\documentclass[nofootinbib,floatfix,a4paper]{revtex4-1}

\usepackage{graphicx}
\usepackage{times}

\usepackage{amssymb}
\usepackage{amsmath}
\usepackage{mathtools}
\usepackage{mathrsfs}
\usepackage{gensymb}
\usepackage[normalem]{ulem}

\usepackage{hyperref}
\hypersetup{
  colorlinks=true,        
  linkcolor=blue,         
  citecolor=cyan,         
}

\newcommand{\lrn}[1]{\textcolor{black}{#1}}

\newcommand{\cf}{cf.,~}
\newcommand{\ie}{i.e.,~}
\newcommand{\eg}{e.g.,~}

\begin{document}

\title{Binary neutron-star mergers: a review of Einstein's richest
  laboratory}

\author{Luca Baiotti} 
\affiliation{Graduate School of Science, Osaka
  University, Toyonaka, 560-0043, Japan}

\author{Luciano Rezzolla}
\affiliation{Institute for Theoretical Physics, Max-von-Laue-Str. 1,
  60438 Frankfurt, Germany}
\affiliation{Frankfurt Institute for Advanced Studies,
  Ruth-Moufang-Str. 1, 60438 Frankfurt, Germany}

\begin{abstract}
The merger of binary neutron-stars systems combines in a single process:
extreme gravity, copious emission of gravitational waves, complex
microphysics, and electromagnetic processes that can lead to
astrophysical signatures observable at the largest redshifts. We review
here the recent progress in understanding what could be considered
Einstein's richest laboratory, highlighting in particular the numerous
significant advances of the last decade. Although special attention is
paid to the status of models, techniques, and results for fully
general-relativistic dynamical simulations, a review is also offered on
initial data and advanced simulations with approximate treatments of
gravity. Finally, we review the considerable amount of work carried out
on the post-merger phase, including: black-hole formation, torus
accretion onto the merged compact object, connection with gamma-ray burst
engines, ejected material, and its nucleosynthesis.
\end{abstract}


 

\maketitle

\tableofcontents 

\newpage
\section{Introduction}
\label{sec:Introduction}

Neutron stars are believed to be born in supernova explosions triggered
by the collapse of the iron core in massive stars. Many astronomical
observations have revealed that binary neutron stars\footnote{\lrn{With
    \textit{``binary neutron-star''} systems we here refer to binary
    systems composed of two neutron stars; in astronomy, such systems are
    often called \textit{``double neutron-star''} systems in order to
    distinguish them from binary systems in which one star is a neutron
    star and the other a white dwarf.}} (BNSs) indeed exist
\cite{Kramer04} and the most important physical properties of all known
such systems are collected in Table \ref{tab:observedNS}. Despite this
observational evidence of existence, the formation mechanisms of BNS
systems are not known in detail. The general picture is that in a binary
system made of two massive main-sequence stars \lrn{of masses between
  approximately $8$ and $25\,M_\odot$}, the more massive one undergoes a
supernova explosion and becomes a neutron star. This is followed by a
very uncertain phase in which the neutron star and the main-sequence star
evolve in a ``common envelope'', that is, with the neutron star orbiting
in the extended outer layers of the secondary star
\cite{Kiziltan:2010a,Ivanova2013,Ozel2016}. At the end of this stage,
also the second main-sequence star undergoes a supernova explosion and,
if the stars are still bound after the explosions, a BNS system is
formed. The common-envelope phase, though brief, is crucial because in
that phase the distance between the stars becomes much smaller as a
result of drag, and this allows the birth of BNS systems that are compact
enough to merge within a Hubble time, following the dissipation of their
angular momentum through the emission of gravitational radiation. \lrn{It
  is also possible that during the common-envelope phase the neutron star
  collapses to a black hole, thus preventing the formation of a BNS.}
Another possible channel for the formation of BNS systems may be the
interaction of two isolated neutron stars in dense stellar regions, such
as globular clusters, in a process called ``dynamical capture''
\cite{Oleary2009, Lee2010, Thompson2011}. Dynamically formed binary
systems are different from the others because they have higher
ellipticities (see Sect. \ref{sec:dynamical-capture}). It is presently
not known what fraction of BNS systems would originate from dynamical
capture, but it is expected that these binaries are only a small part of
the whole population.

\begin{table*}[ht]
  \caption{\label{tab:observedNS} Observational data of neutron stars in
    binary neutron-star systems containing a pulsar. Reported in the
    various columns are: the name of the binary, the total
    (gravitational) mass $M_{\rm tot}$, the (gravitational) masses of the
    pulsar and that of its neutron-star companion $M_{{\rm A}}, M_{_{\rm
        B}}$, the mass ratio $q \leq 1$, the orbital period $T_{\rm
      orb}$, the projected semi-major axis of the orbit $R$ (\ie the
    projection of the semi-major axis onto the line of sight), the
    orbital eccentricity $e_{\rm orb}$, the distance from the Earth, the
    barycentric rotation frequency $f_{\rm s}$, and the inferred surface
    magnetic dipole field $B_{\rm surf}$. The data are taken from the
    respective references and truncated to four significant digits for
    the masses and to two significant digits for the rest. Note that in
    the case of the "double pulsar" system J0737-3039 (the only double
    system where both neutron stars are detectable as pulsars), the
    magnetic field of the second-formed pulsar (not reported in this
    table) is estimated to be 1.59E+12 G.}
\begin{tabular}[t]{lrrrrrrrrrr}
\hline
\hline
      Name  & 
      $M_{\rm tot}$  & 
      $M_{_{\rm A}}$  & 
      $M_{_{\rm B}}$  & 
      $q$  & 
      $T_{\rm orb}$  & 
      $R$  & 
      $e_{\rm orb}$  & 
      $D$  & 
      $f_{\rm s}$  & 
      $B_{\rm surf}$ \\ [0.5ex]
       & 
      $[{M_{\odot}}]$  & 
      $[{M_{\odot}}]$  & 
      $[{M_{\odot}}]$  & 
       & 
      [days]  & 
      [light s]  & 
       & 
      [kpc]  & 
      [Hz]  & 
      [G] \\ [0.5ex] 
\hline
J0453+1559~\cite{Martinez2015}   &  2.734    &  1.559   &  1.174   &  0.75      &  4.1  &  14  &  0.11  &  1.8  &  22  &  9.3E+09          \\ [0.5ex]
J0737-3039~\cite{Kramer2006}~~~~~~~~~~  &  2.587  &  1.338   &  1.249   &  0.93   &  0.10 &  1.4  &  0.088  &  1.1  &  44 & ~~~6.4E+09  \\ [0.5ex]
J1518+4904~\cite{Janssen2008}   &  2.718    &  $\,\,\,\, <$1.766   &  $\,\,\,\, >$0.951   &  $\,\,\,\, >$0.54   &  8.6  &  20  &  0.25  &  0.7  &  24  &  9.6E+08    \\ [0.5ex]
B1534+12~\cite{Fonseca2014}     &  2.678    &  1.333   &  1.345   &  0.99   &  0.42  &  3.7  &  0.27  &  1.0  &  26  &  9.6E+09    \\ [0.5ex]
J1753-2240~\cite{Keith2009}     &  --    &  --   &  --   &  --   &  14  &  18  &  0.30  &  3.5  &  10  &  9.7E+09           \\ [0.5ex]
J1756-2251~\cite{Ferdman2014}   &  2.577    &  1.341  &  1.23   &  0.92    &  0.32  &  2.8  &  0.18  &  0.73  &  35  &  5.4E+09    \\ [0.5ex]
J1807-2500B~\cite{Lynch2012}  &  2.571  &  1.366  &  1.21 &  0.89   & 1.0  &  29  &  0.75  &  --  &  239  &  $\,\,\,\, \leq$9.8E+08    \\ [0.5ex]
J1811-1736~\cite{Corongiu2007}   &  2.571    &  $\,\,\,\, <$1.478   &  $\,\,\,\, >$1.002   &  $\,\,\,\, >$0.68    &  19  &  35 &  0.83  &  5.9  &  9.6  &   9.8E+09    \\ [0.5ex]
J1829+2456~\cite{Champion2004}   &  2.59\phantom{0} &  $\,\,\,\, <$1.298  &  $\,\,\,\, >$1.273  &  $\,\,\,\, >$0.98  &  1.2  &  7.2  &  0.14  &  0.74  &  24  &  1.5E+09    \\ [0.5ex]
J1906+0746~\cite{vanLeeuwen2015}   &  2.613    &  1.291   &  1.322   &  0.98    &  0.17  &  1.4  &  0.085  &  7.4  &  6.9  &  1.7E+12     \\ [0.5ex]
J1913+1102~\cite{Lazarus2016}  &   2.875  &  $\,\,\,\, <$1.84\phantom{0}  & $\,\,\,\, >$1.04\phantom{0}  &  $\,\,\,\, >$0.56   &  0.21 &  1.8  &  0.090 &  13  &  1.1  &  2.1E+09       \\  [0.5ex]
B1913+16~\cite{Weisberg2016}     &  2.828    &  1.449   &  1.389   &  0.96   &  0.32  &  2.3  &  0.62  &  7.1  &  17  &  2.3E+10    \\ [0.5ex]
J1930-1852~\cite{Swiggum2015}   &  2.59\phantom{0} &  $\,\,\,\, <$1.199  &  $\,\,\,\, >$1.363   &  $\,\,\,\, >$0.88  &  45  &  87  &  0.40  &  2.3  &  5.4  &   6.0E+10   \\ [0.5ex]
B2127+11C~\cite{Jacoby2006}    &  2.713    &  1.358  &  1.354   &  1.0\phantom{0}    &  0.34  &  2.5  &  0.68  &  13  &  33  &  1.2E+10    \\ [0.5ex]
\hline
\hline
\end{tabular}
\end{table*}

This is undoubtedly an exciting and dynamical time for research on BNS
mergers, when many accomplishments have been achieved (especially since
2008), while many more need to be achieved in order to describe such
fascinating objects and the related physical phenomena. The first direct
detection through the advanced interferometric LIGO detectors
\cite{Harry2010} of the gravitational-wave signal from what has been
interpreted as the inspiral, merger and ringdown of a binary system of
black holes \cite{Abbott2016a} marks, in many respects, the beginning of
gravitational-wave \lrn{astronomy; a second detection was made a few
  months later \cite{Abbot2016g}}. Additional advanced detectors, such as
Virgo \cite{Accadia2011_etal}, KAGRA \cite{Aso:2013} \lrn{and LIGO India
  (see \eg \cite{Fairhurst2014})}, are going to become operational in the
next few years, and we are likely to witness soon also signals from the
inspiral and post-merger of neutron-star binaries or
neutron-star--black-hole binaries, with a detection rate that has an
uncertainty of three orders of magnitude, but is expected to be of
several events per year \cite{Abadie:2010_etal}.

BNS mergers are rather unique objects in the landscape of relativistic
astrophysics as they are expected to be at the origin of several and
diverse physical processes, namely: {(i)} to be significant sources of
gravitational radiation, not only during the inspiral, but also during
and after the merger; {(ii)} to be possible progenitors for
short-gamma-ray bursts (SGRBs); {(iii)} to be the possible sources of
other electromagnetic and neutrino emission; {(iv)} to be responsible for
the production of a good portion of the very heavy elements in the
Universe. When viewed in this light, BNS mergers naturally appear as
Einstein's richest laboratory, where highly nonlinear gravitational
effects blend with complex microphysical processes and yield astonishing
astrophysical phenomena.

As we will discuss in more detail in the following Section, the typical
scenario leading to SGRBs assumes that a system composed of a rotating
black hole and a surrounding massive torus is formed after the merger
\cite{Narayan92,Eichler89}. A large number of numerical simulations
\cite{Shibata99d, Baiotti08, Anderson2007, Liu:2008xy, Bernuzzi2011} have
confirmed that this scenario can be attained through BNS mergers unless
the progenitor stars have very small masses [smaller than half of the
  maximum allowed mass for neutron stars with a given equation of state
  (EOS)], \lrn{or when the merged object collapses to a black hole as a
  uniformly rotating neutron star in vacuum
  \cite{Margalit2015}}. Furthermore, if sufficiently massive, the torus
could provide the large amount of energy observed in SGRBs, either
through neutrino processes or by extracting the rotational energy of the
black hole via magnetic fields \cite{Paczynski86,
  Eichler89}. Furthermore, if the neutron stars in the binary have
relatively large magnetic fields and extended magnetospheres, the
inspiral could also be accompanied by a precursor electromagnetic signal
\cite{Palenzuela2013a}, while after the merger magnetically confined jet
structures may form once a torus is present around the black hole
\cite{Rezzolla:2011, Paschalidis2014, Dionysopoulou2015, Ruiz2016}.

Possible evidence that a BNS merger can be behind the phenomenology
associated with SGRBs has emerged recently from the infrared excess in
the afterglow curve of Swift's short gamma-ray burst SGRB 130603B
\cite{Berger2013, Tanvir2013}, which has been interpreted as a
``macronova'' emission \cite{Li1998, Kulkarni2005_macronova-term}
(sometimes also referred to as ``kilonova'' \cite{Metzger:2010}), \ie as
due to the radioactive decay of by-products of the $r$-processed matter
from the material ejected in the merger\footnote{We will discuss this
  further in Section \ref{sec:bph}, but we briefly recall here that $r$
  (or rapid) processes are nucleosynthetic processes involving the rapid
  capture of neutrons.}. Other macronova candidates, \lrn{\eg GRB 060614 and
GRB 050709 \cite{Yang2015,Jin2016},} are presently being considered. For
instance, strong evidence for a macronova component has been found
recently in the peculiar long-short event GRB 060614 \cite{Yang2015} and
in its afterglow \cite{Jin2015}, while a careful re-examination of the
afterglow of SGRB 050709, the first short event with an identified
optical afterglow, has highlighted a macronova component \cite{Jin2016}.

The observations of the infrared transient in these afterglows are
important not only because they provide a potential observational link
between two distinct phenomena (\ie a SGRB explosion and a radioactive
decay), but also because they suggest that BNSs can be the site of active
and intense nucleosynthesis. Additional evidence in this direction is
offered by the Solar system abundance of $^{244}$Pu \cite{wallner:15,
  Hotokezaka:2015b} and recent observations of $r$-process enriched stars
in a metal-poor ultra-faint dwarf galaxy \cite{Ji:15}. Both of these
observations suggest that $r$-process elements might be preferentially
produced in rare/high-yield events such as mergers instead of
common/low-yield occurrences such as core-collapse supernovae.

Given the complex nonlinear nature of merging BNSs, it is inevitable that
fully three-dimensional numerical simulations are the only tool available
for studying these processes accurately and with a sufficient degree of
realism. At present, there are about a dozen numerical codes in groups
across the world that are able to produce meaningful results about BNS
mergers. Most of these codes solve the full Einstein equations without
approximations, together with the equations of relativistic hydrodynamics
and/or (resistive) magnetohydrodynamics (MHD) equations. However, there
are also codes that treat matter with smoothed-particle-hydrodynamics
(SPH) methods and with some approximate treatment of gravity, which is
however balanced by more advanced treatments in the microphysical
processes.

Each of these codes represents a complex computational infrastructure
built over the last decade (if not more) and that in most cases already
provides, together with an accurate description of the bulk motion of
matter (before and after the merger), also an approximate representation
of the microphysical aspects related to the EOS, to the neutrino
radiation transport, to the nuclear reactions taking place in the ejected
matter, and, ultimately, to the electromagnetic signal from merging
BNSs. Such computational infrastructures are being continuously updated
and improved, either through the use of more advanced numerical methods,
through the development of novel formulations of the equations, or
through the introduction of new and more refined levels of microphysical
description. Finally, all of these codes also share common scientific
goals: a faithful representation of the gravitational-wave signal
produced before and after the merger, as well as an interpretative and
predictive description of the phenomenology behind SGRBs. This Report is
meant to provide a general but possibly detailed description of the
progress achieved in the numerous areas touched up by investigations of
BNS mergers and hence to provide a snapshot of the status of the field
and of the challenges and goals that lay ahead.
 
\smallskip
The Report is organised as follows: we start in Section \ref{sec:bbp}
with a brief overview of the basic features of the inspiral, merger and
post-merger of binary systems of neutron stars. This is then followed in
Section \ref{sec:ms} by a succinct reminder of the most common
formulation of the set of equations needed to simulate the dynamics of
BNSs, while the problem of computing initial data is reviewed in Section
\ref{sec:ID}. With Section \ref{sec:ph} we will start our review of the
progress in simulations in pure hydrodynamics, leaving treatments that
include magnetic fields and neutrino transport to Section
\ref{sec:bph}. There, special attention is given to the ejecta, which are
thought to produce heavy elements and electromagnetic emission in terms
of a macronova signal. Finally, Section \ref{sec:atas} is dedicated to
the discussion of more advanced techniques and scenarios, which include:
high-order numerical methods, the dynamics of BNSs in alternative
theories of gravity, as well as the dynamics of binary neutron stars in
relativistic collisions. A concluding Section \ref{sec:sao} will
summarise the status of research and its future prospects.

\medskip
We here use a spacelike signature $(-,+,+,+)$ and a system of units in
which $c=G=M_\odot=1$ (unless explicitly shown otherwise for
convenience). Greek indices are taken to run from $0$ to $3$, Latin
indices from $1$ to $3$ and we adopt the standard convention for the
summation over repeated indices. \lrn{Finally, reported below is also a
  quick list of the acronyms adopted in the paper:
\begin{tabbing}
\hglue 0.15truecm \= \hglue 2.1truecm \= hglue 2.0truecm  \kill
\> {ADM}: 	\>  Arnowitt, Deser, Misner\\
\> {AMR}: 	\>  adaptive mesh refinement\\
\> {BNS}: 	\>  binary neutron stars\\
\> {BSSNOK}: 	\>  Baumgarte, Shapiro, Shibata, Nakamura, Oohara, Kojima\\
\> {CCZ4}: 	\>  conformal and covariant Z4\\
\> {EOB}: 	\>  effective one body\\
\> {EOS}: 	\>  equation of state\\
\> {ET}: 	\>  Einstein Telescope\\
\> {HMNS}: 	\>  hypermassive neutron star\\
\> {HRSC}: 	\>  high resolution shock capturing\\
\> {IMHD}: 	\>  ideal magnetohydrodynamics\\
\> {KHI}: 	\>  Kelvin-Helmholtz instability\\
\> {LIGO}: 	\>  Laser Interferometer Gravitational-Wave Observatory\\
\> {MHD}: 	\>  magnetohydrodynamics\\
\> {MRI}: 	\>  magnetorotational instability\\
\> {PSD}: 	\>  power spectral density\\
\> {RMHD}: 	\>  resistive magnetohydrodynamics\\
\> {SGRB}: 	\>  short gamma-ray burst\\
\> {SMNS}: 	\>  supramassive neutron star\\
\> {SNR}: 	\>  signal-to-noise ratio\\
\> {TOV}: 	\>  Tolman, Oppenheimer, Volkoff\\  
\end{tabbing}
}

\newpage
\section{Broadbrush Picture}
\label{sec:bbp}
 
\begin{figure*}
\begin{center}
  \includegraphics[width=1.0\columnwidth]
                  {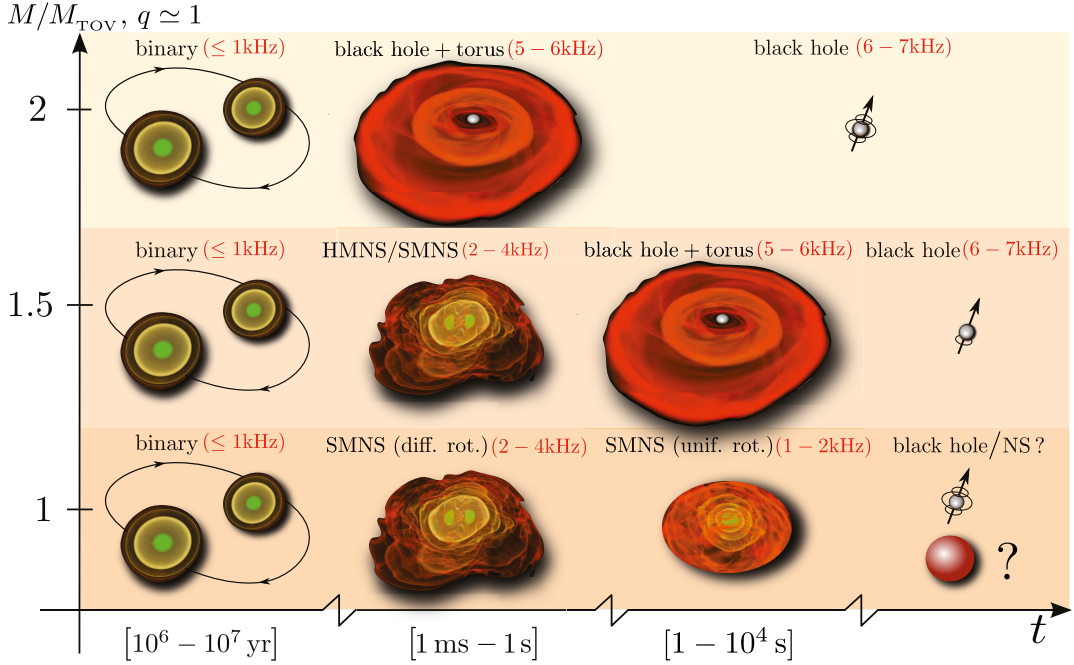}
\end{center}
   \caption{Schematic diagram illustrating the various stages in the
     evolution of an equal-mass binary system of neutron stars as a
     function of the initial mass of the binary. Depending on the initial
     total mass of the binary \lrn{$M$, and on how it relates to the
       maximum mass of a nonrotating neutron star $M_{_{\rm TOV}}$,} the
     binary can either collapse promptly to a black hole surrounded by a
     torus \lrn{(top row)}, or give rise to an HMNS \lrn{(or an SMNS)
       that} ultimately collapses to a black hole and torus \lrn{(middle
       row)}, or \lrn{even} lead to a \lrn{SMNS (first differentially and
       subsequently uniformly rotating) neutron star that} eventually
     yields \lrn{a black hole or a nonrotating neutron star (bottom
       row)}. Also indicated in red are the typical frequencies at which
     gravitational waves are expected to be emitted [Adapted from
       Ref. \cite{Rezzolla_book:2013} by permission of Oxford University
       Press \href{http://www.oup.com}{www.oup.com}].}
   \label{fig:Rezzolla_book:2013}
\end{figure*}

Possibly the best way to summarise the complex sequence of events that is
expected to accompany the evolution of a binary system of neutron stars
is by using a broadbrush picture such as the one illustrated
schematically in Fig. \ref{fig:Rezzolla_book:2013}. More specifically,
the diagram shows on the horizontal axis the progress of time during the
evolution of the system (the intervals in square brackets indicate the
expected duration range of each stage), while on the vertical axis it
displays the ratio of the total \lrn{(gravitational)} mass of the binary
\lrn{(\ie the sum of the gravitational masses of the stars composing the
  system)}, $M$, to the maximum mass of an isolated nonrotating
star,\footnote{An isolated nonrotating neutron star is the solution of
  the Tolman-Oppenheimer-Volkoff (TOV) equation \cite{Tolman39,
    Oppenheimer39b} and so it is often called a ``TOV'' star.} $M_{_{\rm
    TOV}}$. \lrn{Because the EOS describing neutron stars is still
  unknown, the precise value of $M_{_{\rm TOV}}$ cannot be
  determined. However, astronomical observations indicate that it should
  be larger than about two solar masses, since there are two different
  systems that have been measured to have masses in this range: PSR
  J0348+0432 with $M=2.01 \pm 0.04\,M_{\odot}$ \cite{Antoniadis2013}, and
  PSR J1614-2230 with $M=1.97 \pm 0.04\,M_{\odot}$ \cite{Demorest2010}.}

Also indicated \lrn{in the various snapshots of
  Fig. \ref{fig:Rezzolla_book:2013}} are the typical frequencies at which
the corresponding gravitational waves are expected to be emitted. Note
that in all cases, the binary system evolves on the radiation-reaction
timescale, \ie on the timescale set by the loss of energy and angular
momentum via gravitational radiation; this stage lasts for millions of
years at the separations at which BNSs are presently observed (\cf Table
\ref{tab:observedNS}). As can be deduced from Table \ref{tab:observedNS},
the total gravitational masses of the known galactic neutron-star
binaries are in the narrow range $2.57-2.88\,M_{\odot}$; in addition the
masses of the two stars are nearly equal, with differences that are
$10\%$ in general and of $30\%$ at most. Under these conditions, the
stars will inspiral down to very small distances (\ie few tens of
kilometres) without suffering tidal disruptions and hence with a
rest-mass prior to the merger which is essentially the same as the
initial one (the amount of matter lost during the inspiral has been
estimated to be $\ll 10^{-4}\,M_{\odot}$ \cite{Rezzolla:2010}).

As the binary reaches a separation small enough that the changes in the
orbits take place on a timescale of a few seconds only, finite-size
effects such as the tidal deformability of the stars become important and
produce non-negligible changes in the orbits. Numerical-relativity
simulations are the only effective tools to describe the dynamics of the
system in detail from this point onward.

The simplest scenario to illustrate is the one of ``very high-mass''
systems, that is, binaries in which the two component neutron stars have
very large masses, \ie \lrn{$M/M_{_{\rm TOV}} \sim 1.5-2.0$} (top
\lrn{row} of Fig. \ref{fig:Rezzolla_book:2013}). In this case, which is
not expected to be statistically very frequent, the merger will be
accompanied by the ``prompt'' collapse of the binary-merger
product\footnote{We define as `` binary-merger product'' the generic
  object produced after the merger, which can actually change its nature
  over time. This definition is intentionally vague since we want to
  include a multiplicity of possibilities. In fact, depending on the
  total mass and mass ratio of the binary, the EOS, and the time after
  the merger under consideration, the binary-merger product can either be
  a \emph{stable} object, \ie a black hole or a neutron star, or an
  \emph{metastable} one, \ie an object that will eventually reach one of
  the two stable states mentioned above on timescales that can be much
  larger than the dynamical timescale. \label{footnote_bmp}} to a
rotating black hole of dimensionless spin $J/M^2 \simeq 0.7-0.8$,
surrounded by a hot accretion torus with mass $M_{\rm torus} \sim 0.01 -
0.1\, M_{\odot}$, depending on the mass ratio and EOS. The torus will
ultimately accrete onto the black hole on a timescale set by the most
efficient process removing angular momentum, \ie gravitational radiation,
magnetic fields or viscous processes, ultimately leading to an isolated
rotating black hole in vacuum. For any of the mentioned processes, the
timescale can be roughly estimated to be of the order of $1-10\,{\rm s}$.

A second scenario to be considered is the one in which the two component
neutron stars have masses that are not very large, but above the maximum
mass of nonrotating stars, \ie \lrn{$M/M_{_{\rm TOV}} \sim 1.3-1.5$}
(middle \lrn{row} of Fig. \ref{fig:Rezzolla_book:2013}). In this case,
which is expected to be statistically rather frequent \lrn{if $M_{_{\rm
      TOV}} \sim 2.0-2.1\,M_{\odot}$}, the binary-merger product is
expected to be initially a \emph{hypermassive neutron star (HMNS)}, \ie a
neutron star with \lrn{a} mass above the limit for uniformly rotating
neutron stars $M_{\rm max}$\footnote{We recall that a recent
  investigation exploiting universal relations has shown that for any EOS
  it is possible to relate $M_{_{\rm TOV}}$ to the maximum mass that can
  be supported by uniform rotation $M_{\rm max}$ simply as $M_{\rm max}
  \simeq \left(1.203 \pm 0.022\right) M_{_{\rm TOV}}$
  \cite{Breu2016}.}. Because of its large angular momentum and shear, the
HMNS is dynamically unstable to nonlinear instabilities leading to a
barmode deformation \cite{Shibata:2000jt, Baiotti06b, Franci2013,
  Siegel2013, Kiuchi2014}, and it could even be subject to an $m=1$ shear
instability\footnote{As widely known, stellar deformations can be
  described decomposing the linear perturbations of the energy or
  rest-mass density as a sum of quasi-normal modes that are characterized
  by the indices ($\ell,m$) of the spherical harmonic functions. Then the
  $m$ mentioned in the text is the dominant term of such expansion. $m=0$
  is a spherical spherical perturbation, $m=1$ is a one-lobed
  perturbation, $m=2$ is a bar-shaped perturbation. See
  Ref. \cite{Stergioulas03} for a review.}  \cite{Ou06, Corvino:2010,
  Anderson2008, Paschalidis2015, East2016, Radice2016, Lehner2016a}, if a
corotation frequency develops within the HMNS.

Indeed, the collapse of the HMNS to a rotating black hole is temporarily
prevented by its differential rotation, but a number of dissipative
effects, such as magnetic fields, viscosity, or gravitational-wave
emission, will act so as to remove the differential rotation. This will
bring the HMNS towards a configuration that is either still
differentially rotating but unstable to gravitational collapse, or to a
configuration that is uniformly rotating but spinning down because of
angular-momentum loss via, say, electromagnetic emission or neutrino
losses. In the first case, the HMNS will collapse on a dynamical
timescale producing a black-hole--torus configuration as the one
discussed above for the ``prompt'' collapse. In the second case, instead,
the HMNS, by loosing its differential rotation\footnote{This process will
  reduce the gravitational mass of the HMNS, even though its rest mass
  remains essentially constant, apart from the small losses due to the
  emission of winds \lrn{and that should not remove more than a few
    percent in rest mass}.} will evolve into a so-called
\textit{supramassive neutron star} (SMNS), \ie an axisymmetric rotating
\lrn{(either differentially or uniformly)} neutron star with mass
exceeding the limit for nonrotating neutron stars, \ie with mass
$M_{_{\rm TOV}} \leq M \leq M_{\rm max}$. Eventually, when slowed down
sufficiently, the SMNS will reach the stability line to gravitational
collapse, producing, again, a rotating black hole and an accretion torus.
\lrn{Because an SMNS can also be differentially rotating, depending on the
  actual value of $M_{\rm max}$ it is possible that the merger will never
  lead to an HMNS but directly to an SMNS, which then follows the evolution
  described above (see also the bottom row of
  Fig. \ref{fig:Rezzolla_book:2013})}. We should also mention that it has
been pointed out recently by Margalit et al. \cite{Margalit2015} that the
collapse of an SMNS is unlikely to yield a torus as the specific angular
momentum of the SMNS matter is below the one corresponding to stable
circular orbits for the newly produced black hole. While this is correct,
present simulations reveal that the SMNS is also surrounded by a certain
amount of matter with essentially Keplerian angular velocities and which
effectively behaves like a ``disk'' around the central core
\cite{Kastaun2014}. It is therefore reasonable, although it has not been
proven, that this material will remain on stable orbits when the SMNS
collapses, hence leading again to an accretion torus around a rotating
black hole \cite{Rezzolla2014b}.

The final scenario that can take place arises for ``very low-mass''
systems, that is, binaries in which the two component neutron stars have
rather small masses, \ie \lrn{$M/M_{_{\rm TOV}} \sim 0.9-1.2$} (bottom
\lrn{row} of Fig. \ref{fig:Rezzolla_book:2013}). In this case, which is
also expected to be statistically rather rare, the binary-merger product
is unlikely to be a black hole from the beginning. It will instead be a
differentially \lrn{SMNS}, which will lose its angular momentum \lrn{and
  differential rotation} to produce a uniformly rotating star, either
supramassive or not. In the first case, the SMNS will follow the
evolutionary track described above, \ie eventually collapsing to produce
a black-hole--torus system. In the second case, however, the
binary-merger product will finally evolve into a stable, nonrotating
neutron star.

Although the broadbrush picture described above is now well established
and supported by a number of numerical simulations carried out by several
groups, the details of the picture are still far from being clear. In
particular, the ``delay'' between the merger and the collapse to black
hole of the HMNS\lrn{/SMNS} (which we refer to as the ``lifetime'' of the
binary-merger product), depends nonlinearly on a number of factors (\ie
EOS, mass ratio, strength of the magnetic field, efficiency of radiative
losses) and is in general rather difficult to estimate (see
Sects. \ref{sec:hydro_merger_post-merger}, \ref{sec:HD_MHD}, and
\ref{sec:hd_nus} for a discussion). Ravi and Lasky \cite{Ravi2014},
combining in a simplified model a number of channels in which either the
mass or the angular momentum can be lost by the binary-merger product,
have roughly estimated the upper limit of the lifetime of the
binary-merger product to be $\sim 10^4\,{\rm s}$. Clearly, ranging
between a few milliseconds and a few hours, the lifetime of the
binary-merger product represents one of the largest uncertainties in the
post-merger dynamics of BNSs and is likely to remain as such in view of
the difficulties of performing accurate numerical simulations over such
long timescales.

What is clear, however, is that the highly nonlinear regimes encountered
during the merger and after the merger leave numerical solutions as the
only option to investigate these scenarios with sufficient precision. As
a result, several groups worldwide have developed numerical codes able to
solve the equations of relativistic hydrodynamics and
magnetohydrodynamics, together with the Einstein equations, to model the
coalescence and merger of neutron-star binaries. We will describe the
most commonly used numerical methods in the following Section.

\newpage
\section{Mathematical Setup}
\label{sec:ms}

The science of solving equations numerically often has to start from the
formulation of the equations themselves, since different systems of
equations that are mathematically equivalent normally possess different
stability and accuracy properties. In addition, also the numerical
techniques employed to solve the equations obviously have a fundamental
importance in the accuracy and physical consistency of the
solutions. Therefore, we start with a brief review of the mathematical
formulation of the equations. For additional details we refer to a series
of textbooks, where these issues are presented in much greater care and
length \cite{Alcubierre:2008, Bona2009, Baumgarte2010, Gourgoulhon2012,
  Rezzolla_book:2013, Shibata_book:2016}.

\subsection{Spacetime evolution: BSSNOK formulation}
\label{sec:BSSNOK}

One common way to numerically solve the Einstein equations consists in
evolving a conformal-traceless ``$3+1$'' formulation\footnote{The
  ``$3+1$'' formulation, which assumes that spacetime is foliated into a
  family of three-dimensional spacelike hypersurfaces, labelled by their
  time coordinate and with a set of coordinates for each slice, was
  proposed by Arnowitt, Deser, and Misner \cite{Arnowitt59,Arnowitt62}
  and is known as the ADM formulation.} of the Einstein equations called
BSSNOK \cite{Nakamura87, Shibata95, Baumgarte99, Alcubierre99d}, in which
the spacetime is decomposed into three-dimensional spacelike slices,
described by a metric $\gamma_{ij}$, its embedding in the full spacetime,
specified by the extrinsic curvature $K_{ij}$, and the gauge functions
$\alpha$ (lapse) and $\beta^i$ (shift) that specify a coordinate
frame. The three-metric $\gamma_{ij}$ is conformally transformed via
\begin{equation}
  \label{eq:def_g}
  \tilde{\gamma}_{ij} = \phi^2 \gamma_{ij}\,,
\end{equation}
and the conformal factor\footnote{\lrn{Note that the form of the
    conformal factor is arbitrary; for example, other authors, like in
    Refs. \cite{Alcubierre:2008, Baumgarte2010}, set the conformal
    transformation to be $\tilde{\gamma}_{ij} = e^{-4\phi}
    \gamma_{ij}$.}}  $\phi$ is evolved as an independent variable,
whereas $\tilde{\gamma}_{ij}$ is subject to the constraint $\det
(\tilde{\gamma}_{ij}) = 1$ in Cartesian coordinates. The extrinsic
curvature also undergoes the same conformal transformation and its trace
$K$ is evolved as an independent variable. That is, in place of $K_{ij}$,
the following quantities are evolved
\begin{equation}
  \label{eq:def_K}
  K \coloneqq \mathrm{tr} K_{ij} = g^{ij} K_{ij}\,, \qquad
  \tilde{A}_{ij} =
  \phi^2 (K_{ij} - \frac{1}{3}\gamma_{ij} K)\,,
\end{equation}
with $\tilde{A}_{ij}$ being the traceless conformal extrinsic curvature,
\ie $\mathrm{tr} \tilde{A}_{ij}=0$. Finally, new evolution variables are
introduced, defined in terms of the Christoffel symbols
$\tilde{\Gamma}^i_{jk}$ of the conformal three-metric
\begin{equation}
  \label{eq:def_Gamma}
  \tilde{\Gamma}^i \coloneqq \tilde{\gamma}^{jk}\tilde{\Gamma}^i_{jk}\,.
\end{equation}

The Einstein equations are then written as a set of evolution equations
for the listed variables and are given by
\begin{align}
\label{eq:bssn_1}
& \partial_t \tilde{\gamma}_{ij}  = -2 \alpha \tilde{A}_{ij} 
                  + 2 \tilde{\gamma}_{k(i}\partial_{j)}\beta^k
                  - \frac{2}{3}\tilde{\gamma}_{ij} \partial_k\beta^k
                  + \beta^k\partial_k\tilde{\gamma}_{ij} \,,
		  \\ 
\label{eq:bssn_2}
& \partial_t \tilde{A}_{ij}  = \phi^{2} \Big [-D_iD_j\alpha 
                  + \alpha \Big({}^{(3)\!} R_{ij} - 8 \pi S_{ij}\Big)\Big ]^{\!\rm TF\,} 
                  + \beta^k\partial_k\tilde{A}_{ij}
                  + 2\tilde{A}_{k(i}\partial_{j)}\beta^k
                  \nonumber \\
                  & \hskip 4.0cm
                  + \alpha (\tilde{A}_{ij} K  
                  - 2 \tilde{A}_{ik} \tilde{A}^{k}_{\ j})
                  - \frac{2}{3}\tilde{A}_{ij}\partial_k\beta^k\,,
                  \\ 
\label{eq:bssn_3}
& \partial_t \phi    = \frac{1}{3} \phi\, \alpha K 
                  - \frac{1}{3} \phi\,\partial_i\beta^i
                  + \beta^k \partial_k\phi \,,
                  \\ 
\label{eq:bssn_4}
& \partial_t K       = -D_iD^i \alpha + 
		  \alpha \Big [\tilde A_{ij} \tilde A^{ij} 
                  + \frac{1}{3} K^2 + 4\pi (E+S)\Big ] 
                  + \beta^i\partial_iK\,,
                  \\ 
\label{eq:bssn_5}
& \partial_t \tilde{\Gamma}^i    = \tilde{\gamma}^{jk}\partial_j\partial_k\beta^i 
                  + \frac{1}{3} \tilde{\gamma}^{ik}\partial_k\partial_j\beta^j
                  + \frac{2}{3}\tilde{\Gamma}^i \partial_j\beta^j
                  - \tilde{\Gamma}^j\partial_j\beta^i
                  - 2\tilde{A}^{ij} \partial_j\alpha + 
                  \beta^j \partial_j \tilde{\Gamma}^i
                  \nonumber \\
                  & \hskip 2.0cm
                  + 2\alpha \Big(\tilde{\Gamma}^i_{jk} \tilde{A}^{jk} 
                  - 3 \tilde{A}^{ij} \partial_j \ln\phi
                  - \frac{2}{3} \tilde{\gamma}^{ij}\partial_j K\Big)
                  - 16 \pi \alpha \tilde{\gamma}^{ij} S_j\,,
\end{align}
where $^{(3)\!}R$ is the Ricci scalar on a three-dimensional timeslice,
$D_i$ is the covariant derivative with respect to the physical metric
${\gamma}_{ij}$, the index ``TF'' indicates that the trace-free part of
the bracketed term is used, and $E$, $S_j$, and $S_{ij}$ are the matter
source terms defined as
\begin{align}
\label{3+1 matter-terms E}
E&\coloneqq n_\alpha n_\beta T^{\alpha\beta}\,, \\ 
\label{3+1 matter-terms S_i}
S_i&\coloneqq -\gamma_{i\alpha}n_{\beta}T^{\alpha\beta}\,, \\
\label{3+1 matter-terms S_ij}
S_{ij}&\coloneqq \gamma_{i\alpha}\gamma_{j\beta}T^{\alpha\beta}\,, 
\end{align}
where $n_\alpha$ is the future-pointing four-vector orthonormal to the
spacelike hypersurface and $T^{\alpha\beta}$ is the energy-momentum
tensor. The Einstein equations also lead to a set of time-independent
constraint equations that are satisfied within each spacelike slice
\begin{align}
\label{eq:einstein_ham_constraint}
\mathcal{H} &\coloneqq  ^{(3)\!}R + K^2 - K_{ij} K^{ij} - 16\pi E = 0\,, \\
\label{eq:einstein_mom_constraints}
\mathcal{M}^i &\coloneqq D_j(K^{ij} - \gamma^{ij}K) - 8\pi S^i = 0\,,
\end{align}
which are usually referred to as the Hamiltonian ($\mathcal{H}$) and
momentum ($\mathcal{M}^i$) constraints, respectively.

The most commonly used gauges in the BSSNOK formulation (but valid also
for the CCZ4 and Z4c formulation presented in the following Section) are
the hyperbolic singularity-avoiding ``1+log''slicing conditions of the
form
\begin{equation}
(\partial_t - \beta^i\partial_i) \alpha = - f(\alpha) \;
\alpha^2 (K-K_0)\,,
\label{eq:BMslicing}
\end{equation}
with $f(\alpha)>0$ and $K_0 \coloneqq K(t=0)$ \cite{Alcubierre99d,
  Alcubierre01a}, and the ``Gamma-driver'' shift conditions proposed in
\cite{Alcubierre01a,Alcubierre02a}, which essentially act so as to drive
the $\tilde{\Gamma}^{i}$ to be constant
\begin{equation}
\partial^2_t \beta^i = F \, \partial_t \tilde\Gamma^i - \eta \,
\partial_t \beta^i\,,
\label{eq:hyperbolicGammadriver}
\end{equation}
where $F$ and $\eta$ are, in general, positive functions of space and
time. Most often they are implemented in their first-order form:
\begin{align}
\label{shift_evol}
  \partial_t \beta^i - \beta^j \partial_j  \beta^i & =  
\frac{3}{4} \alpha B^i\,,  \\
  \partial_t B^i - \beta^j \partial_j B^i & =  \partial_t \tilde\Gamma^i 
    - \beta^j \partial_j \tilde\Gamma^i - \eta B^i\,,
\end{align}
where $B^i$ is simply an auxiliary variable. The parameter $\eta$ acts as
a damping coefficient and is crucial to avoid strong oscillations in the
shift \cite{Alcubierre01a, Alcubierre02a, Mueller09, Alic:2010}. Overall,
both the 1+log slicing condition and the Gamma-driver shift conditions
satisfy the following basic requirements: (i) if singularities are
present in the spacetime under considerations, these are avoided; (ii) if
coordinate distortions take place on the spatial grid as a result of the
development of large spatial curvatures, these are counteracted.

\subsection{Spacetime evolution: CCZ4 and Z4c formulations}
\label{sec:CCZ4}

Another formulation that has been developed recently and is becoming
increasingly popular is the so-called \emph{CCZ4 formulation}
\cite{Alic2013}. It combines the advantages of a conformal decomposition,
such as the one used in the BSSNOK formulation (\ie well-tested
hyperbolic gauges, no need for excision, robustness to imperfect boundary
conditions), with the advantages of a constraint-damped formulation, such
as the generalized harmonic one (\ie exponential decay of constraint
violations when these are produced). Another conformal formulation of the
Z4 system, the so-called \emph{Z4c formulation}, has also been proposed
recently by \cite{Bernuzzi:2009ex}. This does not include all the
non-principal terms coming from the covariant form of the Z4 equations,
but aims at a system which is as close as possible to BSSNOK. The
resulting set of equations has been applied with success to both
spherically symmetric non-vacuum spacetimes, where it has shown its
ability to damp and propagate away the violations of the constraints
\cite{Bernuzzi:2009ex, Weyhausen:2011cg}, and to generic spacetimes
\cite{Cao:2012,Hilditch2012}.

In essence, the CCZ4 formulation is the conformal and covariant
representation of the original Z4 formulation of the Einstein equations
\cite{Bona:2003fj}, where the original elliptic constraints are converted
into algebraic conditions for a new four-vector $Z_{\mu}$. This
formulation can be derived from the covariant Lagrangian
\begin{equation}\label{action}
{\cal L} = g^{\mu\nu}~[R_{\mu\nu} + 2 \nabla_{\mu} Z_{\nu}]\,,
\end{equation}
by means of a Palatini-type variational principle \cite{Bona:2010is}.
The vector $Z_{\mu}$ measures the deviation from the Einstein field
equations. The algebraic constraints $Z_{\mu}=0$ amount therefore to the
fulfilling of the standard Hamiltonian and momentum constraints. In order
to control these constraints, the original system is supplemented with
damping terms such that the true Einstein solutions (\ie the ones
satisfying the constraints) become an attractor of the enlarged set of
solutions of the Z4 system \cite{Gundlach2005:constraint-damping}. The Z4
damped formulation can be written in covariant form as
\begin{equation}
\label{Z4_covariant}
     R_{\mu\nu} + \nabla_{\mu} Z_{\nu} + \nabla_{\nu} Z_{\mu}  +
     \kappa_1 [n_{\mu} Z_{\nu} + n_{\nu} Z_{\mu} - (1 + \kappa_2)
     g_{\mu\nu} n_{\sigma} Z^{\sigma}] = 
8 \pi (T_{\mu\nu} - \tfrac{1}{2} g_{\mu\nu} T)\,,
\end{equation}
where $n_{\mu}$ is the unit normal to the time slicing, $T_{\mu\nu}$ the
energy-momentum tensor and $T$ its trace.

In the CCZ4 formulation, the energy-momentum constraints become evolution
equations for $Z_{\mu}$, modifying the principal part of the ADM system
and converting it from weakly to strongly hyperbolic
\cite{Bona:2003qn}. The ``$3+1$'' decomposition of the Z4 formulation
including the damping terms, \ie the CCZ4 formulation, reads
\begin{eqnarray}
\label{Z4-dtgamma}
  &&(\partial_t - {\cal L}_{\vec{\beta}}) \gamma_{ij}
  = - {2\alpha}K_{ij} \,,\\
\label{Z4-dtK}
   &&(\partial_t - {\cal L}_{\vec{\beta}})K_{ij} = -\nabla_i\alpha_j
    + \alpha  \left[R_{ij} + 2 \nabla_{(i} Z_{j)}+
    - 2{K_i}^{l} K_{lj}+(K-2\Theta)K_{ij}
    -\kappa_1(1+\kappa_2)\Theta\gamma_{ij}\right] \nonumber \\
    && \hskip 2.5cm -  8\pi\alpha \left[S_{ij}-\frac{1}{2}(S -
      E)\gamma_{ij}\right] \,, \\
\label{Z4-dtTheta} 
&&(\partial_t - {\cal L}_{\vec{\beta}}) \Theta =
\frac{\alpha}{2}\; \left[ R + 2 \nabla_j Z^j + (K -
2\Theta) K
 - K^{ij} K_{ij} - 2 \frac{Z^j {\alpha}_j}{\alpha}
 -2 \kappa_1 (2+\kappa_2)\Theta- 16\pi E\right] ,\nonumber \\
\\
\label{Z4-dtZ}
 &&(\partial_t - {\cal L}_{\vec{\beta}})Z_i =
 \alpha \left[\nabla_j({K_i}^j
  -{\delta_i}^j K) + \partial_i \Theta
  - 2 {K_i}^j Z_j  -  \Theta \frac{{\alpha}_i}{\alpha}
  -\kappa_1 Z_i- 8\pi S_i\right]\,,
\end{eqnarray}
where ${\cal L}_{\vec{\beta}}$ is the Lie derivative along the shift
vector $\vec{\beta}$, $\Theta$ is the projection of the Z4 four-vector
along the normal direction, $\Theta \coloneqq n_{\mu} Z^{\mu} = \alpha
Z^0$, and the matter-related quantities $E$, $S_i$ and $S_{ij}$ are
defined in equations \eqref{3+1 matter-terms E}--\eqref{3+1 matter-terms
  S_ij}.

Equations \eqref{Z4-dtgamma}--\eqref{Z4-dtZ} must be complemented with
suitable gauge conditions that determine the system of coordinates used
during the evolution. Of all the possible options, the most interesting
ones are those which preserve the hyperbolicity of the full evolution
system, such as the $1+\log$ family and the Gamma-driver shift condition,
which was introduced above \cite{Bona:2004yp, Alic:2009}.

\subsection{Spacetime evolution: generalized harmonic formulation}
\label{sec:Harmonic}

Although not widely used, another method of writing the field equations
that has proven very useful and that has lead to the first evolutions of
binary black-hole systems \cite{Pretorius:2004jg}, is the so-called
\emph{generalized harmonic formulation} \cite{Garfinkle02,
  Lindblom:2005gh, Szilagyi:2006qy, Sorkin2010, Brown2011, East2012b}.
As the name suggests, the generalised harmonic formulation of the
Einstein equations is derived by imposing the {\em harmonic coordinate
  condition}, where the four spacetime coordinates $x^\mu$ are chosen to
individually satisfy wave equations: $\Box x^\mu=0$
\cite{Garfinkle02,Pretorius:2004jg}, where $\Box$ is the d'Alambertian
operator. When imposing this condition, the Einstein equations take on a
mathematically appealing form, where the principal part of the evolution
equations for the four-metric is simply given by a wave
equation. Furthermore, to avoid the inconvenient consequences of a strict
harmonic set of coordinates (\eg coordinate focussing and caustics), it
is useful to introduce a set of four source functions $H^\mu$
\begin{equation}
H^\mu \coloneqq \Box x^\mu =
\frac{1}{\sqrt{-g}}\partial_\alpha\left(\sqrt{-g} 
g^{\alpha\beta} \partial_{\beta} x^\mu \right)
= \frac{1}{\sqrt{-g}}\partial_\alpha\left(\sqrt{-g} g^{\alpha\mu} \right)\,,
\end{equation}
or, equivalently, $H_\mu \coloneqq g_{\mu\nu} H^\nu$. In this way one
obtains that
\begin{equation}
\label{hdef}
H_\mu = \partial_\mu\left(\ln\sqrt{-g}\right) -
g^{\alpha\nu}\partial_\alpha g_{\nu\mu}\,.
\end{equation}
and that the symmetric gradient of $H_\mu$ is given by
\begin{align}
\label{dhdef}
\partial_{(\nu}H_{\mu)} = \partial_\mu \partial_\nu\left(\ln\sqrt{-g}\right) -
\frac{1}{2}\left(\partial_\nu g^{\alpha\beta}\partial_\alpha
g_{\mu\beta}+\partial_\mu g^{\alpha\beta}\partial_\alpha
g_{\nu\beta}\right) -
\frac{1}{2}g^{\alpha\beta}\left(\partial_\nu\partial_\alpha g_{\beta\mu}+\partial_\mu\partial_\alpha g_{\beta\nu}\right)\,.
\end{align}
The generalized harmonic decomposition involves replacing particular
combinations of first and second derivatives of the metric in the Ricci
tensor by the equivalent quantities in equations
\eqref{hdef}--\eqref{dhdef} and then promoting the source functions
$H_{\mu}$ to the status of independent quantities. More specifically, it
is possible to rewrite the field equations as
\begin{eqnarray}
g^{\delta\gamma}\partial_\gamma\partial_\delta g_{\alpha\beta} 
+ \partial_\beta g^{\gamma\delta}\partial_\gamma g_{\alpha\delta}
+ \partial_\alpha g^{\gamma\delta}\partial_\gamma g_{\beta\delta}  
+ 2 \partial_{(\beta}H_{\alpha)} 
- 2 H_\delta \Gamma^\delta_{\alpha\beta} 
+2 \Gamma^\gamma_{\delta\beta}\Gamma^\delta_{\gamma\alpha} 
&=& \nonumber\\
&\phantom{=}& \hskip -2.5cm
- 8\pi\left(2 T_{\alpha\beta}-g_{\alpha\beta} \label{efe_h}T\right)\,.
\nonumber \\
\end{eqnarray}
As the $H_{\mu}$ are now four independent functions, it is necessary to
specify four additional, independent differential equations for them. One
can think of the functions $H_\mu$ as representing the four coordinate
degrees of freedom available in general relativity. There are many
conceivable ways of choosing $H_\mu$ (see, \eg
Refs. \cite{Pretorius:2004jg, Haas2016}, for some commonly used options)
and ADM-style gauge conditions can also be used within the harmonic
decomposition.

\subsection{Spacetime evolution: conformally flat approximation}
\label{sec:eq-cf}

Some codes use an approximated formulation of the Einstein equations that
assumes that the three-metric is conformal and flat:
\begin{equation}
\gamma_{ij} = \phi^4  \hat \gamma_{ij}\,
\label{conftrans}
\end{equation}
and
\begin{equation}
\hat \gamma_{ij} = \delta_{ij}\,,
\label{flat}
\end{equation}
where $\phi$ is the conformal factor and $\delta_{jk}$ is the Kronecker
delta. This approximation is also referred to as the \emph{conformally
  flat condition} (or Isenberg-Wilson-Mathews)
\cite{Isenberg08,Wilson89}.

Apart from reducing the complexity of the hydrodynamics and metric
equations, this approach also exhibits numerical stability for long
evolution times, as it solves all constraint equations and thus cannot
violate them by definition. In fact, the constraint equations reduce to
effective flat-space elliptic equations, which are then solved with
standard techniques. For example, the Hamiltonian constraint is combined
with the maximal slicing condition ${\rm tr}\ K_{ij}=0$ and becomes
\begin{equation}
\nabla^2 \phi = - {\phi^5 \over 8} \biggl[ 16 \pi E 
+ K_{ij} K^{ij} \biggr]\,,
\end{equation}
where $E$ is defined in equation \eqref{3+1 matter-terms E}. At each time
slice, a static solution to the exact general-relativistic field
equations is obtained and thus devoid of gravitational radiation. The
modifications to the orbital dynamics from one time slice to the next are
then obtained through some approximate prescription, \eg post-Newtonian
dynamics.

\subsection{Matter evolution: relativistic hydrodynamics}
\label{sec:me_rhd}

The general-relativistic hydrodynamics equations, as given by the
conservation equations for the energy-momentum tensor $T^{\mu\nu}$ and
for the matter current density $J^\mu$, are normally solved numerically
after being recast into a \textit{flux-conservative formulation}
\cite{Marti91, Banyuls97, Ibanez01, Rezzolla_book:2013,
  Shibata_book:2016}. Indeed, a conservative formulation is a necessary
condition to guarantee correct evolution in regions of sharp entropy
generation (\ie shocks). In essence, the conservation equations
\begin{equation}
\label{hydro eqs}
\nabla_\mu J^\mu = 0\,, \,\,\,\,\,\,\,\,\,\,\,\,\,\,\,\,\,\,\,\,
\nabla_\mu T^{\mu\nu} = 0\,,
\end{equation}
are written in a hyperbolic, first-order, flux-conservative form of the
type
\begin{equation}
\label{eq:consform1}
\partial_t \boldsymbol{U} + 
        \partial_i \boldsymbol{F}^{(i)} (\boldsymbol{U}) = 
        \boldsymbol{S} (\boldsymbol{U})\,,
\end{equation}
where $\boldsymbol{F}^{(i)} (\boldsymbol{U})$ and
$\boldsymbol{S}(\boldsymbol{U})$ are the flux vectors and source terms,
respectively \cite{Font03}. Note that the right-hand side (the source
terms) does not depend on derivatives of the energy-momentum tensor.

The fluxes $\boldsymbol{F}^{(i)}$ and the relations between the
\emph{conserved} variables $\boldsymbol{U}$ and the \emph{primitive} (or
physical) variables (\ie the rest-mass density $\rho$, the Lorentz factor
$W$, the specific enthalpy $h$, the three-velocity $v_j$)
are\footnote{Because much of the development of this formulation has
  taken place at the University of Valencia, through the work of
  Ib\'a\~nez and of his collaborators \cite{Marti91, Banyuls97,
    Ibanez01}, this formulation is also known as the \emph{``Valencia
    formulation''} of the relativistic-hydrodynamics equations.}
\cite{Marti91,Banyuls97,Ibanez01}
\begin{align}
\label{eq:consvar_Valencia}
& \boldsymbol{U} =
\left(\begin{array}{c}
D \\ \\
S_j \\ \\
\tau
\end{array}\right)
\coloneqq 
\left(\begin{array}{c}
\rho W \\ \\
\rho h W^2 v_j \\ \\
\rho h W^2 - p - D
\end{array}
\right)\,, 
&&\boldsymbol{F}^i
= \left(\begin{array}{c}
\alpha v^i D-\beta^i D \\  \\
\alpha S^i_{\ j}-\beta^i S_j \\  \\
\alpha (S^i-Dv^i)-\beta^i \tau
\end{array}\right) \, .
\end{align}
The source vector has instead components
\begin{align}
\boldsymbol{S} \coloneqq 
\sqrt{\gamma}\left(
\begin{array}{c}
\label{source_terms}
0 \\   \\
\frac{1}{2}\alpha S^{ik}\partial_j\gamma_{ik}+
S_i\partial_j\beta^i-E\partial_j\alpha \\  \\
\alpha S^{ij}K_{ij} -S^j\partial_j\alpha
\end{array}\right) \,,
\end{align}
where $\gamma$ is the determinant of $\gamma_{ij}$. Note that in order to
close the system of equations for the hydrodynamics an EOS must be
specified to relate the pressure to the rest-mass density, the energy
density and other properties of the fluid (\eg the composition or the
electron fraction).

The general-relativistic hydrodynamics equations are usually solved
making use of high-resolution shock-capturing (HRSC) methods (see Chapter
8 of Ref. \cite{Rezzolla_book:2013} or Chapter 4 of
Ref. \cite{Shibata_book:2016} for a brief overview). One of the most
delicate procedures in solving the system \eqref{eq:consform1} is the
conversion of the conserved variables back to the primitive variables,
because there is no analytical expression for it and regions of very low
rest-mass density may incur into numerical failures. The treatment of
vacuum or low-density regions is indeed one of the most delicate aspects
of modelling numerically BNSs since regions of zero rest-mass density are
not allowed within the HRSC methods normally employed to solve the system
of equations (\ref{eq:consform1}). As a result, an artificial
\emph{atmosphere} is introduced in regions supposed to be vacuum. This
creates inaccuracies, especially where the physical density is low, as in
the material ejected from the BNS system; we will see in
Sect. \ref{sec:hono} how this issue has recently been addressed in an
alternative way \cite{Radice2013c}.

\subsection{Matter evolution: relativistic magnetohydrodynamics}
\label{sec:me_rmhd}

When the presence of electromagnetic fields cannot be ignored, the
conservation equations \eqref{hydro eqs} need to be coupled to the
solution of the Maxwell equations. When simulating BNS mergers, the
latter are normally solved in the so-called ideal magnetohydrodynamics
(MHD) approximation, which assumes that the fluid has zero resistivity
and is therefore a perfect conductor. Since this assumption drastically
simplifies the equations, ideal MHD (IMHD) was adopted in the first
numerical-relativity simulations including magnetic fields and it is
still widely used today, being considered a very good approximation at
least before the actual merger. At the same time, there are several
processes involving compact objects where resistive effects could play an
important role and preliminary attempts to go beyond IMHD and towards
resistive MHD (in particular for studying the region outside the compact
objects) have been first developed in Ref. \cite{Palenzuela:2008sf} and
then applied in Refs. \cite{Dionysopoulou:2012pp, Palenzuela2013,
  Dionysopoulou2015} (see the end of this Section for some more
details). In what follows we summarise very briefly the basic equations
employed in general-relativistic MHD simulations, focusing on the form
they take in the IMHD approximation.

The two pairs of Maxwell equations can be written as
\cite{Anton06,Giacomazzo:2007ti}
\begin{eqnarray}
\nabla_\nu\, ^{*\!\!}F^{\mu \nu} &=& 0 \,,\\ \nabla_\nu F^{\mu \nu} &=& 4
\pi \mathcal{J}^\mu \,,
\end{eqnarray}
where $F^{\mu \nu}$ is the Faraday (or electromagnetic) tensor,
${\boldsymbol {\mathcal J}}$ is the charge current four-vector and
$\,^{*\!\!}{\boldsymbol F}$ is the dual of the electromagnetic tensor
defined as
\begin{equation}
\,^{*\!\!}F^{\mu \nu} \coloneqq 
\frac{1}{2} \eta^{\mu \nu \lambda \delta} F_{\lambda \delta} \,,
\end{equation}
$\eta^{\mu \nu \lambda \delta}$ being the Levi-Civita pseudo-tensor.
The charge current four-vector ${\boldsymbol {\mathcal J}}$ can be in
general expressed as
\begin{equation}
\mathcal{J}^\mu = q u^\mu + \sigma F^{\mu \nu} u_\nu \,,
\end{equation}
where $u^\mu$ is the fluid four-velocity, $q$ is the proper charge
density, and $\sigma$ is the electric conductivity. The IMHD limit
($\sigma \rightarrow \infty$) requires that the electric field measured
by the comoving observer is zero, \ie $F^{\mu \nu} u_\nu=0$. In this
limit, the electromagnetic tensor and its dual is described completely by
the magnetic field ${\boldsymbol b}$ measured in the comoving frame
\begin{equation}
F^{\nu \sigma} = \eta^{\alpha\mu\nu\sigma}b_\alpha u_\mu \,,
\hskip 1.0cm
\,^{*\!\!}F^{\mu\nu} = b^\mu u^\nu - b^\nu u^\mu \,,
\end{equation}
and the Maxwell equations take the simple form
\begin{equation}
\nabla_\nu \,^{*\!\!}F^{\mu\nu} = \frac{1}{\sqrt{-g}} 
\partial_\nu \left[ \sqrt{-g}
\left(b^\mu u^\nu - b^\nu u^\mu \right)\right] = 0 \,,
\label{eq:maxwell}
\end{equation}
The relation between the magnetic field seen by the comoving observer,
${\boldsymbol b}$, and that seen by an Eulerian observer, ${\boldsymbol
  B}$, can be found by using the projection operator $P_{\mu\nu}\coloneqq
g_{\mu\nu}+u_\mu u_\nu$ orthogonal to ${\boldsymbol u}$. Applying this
operator to the definition of the magnetic field ${\boldsymbol B}$, one
can derive the following relations
\begin{equation}
b^0 = \frac{W B^i v_i}{\alpha} \,,\;\;\;  \;\;\;\;
b^i = \frac{B^i + \alpha b^0 u^i}{W} \,,\;\;\;\;\;\;\;
b^2 \coloneqq b^\mu b_\mu = \frac{B^2 + \alpha^2 (b^0)^2}{W^2} \,, 
\label{eq:b0b2}
\end{equation}
where $B^2\coloneqq B^i B_i$.
The time component of equations \eqref{eq:maxwell} provides the
divergence-free condition
\begin{equation}
\partial_i \tilde{B}^i =0 \,,
\label{eq:divergence}
\end{equation}
where $\tilde{B}^i\coloneqq \sqrt{\gamma}B^i$ \lrn{(enforcing this
  condition in numerical simulations is a very active field of research
  and we will comment on this in Section \ref{sec:bph})}. The spatial
components of equations \eqref{eq:maxwell}, on the other hand, yield the
induction equations for the evolution of the magnetic field
\begin{equation}
\partial_t \tilde{B}^i = 
\partial_j(\tilde{v}^i\tilde{B}^j-\tilde{v}^j\tilde{B}^i) \,,
\label{eq:induction}
\end{equation}
where $\tilde{v}^i\coloneqq \alpha v^i-\beta^i$. Also in the case of the
solution of the IMHD equations, the large majority of research groups
makes use of HRSC methods and recast the equations in a flux-conservative
form as first proposed by Ref. \cite{Anton06}, although other (small)
variants have been developed and used
\cite{Duez05MHD0,Giacomazzo:2007ti}.

Finally, we briefly mention the challenges of resistive MHD (RMHD), which
is important to describe realistic plasma instabilities and magnetic
reconnection. In case of finite conductivity, in fact, a relation for the
current as a function of the other fields is needed in order to close the
system, and this is provided, at least in principle, by Ohm's law. In
practice, however, the poor knowledge of the non-ideal microphysical
properties of the matter at the merger of neutron-star binaries makes the
implementation of Ohm's law very delicate, leaving ample room for
phenomenological exploration. In addition to the microphysical
uncertainties, including Ohm's law in the evolution equations changes
their mathematical nature, introducing terms that can become stiff terms
in regions of high (but finite) conductivity. Since such terms reduce
considerably the timestep, making the evolution with explicit time
integrators nearly impossible, successful codes \cite{Palenzuela:2008sf,
  Dionysopoulou:2012pp, Palenzuela2013, Dionysopoulou2015} have
implemented schemes that apply an implicit discretization to the stiff
terms and an explicit one to the non-stiff terms, \ie implicit-explicit
(IMEX) Runge-Kutta methods \cite{pareschi_2005_ier, Rezzolla_book:2013}.

\newpage
\section{Initial data}
\label{sec:ID}

Obviously, numerical simulations involving time evolution must start from
given initial data representing essentially a snapshot of all independent
evolved variables. In the case of general relativity, such initial data
should be a solution of the general-relativistic
\lrn{(magneto)}hydrodynamics equations and of the Einstein equations, in
particular, of the Hamiltonian and momentum constraints, when the
``$3+1$'' ADM decomposition is used. Other constraints could also be
present if additional fields (such as electromagnetic fields) are
included.

We recall that the Hamiltonian and momentum constraints are four coupled
second-order elliptic partial differential equations (PDEs) that are
solved numerically through some iterative procedure that starts with an
initial guess and finds successive solutions, correcting the fields until
some predetermined accuracy criteria are reached. The four constraint
equations, however, are not enough to determine the ten independent
components of the spacetime metric. Suggestions for determining the
remaining degrees of freedom were made long ago, and a widely used
approach is that of employing a conformally flat geometry, also known as
the Isenberg-Wilson-Mathews approach \cite{Isenberg08,Wilson89}.

The popularity of this approach, which by construction suppresses any
gravitational radiation content, stems from the fact that the
restrictions of the conformal flatness of the three-metric and of maximal
slicing simplify the set of equations. 
Furthermore, the condition of helical symmetry applied to the
conservation of the stress energy tensor, that is, the requirement of the
fluid fields to be time independent in the frame that corotates with the
binary, is aimed at enforcing the circularity of the orbit of the BNS
system.
The existence of such a frame is mathematically equivalent to the
existence of a ``helical'' Killing field, and several codes for the
calculation of BNS initial data have been built adopting this assumption
\cite{Wilson95, Bonazzola97, Marronetti98, Baumgarte97, Marronetti99,
  Bonazzola98b, Uryu00, Uryu00a, Usui2000,
  Gourgoulhon-etal-2000:2ns-initial-data, Taniguchi01, Taniguchi02b,
  Taniguchi03, Bejger05, Tsokaros2007, Kiuchi2009, Taniguchi2010,
  Tsokaros2012, Tsokaros2015}\footnote{Particular attention deserves the
  \texttt{LORENE} code \cite{Gourgoulhon-etal-2000:2ns-initial-data,
    lorene}, because it is distributed publicly.}.

\subsection{Irrotational binaries}
\label{sec:ibs}

BNS systems that have evolved without close interaction with other stars
are thought to be on quasi-circular orbits and with minute orbital
eccentricities, \ie $\lesssim 0.01$, having lost any initial eccentricity
through the emission of gravitational radiation \cite{Peters:1963ux}. BNS
formed by dynamical capture are expected to have higher eccentricities by
the time they merge\footnote{Since their detection event rate is likely
  to be much smaller than the one corresponding to the binary evolution
  channel, only a few studies have been performed on these binaries
  \cite{Turner1977a, Turner1977b, East2012c, Gold2012, Rosswog2013,
    Radice2016}. See also Sect. \ref{sec:dynamical-capture}.}. While the
helical symmetry condition demands exact circular orbits, configurations
produced in this framework actually have eccentricities larger than those
thought to be common in old BNS systems by a factor of several
\cite{Tsatsin2013, Miller03b, Miller03c}. Reducing the orbital
eccentricity is an important task in numerical relativity, especially in
order to compute accurate gravitational waveforms during the inspiral. In
fact, the presence of spurious eccentricity complicates comparisons
between waveforms obtained from numerical simulations and those derived
through analytical methods \cite{Baiotti2011, Bernuzzi2012,
  Hotokezaka2013b}. The eccentricity also affects the construction of the
phenomenological hybrid templates, namely templates constructed from the
matching of analytical and numerical waveforms \cite{Ajith:2007qp,
  Ajith:2007kx:longal, Bernuzzi2011, Bernuzzi2015}. These are necessary
because the extraction of neutron-star parameters and deformability
properties, which will place constraints \lrn{on} the neutron-star EOS,
from gravitational-wave observations relies on, and is sensitive to, the
accuracy of the templates \cite{Read2013, Favata2014, Yagi2014,
  Wade2014}.

The spurious eccentricity originates from two main causes, the first of
which being that quasi-equilibrium initial data for BNS assuming a
helical Killing field ignores the radial component of the velocity of the
orbiting stars. This error may be reduced by adding a radial velocity
(determined empirically from time evolutions or calculated from
post-Newtonian expressions of inspiraling point masses) to minimise the
oscillations around the inspiral orbit \cite{Uryu2006, Uryu:2009ye}. The
second cause of spurious eccentricity is in the assumption of conformal
flatness for the three-geometry of the initial hypersurface, which
obviously is only an approximation as it implies the lack of any
gravitational-wave component in the spacetime, which cannot be correct
for a binary that has since long been producing gravitational waves. The
inaccuracy of the binary orbit arising from spatial conformal flatness
can be removed to a high degree if one solves the full Einstein equations
for all metric components, including the non--conformally-flat part of
the spatial metric \cite{Isenberg-1979, Isenberg08, Wilson95,
  Marronetti99, Uryu2006, Uryu:2009ye,Uryu2012}. For example, in what is
called the ``waveless formulation'', the field equations for the metric
components become elliptic equations on an initial slice, and they yield
an asymptotically flat metric. Waveless solutions may determine phase and
frequency of the binary with significantly greater accuracy, particularly
if one first calibrates to the frequencies of a set of quasi-equilibrium
sequences in order to overcome errors in the radial motion, as seen above
\cite{Uryu2006,Uryu:2009ye}.

Both the conformally flat and the waveless approximation do not constrain
the dynamics of the fluid, which is instead specified by the solution of
the equations of hydrostatic equilibrium. To obtain such a solution it is
necessary to specify a condition on the velocity field of the matter in
the star and this condition is, to a large extent, arbitrary. The
simplest option is to consider the two stars as tidally locked or
\emph{corotating} \cite{Baumgarte97, Baumgarte98b}, so that the fluid in
each star is static in the frame that rotates with the
binary\footnote{The corotating configuration is the only rotation state
  that is fully compatible with the helical-symmetry condition and is
  quite common in binary systems, with the closest example being offered
  by the Earth-Moon system.}. While this condition is easy to implement
numerically as it leads to a number of mathematical simplifications,
corotating BNSs are unlikely to exist in nature because neutron-star
viscosities are thought to be insufficient to lead to locking
\cite{Bildsten92}, and therefore different prescriptions have been
suggested. Another choice that simplifies the equations consists in
considering the fluid motion inside the stars to have zero vorticity. The
binaries in this case are referred to as \emph{irrotational}
\cite{Bonazzola97, Asada1998, Teukolsky98, Shibata98} and the neutron
stars have a very small spin angular momentum which is counter-aligned
with the orbital angular momentum. Irrotational binaries have
represented, by far, the most common type of initial data for BNSs and
are still widely used.

Building on the large bulk of work already carried with irrotational
binaries \cite{Bonazzola98b, Uryu00, Uryu00a, Usui2000,
  Gourgoulhon-etal-2000:2ns-initial-data, Taniguchi01, Taniguchi02b,
  Taniguchi03, Taniguchi2010, Bejger05}, a significant effort has been
invested recently in the construction of initial data for BNS that
focuses on reducing the eccentricity of the orbit. The basic approach for
eccentricity reduction developed and employed by Kyutoku et
al. \cite{Kyutoku2014} is similar to the method for eccentricity
reduction for binary black holes of Pfeiffer et
al. \cite{Pfeiffer:2007yz}. As a start, standard quasi-circular initial
data are computed assuming helical symmetry, and these are evolved for
about three orbits. Then, appropriate corrections to the orbital angular
velocity and to the radial velocity are estimated through a fit of the
time derivative of the coordinate orbital angular velocity to an analytic
function. Such corrections are finally applied to the initial data, and
this process is repeated until the eccentricity is reduced below the
desired value. This eccentricity-reduction procedure was tested with
simulations of BNSs and revealed that three successive iterations allowed
for an eccentricity decrease from about $0.01$ to about $0.001$,
\lrn{which is below the threshold set for binary black-hole waveforms
  \cite{Hinder2013}}. As expected, low-eccentricity initial data allow
for smaller modulations in the evolution of the orbital separation, and
the gravitational-wave amplitude and frequency. With these
low-eccentricity initial data, the accuracy of gravitational waves is
limited mostly by the truncation error of the numerical scheme adopted
(which acts as an effective viscosity) and by the boundary conditions at
the outer boundary \cite{Kyutoku2014}, if the latter is causally
connected with the location where the gravitational waves are extracted.

Still concentrating on eccentricity, but with a different goal in mind,
Moldenhauer et al. \cite{Moldenhauer2014} put forth a new method that
extends the notion of helical symmetry to eccentric orbits, by
approximating the elliptical orbit of each companion as instantaneously
circular. This allows one to generate consistent initial data for BNS
systems with high eccentricity, thought to form through dynamical capture
(see Sect. \ref{sec:dynamical-capture}). It was found that the spurious
stellar oscillations seen in simulations of elliptic BNS with less
accurate initial data are reduced by at least an order of magnitude.

\subsection{Spinning binaries}
\label{sec:sbs}

As a way to increase the realism of the initial fluid configuration and
move away from the irrotational condition, an effort has been invested
recently in considering initial data for spinning neutron stars, \ie
\emph{spinning} binaries. In principle, neutron stars in accreting binary
systems could reach very high spin rates and hence it is possible that
also in binary systems the neutron stars retain high spins up to the
merger \cite{Oslowski2011, Kiel2010}. Large spins are expected to have a
noticeable effect on the inspiral and merger of the binary if the
rotation period is within an order of magnitude of the orbital period
\cite{Tichy11}. However, finding ways to construct initial data for
binaries with spinning neutron stars has been a difficult problem. There
have been attempts to produce initial data for spinning BNSs with
different methods. Some of them are also based on a conformally flat
slice in the presence of a helical Killing vector and employ advanced
computationally intensive iterative algorithms \cite{Marronetti03,
  Baumgarte:2009, Tichy11, Tichy12}, while others introduce the spin in
cruder manners \cite{East2012d, Kastaun2013, Tsatsin2013}.

Indeed, before constraint-satisfying solutions for spinning BNS systems
were derived, a number of authors explored the effects of spin with less
accurate initial data. In all cases, the guiding principle was that of
starting from some sort of the constraint-satisfying solutions, \eg of
irrotational binaries or of isolated rotating neutron stars, and to
``perturb'' the system either by introducing some degree of rotation or
by superposing the rotating solutions and suitably boosting them. The
initial data constructed in this way is obviously constraint violating,
but as long as the perturbations are small, the evolution would wash away
the violations (especially if constraint-damping formulations of the
equations are employed) yielding a consistent evolution. Following this
spirit, Kastaun et al. \cite{Kastaun2013} have explored the effect of the
neutron-star spin on the black hole formed after the merger. Their
initial data consisted of irrotational binaries to which various amounts
of rotation in the direction orthogonal to the orbital plane were added
a-posteriori on the irrotational solution to increase the total angular
momentum. Although the initial data violated the constraint equations and
caused oscillations in the initial stages of the evolution, the use of
the constraint-damping CCZ4 formulation (see \ref{sec:CCZ4} and
Refs. \cite{Bernuzzi:2009ex,Alic2013}) led to acceptable constraint
violations during evolution, with on average L2 norms of the Hamiltonian
constraint one order of magnitude smaller than those obtained with other
methods.

Following an alternative route, Tsatsin and Marronetti \cite{Tsatsin2013}
presented a more general method for producing initial data corresponding
to spinning BNS that also allows for arbitrary orbital and radial
velocities. This freedom gives more control also over the orbital
eccentricity. In this work, they also avoid solving the Hamiltonian and
momentum constraints and so sidestep the requirement of finding numerical
solutions of elliptic equations, thus simplifying the
implementation. Also in this study it was found that the constraint
violations were reasonably small and comparable to those obtained with
other methods after some orbits of evolution. They additionally showed
that their method can produce initial-data sets that exhibit
eccentricities smaller than those resulting from evolving helically
symmetric initial-data sets and that possess less spurious radiation of
numerical origin than that found in standard sets.

\begin{figure*}
\begin{center}
\raisebox{0.8cm}{\includegraphics[width=0.48\columnwidth]{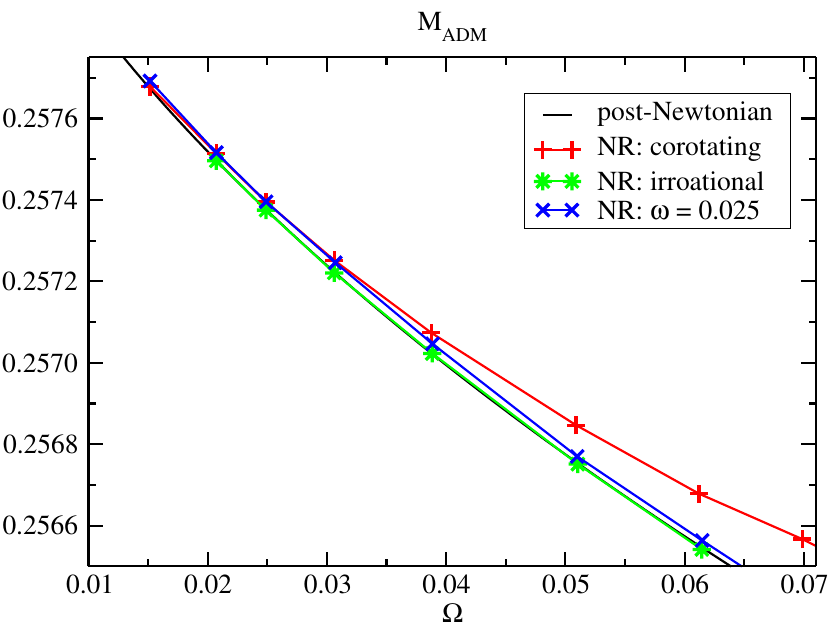}}
\hskip 0.25cm
  \includegraphics[width=0.48\columnwidth]{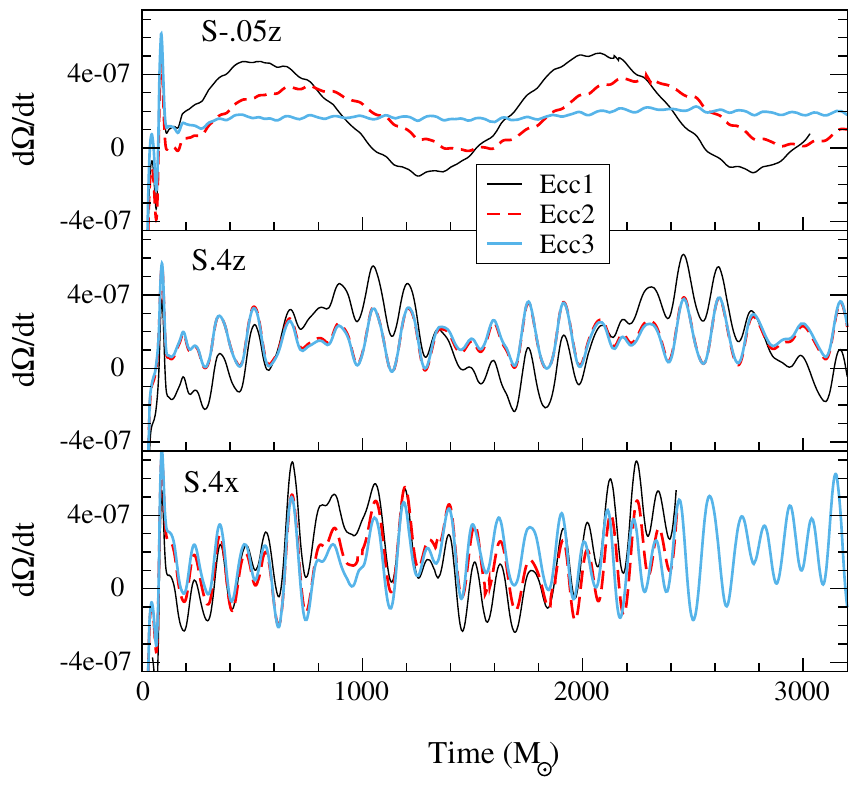}
\end{center}
\caption{\textit{Left panel:} The gravitational mass for a binary as a
  function of orbital angular velocity. Shown are results for
  post-Newtonian point particles (solid black line), and three different
  numerical results (NR) for corotating stars (marked by pluses),
  irrotational stars (marked by stars), and a case where both stars have
  spin (crosses); \lrn{note that the post-Newtonian results overlap
    closely those for irrotational stars} [Reprinted with permission from
    Ref. \protect{} \cite{Tichy12}. \copyright~(2012) by the American
    Physical Society.]  \textit{Right panel:} The derivative of the
  binary orbital frequency as a function of time for different levels of
  eccentricity reduction.  [Reprinted with permission from
    Ref. \protect{} \cite{Tacik15}. \copyright~(2015) by the American
    Physical Society.]}
   \label{fig:initialdata}
\end{figure*}

These heuristic, but overall effective approaches, were paralleled by the
more rigorous work of Tichy \cite{Tichy11,Tichy12}, who developed a
formalism for the calculation of initial data of arbitrarily spinning
neutron stars in binaries, once again adopting the conformally flat
approximation. Tichy's approach starts from the formulation of Shibata
\cite{Shibata98} and makes simplifying assumptions, like that the spins
of each star are small and remain approximately constant, as expected
from the fact that the viscosity of the stars is insufficient for tidal
coupling \cite{Bildsten92}. The result is a system of elliptic equations,
which reduces to the known cases of irrotational and corotating binaries
when appropriate parameters are chosen and has the correct Newtonian
limit. By solving such a system, the neutron stars can be given arbitrary
spin by choosing a (rigid or differential) rotational velocity for each
star \cite{Tichy12}. An example of the results of this formulation are
shown in the left panel of Fig. \ref{fig:initialdata}, which displays the
gravitational mass of the binary as a function of orbital angular
velocity. Shown are results for post-Newtonian point particles (solid
line), and three different numerical results (NR) for corotating stars
(pluses), irrotational stars (marked by stars), and a case where both
stars have spin (crosses)

Since its derivation, the formalism developed by Tichy
\cite{Tichy11,Tichy12} has been adopted in a few works
\cite{Tsokaros2015, Tacik15}. More specifically, after adapting Tichy's
formulation \cite{Tichy11} to a different numerical infrastructure,
Tsokaros et al. \cite{Tsokaros2015} presented an extension of the
\texttt{COCAL} code \cite{Uryu2012} to compute general-relativistic
initial data for symmetric binary compact-star systems with a
nuclear-physics EOS. The new code is able to describe also BNSs made of
arbitrarily (slowly) spinning stars and was tested by comparing it to the
open-source \texttt{LORENE} code
\cite{Gourgoulhon-etal-2000:2ns-initial-data, lorene} in the case of
irrotational and corotating binaries, finding full equivalence in the
solution and in the overall error.

In a separate development, Tacik et al. \cite{Tacik15} also followed the
formalism introduced by Tichy \cite{Tichy11,Tichy12} and solved the
initial-data equations for arbitrary spinning BNS systems with the
multi-domain pseudospectral elliptic solver developed in
Refs. \cite{Pfeiffer:2002wt,Foucart2008}, where the iterative procedure
of Ref. \cite{Foucart2011} is used. Tacik et al. \cite{Tacik15} also
showed how they reduce orbital eccentricity in the initial data,
following an iterative procedure first introduced for binaries containing
black holes \cite{Pfeiffer:2007yz,Boyle:2007ft,Foucart2008,Buonanno2011}
and consisting in first evolving an initial-data set for a few orbits and
then adjusting the initial-data parameters related to eccentricity
according to the analysis of the orbital dynamics. This is summarised in
the right panel of Fig. \ref{fig:initialdata}, which shows the evolution
of the time derivative of the binary orbital frequency for different
levels of eccentricity reduction.

Two remarks before concluding this Section on initial data. The first one
is that recently consistent initial data were produced also for BNS
systems in scalar-tensor theories of gravity \cite{Taniguchi2015} (see
Sect. \ref{sec:atog} for more details). The second, and alarming, remark
is about the recent work of Tsokaros et al. \cite{Tsokaros2016}, who
presented a comparative analysis of the gravitational waveforms relative
to essentially the same physical binary configuration when computed with
two different initial-data codes and then evolved with the same evolution
code. In particular, Tsokaros et al. \cite{Tsokaros2016} have considered
the evolution of irrotational neutron-star binaries computed either with
the pseudo-spectral code \texttt{LORENE}
\cite{Gourgoulhon-etal-2000:2ns-initial-data, lorene} or with the newly
developed finite-difference code \texttt{COCAL} \cite{Tsokaros2015}; both
sets of initial data have been subsequently evolved with the high-order
evolution code \texttt{WhiskyTHC} \cite{Radice2013b, Radice2013c,
  Radice2015}. Despite the initial data showed global (local) differences
that were $\lesssim 0.02\%\ (1\%)$, the difference in the
gravitational-wave phase at the merger time was rather large, reaching
$\sim 1.4$ radians at the merger time, after about $3$ orbits. These
results are a warning signal about the highly nonlinear impact that
errors in the initial data can have on the subsequent evolution and about
the importance of using exactly the same initial data when comparative
studies are done; \lrn{at the same time, they call for the importance of
  sharing initial data as an effective way to quantify the error budget.}

\newpage
\section{Pure-hydrodynamic simulations}
\label{sec:ph}

General-relativistic hydrodynamical simulations of BNSs started being
performed in Japan almost 20 years ago \cite{Nakamura99a, Oohara:1999,
  Shibata99c, Shibata99d, Shibata02a, Shibata:2003ga}. Even if nowadays
many state-of-the-art codes are able to solve more complex sets of
equations (\eg for the evolution of magnetic fields, neutrino emission,
etc.), simulations involving only general-relativistic hydrodynamics are
still the benchmark for any new code and the necessary testbed for more
advanced codes. Furthermore, in many cases, results obtained with pure
hydrodynamics, most notably, gravitational waveforms, provide already a
wealth of information on BNS systems, especially during the inspiral. In
many respects, the inspiral may be considered the {\it easiest} part of
the problem, in which the stars spiral towards each other as a result of
gravitational-radiation losses, being scarcely or not at all affected by
magnetic fields or neutrinos. Its simplicity notwithstanding, this
problem is still the object of continuous efforts and improvements, which
are often carried out through the synergy of numerical simulations and
analytical calculations based on post-Newtonian expansions or other
approximation schemes. We describe progress on this topic in
Sect. \ref{sec:EOB}. The inspiral has also recently attracted renewed
attention with the first simulations of arbitrarily spinning BNS systems
(see Sect. \ref{sec:hydro_inspiral}).

\begin{figure}
\begin{center}
  \includegraphics[width=6.45cm]{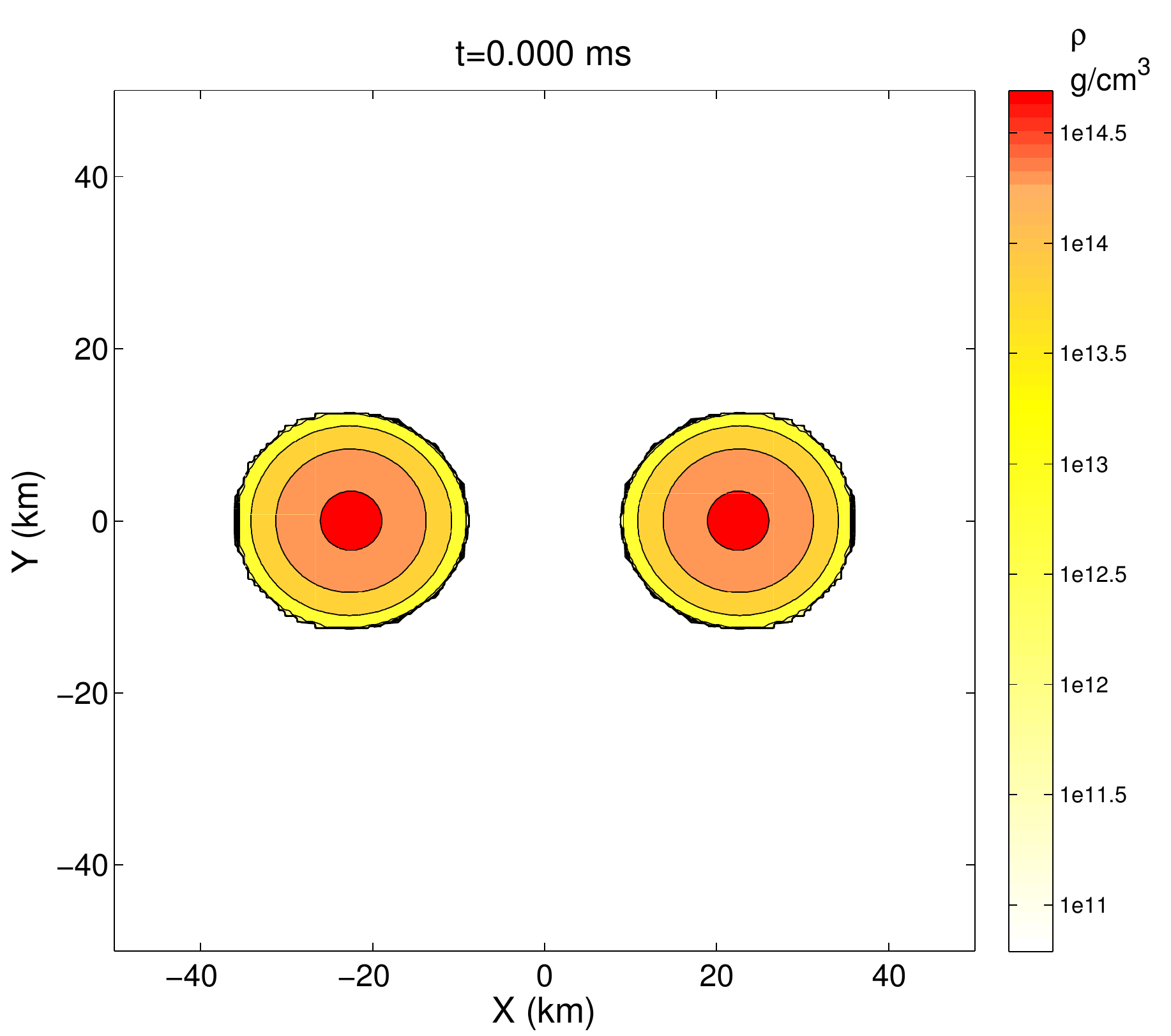} 
  \hskip 0.1cm
  \includegraphics[width=6.45cm]{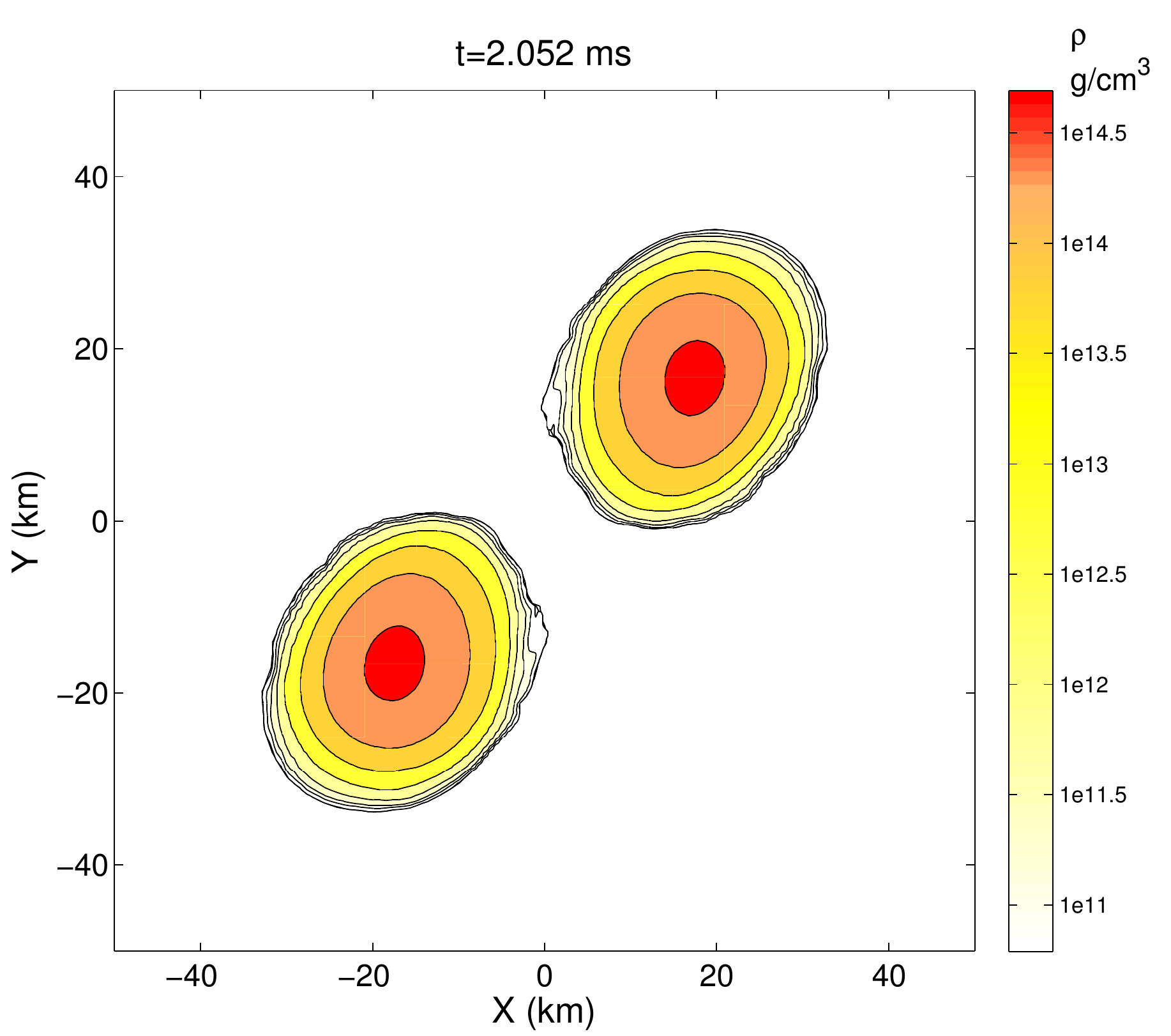} 
  \includegraphics[width=6.45cm]{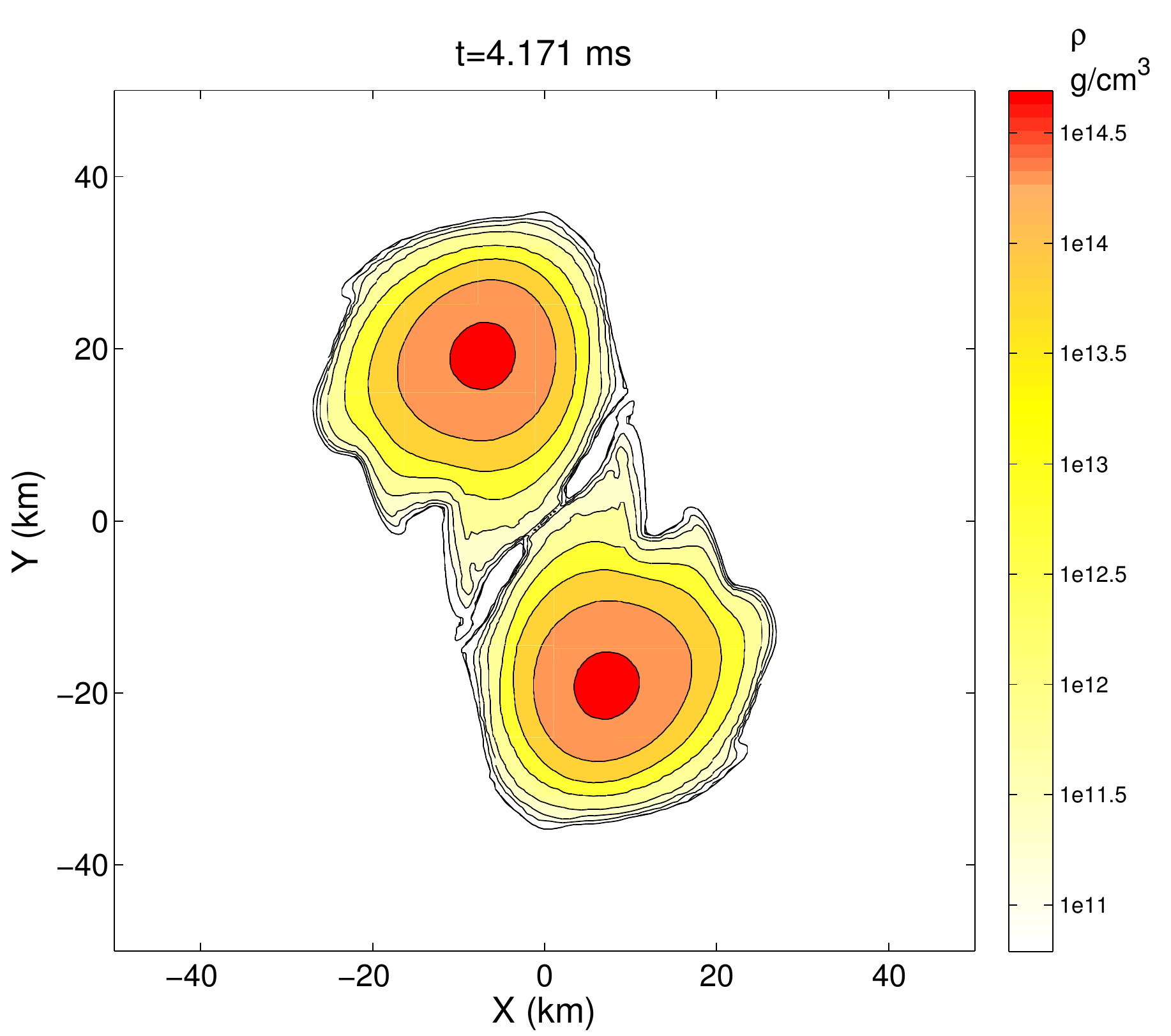} 
  \hskip 0.1cm
  \includegraphics[width=6.45cm]{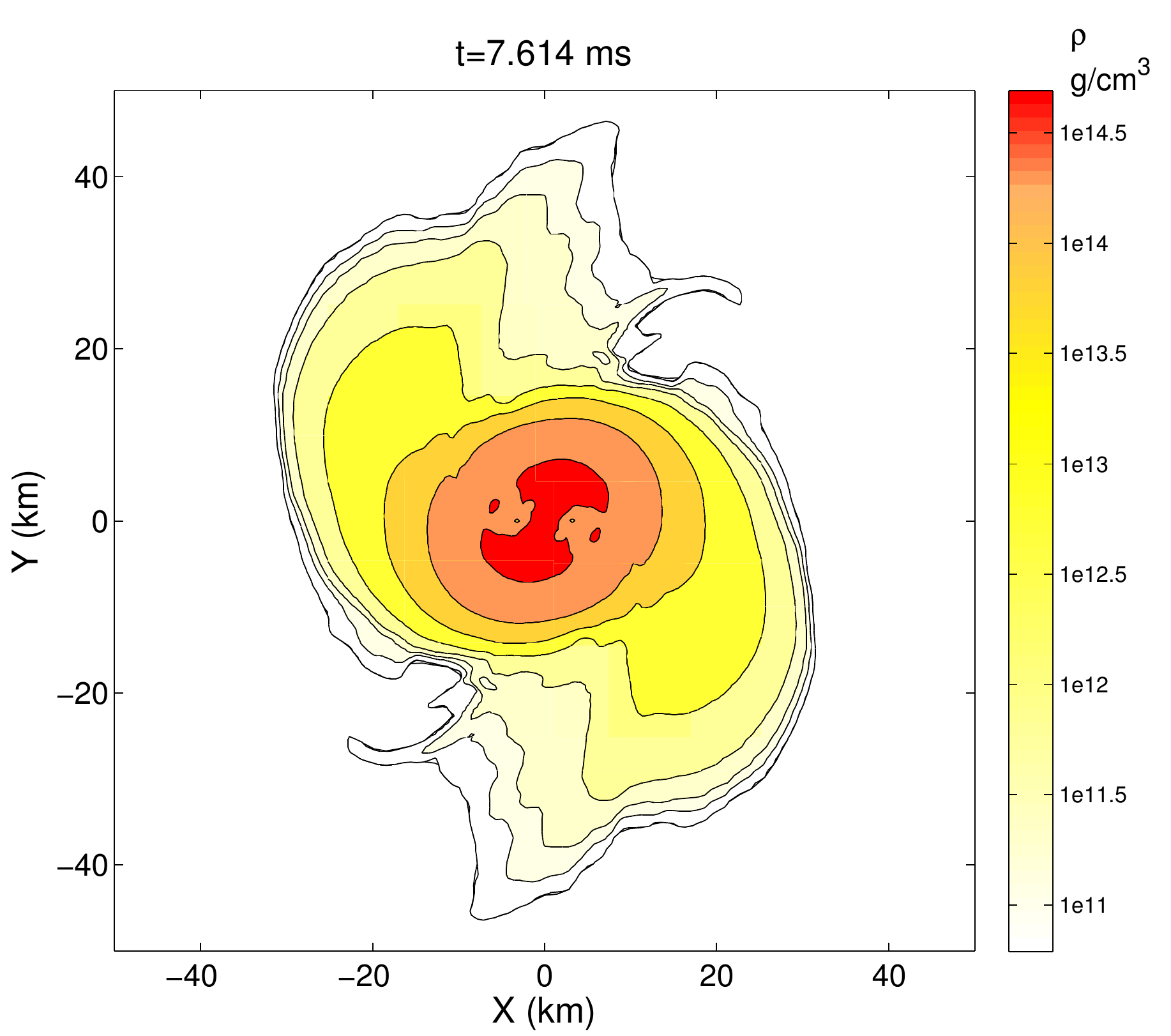} 
  \includegraphics[width=6.45cm]{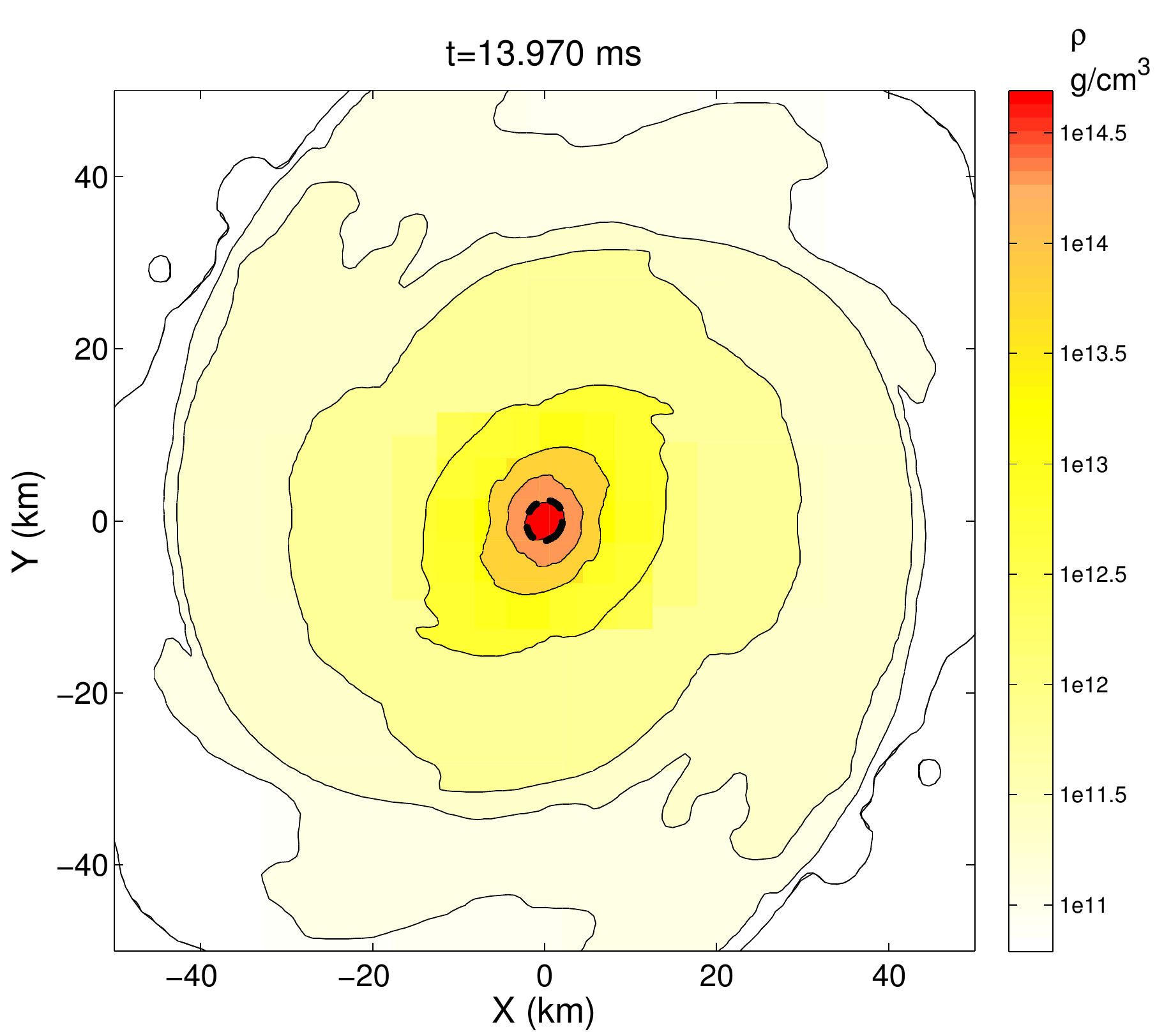} 
  \hskip 0.1cm
  \includegraphics[width=6.45cm]{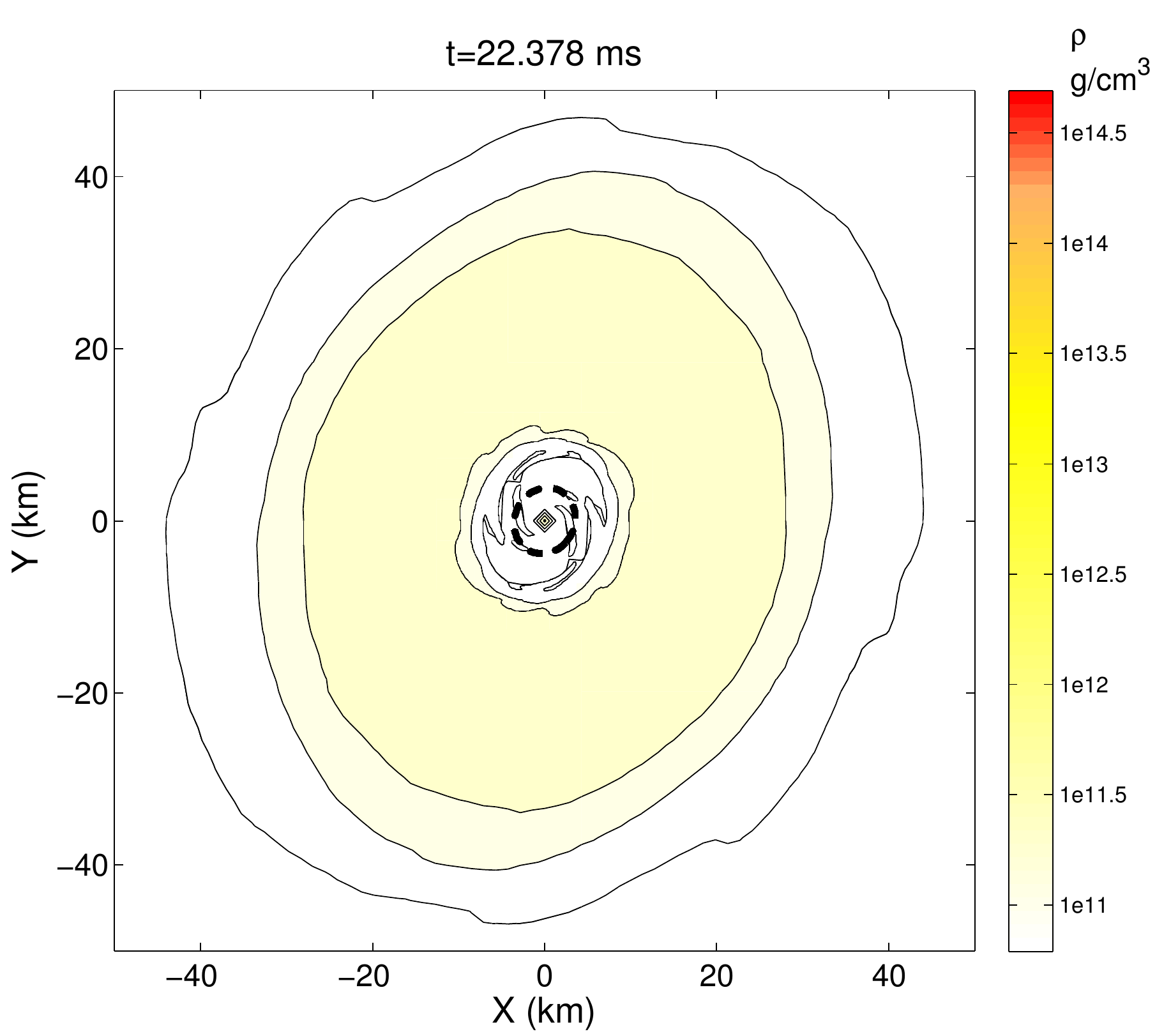} 
\end{center}
   \caption{Isodensity contours in the $(x,y)$ plane for the evolution of
     a high-mass (individual stellar rest mass $1.625 M_\odot$) binary
     with an ideal-fluid EOS. The thick dashed lines in the lower panels
     show the location of the apparent horizon. [Reprinted with
       permission from Ref. \protect{}\cite{Baiotti08}. \copyright~(2008)
       by the American Physical Society.]}
   \label{fig:blackhole_IF_short_by}
\end{figure}

In what follows, we give a general description of the BNS dynamics using
the figures of Ref. \cite{Baiotti08}, which was one of the first to
provide complete and accurate evolutions. Our description is here
intentionally qualitative, as we focus on those aspects that are robust
and independent of the EOS.

For millions of years a comparatively slow inspiral progressively speeds
up until the two neutron stars become so close that tidal waves produced
by the (tidal) interaction start appearing on the stellar surface (these
are clearly visible in the second and third panels of
Fig. \ref{fig:blackhole_IF_short_by}). Such waves are accompanied by
emission of matter stripped from the surface and by shocks that represent
the evolution of small sound waves that propagate from the central
regions of the stars, steepening as they move outwards in regions of
smaller rest-mass density \cite{Stergioulas04,Nagakura2014}.

At the merger, the two stars collide with a rather large impact
parameter. A {\it vortex sheet} (or {\it shear interface}) develops,
where the tangential component of the velocity exhibits a discontinuity.
This condition is known to be unstable to very small perturbations and it
can develop Kelvin-Helmholtz instability (KHI), which curls the interface
forming a series of vortices at all wavelengths \cite{Chandrasekhar81,
  Bodo1994}. Even if this instability is purely hydrodynamical and it is
likely to be important only for binaries with very similar masses, it can
have strong consequences if the stars possess magnetic fields (see
Sect. \ref{sec:HD_MHD}). It has in fact been shown that, in the presence
of an initially poloidal magnetic field, this instability may lead to an
exponential growth of the toroidal component \cite{Price06,
  Giacomazzo2011b, Rezzolla:2011, Neilsen2014, Kiuchi2014}. Such a growth
is the result of the exponentially rapid formation of vortices that curl
magnetic-field lines that were initially purely poloidal. The exponential
growth caused by the KHI leads to an overall amplification of the
magnetic field of about three orders of magnitude
\cite{Kiuchi2014}. 
\lrn{At the same time, high-resolution simulations in core-collapse
  supernovae find that parasitic instabilities quench the MRI, with a
  magnetic-field amplification factor of 100 at most, independently of
  the initial magnetic field strength \cite{Rembiasz2016}. Of course, KHI
  and MRI are two different instabilities, but the lesson these
  simulations provide is that parasitic instabilities may also appear
  during the development of the KHI and limit the overall magnetic-field
  amplification; such parasitic instabilities are at present not yet
  apparent because of the comparatively small resolutions employed when
  modelling BNS mergers.}

\begin{figure}
\begin{center}
   \includegraphics[width=6.9cm]{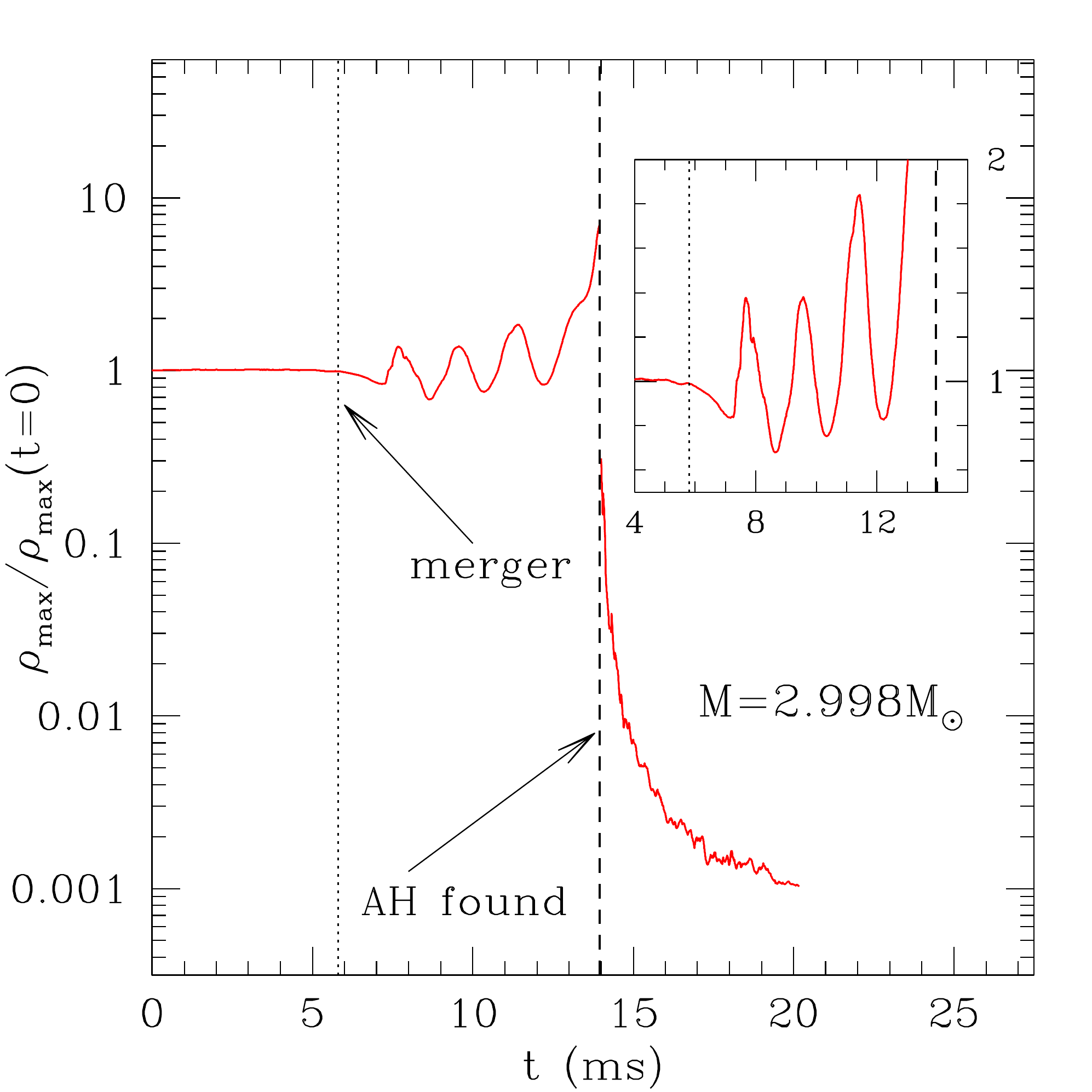}
    \hskip 0.0cm
   \includegraphics[width=6.9cm]{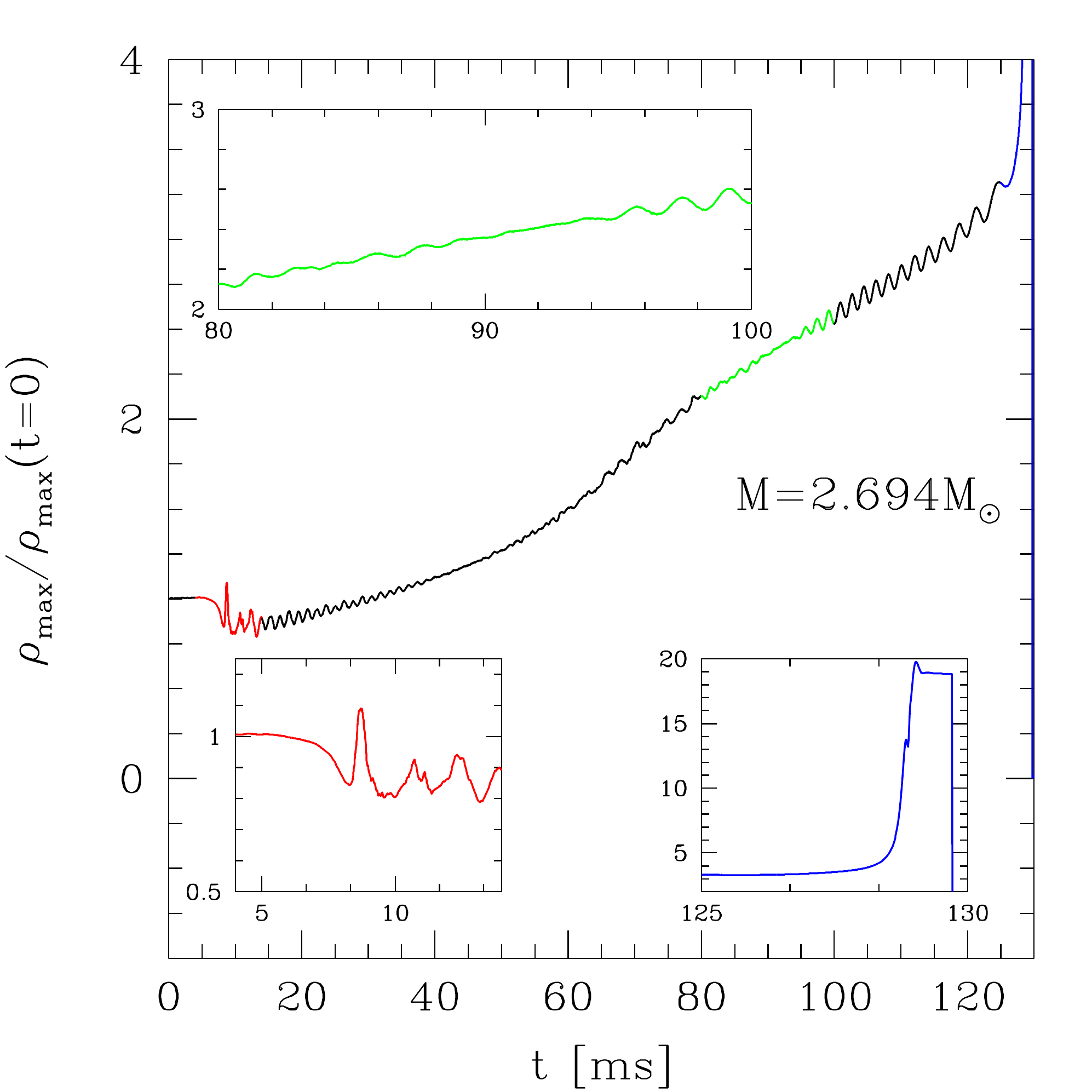}
\end{center}
   \caption{\textit{Left panel:} Evolution of the maximum rest-mass
     density normalized to its initial value for a high-mass (individual
     stellar rest mass $1.625 M_\odot$; gravitational mass of the system
     $2.998 M_\odot$) binary using an ideal-fluid EOS. Indicated with a
     dotted vertical line is the time at which the binary merges, while a
     vertical dashed line shows the time at which an apparent horizon is
     found. After this time, the maximum rest-mass density is computed in
     a region outside the apparent horizon [from
       Ref. \protect{}\cite{Baiotti08}. \copyright~(2008) by the American
       Physical Society] \textit{Right panel:} The same as in the left
     panel but for a low-mass binary (individual stellar rest mass $1.456
     M_\odot$; gravitational mass of the system $2.694 M_\odot$). Note
     that the evolution is much longer in this case and that different
     colours are used to denote the different parts of the evolution (see
     insets). [Adapted from
       Ref. \cite{Rezzolla:2010}. \copyright\protect{} IOP
       Publishing. Reproduced with permission. All rights reserved.]}
  \label{fig:if-rho-high}
\end{figure}

The HMNS produced from the merger may not collapse promptly to a black
hole, but rather undergo large oscillations with variations \lrn{such
  that the the maximum of the rest-mass density may grow to be twice as
  large (or more) as the value in the original stars} (see the right
panel of Fig. \ref{fig:if-rho-high}). These oscillations have a dominant
$m=2$ non-axisymmetric character \cite{Stergioulas2011b} and will be
discussed in detail in Sect. \ref{sec:hydro_merger_post-merger}. As
mentioned earlier, the formation and duration of the HMNS depends on the
stellar masses, the EOS, the effects of radiative cooling, magnetic
fields \cite{Ravi2014} and even on the development of gravitational-wave
driven instabilities \cite{Doneva2015}.

In essentially all cases when a black hole is formed, some amount of
matter remains outside of it, having sufficient angular momentum to stay
orbiting around the black hole on stable orbits. In turn, this leads to
the formation of an accretion torus that may be rather dense ($\rho \sim
10^{12}- 10^{13}\,{\rm g\ cm}^{-3}$) and extended horizontally for tens
of kilometres and vertically for a few tens of kilometres. Also this
point will be discussed in more detail in
Sect. \ref{sec:hydro_merger_post-merger}.

The dynamics of the inspiral and merger of a reference equal-mass binary
system is summarised in Fig. \ref{fig:if-rho-high}, whose panels show the
evolution of the maximum rest-mass density normalized to its initial
value (after the formation of the apparent horizon, the curve shows the
maximum rest-mass density in the region outside the apparent
horizon). Note that together with the large oscillations, the rest-mass
density also experiences a secular growth and the increased compactness
eventually leads to the collapse to a rotating black hole. The
differences in the two panels are essentially related to the initial mass
of the system (\ie $M=2.998\,M_{\odot}$ in the left panel and
$M=2.694\,M_{\odot}$ in the right panel) and it can be seen that, for a
given EOS (even a very simple one like the ideal-fluid EOS used in this
case) smaller masses will yield systematically longer-lived HMNSs.

Of course, the matter dynamics described so far in the various stages of
the evolution of a BNS system are imprinted in the gravitational-wave
signal, which then can be used to extract important information on the
properties of the neutron stars. Different parts of the evolution will
provide distinct pieces of information and with different overall
signal-to-noise (SNR) ratios. For example, the post-merger signal would
provide rather clear signatures but at such high frequencies that it may
be difficult to measure them with present detectors. On the other hand,
as we will discuss in detail in the following section, the inspiral
signal does depend on the EOS much more weakly, but in a way that is
still measurable because it comes at frequencies where the detectors are
more sensitive.

\subsection{Inspiral and merger dynamics}
\label{sec:hydro_inspiral}
\label{sec:EOB}

\begin{figure*}
  \begin{center}
\includegraphics[width=0.48\columnwidth]{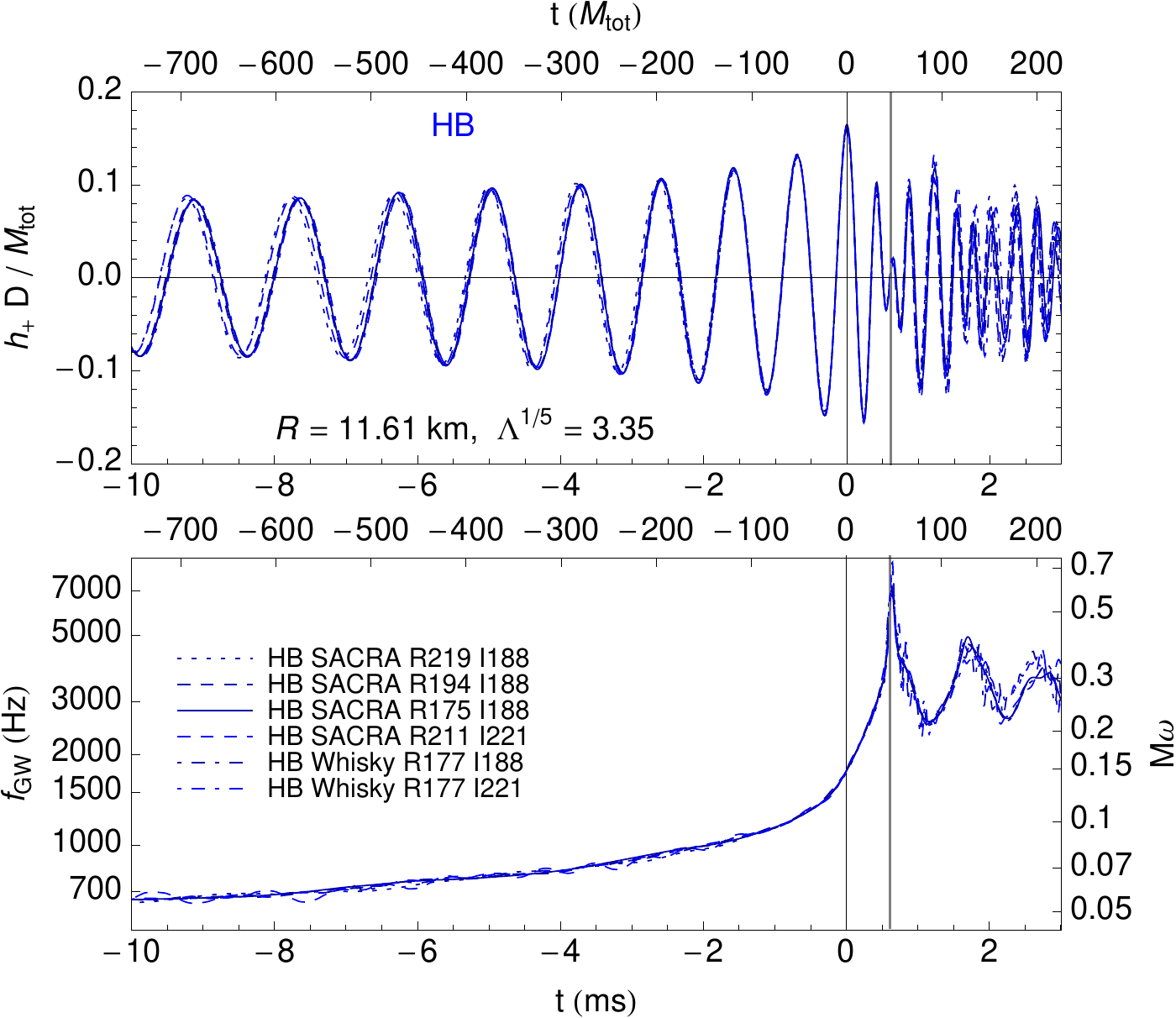}
\hskip 0.2cm
\raisebox{0.8cm}{\includegraphics[width=0.48\columnwidth]{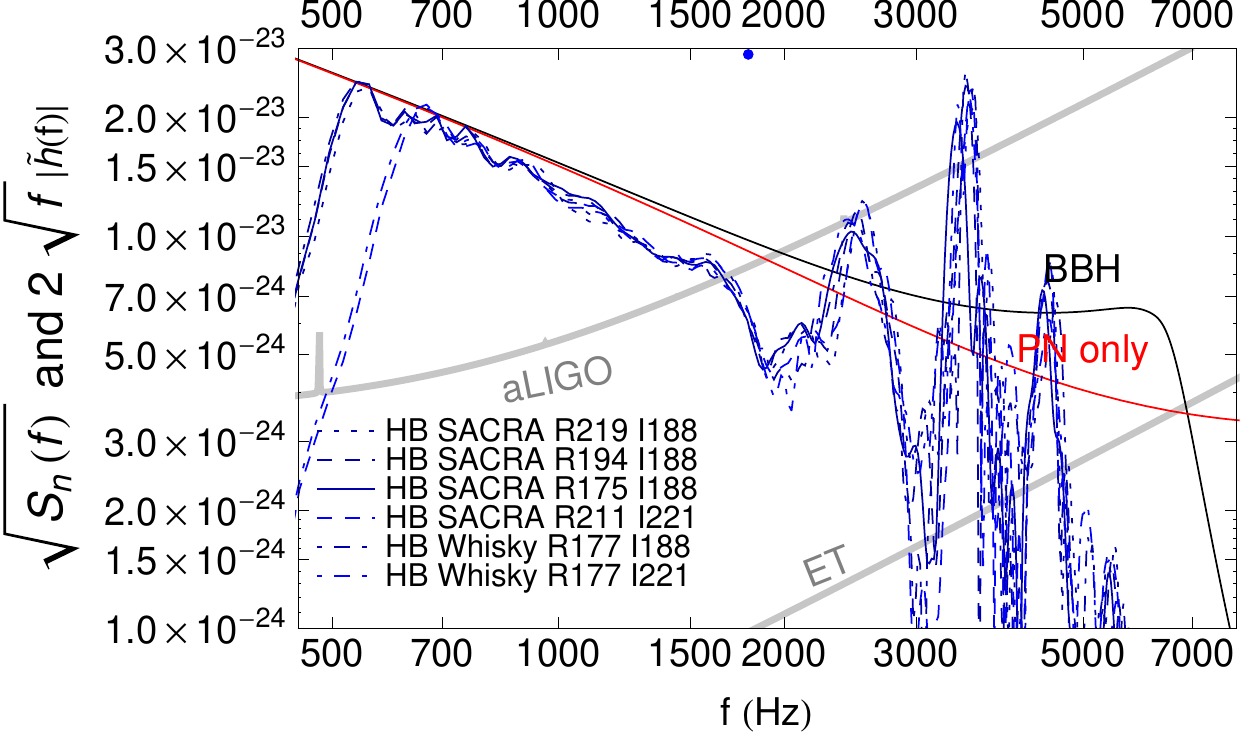}}
\end{center}
\caption{\emph{Left panel:} Waveforms and frequency evolution near
  merger, for model ``HB'' of the set of simulations of Read et
  al. \cite{Read2013}. The time of maximum amplitude is shown with a
  vertical black line and marks the ``merger time'' ($t=0$), while a grey
  vertical line marks the time of maximum frequency. \emph{Right panel:}
  Fourier spectra of the numerical waveforms of the left panel. Example
  noise spectra are indicated by thick grey lines for the Advanced LIGO
  high power noise~\cite{Shoemaker2009} and the Einstein Telescope ET-D
  noise~\cite{Punturo:2010}. The starting frequency depends on the
  initial orbital separation in the simulation. The pre-merger waveform
  gives a roughly monotonically decreasing amplitude, while post-merger
  oscillations contribute spikes at high frequency ($3000-5000\,{\rm
    Hz}$). Black curves indicate the phenomenological
  black-hole--black-hole waveform model of Santamar{\'{\i}}a et
  al. \cite{Santamaria2010} for the same mass parameters and red curves
  indicate the stationary phase approximation of a point-particle
  post-Newtonian inspiral \cite{Cutler94} that includes known terms up to
  3.5 post-Newtonian orders. The frequency of the first peak in the
  amplitude evolution is indicated by a coloured dot on the upper
  axis. [Reprinted with permission from Ref.
    \protect{}\cite{Read2013}. \copyright~(2013) by the American Physical
    Society.]}
\label{fig:Read2013_Fig3-6}
\end{figure*}

In 2013, Read et al. \cite{Read2013} have conducted the first systematic
investigation of the inspiral part of the signal making use of a number
of waveforms from various groups and with the ultimate goal of improving
data-analysis estimates of the measurability of matter effects in BNSs
(see Fig.~\ref{fig:Read2013_Fig3-6}). It was estimated that with a single
close-by source (at a distance of $\sim 100$ Mpc), the neutron star
radius or the dimensionless quadrupole tidal deformability (also called
polarizability coefficient)
\begin{equation}
\Lambda \coloneqq \frac{2}{3} k_2 \left(\frac{R}{M}\right)^5 \,,
\label{eq:Lambda}
\end{equation}
where $k_2$ is the quadrupole Love number, could be constrained to about
10\%. Later, it was confirmed with a more sophisticated statistical
analysis that for neutron-star binaries with individual masses of
$1.4\,M_\odot$, the dimensionless tidal deformability $\Lambda$ could be
determined with about 10\% accuracy by combining information from about
20-100 sources, depending on assumptions about BNS population parameters
(in particular, assuming nonzero spins for the initial neutron stars
shifts the necessary number of sources to higher values)
\cite{DelPozzo2013, Wade2014, Lackey2015, Agathos2015}.

\begin{figure*}
\begin{center}
\raisebox{0.8cm}{\includegraphics[width=0.48\columnwidth]{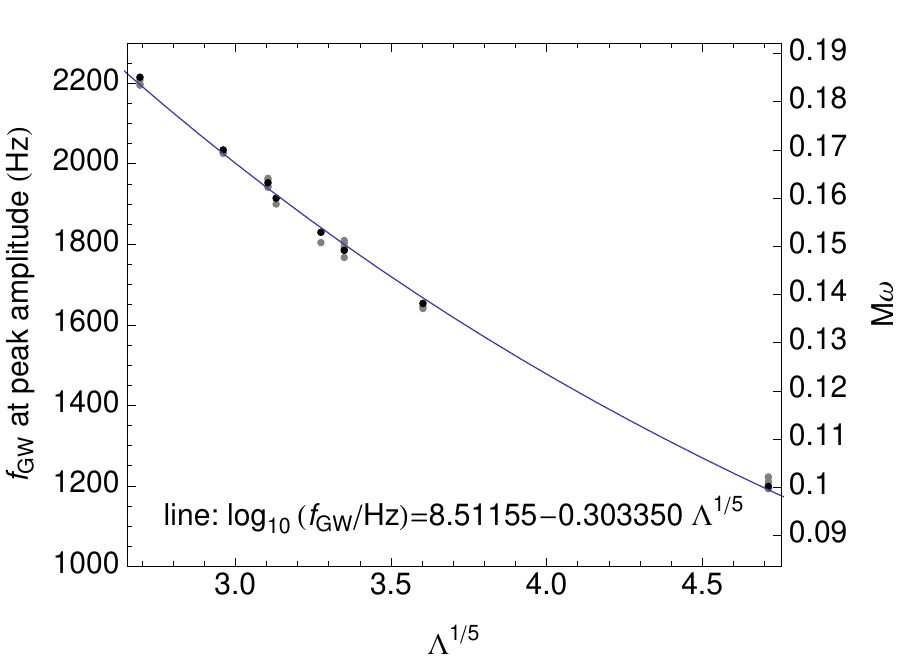}}
\hskip 0.2cm
\includegraphics[width=0.48\columnwidth]{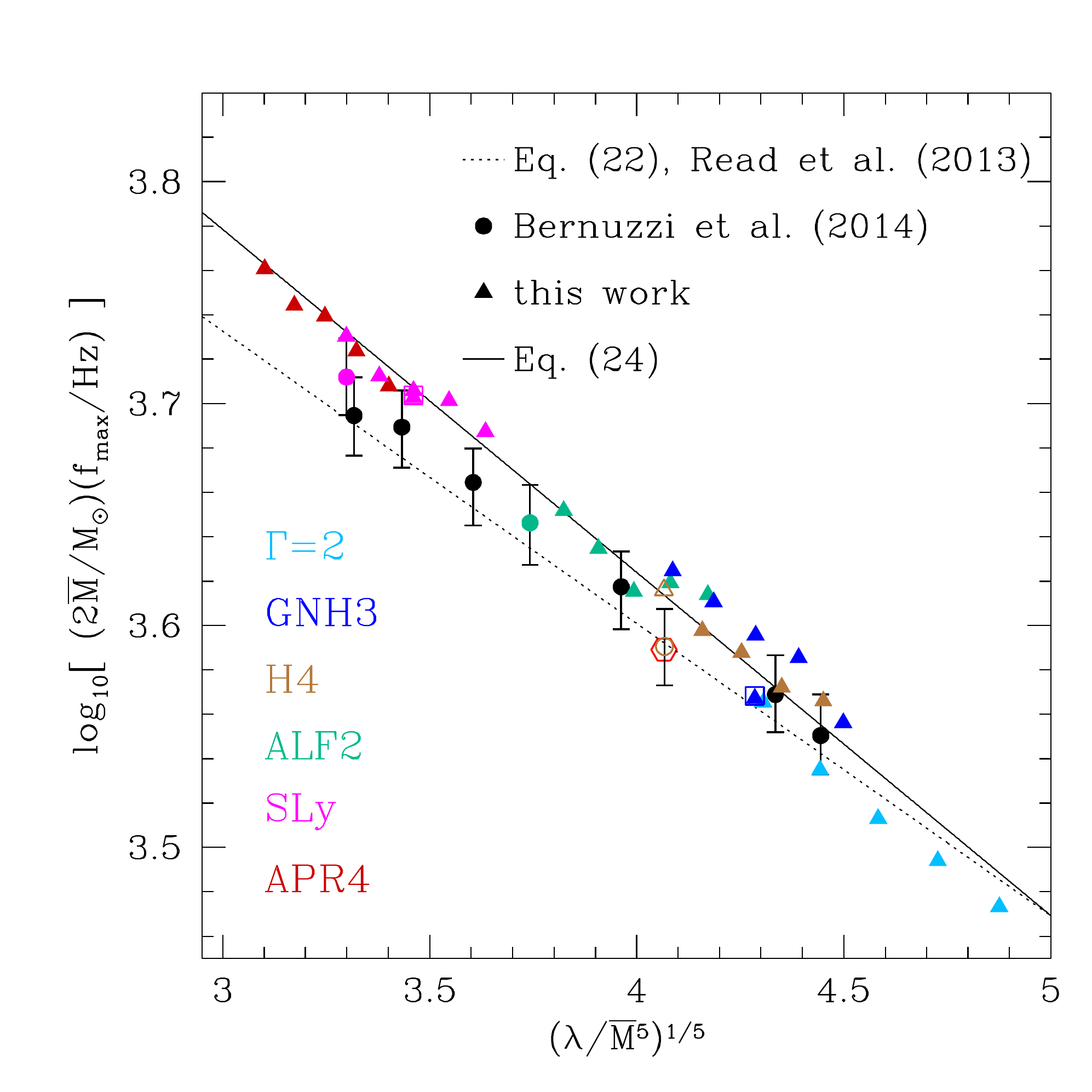}
\end{center}
\caption{\emph{Left panel:} Instantaneous gravitational-wave frequency at
  the point of peak amplitude, as a function of the dimensionless tidal
  deformability $\Lambda^{1/5}$. For each model, the highest-resolution
  simulation for a given EOS is plotted in black, lower-resolution
  simulations in grey. An empirical fit using $\Lambda^{1/5}$ is
  shown. [Reprinted with permission from
    Ref. \protect{}\cite{Read2013}. \copyright~(2013) by the American
    Physical Society.] \emph{Right panel:} Like in the left panel, the
  mass-weighted frequencies at amplitude maximum $f_\mathrm{max}$ are
  shown as a function of the dimensionless tidal deformability
  $\lambda/\bar{M}^5 = \Lambda$, where $\bar{M}$ is the average of the
  initial gravitational mass of the two stars. Filled circles refer to
  the data from Ref. \cite{Bernuzzi2014}, while coloured triangles
  indicate the data from Ref. \cite{Takami2015}, where triangles in boxes
  refer to the unequal-mass binaries. As in the left panel, the dotted
  line shows the relation suggested in Ref. \cite{Read2013}, while the
  solid line represents the best fit to the data (\ie Eq. (24) of
  Ref. \cite{Takami2015}). Note that the systematic differences between
  circles and triangles are due the small differences in the definition
  of the time of merger. [Reprinted with permission from Ref. \protect{}
    \cite{Takami2015}. \copyright~(2015) by the American Physical
    Society.] }
\label{fig:Read2013_Fig4}
\end{figure*}

Read et al. \cite{Read2013} were also the first to find a universal
relation between the frequency of the merger and the tidal deformability
$\Lambda$ of the neutron stars in an equal-mass binary (see
Fig. \ref{fig:Read2013_Fig4}). Here, the frequency of the merger is
defined as the instantaneous gravitational-wave frequency at the time
when the amplitude reaches its first peak. The relation is said to be
universal because it is valid for all the EOSs tried, which were
approximated by piecewise polytropes \cite{Read:2009a} and include a
large range of compactnesses. In a later work, Bernuzzi et
al. \cite{Bernuzzi2014} proposed that the tidal polarizability parameter
$\kappa_2^T$ is a more general choice for the parameter to be related to
the EOS, also because it is extensible to unequal-mass binaries. It is
defined as (see, \eg \cite{Bernuzzi2014})
\begin{equation}
\label{kappa_Bernuzzi}
\kappa_2^{^T} \coloneqq
2\left[
         q \left(\frac{X_{_A}}{C_{_A}}\right)^5k^{^A}_2 +
\frac{1}{q}\left(\frac{X_{_B}}{C_{_B}}\right)^5k^{^B}_2\right]\,,
\end{equation}
where $A$ and $B$ refer to the primary and secondary stars in the binary
$q \coloneqq {M_{_B}}/{M_{_A}} \leq 1$, $X_{_{A,B}} \coloneqq
{M_{_{A,B}}}/(M_{_A}+M_{_B})$, while $k_2^{^{A,B}}$ are the $\ell=2$
dimensionless tidal Love numbers, and $\mathcal{C}_{_{A,B}} \coloneqq
M_{_{A,B}}/R_{_{A,B}}$ are the compactnesses. In the case of equal-mass
binaries, $k_2^{A}=k_2^{B}=\bar{k}_2$, and expression
\eqref{kappa_Bernuzzi} reduces to
\begin{equation}
\kappa_2^{^T} \coloneqq \frac{1}{8}\bar{k}_2
\left(\frac{\bar{R}}{\bar{M}}\right)^5 = \frac{3}{16}\Lambda =
\frac{3}{16} \frac{\lambda}{\bar{M}^5} \,,
\end{equation}
where the quantity $\lambda \coloneqq \tfrac{2}{3} \bar{k}_2
{\bar{R}}^5$, is another commonly employed way of expressing the tidal
Love number for equal-mass binaries \cite{Read2013}, while $\Lambda
\coloneqq \lambda/\bar{M}^5$ is its dimensionless counterpart and was
employed in \cite{Takami2015} \lrn{[all barred quantities reported above
    are meant as averages, \ie $\bar{\Psi} \coloneqq (\Psi_{_A} +
    \Psi_{_B})/2$]}.

Read et al. \cite{Read2013} also investigated the magnitude of systematic
errors arising from numerical uncertainties and hybrid construction and
estimated the frequency at which such effects would interfere with
template-based searches. After defining the \emph{distinguishability} in
terms of the power-spectrum-density-weighted inner product for waveforms
differing in their parameter values (in this case deformability
$\Lambda$), it was concluded that the dependence of the
distinguishability on changes in $\Lambda$ is very similar for
post-Newtonian approximants and for hybridised numerical waveforms. Thus
it was anticipated that post-Newtonian approximants will allow one to
predict how well one will be able to infer the EOS from
gravitational-wave measurements when sufficiently accurate waveforms will
eventually become available.

Very recently, Hotokezaka et al. \cite{Hotokezaka2016} have
quantitatively improved the computations and estimations of Read et
al. \cite{Read2013} in two principal directions. First, they employed new
numerical-relativity simulations of irrotational binaries with longer
inspirals (\ie $14-16$ orbits) and higher accuracy both in the
initial-data setup (\ie residual eccentricity of $\sim 10^{-3}$) and the
evolution (see Ref. \cite{Hotokezaka2015} for details)\footnote{In
  analogy with what presented in Ref. \cite{Hotokezaka2013b}, also in
  Refs. \cite{Kyutoku2014, Hotokezaka2015, Hotokezaka2016} the time
  coordinate and phase of the waveforms are suitably adjusted to yield
  the desired convergence order, but the mathematical grounds for the
  necessity of this operation have not been investigated. Other codes can
  achieve convergence without such rescalings \cite{Thierfelder2011,
    Radice2013b}.}. Second, they included in the analysis lower
frequencies, down to 30 Hz, to which ground-based detectors like Advanced
LIGO are more sensitive. They also adopted EOSs developed more recently
\cite{Steiner2013,Banik2014,Hempel2012}. With these improvements,
Hotokezaka et al. \cite{Hotokezaka2016} found results very similar to
those of Read et al. \cite{Read2013}, namely that deformability $\Lambda$
and radius can be determined to about 10\% accuracy for sources at
$200\,{\rm Mpc}$, and they explain that this is because their
improvements drew the detectability in opposite directions: Increasing
the frequency range increases detectability, while better
numerical-relativity simulations apparently show smaller tidal effects
and so decrease the detectability. They conclude that if the EOS of
neutron stars is \lrn{stiff} (radius around $13\,{\rm km}$), it could be
pinned down by measurements of the radii obtained with this method, but
if the EOS of neutron stars is soft (smaller radii), a single EOS cannot
be identified with this method (unless the signal is very strong, in case
the source is very close).

A considerable portion of the recent research on the inspiral phase is
currently conducted with post-Newtonian expansions of the Einstein
equations (see, \eg Refs. \cite{Blanchet06,Poisson2014}), coupled to the
results of general-relativistic simulations. Another promising method
based on an analytical approximation is the Effective-One-Body (EOB)
formalism \cite{Buonanno:1998gg, Buonanno00a, Damour:1999cr,
  Damour:2001tu}. The high nonlinearity of the Einstein equations makes
it impossible to have satisfactory analytical approximations during the
most nonlinear phases of the merger, but the inspiral part can be very
well approximated, even up to very close before the merger
\cite{Bernuzzi2015}. We recall that the EOB model is a relativistic
generalisation of the Newtonian property that the relative motion of a
two-body system is equivalent to the motion of a particle of mass $\mu
\coloneqq M_{_A} M_{_B} / (M_{_A}+M_{_B})$ in the two-body potential
$V(r)$ of star $A$ and $B$. In the EOB formalism the effective
relativistic radial potential is $W_{\rm
  eff}=\sqrt{A(r)[\mu^2+(P_\phi/r)^2]}$, where $A(r)$ is the main radial
potential and $P_\phi$ is the centrifugal potential
\cite{Bernuzzi2015}. In the test-mass limit, $A(r)$ is equal to the
Schwarzschild potential $A_{\rm Schw}= 1 - 2M/r$ (where $M \coloneqq
M_{_A} + M_{_B}$). For extended bodies $A(r)$ changes because of two
physical effects: (i) mass-ratio effects, parameterized by $\nu\coloneqq
\mu/M$; (ii) tidal effects, parameterized by the relativistic tidal
polarizability parameters $\kappa_{_A}^{(\ell)}$ \cite{Hinderer08,
  Damour:2009, Binnington:2009bb, Hinderer09}, the most important of
which is the quadrupolar (\ie $\ell=2$) combination
$\kappa_2^T=\kappa_{_A}^{(2)}+\kappa_{_B}^{(2)}$ (see equation
\ref{kappa_Bernuzzi}). More specifically, the tidal interactions are
incorporated by a radial potential of the form
$A=A^0(r;\nu)+A^T(r)\kappa_{_A}^{(2)}$, where $A^0(r)$ is the EOB radial
potential for binary black holes and $A^T(r)$ is an additional piece
describing tidal interaction \cite{Damour:2012, Damour:2009wj,
  Bernuzzi2015}. Recently, a further improvement of the EOB formalism has
been proposed by Bernuzzi et al. \cite{Bernuzzi2015}. It is based on a
resummed version of the tidal EOB model that incorporates recent advances
in the relativistic theory of tidal interactions
\cite{Bini2014a,Bini2014b,Dolan2015}. Differently from previous works, no
fitting parameters are introduced for the description of tidal
interaction.

In Bernuzzi et al. \cite{Bernuzzi2015}, the new EOB description was
compared with data from numerical simulations performed with the
\texttt{BAM} code \cite{Thierfelder2011}, which solves the Z4c
formulation of the Einstein equations (see Sect. \ref{sec:CCZ4} and
Refs. \cite{Bernuzzi:2009ex,Alic2013}). The comparison was done mainly
through the gauge-invariant relation between the binding energy and the
orbital angular momentum \cite{Bernuzzi2012,Bernuzzi2014}, and an
agreement within the uncertainty of the numerical data was found for all
models. Previous results \cite{Baiotti:2010, Baiotti2011,
  Hotokezaka2013b, Hotokezaka2015}, obtained with older versions of the
EOB model not including resummation, had also found good agreement
between the phases of gravitational waves simulated in numerical
relativity and those predicted by EOB calculations for most of the
inspiral stage except for the tidally dominated, final inspiral stage, in
which the old EOB results underestimated the tidal effects. The
accumulated phase difference was at most 1-3 radians in the last 15
cycles. The good performance of the EOB formalism has been recently shown
for several cases also by Hotokezaka et al. \cite{Hotokezaka2016}, who
compared the EOB inspiral waveform model to the ones obtained with the
currently available versions of the Taylor-T4 and Taylor-F2 approximants
(see, \eg Ref. \cite{Wade2014}). By computing the distinguishability
between numerical-relativity waveforms and post-Newtonian waveforms, it
appears that the Taylor approximants are worse by a factor of at most
two. The authors suggested that the absence of higher-order
post-Newtonian terms in the Taylor approximants is the likely source for
inaccuracy, especially in the late inspiral.

In another recent development, Hinderer et al. \cite{Hinderer2016}
proposed further improvements of the EOB formalism that include a
treatment of dynamical tides, in addition to the dominant adiabatic
tides. The latter describe distorted neutron stars as remaining in
hydrostatic equilibrium, namely, adjusting instantaneously to the
companion's tidal force which varies because of the orbital motion. The
former, instead, arise only when the frequency of the tidal forcing
becomes close to an eigenfrequency of the stellar normal modes of
oscillation and result in an enhanced and more complex tidal response
than in the adiabatic case. Hinderer et al. \cite{Hinderer2016} validated
their improved EOB gravitational waveforms by comparing with numerical
simulations and found that in some cases the contribution of dynamical
tides to the total deformation can be as large as $30\%$.

A post-Newtonian approach was also used for the determination of the
redshift of gravitational-wave signals from BNSs. Messenger et
al. \cite{Messenger:2011}, in particular, showed that the determination
of the redshift of gravitational waves can be attained solely from
gravitational-wave data, namely without assuming concurrent
electromagnetic observations. The degeneracy in post-Newtonian waveforms
between the mass parameters and the redshift was shown to be broken by
tidal effects, computing which it was estimated that for a range of
representative EOSs the redshift of BNS systems can be determined to an
accuracy of $8\% - 40\%$ for $z < 1$ and of $9\% - 65\%$ for $1 < z <
4$. Other ways to estimate the redshift of BNS sources involve the
post-merger signal \cite{Messenger2013} and will be treated in
Sect. \ref{sec:hydro_merger_post-merger}.

Several works are also using post-Newtonian approximation of waveforms to
investigate what can reasonably be deduced about EOSs from multiple
gravitational-wave observations of BNS inspirals with advanced detectors
\cite{Read:2009b, Hinderer09, Damour:2012, Read2013, DelPozzo2013,
  Wade2014, Lackey2015, Agathos2015}. Hinderer et al. \cite{Hinderer09}
and Damour et al. \cite{Damour:2012} performed Fisher-matrix calculations
(the latter using EOB waveforms) and found that it might be possible to
gain information about the EOSs with advanced detectors. An important
conclusion reached in Ref. \cite{Damour:2012} was that $\Lambda$ can be
measured at the 95\% confidence level through EOB-based merger templates
in match filtered searches by the Advanced LIGO-Virgo-KAGRA detector
network using gravitational-wave signals with a reasonable SNR (\ie
SNR=16) \cite{Damour:2012}.

Recently, Barkett et al. \cite{Barkett2016} proposed a new method for
computing inspiral waveforms for BNS systems by adding post-Newtonian
tidal effects to full numerical simulations of binary black holes. Since
black-hole vacuum simulations are faster than simulations containing
matter, this method would allow to produce gravitational waveforms
faster, but its accuracy is yet to be verified.

All of the results reviewed above referred to binaries that are initially
irrotational. However, as mentioned in Sect. \ref{sec:ID}, progress has
also been achieved recently in simulating the inspiral and merger of
spinning binaries, either from constraint-violating or
constraint-satisfying initial data. More precisely, the first simulations
of spinning BNS mergers were performed by Kastaun et
al. \cite{Kastaun2013}, who considered mergers of equal-mass binaries
with spins of different amplitude and aligned with the orbital angular
momentum. The ultimate goal of these simulations was to determine
whether, as the initial spin of the two stars is increased, the black
hole produced by the merger attains a maximum spin or it increases
indefinitely, leading to a naked singularity. Indeed, the simulations
indicated that it is the first of the two scenarios that takes place (see
Sect. \ref{sec:hydro_merger_post-merger} for more details) but also that
the inspiral lasts longer as the two neutron stars need to shed the
additional angular momentum in the system before merging. This
``hang-up'' effect has been observed in black-hole binaries
\cite{Campanelli:2006uy, Pollney:2007ss, Hannam:2007wf} and is a basic
prediction of the post-Newtonian equations. Nearly at the same time, also
Tsatsin and Marronetti \cite{Tsatsin2013} simulated binaries with spins
aligned and antialigned with the orbital angular momentum, using their
own approach to the construction of spinning initial data (see
Sect. \ref{sec:ID}). Also in this case, and in analogy with dynamics
already studied in binary black holes, the merger of the binary with
antialigned spins lead to a prompt collapse to a black hole, while the
aligned-spin binary lead to the formation of a centrifugally supported
HMNS that survived for several dynamical times before collapsing. Later,
also Kastaun and Galeazzi \cite{Kastaun2014} employed
constraint-violating spinning initial data to investigate how the
dynamical and gravitational-wave spectral properties are influenced by
the presence of rotation.

However, it was not until the work of Bernuzzi et al. \cite{Bernuzzi2013}
that general-relativistic evolutions of constraint-satisfying initial
data of spinning BNS with spins aligned and antialigned to the orbital
angular momentum were performed. The initial data was produced with the
constant-rotational-velocity approach of Tichy \cite{Tichy11,Tichy12},
which provides a more precise measurements of the changes in the phase
evolution during the inspiral due to spin. In particular, it was found
that the orbital motion can be significantly affected by spin-orbit
interactions which delay the merger, at least for high enough
spins. During their three-orbit evolution, they observed accumulated
phase differences of up to $0.7$ gravitational-wave cycles between the
irrotational configuration and the spinning ones. Hence they concluded
that precise modelling of the late-inspiral-merger waveforms needs to
include spin effects even for moderate magnitudes (see also Agathos et
al. \cite{Agathos2015} for the gravitational-wave data-analysis results
taking into account stellar spins). In addition, they studied the shift
to higher frequencies in the main emission mode of the HMNS produced by
spinning neutron stars and confirmed the results of Kastaun et
al. \cite{Kastaun2013} about the maximum spin attained by the black hole.
\lrn{While correct, all of the considerations above are hindered by the
  fact that a realistic estimate of the dimensionless spin parameter
  before the merger is not yet known and it is therefore difficult to
  assess whether spin effects are really going to be present in the
  gravitational-wave signal}.

\subsection{Post-merger dynamics}
\label{sec:hydro_merger_post-merger}

Research on the post-merger phase has been undergoing intense development
over the last few years because of its importance for linking numerical
simulations and astrophysical observations. The (early) post-merger is
also the phase in which most of the energy in gravitational waves is
emitted, as pointed out in Ref. \cite{Bernuzzi2015b}, even though the
gravitational waves emitted in this stage are not those that give the
largest signal-to-noise ratio, because their frequency range is not in
the best sensitivity zone of current interferometric detectors. The
numerical description of this stage is far more challenging than the
inspiral one because of the highly nonlinear dynamics and of the
development of strong, large-scale shocks that inevitably reduce the
convergence order, thus requiring far higher resolutions than the ones
normally employed. As a result, the accuracy of some quantities computed
after the merger is sometimes only marginal. The most notable example of
these quantities is the lifetime of the remnant \lrn{(be it an HMNS or an
  SMNS)} before its collapse to black hole; since this object is only in
metastable equilibrium, even small differences in resolution or even grid
setup are sufficient to change its dynamical behaviour, accelerating or
slowing down its collapse to a black hole. Fortunately, other quantities,
such as the spectral properties of the gravitational-wave post-merger
emission appear far more robust and insensitive to the numerical details;
we will discuss them later in this section.

Since the first general-relativistic simulations of BNS mergers, several
works have studied the nature (neutron star or black hole) of the objects
resulting from the mergers \cite{Shibata99d, Shibata02a, Shibata05c,
  Shibata06a, Yamamoto2008,Baiotti08, Anderson2008, Giacomazzo2011b}. It
is of course important to establish whether a black hole forms promptly
after the merger or instead an HMNS forms and lives for long times (more
than $0.1\,{\rm s}$), because the post-merger gravitational-wave signal
in the two cases is clearly different. Anderson et
al. \cite{Anderson2008} and Giacomazzo et al. \cite{Giacomazzo2011b}
started investigating the dependence of the lifetime of the HMNS on the
magnitude of the initial magnetic field in the case of magnetised
binaries. However, as mentioned above, such investigations are extremely
delicate since it is not straightforward to completely remove the
influence of numerical artefacts on the lifetime of the remnant even in
the absence of magnetic fields, at least with present resolutions.

In an alternative approach, Kaplan et al. \cite{Kaplan2013} have
investigated the role of thermal pressure support in hypermassive merger
remnants by computing sequences of axisymmetric uniformly and
differentially rotating equilibrium solutions to the general-relativistic
stellar structure equations and found that this too is a subtle issue:
the role of thermal effects on the stability and lifetime of a given
configuration depends sensitively and in a complicated way on its
details, like central or mean rest-mass density, temperature
distribution, degree of differential rotation and rotation rate.

Clearly, the issue of the precise lifetime of the binary-merger product
before it reaches its asymptotic state (\cf footnote \ref{footnote_bmp}),
especially when its equilibrium is mediated by the generation of magnetic
fields or radiative losses is far from being solved and will require
computational resources and/or methods not yet available (\cf Sect
\ref{sec:HD_MHD}).

Recently, Paschalidis, East and collaborators \cite{Paschalidis2015,
  East2016} pointed out that a one-arm spiral instability
\cite{Centrella:2001xp, Watts:2003nn, Baiotti06b, Corvino:2010} can
develop in HMNSs formed by dynamical-capture and that the $m=1$ mode
associated with this instability may become the dominant oscillation mode
if the HMNS persists for long enough\footnote{The $m=1$ mode had been
  studied previously together with the other modes, but it had never been
  found to become dominating (see, e.g., \cite{Dietrich:2015b}).}; this
instability has been subsequently studied also in quasi-circular BNSs
\cite{Radice2016a, Lehner2016a}. The instability, is reminiscent of the
shear instability that has been studied in detail for isolated stars
\cite{Baiotti06b, Corvino:2010, Camarda:2009mk, Franci2013b,
  Muhlberger2014} and seems to be correlated with the generation of
vortices near the surface of the HMNS that form due to shearing at the
stellar surface. These vortices then spiral in toward the center of the
star, creating an underdense region near the center. The growth of the
$m=1$ mode and so of the instability, could be related to the fact that
the maximum density does not reside at the center of mass of the star
\cite{Saijo2003}, or to the existence of a corotation band
\cite{Balbinski85b, Luyten:1990, Watts:2003nn}. The instability has an
imprint on the gravitational-wave signal, but the prospects of detection
are not encouraging, because of the small emitted power
\cite{Radice2016a}.

\subsubsection{Influence of the equation of state}

Despite these difficulties, many researchers have taken up the challenge
of studying the properties of the binary-merger product, because this may
give indications on the ultra-high density EOS, the origin of SGRBs, and
even the correct theory of gravity (see also Sects. \ref{sec:HD_MHD},
\ref{sec:atog}). In what follows we will focus in particular on the
determination of the EOS. While detectable differences between
simulations that employed different EOSs already appear during the
inspiral (see Sect. \ref{sec:EOB}), the post-merger phase depends more
markedly on the EOS \cite{Bauswein2014, Takami:2014, Bernuzzi2014,
  Takami2015, Bernuzzi2015, Bernuzzi2015a, Maione2016}. A note of caution
is necessary here to say that post-merger waveforms are at rather high
frequencies and thus probably only marginally measurable by detectors
like Advanced LIGO. Third-generation detectors, such has ET
\cite{Punturo:2010}, may provide the first realistic opportunity to use
gravitational waves to decipher the stellar structure and EOS
\cite{Andersson:2009yt}.

The first attempts to single out the influence of the EOS on the
post-merger dynamics were done in Refs. \cite{Shibata05c, Shibata06a,
  Yamamoto2008, Baiotti08}. These works focused mostly on the dynamics of
equal-mass binaries, as these are thought to be the most common
\cite{Oslowski2011} and are easier and faster to compute, since
symmetries of the configuration can be exploited to save computational
resources. The study of the effect of realistic EOSs in
general-relativistic simulations has been subsequently brought forward by
many groups. Kiuchi et al. \cite{Kiuchi2009} made use of the
Akmal-Pandharipande-Ravenhall (APR) EOS \cite{Akmal1998a}\footnote{The
  APR EOS, and many of the proposed EOSs, were later found to violate the
  light-speed constraint at very high densities and phenomenological
  constraints \cite{Taranto2013}, but no strong conclusions can be made
  to rule out such EOSs on this basis because the constraints themselves
  are affected by errors.}. This nuclear-physics EOS describes matter at
zero temperature and so during the simulation it needs to be combined
with a ``thermal'' part that accounts for the energy increase due to
shock heating (this is mostly done through the addition of an ideal-fluid
part to the EOS; see Ref. \cite{Rezzolla_book:2013} for a
discussion). The resulting \emph{``hybrid EOS''} appears to be
appropriate for studying the inspiral and merger, but may not be
satisfactory for studying the remnant formation and the evolution of the
accretion disc around the formed black hole, because for such cases,
effects associated with the thermal energy (finite temperature), neutrino
cooling, and magnetic fields are likely to play an important role (see
Sects. \ref{sec:HD_MHD}, \ref{sec:hd_nus}). In another work of the same
group \cite{Hotokezaka2011}, the dependence of the dynamical behavior of
BNS mergers on the EOS of the nuclear-density matter with
piecewise-polytropic EOSs \cite{Read:2009a} was studied. Table I in
Ref. \cite{Hotokezaka2011} usefully summarizes the piecewise polytropic
parameters of several realistic EOSs.

One family of EOSs that has received special attention in the past years
is that describing strange matter, namely matter containing hyperons,
which are nucleons containing strange quarks. The strange-matter
hypothesis \cite{Witten84} considers the possibility that the absolute
ground state of matter might not be formed by iron nuclei but by strange
quark matter: a mixture of up, down, and strange quarks. This hypothesis
introduced the possibility that compact stars could be stars made also of
strange-quark matter, or strange stars \cite{Haensel86,Alcock86}. One of
the astrophysical consequences of this is the possibility that collision
events of two strange stars lead to the ejection of strangelets, namely
small lumps of strange quark matter.

Although the occurrence of hyperons at very large nuclear densities is
rather natural, hyperonic EOSs are generally very soft and currently
disfavoured by the observation of a $2 M_\odot$ star
\cite{Antoniadis2013, Demorest2010}, which they can hardly reproduce,
except by fine tuning of the parameters (see, \eg \cite{Alford2005,
  RikovskaStone:2006ta, Weissenborn:2011qu}). This basic inconsistency
between the expectations of many nuclear physicists and the observational
evidence of very massive neutron stars is normally referred to as the
``hyperon puzzle''; those supporting the use of hyperonic EOSs also state
that the existence of exotic phases in strange stars remains
unconstrained and could lead to higher masses
\cite{Bhowmick:2014pma}. Additional work is needed to settle this
``hyperon puzzle'' and we will present results on strange-star
simulations setting these doubts aside.

The first investigations of binary strange stars were those of Bauswein
et al. \cite{Bauswein2009,Bauswein2010}, who employed the MIT bag model
\cite{Farhi1984}. Within this model, quarks are considered as a free or
weakly interacting Fermi gas and the nonperturbative QCD interaction is
simulated by a finite pressure of the vacuum, the {\it bag constant} $B$.
Three-dimensional general-relativistic simulations with conformally flat
gravity of the coalescence of strange stars were performed and the
possibility to discriminate on the strange matter hypothesis by means of
gravitational-wave measurements was explored. The dynamics of mergers of
strange stars, which are usually more compact, is different from those of
neutron-star mergers, most notably in the tidal disruption during the
merger. Furthermore, instead of forming dilute halo-structures around the
binary-merger product, as in the case of neutron-star mergers, the
coalescence of strange stars results in a differentially rotating
hypermassive object with a sharp surface layer surrounded by a
geometrically thin, clumpy high-density strange-quark-matter disc. It was
found that in some cases (some types of EOS and stellar properties) the
analysis of the gravitational-wave signals emitted by strange-star
mergers showed that it may be possible to discern whether strange-star or
neutron-star mergers produced the emission. In particular, it was found
that the maximal frequency during the inspiral and the frequency of the
oscillations of the post-merger remnant are in general higher for
strange-star mergers than for neutron-star mergers. In other cases,
however, there remains a degeneracy among different models, and a
conclusion about the strange-matter hypothesis could be reached only if
other types of observations (\eg of cosmic rays) were available.

Strange-matter EOSs were later studied with a fully general-relativistic
code in a series of articles by Sekiguchi, Kiuchi and collaborators
\cite{Sekiguchi2011b, Kiuchi2012, Sekiguchi2012, Kiuchi2012a}, who showed
results of simulations performed by incorporating both nucleonic and
hyperonic finite-temperature EOSs (and neutrino cooling as well, \cf
Sect. \ref{sec:hd_nus}). It was found that also for the hyperonic EOS, an
HMNS is first formed after the merger and subsequently collapses to a
black hole. The radius of such an HMNS decreases in time because of the
increase of the mass fraction of hyperons and the consequent decrease in
pressure support. Such a shrinking is noticeably larger than the one
simply due to angular-momentum loss through gravitational-wave emission
that is present also in nucleonic EOSs. These differences in the dynamics
are clearly visible in the gravitational-wave signal, whose
characteristic peak frequency has an increase of $20\%-30\%$ during the
HMNS evolution. By contrast, for nucleonic EOSs, the peak
gravitational-wave frequency in the HMNS phase is approximately constant
on the timescales considered. It was also stressed that these results
raise a warning about using the peak frequency of the gravitational-wave
spectrum to extract information of the neutron-star matter (see below),
because it may evolve and so make the relation of the peak frequency with
the HMNS structure ambiguous. Finally, it was found that the torus mass
for the hyperonic EOS is smaller than that for nucleonic EOSs, thus
making hyperonic EOSs less favourable for the description of SGRBs.

\begin{figure*}
\begin{center}
\includegraphics[width=0.60\columnwidth]{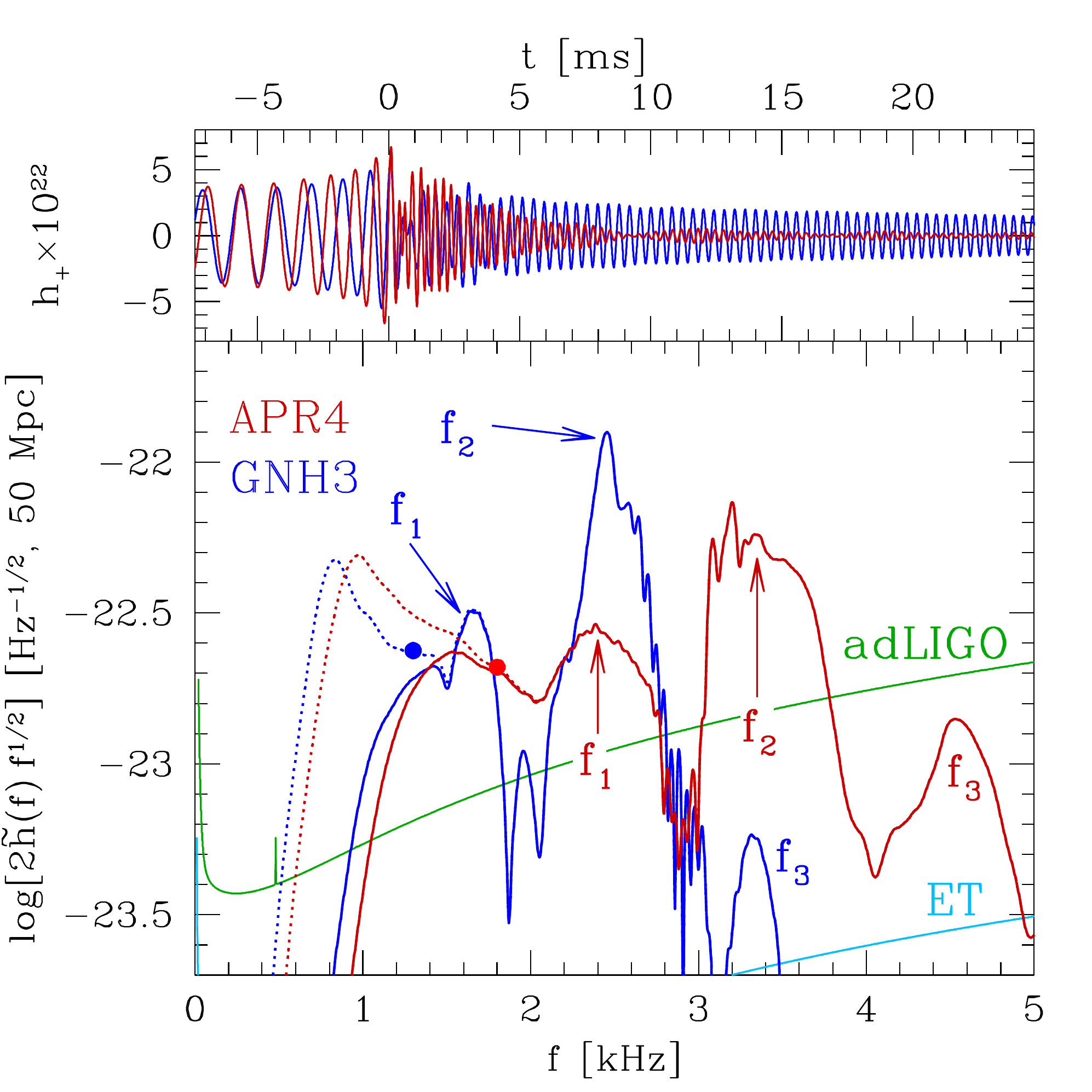}
\caption{Top sub-panel: evolution of $h_+$ for binaries with the APR4 and
  GNH3 EOSs (dark-red and blue lines, respectively) for sources at a
  polar distance of 50 Mpc. Bottom sub-panel: spectral density $2
  \tilde{h}(f) f^{1/2}$ windowed after the merger for the two EOSs and
  sensitivity curves of Advanced LIGO (green line) and ET (light-blue
  line); the dotted lines show the power in the inspiral, while the
  circles mark the contact frequency [Reprinted with permission from
    Ref. \protect{} \cite{Takami:2014}. \copyright~(2014) by the American
    Physical Society.]}
\label{fig:Takami2014_Fig1}
\end{center}
\end{figure*}

\begin{figure*}
\begin{center}
\includegraphics[width=0.90\columnwidth]{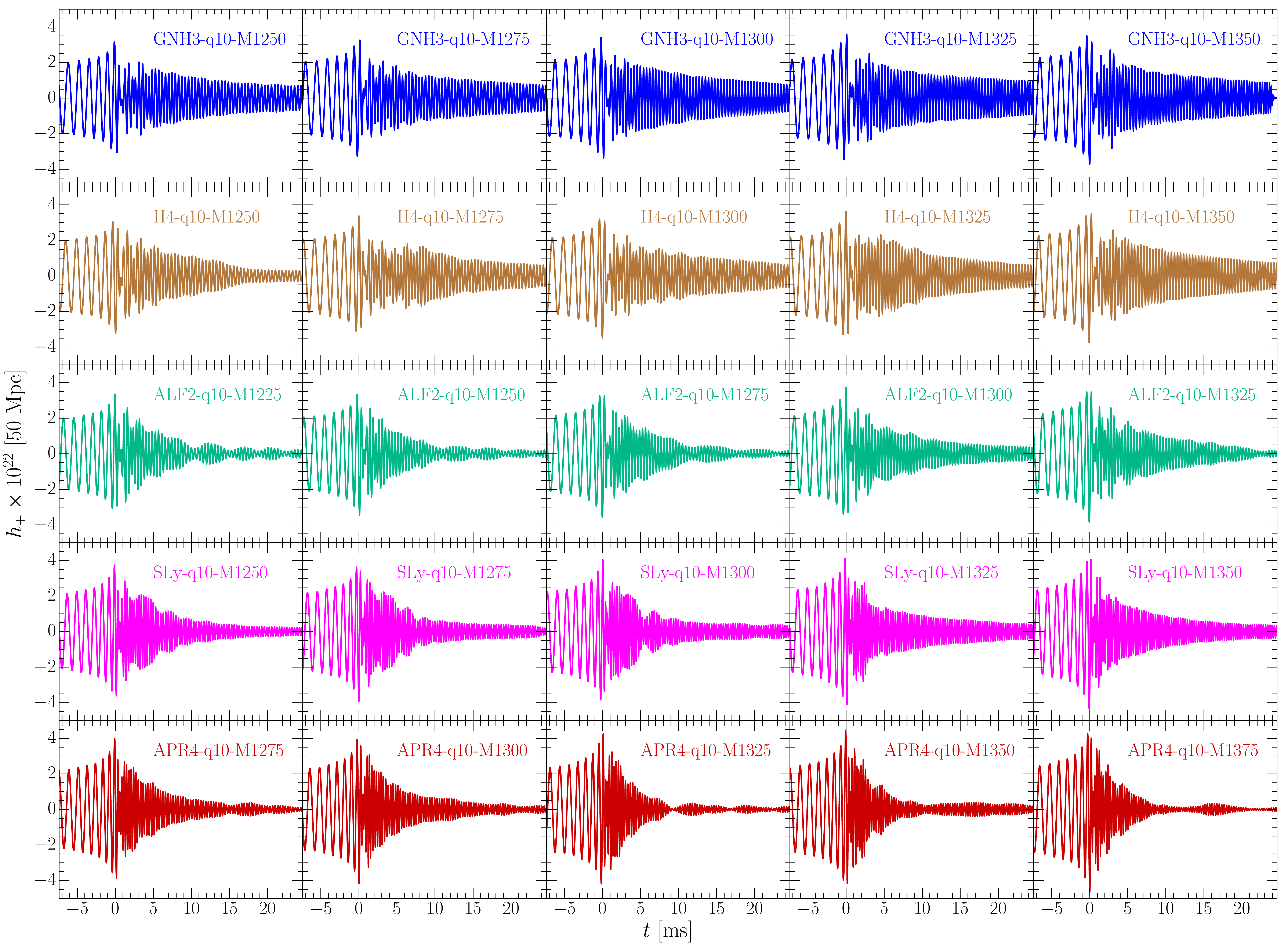}
\caption{Gravitational waveforms for equal-mass binaries with nucleonic
  EOSs, shown in different colours. Each column refers to a given initial
  gravitational mass. [Reprinted with permission from Ref. \protect{}
    \cite{Takami2015}. \copyright~(2015) by the American Physical
    Society.]}
\label{fig:Takami2015_Fig7a}
\end{center}
\end{figure*}

\subsubsection{Spectral properties of the gravitational-wave signal}

In addition to simulating BNS mergers with various EOSs, it is important
to find ways to connect future gravitational-wave observation with the
EOS of the neutron stars. Recently there have been several suggestions on
how to achieve this, based either on the signature represented by the
tidal corrections to the orbital phase or on the power spectral density
(PSD) of the post-merger gravitational waveforms or on the frequency
evolution of the same. The first approach, described in
Sect. \ref{sec:EOB}, is reasonably well understood analytically
\cite{Flanagan08, Baiotti:2010, Bernuzzi2012, Read2013, Bernuzzi2015a}
and can be tracked accurately with advanced high-order numerical codes
\cite{Radice2013b,Radice2013c}. Here we describe works on the post-merger
approach in some detail.

Hotokezaka et al. \cite{Hotokezaka2013c} used their adaptive
mesh-refinement (AMR) code\footnote{In previous works by this group,
  described above, a different code with a uniform grid had been used.}
\texttt{SACRA} \cite{Yamamoto2008} to perform a large number of
simulations with a variety of mass ranges and EOSs (as done before,
approximate finite-temperature effects were added to the cold EOSs
through an additional ideal-fluid term), in order to find universal
features of the frequency evolution of gravitational waves emitted by the
HMNS formed after the merger. In their analysis they found it convenient
to decompose the merger and post-merger gravitational-wave emission in
four different parts: (i) a peak in frequency and amplitude soon after
the merger starts; (ii) a decrease in amplitude during the merger and a
new increase when the HMNS forms; (iii) a damped oscillation of the
frequency during the HMNS phase lasting for several oscillation periods
and eventually settling to an approximately constant value (although a
long-term secular change associated with the change of the state of the
HMNSs is always present); (iv) a final decrease in the amplitude during
the HMNS phase, either monotonical or with modulations. Based on this,
they find an optimal 13-parameters fitting function, using which it may
be possible to constrain the neutron star radius with errors of about $1
\,{\rm km}$ \cite{Hotokezaka2013c}.

In contrast with this multi-stage, multi-parameter description of
Hotokezaka et al. \cite{Hotokezaka2013c}, other groups have concentrated
on the analysis of the full PSD of the post-merger signal, isolating
those spectral features (\ie peaks) that could be used to constrain the
properties of the nuclear-physics EOSs. As a reference, we show in
Fig. \ref{fig:Takami2014_Fig1} the PSDs of some representative
gravitational waves when compared with the sensitivity curves of current
and future gravitational-wave detectors \lrn{\cite{Takami:2014}}. More
specifically, two examples are presented in
Fig. \ref{fig:Takami2014_Fig1}, which refers to two equal-mass binaries
with APR4 and GNH3 EOSs, and with individual gravitational masses at
infinite separation of $\bar{M}/M_{\odot}=1.325$, where $\bar{M}$ is the
average of the initial gravitational mass of the two stars. The top
sub-panel shows the evolution of the $\ell=m=2$ plus polarization of the
strain ($h_+ \sim h_{+}^{22}$), aligned at the merger for sources at a
polar distance of 50 Mpc (dark-red and blue lines for the APR4 and GNH3
EOSs, respectively). The bottom panel, on the other hand, shows the
spectral densities $2 \tilde{h}(f) f^{1/2}$ windowed after the merger for
the two EOSs, comparing them with the sensitivity curves of Advanced LIGO
\cite{url:adLIGO_Sh_curve} (green line) and of the Einstein Telescope
\cite{Punturo2010b, Sathyaprakash:2009xs} (ET; light-blue line). The
dotted lines refer to the whole time series and hence, where visible,
indicate the power during \lrn{the last phase of} the inspiral, while the
circles mark the ``contact frequency'' $f_{\rm
  cont}=\mathcal{C}^{3/2}/(2\pi \bar{M})$ \cite{Damour:2012}, where
$\mathcal{C} \coloneqq \bar{M}/\bar{R}$ is the average compactness,
$\bar{R} \coloneqq (R_1+R_2)/2$, and $R_{1,2}$ are the radii of the
nonrotating stars associated with each binary.

\begin{figure*}
\begin{center}
\includegraphics[width=0.90\columnwidth]{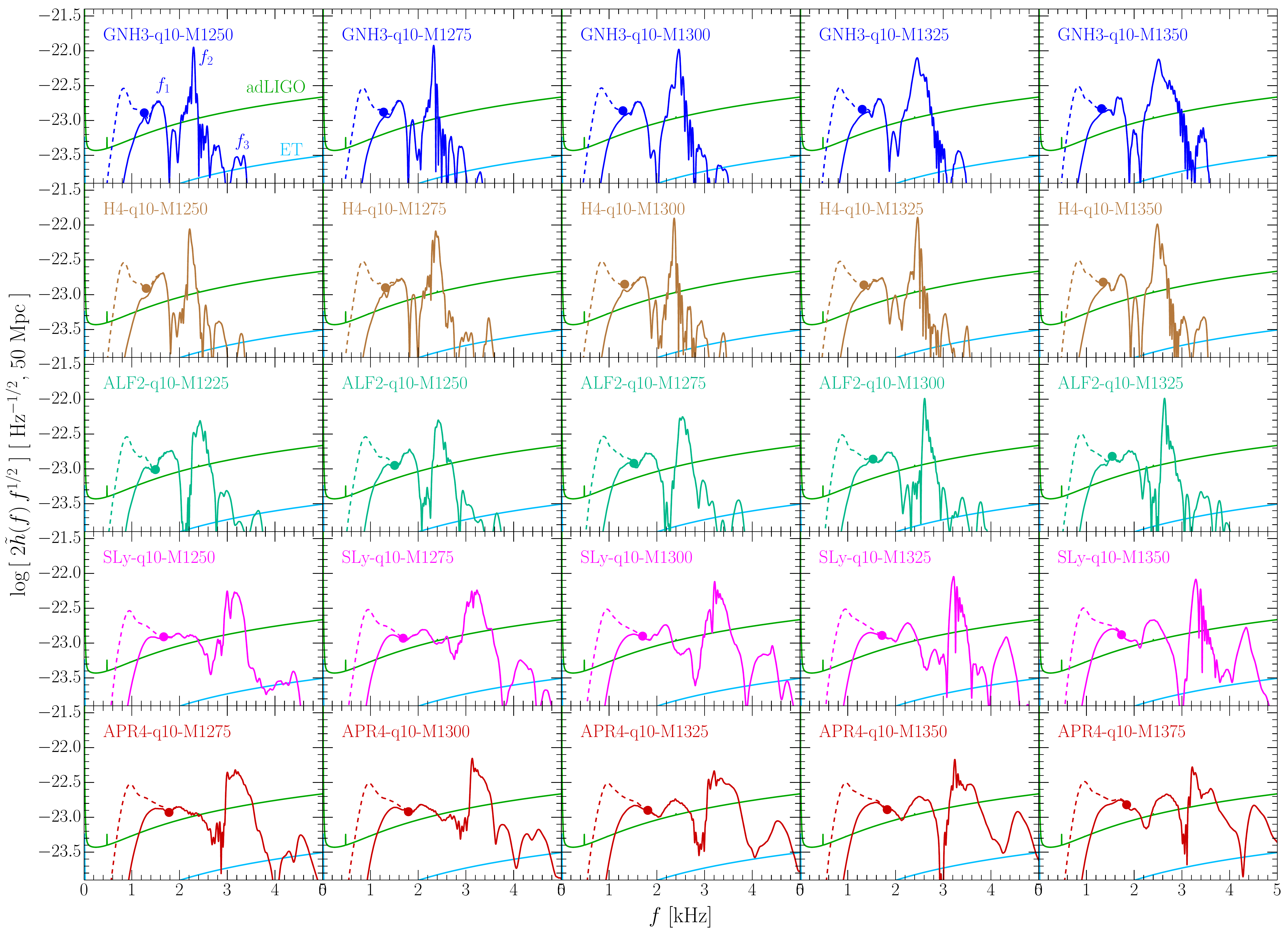}
\caption{PSDs $2 \tilde{h}(f) f^{1/2}$ for the equal-mass binaries with
  nucleonic EOSs shown in Fig. \ref{fig:Takami2015_Fig7a}. Solid
  lines of different colours refer to the high-passed waveforms, while
  the dashed lines refer to the full waveforms. Indicated with coloured
  circles are the various contact frequencies $f_{\rm cont}$, while the
  curves of Advanced LIGO and ET are shown as green and light-blue lines,
  respectively [Reprinted with permission
    from Ref. \protect{} \cite{Takami2015}. \copyright~(2015) by the
    American Physical Society.]}
\label{fig:Takami2015_Fig7b}
\end{center}
\end{figure*}

Note that besides the peak at low frequencies corresponding to the
inspiral (\cf dashed lines), there is one prominent peak and several
other of lower amplitudes. These are related to the oscillations of the
HMNS and would be absent or much smaller if a black hole forms promptly,
in which case the gravitational-wave signal would terminate abruptly with
a cutoff corresponding to the fundamental quasi-normal-mode frequency of
the black hole \cite{Kokkotas99a}. The behaviour summarised in
Fig. \ref{fig:Takami2014_Fig1} is indeed quite robust and has been
investigated by number of authors over the last decade \cite{Oechslin07b,
  Stergioulas2011b, Bauswein2012a, Bauswein2012, Hotokezaka2013c,
  Bauswein2014, Takami:2014, Clark2014, Kastaun2014, Takami2015,
  Bernuzzi2015a, Bauswein2015, Dietrich2015, Foucart2015, DePietri2016,
  Bauswein2015b, Rezzolla2016, Maione2016}. While some of the details of
the picture that emerges from the analysis of the PSDs are still fuzzy,
there are also aspects that are commonly accepted. In particular, there
is a widespread consensus that: (i) the post-merger gravitational-wave
signal possesses spectral features that are robust and that emerge
irrespective of the EOS or the mass ratio; (ii) the frequencies of the
peaks in the post-merger PSD can be used to obtain important information
on the stellar properties (\ie mass and radius) and hence represent a
very good proxy to deduce the EOS. This is summarised in
Fig. \ref{fig:Takami2015_Fig7b}, which shows the PSDs for the equal-mass
binaries with nuclear-physics EOSs reported in
Fig. \ref{fig:Takami2015_Fig7a}. Solid lines of different colours refer
to the high-passed waveforms, while the dashed lines refer to the full
waveforms. Indicated with coloured circles are the various contact
frequencies $f_{\rm cont}$, while the curves of Advanced LIGO and ET are
shown as green and light-blue lines, respectively \cite{Takami2015}.

The first detailed description of a method for extracting information
about the EOS of nuclear matter by carefully investigating the spectral
properties of the post-merger signal was provided by Bauswein et
al. \cite{Bauswein2012a, Bauswein2012}. After performing a large number
of simulations using their conformally-flat SPH code, they pointed out
that the largest peak in the PSD (whose frequency is dubbed there $f_{\rm
  peak}$) correlates with the properties of the EOS, \eg with the radius
of the maximum-mass nonrotating star for the given EOS. The correlation
found was rather tight, but this was partly due to the fact that their
sample was restricted to binaries having all the same total mass (\ie
$2.7\,M_{\odot}$ in the specific case). It was shown that such a
correlation can be used to gain information on the high-density EOS
through gravitational-wave measurements, if the masses of the neutron
stars forming the binaries are known. Additionally, it was recognized
that $f_{\rm peak}$ corresponds to a fundamental fluid mode of the HMNS
with $\ell=2=m$ \cite{Bauswein2012a, Stergioulas2011b} and that the value
of this frequency could also be used to set constraints on the maximum
mass of the system and hence on the EOS \cite{Bauswein2013,
  Bauswein2014}. Subsequent analyses were performed by a number of groups
with general-relativistic codes \cite{Hotokezaka2013c, Takami:2014,
  Takami2015, Dietrich2015, Foucart2015, DePietri2016, Rezzolla2016,
  Maione2016}, which confirmed that the conformally flat approximation
employed by Bauswein and collaborators provided a rather accurate
estimate of the largest peak frequencies in the PSDs.

Takami et al. \cite{Takami:2014, Takami2015} presented a more advanced
method to use detected gravitational waves for determining the EOS of
matter in neutron stars. They used the results of a large number of
accurate numerical-relativity simulations of binaries with different EOSs
and different masses and identified two distinct and robust main spectral
features in the post-merger phase. The first one is the largest peak in
the PSD (whose frequency was called there $f_2$ and essentially coincides
with the $f_{\rm peak}$ of Refs. \cite{Bauswein2012a, Bauswein2012}). The
functions describing the correlations of $f_2$ with the stellar
properties (\eg with the quantity $(\bar{M}/R^3_{\mathrm{max}})^{1/2}$,
where $R_{\mathrm{max}}$ is the radius of the maximum-mass nonrotating
star), which were first proposed by Bauswein et al. \cite{Bauswein2012a},
are not universal, in the sense that different (linear) fits are
necessary for describing the $f_2$-correlations for binaries with
different total masses. This conclusion can be evinced by looking at
Figs. 22--24 of Bauswein et al. \cite{Bauswein2012}, but the different
linear correlations were first explicitly computed by Takami et
al. \cite{Takami:2014, Takami2015} (see also
Ref. \cite{Hotokezaka2013c}).

The second feature identified in all PSDs analysed by Takami et
al. \cite{Takami:2014, Takami2015} is the second-largest peak, which
appears at lower frequencies and was called $f_1$ there. Clear
indications were given about this low-frequency peak being related to the
merger process (\ie the first $\approx 3$ ms after the merger). This was
done by showing that the power in the peak is greatly diminished if the
first few ms after the merger were removed from the waveform. Furthermore,
a simple mechanical toy model was devised that can explain rather
intuitively the main spectral features of the post-merger signal and
therefore shed light on the physical interpretation of the origin of the
various peaks. Despite its crudeness, the toy model was even able to
reproduce the complex waveforms emitted right after the merger, hence
possibly opening the way to an analytical modelling of a part of the
signal \cite{Takami2015}.

\begin{figure*}
\begin{center}
\includegraphics[width=0.90\columnwidth]{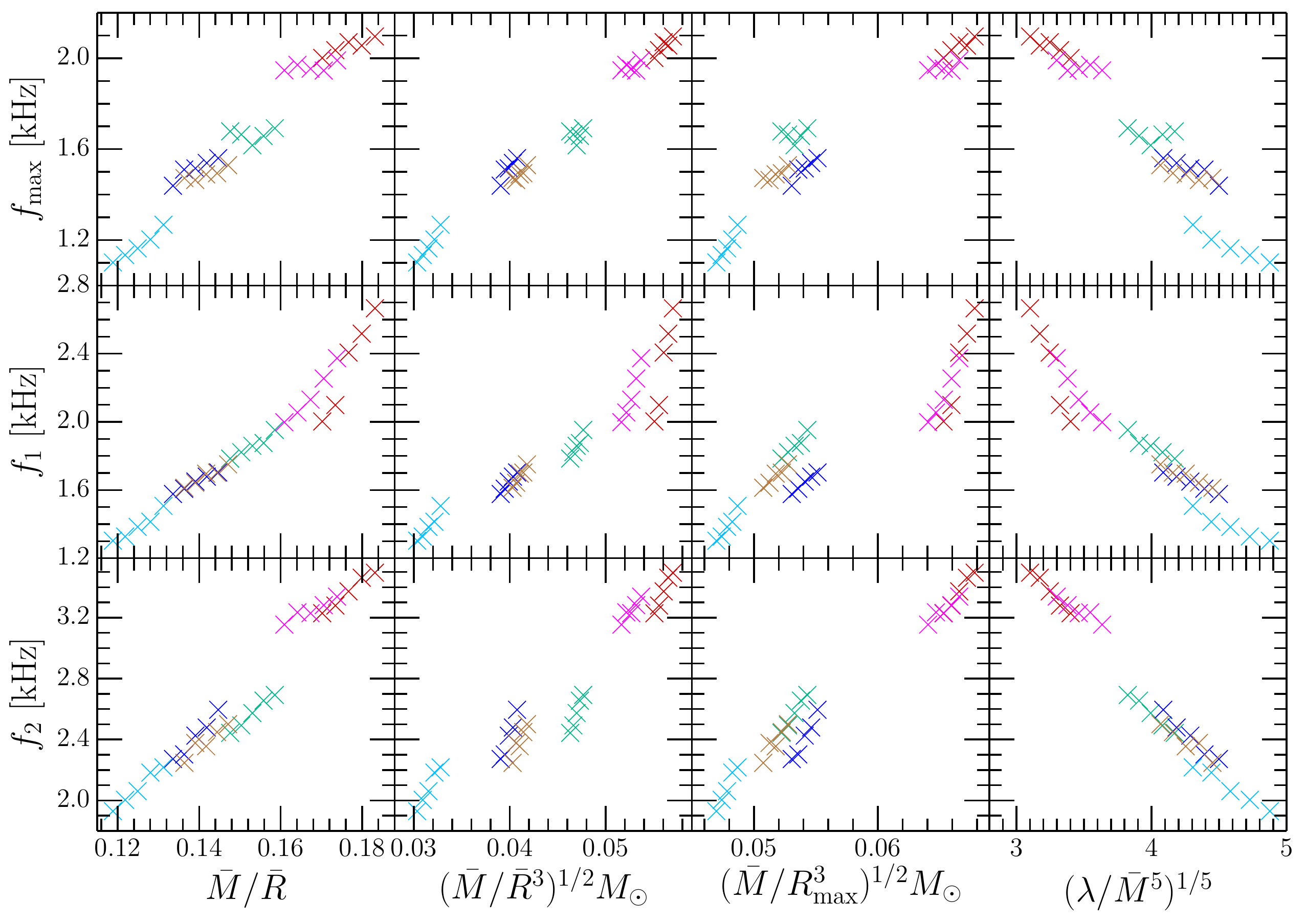}
\caption{Empirical correlations between the $f_{\rm max}, f_1$ and $f_2$
  frequencies and the physical quantities of the binary
  system. [Reprinted with permission from Ref. \protect{}
    \cite{Takami2015}. \copyright~(2015) by the American Physical
    Society.]}
\label{fig:Takami2015_Fig7c}
\end{center}
\end{figure*}

\lrn{More importantly, it was shown that the potential measurement of the
  $f_1$ frequency could reveal the EOS of the merging objects, since a
  correlation was found between the $f_1$-frequency and the average
  compactness of the two stars in the binary. Interestingly, this
  relation appears to be universal, that is, essentially valid for all
  EOSs and masses, and could therefore provide a powerful tool to set
  tight constraints on the EOS \cite{Takami2014, Takami2015}.} Indeed, an
analytic expression was suggested in Ref. \cite{Takami2015} to express
the $f_1$ frequency via a third-order polynomial of the (average) stellar
compactness, which reproduces reasonably well the numerical results. In
addition to the correlations described above, Takami et
al. \cite{Takami:2014, Takami2015} also discussed additional correlations
(24 in all), some of which had been already presented in the literature,
\eg in Refs. \cite{Read2013, Bernuzzi2014}, and some of which are
presented there for the first time. Examples of these correlations are
reported in Fig. \ref{fig:Takami2015_Fig7c} and refer to the $f_{\rm
  max}, f_1$ and $f_2$ frequencies and the physical quantities of the
binary system, \eg the average compactness $\bar{M}/\bar{R}$, the average
density $(\bar{M}/\bar{R}^3)^{1/2}$, the pseudo-average rest-mass density
$(\bar{M}/R^3_{\rm max})^{1/2}$, or the dimensionless tidal deformability
$(\lambda/\bar{M}^5)^{1/5}$ (\cf also Fig. 15 of
Ref. \cite{Takami2015}). In confirmation of the accuracy of the computed
frequencies, very similar values for the $f_1$ frequencies were also
found by Dietrich et al. \cite{Dietrich2015} in a distinct work aimed at
determining the impact that conservative mesh-refinement techniques have
on the accuracy of the post-merger dynamics.

Even though the toy model proposed by Takami et al. \cite{Takami2015}
provides a simple and convincing explanation of the power associated to
the $f_1$ frequency peak, alternative interpretations of the
low-frequency part of the PSD have also been suggested. More
specifically, Bauswein et al. \cite{Bauswein2015} claimed that the
lower-frequency peak (\ie the $f_1$ peak in Refs. \cite{Takami:2014,
  Kastaun2014, Takami2015}) is actually made of two separate peaks
originating from \lrn{different} processes. One of these peaks is said to
be produced by a nonlinear combination between the dominant quadrupolar
oscillation ($f_{\rm peak}$ or $f_2$) and the quasi-radial oscillation of
the remnant and is named $f_{20}$ \cite{Stergioulas2011b}, while the
other is said to be caused by a strong deformation initiated at the time
of the merger, the pattern of which then rotates (in the inertial frame)
slower than the inner cores of the remnant and lasts for a few rotational
periods, while diminishing in amplitude. The gravitational-wave emission
associated with this motion then powers a peak that was named $f_{\rm
  spiral}$ in Ref. \cite{Bauswein2015}. The connection between the
$f_{\rm spiral}$ peak and the deformation was supported by showing that
only PSDs computed from time intervals of the gravitational waveform that
contain the deformation have the $f_{\rm spiral}$ peak. It was also
claimed that the $f_{\rm spiral}$ peak can be roughly reproduced in a toy
model, where two bulges orbit as point particles around the central
double-core structure for a duration of few milliseconds, but no details
were given in Ref. \cite{Bauswein2015}.

In their analysis, Bauswein et al. \cite{Bauswein2015} also proposed an
explanation for the low-frequency modulations seen in quantities like the
lapse function at the stellar center, the maximum rest-mass density, and
the separation between the two cores of the remnant. Such quantities are
modulated according to the orientation of the antipodal bulges of the
deformation with respect to the double central cores: the compactness is
smaller, the central lapse function larger, and the gravitational-wave
amplitude maximal when the bulges and the cores are aligned, and
viceversa.

Making use of a large set of simulations, Bauswein et
al. \cite{Bauswein2015} were able to obtain empirical relations for both
types of low-frequency peaks in terms of the compactness of nonrotating
individual neutron stars. Different relations, however, were found for
different sequences of constant total mass of the binary, in contrast
with what found in Takami et al. \cite{Takami:2014, Takami2015}, where a
different definition for the low-frequency peak was used. As discussed by
Bauswein et al. \cite{Bauswein2015}, the different behaviour could be due
to the fact that the results of Takami et al. \cite{Takami:2014,
  Takami2015} were based on a limited set of five EOSs of soft or
moderate stiffness (with corresponding maximum masses of nonrotating
neutron stars only up to $2.2\,M_\odot$), as well as on different chosen
mass ranges for each EOS with a spread of only $0.2\,M_\odot$ in the
total mass of the binary. In Bauswein et al. \cite{Bauswein2015}, on the
other hand, ten EOSs (including stiff EOSs with maximum masses reaching
up to $2.8\,M_\odot$) and a larger mass range of $2.4-3.0\,M_\odot$ were
used. Overall, the differences between the results of the two groups are
significant only for very low-mass neutron stars (\ie
$M=1.2\,M_{\odot}$), which Takami et al. \cite{Takami:2014,Takami2015}
had not included in their sample because of the low statistical incidence
they are thought to have (see also below).

One important consideration to bear in mind is that measuring the $f_{\rm
  spiral}$ frequencies through the motion of matter asymmetries via
gauge-dependent quantities such as the rest-mass density is essentially
impossible in genuine numerical-relativity calculations. This is because
the spatial gauge conditions can easily distort the coordinate appearance
of mass distributions and even the trajectories of the two stars during
the inspiral (see Appendix A 2 of Ref. \cite{Baiotti08} for some dramatic
examples). In an attempt to clarify the different interpretations
suggested in Refs. \cite{Takami2014, Takami2015, Bauswein2015} and to
bring under a unified framework the spectral properties of the
post-merger gravitational-wave signal, Rezzolla and Takami
\cite{Rezzolla2016} have recently presented a comprehensive analysis of
the gravitational-wave signal emitted during the inspiral, merger and
post-merger of 56 neutron-star binaries \cite{Rezzolla2016}. This sample
of binaries, arguably the largest studied to date with realistic EOSs,
spans across five different nuclear-physics EOSs and seven mass values,
including the very low-mass binaries (\eg with individual neutron-star
masses of $1.2\,M_{\odot}$) that were suggested by
Ref. \cite{Bauswein2015} to be lacking in the previous analysis of
Ref. \cite{Takami2015}. After a systematic analysis of the complete
sample, it was possible to sharpen a number of arguments on the spectral
properties of the post-merger gravitational-wave signal. Overall it has
found that: (i) for binaries with individual stellar masses differing no
more than $20\%$, the frequency at the maximum of the gravitational-wave
amplitude is related quasi-universally with the tidal deformability of
the two stars; (ii) the spectral properties vary during the post-merger
phase, with a transient phase lasting a few milliseconds after the merger
and followed by a quasi-stationary phase; (iii) when distinguishing the
spectral peaks between these two phases, a number of ambiguities in the
identification of the peaks disappear, leaving a simple and robust
picture; (iv) using properly identified frequencies, quasi-universal
relations are found between the spectral features and the properties of
the neutron stars; (v) for the most salient peaks analytic fitting
functions can be obtained in terms of the stellar tidal deformability or
compactness. Overall, the analysis of Rezzolla and Takami
\cite{Rezzolla2016} supports the idea that the EOS of nuclear matter can
be constrained tightly when a signal in gravitational waves from BNSs is
detected.

\begin{figure*}
\begin{center}
  \includegraphics[width=0.90\columnwidth]{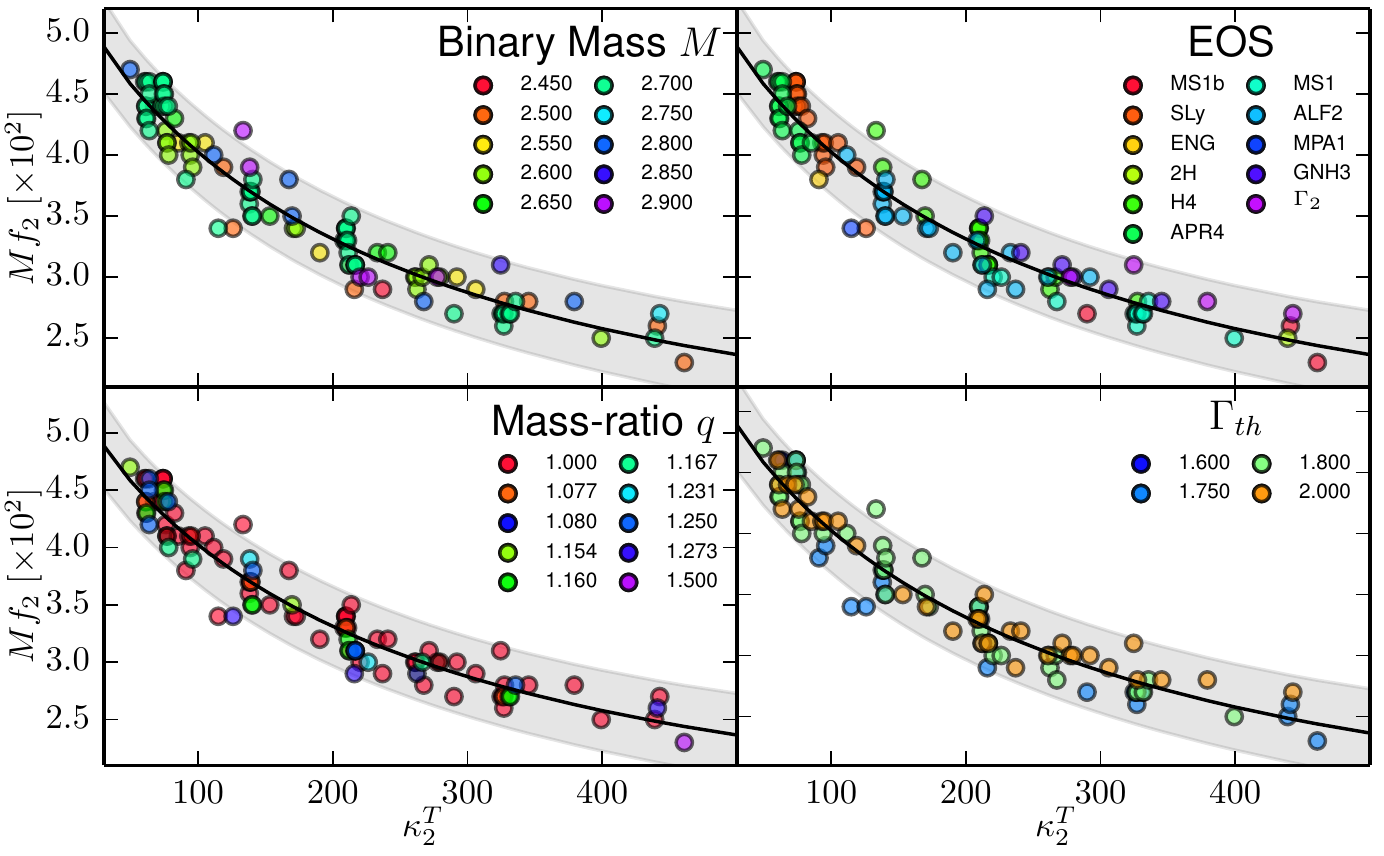}
\end{center}
   \caption{Dimensionless frequency ($Mf_2$) as a function of the tidal
     coupling constant $\kappa_2^T$. Each panel shows the same data set;
     the colour code in each panel indicates the different values of
     binary mass (top left), EOS (top right), mass-ratio (bottom left),
     and $\Gamma_\text{th}$ (bottom right). The black solid line is the
     fit, while the grey area marks the 95\% confidence
     interval. [Reprinted with permission from Ref. \protect{}
       \cite{Bernuzzi2015a}. \copyright~(2015) by the American Physical
       Society.]}
   \label{fig:Bernuzzi2015a}
\end{figure*}

An interesting extension of the work of Takami et
al. \cite{Takami:2014,Takami2015} was suggested by Bernuzzi et
al. \cite{Bernuzzi2015a}, who expressed the correlation between the peak
frequencies $f_2$ with the tidal coupling constant $\kappa_2^T$ instead
of the tidal deformability parameter $\Lambda$, as done in
Refs. \cite{Takami:2014, Takami2015}. As found in previous works by
Bernuzzi et al. \cite{Bernuzzi2014, Bernuzzi2015} (see
Sect. \ref{sec:EOB}), the dimensionless gravitational-wave frequency
depends on the stellar EOS, binary mass, and mass ratio only through the
tidal coupling constants $\kappa_2^T$ and thus this is a better choice of
parameter, also because it can be extended more straightforwardly to the
case of unequal-mass binaries. The relation $f_2(\kappa_2^T)$ was found
in Ref. \cite{Bernuzzi2015a} to be very weakly dependent on the binary
total mass, mass ratio, EOS, and thermal effects (through the ideal-fluid
index $\Gamma_\text{th}$). Relevant dependence on the stellar spins was
instead found. This is shown in Fig. \ref{fig:Bernuzzi2015a}, which
reports the dimensionless frequency $Mf_2$ as a function of the tidal
coupling constant $\kappa_2^T$. Each panel shows the same data set; the
colour code in each panel indicates the different values of binary mass
(top left), EOS (top right), mass-ratio (bottom left), and
$\Gamma_\text{th}$ (bottom right). The black solid line is the fit
obtained by Bernuzzi et al. \cite{Bernuzzi2015a}, while the grey area
marks the 95\% confidence range.

Although not explicitly stated, all of the considerations made so far
about the spectral properties of the post-merger signal refer to binaries
that are initially irrotational. It is therefore natural to ask how does
the dynamics change, both qualitatively and quantitatively, when spinning
binaries are considered (a discussion on the modifications of the
post-merger dynamics introduced by magnetic fields will be discussed in
Sect. \ref{sec:m_pmd}). This was done in part by Bernuzzi et
al. \cite{Bernuzzi2013} and by Kastaun and Galeazzi \cite{Kastaun2014}.
The first work considered in particular whether the main-peak frequency
$f_2$ is influenced by the initial state of rotation and found that this
is indeed the case at least for very rapidly rotating neutron stars,
suggesting that spin effects may be more important than found by
Ref. \cite{Bauswein2012a}. Kastaun and Galeazzi \cite{Kastaun2014}, on
the other hand, analysed the spectral changes induced by the initial spin
and showed that the low-frequency peak $f_1$ is due to the
gravitational-wave signal during the plunge and the first bounces. They
also studied in detail the Fourier decomposition of the rest-mass density
of the binary-merger product and its rotational profile, which is
important for determining its lifetime, especially in view of the
amplification of the magnetic field. A problem that needed to be tackled
in their analysis is that of potential gauge artefacts. We recall, in
fact, that rest-mass density distributions are gauge-dependent quantities
and even when the system approaches an axisymmetric state after the
merger, the spatial coordinates may not reflect this, because the gauge
conditions employed in the evolution introduce local and global
deformations. In order to exclude such systematic gauge effects, Kastaun
and Galeazzi \cite{Kastaun2014} introduced a different coordinate system,
used just for post-processing. In this new coordinate system, they found
that, the Fourier decomposition is far more regular than in the
coordinate system normally used in the evolution. Furthermore, they
showed that, contrary to common assumptions, the binary-merger product
law of differential rotation consists of a slowly rotating core with an
extended and massive envelope rotating close to Keplerian velocity. The
latter result has been confirmed recently also for the binary-merger
product produced by the merger of unequal-mass magnetised binaries
\cite{Endrizzi2016}.

Before concluding this discussion on the post-merger gravitational-wave
signal we should also make two important remarks. The first one is that
gravitational-wave measurements at the expected frequencies and
amplitudes are very difficult, namely limited to sources within $\sim 20
\,{\rm Mpc}$. This number can be easily estimated with
back-of-the-envelope calculations, but it was confirmed through detailed
analysis of the detectability of the dominant oscillation frequency in
Ref. \cite{Clark2014} via a large-scale Monte Carlo study in which
simulated post-merger gravitational-wave signals are injected into
realistic detector data that matches the design goals of Advanced LIGO
and Advanced Virgo. The second consideration to bear in mind is that the
post-merger frequencies evolve in time, albeit only slightly. Hence, the
spectral properties of the gravitational-wave signal can be asserted
reliably only when the signal-to-noise ratio is sufficiently strong so
that even these changes in time can be measured in the evolution of the
PSDs \cite{Kiuchi2012, Hotokezaka2013c, Takami2014, Shibata2014}.

In light of these considerations, the prospects for high-frequency
searches for the post-merger signal are limited to rare nearby
events. Yet, if such detections happen, the error in the estimate of the
neutron star radius will be of the order of a few hundred metres
\cite{Clark2014, Clark2016}.

\subsubsection{Black-hole--torus system}

As mentioned a number of times already, the collapse of the binary-merger
product is expected to lead quite generically to the formation of a
black-hole--torus system, with the black hole having a dimensionless spin
of $J/M^2 \sim 0.7-0.8$ and the torus a mass of $\gtrsim
10^{-3}\,M_{\odot}$. An accurate determination of the final spin produced
in the merger of BNSs is not a mere academic question as the rapid
rotation of the black hole is a key ingredient in all of the models in
which the BNS merger is thought to lead to a jet formation and a
SGRB~\cite{Nakar:2007yr, Lee:2007js, Rezzolla:2011}. Because the
energetics of the emission will depend sensitively on the black-hole
spin, an accurate measure of the maximum value attainable can help set
upper limits on the efficiency of the process. In view of this, Kastaun
et al. \cite{Kastaun2013} have focused on the accurate computation of the
spin of the black hole originated in BNS mergers. Their initial data
consisted of irrotational equal-mass binaries to which various amounts of
rotation (parallel to the orbital angular momentum) are added to increase
the total angular momentum (see Sect. \ref{sec:ID}). Although the initial
data violated the constraint equations, the use of the constraint-damping
CCZ4 formulation (see Sect. \ref{sec:CCZ4}) yielded evolutions with
violations smaller than those with irrotational initial data and
non-constraint-damping formulations.

Overall, Kastaun et al. \cite{Kastaun2013} found that a limit of
$J/M^2\approx 0.89$ exists for the dimensionless spin of the black hole
and that any additional angular momentum present in the binary ends up in
the torus rather than in the black hole, thus providing also another
nontrivial example supporting the cosmic censorship hypothesis. The
material in the torus will eventually be accreted onto the black hole,
transferring angular momentum and further increasing the black-hole
spin. However, this will happen on dissipative time scales which are
longer and thus not relevant for the central engine of gamma-ray bursts,
which should be ignited on a dynamical time scale after the merger.

\subsubsection{Spectral properties and the mass-redshift degeneracy}

Besides providing information on the EOS, the spectral properties of the
gravitational-wave post-merger signal can also \lrn{be} used in a completely
different manner, namely, to remove the degeneracy in the determination
of redshift and mass for cosmological investigations. Indeed, a
well-known problem of the detection of gravitational waves from
compact-object binaries at cosmological distances is the so-called
``mass-redshift degeneracy''. More precisely, given a source of
(gravitational) mass $M$ at a cosmological redshift $z$, a direct
gravitational-wave observation provides information only on the combined
quantity $M(1+z)$, so that it is not possible to have an independent
measurement of $M$ and of $z$. The standard solution to this problem is
to detect an electromagnetic counterpart to the gravitational-wave
signal, so as to measure $z$ and hence the mass $M$. While this is still
possible, it may be not easy in the case of BNSs, as the electromagnetic
counterpart, may not be directed towards us if in the form of the
collimated jet of a SGRB.

\begin{figure*}
\begin{center}
  \includegraphics[width=0.60\columnwidth]{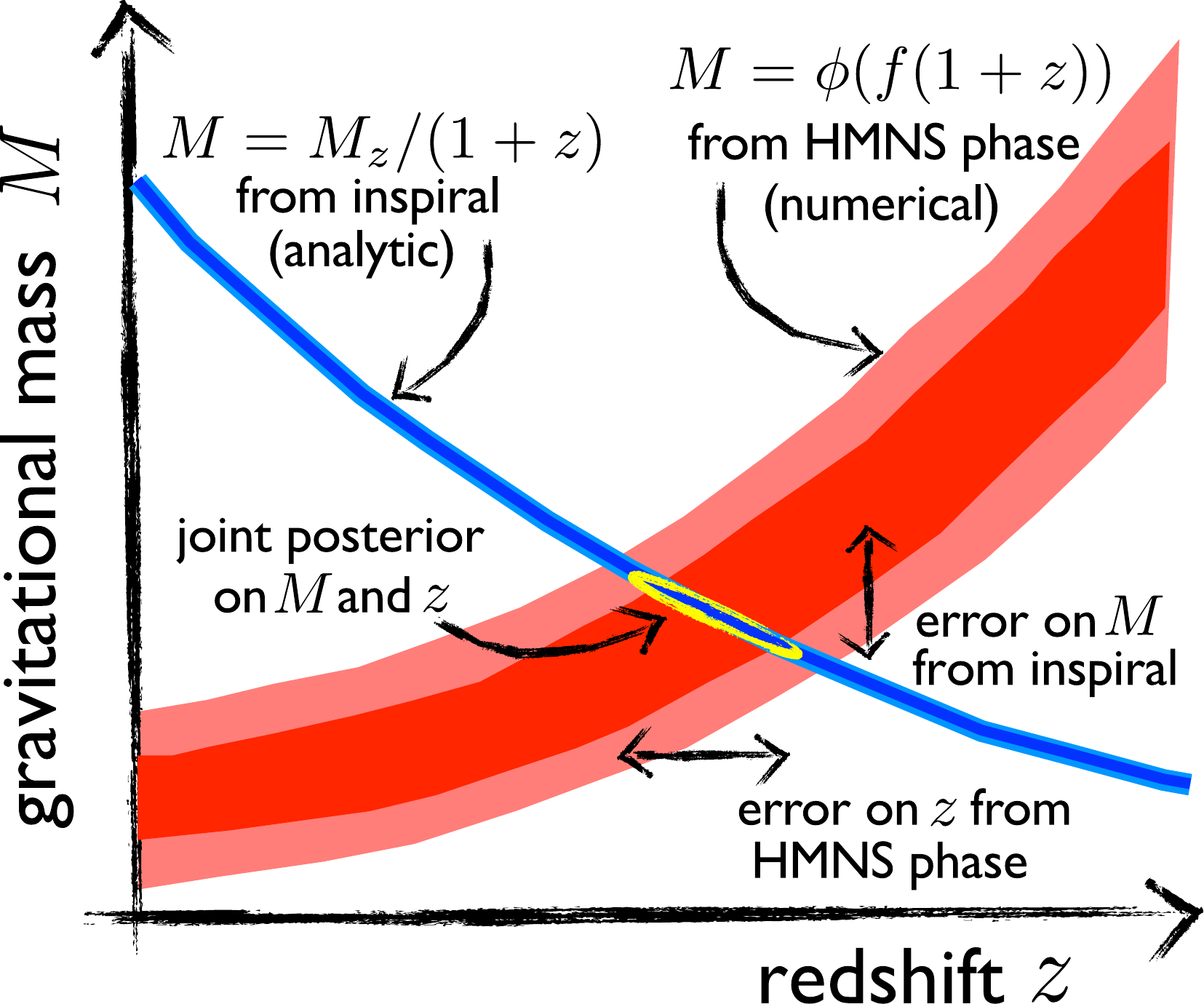}
\end{center}
\caption{A cartoon illustrating how the mass-redshift degeneracy is
  broken through the use of information from the inspiral and HMNS stages
  of a BNS merger event. Cross-correlating the information on the
  redshifted mass as a function of the redshift (blue stripe) with the
  information from the spectral properties of the {HMNS} phase (red
  stripe) will provide a localised range in mass and redshift, breaking
  the degeneracy. [Reprinted with permission from Ref. \protect{}
    \cite{Messenger2013}. \copyright~(2014) by the American Physical
    Society.]}
\label{fig:messenger+}
\end{figure*}

In a recent investigation, Messenger et al. \cite{Messenger2013}
described how this degeneracy can be broken when exploiting information on
the spectral properties of the post-merger gravitational-wave
signal. More specifically, making use of numerically generated BNS
waveforms, it was shown that it is possible to construct frequency-domain
power-spectrum reference templates that capture the evolution of two of
the primary spectral features in the post-merger stage of the waveforms
as a function of the total gravitational mass. 

This is summarised via a cartoon in Fig. \ref{fig:messenger+}, which
shows how the information on the redshifted mass as a function of the
redshift (blue stripe) can be correlated with complementary information
from the spectral properties of the HMNS phase. The overlap will provide
a localised range in mass and redshift, breaking the degeneracy. A
Bayesian inference method was then used to test the ability of the
Einstein Telescope \cite{Punturo2010b} to measure the characteristic
frequencies in the post-merger stage of the signal, finding that redshift
and gravitational mass can be determined separately, with uncertainties
in the redshift of sources at $z = 0.01-0.04$ of $10\% - 20\%$ and in the
gravitational mass of $< 1\%$ in all cases (see also Sect. \ref{sec:EOB}
for a method to determine the redshift based on tidal deformation in the
inspiral of BNS systems).

\begin{figure}
\begin{center}
  \includegraphics[width=6.45cm]{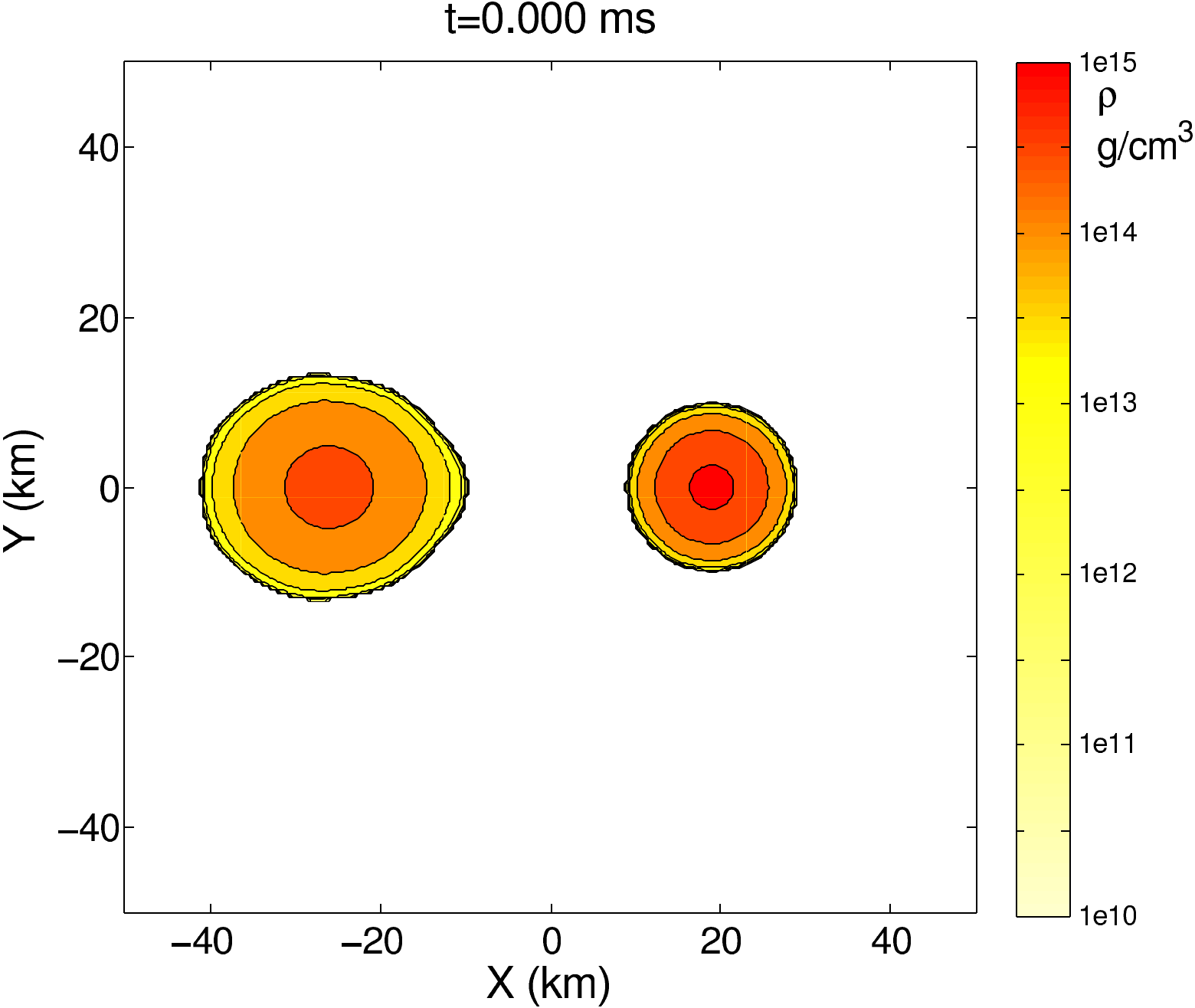}
  \hskip 0.1cm
  \includegraphics[width=6.45cm]{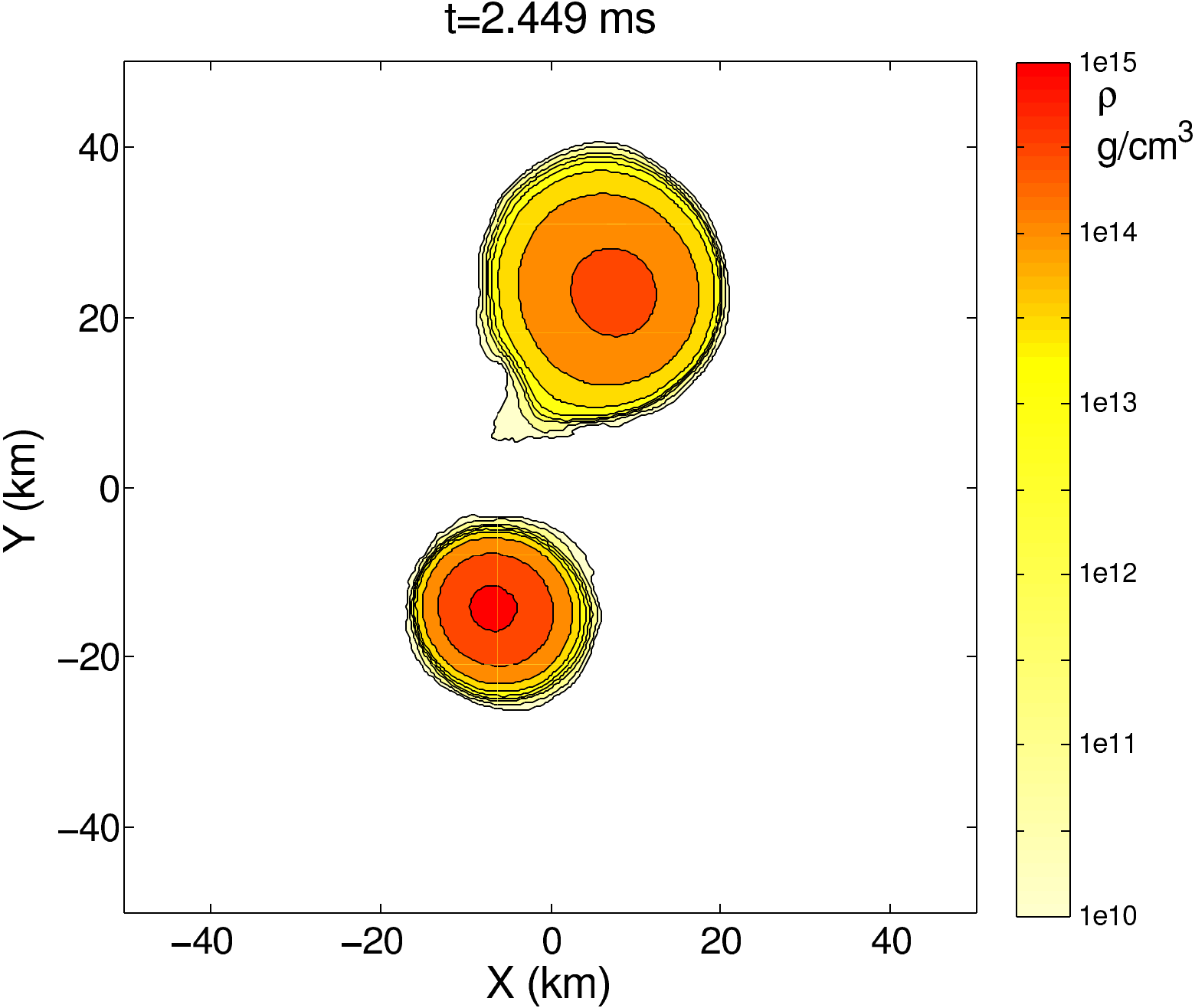}
  \vskip 0.25cm
  \includegraphics[width=6.45cm]{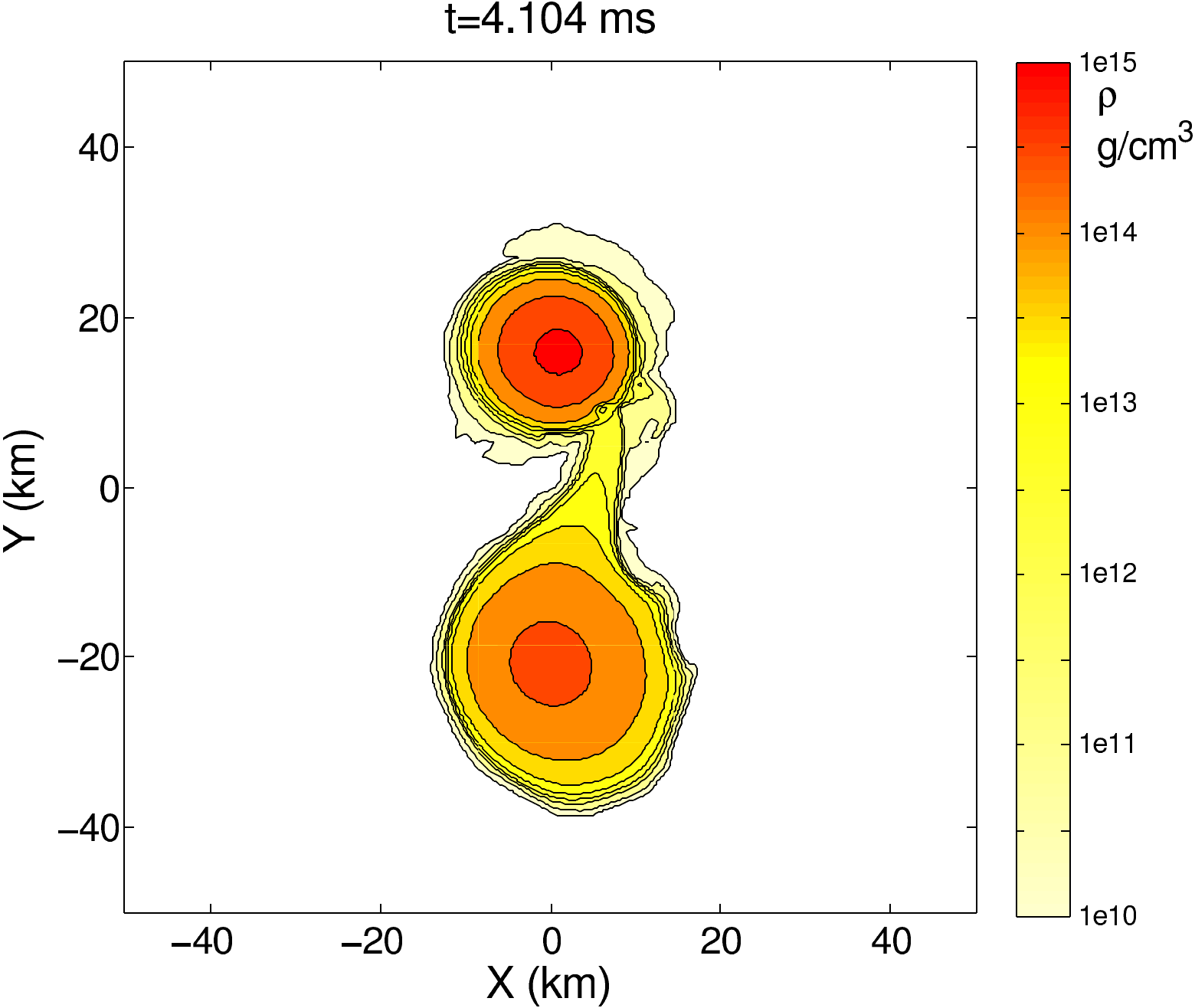}
  \hskip 0.1cm
  \includegraphics[width=6.45cm]{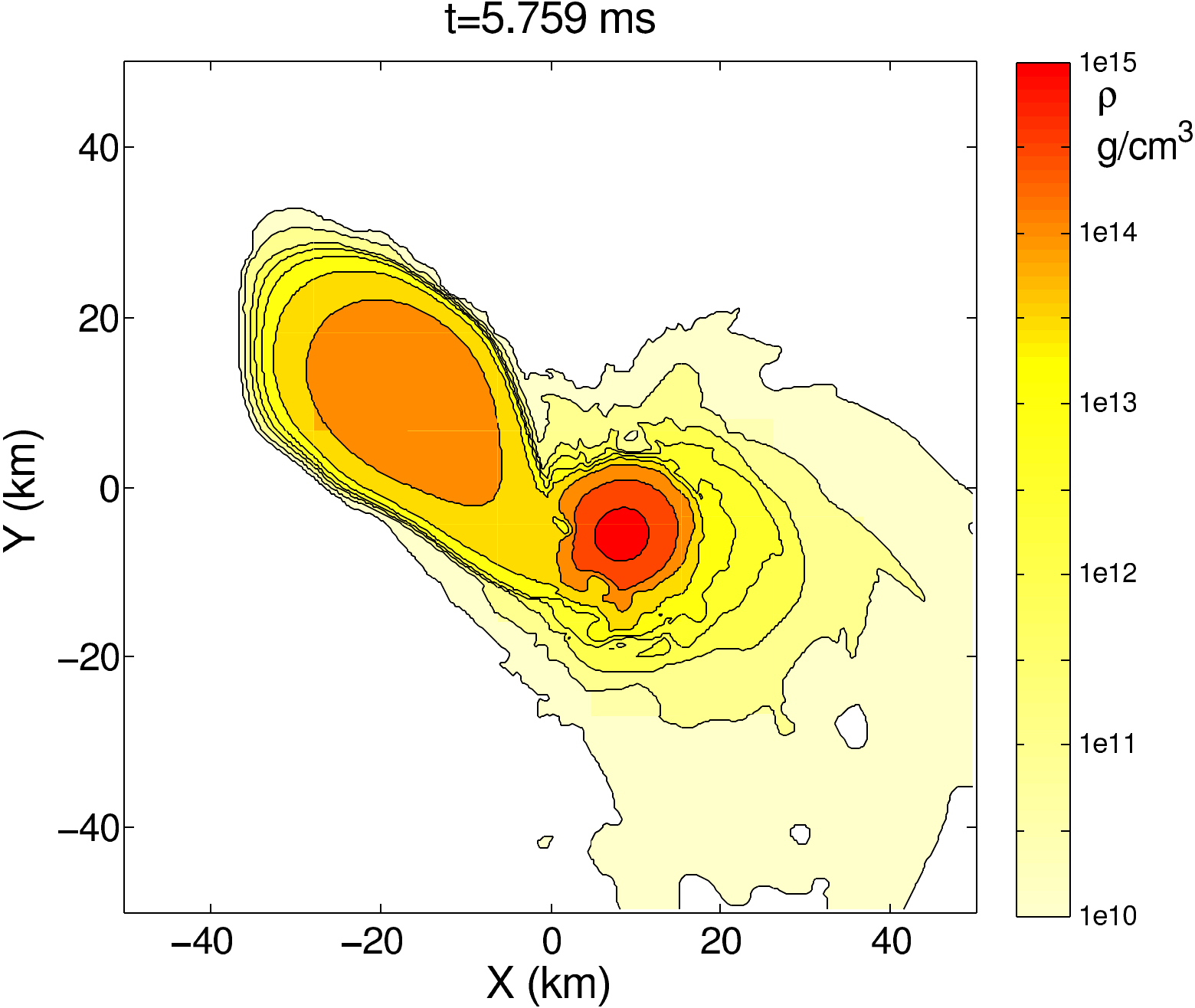}
  \vskip 0.25cm
  \includegraphics[width=6.45cm]{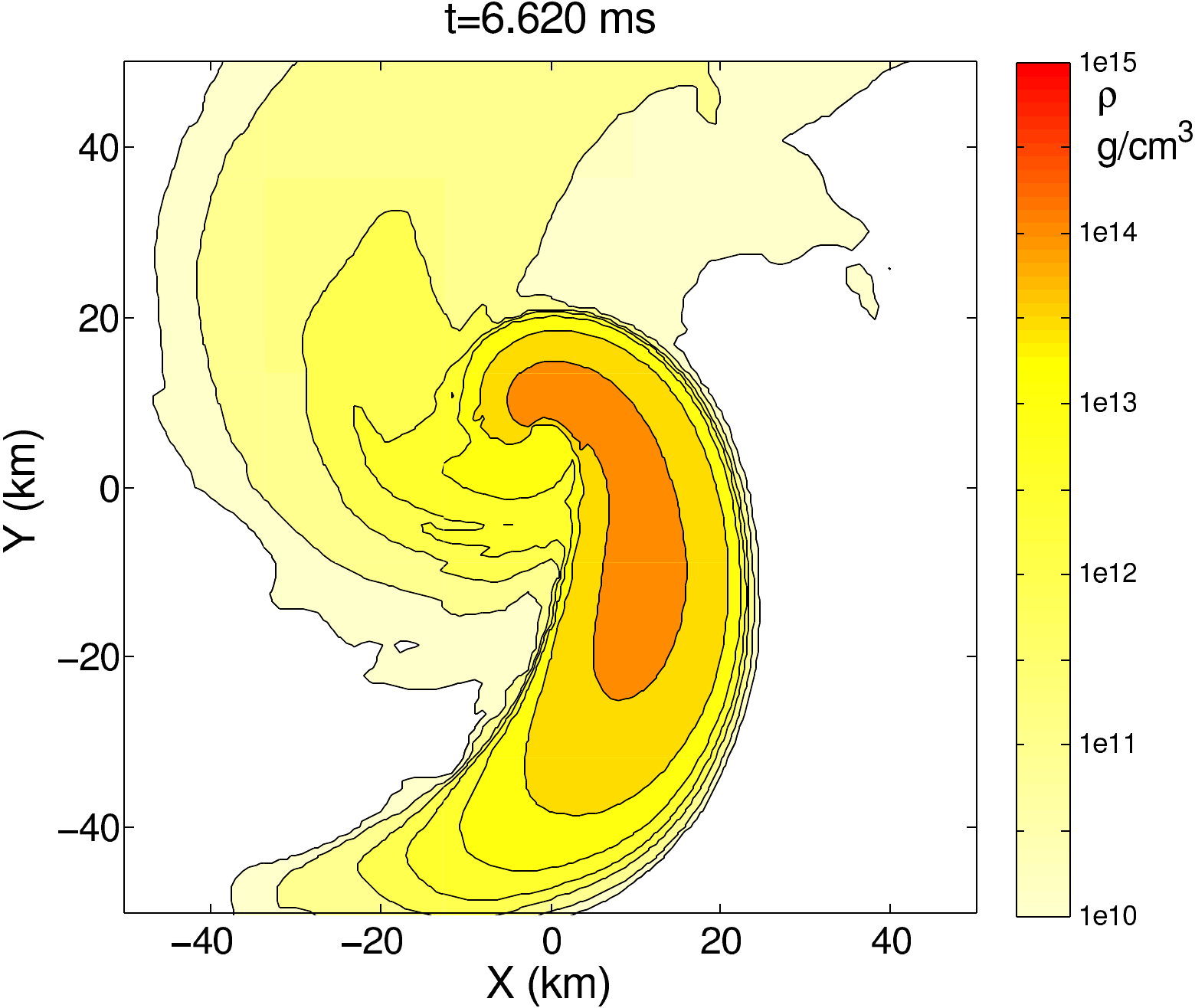}
  \hskip 0.1cm
  \includegraphics[width=6.45cm]{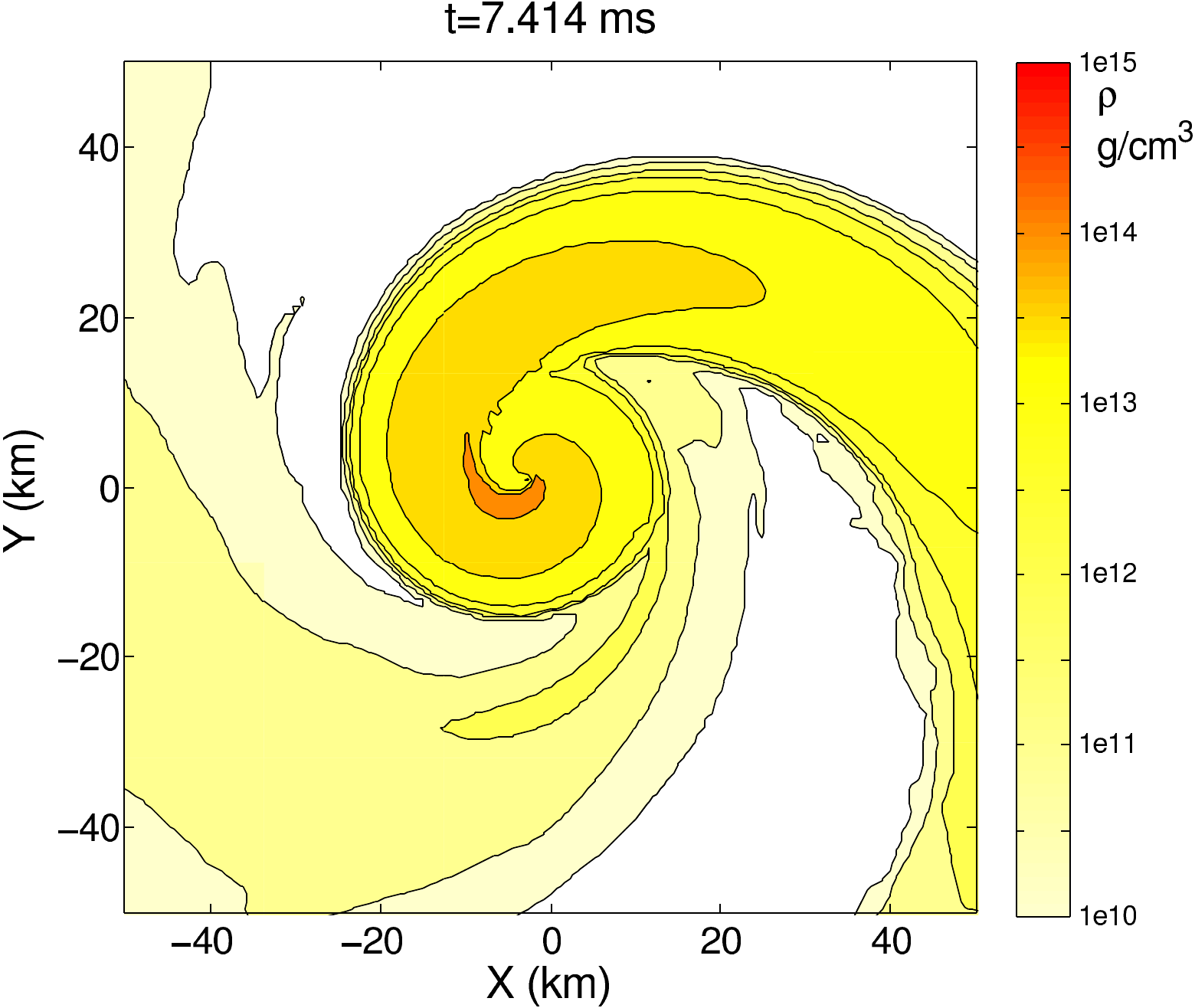}
\end{center}    
\caption{Isodensity contours in the $(x,y)$ plane for the inspiral and
  merger of a binary with mass ratio $q=0.7$ and total gravitational mass
  $M_{\rm tot} = 3.07\,M_{\odot}$. The third frame shows the onset of the
  merger and the last two frames show the behaviour of the system during
  the collapse to a black hole. [From Ref. \cite{Rezzolla:2010}
    \copyright \protect{} IOP Publishing. Reproduced with permission. All
    rights reserved.]}
\label{fig:density-xy-um}
\end{figure}

\subsection{Unequal-mass binaries}
\label{sec:hydro_blackhole_torus}

Neutron stars in binaries are expected to be produced mostly with masses
that are very similar to each other and thus with mass ratios in the
range $q \coloneqq M_{_B}/M_{_A} \in [0.7-1.0]$ \cite{Oslowski2011};
although the number of observed BNS systems is rather limited with only a
dozen of known systems \cite{Lattimer2012rev}, the mass-ratio
distribution is peaked around $q \simeq 0.9-0.95$. Although these mass
differences may appear small and indeed are so when comparing with the
mass ratios expected for binaries of stellar-mass black holes and even
more of supermassive black holes, these mass asymmetries are sufficient
to produce significantly different dynamics.

The first general-relativistic investigations of unequal-mass BNS systems
were performed in Refs. \cite{Shibata:2003ga, Rezzolla:2010}, although
these systems have been studied also earlier in various approximations
(see, \eg Refs. \cite{Rosswog2000,Oechslin06}). A representative example
of the dynamics of an unequal-mass binary with an ideal-fluid EOS is
shown in Fig. \ref{fig:density-xy-um}, which shows isodensity contours in
the equatorial plane for the inspiral and merger of a binary with a mass
ratio $q = M_{_B}/M_{_A} = 0.7$ and a total gravitational mass $M_{\rm
  tot} = M_{_B} + M_{_A} = 3.07\,M_{\odot}$. During the inspiral phase,
the heavier and more compact star is only slightly affected by its
companion, whereas the latter is decompressed rapidly while being
accreted onto the heavier star (see the three intermediate panels of
Fig. \ref{fig:density-xy-um}).

In addition to a different dynamics during the inspiral, unequal-mass
mergers also lead to substantial differences in the tori formed after the
merger and surrounding the black hole. More specifically, while
equal-mass mergers produce a highly symmetric disc, unequal-mass mergers
produce initially asymmetric discs, because of the presence of a large
spiral arm. The tidal disruption of the smaller-mass neutron star results
in an extended tail (extending well beyond the domain shown in
Fig. \ref{fig:density-xy-um}), which, unlike what happens in the
equal-mass case, transfers angular momentum outwards very efficiently.

There has been a lot of research on general-relativistic simulations of
the black-hole--torus system emerging from BNS mergers of unequal-mass
binaries. Kiuchi et al. \cite{Kiuchi2010} tried to schematise the
relation between the possible formation process of the central engine of
SGRBs (namely a system composed of a black hole surrounded by a disc) and
the gravitational waveforms that it generated. In particular, it was
pointed out that the gravitational-wave spectra of different models are
all qualitatively similar and that the various features of such spectra
are connected qualitatively with the different phases of the merger (\eg
the transition between a merged object with two separate density maxima,
reminiscent of the individual stars, and a merged object with a single
density maximum; or the formation of spiral arms; or the ringdown of the
black hole). It was also found that the torus mass has a positive
correlation with the frequency of the highest peak in the spectra, with
the total mass, and with the mass ratio.

More quantitative estimates of the disc mass and spin were given by
Rezzolla et al. \cite{Rezzolla:2010}, who also measured recoil velocities
of the black hole. In case of unequal-mass mergers, in fact, the formed
black hole may recoil as a result of the asymmetrical emission of
gravitational radiation in the final stages of the inspiral. Velocities
of $\lesssim 100\,{\rm km}\,{\rm s}^{-1}$, much smaller than those
observed in black hole simulations (see, \eg \cite{Gonzalez:2006md,
  Koppitz-etal-2007aa, Healy2014}), were calculated. Yet, such a recoil
could yield astrophysically interesting results, being comparable to or
larger than the escape velocity from the core of a globular cluster, that
is, $v_{\rm esc}\sim 50\,{\rm km}\,{\rm s}^{-1}$ \cite{Webbink:1985}.

Using the results of numerous simulations and some simple physical
considerations, Rezzolla et al. \cite{Rezzolla:2010} also built a
phenomenological expression that reproduces reasonably well the mass
distribution of tori produced in the merger of unequal-mass binaries, \ie
\begin{equation}
\label{eq:fitNSs}
M_{b, {\rm tor}} = [c_1 (1-q) + c_2] [c_3 (1+q)M_{_{\rm TOV}}-M_{\rm tot}]
\,,
\end{equation}
where $M_{\rm tot}$ and $M_{_{\rm TOV}}$ are the gravitational mass of
the binary and the maximum mass for an isolated neutron star with the
same EOS, respectively. The coefficients $c_1=2.974 \pm 3.366$,
$c_2=0.11851 \pm 0.07192$, and $c_3 = 1.1193 \pm 0.1579$ were determined
by fitting equation (\ref{eq:fitNSs}) to the results of the fully
general-relativistic simulations of Refs. \cite{Baiotti08} and
\cite{Rezzolla:2010}, but rescaled to allow for a value of $M_{_{\rm
    TOV}}=2.20\,M_{\odot}$ to be more consistent with current
observations of neutron star masses\footnote{\lrn{For completeness we
    also recall the coefficients originally reported in
    \cite{Rezzolla:2010}, that are: $c_1 = 1.115 \pm 1.090, c_2 = 0.039
    \pm 0.023, c_3 = 1.139 \pm 0.149$.}}. We expect that the values of
the coefficients will depend on the EOS considered (in
\cite{Rezzolla:2010} a simple ideal-fluid EOS with $\Gamma=2$ was used),
but also that the functional dependence suggested in \eqref{eq:fitNSs}
will be the same for all EOSs.

Overall, expression \eqref{eq:fitNSs} indicates that: (i) the mass of the
torus increases with the asymmetry in the mass ratio; (ii) such an
increase is not monotonic and for sufficiently small mass ratios the
tidal disruption leads to tori that have a smaller mass for binaries with
the same total mass; (iii) tori with masses up to $\sim 0.35\,M_{\odot}$
are possible for mass ratios $q\sim 0.75-0.85$. This information has also
been used to derive a relation between observations of the energy emitted
by SGRBs and the mass of BNSs that produce the tori that are presumed to
feed them \cite{Giacomazzo2012b}. In particular, after comparing the
masses of the tori with the results of the simulations it was possible to
infer the properties of the binary progenitors that yield SGRBs. By
assuming a constant efficiency in converting torus mass into jet energy,
(\ie an efficiency of $\sim 10\%$), it was found that most of the tori
would have masses smaller than $0.01M_{\odot}$, favouring ``high-mass''
binary NSs mergers, \ie binaries with total masses $\gtrsim1.5$ times the
maximum mass of an isolated NS. This has important consequences for the
gravitational-wave signals that may be detected in association with
SGRBs, since ``high-mass'' systems do not form a long-lived HMNS after
the merger. Furthermore, the analysis carried out by Giacomazzo et
al. \cite{Giacomazzo2012b} suggested that although binary systems
comprising a black hole and a neutron star cannot be excluded as the
engine of at least some of the SGRBs, the black hole would need to have
an initial dimensionless spin of $J/M^2 \sim 0.9$, or higher.

A more recent analysis has been carried out by Dietrich et
al. \cite{Dietrich2015}, who have presented results of a systematic
investigation of unequal-mass binaries using four different
nuclear-physics EOSs and significantly different stellar masses, \ie with
$q \simeq 0.66-0.86$. The binary with very small ratio is probably at the
edge of what is realistic to expect from an astrophysical scenario, but
is a useful reference to explore the dynamics of these systems in the
most extreme conditions. Indeed, for $q \simeq 0.66$ and a stiff EOS, the
secondary star is highly deformed during the inspiral, and its tidal
disruption ejects substantial amounts of unbound matter, \ie
$0.03\,M_\odot$, which could have a significant impact on the subsequent
nucleosynthesis (\cf Sect. \ref{sec:ejecta}). Also quite large is the
mass of the torus produced around the black hole, which has been measured
to be $\sim 0.3\,M_{\odot}$ and that would obviously lead to a powerful
electromagnetic counterpart

A number of very recent works have considered unequal-mass binaries as
complementary cases within more systematic investigations, such as those
involving microphysical nuclear EOSs and neutrino effects
\cite{Lehner2016,Sekiguchi2016}, the spectral properties of the
post-merger signal \cite{Rezzolla2016} or the influence of magnetic
fields \cite{Endrizzi2016}. In general, it was found that the results do
not differ considerably from the equal-mass binary simulations of the
same kind as long as the difference in the masses is 10\% or less. One
aspect that may show larger differences is the composition of the ejecta
\cite{Sekiguchi2016}.

\subsection{Dynamically captured binaries}
\label{sec:dynamical-capture}

Dynamical-capture compact-object binaries are binaries that form through
the close interaction (\ie {\it ``collision''}) of compact objects, as
opposed to the more standard compact-object binaries, whose components
were originally in the binary, \emph{before} becoming compact
objects. Recent studies have suggested that there may be a significant
population of compact-object binaries formed via dynamical captures in
dense stellar environments \cite{Oleary2009,Lee2010,Thompson2011}, such
as globular clusters. These binaries could contribute to the observed
SGRB rate, even if it is difficult to quantify what fraction of SGRBs may
be attributed to them \cite{East2013}.

\begin{figure*}
\begin{center}
\raisebox{0.3cm}{\includegraphics[width=0.55\columnwidth]{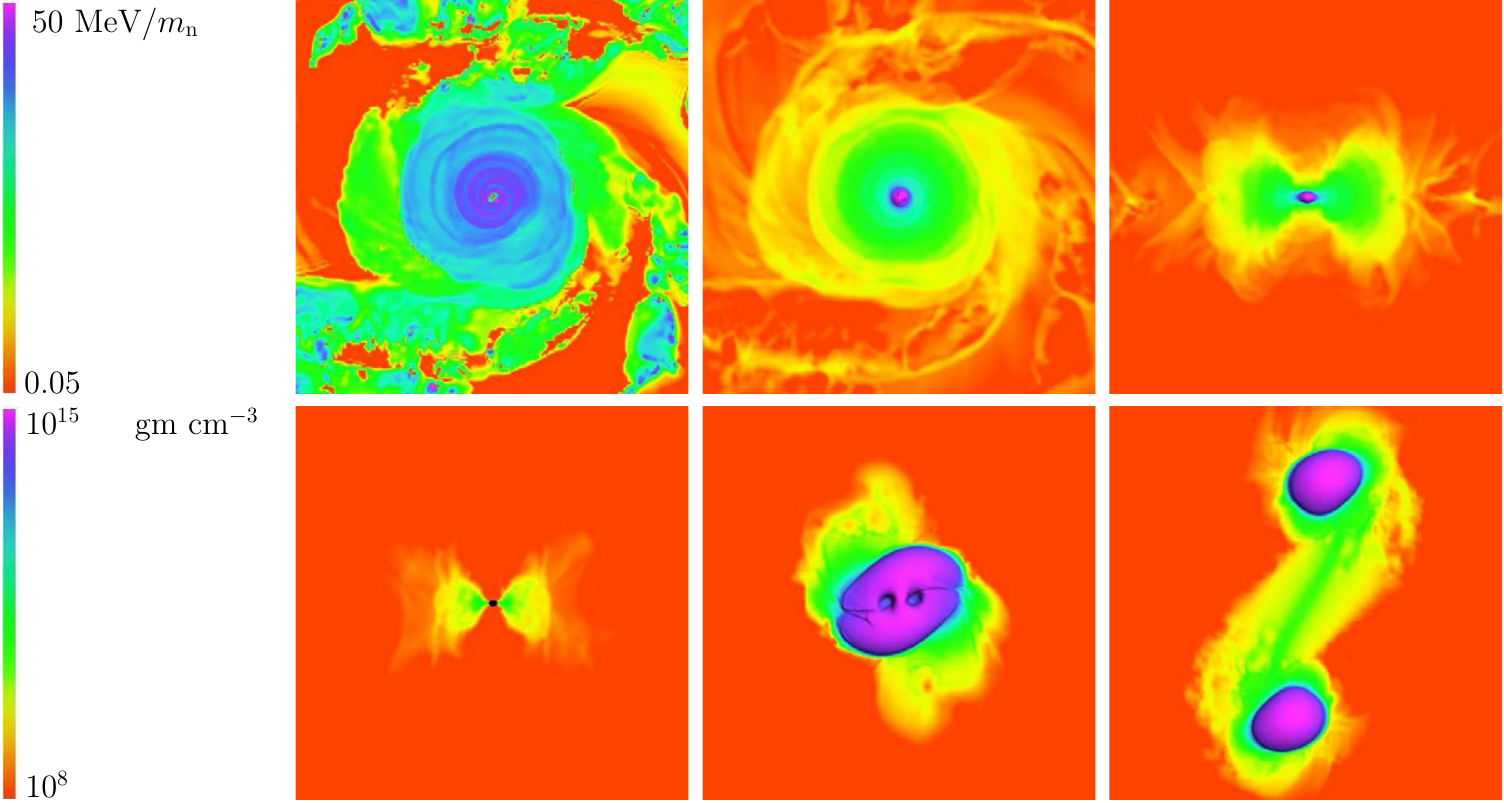}}
  \hskip 0.5cm
  \includegraphics[width=0.40\columnwidth]{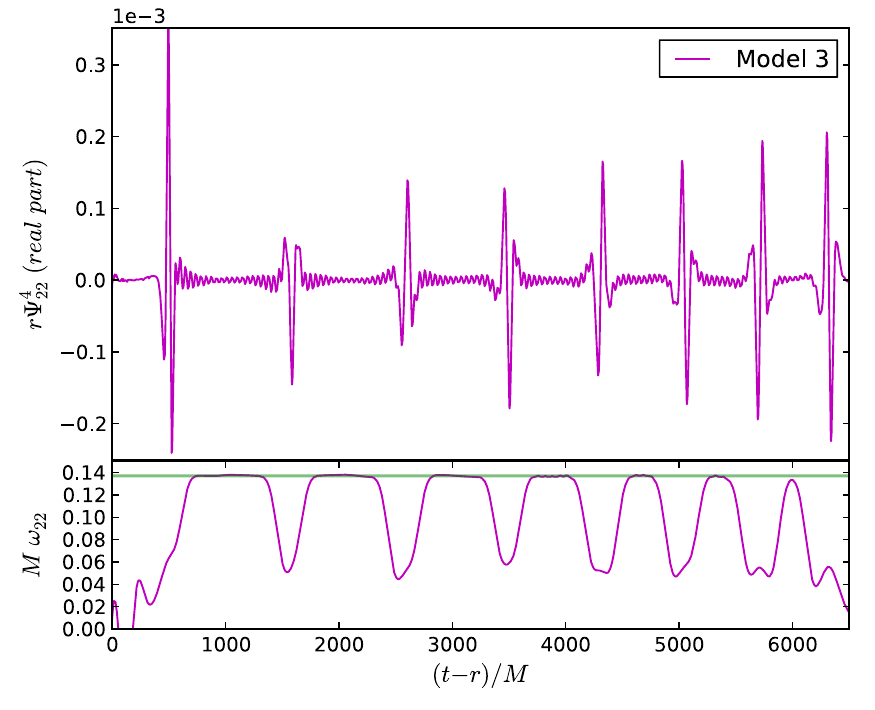}
\end{center}
\caption{{\it Left panel:} Snapshots of thermal specific energy (top left
  sub-panel) and rest-mass density (other five sub-panels). The top left
  and top middle, and bottom middle and bottom right sub-panels show the
  equatorial plane, while the other two show a perpendicular plane
  through the center of mass. The top sub-panels show an HMNS with
  surrounding disc and unbound material. The bottom sub-panels show, from
  left to right, a black hole and a surrounding disc; neutron stars with
  excited f-mode perturbations after a close encounter. The first four
  sub-panels have the same distance scale, where the coordinate radius of
  the HMNS and black hole are $\approx 13$ and $\approx 6\,{\rm km}$,
  respectively. The last two sub-panels share a second distance scale;
  the coordinate separation between the NSs in the last panel is $\approx
  73\,{\rm km}$. [Adapted from Ref. \cite{East2012c} with permission by
    the authors.]
  {\it Right panel:} Real part of the waveform (upper panel) and
  instantaneous gravitational wave frequency (lower panel) as a function
  of retarded time, for a model of Ref. \cite{Gold2012} that shows
  several elliptical orbits. [Reprinted with permission from
    Ref. \protect{} \cite{Gold2012}. \copyright~(2012) by the American
    Physical Society.]}
\label{fig:gold+}
\end{figure*}

Dynamically formed binary systems are born at small orbital separations
and with large eccentricities, so that their orbits are likely to remain
eccentric up to merger, since circularisation due to gravitational-wave
emission requires longer timescales. As a result, their close orbital
dynamics and gravitational-wave signal may differ substantially from
those of standard BNS systems in quasi-circular orbits. Initially,
dynamically formed binary systems may produce at periastron a series of
well-separated, repeated gravitational-wave bursts that last for minutes
to days (see right panel of Fig. \ref{fig:gold+}. This sequence of
  bursts gradually transforms into the final chirp inspiral signal of an
  eccentric binary system.

Although the integrated energy released in gravitational waves is
comparable to that of a quasi-circular inspiral, it will be more
difficult to measure it in the advanced interferometric detectors that
are coming into operation because a larger part of the radiation from
dynamical-capture binaries is emitted at the close periastron and at
rather high frequencies, which are mostly outside the best sensitivity
range of the detectors \cite{East2013}.

As for the matter dynamics, each close encounter may launch a tidal tail
and the neutron stars may be spun up to rotation frequencies close to
breakup. The final remnant is then, like in the standard merger case, a
compact object surrounded by a disc, but the disc is more likely to be
externally fed by the tidal tails that are produced at each close
encounter. A good fraction of the disrupted mass that orbits at some
distance from the merged object is still gravitationally bound and will
fall back on the central remnant on timescales substantially exceeding
the dynamical timescale of the central object, but within seconds at
most. Some snapshots of a simulation are reproduced from
Ref. \cite{East2012c} in the left panel of Fig. \ref{fig:gold+}.

Newtonian calculations of eccentric binaries were performed long ago
\cite{Turner1977a,Turner1977b}, but general-relativistic simulations
appeared in the literature only in 2012, with the work of Gold et
al. \cite{Gold2012}. One important result of the Newtonian studies
\cite{Turner1977b} is that the eccentric orbits lead to tidal
interactions that can excite oscillations of the stars. Such
oscillations, in turn, generate an additional and characteristic
gravitational-wave signal. In some cases, the gravitational signal can be
dominated by these non-orbital contributions.

\begin{figure*}
  \includegraphics[width=\textwidth]{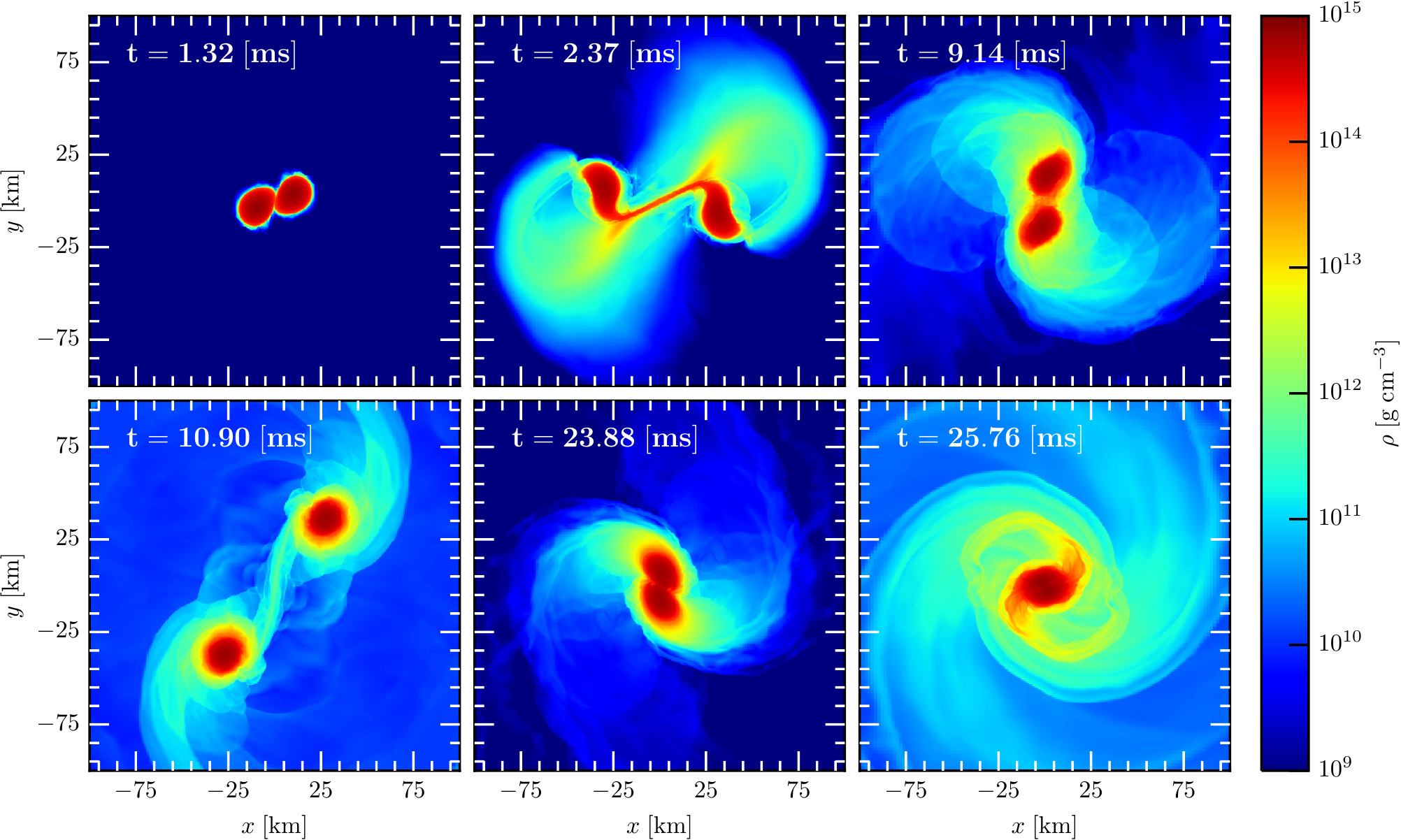}
  \caption{Rest-mass density in the orbital plane for a
    parabolic-encounter simulation at six different times. The neutron
    stars undergo three close encounters before merging. The panels show
    snapshots of the two stars immediately before and after each
    encounter. Tidal torques at the periastron result in large mass
    ejection and trigger oscillations in the neutron stars. [Adapted from
      Fig. 1 of Ref. \cite{Radice2016} with permission of authors.]}
  \label{fig:Radice2016-parabolic-rho_xy}
\end{figure*}

The general-relativistic simulations Gold et al. \cite{Gold2012} started
from initial-data sets created by superposing two boosted nonrotating
equal-mass stars without solving the constraint equations. They found,
however, that the constraint violation was at the level of the truncation
error of the evolution scheme. The gravitational-wave signal computed in
Ref. \cite{Gold2012} (and reproduced in the right panel of
Fig. \ref{fig:gold+}) was obviously much more accurate than that from
Newtonian calculations and was indeed interpreted as produced by
oscillations of the individual stars that are tidally induced by the
companion. Such a clear signal from orbit-induced stellar oscillations
could be a very interesting source for third-generation
gravitational-wave detectors such as the Einstein Telescope
\cite{Punturo2010b}, but not for second-generation ones (except for very
nearby events), because of the high frequency and short duration of the
merger or of the quasi-periodic signals from the HMNS.

Although they employed a simple polytropic EOS with $\Gamma=2$, Gold et
al. \cite{Gold2012} also pointed out that the neutron-star crusts may
disrupt or fracture during the inspiral under such strong tidal
deformations. This point was later put in evidence also by East and
Pretorius \cite{East2012c}, who performed improved general-relativistic
simulations that included initial data produced by solving the constraint
equations for colliding neutron stars (with equal and unequal masses)
with a given impact parameter and piecewise polytropic EOS models
\cite{Read:2009a}. It was then shown that dynamical-capture mergers can
produce accretion discs with masses $\lesssim 0.1\,M_{\odot}$ and eject
unbound material of up to a few percent of a solar mass.

Indeed, the copious ejection of matter is one of the aspects of the
dynamics of dynamically captured binaries that makes them particularly
interesting. Later on, in Sect. \ref{sec:ejecta}, we will discuss in more
detail the work done in this context and will also review the impact that
the ejected material has on the electromagnetic signal and on the
nucleosynthesis. However, it is worth mentioning here three works that
represent in many respects the state of the art in the modelling of
dynamically captured BNSs. The first one is by Rosswog et al.
\cite{Rosswog2013}, who studied dynamical-capture binaries with a code
that treats gravity in a Newtonian fashion but that treats microphysics
(neutrinos, radioactive decay) in very sophisticated ways. In addition to
finding results similar to the ones in Refs. \cite{Gold2012, East2012c}
for the bulk dynamics and for the amount of ejected material, Rosswog et
al. \cite{Rosswog2013} also computed neutrino luminosities from dynamical
collisions and found them to be at least comparable to those from the
merger of BNSs in quasi-circular orbits. The second work is by Radice et
al. \cite{Radice2016}, which is more recent and has considered the
dynamics of BNSs in parabolic orbits in full general relativity, with
high-order numerical methods and with different treatments of the
radiative effects. It was then found that eccentric binaries can eject
significantly more material than quasi-circular binaries and that the
outflow is composed of a combination of tidally- and shock-driven ejecta,
mostly distributed over a broad $\sim 60^\circ$ angle from the orbital
plane, and, to a lesser extent, by thermally driven winds at high
latitudes \cite{Radice2016}. Ejecta from such eccentric mergers were also
found to be more neutron rich than those of quasi-circular mergers. This
is due to the strong tidal torques exerted on the neutron stars during
their periastron passages and that lead to the ejection of cold,
neutron-rich material. The third work is by East et al. \cite{East2016}
and focused on the influence of the initial spin of the neutron stars in
eccentric orbits. It was found that even moderate spins can significantly
increase the amount of ejected material, including the amount unbound
with velocities greater than half the speed of light. This would predict
brighter electromagnetic signatures (cf. Sect. \ref{sec:ejecta}).

The dynamics of a representative parabolic-encounter simulation from
Ref. \cite{Radice2016} is shown in
Fig. \ref{fig:Radice2016-parabolic-rho_xy}, where the rest-mass density
is shown in the orbital plane at representative times during the
evolution before and after each close encounter. During the periastron
passage strong tidal torques and shocks result in episodic outflow
events. Part of the ejected neutron-rich matter is unbound from the
system, while the rest settles in a thick atmosphere around the neutron
stars.

\newpage
\section{Beyond pure hydrodynamic simulations}
\label{sec:bph}

\subsection{Ideal and resistive magnetohydrodynamics simulations}
\label{sec:HD_MHD}

Very strong magnetic fields\footnote{Quantum electrodynamics effects may
  become important and would need to be accounted for if the magnetic
  fields are sufficiently strong and the rest-mass density sufficiently
  low. For example, the so-called Landau quantisation (see, \eg
  Ref. \cite{Harding2006}) may occur if the magnetic fields exceed a
  critical value of $B_{\rm crit} = 4.414 \times 10^{13}$ G and the
  rest-mass density is below a critical value $\rho_{\rm crit, B} =
  7.04\times 10^{10}(0.1/Y_e)(B/10^{16} {\mathrm G})^{3/2} {\mathrm
    g}/\mathrm{cm}^3$, where $Y_e$ is the electron fraction per
  baryon. Kiuchi et al. \cite{Kiuchi2015a} have concluded that Landau
  quantisation is not important during the merger of magnetised BNSs
  because the regions where the QED limit is exceeded are at rest-mass
  densities above the critical one. Other effects have not been
  investigated and are routinely neglected in general-relativistic MHD
  simulations.}, in the range of $10^{12}-10^{15}\,{\rm G}$, are known to
endow neutron stars. In addition to modifying the dynamics of the matter,
such strong magnetic fields are likely related to observable
electromagnetic emissions (see Sect. \ref{sec:EM_counterparts}), which
will yield important information on these systems when observed through
the electromagnetic counterparts they can lead to. We should note that
hereafter we will make
the distinction between \lrn{\emph{``prompt''} electromagnetic
  counterparts and \emph{``delayed''} electromagnetic counterparts, or
  \emph{``afterglows''}}.
The former is expected to take place in a window in time that starts just
before (\ie a few seconds) the merger and ends a few hours after the
merger, while the latter is instead expected to take place from days to
years after the actual merger. Hence, while we discuss \lrn{prompt}
electromagnetic counterparts in Sections \ref{sec:magn-inspiral},
\ref{sec:m_pmd}, and \ref{sec:EM_counterparts}, we will discuss
\lrn{delayed electromagnetic counterparts} in Section \ref{sec:ejecta}.

Although neutron stars can in principle be endowed with very large magnetic
fields, in practice, such large fields are not expected to be present in
old neutron stars right before merger, as they must have decayed
considerably, reaching values that are instead in the range \ie
$10^{8}-10^{10}\,{\rm G}$. Yet, even comparatively weak initial magnetic
fields are expected to be amplified during and after the merger through
different mechanisms and instabilities, which have received particular
attention over the years. In particular, five different and quite general
amplification mechanisms have been so far explored with BNS mergers: (i)
magnetic compression; (ii) turbulent amplification, (iii) the
Kelvin-Helmholtz instability (KHI), either at the merger or later on in
the dynamics \cite{Dionysopoulou2015} (\cf Sect. \ref{sec:ph}); (iv)
magnetic winding by differential rotation; (v) the magneto-rotational
instability (MRI) \cite{Velikhov1959, Chandrasekhar1960, Balbus1991,
  BalbusHawley1998} in the binary-merger product and/or the massive disc
around it or around the black hole after the collapse has taken place.

In view of the considerations above, the inclusion of magnetic fields in
simulations of BNS mergers, although far from being trivial, is a
necessity, and a number of codes have been developed to solve the
equations of relativistic ideal MHD (IMHD; see
Sect. \ref{sec:me_rmhd}). While several of the codes adopt a fixed curved
background spacetime (see, \eg Refs. \cite{Komissarov1999, Koide98,
  DelZanna2003, Gammie03, Anninos05c, Anton06, DelZanna2007, Zink2011}),
considerable development has also been made with codes that can handle
arbitrary and dynamical spacetimes (see, \eg Refs. \cite{Duez05MHD0,
  Shibata05b, Neilsen2005, Giacomazzo:2007ti, Farris08,
  Moesta13_GRHydro}). In addition, an effort has been dedicated to going
beyond IMHD and thus to including resistive effects as a way to better
model resistive contributions to the energy losses from the system. The
importance of resistivity effects can be easily deduced from observing
the evolution of a current sheet in high but finite conductivity. Under
these conditions, several instabilities can take place in the plasma and
release substantial amounts of energy via magnetic
reconnection~\cite{Biskamp1986}. Because the study of reconnection in
relativistic phenomena is important to try to explain the origin of
flares in relativistic sources, it is not surprising then that several
groups have developed in recent years numerical codes to solve the
equations of special and general-relativistic RMHD~\cite{Komissarov2007,
  Palenzuela:2008sf, Dumbser2009, Zenitani2010, Takamoto2011b,
  Zanotti2011b, Bucciantini2012, Dionysopoulou:2012pp, Palenzuela2013,
  Palenzuela2013a, Palenzuela2013b, Ponce2014, Dionysopoulou2015}.

\begin{figure*}
\begin{center}
\raisebox{0.8cm}{\includegraphics[width=0.49\columnwidth]{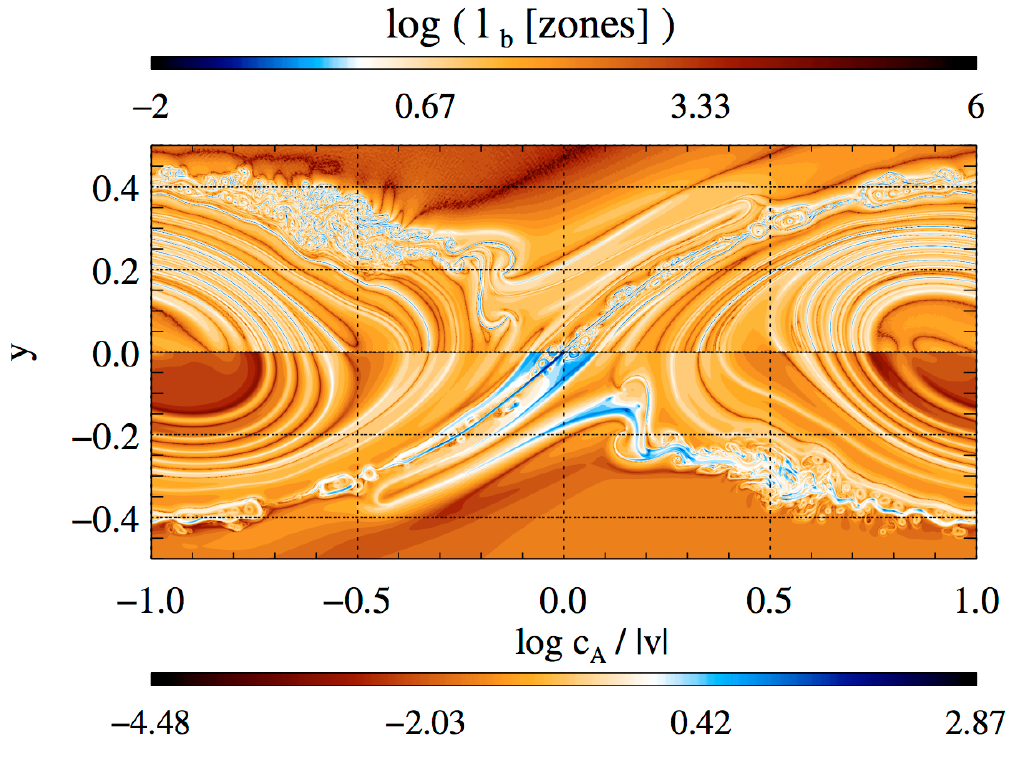}}
  \hskip 0.2cm
  \includegraphics[width=0.49\columnwidth]{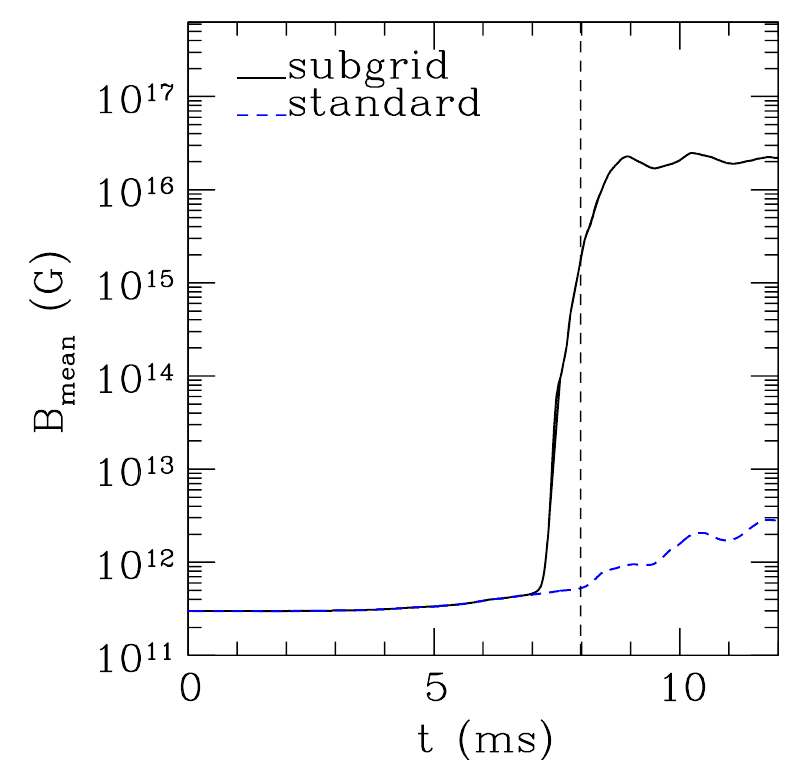}
\end{center}
\caption{{\it Left panel:} Snapshot of the structure of a model studied
  in Ref. \cite{Obergaulinger10}, taken shortly after the termination of
  the kinematic amplification phase. The top half shows the logarithm of
  the characteristic length scale of the magnetic field in units of the
  zone size. The bottom half shows the logarithm of the ratio of the
  Alfv\'en velocity and the modulus of the fluid velocity, bluish and
  reddish colours denoting strongly and weakly magnetized regions,
  respectively. [Reproduced with permission from
    \cite{Obergaulinger10}. \copyright~ESO.] {\it Right panel:} Evolution
  of the mean value of the magnetic field when a subgrid model is
  implemented (black solid line) and when it is not (blue dashed
  line). The vertical dashed line shows the time of merger (when the
  neutron-star cores collide). In the simulation implementing the subgrid
  model the magnetic field grows by five orders of magnitude up to $\sim
  10^{16}$ G and saturates when reaching equipartition with the kinetic
  energy of the fluid in the turbulent regions. In the simulation without
  the subgrid model the magnetic field grows by only one order of
  magnitude. [Reprinted with permission from Ref. \protect{}
    \cite{Giacomazzo:2014b}. \copyright~(2015) by the American Physical
    Society.}
\label{fig:Obergaulinger+2010}
\end{figure*}

Among the instabilities expected to develop, the KHI and MRI in
particular are those that give the fastest, exponential growths, and
therefore particular attention has been concentrated on them.  \lrn{In
  contrast to the KHI, where a numerical triggering is necessary when
  dealing with inviscid fluids \cite{Radice2012a}, the MRI develops also
  in inviscid plasmas, but
requires the highest grid resolutions to be resolved for realistic
magnetic-filed strengths, and so is effectively the most difficult to
investigate.} Local simulations with sufficient resolutions are possible,
as nicely displayed in the left panel of
Fig. \ref{fig:Obergaulinger+2010}, which shows a snapshot of a developed
KHI, but it is believed that in the near future computational resources
will not be sufficient to study the MRI quantitatively in BNS merger
simulations, despite some impressively expensive attempts having been
made \cite{Kiuchi2014, Kiuchi2015}. As a result, current attempts are
moving towards the use of models of the magnetic-field amplification at
the subgrid level \cite{Giacomazzo:2014b, Palenzuela2015}, as done in
other research fields where turbulent flows need to be modeled. In
practice, these approaches aim at including the contribution to the bulk
flow from the unresolved turbulent dynamics by introducing suitably tuned
corrections to the equations (most notably the induction
equation). Subgrid approaches make computationally feasible the inclusion
of effects that are prohibitively expensive to be included via direct
simulations, but the results they provide are dependent on the choice
made for the subgrid dynamics. Therefore, although subgrid modelling
appears the only convincing way to proceed in the near future,
computations based on direct simulations under reasonably realistic
conditions or first-principle studies of the properties of relativistic
turbulence (see, \eg \cite{Obergaulinger10, Zrake2011, Zrake2011a,
  Radice2012b}) are definitively still very important.

In addition to the challenges related to the abundance of physical
instabilities, IMHD and RMHD simulations are also made difficult by the
necessity of conserving during the evolution the divergence-free
condition of the magnetic field. This is an obvious requirement to avoid
the spurious generation of magnetic monopoles which would act as energy
sinks, but is particularly challenging to enforce when adopting
mesh-refined grids. This issue has been addressed in general-relativistic
MHD simulations mainly in three ways: (i) through the \emph{``constrained
  transport''} approach \cite{Evans1988, Balsara99, Toth2000,
  Balsara2001, Balsara2009}; (ii) through the \emph{``hyperbolic
  divergence cleaning''} method \cite{Dedner:2002}; (iii) through the use
of the vector potential as evolution variable in place of the magnetic
field \cite{Etienne:2010ui, Giacomazzo2011b, Etienne2012a,
  Farris2012}. While we refer to Ref. \cite{Toth2000} for a more detailed
review of these methods, we here simply recall that the
constraint-transport approach preserves the constraint to machine
accuracy but requires special interpolation at refinement-level
boundaries in order to preserve the constraint. Hyperbolic divergence
cleaning, on the other hand, may require a tuning of the parameters it
uses to damp-out any violation of the constraints and hence its
efficiency is problem dependent but generally worse than the constraint
transport approach. Finally, the vector-potential formulation is the one
that seems to work best in AMR, since the constraint is preserved by
construction with the vector potential, even though problems may arise
with the restriction and prolongation operations of AMR grids. Overall,
the last two approaches, although not optimal, are easier to implement
and hence have become the standard choice in many general-relativistic
IMHD and RMHD codes.

\subsubsection{Inspiral and merger dynamics}
\label{sec:magn-inspiral}

As in the case of pure-hydrodynamics simulations, also in IMHD, the study
of the inspiral phase is the simplest to calculate as no instabilities
are expected to develop and hence has been the focus of the early
calculations. Furthermore, given that the inspiral may be the only part
of the signal that can be detected by present gravitational-wave
detectors for sources that have a small signal-to-noise ratio, it is
natural to ask whether the detection of the inspiral can be used to
measure the strength of the magnetic field in the neutron stars prior to
the merger. The answer to this question can be easily worked out
analytically and it is not difficult to conclude that, for realistic
initial magnetic fields (\ie $B_{0} \lesssim 10^{10}\,{\rm G}$) the
magnetic energy in the two stars is several orders of magnitude smaller
than the binding energy of the binary and hence magnetic fields can only
provide very small corrections to the orbital dynamics. While correct,
this estimate does not address the problem of whether an advanced
detector would be able to measure such a small correction.

Hence, a number of groups have undertaken the task of providing a
quantitative answer to this question by measuring the orbital corrections
induced in BNS systems with variable magnetic-field strengths
\cite{Anderson2008, Liu:2008xy, Giacomazzo:2009mp, Giacomazzo2011b}. A
common feature of these simulations is that they all employed very
simplified EOSs (\ie an ideal-fluid EOS with $\Gamma=2$) and all used
fully buried magnetic fields that were added to the pure hydrodynamical
solutions of irrotational binaries. More specifically, since no
self-consistent solution was (and still is!) available for magnetized
binaries, a poloidal magnetic field was added a posteriori using a
suitable prescription for the vector potential. The disappointing, but
not surprising, answer coming from these simulations is that for
realistic magnetic-field strengths the changes introduced in the inspiral
waveforms are too small to be detected by present and possibly future
detectors such as the Einstein Telescope \cite{Giacomazzo:2009mp,
  Giacomazzo2011b}. On the other hand, if unexpectedly high magnetic
fields, \ie $B_{0} \gtrsim 10^{10}\,{\rm G}$ are present in the stars
before the merger, these would leave a sufficiently strong imprint to be
detected \cite{Giacomazzo:2009mp}.

\begin{figure*}
\begin{center}
  \includegraphics[width=0.49\columnwidth]{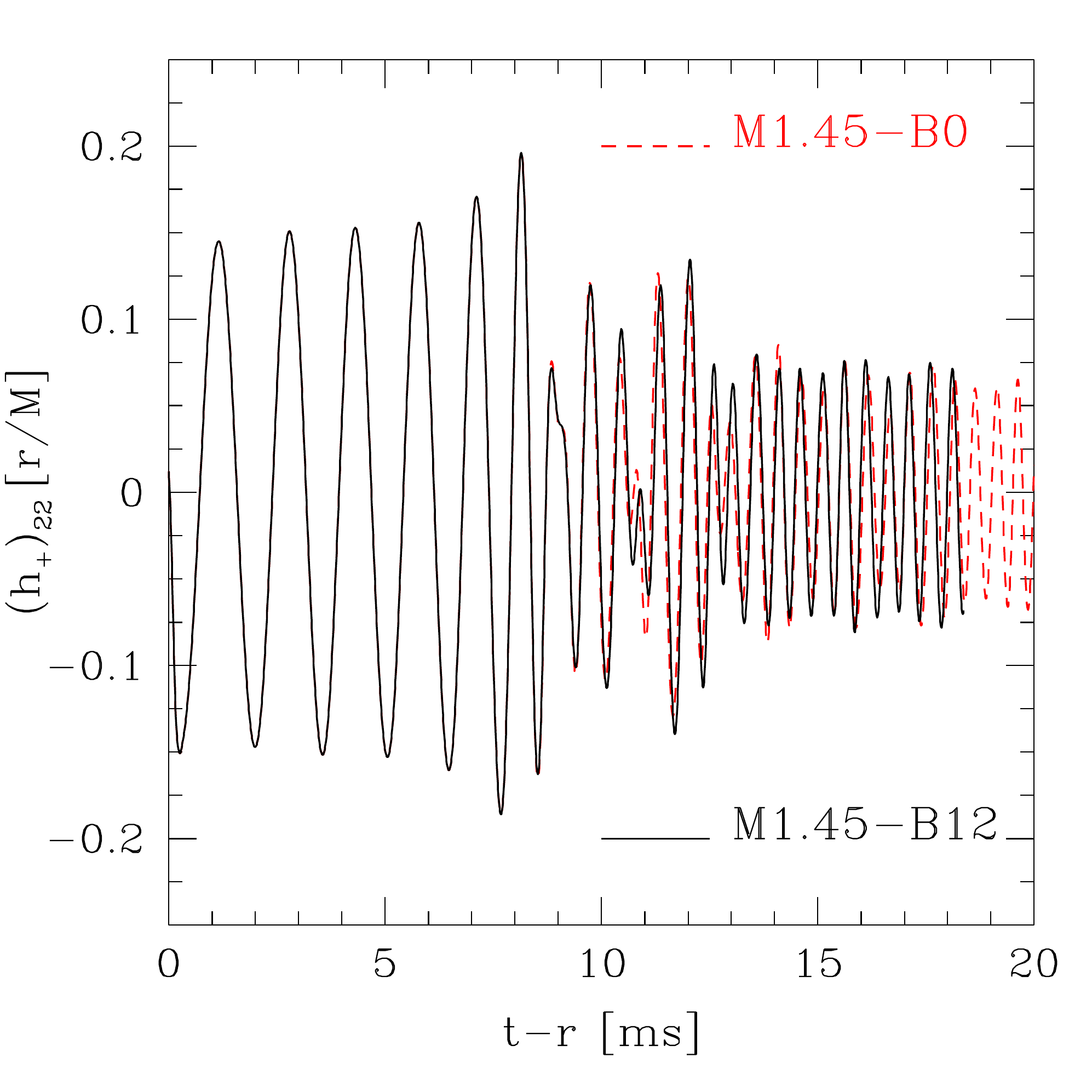}
  \hskip 0.2cm
  \includegraphics[width=0.49\columnwidth]{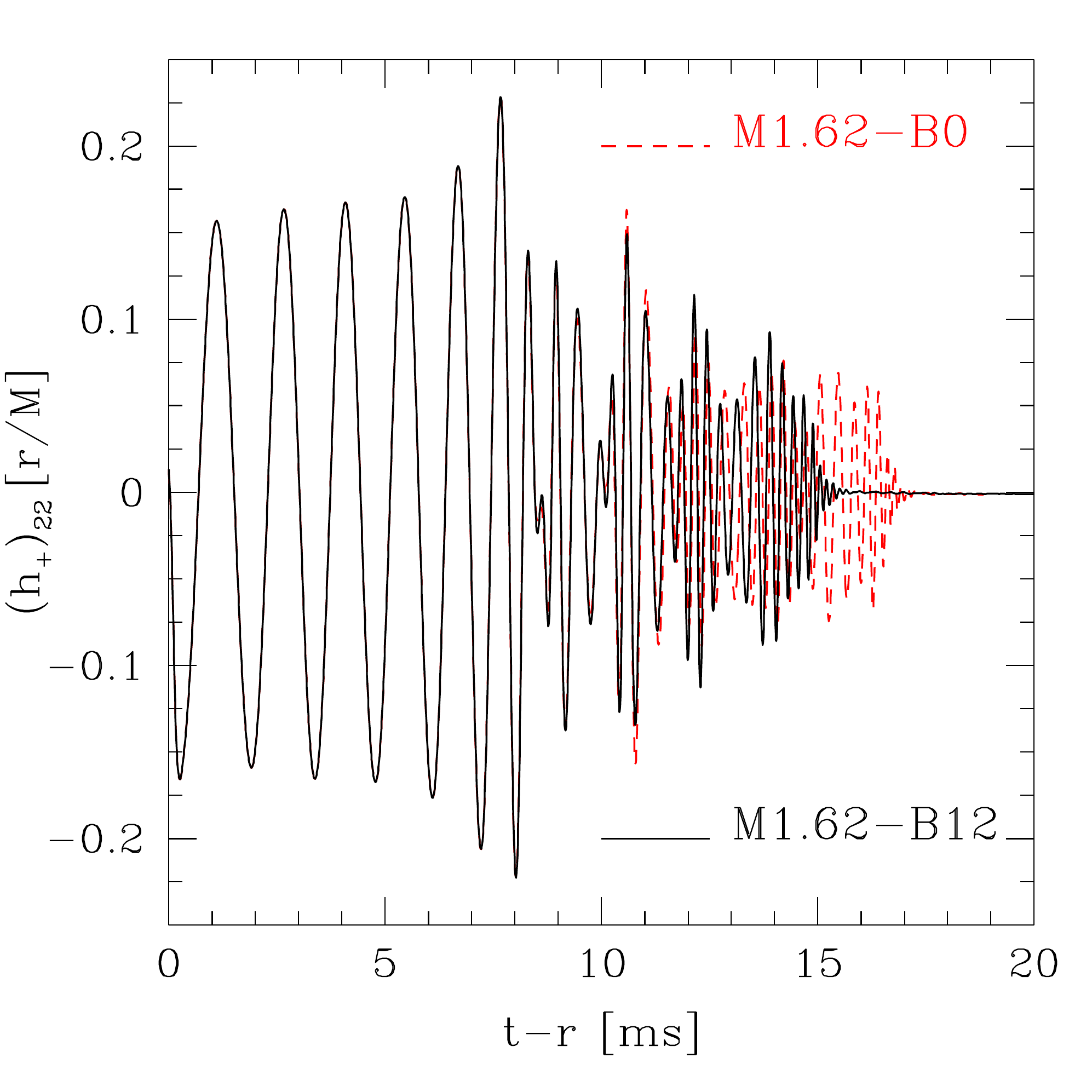}
\end{center}
\caption{Gravitational waves for a low-mass (individual stellar rest mass
  $1.445 M_\odot$) binary (left panel) and a high-mass (individual
  stellar rest mass $1.625 M_\odot$) binaries (right panel) as a function
  of the retarded time $t-r$ in ms. Shown with red dashed lines are the
  corresponding models with zero magnetisation. Note that the differences
  become appreciable only after the merger. [Reprinted with permission
    from Ref. \protect{} \cite{Giacomazzo2011b}. \copyright~(2011) by the
    American Physical Society.]}
\label{fig:giacomazzo+}
\end{figure*}

To reach these conclusions, Giacomazzo et al. \cite{Giacomazzo:2009mp}
have computed the overlap, ${\cal O}$, between the waveforms produced
during the inspiral by magnetized and unmagnetized binaries, finding that
for initial magnetic-field strength $B_{0}\lesssim 10^{14}\,{\rm G}$ the
overlap during the inspiral is ${\cal O}\gtrsim 0.999$ and is quite
insensitive to the mass of the neutron stars. Only for unrealistically
large magnetic fields like $B_{\rm 0}\simeq 10^{17}\,{\rm G}$, the
overlap does decrease noticeably, becoming ${\cal O}\lesssim
0.76\,(0.67)$ for stars with rest-mass masses $M_{b} \simeq
1.4\,(1.6)\,M_{\odot}$, respectively. On the other hand, in agreement
with what found by Refs. \cite{Anderson2008, Liu:2008xy}, Giacomazzo et
al. \cite{Giacomazzo2011b} also find that magnetic fields do have an
impact after the merger. Some representative waveforms of BNS systems
with different masses and magnetic-field strengths are shown in
Fig. \ref{fig:giacomazzo+}, which displays the waveforms for a low-mass
(left panel) and a high-mass binary (right panel) as a function of the
retarded time. Shown with red dashed lines are the corresponding models
with zero magnetisation. It is clear how the black and red lines become
discernible only after the merger.

Already the early works on the inspiral and merger of magnetised BNSs,
\eg Refs. \cite{Anderson2008, Liu:2008xy, Giacomazzo:2009mp,
  Giacomazzo2011b}, pointed out that the KHI develops at the beginning of
the merger in the shear layer where the two stars enter into contact,
causing an amplification of the magnetic-field strength. Because these
works used rather coarse spatial resolutions, the highest being that of
Ref. \cite{Giacomazzo2011b}, $220\,{\rm m}$ on the finest grid, the
amplification reached was rather modest: of one order of magnitude at
most \cite{Giacomazzo2011b}. Hence, it was not possible to confirm the
results of Price and Rosswog \cite{Price06} and assess whether a KHI
fully develops at the merger.

\begin{figure*}
\begin{center}
  \includegraphics[width=0.49\columnwidth]{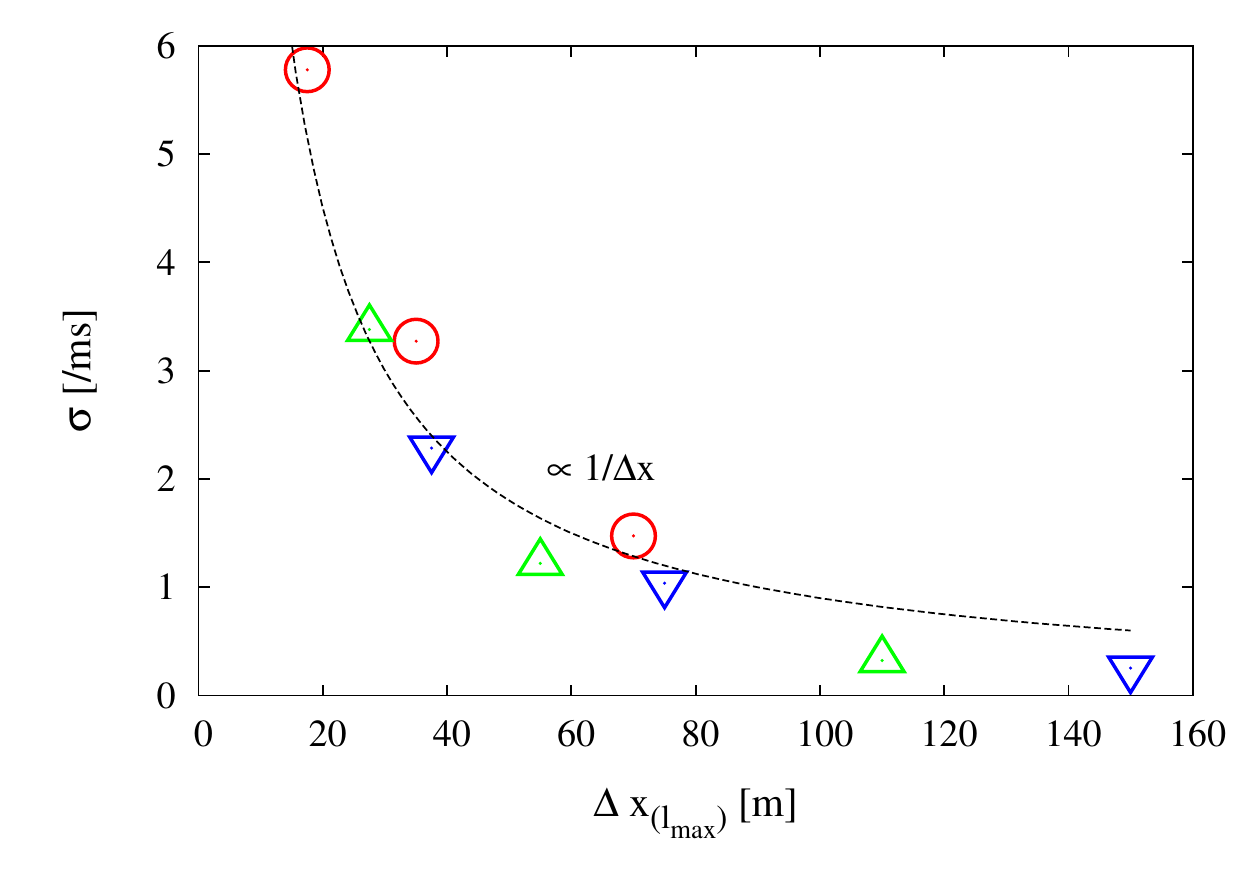}
  \hskip 0.2cm                           
  \includegraphics[width=0.49\columnwidth]{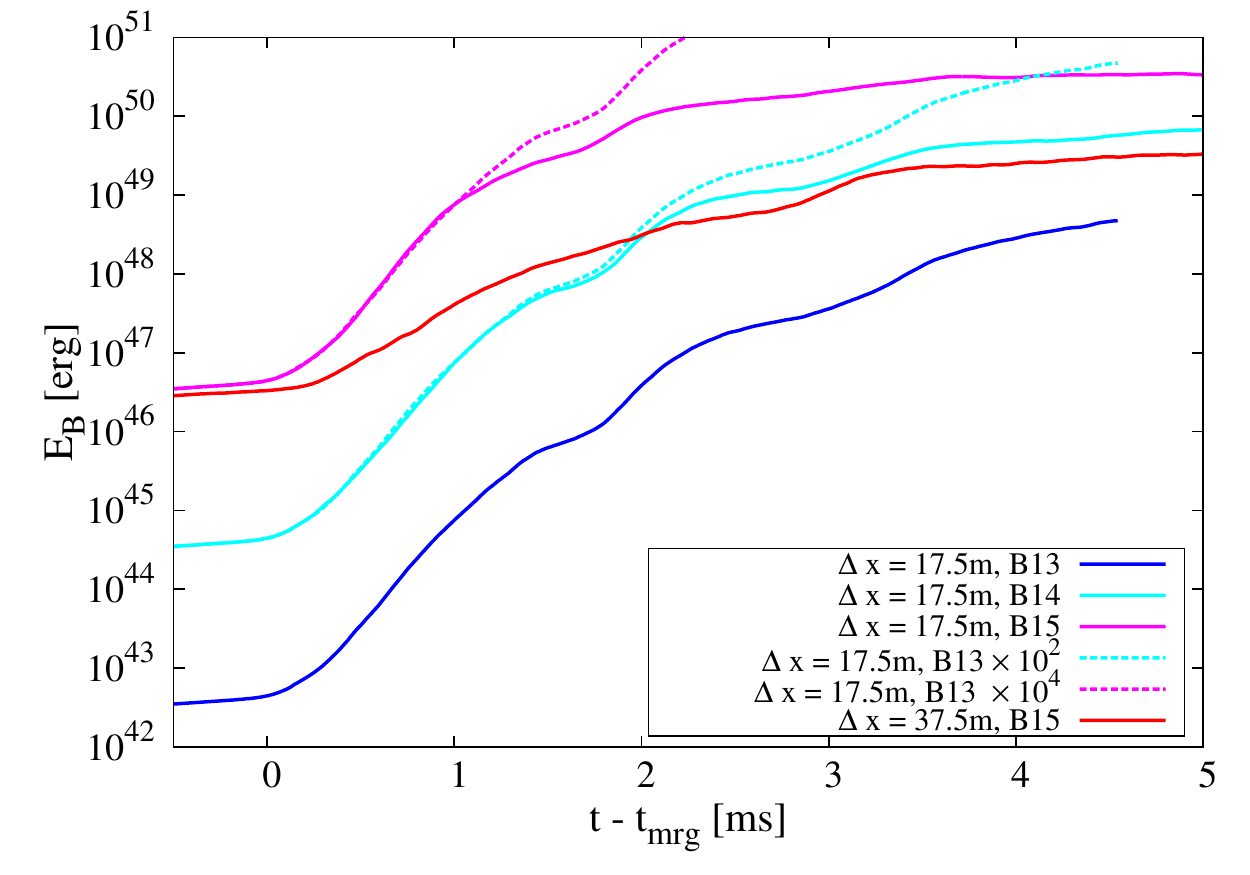}
\end{center}
\caption{\emph{Left panel:} Growth rate of the magnetic-field energy as a
  function of the final grid resolution. \emph{Right panel:} Evolution of
  the magnetic-field energy for a number of different simulations with
  different initial magnetic fields (see legend). The cyan and magenta
  dashed curves show the evolution when the initial magnetic field of
  $B_0 \sim 10^{13}\,{\rm G}$ has been magnified by a factor of $10^2$
  and $10^4$, respectively. [Reprinted with permission from
    Ref. \protect{} \cite{Kiuchi2015a}. \copyright~(2015) by the American
    Physical Society.]}
\label{fig:kiuchi_15}
\end{figure*}

However, simulations with very high resolution were carried out by Kiuchi
et al. \cite{Kiuchi2014}, who used the code already employed and tested
in Refs. \cite{Kiuchi2012b,Kiuchi2013}. With a grid spacing of $70\,{\rm
  m}$ on the finest grid and an evolution time of $\sim 100\,{\rm ms}$,
these computations were the most accurate and computationally expensive
simulations of magnetised BNS mergers, but have been further improved in
even more recent (and expensive) simulations by Kiuchi et
al. \cite{Kiuchi2015a}, where an impressive resolution of only
$17.5\,{\rm m}$ on the finest grid was used. In both works
\cite{Kiuchi2014, Kiuchi2015a}, the simulations used a nuclear-physics
EOS (\ie the stiff EOS H4 \cite{GlendenningMoszkowski91}, which is based
on relativistic mean field theory with hyperon effects). The study of the
various amplification mechanisms observed in these simulations confirmed
and improved previous results. In particular, the development of the KHI
was studied with great care and, thanks to the high resolution employed,
Kiuchi et al. \cite{Kiuchi2014, Kiuchi2015a} were able to show that the
amplification factor depends on the grid resolution but not on the
initial magnetic-field strength \cite{Obergaulinger10, Giacomazzo2011b,
  Rezzolla:2011}; this is nicely summarised in the left panel of
Fig. \ref{fig:kiuchi_15}, which shows that the growth rate actually
scales inversely with resolution \cite{Kiuchi2015a}. The figure also
shows a divergent behaviour, so that even with resolutions one order of
magnitude higher than other groups, Kiuchi et al. \cite{Kiuchi2015a} were
not able to observe the saturation that is however expected to take place
under more realistic conditions. This behavior can be understood in terms
of the basic properties of the KHI, whose cut-off wavenumber can only
increase with grid resolution unless some other cut-off is introduced by
other physical processes, \eg via viscosity or resistivity. These results
seem to suggest that, unless a better microphysical treatment is
introduced, \eg through a realistic treatment of viscosity and
resistivity, even resolutions orders of magnitude higher than those
commonly used nowadays may be insufficient to capture the realistic
development of the KHI in magnetised BNS mergers.

Another important aspect explored by the simulations of Kiuchi et
al. \cite{Kiuchi2012b,Kiuchi2015a} is that of the amplification of the
magnetic field once the KHI has fully developed. This is reported in the
right panel of Fig. \ref{fig:kiuchi_15}, which shows the evolution of the
magnetic-field energy for a number of different simulations with
different initial magnetic fields (see legend). The cyan and magenta
dashed curves show the evolution when the initial magnetic field of $B_0
\sim 10^{13}\,{\rm G}$ has been magnified by a factor of $10^2$ and
$10^4$, respectively. The data refers to the simulations reported in
Ref. \cite{Kiuchi2015a} and shows that amplifications in the magnetic
field of about three orders of magnitude are possible; these results also
show how the amplification of the magnetic field in such direct
simulations depends sensitively on the grid resolution, since Kiuchi et
al. \cite{Kiuchi2014} found a much smaller amplification (of one order of
magnitude only) when employing a resolution of $70\,{\rm m}$.

In summary, despite the computationally impressive efforts carried out
recently by Kiuchi et al. \cite{Kiuchi2014, Kiuchi2015a}, the issue of
the final amplification of the magnetic field as a result of the
development of the KHI at the merger of magnetised BNS systems remains
open. It is clear that the magnetic-field energy is amplified by at least
about three orders of magnitude after the merger and also that the
saturation energy of the magnetic-field is likely to be $\gtrsim 0.1\%$
of the bulk kinetic energy, \ie $\gtrsim 4 \times 10^{50}\,{\rm erg}$. It
is yet unclear, however, whether these results will continue to hold if
one considered lower but more realistic initial magnetic field, \eg $B_0
\sim 10^{8}-10^{10}\,{\rm G}$, and whether the newly produced
magnetic-field energy can actually saturate near equipartition. We should
however remark that, albeit almost prohibitively expensive, such direct
simulations are essential, since they can provide the correct physical
input for the subgrid modelling that may be necessary for less expensive
simulations investigating large parameter spaces.

Exploring a computationally less expensive approach to the development of
the KHI and hence to the amplification of the magnetic fields at the
merger, Giacomazzo et al. \cite{Giacomazzo:2014b} were the first to study
magnetic-field amplification in the KHI via a subgrid model for the
magnetic-field amplification. The resolution on the finest grid was about
$225\,{\rm m}$, but they introduced a subgrid model that served to
include within global direct simulations some effects of the small-scale
amplification of the magnetic field caused by turbulence. More in detail,
this subgrid model was intended to account for the electromotive forces
arising from unresolved fluctuations in the magnetic field and bulk fluid
velocity. The magnetic field evolution was modified according to the
strength of the fluid vorticity, the rest-mass density and the original
magnetic field, and tuning the magnitude of the magnification through the
results of high-resolution local special-relativistic simulations of
driven relativistic MHD turbulence made by Zrake et
al. \cite{Zrake2013b}. Among other results, Giacomazzo et
al. \cite{Giacomazzo:2014b} found that the magnetic field reaches
saturation very rapidly after the merger (cf. the right panel of
Fig.~\ref{fig:Obergaulinger+2010}), with different resolutions giving the
same amplification. While this is the first time that a similar result
has been obtained in BNS simulations, it is also a natural consequence of
the recipe adopted for the subgrid modelling, which effectively quenches
the magnetic-field growth at equipartition.

In a series of recent works, Neilsen et al. \cite{Neilsen2014} and
Palenzuela et al. \cite{Palenzuela2015} have described simulations from a
code that implements general-relativistic MHD with tabulated EOSs and a
neutrino leakage scheme to account for cooling via neutrino
emission\footnote{A more complete discussion of results about neutrino
  emission will be presented more in detail in
  Sect. \ref{sec:hd_nus}.}. The resolution of the finest grid was of
$460\,{\rm m}$ in Ref. \cite{Neilsen2014} and of $230\,{\rm m}$ in
Ref. \cite{Palenzuela2015}, which also adopted a subgrid modelling of the
magnetic field in terms of the fluid vorticity. Overall, these works
confirmed that the effects of the magnetic field and of neutrino
production and cooling play a subleading role in the dynamics of the
binary during the inspiral, and that during the merger the magnetic field
strength can increase to $10^{15}-10^{16}\,{\rm G}$ through compression,
the KHI and turbulent amplification.

\begin{figure*}
\begin{center}
  \raisebox{0.3cm}{\includegraphics[width=0.425\columnwidth]{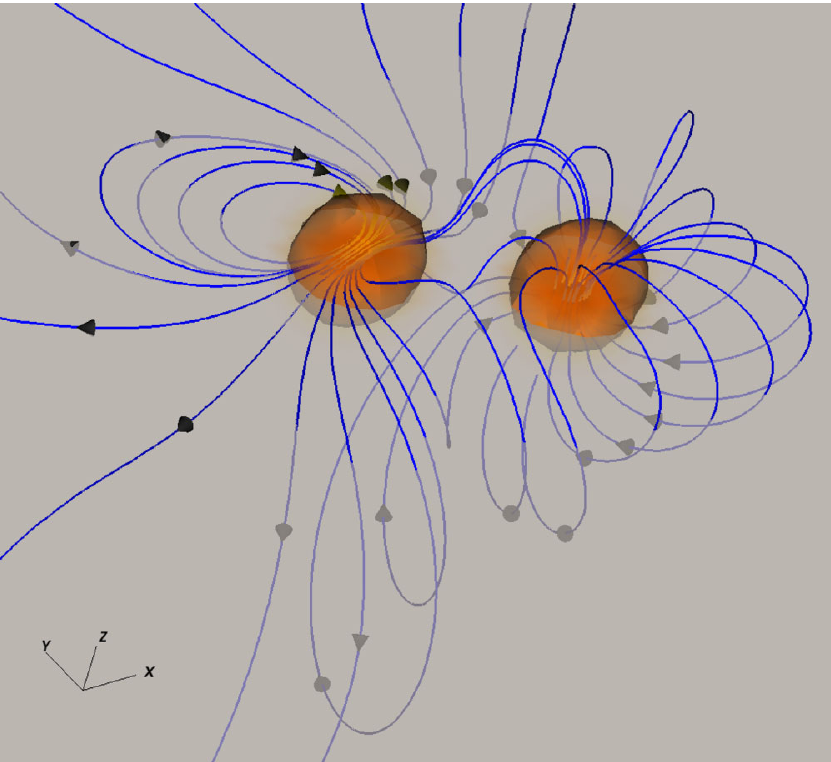}}
  \hskip 0.5cm                           
  \includegraphics[width=0.525\columnwidth]{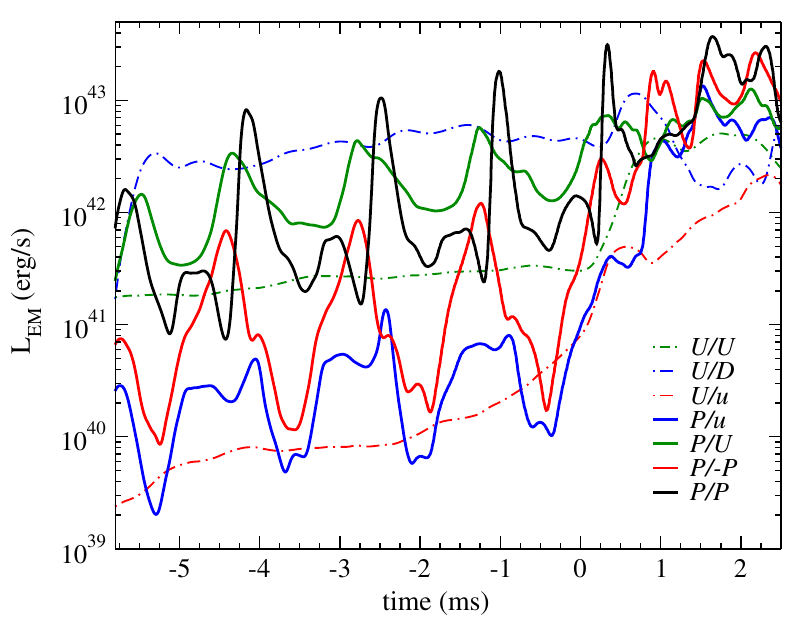}
\end{center}
\caption{\emph{Left panel:} Three-dimensional arrangement of
  magnetic-field lines, roughly 3 ms before merger, for an evolution
  starting with one star having a dipole magnetic field perpendicular to
  the orbital angular momentum and the other having it parallel (the
  \lrn{``$P/U$} case'' in the nomenclature of Ref. \cite{Ponce2014}). The
  orbital plane was made translucent (displayed in grey) to aid in
  distinguishing sections of the field lines lying above from those below
  the $z=0$ plane (the latter in less vivid colours). Notice that field
  lines naturally venture off the orbital plane. \emph{Right panel:}
  Luminosities for different magnetic-dipole configurations in a binary
  neutron-star system, shifted in time so that $t=0$ denotes first
  contact between the stars. Solid lines refer to configurations studied
  in Ref. \cite{Ponce2014}, dot-dashed lines to those studied in
  Refs. \cite{Palenzuela2013a, Palenzuela2013b}. The legend refers to the
  initial alignment of the initial dipole magnetic field of each star in
  the binary: $U$ means parallel to the orbital angular momentum, $u$
  means parallel to the orbital angular momentum but with a smaller
  intensity, $D$ means antiparallel to the orbital angular momentum, $P$
  means perpendicular to the orbital angular momentum. [Reprinted with
    permission from Ref. \protect{} \cite{Ponce2014}. \copyright~(2014)
    by the American Physical Society.]}
\label{fig:ponce_14}
\end{figure*}

A distinct, but equally interesting series of studies about the inspiral
part of magnetized BNS simulations has been focusing on the interaction
of the stellar magnetospheres just before the merger, exploring whether
the resulting electromagnetic radiation can represent a counterpart to
the gravitational-wave emission. A number of mechanisms have been
suggested in this regard \cite{Palenzuela2013a, Palenzuela2013b}, \eg the
interaction with the interstellar medium in the form of synchrotron radio
emission, the emission of flares induced by resonant excitations of
neutron-star modes by tides that could induce crust cracking, and
unipolar induction \cite{Goldreich1969a}, the extraction of stellar
kinetic energy through the interaction of the stellar magnetosphere with
an external magnetic field. This last process can be modeled as a perfect
conductor moving through an ambient magnetic field, which induces a
charge separation on its surface and drives electrical currents. The
kinetic energy from the moving conductor is extracted in the form of MHD
waves propagating along the magnetic field lines, or of particle
acceleration as in ordinary pulsars. In terms of simulations, a
systematic investigation has been carried out by Palenzuela et
al. \cite{Palenzuela2013a, Palenzuela2013b}, who studied in RMHD the
electromagnetic emission originating a few orbits before the merger of a
BNS\footnote{Although with an admittedly coarse resolution of about
  $300\,{\rm m}$, the system was also followed up to a few milliseconds
  after the merger, thus including the dynamics of the magnetic fields
  during the formation of the HMNS.}. By using the Poynting flux as a
first approximation to the energetics, Palenzuela et
al. \cite{Palenzuela2013a,Palenzuela2013b} found that the emitted power
can outshine pulsars in binaries, that it displays a distinctive angular-
and time-dependent pattern, and that it radiates within large opening
angles. These properties suggest that some BNS mergers could yield
interesting \lrn{prompt} electromagnetic counterparts to
gravitational-wave events, although the comparatively low luminosity of
$\sim 10^{40}-10^{42}\,{\rm erg/s}$ and the large error box of
gravitational-wave detectors would make the identification rather
challenging.

The systematic work of Refs. \cite{Palenzuela2013a, Palenzuela2013b} was
completed in a follow-up analysis \cite{Ponce2014a}, where the same group
studied the dependence of the electromagnetic luminosity on the
inclination of the dipolar magnetic fields relative to the orbital plane;
in principle, the dipoles are expected to be arbitrarily oriented, while
they were assumed perpendicular to the orbital plane in the previous
works. In this way it was pointed out that indeed there is a strong
dependence on the dipole orientations, as it can be seen in the right
panel of Fig. \ref{fig:ponce_14}. This dependence can be linked to the
reconnection and redistribution of the magnetic field (the shape of the
magnetic-field lines can become rather complex, as shown in the left
panel of Fig. \ref{fig:ponce_14}) as the stars interact. In particular,
the luminosities are roughly periodic, accompanied by an overall increase
as the orbit tightens, and the local maxima of the luminosity occur when
the magnetic fields emanating from each star reconnect. The Poynting flux
was found to be not strongly collimated and thus producing isotropic
emissions. Finally, as we will further discuss in Sect. \ref{sec:atog},
Ponce et al. \cite{Ponce2014a} also examined whether the characteristics
of the electromagnetic counterparts can provide an independent way to
test gravity in the strong regime. They found that in some cases the
electromagnetic flux emitted by binaries in scalar-tensor theories may
show small but potentially measurable deviations from the prediction of
general relativity.

\subsubsection{Post-merger dynamics: short-lived merger product and
  black-hole--torus system}
\label{sec:m_pmd}

We now shift to describing the progress of MHD simulations relative to
the post-merger phase when the binary-merger product is short lived. We
recall that we define the binary-merger product as whatever object is
produced after the merger, bearing in mind that this object can actually
change its nature over time. In fact, depending on the total mass and
mass ratio of the binary, the EOS, and the time after the merger, this
can be a \emph{stable} object, \ie a black hole or a neutron star, or a
\emph{metastable} object that will eventually reach, on timescales that
can be much larger than the dynamical timescale, one of the two stable
states mentioned above. In the phase in which the binary-merger product
is a metastable object (if this phase exists at all), the binary-merger
product will either be an SMNS (\ie a uniformly rotating star with mass
above the maximum mass for nonrotating star $M_{_{\rm TOV}}$, but below
the maximum mass for uniformly rotating stars $M_{\rm max}$, with $M_{\rm
  max} \simeq 1.20\,M_{_{\rm TOV}}$ \cite{Breu2016}), or an HMNS (\ie a
differentially rotating star more massive than an SMNS). When talking
about a {\it short/long-lived} binary-merger product we refer only to the
metastable phase of the binary-merger product before its collapse to
black hole, hence considering whether the metastable object lives for a
{\it short} or {\it long} time before collapsing to a black hole. In this
Section, in particular, we will concentrate on those scenarios where the
progenitors neutron stars have masses sufficiently large so that the
binary-merger product collapses rather rapidly (\ie within $\sim 1\,{\rm
  s}$) to a rotating black hole, producing a black-hole--torus system
(see Sect. \ref{sec:EM_counterparts} for a discussion of long-lived
binary-merger products).

Before dwelling on the details of this section, it may be useful to
remark right at the outset that simulating this stage of the binary
evolution is particularly challenging and possibly the most difficult
aspect of the simulation of merging BNSs. This is because after the
merger convergence order of any known numerical method reduces to one or
less, while ultra-high accuracy is needed to resolve those MHD
instabilities as the MRI that are supposed to play a crucial role. It
will probably be difficult to resolve such instabilities in current and
near-future simulations. Important additional factors that make
post-merger simulations more challenging are related to both missing
physical input, such as accurate neutrino treatment, and numerical
issues, such as the dependence of the HMNS lifetime on the grid
resolution, the AMR grid setup, the boundary location and conditions, and
the unresolved numerical resistivity.

The first, pioneering works in IMHD, reported in
Refs. \cite{Anderson2008, Liu:2008xy, Giacomazzo:2009mp}, which used
extremely high and unrealistic initial magnetic fields of $B_0 =
10^{15}-10^{17}\,{\rm G}$, explored the effects that magnetic fields have
on the lifetime of the HMNS created after the merger. Later on,
Giacomazzo et al. \cite{Giacomazzo2011b} reported a study with more
realistic initial magnetic fields and pointed out that the amplified
magnetic field in the HMNS can redistribute the angular momentum,
transporting it outwards and reducing the amount of differential rotation
that is essential in supporting the HMNS against gravitational
collapse. As a result, if the magnetic tension is sufficiently strong to
be comparable to or larger than the matter pressure gradients (\ie $B_{0}
\gtrsim 10^{8}\,{\rm G}$), magnetic fields can contribute to causing the
collapse of the HMNS \cite{Giacomazzo2011b, Rezzolla_book:2013}. For
larger initial magnetic fields (\ie $B_0 \sim 10^{12}-10^{17}\,{\rm G}$),
however, the amplified magnetic fields will also introduce a significant
magnetic pressure, which provides additional pressure support and thus
either compensates or even dominates the angular-momentum redistribution,
with an overall delay of the collapse \cite{Giacomazzo2011b}.

\begin{figure*}
  \begin{center}
     \includegraphics[angle=0,width=4.4cm]{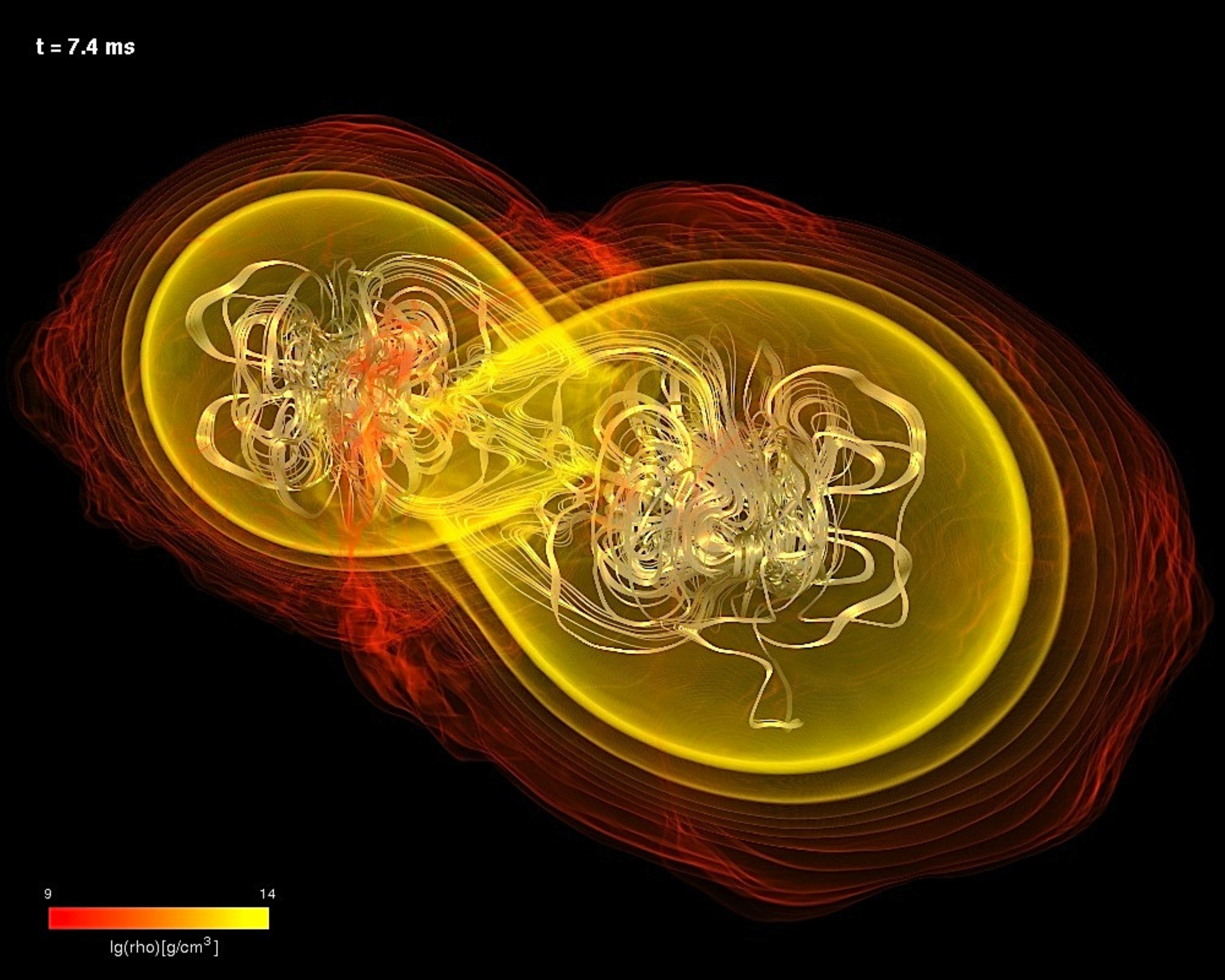}
     \includegraphics[angle=0,width=4.4cm]{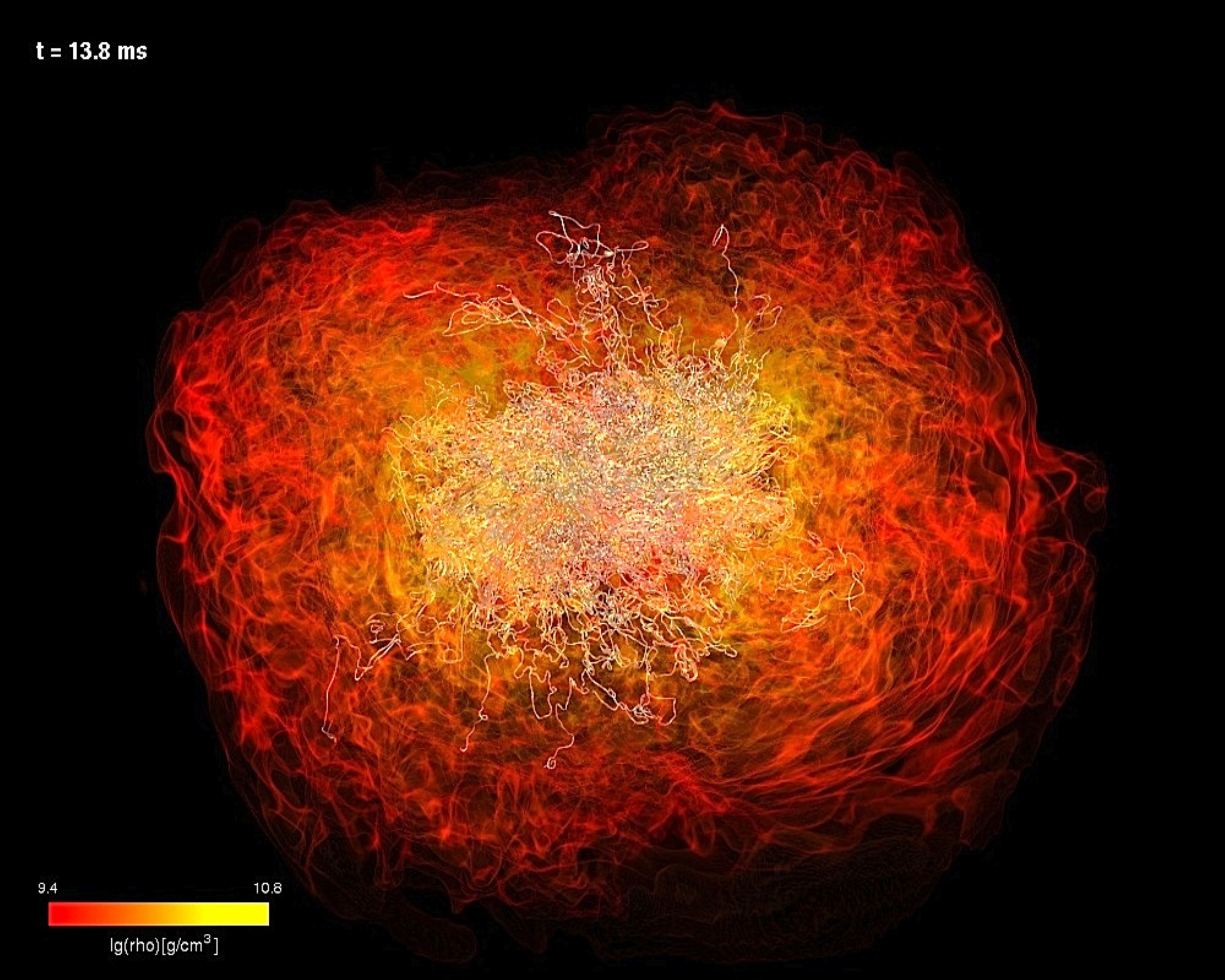}
     \includegraphics[angle=0,width=4.4cm]{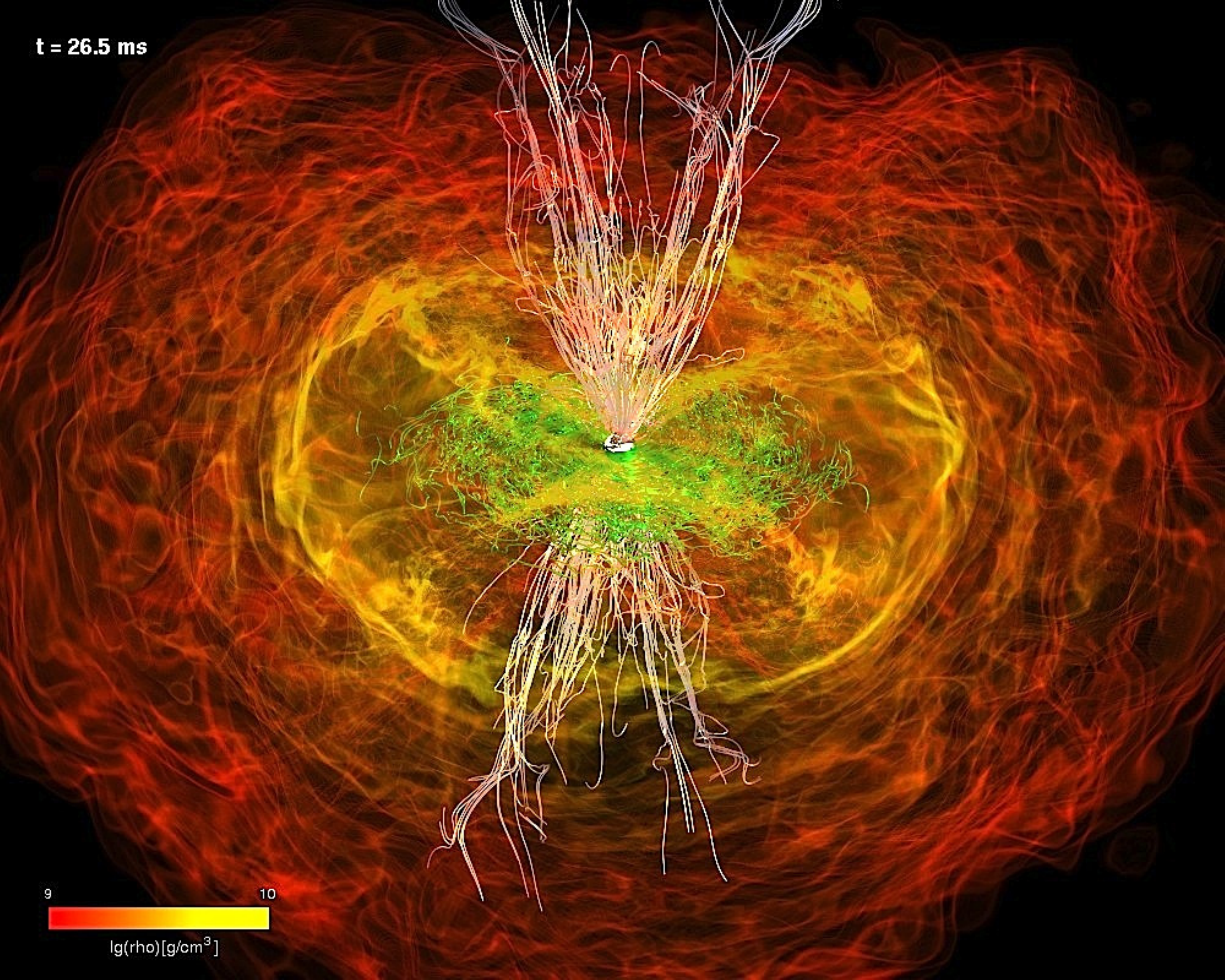}
  \end{center} 
  \caption{Snapshots at representative times of the evolution of the
    binary and of the formation of a large-scale ordered magnetic
    field. Shown with a colour-code map is the rest-mass density, over
    which the magnetic-field lines are superposed. The panels refer to
    the binary during the merger ($t=7.4\,{\rm ms}$), \textit{before} the
    collapse to black hole ($t=13.8\,{\rm ms}$), and \textit{after} the
    formation of the black hole ($t=26.5\,{\rm ms}$). Green lines sample
    the magnetic field in the torus and on the equatorial plane, while
    white lines show the magnetic field outside the torus and near the
    black-hole spin axis. The inner/outer part of the torus has a size of
    $90$ to $170\,{\rm km}$, while the horizon has a diameter of
    $\simeq9\,{\rm km}$. [Adapted from Ref. \cite{Rezzolla:2011} with
      permission by the authors.]} \label{fig:missinglink}
\end{figure*}

The study of Giacomazzo et al. \cite{Giacomazzo2011b} was then improved
by Rezzolla et al. \cite{Rezzolla:2011}, whose \lrn{analysis of the
  post-merger evolution was extended to much longer times, thus providing
  the first evidence of} the connection between the merger of magnetized
BNSs and SGRBs (see also Sect. \ref{sec:hydro_merger_post-merger}). In
particular, it was shown for the first time that, after black-hole
formation, a poloidal component of the magnetic field around the
black-hole rotation axis is generated in addition to the predominant
toroidal component that persists in the accreting torus. Representative
snapshots of the rest-mass density and of the magnetic-field lines are
shown in Fig. \ref{fig:missinglink}. The low-matter-density ``funnel'' or
``magnetic-jet'' structure \lrn{(\ie a structure characterized by a
  large-scale ordered poloidal magnetic field along the black-hole spin
  axis)} may be arguably connected to the launch of jets related to
SGRBs, even if it should be remarked that \emph{no ultrarelativistic
  outflow} was observed in the funnel produced in those simulations. An
intense outflow of matter was instead seen from the torus and, in
particular, in the regions that surround the magnetic jet. Within the
IMHD approximation, this outflow was able to stretch the magnetic field
lines and lead to the formation of the funnel with predominantly poloidal
magnetic field.

\begin{figure}
  \begin{center}
    \includegraphics[width=\columnwidth]{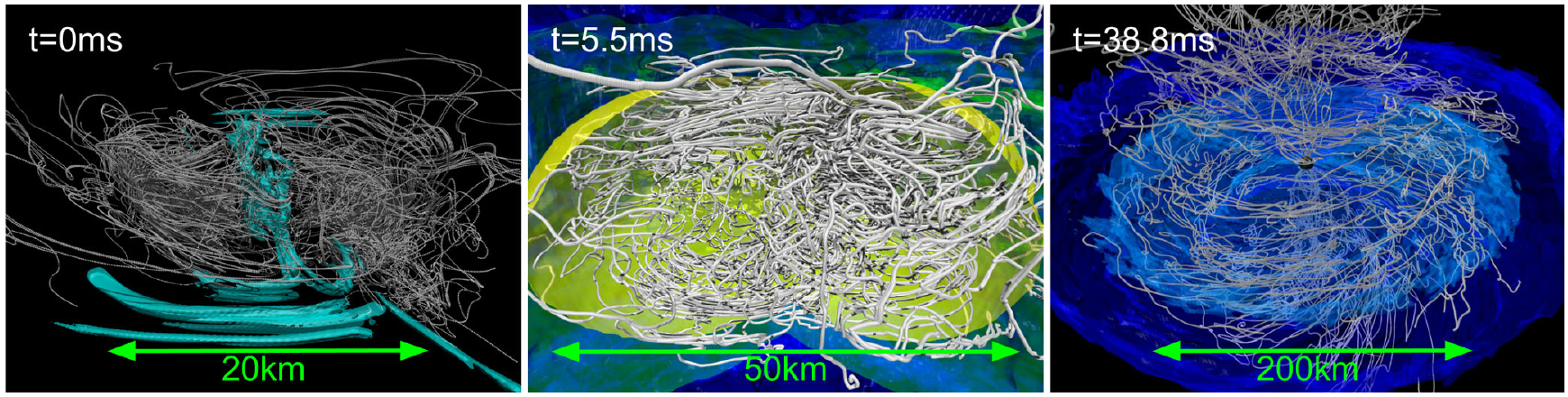} 
  \end{center}
  \caption{Snapshots of the density, magnetic-field strength and
    magnetic-field lines. The left, middle, and right panels show the
    configuration just after the onset of the merger, during the HMNS
    phase, and when a black hole formed, surrounded by an accretion
    torus, respectively. In each panel, the white curves are the
    magnetic-field lines. In the left panel, the cyan colour represents
    magnetic fields stronger than $10^{15.6}$ G. In the middle panel, the
    yellow, green, and dark blue surfaces represent the density
    iso-surfaces of $10^{14}$, $10^{12}$, and $10^{10}\, {\rm g/cm}^{3}$,
    respectively. In the right panel, the light and dark blue surfaces
    are the density iso-surfaces of $10^{10.5}$ and $10^{10}\, {\rm
      g/cm}^{3}$, respectively. [Reprinted with permission from
      Ref. \protect{} \cite{Kiuchi2014}. \copyright~(2014) by the
      American Physical Society.]}
   \label{fig:Kiuchi2014_Fig1}
\end{figure}
\begin{figure*}
  \begin{center}
     \includegraphics[angle=0,width=4.4cm]{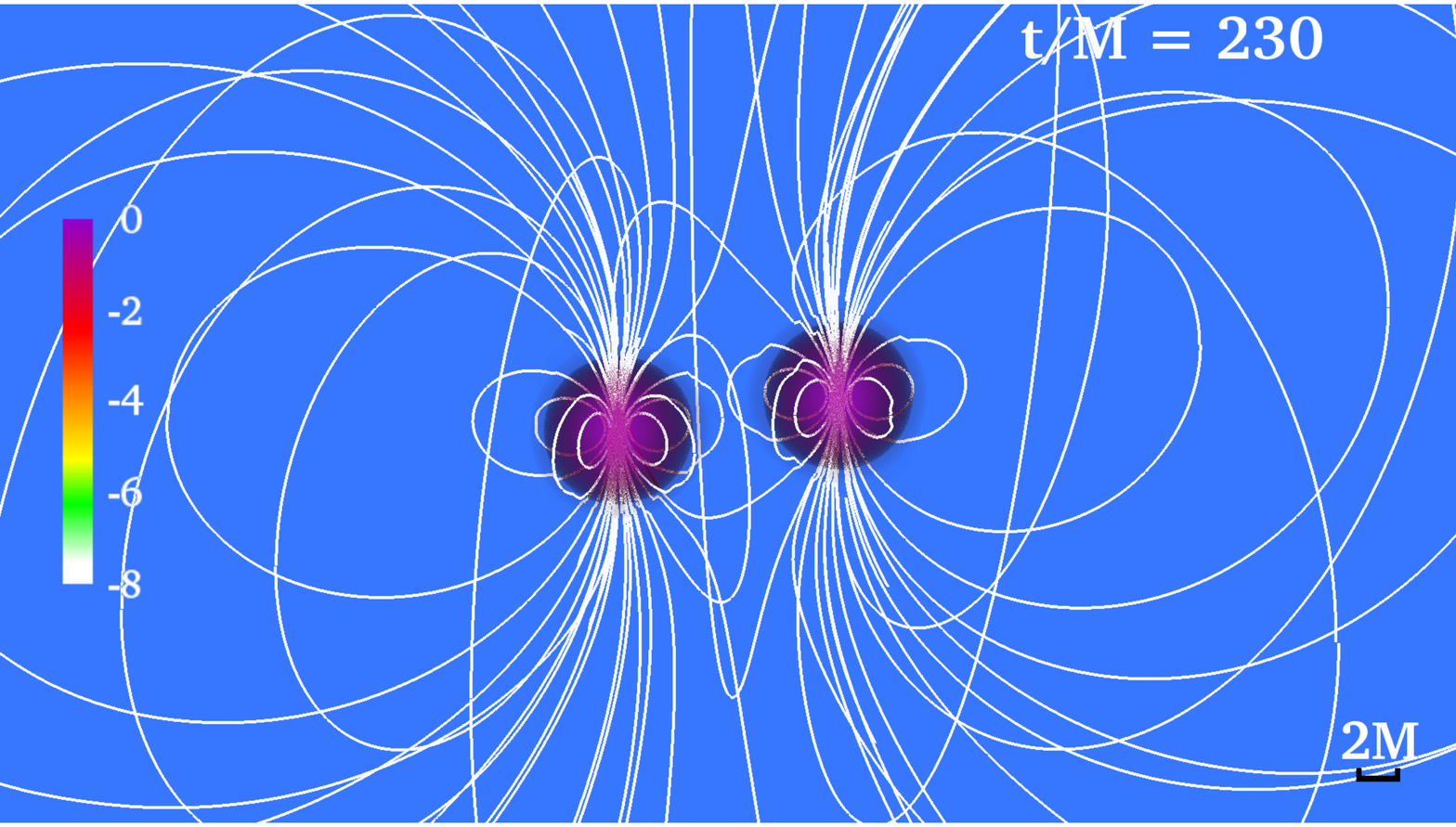}
     \includegraphics[angle=0,width=4.4cm]{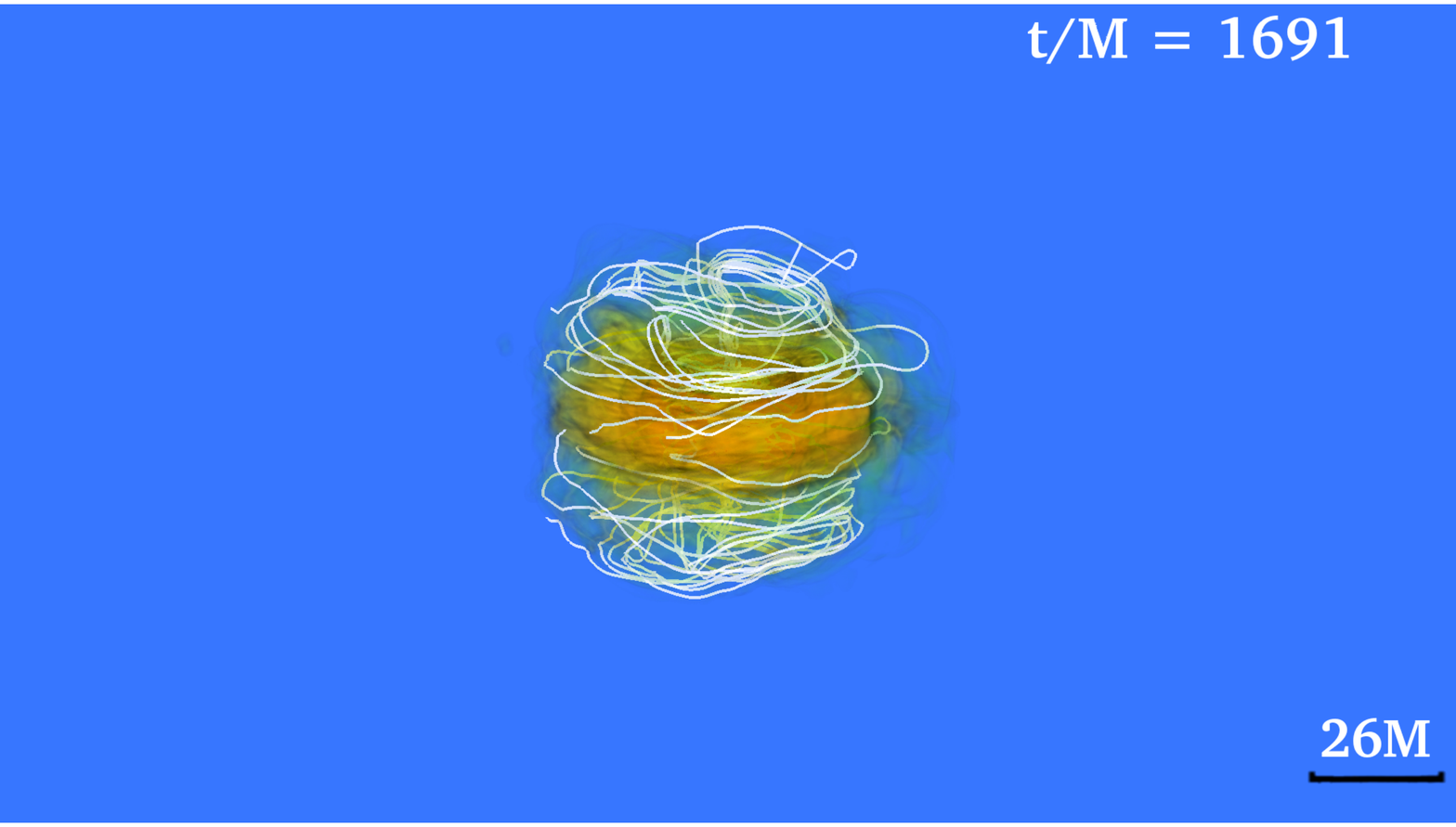}
     \includegraphics[angle=0,width=4.4cm]{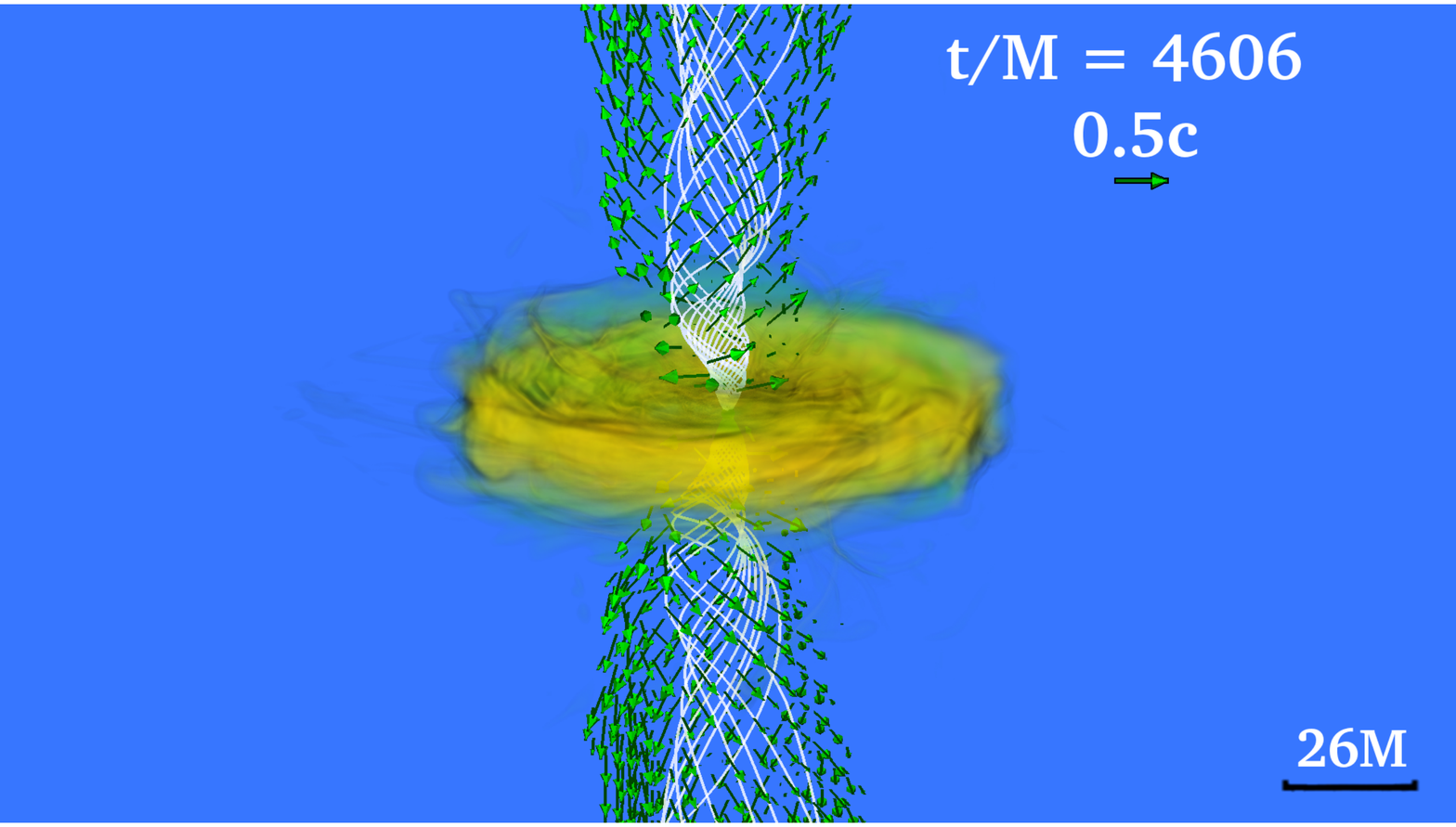}
  \end{center} 
  \caption{Snapshots of the rest-mass density, normalized to its initial
    maximum value $\rho_{0,\text{max}}= 5.9\times 10^{14}\, {\rm
      g/cm}^{3}$ shown in a logarithmic scale at selected times.  Arrows
    indicate plasma velocities, and white lines show the magnetic-field
    structure. [Adapted from Ref. \cite{Ruiz2016} with permission by
      authors.]}
  \label{fig:ruiz2016}
\end{figure*}

In a related work, Aloy et al. \cite{Aloy2012}, studied magnetic
instabilities in BNS mergers comparing the results of two different
numerical approaches: the above-mentioned global numerical simulations of
Rezzolla et al. \cite{Rezzolla:2011} and the local numerical simulations
of Obergaulinger et al. \cite{Obergaulinger10}, which can be run with
numerical resolution much higher than that of global simulations and
adequate to capture the fastest growing modes. While this is a very
interesting approach, the results of Aloy et al. \cite{Aloy2012} found
significant differences between local and global simulations. These could
be due to a number of causes and primarily to the fact that the
development of the MRI in BNSs may be influenced by the varying
background flow that is produced by the colliding stellar cores and that
cannot be reproduced in the local simulations.

The scenario of magnetic-field amplification in IMHD simulations has also
been investigated with the high-resolution simulations of Kiuchi et
al. \cite{Kiuchi2014, Kiuchi2015a}, who describe in detail how the growth
of the magnetic-field energy in the HMNS phase is attributable to
nonaxisymmetric MRI in the low-density regions, while magnetic winding
contributes to the growth of the toroidal magnetic-field energy as
well. Figure \ref{fig:Kiuchi2014_Fig1} shows some representative
snapshots of the rest-mass density and magnetic-field lines as obtained
from the simulation of Ref. \cite{Kiuchi2014}. The last panel in
particular seems to show the formation of a poloidal component, though
not as clear as in the simulations \cite{Rezzolla:2011}. The different
behaviour may be due to the different EOSs employed in the simulations,
\lrn{or to the different visualisation of the magnetic-field lines}. In
particular, Kiuchi et al. \cite{Kiuchi2014} use a much stiffer EOS (H4
\cite{GlendenningMoszkowski91}), which leads to a smaller matter outflow
and, consequently, to a weaker poloidal-field component; the latter, we
recall, is generated by the motion of matter along the edges of the torus
near the rotation axis since in IMHD the magnetic-field lines are
advected by the fluid.

Indeed, the generation of a coherent poloidal magnetic field at late
times was also observed in other simulations employing the ideal-fluid
EOS, such as those of black-hole--neutron-star binaries of Etienne et
al. \cite{Etienne2012b} or, more recently, of Paschalidis et
al. \cite{Paschalidis2014} for neutron-star--black-hole systems, and of
Ruiz et al. \cite{Ruiz2016} for BNS systems. Indeed, it is worth noting
that the most recent work of Ruiz et al. \cite{Ruiz2016} is very similar
to the one by Rezzolla et al. \cite{Rezzolla:2011} in terms of EOS and
stellar properties, but uses higher initial magnetic fields and has a
different treatment of the ``atmosphere'', which could explain why, in
addition to a coherent magnetic structure, Ruiz et al. \cite{Ruiz2016}
\lrn{find that the funnel becomes magnetically dominated at the end, and}
are able to measure a sustained outflow (although only mildly
relativistic). This is shown in the representative snapshots reported in
Fig. \ref{fig:ruiz2016}.

Finally, as an additional confirmation of the robustness of the process,
the formation of a coherent magnetic-jet structure was observed also in
the work of Dionysopoulou et al. \cite{Dionysopoulou2015}, who have
evolved BNSs in RMHD (see discussion below); also in this latter case,
however, only a magnetic structure was formed and no sustained
ultra-relativistic outflow was observed\footnote{It is presently unclear
  whether the jet acceleration mechanism is the result of some energy
  extraction from the accreting black hole, as in the Blandford-Znajek
  mechanism \cite{Blandford1977}, or of some other acceleration
  mechanism, \eg as in the Aloy-Rezzolla booster
  \cite{Aloy:2006rd}.}. Overall, the results of
Refs. \cite{Rezzolla:2011, Dionysopoulou2015, Kiuchi2015, Ruiz2016} are
interesting because they provide proof-of-principle evidence that the
merger of magnetised BNSs could provide the basic physical processes
necessary to explain the phenomenology invoked to explain the
observations of SGRBs. Hence, they serve as a link between the
theoretical modelling of BNSs and the observations of SGRBs; however only
an electromagnetic counterpart to a BNS merger can cast this link on an
undisputed ground.

Resistive MHD codes are harder to build and use because of the increased
complexity of the equations, because of the additional difficulties posed
by their numerical solution (the equations easily become stiff in regions
of high conductivity), and because the choice of realistic values for the
resistivity is far from trivial (see, \eg \cite{Uzdensky2011}), as
experiments and current astronomical observations do not set any
stringent constraint on its values at the temperatures, rest-mass
densities, and magnetic fields appearing in BNS mergers. The resistive
MHD code \texttt{WhiskyRMHD} of Dionysopoulou and collaborators
\cite{Dionysopoulou2015} matches the highly conducting stellar interior
to an electrovacuum exterior. This is different from and complementary to
the approach of Refs. \cite{Palenzuela2013a, Palenzuela2013b, Ponce2014},
in which the resistive description is matched to a force-free one in
order to study the interaction of the two stellar magnetospheres before
the merger (see Sect. \ref{sec:EM_counterparts} for more details). This
choice was made because the focus of Dionysopoulou et
al. \cite{Dionysopoulou2015} was the dynamics of post-merger objects,
which are surrounded by matter emitted through the large baryonic winds
produced after the merger, and to ensure that in the low-density regions
the electromagnetic fields behave as if in vacuum.

\begin{figure*}
  \begin{center}
     \includegraphics[width=0.49\columnwidth]{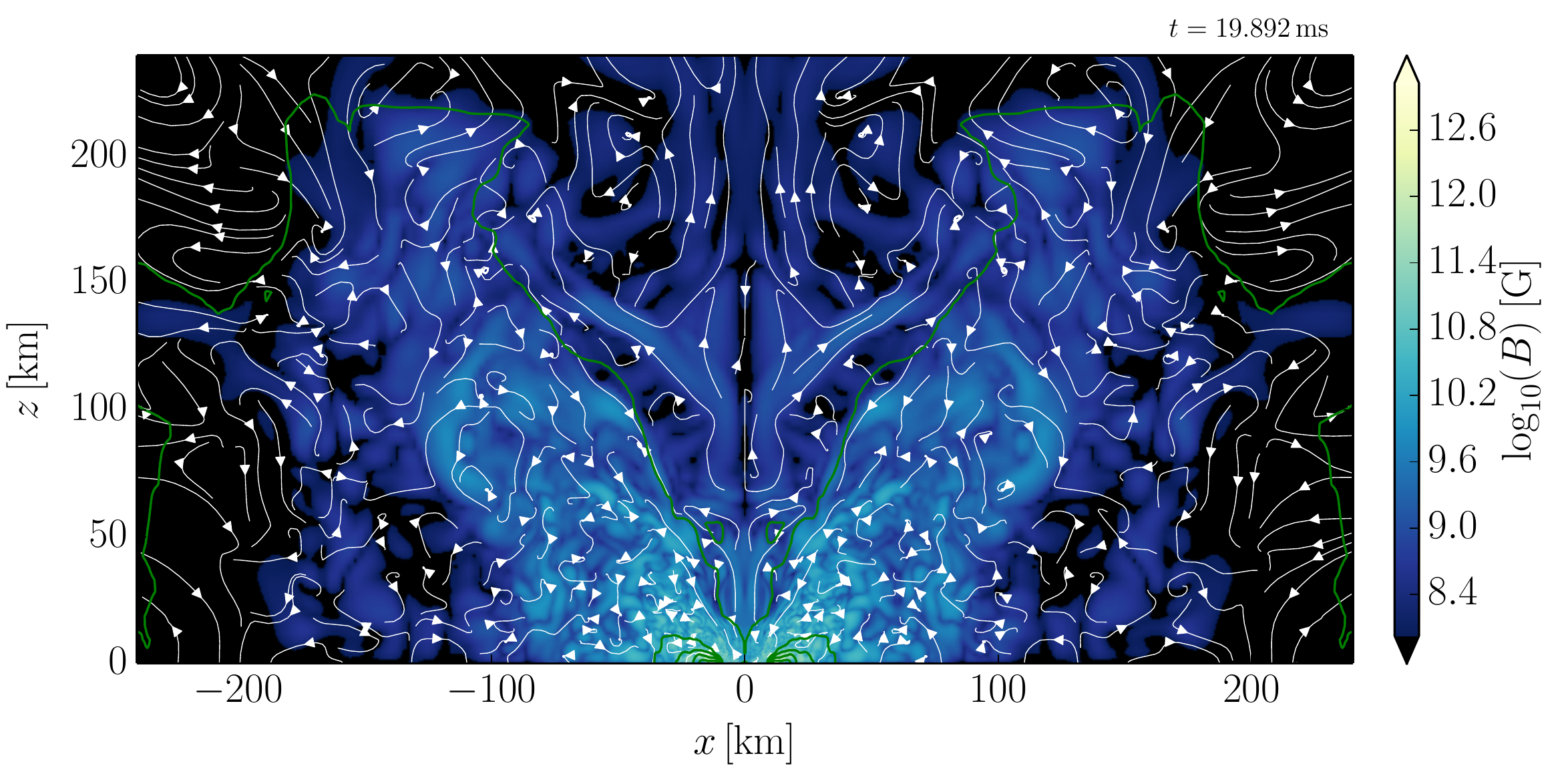}
     \hskip 0.2cm
     \includegraphics[width=0.49\columnwidth]{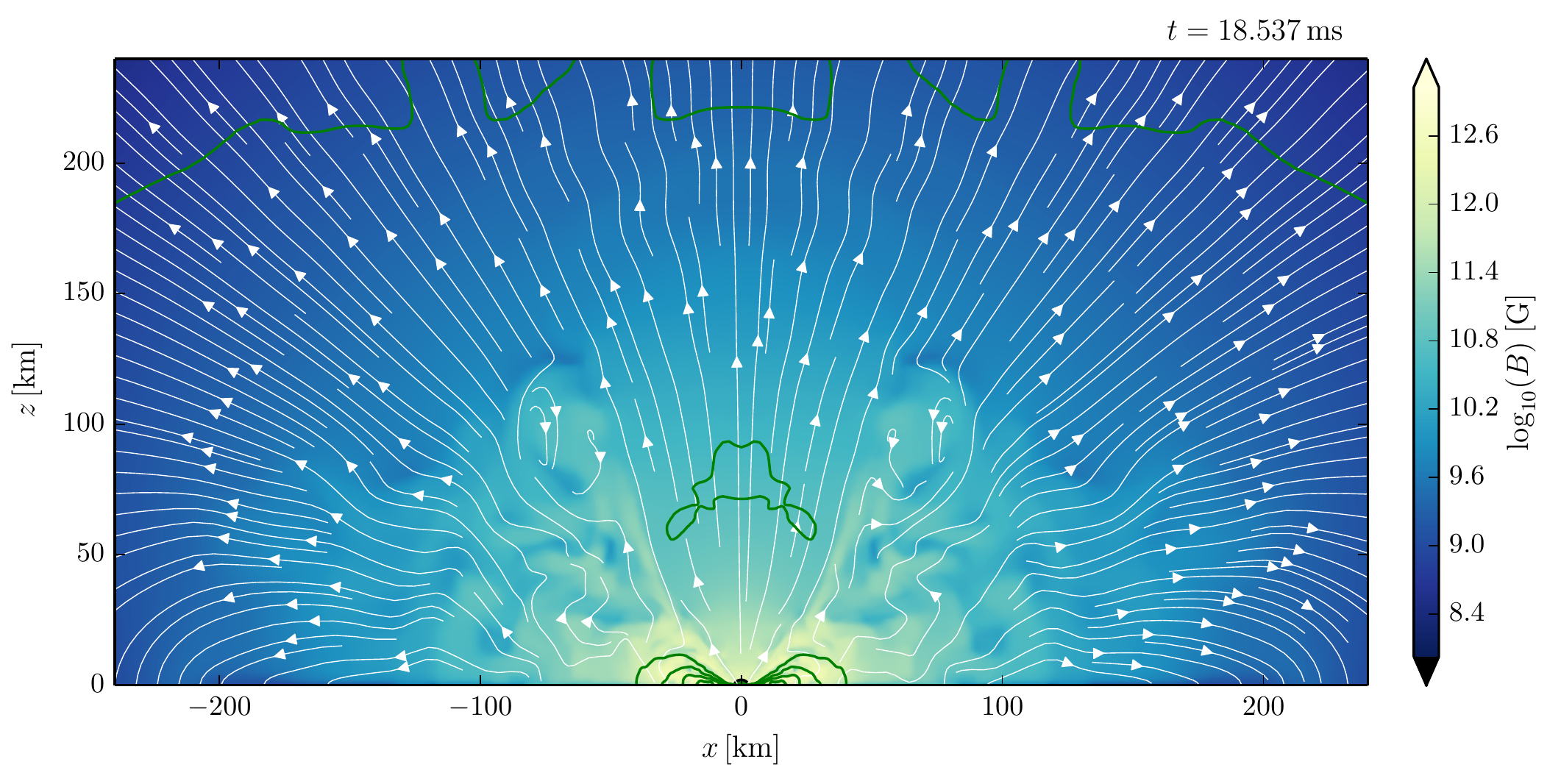}
  \end{center} 
  \caption{Large-scale two-dimensional snapshots on the $(x,z)$ planes of
    the magnetic field in the case of a simulation in IMHD (left panel)
    or RMHD (right panel). The snapshots refer to slightly different
    times but in both cases to when a black hole has already formed. Note
    again the formation of a magnetic-jet structure around the black-hole
    rotation axis, which extends on scales that are much larger than
    those of the accreting torus. Note also that in the RMHD case, the
    magnetic field shows a coherence on the largest scales of the
    system. [Reprinted with permission from Ref. \protect{}
      \cite{Dionysopoulou2015}. \copyright~(2015) by the American
      Physical Society.}
\label{fig:RMHD}
\end{figure*}

Using such a resistive code, Dionysopoulou et
al. \cite{Dionysopoulou2015} performed a systematic comparison of the
dynamics of equal-mass magnetized BNSs when simulated either in IMHD or
in RMHD. One of the most important differences found is that, since in
RMHD the magnetic field is not perfectly locked with the plasma,
resistive simulations show a less efficient redistribution of the angular
momentum in the HMNS. This, in turn, causes an increase in the lifetime
of the HMNS. Another difference is that the modulus of the magnetic field
along the rotation axis is about two orders of magnitude larger than in
the IMHD simulation. This is due mainly to the intense currents produced
by the rapidly rotating torus and also to the magnetic-field diffusion of
the strong magnetic field in the torus across the walls of the funnel. As
mentioned above, Dionysopoulou et al. \cite{Dionysopoulou2015}, also
confirmed the formation of a low-density funnel produced by a
predominantly poloidal magnetic field along the black-hole rotation axis,
although also in this case no ultra-relativistic outflow was observed in
the simulation. The magnetic-jet structure that forms can be viewed as a
quasi-stationary structure that confines the tenuous plasma in the funnel
away from the dense plasma in the torus. However, the plasma in the
funnel does not have sufficient internal energy to launch a relativistic
outflow. It is also possible that reconnection processes or neutrino pair
annihilation, not treated in their work, could generate the energy
required for launching a relativistic outflow along the baryon-poor
funnel. Two representative snapshots of these simulations are presented
in Fig. \ref{fig:RMHD}, which displays large-scale two-dimensional views
on the $(x,z)$ planes of the magnetic field for an IMHD simulation (left
panel) or an RMHD one (right panel). The snapshots refer to when a black
hole has already formed. It should be noted that the formation of a
magnetic-jet structure around the black-hole rotation axis extends on
scales that are much larger than those of the accreting torus. Note also
that in the RMHD case, the magnetic field shows a coherence on the
largest scales of the system.

Finally, Dionysopoulou et al. \cite{Dionysopoulou2015} noted the
potentially very interesting fact that a shearing boundary layer is
present at the interface between the magnetic-jet structure and the
torus. A KHI is expected to form there, but it was not investigated
because of the insufficient resolution.

In summary, a number of simulations suggest that when the binary-merger
product from the merger of magnetised BNSs collapses to a rotating black
hole, the matter in the region along the rotation axis does not have much
centrifugal support and accretes onto the black hole, leaving behind a
funnel of low-density matter. Both in IMHD and in RMHD simulations, the
magnetic fields in this region are stretched into a poloidal component,
either because of the inflow onto the black hole, or because of the
outflow along the torus edges, or both. While reasonable and
fundamentally what is expected, many details of this picture still need
to be cast on firmer grounds.

\subsubsection{Post-merger dynamics: long-lived merger product and extended 
  X-ray emission}
\label{sec:EM_counterparts}

We next shift to describing the progress of MHD simulations relative to
the post-merger phase, concentrating on those scenarios in which the
binary-merger product is long-lived (\ie does not collapse on a timescale
of $\sim 10^3-10^4\,{\rm s}$) and could be used to explain otherwise
puzzling astronomical observations. We recall, in fact, that the
\emph{Swift} satellite~\cite{Gehrels_etal2004} has revealed phases of
roughly constant luminosity in the X-ray afterglows of a large subclass
of SGRBs (\ie $\gtrsim 25\%$ of the full set of SGRBs). These are
referred to as ``X-ray plateaus'' (see, \eg Refs. \cite{Rowlinson2013,
  Gompertz2013}) and last $10-10^4\,\mathrm{s}$. The riddle is then in
the timescales involved, which are too long if the X-ray emission is
really an afterglow. Making the standard assumption that the gamma-ray
emission is associated to an ultra-relativistic jet launched by the black
hole as it accretes matter from the torus and since the torus mass is
$\lesssim0.1\,M_{\odot}$, with accretion rates $\sim
10^{-3}-10^{-2}\,M_{\odot}\,{\rm ms}^{-1}$~\cite{Rezzolla:2010,
  Hotokezaka2013}, the accretion timescale is at most
$\sim1\,{\rm~s}$. This is three or more orders of magnitude smaller than
the observed timescale for the \emph{sustained} X-ray emission.

A way out from this riddle is in principle available and involves the
presence a long-lived ``proto-magnetar'', that is, a uniformly rotating
object formed in the merger that powers the X-ray emission through
standard dipolar radiation and spin-down~\cite{Zhang2001, Gao2006,
  Fan2006, Metzger2008, Metzger:2011, Bucciantini2012}. By performing
general-relativistic MHD simulations of BNS mergers, Giacomazzo et
al. \cite{Giacomazzo2013} showed for the first time that the end result
of the merger may be a stable magnetar, surrounded by an extended
disc. This result is not particularly surprising given that the total
mass of the binary was chosen to be below the maximum mass of the
corresponding nonrotating star, but it was nevertheless useful to remark
that the violent dynamics at the merger is not sufficient to induce the
collapse of a binary with a subcritical mass. Giacomazzo et
al. \cite{Giacomazzo2013} also estimated the proto-magnetar typical
periods of the order of a few milliseconds, magnetic field strengths in
the range $B \sim 10^{15}-10^{16}\,{\rm G}$, and, under the assumption of
energy loss by pure dipole radiation, luminosities of $\sim
10^{46}-10^{49}\,{\rm erg/s}$. In a follow-up work, Dall'Osso et
al. \cite{DallOsso2014} studied the gravitational-wave emission from the
proto-magnetars deformed because of the large magnetisation
\cite{Ciolfi2010, Frieben2012, Ciolfi2013}, finding, as expected, that
such radiation is dependent on the EOS and yields a potential detection
rate of $0.1-1\,{\rm yr}^{-1}$ events with advanced detectors.

Although the formation of a proto-magnetar and its dipolar radiation can
help explain the long timescale over which the sustained X-ray emission
is observed in some SGRBs, it also introduces a different riddle. What is
difficult to explain in this case is the timing of the gamma- and X-ray
emissions. If the X-ray emission is produced by the binary-merger
product, then it \emph{cannot} follow the gamma-ray emission, which seems
to require a jet and hence a black hole. Indeed, none of the simulations
to date indicates the generation of a collimated jet by the binary-merger
product~\cite{Price06, Liu:2008xy, Giacomazzo:2010, Palenzuela2013a,
  Kiuchi2014, Giacomazzo:2014b, Kiuchi2015a}, which instead appears after
the formation of a black hole~\cite{Rezzolla:2011, Dionysopoulou2015,
  Ruiz2016}. To resolve this riddle a scenario has been developed that
involves a delay in the detection of the X-ray emission which is trapped
in the material ejected after the merger. The scenario has been suggested
simultaneously by two groups, \ie as the ``two-winds'' scenario
\cite{Rezzolla2014b} or as the ``time-reversal'' scenario
\cite{Ciolfi2014}; while both scenarios stem from initial joint
discussions among the authors, they have been developed entirely
independently and present slight differences\footnote{\lrn{More
    specifically, the model of Ref. \cite{Rezzolla2014b}, provides a
    detailed calculation of the resulting X-ray lightcurves and suggests
    as signature the presence of inverse-Compton scattered thermal
    cocoon photons that should show up at energies $> 10\,{\rm MeV}$ with
    a luminosity $\sim 10^{50}\,{\rm erg/s}$. The model of
    Ref. \cite{Ciolfi2014}, on the other hand, provides a more systematic
    analysis that the time delay between the prompt SGRB and the
    long-lasting X- ray signal can be large enough to explain the
    observed X-ray emission.}}. In what follows we concentrate on the
former, but details on the latter can be found in
Refs. \cite{Ciolfi2014}.

\begin{figure*}
  \begin{center}
    \includegraphics[width=6.0cm,height=7.0cm]{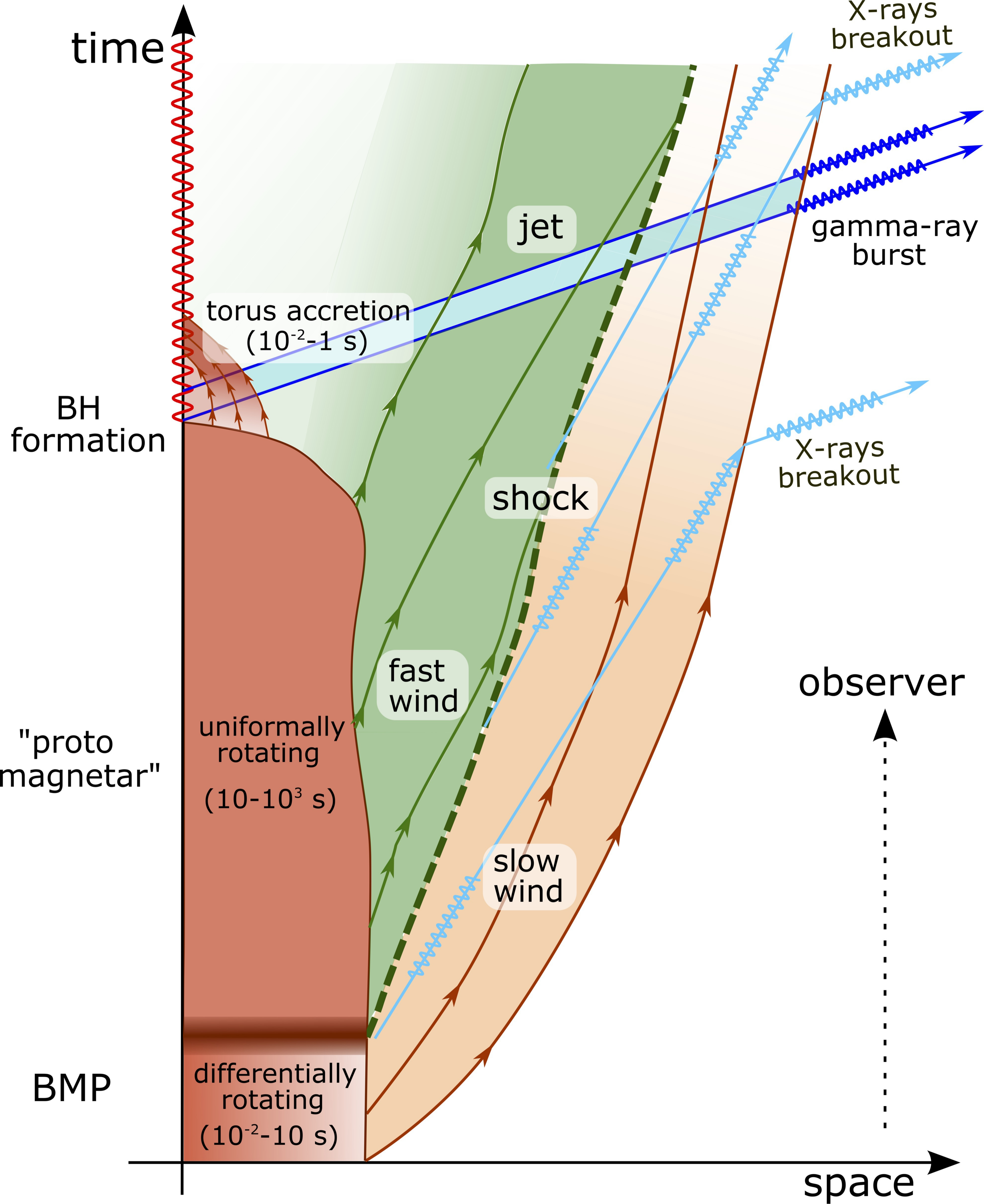}
    \hskip 1.0cm
    \raisebox{2.0cm}{\includegraphics[width=6.0cm]{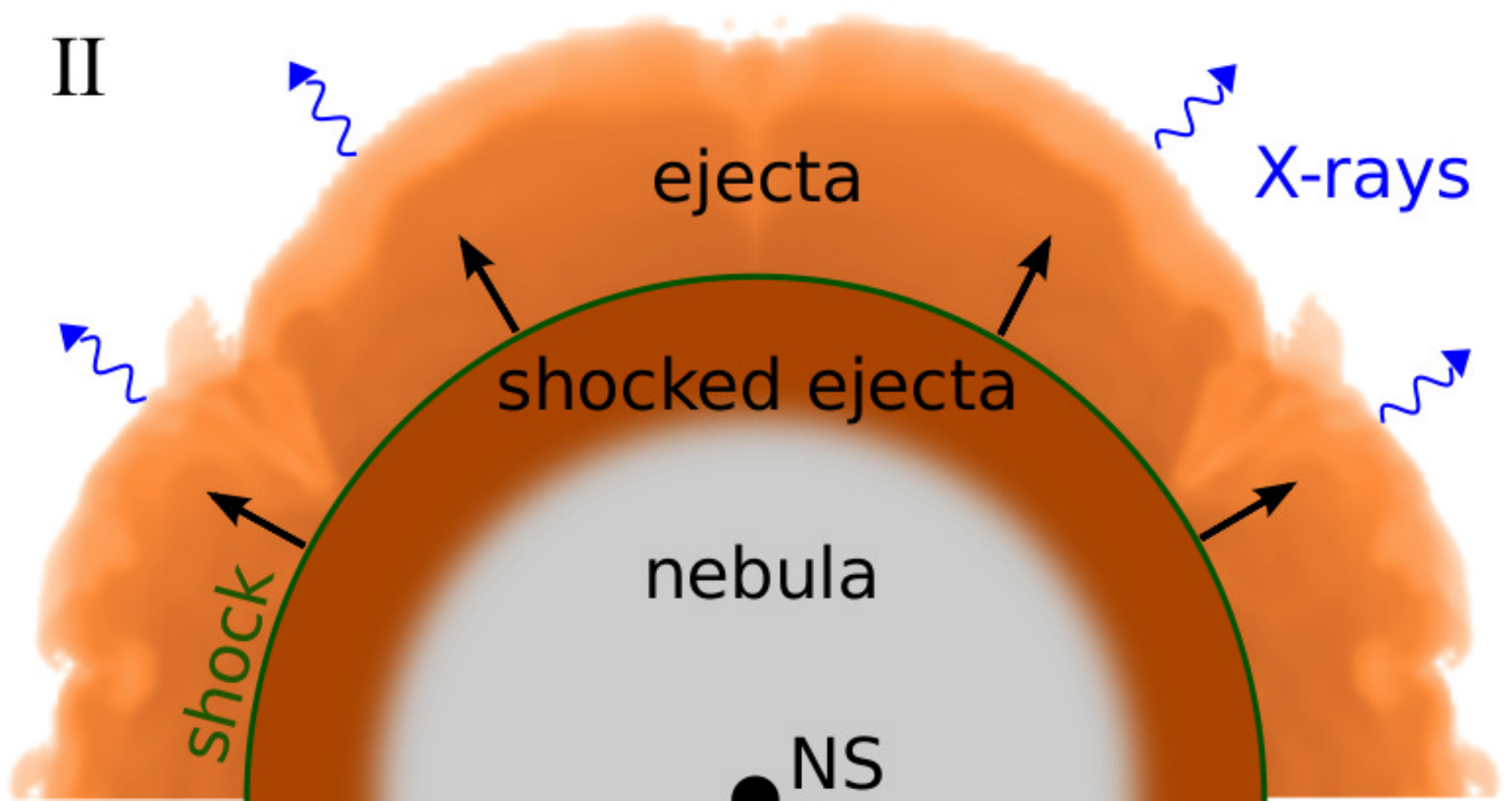}}
  \end{center}
  \caption{\textit{Left panel:} Schematic spacetime diagram showing in
    red the region occupied by the binary-merger product that eventually
    collapses leading to a black-hole--torus system. Shown in brown and
    green are the regions occupied by the magnetically driven slow wind
    and by the dipole-driven fast wind. The interaction of the two winds
    generates a shock and the sustained X-ray emission, while a jet is
    produced by the black-hole--torus system. [Reproduced from
      Ref. \cite{Rezzolla2014b} with permission by authors.]
    \textit{Right panel:} Schematic representation at a given time of the
    shock structure. [Reproduced from \cite{Ciolfi2014} with permission
      by authors.]}
\label{fig:twowinds}
\end{figure*}

The easiest way to describe this scenario is to discuss it via a
spacetime diagram as the one shown in the left panel of
Fig.~\ref{fig:twowinds}. There, shown as red-shaded is the region
occupied by the binary-merger product, which eventually collapses to
produce a rapidly rotating black hole surrounded by an accreting
torus. The binary-merger product is expected to rotate differentially for
an Alfv\'en timescale, \ie $\lesssim1-10\,{\rm s}$ and if it does not
collapse to a black hole when differential rotation is lost, it rotates
uniformly for considerably longer, \ie $\lesssim 10^3 - 10^4\,{\rm
  s}$. Shown instead as brown-shaded is the region occupied by a
\textit{slow} and baryon-rich wind, whose geometry is approximately
spherical and which moves at bulk speeds of $\sim0.01-0.1\,c$, then
progressively slowing down as it loses part of its kinetic energy to
climb out of the gravitational potential. There are a number of ways in
which this wind can be driven, possibly all acting at the same time: via
shock heating~\cite{Hotokezaka2013}, via magnetic fields and differential
rotation~\cite{Kiuchi2012b, Franci2013b, Siegel2014}, or via
neutrinos~\cite{Metzger2014, Perego2014}. In all cases, the duration of
the slow wind is $\lesssim 1 - 10\,{\rm s}$, and in the first two
scenarios the wind is isotropic for realistic magnetic-field
topologies~\cite{Siegel2014}.

Once differential rotation is lost, the magnetically driven wind is
quenched and the uniformly rotating and magnetized binary-merger product
will emit a \emph{fast} and baryon-poor wind (green-shaded area) moving
with bulk speeds of $\sim 0.3 - 0.5$. The binary-merger product can
provide a continuous source of dipole radiation over a timescale set by
the stability of the binary-merger product, \ie $\sim1\,{\rm
  s}-10^3\,{\rm s}$.

Since the slow and fast winds have different velocities, the latter
catches up with the former, producing a shock which heats the matter
locally and leads to an X-ray emission. At the same time, because the
matter of the slow wind is baryon rich and optically thick, the X-ray
photons cannot propagate freely, but rather diffuse through the slow-wind
material till reaching a photospheric radius from which they travel
directly towards the observer. The effective speed of propagation of the
X-ray photons is $\sim\,c/\tau$, where $\tau\gg1$ is the optical depth of
the slow wind where photons are produced, and the shock front moves
through the wind with a relative speed of $\sim\,c/5$, therefore X-ray
diffusion can be ignored until the shock is close to the photosphere. A
snapshot of the expanded wind structure is shown in the right panel of
Fig. \ref{fig:twowinds} and is taken from Ref. \cite{Ciolfi2014}.

As the X-ray photons ``slowly'' diffuse in the slow wind, the
binary-merger product will spin down via dipolar emission to a
sufficiently slow rate to collapse to a black hole surrounded by a hot
dense torus\footnote{As mentioned in Sect. \ref{sec:bbp}, it has been
  pointed out that the collapse of the uniformly rotating SMNS cannot
  lead to a torus surrounding the black hole as too little angular
  momentum is available in outer layers of the star
  \cite{Margalit2015}. At the same time, it has been noted that the
  binary-merger product prior to collapse is still expected to be
  surrounded by matter in quasi-Keplerian orbits, so that a torus is
  likely to be formed \cite{Rezzolla2014b}. This conjecture has still to
  be proven via simulations.}. magnetic instabilities will develop in
the torus, amplifying the magnetic field \cite{Rezzolla:2011, Kiuchi2014}
and leading to the construction of a jet-like magnetic structure
\cite{Rezzolla:2011, Dionysopoulou2015, Ruiz2016}. This magnetic funnel
can collimate the low-density material in its interior, which could be
heated either by the neutrinos emitted from the torus \cite{Ruffert99b},
or via magnetic reconnection. In addition, the matter ejected with the
slow wind can further confine the propagation of the jet \cite{Aloy:2005,
  Murguia-Berthier2014, Nagakura2014}. As a result, an ultrarelativistic
jet could be launched propagating with Lorentz factors
$\Gamma\sim100-1000$ (light-blue shaded area). Clearly, within this
scenario the dynamics of the jet across the winds material is similar to
the one envisaged for long GRBs, so that a burst of gamma rays is
expected to be produced as the jet breaks out, with luminosities of $L
\simeq 10^{50} - 10^{51}\,{\rm erg/s}$, over the timescale of the
duration of the accreting torus, \ie $0.01-1\,{\rm s}$.

Although still largely qualitative, the two-winds/time-reversal models
solve both the X-ray timescale riddle (the emission is produced by the
binary-merger product, which can survive up to $10^4\,{\rm s}$) and the
timing riddle (the X- and gamma-ray emission are produced at different
times, locations and propagate at different speeds). In order to
reproduce self-consistently this scenario one would require numerical
simulations on timescales of hours and these are still too expensive to
be feasible, even in two dimensions. However, there are several
observational features that can be used to confirm or rule out this novel
paradigm. First, it is clear that the launching of the jet will take
place considerably after the actual merger of the two neutron stars,
which is also when the gravitational-wave amplitude reaches its first
maximum. Hence, the observation of a SGRB which is seen to take place
$10^3 - 10^4\,{\rm s}$ after the maximum gravitational-wave emission
would be a confirmation of the validity of this scenario for SGRBs with
extended X-ray emission. Second, future observations could test this
scenario by looking for inverse-Compton scattered thermal cocoon photons
that should show up at energies $>10\,{\rm MeV}$ with a luminosity $\sim
10^{50}\,{\rm erg\,s}^{-1}$ lasting for about a second. Finally, the
detection of an X-ray emission anticipating the SGRB, such as the
precursor signals in some SGRBs \cite{Troja2010} would also represent a
strong validation of this model.

\subsection{Inclusion of radiative losses}
\label{sec:hd_nus}

Radiative effects, and in particular those associated with the emission
of neutrinos, can influence significantly the evolution of the HMNS and
the disc produced after the merger, and possibly play an important role
in the mechanisms that generate SGRBs. We recall that before the merger,
each neutron star can be considered ``cold'' (namely, such that the
thermal energy of constituent nucleons is much smaller than the Fermi
energy); at the merger, however, strong shocks are generated, which heat
the merged stellar object up to temperatures of $\sim 30-50\,{\rm
  MeV}$. This increased internal energy can be lost very efficiently via
copious emission of neutrinos, which may reach luminosities of the order
of $10^{53}\,{\rm erg\ s}^{-1}$, which can be deposited along the
baryon-free axis of rotation of the black hole via neutrino pair
annihilation. This energy injection can therefore play an important role,
or even be entirely responsible for powering the relativistic jets needed
for the beamed emission of a SGRB \cite{Narayan92, Piran:2004ba,
  Nakar:2007yr} (see however Ref. \cite{Just2016} for a different
conclusion). Neutrinos are also expected to play an important role in the
ejection of matter from the merged object and such matter contributes to
heavy-element generation and macronovae/kilonovae phenomena (see
Sects. \ref{sec:EM_counterparts} and \ref{sec:ejecta}).

For all of these reasons, incorporating the energy and momentum losses
via neutrinos in simulations of merging BNS is universally recognised as
very important. Yet, a complete treatment of the microphysics of these
systems would generally require solving the full transport problem (the
Boltzmann equation, including the absorption, emission, and scattering
source terms) in six dimensions (plus time): three for the spatial
components and an additional three for the momentum components. Even if
Shibata et al. \cite{Shibata2014a} have derived a concise and general
formulation for the conservative form (very important for numerical
simulations, as mentioned in Sect. \ref{sec:me_rhd}) of the Boltzmann
equation in general relativity, such simulations are out of reach for
current (and near-future) computational resources (except for
one-dimensional simulations, like in Ref. \cite{Abdikamalov12}).

As a result, a number of approximations, either in the gravitational
sector or in the radiation one, need to be adopted to make some progress.
A simple and yet reasonably robust approach to approximate neutrino
radiative losses is the so-called grey (energy-averaged) ``leakage
scheme'', which is essentially a parameterized neutrino cooling
scheme. This scheme was initially developed by van Riper and Lattimer
\cite{vanRiper1981} and has been widely used for both core-collapse
supernovae and BNS simulations \cite{vanRiper1981, Ruffert96b,
  Rosswog:2003b, OConnor10, Galeazzi2013, Neilsen2014, Perego2014}. Such
schemes estimate only the local changes in the lepton number and the
associated energy losses via neutrino emission, but this is probably
sufficient when simulating the late stages of compact binary mergers,
because they usually evolve on short timescales and thus the details of
radiation transport are expected not to dominate the bulk dynamics of the
system.

Furthermore, neutrino leakage schemes provide a good approximation at
those rest-mass densities in which neutrinos are either mostly trapped or
almost free streaming, namely at rest-mass densities larger than $\sim
10^{12}\,{\rm g/cm}^{3}$ and temperatures around $\sim 10\,{\rm MeV}$
(neutrinos are trapped because their scattering off baryons is so
efficient that they quickly reach a thermal equilibrium with the nuclear
matter), or at rest-mass densities smaller than $\sim 10^{11}\,{\rm
  g/cm}^{3}$ and with energies below $\sim 10\,{\rm MeV}$ (neutrinos in
these regimes interact rarely with the nuclear matter and can therefore
be considered as free streaming). Most of the matter in BNS systems
before and in the early stages of the merger is in such ranges, while for
matter outside those ranges interpolation between the two limiting
regimes can be used.

Of course, also more sophisticated approaches to approximating the
radiative transport in general-relativistic simulations have been
proposed over the years. One of them is Thorne's truncated moment
formalism for radiation hydrodynamics \cite{Thorne1981, Rezzolla1994,
  Shibata2011, Shibata2012, Cardall2013, Takahashi2013a,
  Hotokezaka2013d}, whose simplest implementation involves the evolution
of the zeroth moment of the free-streaming neutrino distribution function
on a set of individual radial rays (this is also referred to as the
``M0'' approach). Such a scheme is simpler than the two-moment grey
method (``M1'') \cite{Cardall2013, Wanajo2014, Sekiguchi2015,
  Foucart2015} and, because it tracks both neutrino density and average
energies, it allows one to model a number of important effects, such as
gravitational redshift, velocity dependence and non local-thermodynamical
equilibrium that cannot be easily incorporated into grey schemes. In this
respect, the work of Sekiguchi et al. \cite{Sekiguchi2015,Sekiguchi2016},
\lrn{whose neutrino transport is computed in a leakage-based scheme
  \cite{Sekiguchi2010, Sekiguchi2011a, Sekiguchi2012} and incorporates
  Thorne's moment formalism \cite{Thorne1981}, can currently be
  considered the most sophisticated approaches to account for radiative
  losses in binary-merger simulations in general relativity, together
  with the M1 approach of Foucart et al. \cite{Foucart2015a,
    Foucart2015}.}

Another approach developed recently is the spherical-harmonics method,
based on an expansion of the radiation intensity in angles using
spherical harmonics basis functions \cite{McClarren10, Radice2012b}. Yet
another example are the ``ray-by-ray'', multi-energy-group neutrino
schemes, like multigroup flux-limited diffusion and isotropic diffusion
source approximation \cite{Mezzacappa1993, Scheck2006, Ott08,
  Liebendoerfer2009, Sumiyoshi:12}. These schemes can offer a rather good
approximation under many conditions, but they are still rather
computationally expensive and and have not yet been used in BNS
simulations.

When discussing the actual applications of radiative-transfer schemes to
BNS simulations, we should start by mentioning that the first ones were
obviously made in Newtonian gravity, where neutrino leakage schemes were
initially employed by Ruffert et al. \cite{Ruffert96b, Ruffert97,
  Ruffert99b, Ruffert01} and more recently by Rosswog et
al. \cite{Rosswog:2003b, Rosswog:2003}, also with nuclear EOSs and
magnetic fields. On the other hand, the first works in general relativity
considering some form of neutrino cooling are those of Sekiguchi et
al. \cite{Sekiguchi2011} and have been followed by others
\cite{Sekiguchi2011b, Kiuchi2012, Sekiguchi2012, Kiuchi2012a,
  Galeazzi2013, Neilsen2014, Galeazzi2013, Palenzuela2015, Radice2016,
  Lehner2016}. In addition to the general-relativistic hydrodynamics
equations with realistic hot EOSs, these works generally solve also the
evolution equations for the neutrino, electron, and the total lepton
fractions per baryon, together with weak interaction processes and
electron/positron capture.

Although the mathematical approaches for the inclusion of the radiative
losses from BNS mergers, as well the numerical methods and their
accuracies, vary considerably from simulation to simulation, over the
last few years several groups \cite{Sekiguchi2011b, Kiuchi2012,
  Sekiguchi2012, Kiuchi2012a, Galeazzi2013, Neilsen2014, Galeazzi2013,
  Palenzuela2015, Sekiguchi2015, Radice2016, Lehner2016, Sekiguchi2016}
obtained a number of results that do not depend on such ``details'' and
can therefore be considered robust features of the radiative properties
of BNS merger simulations. The first of such features is the fact the
neutrino luminosity is high in the outer regions of the HMNS, in
particular near the polar cap, where the rest-mass density is also
relatively small. This is a result that supports the merger hypothesis
for the engine of SGRB: pair annihilation of neutrinos and antineutrinos
could supply the required amount of thermal energy necessary to drive a
fireball along the rotation axis. This conclusion is in stark contrast
with the recent work of Just et al. \cite{Just2016}, who have considered
the long-term evolution of the black-hole--torus system using a
multi-energy group M1-type scheme for the neutrino transport. In
particular, Just et al. \cite{Just2016} found that the neutrino emission
of the accreting torus (which is also expanding because of viscous
transport) is too short and too weak to yield the energy necessary for
the outflows to break out from the surrounding ejecta shell as highly
relativistic jets. While the work of Just et al. \cite{Just2016} explores
in detail timescales that are considerably longer than those considered
so far, it is also restricted to two spatial dimensions and, more
importantly, is not fully general-relativistic. Because a number of
examples have shown that the inclusion of general relativity is essential
for a proper assessment of the energetics and efficiencies in black-hole
processes, additional simulations in full general relativity are needed
to assess whether or not BNS mergers can provide neutrino-powered jets.

\begin{figure*}
  \begin{center}
    \includegraphics[width=0.6\columnwidth]{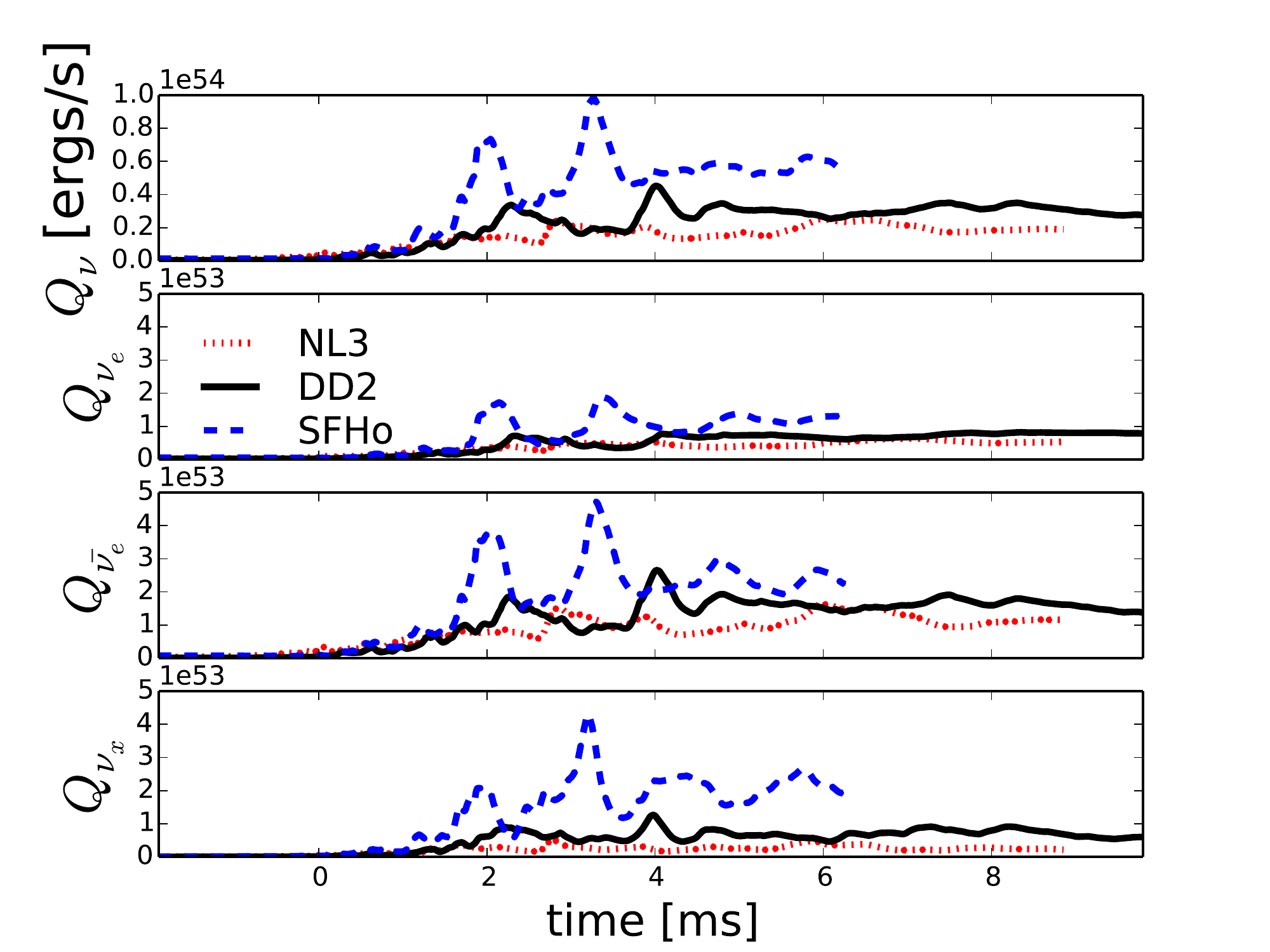}
  \end{center}
  \caption{Neutrino luminosities for the different EOSs used in
    Ref. \cite{Palenzuela2015}. The panels show, from top to bottom, the
    total neutrino luminosity and the luminosities for electron
    neutrinos, electron antineutrinos, and the other four lepton flavours
    combined. [Reprinted with permission from Ref. \protect{}
      \cite{Palenzuela2015}. \copyright~(2015) by the American Physical
      Society.]}
\label{fig:Palenzuela2015}
\end{figure*}

A second robust feature is that the dominant neutrino emission, with
luminosities of a few $10^{54}\,{\rm erg\ s}^{-1}$ consists of
electron antineutrinos (see Fig. \ref{fig:Palenzuela2015}); this is
because the HMNS has a high temperature, and hence, electron-positron
pairs are efficiently produced from thermal photons, but then neutrons
efficiently capture the positrons to emit antineutrinos, whereas
electrons are not captured by protons as frequently as positrons because
the proton fraction is much smaller than the neutron fraction.

A third and somewhat expected result is that the simulations have shown
that the lifetime of the binary-merger product is influenced also by the
timescale of neutrino cooling, as well as by the strength of the magnetic
field and by the EOS. After the eventual collapse to black hole, muon and
tau neutrino luminosities steeply decrease because most of the
high-temperature regions are covered by the event horizon, while the
luminosities of electron neutrinos and antineutrinos decrease only
gradually, because these neutrinos are emitted via charged-current
processes from the accretion disc.

\begin{figure*}
  \begin{center}
    \includegraphics[width=0.6\columnwidth]{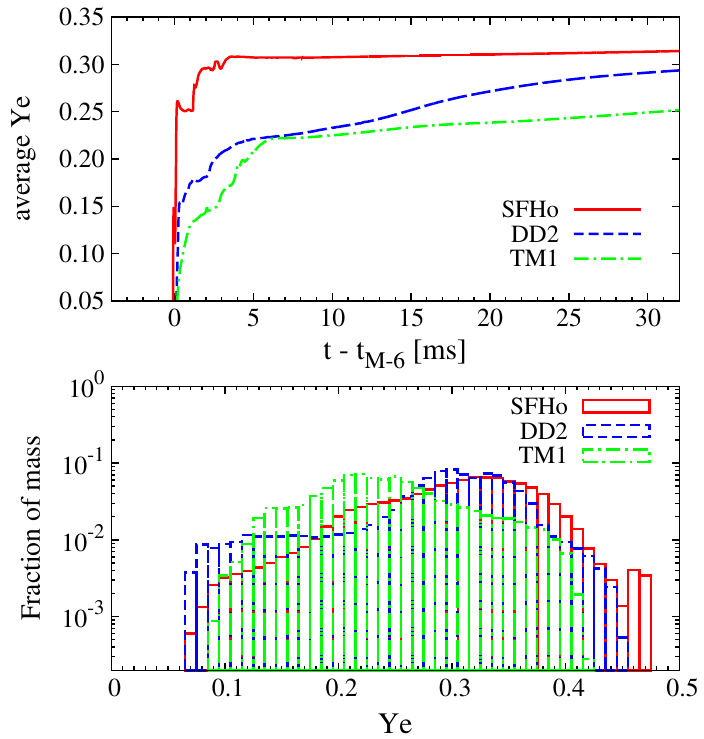}
  \end{center}
  \caption{\emph{Upper panel:} evolution of the averaged value of the
    electron fraction $Y_e$ for different EOSs: SFHo (red solid), DD2
    (blue dashed), and TM1 (green dotted dashed). \emph{Lower panel:}
    mass-distribution histograms of $Y_e$ normalized to the total mass of
    ejecta measured at about 25 ms after the onset of merger for SFHo,
    DD2, and TM1. [Reprinted with permission from Ref. \protect{}
      \cite{Sekiguchi2015}. \copyright~(2015) by the American Physical
      Society.]}
\label{fig:Sekiguchi2015_4}
\end{figure*}

A fourth result, due mostly to Refs. \cite{Sekiguchi2015, Palenzuela2015,
  Radice2016}, is that after the merger, the electron fraction in the
ejected matter has a broad distribution, in a range between 0.05 and 0.45
(see Fig. \ref{fig:Sekiguchi2015_4}). This result is rather robust and
depends very weakly on the grid resolution or the EOS
considered. However, it is very different from what found in previous
studies, which used either Newtonian codes \cite{Korobkin2012,
  Rosswog2014a} or codes with an approximate treatment of general
relativity \cite{Oechslin07a, Bauswein2013b} and which had found that the
distribution of the electron fraction was very narrow, with an average
value $\lesssim 0.1$. There are several possible explanations for these
differences: firstly, the Newtonian simulations of Refs.
\cite{Korobkin2012, Rosswog2014a}, while taking into account neutrino
cooling, underestimated significantly the effect of shock heating and
thus the effect of the positron capture was much weaker. Furthermore,
these simulations did not consider neutrino heating (absorption), which
is expected to play an important role for stiffer EOSs, where positron
capture is relatively less important because of the lower temperature
\cite{Sekiguchi2015}. Secondly, the differences with respect to
approximate general-relativistic simulations \cite{Oechslin07a,
  Bauswein2013b} can be explained by noting that the ejecta in such
simulations remained neutron rich because weak interaction processes were
not taken into account and thus there was no way to change the electron
fraction \cite{Sekiguchi2015}.

Finally, in addition to these differences, Radice et al.
\cite{Radice2016} also found that the ejecta have a bi-modal distribution
in the electron fraction $Y_e$, with maxima around $0.08$ and
$0.16$. This is because part of the outflow, driven by tidal torques, is
cold and neutron rich, while another component, launched by shocks during
merger, has high temperatures and rapid protonisation with values of
$Y_e$ peaking around $0.16$.

As a concluding remark in this section we note that the presence of
magnetic fields and radiative losses in full general relativity has been
considered so far only by Neilsen et al. \cite{Neilsen2014} and by
Palenzuela et al. \cite{Palenzuela2015}. Their results indicate that
magnetic fields do not seem to play a significant role in modifying the
neutrino emission, although the resolutions employed were somewhat lower
than the ones needed to resolve the complex dynamics behind
magnetic-field amplification. Particularly interesting in the work of
Neilsen et al. \cite{Neilsen2014} is the use of a novel method for the
calculation of the optical depth, which simplifies its use with
distributed adaptive mesh refinement and has later been employed also in
Ref. \cite{Radice2016}.

\subsection{Non-bulk dynamics: ejecta, nucleosynthesis and afterglows}
\label{sec:ejecta}

In what follows we will concentrate on two advanced topics that have been
pursued in the modelling of merging BNSs and that revolve about the
dynamics of that \emph{tiny} amount of matter (\ie $\lesssim 0.1\%$ of
the total mass of the system) that does not participate to the bulk
dynamics nor is ultimately accreted onto the black hole, but that leaves
the system because it becomes gravitationally unbound. Over the last few
years a lot of progress has been made in the study of such material
ejected from BNSs during and after the merger. The neutron-rich dense
material that becomes gravitationally unbound and so escapes to larger
distances, undergoes decompression and changes in its electron fraction,
so that rapid neutron capture processes (\ie ``$r$ processes'')
\cite{Arnould2007} and radioactive decays \cite{Li1998, Rosswog05,
  Metzger:2010, Roberts2011} take place. While $r$ processes are
responsible for the nucleosynthesis of heavy elements, possibly playing a
fundamental role in the cosmic chemical evolution, the radioactive decays
could lead to the release of large amounts of electromagnetic radiation
which would become observable as \lrn{a delayed electromagnetic
  counterpart}. Both of these processes are related to two outstanding
issues in astrophysics: the explanation of the observed heavy-element
abundance in the Universe and the modelling of afterglows (of SGRBs or
mergers) that are expected to produce optical and infrared (or even
longer-wavelength) emissions. Although such topics are obviously closely
related, we present them separately in Sects. \ref{sec:emain} and
\ref{sec:maoa}, as this helps us collect the various contributions.

\subsubsection{Ejected matter and nucleosynthesis}
\label{sec:emain}

As mentioned above, that of the construction of the chemical abundance
presently observed in the Universe is a long-standing problem in
astrophysics. It is clear that Big-Bang nucleosynthesis cannot produce
heavy elements and that these need to be produced within stars through
their nuclear evolution. Core-collapse supernovae have been traditionally
considered as the channel through which heavy elements produced via $r$
processes can be distributed across the Universe. This picture, however,
is becoming increasingly difficult to support as the numerical
simulations are showing that the physical conditions necessary to produce
elements with atomic number $A > 90$, namely, high entropy, low electron
fraction, and very rapid expansion, are hard to be met in core-collapse
supernovae. Such conditions, however, are quite natural in BNS mergers,
which have therefore become an interesting alternative for $r$-process
nucleosynthesis and the construction of heavy/very-heavy elements in the
Universe.

We recall that BNS mergers can release neutron-rich matter in at least
four different ways (see also Sect. \ref{sec:EM_counterparts}): (i) in
the matter that is ejected dynamically via gravitational torques (tidal
ejection) \cite{Rezzolla:2010, Roberts2011, Rosswog2013a, Bauswein2013b}
or shocks \cite{Kyutoku2012}; (ii) through neutrino-driven winds
\cite{Dessart2009, Perego2014, Wanajo2014, Just2015, Sekiguchi2015,
  Just2016, Sekiguchi2016}; (iii) through magnetically driven winds (see
Sect. \ref{sec:EM_counterparts} \cite{Shibata2011b, Kiuchi2012b,
  Siegel2014, Rezzolla2014b, Ciolfi2014}; (iv) from shock waves in the
late-time evolution of accretion discs \cite{Goriely2011, Kastaun2014}.
In all cases, the ejected matter is neutron-rich, \lrn{cold if
  dynamically ejected (\ie if not shock heated), and in beta equilibrium;
  however, to} the different channels correspond different amounts of
ejected matter, as well as different distributions in entropy, electron
fraction, and velocity. As a result, they might possibly produce
different nucleosynthetic signatures \cite{Rosswog2013, Piran2013,
  Perego2014, Wanajo2014, Just2015, Sekiguchi2015, Radice2016, Just2016,
  Sekiguchi2016}.

Perhaps in no aspect of the dynamics of BNS mergers other than that of
the ejected matter and nucleosynthesis are the differences between
various approaches larger. On the one side, there are codes making use of
Newtonian (or pseudo-Newtonian) descriptions of gravity and implementing
either SPH techniques \cite{Bauswein2013b, Rosswog2013, Piran2013,
  Rosswog2014a, Grossman2014, Metzger2015} or Eulerian hydrodynamics
approaches \cite{Perego2014, Just2015, Just2016}, which are combined with
very advanced treatments of radiative transfer and nuclear-reaction
networks. In addition, either exploiting symmetries or larger timesteps,
these codes can achieve evolutions of the dynamical ejecta over very
long-term and through direct hydrodynamical simulations lasting up to 100
years and over a density range of roughly 40 orders of magnitude. On the
other side, there are codes in full general relativity, but that
implement simpler approaches to the neutrino radiative losses and are
necessarily restricted to much shorter timescales \cite{Wanajo2014,
  Sekiguchi2015, Foucart2015, Palenzuela2015, Radice2016, Lehner2016,
  Sekiguchi2016}. Inevitably perhaps, these methodological differences
are then responsible for systematic differences in the results from the
two classes of approaches. Fortunately, these differences are mostly
quantitative, leaving behind a qualitative picture which is instead
rather robust.

The most important quantitative difference is in the actual amount of
mass ejected. Codes adopting approximate treatments of gravity, in fact,
tend to overestimate the relative ratio of tidally- to shock-driven
ejecta for quasi-circular binaries \cite{Bauswein2013}. More
specifically, simulations that are not fully general-relativistic show
that even mergers from equal-mass systems tend to develop prominent tidal
arms before the actual contact. At the same time, these simulations tend
to underestimate the strength of the shocks that develop at the merger
and that are able to impart larger kinetic and internal energies to the
shocked material \cite{Goriely2015, Radice2016}. Both of these aspects
are clearly the consequence of ``shallower'' gravitational potentials
than those computed in general relativity, which tend to produce larger
``spills'' of matter during the inspiral and less ``catastrophic''
mergers. These differences apply also to eccentric binaries, where
Newtonian studies tend to overestimate the amount of shocked ejecta,
leading to outflows that are more proton-rich in eccentric binaries than
in quasi-circular mergers \cite{Rosswog2013} (see also
Sect. \ref{sec:dynamical-capture}). In contrast, general-relativistic
eccentric mergers are typically more neutron rich than those of
quasi-circular mergers, the reason being that the ejecta in eccentric
mergers is increasingly dominated by the tidal component
\cite{Radice2016}.

\begin{figure*}
  \begin{center}
    \includegraphics[width=0.6\columnwidth]{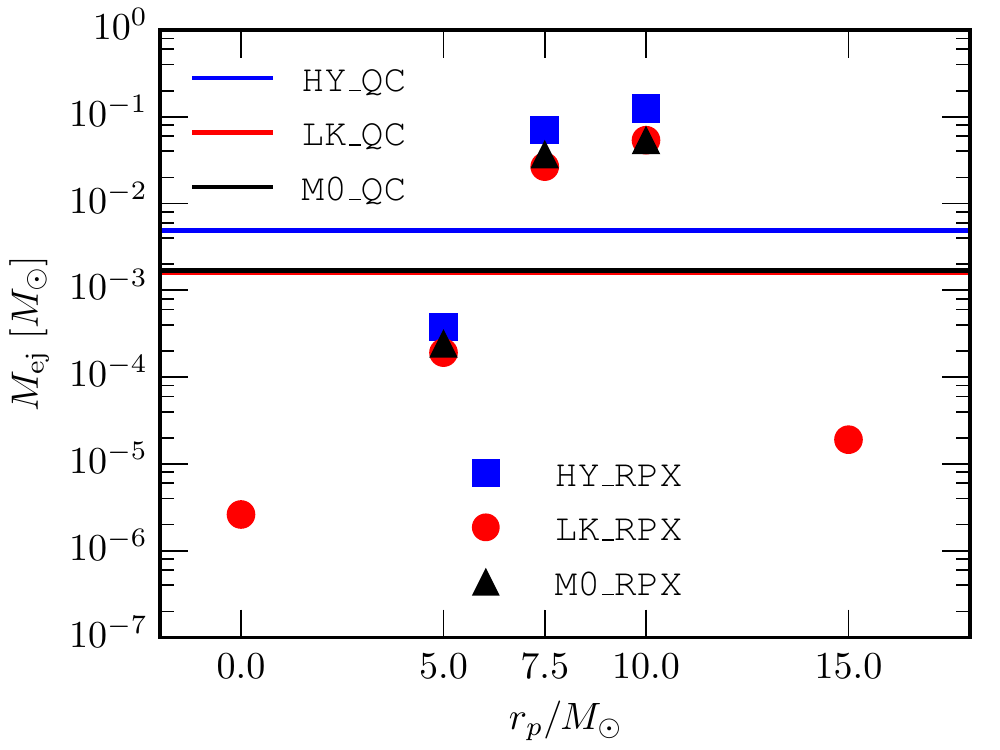}
  \end{center}
  \caption{Dynamically ejected mass for BNS mergers coming either from
    \lrn{dynamical encounters or from quasi-circular inspirals (in this
      latter case since the mass does not depend on $r_p$, it is marked
      as a horizontal line)}. The ejecta mass is computed integrating in
    time the flux of unbound matter (with the covariant time-component of
    the four-velocity $u_t \leq -1$) across the surface of a spherical
    sphere with radius $r \simeq 295\ \mathrm{km}$. Overall, eccentric
    binaries can eject up to $2$ orders of magnitude more mass than
    quasi-circular binaries. [Adapted from Ref. \cite{Radice2016}.]}
\label{fig:ejected}
\end{figure*}

A summary of how much the amount of ejected matter can vary under
different physical conditions, either before the merger (quasi-circular
versus eccentric encounters) or after the merger (in pure hydrodynamics
or with radiative effects taken into account) is shown in
Fig. \ref{fig:ejected}, which is taken from the work of Radice et
al. \cite{Radice2016}. More specifically, shown are the total ejected
rest mass as computed integrating in time the flux of matter with
covariant time-component of the four-velocity $u_t \leq -1$, across a
spherical coordinate surface with radius $\simeq
295\ \mathrm{km}$\footnote{A detailed discussion of possible alternative
  criteria to identify unbound fluid elements has been made by Kastaun
  and Galeazzi \cite{Kastaun2014}.}. Shown with symbols are the results
of simulations of BNS systems on parabolic orbits with varying impact
parameters, while the horizontal lines show the corresponding results
when considering binaries in quasi-circular orbits (the gravitational
masses are essentially the same in the two cases) \cite{Radice2016}.

Note that the eccentric mergers lead to considerably larger amounts of
ejected masses, which is hardly surprising given that the binaries
themselves are in this case only weakly bound. Furthermore, the mass lost
depends sensitively on the impact parameter, showing a local maximum for
BNS systems close to the threshold between prompt merger and
multiple-encounters \cite{East2012b, Radice2016}. Overall, eccentric
encounters can yield up to $\sim 0.1\ M_\odot$ of ejecta, slightly less
than what can be achieved with black-hole--neutron-star mergers
\cite{Foucart2015, Kyutoku2015}, but almost two orders of magnitude
larger than what is ejected by mergers of BNSs in quasi-circular orbits
\cite{Radice2016} and one order of magnitude larger than what is ejected
in unequal-mass BNS mergers in quasi-circular orbits
\cite{Sekiguchi2016}. The ejecta from unequal-mass BNS mergers may also
have a different composition (electron fraction), because the tidal
component is larger for mass ratios more different from one
\cite{Lehner2016,Sekiguchi2016}.

The amount of ejected matter also depends on the inclusion of radiative
losses, which are either absent (blue squares) or accounted via a
neutrino leakage (red circles) or an ``M0'' approach (black triangles),
discussed in Sect. \ref{sec:hd_nus}. The differences in this case are far
smaller, but it is also clear that neglecting neutrino cooling results is
an overestimate of the unbound mass by a factor $\gtrsim 2$. This is
because neutrino losses in the optically thin outflows are rapid and
sufficient to cause part of the material to become gravitationally bound
again by removing part of its internal energy. Finally, we note that the
amount of matter ejected from quasi-circular mergers is larger if the
progenitors are spinning. Kastaun and Galeazzi \cite{Kastaun2014}
measured ejecta masses of a factor of a few larger for
constraint-violating spinning initial data (see discussion in
Sect. \ref{sec:sbs}. More work is needed to further explore ejecta in the
case of realistic spins and constraint-satisfying initial data.

Next, we turn our attention to the nucleosynthesis resulting from the
ejected matter. In this context, when discussing more in detail the
results of a number of simulations performed by various groups using
different approaches and approximations, we will distinguish those works
that focused on the nucleosynthesis of matter that was dynamically
ejected from those studies that instead focused on neutrino-driven
ejecta. For the first class of simulations, the ejecta are normally
followed through the times when radioactively powered transients should
peak (of the order of days) and up to the point when the radio flares
from the interaction with the ambient medium are expected (of the order
of years). Essentially all results, independently of whether performed in
full general relativity \cite{Hotokezaka2013, Tanaka2013, Sekiguchi2015,
  Foucart2015, Palenzuela2015, Radice2016, Lehner2016, Sekiguchi2016} or
with approximate treatments of gravity \cite{Rosswog2013, Rosswog2013a},
conclude that the dynamical ejecta very robustly produce strong $r$-process
elements with $A > 130$ with a pattern that is essentially independent of
the details of the merging system. In particular, the $r$-process peaks
around $A \approx 130$ and $A \approx 195$ are fairly independent from
the details of the initial data used (quasi-circular or eccentric) or the
details of the radiative losses (from pure hydrodynamics simulations over
to those having ``M1'' approaches) \cite{Radice2016}.

Furthermore, the recent simulations by Wanajo et al. \cite{Wanajo2014},
Sekiguchi et al. \cite{Sekiguchi2015}, and Radice et
al. \cite{Radice2016}, using a general-relativistic treatment, advanced
neutrino transport schemes and state-of-the-art finite-temperature EOSs,
also showed that the dynamical and early-merger neutrino-wind ejecta of
BNS mergers can be the dominant origin of all the Galactic $r$-process
nuclei. These general-relativistic simulations \cite{Wanajo2014,
  Sekiguchi2015, Radice2016} have highlighted that shock-heated and
neutrino-processed dynamical ejecta from the HMNS are not made of almost
pure neutron matter (as previously thought), but also contain modestly
neutron-rich matter. This result makes the $r$-process abundance
distribution from the calculations agree with the solar one over the full
range of $A\simeq 90-240$, while all previous works obtained a good
matching without fine tuning only for the production of nuclei with
$A\gtrsim 130$.

Interestingly, Goriely et al. \cite{Goriely2015}, have conducted a
systematic investigation of the potential impact that weak interactions
could have on the electron-fraction evolution in merger ejecta,
suggesting that the relative contributions of matter in the intervals
$A\lesssim 90$, $90\lesssim A \lesssim 140$ and $A\gtrsim 140$ depend
strongly not only on the EOS, but also on whether a large tidal tail
develops after the merger. If this is the case, considerable amounts of
cold, unshocked matter could be centrifugally ejected before the neutrino
luminosities rise and thus before neutrino exposure of these ejecta plays
an important role. Under these conditions, the ejecta will be expelled
highly anisotropically and will carry a dominant fraction of the mass in
the form of $A \gtrsim 140$ nuclei \cite{Goriely2015}.

\begin{figure*}
  \begin{center}
    \includegraphics[width=0.49\columnwidth]{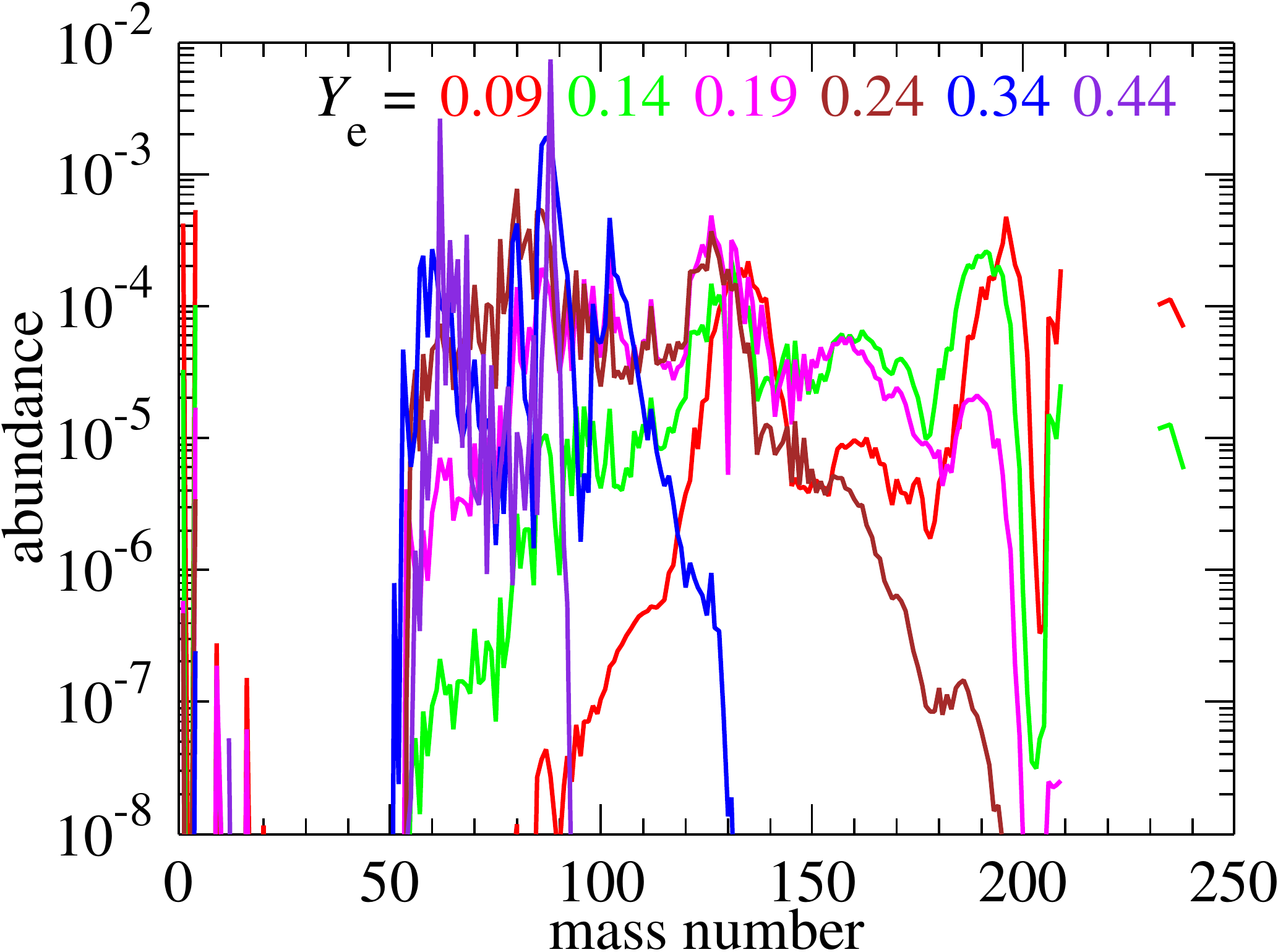}
    \hskip 0.2cm
    \includegraphics[width=0.49\columnwidth]{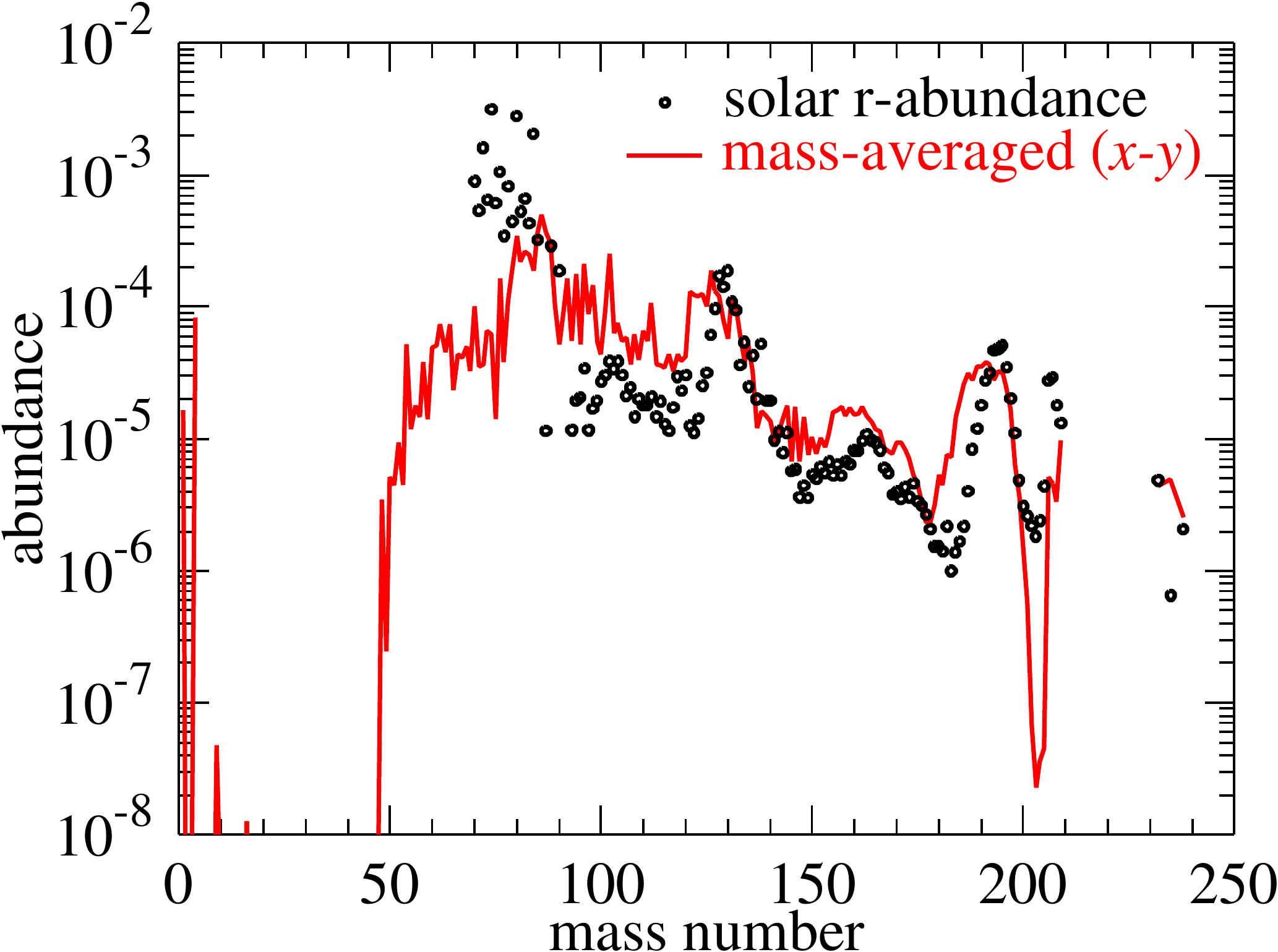}
  \end{center}
  \caption{\emph{Left panel:} Final nuclear abundances for selected
    trajectories with given electron fraction on the orbital
    plane. \emph{Right panel:} Comparison between the solar $r$-process
    abundances and the mass-averaged nuclear abundances obtained by
    weighting the final yields for the representative trajectories in the
    simulations with their electron-fraction mass fractions on the
    orbital plane. [Adapted from Ref. \cite{Wanajo2014} with permission
      by the authors.]}
\label{fig:abundances}
\end{figure*}

A summary of the results relative to the nucleosynthesis of the matter
ejected in BNS mergers is shown in Fig. \ref{fig:abundances}, which is
taken from the work of Wanajo et al. \cite{Wanajo2014}, where
thermodynamical trajectories of particles in the ejecta are traced on the
orbital plane and each representative particle is chosen from a bin
defined by an electron fraction $0.09 \leq Y_e \leq 0.44$. The left
panel, in particular, shows the final nuclear abundances for the selected
trajectories, which reproduce a variety of nucleosynthetic features: the
iron-peak and $A \approx 90$ abundances made in nuclear quasi-equilibrium
for $Y_e \gtrsim 0.4$, light $r$-process abundances for $Y_e \approx
0.2$--$0.4$, and heavy $r$-process abundances for $Y_e \lesssim 0.2$. The
right panel, on the other hand, shows the mass-averaged nuclear
abundances by weighting the final yields for the representative
trajectories with their $Y_e$ mass fractions on the orbital plane. The
agreement with the solar $r$-process abundance distribution over the full
atomic-number range of $A \approx 90-240$ is quite good; an even closer
agreement in the high-$A$ regime, including in particular the second $A
\approx 130$ and third $A \approx 195$ peaks, has been shown recently by
Radice et al. \cite{Radice2016}, by considering full three-dimensional
trajectories, not restricted to the orbital plane.

When considering now those studies that focused on neutrino-driven
ejecta, we should note that in this case, because one needs to explore
timescales that are of the order of seconds (and not of tens of
milliseconds), the simulations all employ some approximation, either in
the treatment of gravity or in the underlying spatial symmetries. With
these considerations in mind, the inclusion of neutrino transport and the
contribution from neutrino-driven winds in long-term simulations do bring
different results for the properties of the dynamical ejecta of the
merger such as total mass, the average electron fraction, and the thermal
energy, and thus for the properties of the nucleosynthesis as well
\cite{Dessart2009, Perego2014, Just2015, Wanajo2014, Sekiguchi2015}. This
is because some fraction of the total ejected material comes through
neutrino dynamics, especially for stiffer EOSs. Furthermore, such ejecta
have the peculiarity of being emitted also at high latitudes, near the
rotation axis, while other types of ejecta are mostly constrained to
lower latitudes. More specifically, Perego et al. \cite{Perego2014} found
that the neutrino winds provide a substantial contribution to the total
mass lost in a BNS merger and that material ejected at higher and lower
latitudes has different properties. It was found that polar ejecta,
characterized by more intense neutrino irradiation, larger electron
fraction, entropies, and expansion velocities, produce $r$-process
contributions from $A \approx 80$ to $A \approx 130$, while the more
neutron-rich, lower latitude parts produce elements up to the third
$r$-process peak, near $A \approx 195$.

An exhaustive analysis has been carried out by Just et
al. \cite{Just2015}, who employed a three-dimensional relativistic SPH
code, using the conformally flat approximation for the spacetime
\cite{Oechslin07a} and different microphysical EOSs for
nonzero-temperature neutron-star matter. The subsequent, seconds-long
evolutions of the black-hole--accretion-torus system were carried out
starting from tori in rotational equilibrium with a Newtonian
hydrodynamics code that includes dynamical viscosity effects,
pseudo-Newtonian gravity for rotating black holes, and an
energy-dependent two-moment closure scheme for the transport of electron
neutrinos and antineutrinos. As a result of their analysis, Just et
al. \cite{Just2015} reiterated that the solar $r$-abundance pattern can
be reproduced well in the whole range $90 < A < 240$ by a mass-weighted
combination of the dynamical ejecta from the BNS merger phase and the
secular ejecta from the black-hole--torus evolution, to which
neutrino-powered winds contribute only at most another 1\%. In
particular, the black-hole--torus outflows are able to reproduce well the
region $A \lesssim 140$, where the prompt merger ejecta under-produce the
nuclear abundances.

In conclusion, research on the ejected matter and the consequent
heavy-element nucleosynthesis from BNS mergers is a comparatively young
field of research that has already seen considerable efforts at multiple
levels of mathematical approximation and physical
sophistication. Overall, the results show that amounts of matter in the
range $10^{-3} - 10^{-1}\,M_{\odot}$ can be lost to spatial infinity in
the course of the merger and post-merger dynamics, and that the details
depend on a rich combination of factors: EOS, mass ratio, pre-merger
dynamics, etc. At the same time, robust results emerge from the nuclear
evolution of the ejected matter, which provides a surprisingly good match
with the chemical abundances observed and promotes BNS mergers to a
primary role in the chemical enrichment of the Universe. These promising
first results also call for further investigations to set on firmer
grounds a number of details that have been neglected so far for a number
of reasons. These include: long-timescale fully general-relativistic
calculations, improved treatments of radiative transfer, and more
advanced numerical approaches for the study of the tiny and tenuous
component of the BNS matter that is ejected at the merger.

\subsubsection{Macronovae and other afterglows}
\label{sec:maoa}

Observations of electromagnetic radiation from astrophysical events that
also emit gravitational radiation will be very important and for several
reasons. Such simultaneous detections of electromagnetic and
gravitational-wave signals, in fact, would significantly increase the
confidence in the detection of gravitational waves \cite{Metzger2012,
  Piran2013}. Furthermore, such delayed electromagnetic counterparts
would contribute considerably to the sky localisation of astrophysical
events emitting gravitational waves. Since gravitational-wave detectors
will have only moderate space localisation capabilities (of the order of
one square degree at best), electromagnetic observations can be crucial
for extracting information about the astrophysical properties of the
source and of its environment (\eg the host galaxy and its redshift).

\begin{figure}
\begin{center}
  \includegraphics[width=0.6\columnwidth]{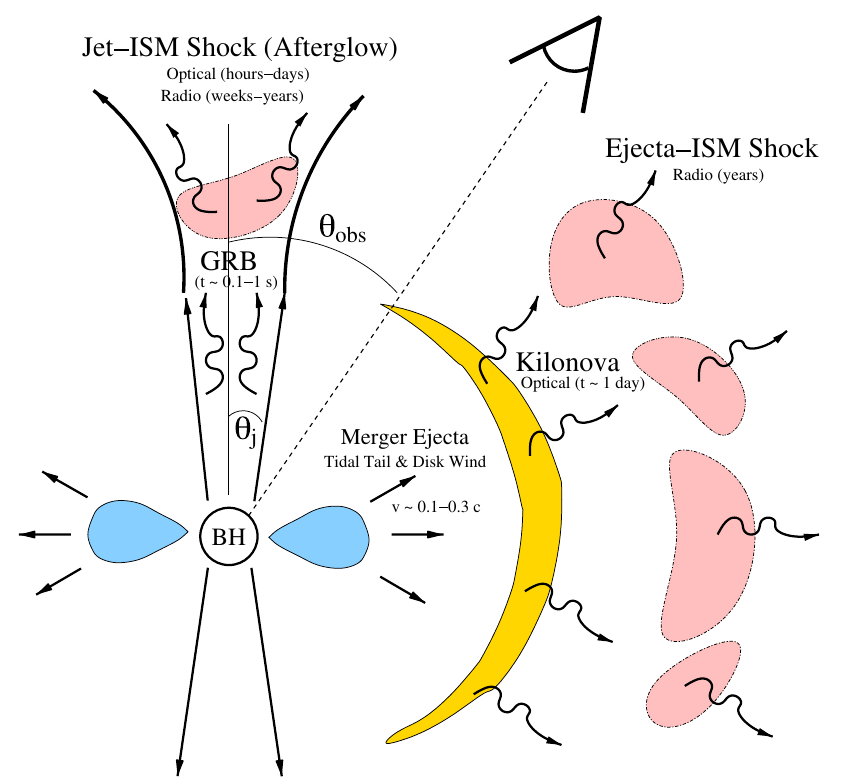} 
\end{center}
\caption{Schematic summary of potential electromagnetic counterparts of
  BNS mergers, as a function of the observer angle,
  $\theta_{\rm{obs}}$. Following the merger, a centrifugally supported
  disc (blue) remains around the central compact object (usually a black
  hole). Rapid accretion lasting $\sim 1$ s powers a collimated
  relativistic jet, which produces a SGRB. Due to relativistic beaming,
  the gamma-ray emission is restricted to observers at small viewing
  angles. Non-thermal afterglow emission results from the interaction of
  the jet with the surrounding circumburst medium (pink). Optical
  afterglow emission is observable on timescales up to days/weeks by
  observers with larger viewing angles. Radio afterglow emission is
  observable from all viewing angles (isotropic) once the jet decelerates
  to mildly relativistic speeds on a timescale of weeks/months, and can
  also be produced on timescales of years from sub-relativistic
  ejecta. Short-lived isotropic optical emission lasting a few days
  (kilonova/macronova; yellow) can also accompany the merger, powered by
  the radioactive decay of heavy elements synthesised in the
  ejecta. [Reproduced from Ref. \cite{Metzger2012} with permission by the
    authors.]}
   \label{fig:Metzger2012_Fig1}
\end{figure}

As discussed in detail in Sect. \ref{sec:hydro_merger_post-merger}, in
addition to SGRBs, which are the most plausible electromagnetic
manifestation of BNS mergers \cite{Narayan92,Eichler89}, and certainly
the best studied electromagnetic emission that such mergers could produce
(either before of after the merger; see Sect. \ref{sec:magn-inspiral}),
there is also another class of delayed electromagnetic counterparts to
BNS mergers that has caught a widespread attention recently. This
electromagnetic emission is the one caused by the radioactive decay
taking place mostly in ejecta and by the interaction of the ejecta with
the interstellar medium, long time after the merger. This is nicely
summarised in the schematic diagram in Fig.\ref{fig:Metzger2012_Fig1},
which illustrates the various potential electromagnetic emissions from
the post-merger phase. In this Section, we will address the latter
processes, among which the most promising are the \emph{macronovae} (or
\emph{kilonovae}) \cite{Li1998, Kulkarni2005_macronova-term,
  Metzger:2010, Berger2013, Tanvir2013, Yang2015, Jin2015, Jin2016}.

We recall that macronovae \cite{Li1998} are electromagnetic emissions
about one to three orders of magnitudes brighter than regular novae
(which are caused by hydrogen fusion explosions on a white dwarf
accreting from a larger companion star), hence the name macronovae
\cite{Kulkarni2005_macronova-term} or kilonovae \cite{Metzger:2010}. In
the standard scenario, macronovae shine days after the merger in
ultra-violet, optical, and infrared bands probably because of radioactive
decay of $r$-process elements \cite{Li1998, Kulkarni2005_macronova-term,
  Metzger:2010} (however, see also \cite{Kisaka2015} for an alternative
explanation). Macronovae are particularly promising delayed
electromagnetic counterparts because their emission is relatively
isotropic, contrary to SGRBs, which are thought to be highly beamed;
furthermore, statistical evidence being presently collected indicates that
they may be ubiquitous \cite{Jin2016}. Hence, the probability of
observing electromagnetic radiation from a given macronova together with
the gravitational waves from the BNS system is larger than the
probability to observe the SGRB together with the gravitational emission.

The results of numerical simulations agree on the fact that the nuclear
decay in ejecta from BNS mergers is consistent with the electromagnetic
radiation observed as macronovae, even if, as pointed out in detail in
Ref. \cite{Radice2016}, different initial data and levels of
microphysical description give intrinsic luminosities and peak times that
are different of factors of a few. The consistency is present
notwithstanding the fact that the codes used for such simulations are
rather diverse, both in the mathematical description and in the numerical
techniques used. Some codes, in particular, make use of Newtonian (or
pseudo-Newtonian) descriptions of gravity and implement either SPH
techniques \cite{Bauswein2013b, Metzger2014} or Eulerian hydrodynamics
approaches \cite{Perego2014, Just2015}, which are combined with very
advanced treatments of radiative transfer and nuclear-reaction networks.
In addition, either exploiting symmetries or larger timesteps, these
codes can achieve evolutions of the dynamical ejecta over very long-term
and through direct hydrodynamical simulations lasting up to 100 years and
over a density range of roughly 40 orders of magnitude. Other codes,
instead, implement full general relativity, but simpler approaches to the
neutrino radiative losses and are necessarily restricted to much shorter
timescales \cite{Wanajo2014, Radice2016}.

Another feature found irrespective of the various approaches used is the
clear dependence of the optical emission on the EOS of the neutron stars,
with softer EOSs yielding brighter transients and peaking on longer
timescales and with a lower effective temperature \cite{Bauswein2013b,
  Tanaka2013, Wanajo2014, Radice2016}. Finally, also a commonly observed
feature is the dependence on the binary mass ratio, with brighter
emissions with increasingly smaller mass ratios \cite{Tanaka2013}. Both
these last two features are rather easy to explain as softer EOSs and
smaller mass ratios generically yield larger ejecta and hence larger
amounts of matter undergoing radioactive decay. 

In view of this, the dependence on the EOS and in particular the
relations between peak luminosity (or peak timescale or effective
temperature at the time of the peak luminosity) and the original stellar
radii suggest the possibility to constrain neutron-star radii and thus
the high-density EOSs from observations of optical transients associated
with BNS mergers \cite{Bauswein2013b, Tanaka2013,
  Hotokezaka2013d}. Following this line of thought, Hotokezaka et
al. \cite{Hotokezaka2013d} were the first to attempt to extract
information on possible progenitor models for the macronova transient
associated with GRB 130603B. Assuming that the electromagnetic transient
was powered by the radioactive decay of $r$-process elements, they
estimated that the central engine of GRB 130603B was a massive torus of
approximately $0.1\,M_\odot$ produced in a BNS merger and accreting onto
a spinning black hole. Soft EOS models were found to be favoured to
explain the observations.

As mentioned in Sects. \ref{sec:hd_nus} and \ref{sec:emain}, the
inclusion of neutrino transport in simulations does bring different
results for the properties of the dynamical ejecta of the merger, such as
total mass, average electron fraction, and thermal energy, and thus for
the properties of the nucleosynthesis as well \cite{Dessart2009,
  Perego2014, Just2015, Wanajo2014, Sekiguchi2015, Lehner2016,
  Radice2016, Sekiguchi2016}. This is because some fraction of the total
ejected material is the result of neutrino emission, in particular for
stiffer EOSs as these produce smaller amounts of dynamical ejecta.

Using a pseudo-Newtonian code that solves the hydrodynamics equations
with dynamical viscosity and in axisymmetry, Metzger and Fern{\'a}ndez
\cite{Metzger2014} have studied the dependence of the ejecta and their
emissions over the lifetime of the HMNS, till its collapse, if this takes
place\footnote{We should note that the use of a viscosity description
  instead of magnetohydrodynamics modelling is a major source of
  uncertainty as it may well influence the nucleosynthetic predictions,
  as shown in Ref. \cite{Just2015}.}. Starting with models of tori in
equilibrium and approximating the HMNS as a reflecting inner boundary at
a fixed radius, they found that the lifetime of the merger remnant may be
directly imprinted in the radioactively powered macronova emission
following the merger. In fact, when black-hole formation is relatively
prompt (\ie when the HMNS survives for $\lesssim 100\,{\rm ms}$),
outflows from the disc are sufficiently neutron rich to form heavy
$r$-process elements, resulting in week-long emission with a spectral peak
in the near-infrared, similar to that produced by the dynamical ejecta,
emitted during the merger. In contrast, if the black-hole formation is
delayed, neutrinos from the HMNS are able to raise the electron fraction
in the polar direction to values such that outflows may be generated. The
lower opacity of this region would then produce a brighter, bluer, and
shorter-lived day-long emission prior to the late peak in the near
infrared from the dynamical ejecta and equatorial wind.

Almost at the same time, Perego et al. \cite{Perego2014}, presented
detailed, three-dimensional viscous-hydrodynamics simulations performed
with a Newtonian Eulerian code with a spectral neutrino leakage scheme
accounting for neutrino absorption (but not for neutrino annihilation in
optically thin conditions, which may further enhance ejecta). The
simulations were started from a direct re-mapping of the matter
distribution of three-dimensional SPH simulations of the merger of two
non-spinning $1.4\,M_\odot$ neutron stars obtained through the code of
Rosswog and Price \cite{Rosswog2007}, covering the time interval up to
about $100\,{\rm ms}$ after the merger and including a spatial domain
extending up to $1500\,{\rm km}$ from the HMNS, which was again treated
as an inner boundary. With this setup, Perego et al. \cite{Perego2014}
found that the collapse timescale has a minor impact on electromagnetic
counterparts and nucleosynthesis yields, in contrast to the results of
Metzger and Fern{\'a}ndez \cite{Metzger2014}. Perego et
al. \cite{Perego2014} also remark that winds from the polar regions
produce ultraviolet and optical transients reaching luminosities up to
$10^{41}\,{\rm erg\ s}^{-1}$ and peaking around one day after emission,
as found by Metzger and Fern{\'a}ndez \cite{Metzger2014} too. Winds from
lower latitude regions, instead, being contaminated with high-opacity
heavy elements, produce dimmer and redder signals, peaking after about
two days in optical and infrared.

More recently, Metzger et al. \cite{Metzger2015} considered the
electromagnetic emissions coming from the decay of free neutrons emitted
by the BNS merger, following a suggestion first pointed out by Kulkarni
\cite{ Kulkarni2005_macronova-term}. Even if most of the material ejected
is so dense that almost all neutrons are captured into heavy nuclei via
$r$ processes, some fast expanding matter ejected during the initial
phases of the merger from the interface between the two stars could
produce a relevant flow of free neutrons. Such free neutrons could alter
the early macronova light curves, since their decay timescale is long and
they release a large quantity of energy per decay as compared to the
majority of the $r$ processes. The signature of free-neutron decay would
be a blue/visual bump that peaks at a luminosity $L_{\rm tot}\sim 10^{41}
{\rm erg \, s}^{-1}$ on a timescale of less than few hours after the
merger. Metzger et al. \cite{Metzger2015} calculated that the observation
of the imprint of free neutrons on the light curve is within reach of
near-future telescopes and that the presence of such imprint would allow
immediate discrimination between a BNS merger and the merger of a neutron
star with a black hole, which would not show such a feature.

When considering fully general-relativistic approaches to the modelling
of the macronova afterglow emission, much less has been done and the
state-of-the-art is represented by the work of Wanajo et
al. \cite{Wanajo2014}, who employed advanced neutrino transport schemes
\cite{Thorne1981,Shibata2011}, finite-temperature EOSs and a detailed
nuclear-reaction network. When the $r$ processes end a few hundreds of ms
after the merger, their simulations followed the $\beta$ decay, fission,
and $\alpha$ decay for a timescale of about 100 days and found that the
total heating rate is dominated at all times by $\beta$ decays from a
small number of species with precisely measured half-lives and that
therefore uncertainties in other nuclear data seem to be irrelevant. More
recent works \cite{Hotokezaka2016b,Barnes2016}, however, found that
realistic thermalisation efficiencies of $\beta$ decay may be lower than
thought before and, if so, contributions from fission and alpha-decays of
transuranic nuclei (whose reaction parameters are not well known) can
actually be important at late times.

We conclude this Section by discussing yet another type of ``afterglow''
electromagnetic transient that could accompany BNS mergers and is related
to the dissipation of the kinetic energy of the ejecta in the ambient
interstellar medium. This dissipation manifests itself through radio
flares that arise from the interaction of sub- to mildly relativistic
outflows with the surrounding matter and lead to non-thermal radio
synchrotron emission \cite{Nakar2011}. The resulting signal is a
potentially detectable, isotropic radio emission that peaks in the radio
band near $1\,{\rm GHz}$ and persists on a detectable level for weeks
\cite{Rosswog2013, Piran2013, Hotokezaka2015MNRAS, Radice2016}. Studies
performed with Newtonian codes \cite{Rosswog2013, Piran2013} suggest that
observations at the sub-milliJanksy level at $1.4\,{\rm GHz}$ are optimal
for the detection of BNS mergers through their post-merger radio
transient. However, their detectability depends strongly on the density
of the medium around the merging binaries, which may be low if the binary
has been ejected from its host galaxy before the merger. More recent
works have considered initial configurations in full general relativity
\cite{Hotokezaka2015MNRAS, Radice2016}. In particular, Radice et
al. \cite{Radice2016} have found that, depending on the initial
configurations and the level of microphysical description, the timescale
for the radio emission varies from about 1 year to almost 30 years and
the radio fluence may vary by more than two orders of magnitude.

\newpage
\section{Advanced Techniques and Alternative Scenarios}
\label{sec:atas}

\subsection{The zoology of binary neutron-star codes}
\label{sec:zoo}

As the study of the merger of BNS systems is transforming into a mature
field of research, a number of numerical codes have been developed to
achieve this result, producing a relative abundance of computational
infrastructures able to solve the Einstein equations together with those
of relativistic hydrodynamics and MHD. Some of the codes used in the last
five years for general-relativistic simulations of BNS were first written
ten or more years ago and have been continuously developed
since\footnote{A number of other general-relativistic hydrodynamics codes
  on fixed spacetimes have been developed over the years for applications
  on astrophysical compact objects. Among such codes are: \texttt{Athena}
  \cite{Stone2008}, \texttt{Athena++} \cite{White2015}, \texttt{Castro}
  \cite{Almgren2010}, \texttt{COSMOS++} \cite{Anninos05c}, \texttt{Enzo}
  \cite{Collins2010}, \texttt{Flash} \cite{Fryxell2000}, and
  \texttt{HARM} \cite{Gammie03}.}. This is the case for the
\texttt{SACRA} code \cite{Yamamoto2008, Shibata2008, Pannarale2013a,
  Kiuchi2014, Sekiguchi2016} (and the other codes developed by the group
now in Kyoto, starting from Refs. \cite{Shibata99d, Shibata99c,
  Shibata99a}), the \texttt{Whisky} code (in its various versions
\cite{Baiotti04, Giacomazzo:2007ti, Dionysopoulou:2012pp, Galeazzi2013,
  Radice2013c}), the \texttt{HAD} code \cite{Anderson2007, Anderson2008,
  Palenzuela2009b, Palenzuela2013, Palenzuela2015}, and the code
developed by the group in Illinois \cite{Yo02a, Duez:2002bn, Faber2007,
  Etienne:2010ui, Etienne2015}. A special note should be made for the
publicly available Einstein Toolkit \cite{loeffler_2011_et, ET2013,
  Moesta13_GRHydro, DePietri2016, Maione2016, einsteintoolkitweb}, whose
routines for hydrodynamics were originally taken from a public version of
the \texttt{Whisky} code \cite{whisky-web} and then were developed in a
independent way, its implementations becoming increasingly different from
the original ones.

In addition to the more ``traditional'' codes mentioned above, new codes
have recently been developed and particular attention is deserved by the
relativistic-hydrodynamics code developed by Thierfelder et
al. \cite{Thierfelder2011} and coupled to the \texttt{BAM} infrastructure
\cite{Bruegmann97, Bruegmann:2003aw, Bruegmann:2006at} code for spacetime
evolution and grid setup. The code implements HRSC schemes on a hierarchy
of mesh refined Cartesian grids with moving boxes and solves the Z4c
formulation of the Einstein equations (see Sect. \ref{sec:CCZ4}).

Another recently developed code is the one by East et
al. \cite{East2012b, East2012b0, East2015, Paschalidis2015, East2016},
where the authors show how they evolve the fluid conservatively using
HRSC methods and a refluxing AMR scheme, while the field equations are
solved in the generalized-harmonic formulation \cite{Garfinkle02,
  Pretorius:2004jg} (see Sect. \ref{sec:Harmonic}) with finite
differences, like the \texttt{HAD} code mentioned above.

Last but certainly not least, is the use of the \texttt{SpEC}
computational infrastructure \cite{Duez:2008rb}, which had been
originally employed to study the inspiral and merger of binary black
holes or of neutron-star--black-hole binaries \cite{Foucart2010,
  Foucart2011, Foucart2012, Foucart2013a, Foucart2015a}, to investigate
BNS mergers \cite{Haas2016}. \texttt{SpEC} employs a mixed approach to
solve the Einstein equations in the generalized harmonic formulation
coupled to matter \cite{Duez:2008rb}, in which the evolution equations
for the spacetime metric are solved using spectral methods as described
in Refs. \cite{Scheel:2008rj, Lovelace2011, Buchman2012, Ossokine2013},
while the fluid equations are solved using HRSC methods as described in
Refs. \cite{Duez:2008rb, Foucart2011, Foucart2013a}. It also uses
comoving coordinates, which minimise the movement of the stars on the
numerical grid during the inspiral, thus decreasing the numerical error
due to interpolation between different refinement levels.

As previously mentioned several times, in addition to Eulerian codes that
solve the full Einstein equations without approximations (except the
truncation error and the errors associated to numerical methods), some
other codes currently employed for BNS merger simulations make
theoretical approximations to the evolution of gravity but are very
advanced in treating microphysical processes and treat matter with
smoothed-particle-hydrodynamics (SPH) methods. In particular, one code of
Rosswog \cite{Korobkin2012, Rosswog2014a} assumes Newtonian gravity and
another code of Rosswog \cite{Rosswog2000, rosswog_2010_csr, Rosswog2013,
  Piran2013} and the code of Bauswein, Janka, Oechslin, and collaborators
\cite{Oechslin02, Oechslin07a, Bauswein:2010dn, Bauswein2013b,
  Goriely2011, Just2015} adopt a conformally flat approximation to
general relativity.

A perhaps disappointing note that however needs to be made is that
despite this abundance of numerical codes, systematic comparisons among
the various codes are done seldom and the only cases we are aware of are
the quantitative comparisons between the \texttt{SACRA} and
\texttt{Whisky} codes \cite{Baiotti:2010ka, Read2013} and between the
\texttt{BAM} and \texttt{SpEC} codes \cite{Haas2016}; such comparisons
are tedious and computationally expensive but important to gauge the
systematic errors that the various code can introduce.

\subsection{High-order numerical methods}
\label{sec:hono}

In order to use results from numerical-relativity simulations to aid and
interpret the analysis of gravitational-wave data from BNSs that will be
soon measured by advanced detectors, it is of fundamental importance to
improve not only the accuracy but also the computational efficiency of
the simulations\footnote{The computational efficiency of the simulations
  can be defined in a number of different ways, but an effective one
  involves the computational cost in terms of CPU hours to obtain a given
  result with a specified truncation error. The smaller the number of CPU
  hours to obtain a result with a given truncation error, the more
  efficient the code.}. Having these goals in mind, a number of groups
have made progress in improving the numerical methods used in simulations
of BNSs, starting with the simpler solution of the pure-hydrodynamics
equations. There are two obvious directions in which developers are
heading through improved numerical methods in numerical-relativity
codes. The first one is a high convergence order of the emitted
gravitational waves; the second one is the ability to limit the growth of
(or even damp) the violation of the constraint equations
\eqref{eq:einstein_ham_constraint}, as discussed in detail in
Sect. \ref{sec:CCZ4}.

In this context, the newly developed \texttt{WhiskyTHC} code by Radice,
Rezzolla and Galeazzi \cite{Radice2013b, Radice2013c, Radice2015}
deserves a special mention as it represents the first
relativistic-hydrodynamics code employing high-resolution
shock-capturing, finite-differencing schemes to go beyond second-order
convergence in the modelling of BNS mergers\footnote{By using standard
  finite-difference methods, Hotokezaka et al. \cite{Hotokezaka2015} also
  reported third- or higher-order convergence in some specific derived
  quantities in their long (\ie $\simeq 16$ orbits),
  low-initial-eccentricity ($\lesssim 10^{-3}$) simulations; however the
  phase evolutions of the gravitational waves from simulations with
  different resolutions could be matched only after a suitable stretching
  of the time coordinate; it is not clear how this affects the
  convergence order.}. Thanks to this increased convergence order, from
the $\simeq 1.8$ of ``traditional'' codes such as \texttt{Whisky} to
$\simeq 3.2$ of \texttt{WhiskyTHC}, Radice et al. \cite{Radice2013b,
  Radice2013c, Radice2015} were able to compute very accurate
gravitational waveforms, including those for higher-compactness binaries,
which are much more challenging to evolve accurately, because numerical
viscosity becomes the leading source of de-phasing from the
point-particle limit, since tidal effects are small. For all of the
simulated BNSs, remarkable agreement was found between
Richardson-extrapolated numerical waveforms and the ones from the tidally
corrected post-Newtonian Taylor-T4 model \cite{Flanagan08, Hinderer09,
  Santamaria2010, Vines:2010ca, Pannarale2011, Maselli2012}.

As the name implies, \texttt{WhiskyTHC} results from the merger of two
codes: \texttt{Whisky} \cite{Baiotti03a} and \texttt{THC}
\cite{Radice2012a}\footnote{THC stand for Templated-Hydrodynamics
  Code. The ``templated'' aspect reflects the fact that the code design is
  based on a modern C++ paradigm called template metaprogramming, in
  which part of the code is generated at compile time. Using this
  particular programming technique it is possible to construct
  object-oriented, highly modular codes without the extra computational
  costs associated with classical polymorphism, because, in the templated
  case, polymorphism is resolved at compile time allowing the compiler to
  inline all the relevant function calls.}. It inherited from
\texttt{THC} the use of state-of-the-art high-order flux-vector splitting
finite-differencing techniques and from \texttt{Whisky} the module for
the recovery of the primitive quantities as well as the new EOS framework
recently introduced in Ref. \cite{Galeazzi2013}. More specifically,
\texttt{WhiskyTHC} employs a flux-vector splitting scheme that uses up to
seventh-order reconstruction in characteristics fields and a novel
entropy-fix prescription for the Roe flux splitting \cite{Radice2013b}.

Among other more traditional algorithms, \texttt{WhiskyTHC} implements
the formally-fifth-order-in-space MP5 scheme\footnote{Other codes also
  implement variations of this scheme.} \cite{suresh_1997_amp} for the
flux reconstruction of the local characteristic variables, using in
particular the explicit expression for the eigenvalues and left- and
right-eigenvectors that can be found in, \eg
Ref. \cite{Rezzolla_book:2013}. The spacetime is evolved via an
implementation of the CCZ4 formulation (see Sect. \ref{sec:CCZ4})
publicly available in the \texttt{Mclachlan} code \cite{mclachlanweb} of
the Einstein Toolkit \cite{loeffler_2011_et, ET2013,einsteintoolkitweb}.

\begin{figure}
\begin{center}
  \includegraphics[width=0.49\columnwidth]{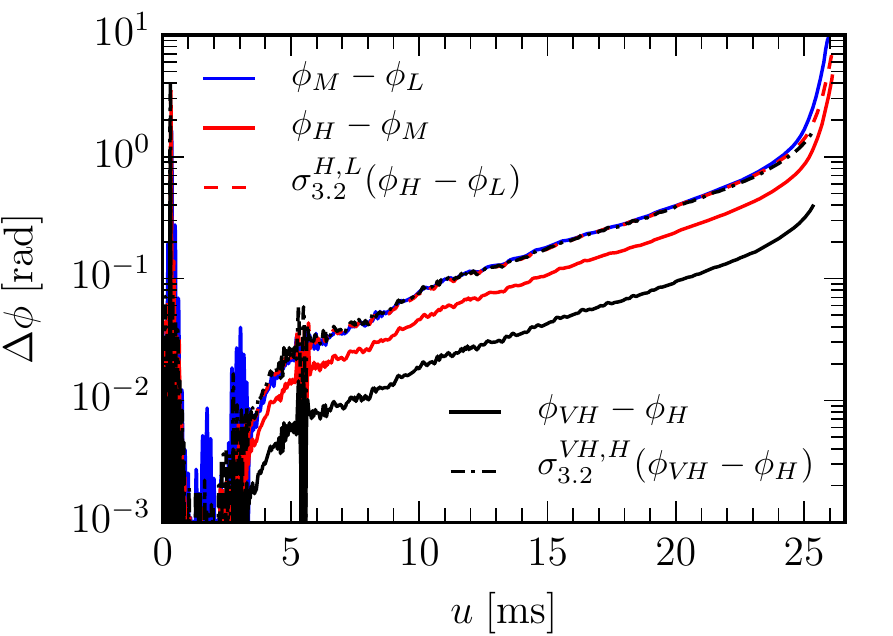} 
  \hskip 0.2cm
  \includegraphics[width=0.49\columnwidth]{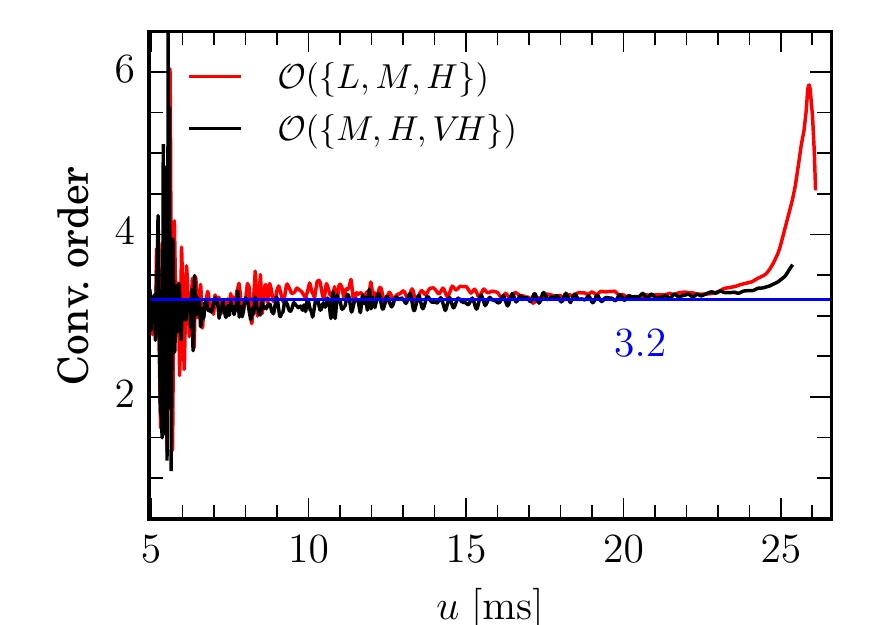} 
\end{center}
   \caption{Accumulate de-phasing (left panel) and estimated order of
     convergence (right panel) for the $\ell = 2, m = 2$ mode of the Weyl
     scalar $\Psi_4$ as extracted at $r = 450\ M_\odot$. The de-phasing
     between high ($H$) and medium ($M$) and very high ($V\!H$) and high
     are also rescaled assuming an order of convergence of $3.2$. The $M,
     H$, and $V\!H$ resolution correspond to spatial spacings on the
     finest grid of $h =0.2,\ 0.133,\ 0.1\,M_\odot \simeq
     295,\ 197,\ 148\,{\rm m}$, respectively. The instantaneous order of
     convergence is estimated separately from the first three resolutions
     $\mathcal{O}(\{L,M,H\})$ and from the last three
     $\mathcal{O}(\{M,H,V\!H\})$. Adapted from Ref. \cite{Radice2015}
   }
   \label{fig:Radice2013c_Fig13}
\end{figure}

As shown in Fig. \ref{fig:Radice2013c_Fig13}, the convergence order of
\texttt{WhiskyTHC} is $3.2$ in the phase (left panel) and the amplitude
of the gravitational waves, without the need to perform any artificial
manipulation of the waveforms \cite{Radice2015}. The measured order of
convergence of $3.2$ is somewhat smaller than the formal order of four of
the scheme, but this is expected because HRSC methods typically reach
their nominal convergence order only at higher resolutions \cite{Shu97,
  Radice2012a}; see also Ref. \cite{Zlochower2012} for a discussion of
other possible sources of errors. Finally,
Fig. \ref{fig:Radice2013c_Fig13} illustrates the fact that the solutions
obtained through \texttt{WhiskyTHC} show a loss of convergence (with
apparent super-convergence) after time $u \gtrsim 24.6\ \mathrm{ms}$, as
also observed with other codes \cite{Bernuzzi2011}. This is roughly the
time when the two neutron stars enter into contact. At this time the
de-phasing between the $H$ and $V\!H$ resolution is $\simeq
0.26\ \mathrm{rad}$; such dephasings over almost seven orbits are
comparable with those obtained for binary black-hole waveforms
\cite{Hinder2013} and highlight that, given sufficiently good initial
data, very accurate waveforms can be computed also for BNS mergers.

The possibility of employing high-order
weighted-essentially-non-oscillatory (WENO) schemes \cite{Shu97,
  Rezzolla_book:2013} in BNS mergers has been first considered by De
Pietri et al. \cite{DePietri2016}, who, \lrn{through a series of
  simulations employing both the BSSNOK and the CCZ4 formulations and
  making use of the public codes within the Einstein Toolkit. Besides
  proving the maturity of such codes, Pietri et al. \cite{DePietri2016}}
have shown the ability of these schemes to yield physically consistent
results even at rather low resolutions\footnote{De Pietri et
  al. \cite{DePietri2016} also remarked that the CCZ4 formulation
  normally requires higher resolutions to obtain consistent results, in
  agreement with what reported by Alic et
  al. \cite{Alic2013}.}. Simulations of BNS mergers with WENO schemes
have also been reported more recently by Bernuzzi and Dietrich
\cite{Bernuzzi2016}, who find that these schemes can be employed robustly
for simulating the inspiral-merger phase and can improve the assessment
of the error budget in the calculation of the gravitational waveforms
when compared to finite-volume methods. The convergence order reported by
Bernuzzi and Dietrich \cite{Bernuzzi2016}, however, is only $\simeq 2$,
mostly because optimal choices for the WENO weights for this type of
problem and in this range of frequencies are not available yet.

Before concluding this Section, we note that an effort has been invested
recently also in the application of discontinuous-Galerkin methods
\cite{Cockburn1989a, Cockburn1990, Cockburn1998, Rezzolla_book:2013}
within general-relativistic codes solving the equations of relativistic
hydrodynamics or MHD. This is very much a growing field and, at present,
these studies are either limited to dynamical curved spacetimes in one
spatial dimension \cite{Radice2011}, to fixed curved spacetimes in three
spatial dimensions \cite{Bugner2015}, or to flat spacetimes, but in
relativistic MHD \cite{Zanotti2015, Zanotti2015b}.

\subsection{Advanced Numerical Techniques}
\label{sec:ant}

Among the advanced numerical techniques recently introduced in
numerical-relativity codes modelling BNS mergers there is one that deals
with the rather delicate treatment of interfaces between fluid regions
and regions supposed to be vacuum. As mentioned above, this is one of the
most challenging problems in Eulerian (relativistic) hydrodynamics codes
(see \eg \cite{galeazzi_master, kastaun_2006_hrs, Millmore2010}). As
mentioned in Section \ref{sec:me_rhd}, the most commonly used approach to
treat vacuum regions is to fill them with a low-density fluid, the {\it
  atmosphere}, such that if the evolution of a fluid element would bring
it to have a rest-mass density below a certain threshold, some
hydrodynamical variables are set to a floor value. This approach works
reasonably well for standard second-order codes and has been adopted,
with some variations, by the vast majority of the
relativistic-hydrodynamics codes, but it is problematic for higher-order
codes \cite{Radice2011}. The reason is that small numerical oscillations
can easily couple with the prescription for the floor, violating the
conservative character of the equations and affecting artificially the
conservation of mass, energy, and momentum. Such oscillations may be
amplified and ultimately cause instabilities in the evolution.

Radice et al. \cite{Radice2013c} have proposed and tested different
solutions to this problem, finding that the positivity-preserving limiter
recently proposed by Hu et al. \cite{Hu2013} is the one that works
best. This method still requires a low-density fluid everywhere and the
enforcement of a floor value, but has the important property of enforcing
the local conservation of the solution up to floating-point precision. In
practice, the floor for the rest-mass density can be arbitrarily small
and does not require any tuning. In contrast, the classical atmosphere
prescriptions usually work only in a limited range floor values. The
final prescription is rather simple: one fills the vacuum with a low
rest-mass density floor at the beginning of a simulation and then lets it
evolve freely, only relying on the positivity preserving limiters to
ensure its proper behaviour. This typically results in the creation of
accretion flows onto the compact objects, but, given the low rest-mass
density that can be chosen for the floor\footnote{In
  Refs. \cite{Radice2013b, Radice2013c, Radice2015}, the atmosphere floor
  is taken to be $16$ orders of magnitude smaller than some reference
  rest-mass density (normally the initial maximum rest-mass density),
  essentially below floating-point precision.}, the effects of this
artificial accretion are completely negligible. Radice et
al. \cite{Radice2013c} tested all their proposed different atmosphere
prescriptions and found that they give identical results during the
inspiral and yield very marginal differences in the merger phase, while
significant differences may appear in the post-merger phase. This
suggests that the treatment of the neutron-star surface is not a leading
source of error in BNS simulations, as far as the inspiral
gravitational-wave signal is concerned.

Also worth discussing as an advanced numerical technique in codes for BNS
mergers is the solution of the problem of small violations of the
conservation of rest-mass that can take place when the fluid flow crosses
a mesh-refinement boundary. These violations are the result of the jump
in resolution across the refinement levels that, because of the average
nature of finite-volume methods, can lead to violations of the
conservative formulation of the hydrodynamic equations. The solution to
this problem is called {\it refluxing} and was originally proposed by
Berger and Colella \cite{Berger89}. As suggested by the name, the
refluxing technique involves a suitable correction of the fluid fluxes
across the mesh-refinement boundaries so as to enforce the conservative
nature of the equations.

\begin{figure*}
\begin{center}
  \includegraphics[width=0.625\columnwidth]{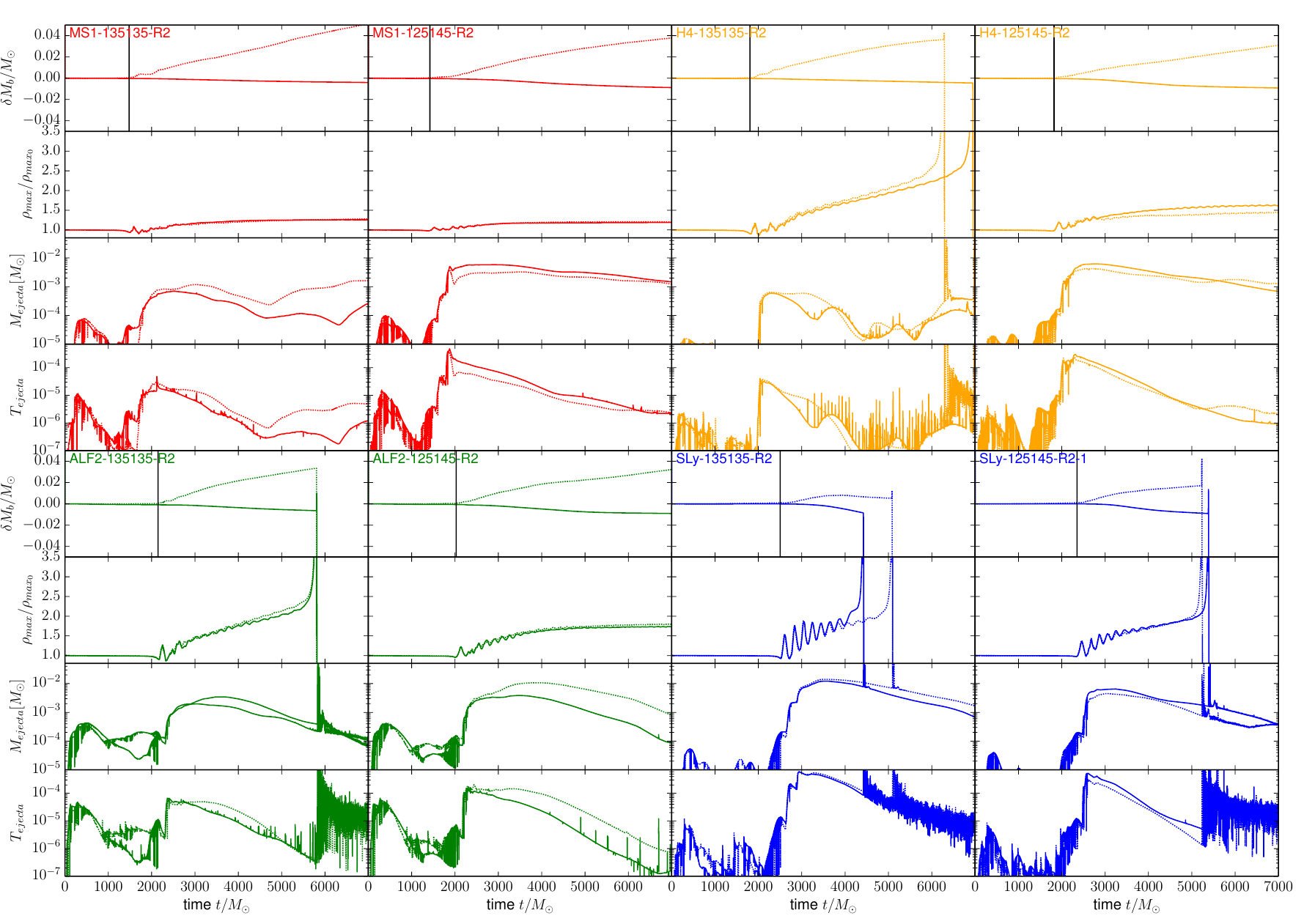} 
  \includegraphics[width=0.365\columnwidth]{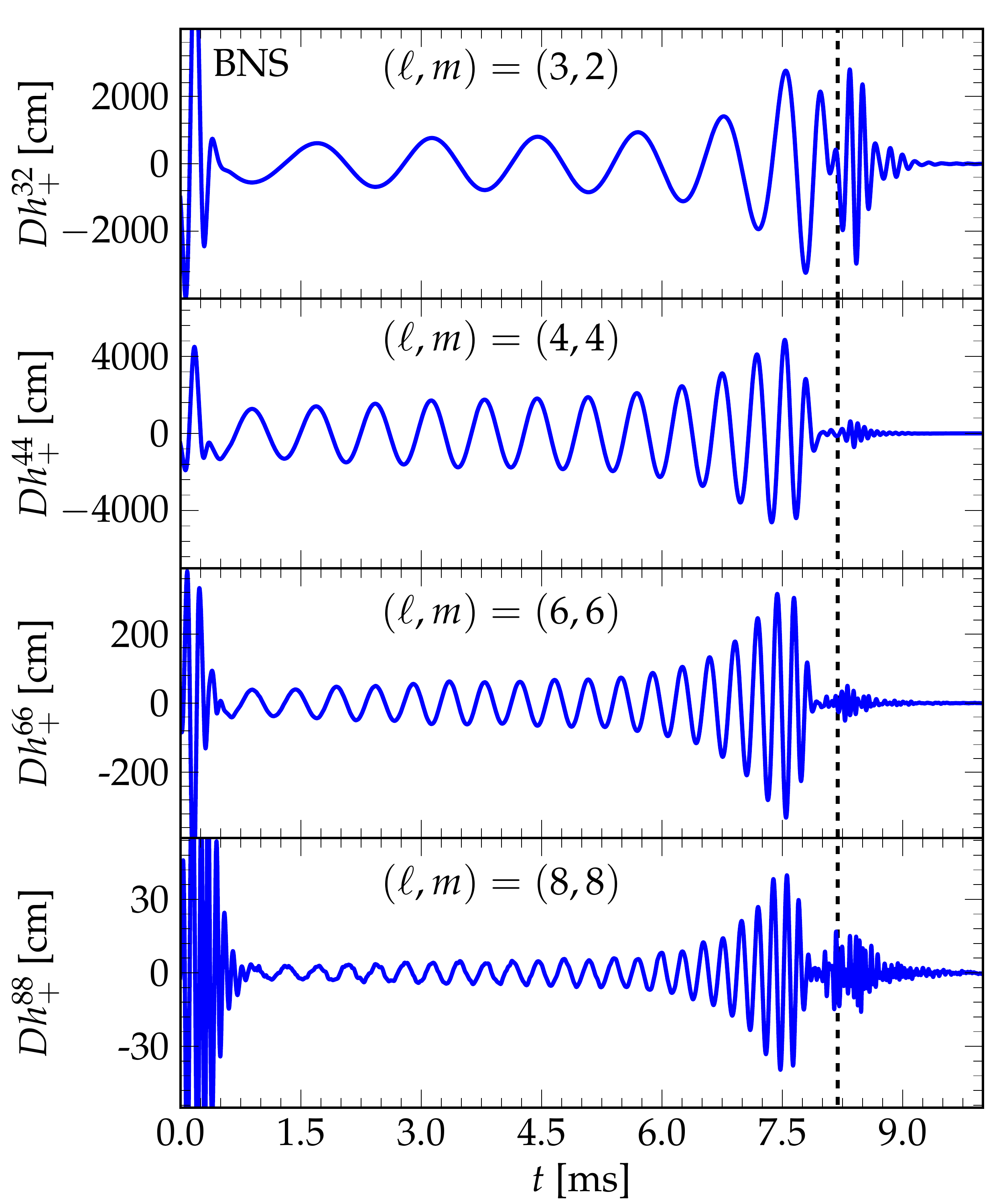} 
\end{center}
\caption{\emph{Left panel:} Evolution of some dynamical quantities for
  simulations of configurations with total mass $M=2.7M_\odot$ and
  initial mass ratios $q=1$ and $q=1.16$. Results for different EOSs are
  in different colours. For each configuration, the panel contains four
  plots. From top to bottom: violation of rest-mass conservation $\delta
  M =M(t) - M(t=0)$ on the intermediate fixed refinement level; maximum
  rest-mass density on the grid scaled to its initial value; rest mass of
  the ejected material; kinetic energy of the ejecta. Results for
  conservative AMR are presented with solid lines, while the
  corresponding results for nonconservative AMR are shown with dashed
  lines. Vertical lines indicate the time of merger, determined by the
  maximum in the gravitational-wave amplitude. [Reprinted with permission
    from Ref. \protect{} \cite{Dietrich2015} \copyright~(2015) by the
    American Physical Society.] \emph{Right panel:} Various
  gravitational-wave modes $(\ell, m)$ from a BNS simulation, obtained
  via Cauchy-characteristic extraction. The vertical line indicates the
  time of appearance of an apparent horizon. [Reprinted with permission
    from Ref. \protect{} \cite{Reisswig2012b}. \copyright~(2013) by the
    American Physical Society.] }
   \label{fig:Dietrich2015_Fig10}
\end{figure*}

A first implementation of this technique within the Einstein Toolkit
\cite{loeffler_2011_et, ET2013, Moesta13_GRHydro, einsteintoolkitweb} was
performed by Reisswig et al. \cite{Reisswig2012b} and was later followed
by Dietrich et al. \cite{Dietrich2015}, who carried out a detailed study
of its efficacy. In particular, as shown in the left panel of
Fig. \ref{fig:Dietrich2015_Fig10}, results obtained with and without the
conservative mesh-refinement algorithm were compared, and it was found
that simulations of the post-merger phase can be affected by systematic
errors if mass conservation across mesh refinements is not enforced. Such
errors may affect estimates of the mass of the ejecta by factors of
several. The computational overhead due to the enforcement of rest-mass
conservation amounts to an increase of the evolution time of $\sim 10\%$
at most, which is quite acceptable in view of the benefits in terms of
accuracy. Overall, it was found that the conservation of rest mass is
improved of a factor $\sim 5$, but also that the variance in other
relevant quantities, such as the total mass of the ejecta and the time of
collapse of the binary-merger product, depends sensibly on other factors
that are far more difficult to control, \eg the specific grid setup or
the treatment of the atmosphere, in addition to the EOS.

The impressive investigation work by Reisswig et
al. \cite{Reisswig2012b}, also contained two additional new contributions
to the use of advanced techniques to study BNS merger. The first one is
the implementation in relativistic hydrodynamics of adapted curvilinear
grids, called {\it multipatches} or {\it multiblocks}, coupled with
flux-conservative, cell-centered adaptive mesh refinement. The idea is to
cover the simulation domain with multiple curvilinear coordinate {\it
  patches}. Each patch is locally uniform, and diffeomorphic mappings
from local to global coordinates enable the representation of a wide
range of grid shapes in different regions of the simulation. The most
common and useful setup here consists of a central Cartesian patch (for
resolving the aspherical region of coalescence and merger) surrounded by
six {\it inflated-cube} spherical grid patches. Multiblocks are useful in
many different ways \cite{Reisswig2012b}: (i) it is possible to set the
outer boundary very far, enough for it to be causally disconnected from
the interior evolution and the gravitational-wave extraction zone, thus
avoiding systematic errors from the approximate and
non--constraint-preserving outer-boundary condition; (ii) it is possible
to track the ejected material out to large radii with relatively high
resolution; (iii) the number of mesh-refinement levels can be decreased
without reducing the resolution, leading to better parallel scaling.

The second important contribution by Reisswig et al. \cite{Reisswig2012b}
is the first application in the context of BNS mergers of
Cauchy-characteristic extraction (see Ref. \cite{Bishop2016} for a
review) for gravitational-wave extraction at future null infinity. The
characteristic formulation takes spacetime as foliated into a sequence of
null cones emanating from a central geodesic. This approach has the
advantage that the Einstein equations can be compactified so that future
null infinity is rigorously represented on a finite grid, and there is no
artificial outer boundary condition. However, it suffers from the
disadvantage that the coordinates are based on light rays, which can be
focused by a strong field to form caustics, which complicate a numerical
computation \cite{Bishop97b}. Cauchy-characteristic extraction is capable
of determining the gravitational radiation content unambiguously and
without finite-radius and gauge errors. The combined use of causally
disconnected outer boundaries and the application of
Cauchy-characteristic gravitational-wave extraction allowed Reisswig et
al. \cite{Reisswig2012b} to extract higher orders than the $\ell=2=m$
gravitational-wave mode, obtaining non-negligible signals up to $\ell =6$
(see the right panel of Fig. \ref{fig:Dietrich2015_Fig10}).

\subsection{Alternative theories of gravity}
\label{sec:atog}

Building up on previous analytical studies within the post-Newtonian
approximation on the dynamics and gravitational-wave emission from BNSs
in scalar-tensor theories (see, \eg Refs. \cite{Eardley1975a, Will92,
  Alsing2012, Mirshekari2013}) a number of recent numerical-relativity
investigations have addressed the problem of BNS mergers in gravitational
theories alternative to general relativity \cite{Barausse2013,
  Shibata2014, Palenzuela2014, Sampson2014, Taniguchi2015,
  Ponce2014a}. The efforts are concentrated on scalar-tensor theories of
gravity \cite{Wagoner1970, Nordtvedt1970, Will92}, where the
gravitational interaction is mediated by a scalar degree of freedom, in
addition to the usual tensor one. Such theories have received much
attention, because the presence of a scalar field in nature is motivated
by, \eg the low-energy limit of string theories, the observation of the
Higgs boson, and cosmological phenomenology (see Refs. \cite{Will92,
  Damour1992} for reviews).

\begin{figure*}
\begin{center}
  \includegraphics[width=0.52\columnwidth]{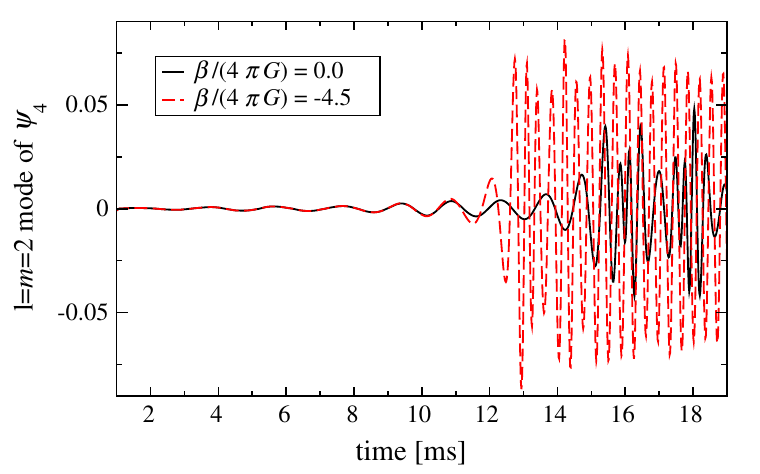} 
  \includegraphics[width=0.47\columnwidth]{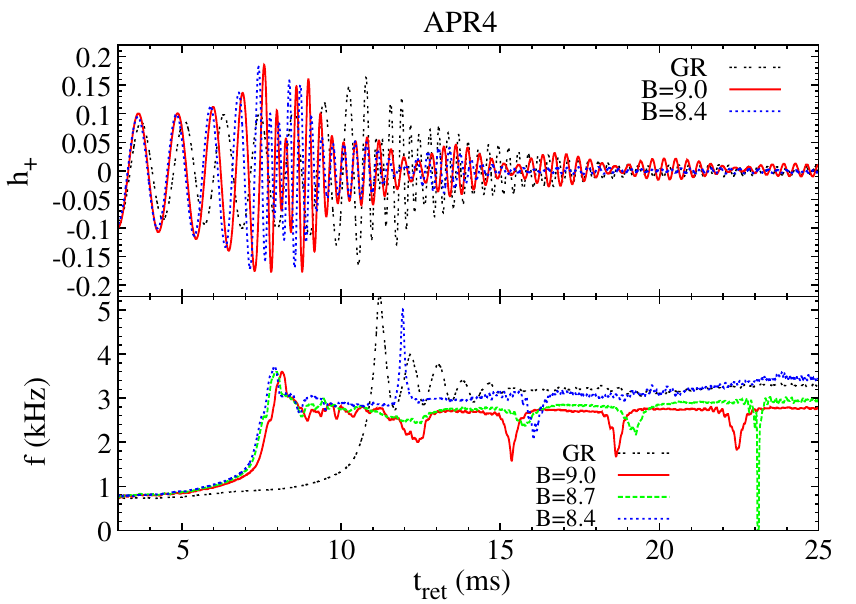} 
\end{center}
\caption{\emph{Left panel:} The dominant mode of the gravitational wave
  for an equal-mass binary with individual rest masses of $1.625
  M_\odot$. [Reprinted with permission from Ref. \protect{}
    \cite{Barausse2013}. \copyright~(2013) by the American Physical
    Society.] \emph{Right panel:} Plus-polarization of gravitational
  waves observed along the axis perpendicular to the orbital plane and
  their frequency as functions of the retarded time for simulations of
  BNSs of total rest mass $M= 2.7M_\odot$ with the APR4 EOS [Reprinted
    with permission from Ref. \protect{}
    \cite{Shibata2014}. \copyright~(2014) by the American Physical
    Society.] In both panels, the results for general relativity are
  compared with the waves obtained in scalar-tensor theory, with
  different values of the free parameter ($\beta$ on the left and $B$ on
  the right; $B=-2\beta$).}
   \label{fig:Shibata_scalar-tensor_Fig4}
\end{figure*}

The parameters of scalar-tensor theories are constrained by solar-system
experiments and binary-pulsars observations \cite{Kramer2009,
  Freire2012}, but even if such constraints are satisfied, significant
differences from the predictions of general relativity could be
observable in stronger gravity regimes. In particular, strong-field
aspects of the dynamics of BNSs in scalar-tensor theories, \eg
late-inspirals, mergers, and post-merger phases, can show large
deviations from general relativity, which cannot be captured accurately
by weak-field analyses \cite{Barausse2013} and cannot be reproduced
within general relativity, \eg via exotic EOSs. Although the scalar field
is only weakly coupled, it carries energy away from the source, exerting
a significant back-reaction that could leave an imprint in the
gravitational waves, detectable with Advanced LIGO/Virgo \cite{Babusci01,
  Barausse2013, Shibata2014, Palenzuela2014, Sampson2014} (see
Fig. \ref{fig:Shibata_scalar-tensor_Fig4}). In practice, for some values
of the parameters of the theory, neutron stars of a given mass and EOS
may merge at significantly lower frequency than in general relativity
\cite{Barausse2013}. A faster orbital decay may happen also because, in
an unequal-mass binary, the system would emit also dipolar radiation due
to the scalar field \cite{Damour1992, Damour1993, Barausse2013}. Such
additional decay, as well as the existence of dipole radiation,
scalarisation, and the magnitude of the parameters of the theory, are
rather stringently constrained by binary-pulsars observations
\cite{Kramer2009, Freire2012}, which however do not probe regions of the
space of parameters relative to fields as strong as those occurring in
the late inspiral of the binary.

Since the scalar charge of a compact object depends on the gravitational
binding energy (or compactness) of the object itself, a neutron star can
undergo a {\it spontaneous scalarisation} if it becomes compact enough
(and for some values of the parameters of the scalar-tensor theory). That
is, it can acquire a scalar field from an initial state where scalar
fields were absent, since the scalarised configuration represents a
lower-energy state \cite{Damour1992,Damour1993}. Furthermore, if the star
is in a binary, the orbital binding energy of the binary system
contributes to the scalarisation, and this process is referred to as {\it
  dynamical scalarisation} \cite{Barausse2013}. Dynamical scalarisation
may also happen after the onset of the merger, if a more compact neutron
star is formed \cite{Shibata2014}. Finally, once a star in a binary
acquires a scalar charge, it can also scalarise its binary companion,
through a process that is referred to as {\it induced scalarisation}
\cite{Barausse2013}).

By modelling the early plunge through a toy model based on a simple EOS
and employing Markov-Chain Monte-Carlo techniques, Sampson et
al. \cite{Sampson2014} studied whether deviations from general relativity
in gravitational-wave signals that are caused by a certain class of
scalar-tensor theories can be detected with advanced detectors and found
that this is the case for the standard projected sensitivity of the
detectors, if the stars are sufficiently compact. However, not all
scalarisation effects are easily detectable, so that dynamical
scalarisation in the late inspiral of BNSs will be difficult to constrain
with gravitational-wave observations, unless the system scalarises at a
low enough orbital frequency (\ie large enough orbital separation) that a
sufficient amount of SNR is accumulated while the scalar-tensor
modifications are active.

The first fully nonlinear numerical simulations of BNSs in scalar-tensor
theories were performed by Barausse et al. \cite{Barausse2013} and were
rapidly followed by those of Shibata et al. \cite{Shibata2014}, who found
similar results, but through more sophisticated and detailed
simulations. Such simulations employed initial conditions from
quasi-equilibrium configurations that consistently include also the
scalar field, realistic EOSs (Barausse et al. \cite{Barausse2013}
employed a polytropic EOS), and free parameters of the theory chosen
taking into account the constraints imposed by the latest observations of
neutron-star--white-dwarf binaries with pulsar timing. We note that while
the simulations performed by Barausse et al. \cite{Barausse2013} employed
the so-called ``Einstein frame'', those of Shibata et
al. \cite{Shibata2014} adopted the so-called ``Jordan frame'' (the
results of Barausse et al. \cite{Barausse2013} are however expressed in
the Jordan frame). 

We should note that in both sets of simulations, the values chosen for
the parameters of the scalar field (and in particular the signs of such
parameters) are compatible with solar system experiments, but only if the
scalar field is assumed to have a mass in some specific range, namely
with a Compton wavelength between kilometres and the Hubble scale
\cite{Sampson2014, Ramazanoglu2016}. This assumption does not affect the
results of simulations which are relative to much smaller lengthscales,
but it remains an assumption.

Shibata et al. \cite{Shibata2014} also extended the study to the merger
and post-merger phases and emphasized that also the evolution of the
remnant neutron star in scalar-tensor theories may be quantitatively
different from that predicted by general relativity, because, when the
remnant is scalarised, the compactness is different from the one computed
in general relativity, and thus the frequency of the quasi-periodic
stellar oscillations and so of the gravitational waves is also
different. Such a difference depends on the EOS. Shibata et
al. \cite{Shibata2014} underline that the possibility (studied in many
works \cite{Bauswein2012a, Bauswein2012, Hotokezaka2013c, Bauswein2014,
  Takami:2014, Clark2014, Takami2015, Rezzolla2016, Maione2016}) to
constrain the EOS of neutron stars by observing the frequency of
post-merger quasi-periodic gravitational waves emitted by the
binary-merger product may be compromised by the degeneracy introduced by
scalar-tensor theories, since the frequency of such gravitational-wave
emission depends also on the degree of scalarisation.

Taniguchi et al. \cite{Taniguchi2015}, then, used quasi-equilibrium
sequences, computing equal-mass, irrotational BNSs in scalar-tensor
theory and with realistic EOSs (approximated as piecewise
polytropes). These quasi-equilibrium sequences were used to compute the
dependence of the binary scalar charge and binding energy on the orbital
angular frequency and found that the absolute value of the binding energy
is smaller than in general relativity, at most by $14\%$. It was also
noted that dynamical scalarisation can yield a different number of
gravitational-wave cycles prior to merger and that such a difference is
much larger than the effect due to tidal interactions, which is of the
order of a few gravitational-wave cycles at most.

There has been some debate on whether in order to compute the dynamics
and the waveforms accurately enough it is necessary to solve consistently
the initial data problem for configurations in which the scalar field is
consistently taken into account as opposed to configurations in which the
scalar field develops dynamically during the evolution (the initial data
for such configurations without scalar field is of course consistently
solved in all of the cited works). On the one side, in
Refs. \cite{Barausse2013, Sampson2014} it is stated that even for such
configurations the error introduced by neglecting the scalar field at the
initial time is negligible; this conclusion is based also on the fact
that the results of Refs.\cite{Barausse2013,Sampson2014} were later
reproduced analytically in the post-Newtonian approximation
\cite{Palenzuela2014,Sennet2016} (see below). On the other side, in
Ref. \cite{Taniguchi2015} it is claimed that inaccurate initial data may
cause artefacts in the dynamical evolution of the binary system, like a
plunge more rapid than in general relativity, as reported in
Refs. \cite{Barausse2013,Sampson2014} and not observed in
Refs. \cite{Shibata2014,Taniguchi2015}. It should also be noted that the
considerations made in Ref. \cite{Taniguchi2015} may be limited by the
fact that they refer to equilibrium configurations and do not involve
evolutions. Further work is needed to clarify the issue.

As mentioned above, Palenzuela et al. \cite{Palenzuela2014} have made
computations with an improved post-Newtonian orbital-evolution technique,
interfaced with a set of nonlinear algebraic equations that provide a
description of the scalar charge (see also Ref. \cite{Sennet2016} for an
improved approach using a resummed post-Newtonian expansion). After
validating this semi-analytical procedure by comparing results to those
of fully general-relativistic simulations, they investigated the behavior
of BNSs prior to merger, exploring large portions of the parameter space
of scalar-tensor theories and including equal-mass, unequal-mass, and
eccentric BNS systems. This simplified model also allows for an efficient
generation of templates of waveforms within scalar-tensor theories, for
the exploration of possible degeneracies, and for the investigation of
the extent to which existing data-analysis techniques can be exploited to
test gravity theories alternative to general relativity.

Finally, Ponce et al. \cite{Ponce2014a} examined whether the
characteristics of the \lrn{prompt} electromagnetic counterparts (see
Sect.  \ref{sec:EM_counterparts}) to BNS systems can provide an
independent way to test gravity in the strong-field regime. In
particular, they found that in some cases the electromagnetic emission
during the inspiral in scalar-tensor theories may show deviations from
the prediction of general relativity. These differences are quite small
and so very accurate measurements are required to differentiate between
general relativity and scalar-tensor theories when using electromagnetic
observations alone; of course, a multimessenger signal would provide far
more stringent constraints.

\subsection{Relativistic collisions}

We conclude this Chapter and review with a topic that still involves
pairs of neutron stars, although under rather unusual circumstances,
namely, through their head-on collision when they are boosted to
relativistic speeds. Hence, this is not quite the type of BNS ``merger''
discussed so far, although it still involves the merger of two neutron
stars.

Much of the interest in the collision of two neutron stars moving at
relativistic speeds obviously stems from the possibility of using this
system to study, in a controlled numerical environment, the conditions
that may lead to black-hole formation, and, in particular, from the
interest in extending to a more dynamical context Thorne's \textit{hoop
  conjecture} \cite{Thorne72a}. We recall that the hoop conjecture casts
in rather loose terms the conditions for black-hole formation by stating
that a black hole will be formed if an object of mass (energy) $M$ is
confined in a volume that in all directions (hence the word
\textit{hoop}) has a radius $R \leq 2M$. While somewhat obvious, this
conjecture does not distinguish the amount of energy in $M$ that could be
of kinetic nature. It is not difficult, in fact, to imagine that the
collision of two compact objects of mass $M/2$ when in isolation, could
or could not lead to the formation of a black hole depending on the
amount of kinetic energy they possess at the time of the collision
\cite{Eardley_2002, Choptuik:2010a}.

These considerations are not entirely academic and were spurred by the
first operations of the Large Hadron Collider at CERN. Indeed, after
applying the conjecture to particles, one finds that the threshold for
black-hole formation occurs at Planck energy scales. However, some
theories of quantum gravity with small or warped extra dimensions
\cite{Antoniadis1998,Argyres_1998,Yoo_2010,Yoshino_2003} predict that the
energy required for black-hole formation might be significantly smaller
than the Planck scale \cite{Argyres_1998}. In this case, proton
collisions at the Large Hadron Collider (LHC) \cite{Dimopoulos_2001}, or
collisions of cosmic rays with the Earth's atmosphere \cite{Feng2002},
may become relevant to the formation of microscopic black holes. To date,
however, no evidence for black-hole formation in these events has been
found \cite{Chatrchyan2013}.

\begin{figure}
\begin{center}
  \includegraphics[width=0.495\columnwidth]{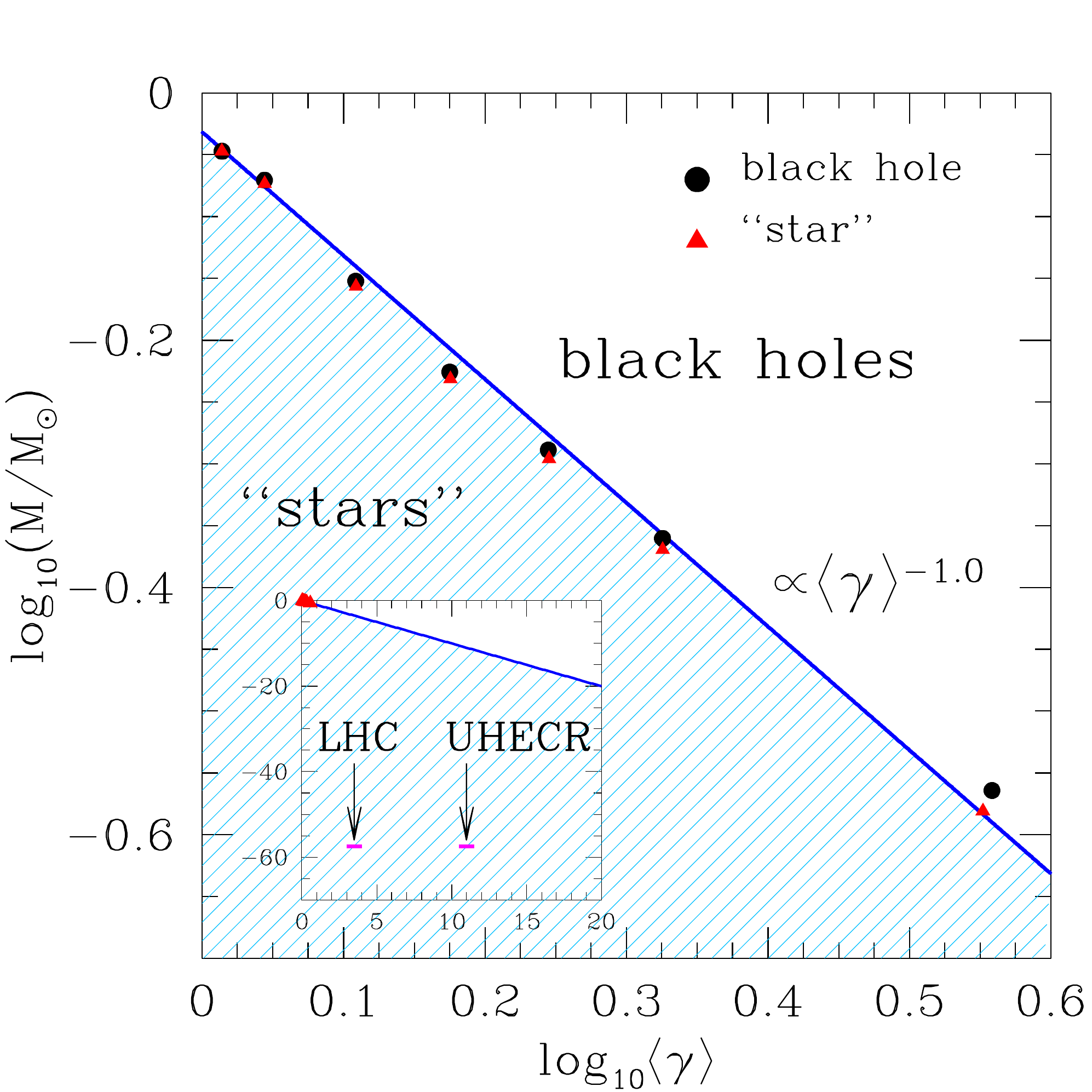}
  \raisebox{0.6cm}{\includegraphics[width=0.495\columnwidth]{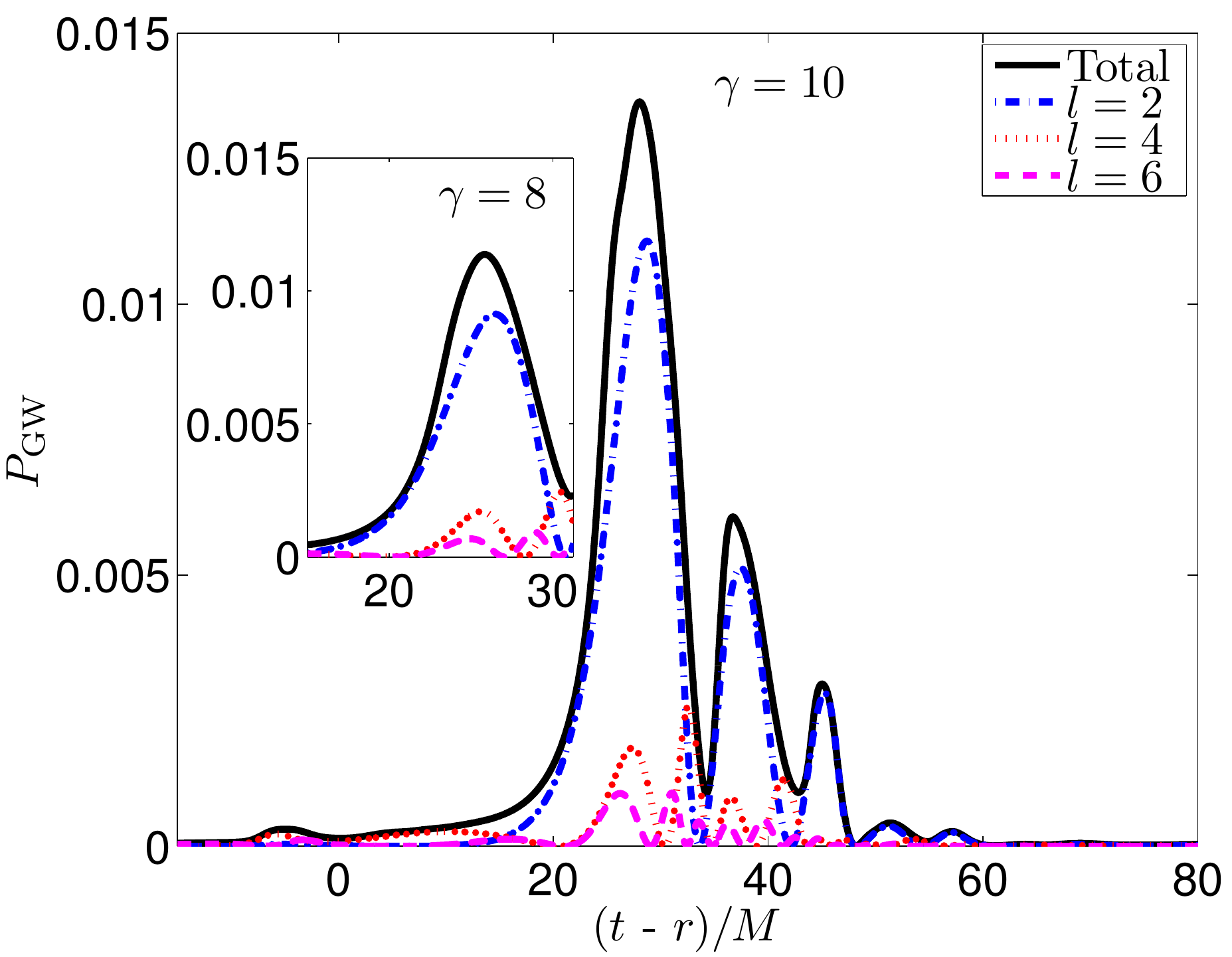}}
\end{center}
   \caption{\emph{Left panel:} Critical line as a function of the
     effective Lorentz factor; here circles indicate black holes, while
     triangles self-gravitating partially bound objects. [From
       Ref. \cite{Rezzolla2013} \copyright \protect{} IOP
       Publishing. Reproduced with permission. All rights reserved.]
     \emph{Right panel:} Total and spin-weight 2 spherical harmonic
     decomposition of the power in gravitational waves from a collision
     with $\langle \gamma \rangle = 10$ (the inset shows a case with
     $\langle \gamma \rangle = 8$). Only the power for $t-r>0$ is
     accounted for in the estimate of the emitted energy. [Reprinted with
       permission from Ref. \protect{} \cite{East2012}. \copyright~(2013)
       by the American Physical Society.]}
   \label{fig:headon}
\end{figure}

Also with this in mind, Rezzolla and Takami \cite{Rezzolla2013},
expanding on their previous work \cite{Kellermann:08a, Kellermann:10,
  Radice:10}, studied black-hole formation from the ultrarelativistic
head-on collision of two neutron stars. They performed simulations that
improve several aspects of previous work \cite{Eardley_2002,
  Choptuik:2010a}. In particular, their colliding objects are not treated
as point particles or scalar fields, but rather as extended and
self-gravitating fluid objects, thus representing a more realistic
description of the baryonic matter involved in ultra-relativistic
heavy-ion collisions. Their results confirmed that the collision of two
self-gravitating objects gives rise to a type-I critical behaviour
\cite{noble_2003_nsr, Noble08a, Kellermann:08a, Kellermann:10, Radice:10,
  Noble2016}\footnote{In analogy with first- and second-order phase
  transitions in statistical mechanics, critical phenomena characterised
  by a finite mass at the black-hole--formation threshold are called type
  I, while the critical phenomena with a power-law scaling of such mass
  are called type II \cite{Gundlach-2003-critical-review}.}, with a black
hole being produced for masses above a critical value $M_{\rm c}$ and a
partially bound object for masses below $M_{\rm c}$.

Interestingly, the critical value depends on the Lorentz factor of the
colliding stars, so that increasingly smaller masses are needed to
produce a black hole when the stars are boosted to larger and larger
velocity. The location of this critical line in the $(M, \langle \gamma
\rangle)$ plane is expressed by the simple relation $M_{\mathrm
  c}/M_{\odot} \approx 0.92 ~ \langle \gamma \rangle^{-1.03}$, where
$\langle \gamma \rangle$ is the average Lorentz factor of the boosted
stars \cite{Rezzolla2013} (see left panel of Fig. \ref{fig:headon}). This
expression essentially states the conservation of energy, but also that
in the limit of zero initial velocities, \ie $\langle \gamma \rangle \to
1$, twice the critical mass is slightly larger than the maximum mass of
the relative spherical-star sequence, \ie $2 M_{\mathrm c} \simeq
1.12\,M_{_{\rm TOV}} \simeq 1.12\times 1.637\,M_{\odot}$ for the EOS
considered. In the opposite limit of $\langle \gamma \rangle \to \infty$,
the expression above predicts that the critical mass will go zero: as the
kinetic energy diverges, no room is left for selfgravitating matter,
which will all be ejected but for an infinitesimal amount that will go
into building the zero-mass critical black hole. While the precise
numbers reported in Ref. \cite{Rezzolla2013} refer to a given choice of
the EOS, it is claimed that the same behaviour should hold qualitatively
for all EOSs.

Returning to the hoop conjecture, Rezzolla and Takami \cite{Rezzolla2013}
showed that the presence of a critical line can also be cast as a
condition on a critical compactness, \ie it can be expressed as $2M_{\rm
  lab}/R \approx 0.16\, \langle \gamma \rangle^{-0.13}$, where
\hbox{$M_{\rm lab} \coloneqq \langle \gamma \rangle M$} is the mass in
the lab frame and $R$ is the largest dimension in that frame being
transverse to the motion. Stated differently, two colliding
self-gravitating systems moving with an average boost of $\langle \gamma
\rangle$ will produce a black hole if the transverse compactness is
$\approx 0.08/\langle \gamma \rangle^{0.13}$. This extends the static
hoop conjecture to a dynamical framework \cite{Rezzolla2013}.

At about the same time when the results above were published, East and
Pretorius \cite{East2012} took further the investigation by exploring
significantly higher-boost collisions where the ratio of kinetic to rest
mass energy is of order $10:1$, although for a more restricted range of
masses. East and Pretorius \cite{East2012} found that the threshold for
black-hole formation is a factor of a few less than the estimate of the
hoop conjecture. A new and interesting phenomenon was also presented,
namely that for boosts slightly above the threshold, two separate
apparent horizons form shortly after the collision and then some time
later are encompassed by a single horizon that rings down to a
Schwarzschild black hole. This is qualitatively explained as due to the
strong focusing of the fluid elements of one star by the boosted
spacetime of the other, and viceversa. Finally, East and Pretorius
\cite{East2012} were able to measure for the first time the
gravitational-wave emission generated by the collision, finding that up
to $16\pm 2\%$ of the total energy of spacetime is radiated via
gravitational waves (see right panel of Fig. \ref{fig:headon}).

\newpage
\section{Summary and outlook}
\label{sec:sao}

As anticipated in the Introduction, there is little doubt that this is a
particularly exciting and highly dynamical time for research on neutron
stars, in general, and on BNS mergers, in particular. In less than 10
years, \ie starting approximately from 2008, a considerable effort by
several groups across the world has obtained numerous important results
about the dynamics of binary systems of neutron stars, employing a large
variety of numerical (in most cases) and analytical (in a few cases)
techniques and exploring this process with different degrees of
approximation and realism.

Altogether, these works have revealed that the merger of a binary system
of neutron stars is a marvellous physical laboratory. Indeed, BNS mergers
are expected to be behind several fascinating physical processes, which
we recall here: {(i)} they are significant sources of gravitational
radiation; {(ii)} they could act as possible progenitors for
short-gamma-ray bursts (SGRBs); {(iii)} they have the potential to
produce electromagnetic and neutrino emission that is visible from
enormous distances; {(iv)} they are likely responsible for the production
of a good portion (if not all) of the very heavy elements in the
Universe. When viewed across this lens, it is quite natural to consider
BNS mergers as Einstein's richest laboratory, binding in the same
environment highly nonlinear gravitational dynamics with complex
microphysical processes and astonishing astrophysical phenomena.

The huge progress accomplished over the last ten years has helped trace a
broadbrush picture of BNS mergers that has several sound aspects, among
which the most robust in our opinion are the following ones:\footnote{In
  this Section we will intentionally omit references to avoid cluttering
  the text; all the relevant references can be found in the various
  Sections covering the topics discussed here.}

\begin{itemize}

\item Independently of the fine details of the EOS, of the mass ratio or
  of the presence of magnetic fields, the merger of a binary system of
  neutron stars eventually leads to a rapidly rotating black hole with
  dimensionless spin $J/M^2 \simeq 0.7-0.8$ surrounded by a hot accretion
  torus with mass \lrn{in the range $M_{\rm torus} \sim 0.001 -
    0.1\,M_{\odot}$.} Only very low-mass progenitors whose total mass is
  below the maximum mass of a (nonrotating) neutron star would not
  produce a black hole. It is unclear whether such progenitors are
  statistically important.

\item The complete gravitational-wave signal from inspiralling and
  merging BNSs can be computed numerically with precision that is smaller
  but overall comparable with that available for black holes.

\item When considering the inspiral-only part of the gravitational-wave
  signal, semi-analytical approximations either in the post-Newtonian or
  EOB approximation, can reproduce the results of numerical-relativity
  calculations essentially up to the merger.

\item The gravitational-wave spectrum is marked by precise frequencies,
  either during the inspiral or after the merger that exhibit a
  ``quasi-universal'' behaviour. \lrn{In other words, while the position
    of the peaks depends on the EOS, it can be easily factored out to
    obtain EOS-independent relations between the frequencies of the peaks
    and the properties of the progenitor stars.}

\item The result of the merger, \ie the binary-merger product, is a
  highly massive and differentially rotating neutron star. The lifetime
  of the binary-merger product depends on a number of factors, including
  the mass of the progenitors, their mass ratio and EOS, as well as the
  role played by magnetic fields and neutrino losses. While sufficiently
  large initial masses can yield a prompt collapse at the merger, smaller
  masses can lead to a binary-merger product surviving hundreds of
  seconds and possibly more.

\item When considering magnetic fields of realistic strengths endowing
  the stars prior to the merger, the correction imprinted by them on the
  gravitational-wave signal during the inspiral are too small to be
  detected from advanced gravitational-wave detectors. Electromagnetic
  signals could be produced before the merger, but these are probably too
  weak to be detected from cosmological distances.

\item Magnetic fields are expected to be amplified both at the merger
  (via Kelvin-Helmholtz instability), after it and before the collapse
  and after the formation of a black-hole--torus system (in all cases via
  a magnetorotational instability or a dynamo action converting small-scale
  fields into large-scale ones). The final and effective amplification
  of the resulting magnetic fields is still uncertain, although it should
  be of at least two-three orders of magnitude.

\item The interaction of amplified magnetic fields and accretion in the
  black-hole--torus system leads to the formation of a magnetically
  confined plasma along the polar directions of the black hole. Under
  suitable conditions, the plasma in this funnel may be launched to
  \lrn{ultrarelativistic} speeds (still unobserved in simulations).

\item Matter is expected to be ejected both at the merger and
  subsequently as a result of a combination of processes: tidal and
  dynamical mass ejection, magnetically driven winds, neutrino-driven
  winds, shock-heating winds. Overall, the matter ejected from binaries
  in quasi-circular orbits amounts to $M_{\rm ejected} \sim 0.001 -
  0.01\,M_{\odot}$, while binaries in eccentric orbits can yield up to
  one order of magnitude more.

\item The ejected and unbound matter is expected to undergo nuclear
  transformations that are mediated by the emission and absorption of
  neutrinos. Rapid neutron-capture processes ($r$-processes) will then
  lead to nucleosynthetic yields that are insensitive to input physics or
  merger type in the regions of the second and third $r$-process peaks,
  matching the Solar abundances surprisingly well. However, first-peak
  elements are difficult to explain without invoking contributions from
  either neutrino and viscously-driven winds operating on longer
  timescales after the merger, or from core-collapse supernovae.

\item The radioactive decay of the ejected matter or its interaction with
  the interstellar medium are likely to yield afterglows in the infrared
  or radio bands that are expected to follow the merger after timescales
  that go from several days to years.

\end{itemize}

Note that many of the aspects listed above are robust but have been
addressed mostly at a rather qualitative level, with precisions that
range from ``a-factor-of-a-few'' up to ``order-of-magnitude''
estimates. Furthermore, these results can be seen as the low-hanging
fruits of a tree that still has a number of results to offer, although
these will require an equal, if not larger, investments of effort,
microphysical and numerical developments, and, of course, of computer
time. Among the most pressing and exciting open issues we certainly list
the following ones:

\begin{itemize}

\item Consistent initial data for magnetised and arbitrarily spinning
  neutron-star binaries.
  
\item Semi-analytical and faithful description of the complete
  gravitational-wave signal, from the inspiral to the formation of a
  black-hole or stable neutron star.

\item Robust and accurate estimate of the critical mass to prompt
  collapse and of the survival time of the binary-merger product.

\item Robust and accurate estimate of the processes mediating the
  accretion of the torus formed around the black hole, hence obtaining
  reliable measurements of the timescale for accretion.

\item Assessment of the role that turbulence and instabilities play in
  the amplification of the progenitor magnetic field and determination of
  the final strengths to be expected.

\item Robust and accurate sub-grid modelling for a realistic simulation
  of the magnetic-field dynamics at the smallest scales.

\item Assessment of a possible dynamo action occurring in a long-lived
  binary-merger product and leading to the generation of an ordered and
  large-scale magnetic field.

\item Determination of the microphysical processes leading to the
  formation and launching of an ultra-relativistic jet.

\item Determination of the acceleration sites of charged particles to
  ultrarelativistic energies and calculation of the energy distribution
  functions. 

\item Determination of the role played by neutrino losses in launching
  the jet and in modifying the chemical composition of the ejected
  material.

\item Determination of the relative importance of dynamical tidal
  torques, magnetic unbalance, neutrino emission, or shock heating, for
  the ejection of matter from the system.

\item Robust and accurate determination of the physical and chemical
  properties of material ejected from the whole merger process.

\item Quantitative and accurate predictions of the electromagnetic signal
  produced by the merger, either directly or indirectly as afterglows.

\end{itemize}

In conclusion, if the first direct detection of the gravitational-wave
signals from binary systems of black holes has officially given birth to
the era of gravitational-wave astronomy and has, once again, emphasised
general relativity as the best theory of gravitation known, the huge
advances that are expected to come in the next few years on the physics
and astrophysics of BNSs will help lift many of the veils that still
cover Einstein's richest laboratory.

\section*{Acknowledgements}

\noindent Writing this review has been an effort we have diluted over a
couple of years, during which it has been continuously updated as new
results were produced and additional progress was made. We are indebted
to the numerous friends and colleagues that have provided useful input
and have helped us obtaining a review that is as complete and detailed as
we could possibly have written. Special thanks go to E. Barausse,
L. Bovard, R. De Pietri, N. Stergioulas, K. Taniguchi, A. Tsokaros, and
S. Wanajo and for useful input on this review and to D. Alic,
K. Dionysopoulou, T. Font, F. Guercilena, M. Hanauske, B. Mundim, and
K. Takami for uncountable discussions.

\noindent Partial support has come from ``NewCompStar'', COST Action
MP1304, from the LOEWE-Program in HIC for FAIR, the European Union's
Horizon 2020 Research and Innovation Programme under grant agreement
No. 671698 (call FETHPC-1-2014, project ExaHyPE), from the ERC Synergy
Grant ``BlackHoleCam - Imaging the Event Horizon of Black Holes'' (Grant
610058), and from JSPS Grant-in-Aid for Scientific Research(C)
No. 26400274.

\newpage
\bibliographystyle{apsrev4-1}
\bibliography{aeireferences}

\begin{thebibliography}{551}%
\makeatletter
\providecommand \@ifxundefined [1]{%
 \@ifx{#1\undefined}
}%
\providecommand \@ifnum [1]{%
 \ifnum #1\expandafter \@firstoftwo
 \else \expandafter \@secondoftwo
 \fi
}%
\providecommand \@ifx [1]{%
 \ifx #1\expandafter \@firstoftwo
 \else \expandafter \@secondoftwo
 \fi
}%
\providecommand \natexlab [1]{#1}%
\providecommand \enquote  [1]{``#1''}%
\providecommand \bibnamefont  [1]{#1}%
\providecommand \bibfnamefont [1]{#1}%
\providecommand \citenamefont [1]{#1}%
\providecommand \href@noop [0]{\@secondoftwo}%
\providecommand \href [0]{\begingroup \@sanitize@url \@href}%
\providecommand \@href[1]{\@@startlink{#1}\@@href}%
\providecommand \@@href[1]{\endgroup#1\@@endlink}%
\providecommand \@sanitize@url [0]{\catcode `\\12\catcode `\$12\catcode
  `\&12\catcode `\#12\catcode `\^12\catcode `\_12\catcode `\%12\relax}%
\providecommand \@@startlink[1]{}%
\providecommand \@@endlink[0]{}%
\providecommand \url  [0]{\begingroup\@sanitize@url \@url }%
\providecommand \@url [1]{\endgroup\@href {#1}{\urlprefix }}%
\providecommand \urlprefix  [0]{URL }%
\providecommand \Eprint [0]{\href }%
\providecommand \doibase [0]{http://dx.doi.org/}%
\providecommand \selectlanguage [0]{\@gobble}%
\providecommand \bibinfo  [0]{\@secondoftwo}%
\providecommand \bibfield  [0]{\@secondoftwo}%
\providecommand \translation [1]{[#1]}%
\providecommand \BibitemOpen [0]{}%
\providecommand \bibitemStop [0]{}%
\providecommand \bibitemNoStop [0]{.\EOS\space}%
\providecommand \EOS [0]{\spacefactor3000\relax}%
\providecommand \BibitemShut  [1]{\csname bibitem#1\endcsname}%
\let\auto@bib@innerbib\@empty
\bibitem [{\citenamefont {Kramer}\ \emph {et~al.}(2004)\citenamefont {Kramer},
  \citenamefont {Lyne}, \citenamefont {Burgay}, \citenamefont {Possenti},
  \citenamefont {Manchester}, \citenamefont {Camilo}, \citenamefont
  {McLaughlin}, \citenamefont {Lorimer}, \citenamefont {D'Amico}, \citenamefont
  {Joshi}, \citenamefont {Reynolds},\ and\ \citenamefont {Freire}}]{Kramer04}%
  \BibitemOpen
  \bibfield  {author} {\bibinfo {author} {\bibfnamefont {M.}~\bibnamefont
  {Kramer}}, \bibinfo {author} {\bibfnamefont {A.}~\bibnamefont {Lyne}},
  \bibinfo {author} {\bibfnamefont {M.}~\bibnamefont {Burgay}}, \bibinfo
  {author} {\bibfnamefont {A.}~\bibnamefont {Possenti}}, \bibinfo {author}
  {\bibfnamefont {R.}~\bibnamefont {Manchester}}, \bibinfo {author}
  {\bibfnamefont {F.}~\bibnamefont {Camilo}}, \bibinfo {author} {\bibfnamefont
  {M.}~\bibnamefont {McLaughlin}}, \bibinfo {author} {\bibfnamefont
  {D.}~\bibnamefont {Lorimer}}, \bibinfo {author} {\bibfnamefont
  {N.}~\bibnamefont {D'Amico}}, \bibinfo {author} {\bibfnamefont
  {B.}~\bibnamefont {Joshi}}, \bibinfo {author} {\bibfnamefont
  {J.}~\bibnamefont {Reynolds}}, \ and\ \bibinfo {author} {\bibfnamefont
  {P.}~\bibnamefont {Freire}},\ }in\ \href@noop {} {\emph {\bibinfo {booktitle}
  {Binary Pulsars}}},\ \bibinfo {editor} {edited by\ \bibinfo {editor}
  {\bibnamefont {Rasio}}\ and\ \bibinfo {editor} {\bibnamefont {Stairs}}}\
  (\bibinfo  {publisher} {PSAP},\ \bibinfo {address} {Chicago},\ \bibinfo
  {year} {2004})\BibitemShut {NoStop}%
\bibitem [{\citenamefont {{Kiziltan}}\ \emph {et~al.}(2013)\citenamefont
  {{Kiziltan}}, \citenamefont {{Kottas}}, \citenamefont {{De Yoreo}},\ and\
  \citenamefont {{Thorsett}}}]{Kiziltan:2010a}%
  \BibitemOpen
  \bibfield  {author} {\bibinfo {author} {\bibfnamefont {B.}~\bibnamefont
  {{Kiziltan}}}, \bibinfo {author} {\bibfnamefont {A.}~\bibnamefont
  {{Kottas}}}, \bibinfo {author} {\bibfnamefont {M.}~\bibnamefont {{De
  Yoreo}}}, \ and\ \bibinfo {author} {\bibfnamefont {S.~E.}\ \bibnamefont
  {{Thorsett}}},\ }\href {\doibase 10.1088/0004-637X/778/1/66} {\bibfield
  {journal} {\bibinfo  {journal} {Astrophys. J.}\ }\textbf {\bibinfo {volume}
  {778}},\ \bibinfo {eid} {66} (\bibinfo {year} {2013})},\ \Eprint
  {http://arxiv.org/abs/1011.4291} {arXiv:1011.4291 [astro-ph.GA]} \BibitemShut
  {NoStop}%
\bibitem [{\citenamefont {{Ivanova}}\ \emph {et~al.}(2013)\citenamefont
  {{Ivanova}}, \citenamefont {{Justham}}, \citenamefont {{Chen}}, \citenamefont
  {{De Marco}}, \citenamefont {{Fryer}}, \citenamefont {{Gaburov}},
  \citenamefont {{Ge}}, \citenamefont {{Glebbeek}}, \citenamefont {{Han}},
  \citenamefont {{Li}}, \citenamefont {{Lu}}, \citenamefont {{Marsh}},
  \citenamefont {{Podsiadlowski}}, \citenamefont {{Potter}}, \citenamefont
  {{Soker}}, \citenamefont {{Taam}}, \citenamefont {{Tauris}}, \citenamefont
  {{van den Heuvel}},\ and\ \citenamefont {{Webbink}}}]{Ivanova2013}%
  \BibitemOpen
  \bibfield  {author} {\bibinfo {author} {\bibfnamefont {N.}~\bibnamefont
  {{Ivanova}}}, \bibinfo {author} {\bibfnamefont {S.}~\bibnamefont
  {{Justham}}}, \bibinfo {author} {\bibfnamefont {X.}~\bibnamefont {{Chen}}},
  \bibinfo {author} {\bibfnamefont {O.}~\bibnamefont {{De Marco}}}, \bibinfo
  {author} {\bibfnamefont {C.~L.}\ \bibnamefont {{Fryer}}}, \bibinfo {author}
  {\bibfnamefont {E.}~\bibnamefont {{Gaburov}}}, \bibinfo {author}
  {\bibfnamefont {H.}~\bibnamefont {{Ge}}}, \bibinfo {author} {\bibfnamefont
  {E.}~\bibnamefont {{Glebbeek}}}, \bibinfo {author} {\bibfnamefont
  {Z.}~\bibnamefont {{Han}}}, \bibinfo {author} {\bibfnamefont {X.-D.}\
  \bibnamefont {{Li}}}, \bibinfo {author} {\bibfnamefont {G.}~\bibnamefont
  {{Lu}}}, \bibinfo {author} {\bibfnamefont {T.}~\bibnamefont {{Marsh}}},
  \bibinfo {author} {\bibfnamefont {P.}~\bibnamefont {{Podsiadlowski}}},
  \bibinfo {author} {\bibfnamefont {A.}~\bibnamefont {{Potter}}}, \bibinfo
  {author} {\bibfnamefont {N.}~\bibnamefont {{Soker}}}, \bibinfo {author}
  {\bibfnamefont {R.}~\bibnamefont {{Taam}}}, \bibinfo {author} {\bibfnamefont
  {T.~M.}\ \bibnamefont {{Tauris}}}, \bibinfo {author} {\bibfnamefont
  {E.~P.~J.}\ \bibnamefont {{van den Heuvel}}}, \ and\ \bibinfo {author}
  {\bibfnamefont {R.~F.}\ \bibnamefont {{Webbink}}},\ }\href {\doibase
  10.1007/s00159-013-0059-2} {\bibfield  {journal} {\bibinfo  {journal}
  {Astronomy and Astrophysics Reviews}\ }\textbf {\bibinfo {volume} {21}},\
  \bibinfo {eid} {59} (\bibinfo {year} {2013})},\ \Eprint
  {http://arxiv.org/abs/1209.4302} {arXiv:1209.4302 [astro-ph.HE]} \BibitemShut
  {NoStop}%
\bibitem [{\citenamefont {{{\"O}zel}}\ and\ \citenamefont
  {{Freire}}(2016)}]{Ozel2016}%
  \BibitemOpen
  \bibfield  {author} {\bibinfo {author} {\bibfnamefont {F.}~\bibnamefont
  {{{\"O}zel}}}\ and\ \bibinfo {author} {\bibfnamefont {P.}~\bibnamefont
  {{Freire}}},\ }\href {\doibase 10.1146/annurev-astro-081915-023322}
  {\bibfield  {journal} {\bibinfo  {journal} {Annual Review of Astronomy and
  Astrophysics}\ }\textbf {\bibinfo {volume} {54}},\ \bibinfo {pages} {401}
  (\bibinfo {year} {2016})},\ \Eprint {http://arxiv.org/abs/1603.02698}
  {arXiv:1603.02698 [astro-ph.HE]} \BibitemShut {NoStop}%
\bibitem [{\citenamefont {{O'Leary}}\ \emph {et~al.}(2009)\citenamefont
  {{O'Leary}}, \citenamefont {{Kocsis}},\ and\ \citenamefont
  {{Loeb}}}]{Oleary2009}%
  \BibitemOpen
  \bibfield  {author} {\bibinfo {author} {\bibfnamefont {R.~M.}\ \bibnamefont
  {{O'Leary}}}, \bibinfo {author} {\bibfnamefont {B.}~\bibnamefont {{Kocsis}}},
  \ and\ \bibinfo {author} {\bibfnamefont {A.}~\bibnamefont {{Loeb}}},\ }\href
  {\doibase 10.1111/j.1365-2966.2009.14653.x} {\bibfield  {journal} {\bibinfo
  {journal} {Mon. Not. R. Astron. Soc.}\ }\textbf {\bibinfo {volume} {395}},\
  \bibinfo {pages} {2127} (\bibinfo {year} {2009})},\ \Eprint
  {http://arxiv.org/abs/0807.2638} {arXiv:0807.2638} \BibitemShut {NoStop}%
\bibitem [{\citenamefont {{Lee}}\ \emph {et~al.}(2010)\citenamefont {{Lee}},
  \citenamefont {{Ramirez-Ruiz}},\ and\ \citenamefont {{van de
  Ven}}}]{Lee2010}%
  \BibitemOpen
  \bibfield  {author} {\bibinfo {author} {\bibfnamefont {W.~H.}\ \bibnamefont
  {{Lee}}}, \bibinfo {author} {\bibfnamefont {E.}~\bibnamefont
  {{Ramirez-Ruiz}}}, \ and\ \bibinfo {author} {\bibfnamefont {G.}~\bibnamefont
  {{van de Ven}}},\ }\href {\doibase 10.1088/0004-637X/720/1/953} {\bibfield
  {journal} {\bibinfo  {journal} {Astrophys. J.}\ }\textbf {\bibinfo {volume}
  {720}},\ \bibinfo {pages} {953} (\bibinfo {year} {2010})},\ \Eprint
  {http://arxiv.org/abs/0909.2884} {arXiv:0909.2884 [astro-ph.HE]} \BibitemShut
  {NoStop}%
\bibitem [{\citenamefont {{Thompson}}(2011)}]{Thompson2011}%
  \BibitemOpen
  \bibfield  {author} {\bibinfo {author} {\bibfnamefont {T.~A.}\ \bibnamefont
  {{Thompson}}},\ }\href {\doibase 10.1088/0004-637X/741/2/82} {\bibfield
  {journal} {\bibinfo  {journal} {Astrophys. J.}\ }\textbf {\bibinfo {volume}
  {741}},\ \bibinfo {eid} {82} (\bibinfo {year} {2011})},\ \Eprint
  {http://arxiv.org/abs/1011.4322} {arXiv:1011.4322 [astro-ph.HE]} \BibitemShut
  {NoStop}%
\bibitem [{\citenamefont {{Martinez}}\ \emph {et~al.}(2015)\citenamefont
  {{Martinez}}, \citenamefont {{Stovall}}, \citenamefont {{Freire}},
  \citenamefont {{Deneva}}, \citenamefont {{Jenet}}, \citenamefont
  {{McLaughlin}}, \citenamefont {{Bagchi}}, \citenamefont {{Bates}},\ and\
  \citenamefont {{Ridolfi}}}]{Martinez2015}%
  \BibitemOpen
  \bibfield  {author} {\bibinfo {author} {\bibfnamefont {J.~G.}\ \bibnamefont
  {{Martinez}}}, \bibinfo {author} {\bibfnamefont {K.}~\bibnamefont
  {{Stovall}}}, \bibinfo {author} {\bibfnamefont {P.~C.~C.}\ \bibnamefont
  {{Freire}}}, \bibinfo {author} {\bibfnamefont {J.~S.}\ \bibnamefont
  {{Deneva}}}, \bibinfo {author} {\bibfnamefont {F.~A.}\ \bibnamefont
  {{Jenet}}}, \bibinfo {author} {\bibfnamefont {M.~A.}\ \bibnamefont
  {{McLaughlin}}}, \bibinfo {author} {\bibfnamefont {M.}~\bibnamefont
  {{Bagchi}}}, \bibinfo {author} {\bibfnamefont {S.~D.}\ \bibnamefont
  {{Bates}}}, \ and\ \bibinfo {author} {\bibfnamefont {A.}~\bibnamefont
  {{Ridolfi}}},\ }\href {\doibase 10.1088/0004-637X/812/2/143} {\bibfield
  {journal} {\bibinfo  {journal} {\apj}\ }\textbf {\bibinfo {volume} {812}},\
  \bibinfo {eid} {143} (\bibinfo {year} {2015})},\ \Eprint
  {http://arxiv.org/abs/1509.08805} {arXiv:1509.08805 [astro-ph.HE]}
  \BibitemShut {NoStop}%
\bibitem [{\citenamefont {{Kramer}}\ \emph {et~al.}(2006)\citenamefont
  {{Kramer}}, \citenamefont {{Stairs}}, \citenamefont {{Manchester}},
  \citenamefont {{McLaughlin}}, \citenamefont {{Lyne}}, \citenamefont
  {{Ferdman}}, \citenamefont {{Burgay}}, \citenamefont {{Lorimer}},
  \citenamefont {{Possenti}}, \citenamefont {{D'Amico}}, \citenamefont
  {{Sarkissian}}, \citenamefont {{Hobbs}}, \citenamefont {{Reynolds}},
  \citenamefont {{Freire}},\ and\ \citenamefont {{Camilo}}}]{Kramer2006}%
  \BibitemOpen
  \bibfield  {author} {\bibinfo {author} {\bibfnamefont {M.}~\bibnamefont
  {{Kramer}}}, \bibinfo {author} {\bibfnamefont {I.~H.}\ \bibnamefont
  {{Stairs}}}, \bibinfo {author} {\bibfnamefont {R.~N.}\ \bibnamefont
  {{Manchester}}}, \bibinfo {author} {\bibfnamefont {M.~A.}\ \bibnamefont
  {{McLaughlin}}}, \bibinfo {author} {\bibfnamefont {A.~G.}\ \bibnamefont
  {{Lyne}}}, \bibinfo {author} {\bibfnamefont {R.~D.}\ \bibnamefont
  {{Ferdman}}}, \bibinfo {author} {\bibfnamefont {M.}~\bibnamefont {{Burgay}}},
  \bibinfo {author} {\bibfnamefont {D.~R.}\ \bibnamefont {{Lorimer}}}, \bibinfo
  {author} {\bibfnamefont {A.}~\bibnamefont {{Possenti}}}, \bibinfo {author}
  {\bibfnamefont {N.}~\bibnamefont {{D'Amico}}}, \bibinfo {author}
  {\bibfnamefont {J.~M.}\ \bibnamefont {{Sarkissian}}}, \bibinfo {author}
  {\bibfnamefont {G.~B.}\ \bibnamefont {{Hobbs}}}, \bibinfo {author}
  {\bibfnamefont {J.~E.}\ \bibnamefont {{Reynolds}}}, \bibinfo {author}
  {\bibfnamefont {P.~C.~C.}\ \bibnamefont {{Freire}}}, \ and\ \bibinfo {author}
  {\bibfnamefont {F.}~\bibnamefont {{Camilo}}},\ }\href {\doibase
  10.1126/science.1132305} {\bibfield  {journal} {\bibinfo  {journal}
  {Science}\ }\textbf {\bibinfo {volume} {314}},\ \bibinfo {pages} {97}
  (\bibinfo {year} {2006})},\ \Eprint {http://arxiv.org/abs/astro-ph/0609417}
  {astro-ph/0609417} \BibitemShut {NoStop}%
\bibitem [{\citenamefont {{Janssen}}\ \emph {et~al.}(2008)\citenamefont
  {{Janssen}}, \citenamefont {{Stappers}}, \citenamefont {{Kramer}},
  \citenamefont {{Nice}}, \citenamefont {{Jessner}}, \citenamefont
  {{Cognard}},\ and\ \citenamefont {{Purver}}}]{Janssen2008}%
  \BibitemOpen
  \bibfield  {author} {\bibinfo {author} {\bibfnamefont {G.~H.}\ \bibnamefont
  {{Janssen}}}, \bibinfo {author} {\bibfnamefont {B.~W.}\ \bibnamefont
  {{Stappers}}}, \bibinfo {author} {\bibfnamefont {M.}~\bibnamefont
  {{Kramer}}}, \bibinfo {author} {\bibfnamefont {D.~J.}\ \bibnamefont
  {{Nice}}}, \bibinfo {author} {\bibfnamefont {A.}~\bibnamefont {{Jessner}}},
  \bibinfo {author} {\bibfnamefont {I.}~\bibnamefont {{Cognard}}}, \ and\
  \bibinfo {author} {\bibfnamefont {M.~B.}\ \bibnamefont {{Purver}}},\ }\href
  {\doibase 10.1051/0004-6361:200810076} {\bibfield  {journal} {\bibinfo
  {journal} {Astron. Astrophys.}\ }\textbf {\bibinfo {volume} {490}},\ \bibinfo
  {pages} {753} (\bibinfo {year} {2008})},\ \Eprint
  {http://arxiv.org/abs/0808.2292} {arXiv:0808.2292} \BibitemShut {NoStop}%
\bibitem [{\citenamefont {{Fonseca}}\ \emph {et~al.}(2014)\citenamefont
  {{Fonseca}}, \citenamefont {{Stairs}},\ and\ \citenamefont
  {{Thorsett}}}]{Fonseca2014}%
  \BibitemOpen
  \bibfield  {author} {\bibinfo {author} {\bibfnamefont {E.}~\bibnamefont
  {{Fonseca}}}, \bibinfo {author} {\bibfnamefont {I.~H.}\ \bibnamefont
  {{Stairs}}}, \ and\ \bibinfo {author} {\bibfnamefont {S.~E.}\ \bibnamefont
  {{Thorsett}}},\ }\href {\doibase 10.1088/0004-637X/787/1/82} {\bibfield
  {journal} {\bibinfo  {journal} {\apj}\ }\textbf {\bibinfo {volume} {787}},\
  \bibinfo {eid} {82} (\bibinfo {year} {2014})},\ \Eprint
  {http://arxiv.org/abs/1402.4836} {arXiv:1402.4836 [astro-ph.HE]} \BibitemShut
  {NoStop}%
\bibitem [{\citenamefont {{Keith}}\ \emph {et~al.}(2009)\citenamefont
  {{Keith}}, \citenamefont {{Kramer}}, \citenamefont {{Lyne}}, \citenamefont
  {{Eatough}}, \citenamefont {{Stairs}}, \citenamefont {{Possenti}},
  \citenamefont {{Camilo}},\ and\ \citenamefont {{Manchester}}}]{Keith2009}%
  \BibitemOpen
  \bibfield  {author} {\bibinfo {author} {\bibfnamefont {M.~J.}\ \bibnamefont
  {{Keith}}}, \bibinfo {author} {\bibfnamefont {M.}~\bibnamefont {{Kramer}}},
  \bibinfo {author} {\bibfnamefont {A.~G.}\ \bibnamefont {{Lyne}}}, \bibinfo
  {author} {\bibfnamefont {R.~P.}\ \bibnamefont {{Eatough}}}, \bibinfo {author}
  {\bibfnamefont {I.~H.}\ \bibnamefont {{Stairs}}}, \bibinfo {author}
  {\bibfnamefont {A.}~\bibnamefont {{Possenti}}}, \bibinfo {author}
  {\bibfnamefont {F.}~\bibnamefont {{Camilo}}}, \ and\ \bibinfo {author}
  {\bibfnamefont {R.~N.}\ \bibnamefont {{Manchester}}},\ }\href {\doibase
  10.1111/j.1365-2966.2008.14234.x} {\bibfield  {journal} {\bibinfo  {journal}
  {Mon. Not. R. Astron. Soc.}\ }\textbf {\bibinfo {volume} {393}},\ \bibinfo
  {pages} {623} (\bibinfo {year} {2009})},\ \Eprint
  {http://arxiv.org/abs/0811.2027} {arXiv:0811.2027} \BibitemShut {NoStop}%
\bibitem [{\citenamefont {{Ferdman}}\ \emph {et~al.}(2014)\citenamefont
  {{Ferdman}}, \citenamefont {{Stairs}}, \citenamefont {{Kramer}},
  \citenamefont {{Janssen}}, \citenamefont {{Bassa}}, \citenamefont
  {{Stappers}}, \citenamefont {{Demorest}}, \citenamefont {{Cognard}},
  \citenamefont {{Desvignes}}, \citenamefont {{Theureau}}, \citenamefont
  {{Burgay}}, \citenamefont {{Lyne}}, \citenamefont {{Manchester}},\ and\
  \citenamefont {{Possenti}}}]{Ferdman2014}%
  \BibitemOpen
  \bibfield  {author} {\bibinfo {author} {\bibfnamefont {R.~D.}\ \bibnamefont
  {{Ferdman}}}, \bibinfo {author} {\bibfnamefont {I.~H.}\ \bibnamefont
  {{Stairs}}}, \bibinfo {author} {\bibfnamefont {M.}~\bibnamefont {{Kramer}}},
  \bibinfo {author} {\bibfnamefont {G.~H.}\ \bibnamefont {{Janssen}}}, \bibinfo
  {author} {\bibfnamefont {C.~G.}\ \bibnamefont {{Bassa}}}, \bibinfo {author}
  {\bibfnamefont {B.~W.}\ \bibnamefont {{Stappers}}}, \bibinfo {author}
  {\bibfnamefont {P.~B.}\ \bibnamefont {{Demorest}}}, \bibinfo {author}
  {\bibfnamefont {I.}~\bibnamefont {{Cognard}}}, \bibinfo {author}
  {\bibfnamefont {G.}~\bibnamefont {{Desvignes}}}, \bibinfo {author}
  {\bibfnamefont {G.}~\bibnamefont {{Theureau}}}, \bibinfo {author}
  {\bibfnamefont {M.}~\bibnamefont {{Burgay}}}, \bibinfo {author}
  {\bibfnamefont {A.~G.}\ \bibnamefont {{Lyne}}}, \bibinfo {author}
  {\bibfnamefont {R.~N.}\ \bibnamefont {{Manchester}}}, \ and\ \bibinfo
  {author} {\bibfnamefont {A.}~\bibnamefont {{Possenti}}},\ }\href {\doibase
  10.1093/mnras/stu1223} {\bibfield  {journal} {\bibinfo  {journal} {Mon. Not.
  R. Astron. Soc.}\ }\textbf {\bibinfo {volume} {443}},\ \bibinfo {pages}
  {2183} (\bibinfo {year} {2014})},\ \Eprint {http://arxiv.org/abs/1406.5507}
  {arXiv:1406.5507 [astro-ph.SR]} \BibitemShut {NoStop}%
\bibitem [{\citenamefont {{Lynch}}\ \emph {et~al.}(2012)\citenamefont
  {{Lynch}}, \citenamefont {{Freire}}, \citenamefont {{Ransom}},\ and\
  \citenamefont {{Jacoby}}}]{Lynch2012}%
  \BibitemOpen
  \bibfield  {author} {\bibinfo {author} {\bibfnamefont {R.~S.}\ \bibnamefont
  {{Lynch}}}, \bibinfo {author} {\bibfnamefont {P.~C.~C.}\ \bibnamefont
  {{Freire}}}, \bibinfo {author} {\bibfnamefont {S.~M.}\ \bibnamefont
  {{Ransom}}}, \ and\ \bibinfo {author} {\bibfnamefont {B.~A.}\ \bibnamefont
  {{Jacoby}}},\ }\href {\doibase 10.1088/0004-637X/745/2/109} {\bibfield
  {journal} {\bibinfo  {journal} {\apj}\ }\textbf {\bibinfo {volume} {745}},\
  \bibinfo {eid} {109} (\bibinfo {year} {2012})},\ \Eprint
  {http://arxiv.org/abs/1112.2612} {arXiv:1112.2612 [astro-ph.HE]} \BibitemShut
  {NoStop}%
\bibitem [{\citenamefont {{Corongiu}}\ \emph {et~al.}(2007)\citenamefont
  {{Corongiu}}, \citenamefont {{Kramer}}, \citenamefont {{Stappers}},
  \citenamefont {{Lyne}}, \citenamefont {{Jessner}}, \citenamefont
  {{Possenti}}, \citenamefont {{D'Amico}},\ and\ \citenamefont
  {{L{\"o}hmer}}}]{Corongiu2007}%
  \BibitemOpen
  \bibfield  {author} {\bibinfo {author} {\bibfnamefont {A.}~\bibnamefont
  {{Corongiu}}}, \bibinfo {author} {\bibfnamefont {M.}~\bibnamefont
  {{Kramer}}}, \bibinfo {author} {\bibfnamefont {B.~W.}\ \bibnamefont
  {{Stappers}}}, \bibinfo {author} {\bibfnamefont {A.~G.}\ \bibnamefont
  {{Lyne}}}, \bibinfo {author} {\bibfnamefont {A.}~\bibnamefont {{Jessner}}},
  \bibinfo {author} {\bibfnamefont {A.}~\bibnamefont {{Possenti}}}, \bibinfo
  {author} {\bibfnamefont {N.}~\bibnamefont {{D'Amico}}}, \ and\ \bibinfo
  {author} {\bibfnamefont {O.}~\bibnamefont {{L{\"o}hmer}}},\ }\href {\doibase
  10.1051/0004-6361:20054385} {\bibfield  {journal} {\bibinfo  {journal}
  {Astron. Astrophys.}\ }\textbf {\bibinfo {volume} {462}},\ \bibinfo {pages}
  {703} (\bibinfo {year} {2007})},\ \Eprint
  {http://arxiv.org/abs/astro-ph/0611436} {astro-ph/0611436} \BibitemShut
  {NoStop}%
\bibitem [{\citenamefont {{Champion}}\ \emph {et~al.}(2004)\citenamefont
  {{Champion}}, \citenamefont {{Lorimer}}, \citenamefont {{McLaughlin}},
  \citenamefont {{Cordes}}, \citenamefont {{Arzoumanian}}, \citenamefont
  {{Weisberg}},\ and\ \citenamefont {{Taylor}}}]{Champion2004}%
  \BibitemOpen
  \bibfield  {author} {\bibinfo {author} {\bibfnamefont {D.~J.}\ \bibnamefont
  {{Champion}}}, \bibinfo {author} {\bibfnamefont {D.~R.}\ \bibnamefont
  {{Lorimer}}}, \bibinfo {author} {\bibfnamefont {M.~A.}\ \bibnamefont
  {{McLaughlin}}}, \bibinfo {author} {\bibfnamefont {J.~M.}\ \bibnamefont
  {{Cordes}}}, \bibinfo {author} {\bibfnamefont {Z.}~\bibnamefont
  {{Arzoumanian}}}, \bibinfo {author} {\bibfnamefont {J.~M.}\ \bibnamefont
  {{Weisberg}}}, \ and\ \bibinfo {author} {\bibfnamefont {J.~H.}\ \bibnamefont
  {{Taylor}}},\ }\href {\doibase 10.1111/j.1365-2966.2004.07862.x} {\bibfield
  {journal} {\bibinfo  {journal} {Mon. Not. R. Astron. Soc.}\ }\textbf
  {\bibinfo {volume} {350}},\ \bibinfo {pages} {L61} (\bibinfo {year}
  {2004})},\ \Eprint {http://arxiv.org/abs/astro-ph/0403553} {astro-ph/0403553}
  \BibitemShut {NoStop}%
\bibitem [{\citenamefont {{van Leeuwen}}\ \emph {et~al.}(2015)\citenamefont
  {{van Leeuwen}}, \citenamefont {{Kasian}}, \citenamefont {{Stairs}},
  \citenamefont {{Lorimer}}, \citenamefont {{Camilo}}, \citenamefont
  {{Chatterjee}}, \citenamefont {{Cognard}}, \citenamefont {{Desvignes}},
  \citenamefont {{Freire}}, \citenamefont {{Janssen}}, \citenamefont
  {{Kramer}}, \citenamefont {{Lyne}}, \citenamefont {{Nice}}, \citenamefont
  {{Ransom}}, \citenamefont {{Stappers}},\ and\ \citenamefont
  {{Weisberg}}}]{vanLeeuwen2015}%
  \BibitemOpen
  \bibfield  {author} {\bibinfo {author} {\bibfnamefont {J.}~\bibnamefont {{van
  Leeuwen}}}, \bibinfo {author} {\bibfnamefont {L.}~\bibnamefont {{Kasian}}},
  \bibinfo {author} {\bibfnamefont {I.~H.}\ \bibnamefont {{Stairs}}}, \bibinfo
  {author} {\bibfnamefont {D.~R.}\ \bibnamefont {{Lorimer}}}, \bibinfo {author}
  {\bibfnamefont {F.}~\bibnamefont {{Camilo}}}, \bibinfo {author}
  {\bibfnamefont {S.}~\bibnamefont {{Chatterjee}}}, \bibinfo {author}
  {\bibfnamefont {I.}~\bibnamefont {{Cognard}}}, \bibinfo {author}
  {\bibfnamefont {G.}~\bibnamefont {{Desvignes}}}, \bibinfo {author}
  {\bibfnamefont {P.~C.~C.}\ \bibnamefont {{Freire}}}, \bibinfo {author}
  {\bibfnamefont {G.~H.}\ \bibnamefont {{Janssen}}}, \bibinfo {author}
  {\bibfnamefont {M.}~\bibnamefont {{Kramer}}}, \bibinfo {author}
  {\bibfnamefont {A.~G.}\ \bibnamefont {{Lyne}}}, \bibinfo {author}
  {\bibfnamefont {D.~J.}\ \bibnamefont {{Nice}}}, \bibinfo {author}
  {\bibfnamefont {S.~M.}\ \bibnamefont {{Ransom}}}, \bibinfo {author}
  {\bibfnamefont {B.~W.}\ \bibnamefont {{Stappers}}}, \ and\ \bibinfo {author}
  {\bibfnamefont {J.~M.}\ \bibnamefont {{Weisberg}}},\ }\href {\doibase
  10.1088/0004-637X/798/2/118} {\bibfield  {journal} {\bibinfo  {journal}
  {\apj}\ }\textbf {\bibinfo {volume} {798}},\ \bibinfo {eid} {118} (\bibinfo
  {year} {2015})},\ \Eprint {http://arxiv.org/abs/1411.1518} {arXiv:1411.1518
  [astro-ph.SR]} \BibitemShut {NoStop}%
\bibitem [{\citenamefont {{Lazarus}}\ \emph {et~al.}(2016)\citenamefont
  {{Lazarus}}, \citenamefont {{Freire}}, \citenamefont {{Allen}}, \citenamefont
  {{Aulbert}}, \citenamefont {{Bock}}, \citenamefont {{Bogdanov}},
  \citenamefont {{Brazier}}, \citenamefont {{Camilo}}, \citenamefont
  {{Cardoso}}, \citenamefont {{Chatterjee}}, \citenamefont {{Cordes}},
  \citenamefont {{Crawford}}, \citenamefont {{Deneva}}, \citenamefont
  {{Eggenstein}}, \citenamefont {{Fehrmann}}, \citenamefont {{Ferdman}},
  \citenamefont {{Hessels}}, \citenamefont {{Jenet}}, \citenamefont
  {{Karako-Argaman}}, \citenamefont {{Kaspi}}, \citenamefont {{Knispel}},
  \citenamefont {{Lynch}}, \citenamefont {{van Leeuwen}}, \citenamefont
  {{Machenschalk}}, \citenamefont {{Madsen}}, \citenamefont {{McLaughlin}},
  \citenamefont {{Patel}}, \citenamefont {{Ransom}}, \citenamefont {{Scholz}},
  \citenamefont {{Seymour}}, \citenamefont {{Siemens}}, \citenamefont
  {{Spitler}}, \citenamefont {{Stairs}}, \citenamefont {{Stovall}},
  \citenamefont {{Swiggum}}, \citenamefont {{Venkataraman}},\ and\
  \citenamefont {{Zhu}}}]{Lazarus2016}%
  \BibitemOpen
  \bibfield  {author} {\bibinfo {author} {\bibfnamefont {P.}~\bibnamefont
  {{Lazarus}}}, \bibinfo {author} {\bibfnamefont {P.~C.~C.}\ \bibnamefont
  {{Freire}}}, \bibinfo {author} {\bibfnamefont {B.}~\bibnamefont {{Allen}}},
  \bibinfo {author} {\bibfnamefont {C.}~\bibnamefont {{Aulbert}}}, \bibinfo
  {author} {\bibfnamefont {O.}~\bibnamefont {{Bock}}}, \bibinfo {author}
  {\bibfnamefont {S.}~\bibnamefont {{Bogdanov}}}, \bibinfo {author}
  {\bibfnamefont {A.}~\bibnamefont {{Brazier}}}, \bibinfo {author}
  {\bibfnamefont {F.}~\bibnamefont {{Camilo}}}, \bibinfo {author}
  {\bibfnamefont {F.}~\bibnamefont {{Cardoso}}}, \bibinfo {author}
  {\bibfnamefont {S.}~\bibnamefont {{Chatterjee}}}, \bibinfo {author}
  {\bibfnamefont {J.~M.}\ \bibnamefont {{Cordes}}}, \bibinfo {author}
  {\bibfnamefont {F.}~\bibnamefont {{Crawford}}}, \bibinfo {author}
  {\bibfnamefont {J.~S.}\ \bibnamefont {{Deneva}}}, \bibinfo {author}
  {\bibfnamefont {H.-B.}\ \bibnamefont {{Eggenstein}}}, \bibinfo {author}
  {\bibfnamefont {H.}~\bibnamefont {{Fehrmann}}}, \bibinfo {author}
  {\bibfnamefont {R.}~\bibnamefont {{Ferdman}}}, \bibinfo {author}
  {\bibfnamefont {J.~W.~T.}\ \bibnamefont {{Hessels}}}, \bibinfo {author}
  {\bibfnamefont {F.~A.}\ \bibnamefont {{Jenet}}}, \bibinfo {author}
  {\bibfnamefont {C.}~\bibnamefont {{Karako-Argaman}}}, \bibinfo {author}
  {\bibfnamefont {V.~M.}\ \bibnamefont {{Kaspi}}}, \bibinfo {author}
  {\bibfnamefont {B.}~\bibnamefont {{Knispel}}}, \bibinfo {author}
  {\bibfnamefont {R.}~\bibnamefont {{Lynch}}}, \bibinfo {author} {\bibfnamefont
  {J.}~\bibnamefont {{van Leeuwen}}}, \bibinfo {author} {\bibfnamefont
  {B.}~\bibnamefont {{Machenschalk}}}, \bibinfo {author} {\bibfnamefont
  {E.}~\bibnamefont {{Madsen}}}, \bibinfo {author} {\bibfnamefont {M.~A.}\
  \bibnamefont {{McLaughlin}}}, \bibinfo {author} {\bibfnamefont
  {C.}~\bibnamefont {{Patel}}}, \bibinfo {author} {\bibfnamefont {S.~M.}\
  \bibnamefont {{Ransom}}}, \bibinfo {author} {\bibfnamefont {P.}~\bibnamefont
  {{Scholz}}}, \bibinfo {author} {\bibfnamefont {A.}~\bibnamefont {{Seymour}}},
  \bibinfo {author} {\bibfnamefont {X.}~\bibnamefont {{Siemens}}}, \bibinfo
  {author} {\bibfnamefont {L.~G.}\ \bibnamefont {{Spitler}}}, \bibinfo {author}
  {\bibfnamefont {I.~H.}\ \bibnamefont {{Stairs}}}, \bibinfo {author}
  {\bibfnamefont {K.}~\bibnamefont {{Stovall}}}, \bibinfo {author}
  {\bibfnamefont {J.}~\bibnamefont {{Swiggum}}}, \bibinfo {author}
  {\bibfnamefont {A.}~\bibnamefont {{Venkataraman}}}, \ and\ \bibinfo {author}
  {\bibfnamefont {W.~W.}\ \bibnamefont {{Zhu}}},\ }\href {\doibase
  10.3847/0004-637X/831/2/150} {\bibfield  {journal} {\bibinfo  {journal}
  {\apj}\ }\textbf {\bibinfo {volume} {831}},\ \bibinfo {eid} {150} (\bibinfo
  {year} {2016})},\ \Eprint {http://arxiv.org/abs/1608.08211} {arXiv:1608.08211
  [astro-ph.HE]} \BibitemShut {NoStop}%
\bibitem [{\citenamefont {{Weisberg}}\ and\ \citenamefont
  {{Huang}}(2016)}]{Weisberg2016}%
  \BibitemOpen
  \bibfield  {author} {\bibinfo {author} {\bibfnamefont {J.~M.}\ \bibnamefont
  {{Weisberg}}}\ and\ \bibinfo {author} {\bibfnamefont {Y.}~\bibnamefont
  {{Huang}}},\ }\href {\doibase 10.3847/0004-637X/829/1/55} {\bibfield
  {journal} {\bibinfo  {journal} {\apj}\ }\textbf {\bibinfo {volume} {829}},\
  \bibinfo {eid} {55} (\bibinfo {year} {2016})},\ \Eprint
  {http://arxiv.org/abs/1606.02744} {arXiv:1606.02744 [astro-ph.HE]}
  \BibitemShut {NoStop}%
\bibitem [{\citenamefont {{Swiggum}}\ \emph {et~al.}(2015)\citenamefont
  {{Swiggum}}, \citenamefont {{Rosen}}, \citenamefont {{McLaughlin}},
  \citenamefont {{Lorimer}}, \citenamefont {{Heatherly}}, \citenamefont
  {{Lynch}}, \citenamefont {{Scoles}}, \citenamefont {{Hockett}}, \citenamefont
  {{Filik}}, \citenamefont {{Marlowe}}, \citenamefont {{Barlow}}, \citenamefont
  {{Weaver}}, \citenamefont {{Hilzendeger}}, \citenamefont {{Ernst}},
  \citenamefont {{Crowley}}, \citenamefont {{Stone}}, \citenamefont {{Miller}},
  \citenamefont {{Nunez}}, \citenamefont {{Trevino}}, \citenamefont
  {{Doehler}}, \citenamefont {{Cramer}}, \citenamefont {{Yencsik}},
  \citenamefont {{Thorley}}, \citenamefont {{Andrews}}, \citenamefont {{Laws}},
  \citenamefont {{Wenger}}, \citenamefont {{Teter}}, \citenamefont {{Snyder}},
  \citenamefont {{Dittmann}}, \citenamefont {{Gray}}, \citenamefont {{Carter}},
  \citenamefont {{McGough}}, \citenamefont {{Dydiw}}, \citenamefont {{Pruett}},
  \citenamefont {{Fink}},\ and\ \citenamefont {{Vanderhout}}}]{Swiggum2015}%
  \BibitemOpen
  \bibfield  {author} {\bibinfo {author} {\bibfnamefont {J.~K.}\ \bibnamefont
  {{Swiggum}}}, \bibinfo {author} {\bibfnamefont {R.}~\bibnamefont {{Rosen}}},
  \bibinfo {author} {\bibfnamefont {M.~A.}\ \bibnamefont {{McLaughlin}}},
  \bibinfo {author} {\bibfnamefont {D.~R.}\ \bibnamefont {{Lorimer}}}, \bibinfo
  {author} {\bibfnamefont {S.}~\bibnamefont {{Heatherly}}}, \bibinfo {author}
  {\bibfnamefont {R.}~\bibnamefont {{Lynch}}}, \bibinfo {author} {\bibfnamefont
  {S.}~\bibnamefont {{Scoles}}}, \bibinfo {author} {\bibfnamefont
  {T.}~\bibnamefont {{Hockett}}}, \bibinfo {author} {\bibfnamefont
  {E.}~\bibnamefont {{Filik}}}, \bibinfo {author} {\bibfnamefont {J.~A.}\
  \bibnamefont {{Marlowe}}}, \bibinfo {author} {\bibfnamefont {B.~N.}\
  \bibnamefont {{Barlow}}}, \bibinfo {author} {\bibfnamefont {M.}~\bibnamefont
  {{Weaver}}}, \bibinfo {author} {\bibfnamefont {M.}~\bibnamefont
  {{Hilzendeger}}}, \bibinfo {author} {\bibfnamefont {S.}~\bibnamefont
  {{Ernst}}}, \bibinfo {author} {\bibfnamefont {R.}~\bibnamefont {{Crowley}}},
  \bibinfo {author} {\bibfnamefont {E.}~\bibnamefont {{Stone}}}, \bibinfo
  {author} {\bibfnamefont {B.}~\bibnamefont {{Miller}}}, \bibinfo {author}
  {\bibfnamefont {R.}~\bibnamefont {{Nunez}}}, \bibinfo {author} {\bibfnamefont
  {G.}~\bibnamefont {{Trevino}}}, \bibinfo {author} {\bibfnamefont
  {M.}~\bibnamefont {{Doehler}}}, \bibinfo {author} {\bibfnamefont
  {A.}~\bibnamefont {{Cramer}}}, \bibinfo {author} {\bibfnamefont
  {D.}~\bibnamefont {{Yencsik}}}, \bibinfo {author} {\bibfnamefont
  {J.}~\bibnamefont {{Thorley}}}, \bibinfo {author} {\bibfnamefont
  {R.}~\bibnamefont {{Andrews}}}, \bibinfo {author} {\bibfnamefont
  {A.}~\bibnamefont {{Laws}}}, \bibinfo {author} {\bibfnamefont
  {K.}~\bibnamefont {{Wenger}}}, \bibinfo {author} {\bibfnamefont
  {L.}~\bibnamefont {{Teter}}}, \bibinfo {author} {\bibfnamefont
  {T.}~\bibnamefont {{Snyder}}}, \bibinfo {author} {\bibfnamefont
  {A.}~\bibnamefont {{Dittmann}}}, \bibinfo {author} {\bibfnamefont
  {S.}~\bibnamefont {{Gray}}}, \bibinfo {author} {\bibfnamefont
  {M.}~\bibnamefont {{Carter}}}, \bibinfo {author} {\bibfnamefont
  {C.}~\bibnamefont {{McGough}}}, \bibinfo {author} {\bibfnamefont
  {S.}~\bibnamefont {{Dydiw}}}, \bibinfo {author} {\bibfnamefont
  {C.}~\bibnamefont {{Pruett}}}, \bibinfo {author} {\bibfnamefont
  {J.}~\bibnamefont {{Fink}}}, \ and\ \bibinfo {author} {\bibfnamefont
  {A.}~\bibnamefont {{Vanderhout}}},\ }\href {\doibase
  10.1088/0004-637X/805/2/156} {\bibfield  {journal} {\bibinfo  {journal}
  {\apj}\ }\textbf {\bibinfo {volume} {805}},\ \bibinfo {eid} {156} (\bibinfo
  {year} {2015})},\ \Eprint {http://arxiv.org/abs/1503.06276} {arXiv:1503.06276
  [astro-ph.HE]} \BibitemShut {NoStop}%
\bibitem [{\citenamefont {{Jacoby}}\ \emph {et~al.}(2006)\citenamefont
  {{Jacoby}}, \citenamefont {{Cameron}}, \citenamefont {{Jenet}}, \citenamefont
  {{Anderson}}, \citenamefont {{Murty}},\ and\ \citenamefont
  {{Kulkarni}}}]{Jacoby2006}%
  \BibitemOpen
  \bibfield  {author} {\bibinfo {author} {\bibfnamefont {B.~A.}\ \bibnamefont
  {{Jacoby}}}, \bibinfo {author} {\bibfnamefont {P.~B.}\ \bibnamefont
  {{Cameron}}}, \bibinfo {author} {\bibfnamefont {F.~A.}\ \bibnamefont
  {{Jenet}}}, \bibinfo {author} {\bibfnamefont {S.~B.}\ \bibnamefont
  {{Anderson}}}, \bibinfo {author} {\bibfnamefont {R.~N.}\ \bibnamefont
  {{Murty}}}, \ and\ \bibinfo {author} {\bibfnamefont {S.~R.}\ \bibnamefont
  {{Kulkarni}}},\ }\href {\doibase 10.1086/505742} {\bibfield  {journal}
  {\bibinfo  {journal} {Astrophys. J. Lett.}\ }\textbf {\bibinfo {volume}
  {644}},\ \bibinfo {pages} {L113} (\bibinfo {year} {2006})},\ \Eprint
  {http://arxiv.org/abs/astro-ph/0605375} {astro-ph/0605375} \BibitemShut
  {NoStop}%
\bibitem [{\citenamefont {{Harry}}\ \emph {et~al.}(2010)\citenamefont {{Harry}}
  \emph {et~al.}}]{Harry2010}%
  \BibitemOpen
  \bibfield  {author} {\bibinfo {author} {\bibfnamefont {G.~M.}\ \bibnamefont
  {{Harry}}} \emph {et~al.},\ }\href {\doibase 10.1088/0264-9381/27/8/084006}
  {\bibfield  {journal} {\bibinfo  {journal} {Class. Quantum Grav.}\ }\textbf
  {\bibinfo {volume} {27}},\ \bibinfo {pages} {084006} (\bibinfo {year}
  {2010})}\BibitemShut {NoStop}%
\bibitem [{\citenamefont {{The LIGO Scientific Collaboration}}\ and\
  \citenamefont {{the Virgo Collaboration}}(2016)}]{Abbott2016a}%
  \BibitemOpen
  \bibfield  {author} {\bibinfo {author} {\bibnamefont {{The LIGO Scientific
  Collaboration}}}\ and\ \bibinfo {author} {\bibnamefont {{the Virgo
  Collaboration}}},\ }\href {\doibase 10.1103/PhysRevLett.116.061102}
  {\bibfield  {journal} {\bibinfo  {journal} {Phys. Rev. Lett.}\ }\textbf
  {\bibinfo {volume} {116}},\ \bibinfo {eid} {061102} (\bibinfo {year}
  {2016})},\ \Eprint {http://arxiv.org/abs/1602.03837} {arXiv:1602.03837
  [gr-qc]} \BibitemShut {NoStop}%
\bibitem [{\citenamefont {{Abbott}}\ \emph {et~al.}(2016)\citenamefont
  {{Abbott}}, \citenamefont {{Abbott}}, \citenamefont {{Abbott}}, \citenamefont
  {{Abernathy}}, \citenamefont {{Acernese}}, \citenamefont {{Ackley}},
  \citenamefont {{Adams}}, \citenamefont {{Adams}}, \citenamefont {{Addesso}},
  \citenamefont {{Adhikari}},\ and\ \citenamefont {et~al.}}]{Abbot2016g}%
  \BibitemOpen
  \bibfield  {author} {\bibinfo {author} {\bibfnamefont {B.~P.}\ \bibnamefont
  {{Abbott}}}, \bibinfo {author} {\bibfnamefont {R.}~\bibnamefont {{Abbott}}},
  \bibinfo {author} {\bibfnamefont {T.~D.}\ \bibnamefont {{Abbott}}}, \bibinfo
  {author} {\bibfnamefont {M.~R.}\ \bibnamefont {{Abernathy}}}, \bibinfo
  {author} {\bibfnamefont {F.}~\bibnamefont {{Acernese}}}, \bibinfo {author}
  {\bibfnamefont {K.}~\bibnamefont {{Ackley}}}, \bibinfo {author}
  {\bibfnamefont {C.}~\bibnamefont {{Adams}}}, \bibinfo {author} {\bibfnamefont
  {T.}~\bibnamefont {{Adams}}}, \bibinfo {author} {\bibfnamefont
  {P.}~\bibnamefont {{Addesso}}}, \bibinfo {author} {\bibfnamefont {R.~X.}\
  \bibnamefont {{Adhikari}}}, \ and\ \bibinfo {author} {\bibnamefont
  {et~al.}},\ }\href {\doibase 10.1103/PhysRevLett.116.241103} {\bibfield
  {journal} {\bibinfo  {journal} {Physical Review Letters}\ }\textbf {\bibinfo
  {volume} {116}},\ \bibinfo {eid} {241103} (\bibinfo {year} {2016})},\ \Eprint
  {http://arxiv.org/abs/1606.04855} {arXiv:1606.04855 [gr-qc]} \BibitemShut
  {NoStop}%
\bibitem [{\citenamefont {{Accadia}}\ \emph {et~al.}(2011)\citenamefont
  {{Accadia}} \emph {et~al.}}]{Accadia2011_etal}%
  \BibitemOpen
  \bibfield  {author} {\bibinfo {author} {\bibfnamefont {T.}~\bibnamefont
  {{Accadia}}} \emph {et~al.},\ }\href {\doibase
  10.1088/0264-9381/28/11/114002} {\bibfield  {journal} {\bibinfo  {journal}
  {Class. Quantum Grav.}\ }\textbf {\bibinfo {volume} {28}},\ \bibinfo {eid}
  {114002} (\bibinfo {year} {2011})}\BibitemShut {NoStop}%
\bibitem [{\citenamefont {{Aso}}\ \emph {et~al.}(2013)\citenamefont {{Aso}},
  \citenamefont {{Michimura}}, \citenamefont {{Somiya}}, \citenamefont
  {{Ando}}, \citenamefont {{Miyakawa}}, \citenamefont {{Sekiguchi}},
  \citenamefont {{Tatsumi}},\ and\ \citenamefont {{Yamamoto}}}]{Aso:2013}%
  \BibitemOpen
  \bibfield  {author} {\bibinfo {author} {\bibfnamefont {Y.}~\bibnamefont
  {{Aso}}}, \bibinfo {author} {\bibfnamefont {Y.}~\bibnamefont {{Michimura}}},
  \bibinfo {author} {\bibfnamefont {K.}~\bibnamefont {{Somiya}}}, \bibinfo
  {author} {\bibfnamefont {M.}~\bibnamefont {{Ando}}}, \bibinfo {author}
  {\bibfnamefont {O.}~\bibnamefont {{Miyakawa}}}, \bibinfo {author}
  {\bibfnamefont {T.}~\bibnamefont {{Sekiguchi}}}, \bibinfo {author}
  {\bibfnamefont {D.}~\bibnamefont {{Tatsumi}}}, \ and\ \bibinfo {author}
  {\bibfnamefont {H.}~\bibnamefont {{Yamamoto}}},\ }\href {\doibase
  10.1103/PhysRevD.88.043007} {\bibfield  {journal} {\bibinfo  {journal} {Phys.
  Rev. D}\ }\textbf {\bibinfo {volume} {88}},\ \bibinfo {eid} {043007}
  (\bibinfo {year} {2013})},\ \Eprint {http://arxiv.org/abs/1306.6747}
  {arXiv:1306.6747 [gr-qc]} \BibitemShut {NoStop}%
\bibitem [{\citenamefont {{Fairhurst}}(2014)}]{Fairhurst2014}%
  \BibitemOpen
  \bibfield  {author} {\bibinfo {author} {\bibfnamefont {S.}~\bibnamefont
  {{Fairhurst}}},\ }\href {\doibase 10.1088/1742-6596/484/1/012007} {\bibfield
  {journal} {\bibinfo  {journal} {Journal of Physics Conference Series}\
  }\textbf {\bibinfo {volume} {484}},\ \bibinfo {eid} {012007} (\bibinfo {year}
  {2014})},\ \Eprint {http://arxiv.org/abs/1205.6611} {arXiv:1205.6611 [gr-qc]}
  \BibitemShut {NoStop}%
\bibitem [{\citenamefont {{Abadie}}\ \emph {et~al.}(2010)\citenamefont
  {{Abadie}} \emph {et~al.}}]{Abadie:2010_etal}%
  \BibitemOpen
  \bibfield  {author} {\bibinfo {author} {\bibfnamefont {J.}~\bibnamefont
  {{Abadie}}} \emph {et~al.},\ }\href {\doibase 10.1088/0264-9381/27/17/173001}
  {\bibfield  {journal} {\bibinfo  {journal} {Class. Quantum Grav.}\ }\textbf
  {\bibinfo {volume} {27}},\ \bibinfo {pages} {173001} (\bibinfo {year}
  {2010})},\ \Eprint {http://arxiv.org/abs/1003.2480} {arXiv:1003.2480
  [astro-ph.HE]} \BibitemShut {NoStop}%
\bibitem [{\citenamefont {{Narayan}}\ \emph {et~al.}(1992)\citenamefont
  {{Narayan}}, \citenamefont {{Paczynski}},\ and\ \citenamefont
  {{Piran}}}]{Narayan92}%
  \BibitemOpen
  \bibfield  {author} {\bibinfo {author} {\bibfnamefont {R.}~\bibnamefont
  {{Narayan}}}, \bibinfo {author} {\bibfnamefont {B.}~\bibnamefont
  {{Paczynski}}}, \ and\ \bibinfo {author} {\bibfnamefont {T.}~\bibnamefont
  {{Piran}}},\ }\href {\doibase 10.1086/186493} {\bibfield  {journal} {\bibinfo
   {journal} {Astrophysical Journal, Letters}\ }\textbf {\bibinfo {volume}
  {395}},\ \bibinfo {pages} {L83} (\bibinfo {year} {1992})},\ \Eprint
  {http://arxiv.org/abs/astro-ph/9204001} {astro-ph/9204001} \BibitemShut
  {NoStop}%
\bibitem [{\citenamefont {{Eichler}}\ \emph {et~al.}(1989)\citenamefont
  {{Eichler}}, \citenamefont {{Livio}}, \citenamefont {{Piran}},\ and\
  \citenamefont {{Schramm}}}]{Eichler89}%
  \BibitemOpen
  \bibfield  {author} {\bibinfo {author} {\bibfnamefont {D.}~\bibnamefont
  {{Eichler}}}, \bibinfo {author} {\bibfnamefont {M.}~\bibnamefont {{Livio}}},
  \bibinfo {author} {\bibfnamefont {T.}~\bibnamefont {{Piran}}}, \ and\
  \bibinfo {author} {\bibfnamefont {D.~N.}\ \bibnamefont {{Schramm}}},\ }\href
  {\doibase 10.1038/340126a0} {\bibfield  {journal} {\bibinfo  {journal}
  {Nature}\ }\textbf {\bibinfo {volume} {340}},\ \bibinfo {pages} {126}
  (\bibinfo {year} {1989})}\BibitemShut {NoStop}%
\bibitem [{\citenamefont {{Shibata}}\ and\ \citenamefont {{Ury{\=
  u}}}(2000)}]{Shibata99d}%
  \BibitemOpen
  \bibfield  {author} {\bibinfo {author} {\bibfnamefont {M.}~\bibnamefont
  {{Shibata}}}\ and\ \bibinfo {author} {\bibfnamefont {K.}~\bibnamefont
  {{Ury{\= u}}}},\ }\href {\doibase 10.1103/PhysRevD.61.064001} {\bibfield
  {journal} {\bibinfo  {journal} {Phys. Rev. D}\ }\textbf {\bibinfo {volume}
  {61}},\ \bibinfo {eid} {064001} (\bibinfo {year} {2000})},\ \Eprint
  {http://arxiv.org/abs/gr-qc/9911058} {gr-qc/9911058} \BibitemShut {NoStop}%
\bibitem [{\citenamefont {{Baiotti}}\ \emph {et~al.}(2008)\citenamefont
  {{Baiotti}}, \citenamefont {{Giacomazzo}},\ and\ \citenamefont
  {{Rezzolla}}}]{Baiotti08}%
  \BibitemOpen
  \bibfield  {author} {\bibinfo {author} {\bibfnamefont {L.}~\bibnamefont
  {{Baiotti}}}, \bibinfo {author} {\bibfnamefont {B.}~\bibnamefont
  {{Giacomazzo}}}, \ and\ \bibinfo {author} {\bibfnamefont {L.}~\bibnamefont
  {{Rezzolla}}},\ }\href {\doibase 10.1103/PhysRevD.78.084033} {\bibfield
  {journal} {\bibinfo  {journal} {Phys. Rev. D}\ }\textbf {\bibinfo {volume}
  {78}},\ \bibinfo {pages} {084033} (\bibinfo {year} {2008})},\ \Eprint
  {http://arxiv.org/abs/0804.0594} {arXiv:0804.0594 [gr-qc]} \BibitemShut
  {NoStop}%
\bibitem [{\citenamefont {{Anderson}}\ \emph
  {et~al.}(2008{\natexlab{a}})\citenamefont {{Anderson}}, \citenamefont
  {{Hirschmann}}, \citenamefont {{Lehner}}, \citenamefont {{Liebling}},
  \citenamefont {{Motl}}, \citenamefont {{Neilsen}}, \citenamefont
  {{Palenzuela}},\ and\ \citenamefont {{Tohline}}}]{Anderson2007}%
  \BibitemOpen
  \bibfield  {author} {\bibinfo {author} {\bibfnamefont {M.}~\bibnamefont
  {{Anderson}}}, \bibinfo {author} {\bibfnamefont {E.~W.}\ \bibnamefont
  {{Hirschmann}}}, \bibinfo {author} {\bibfnamefont {L.}~\bibnamefont
  {{Lehner}}}, \bibinfo {author} {\bibfnamefont {S.~L.}\ \bibnamefont
  {{Liebling}}}, \bibinfo {author} {\bibfnamefont {P.~M.}\ \bibnamefont
  {{Motl}}}, \bibinfo {author} {\bibfnamefont {D.}~\bibnamefont {{Neilsen}}},
  \bibinfo {author} {\bibfnamefont {C.}~\bibnamefont {{Palenzuela}}}, \ and\
  \bibinfo {author} {\bibfnamefont {J.~E.}\ \bibnamefont {{Tohline}}},\ }\href
  {\doibase 10.1103/PhysRevD.77.024006} {\bibfield  {journal} {\bibinfo
  {journal} {Phys. Rev. D}\ }\textbf {\bibinfo {volume} {77}},\ \bibinfo {eid}
  {024006} (\bibinfo {year} {2008}{\natexlab{a}})},\ \Eprint
  {http://arxiv.org/abs/0708.2720} {arXiv:0708.2720 [gr-qc]} \BibitemShut
  {NoStop}%
\bibitem [{\citenamefont {Liu}\ \emph {et~al.}(2008)\citenamefont {Liu},
  \citenamefont {Shapiro}, \citenamefont {Etienne},\ and\ \citenamefont
  {Taniguchi}}]{Liu:2008xy}%
  \BibitemOpen
  \bibfield  {author} {\bibinfo {author} {\bibfnamefont {Y.~T.}\ \bibnamefont
  {Liu}}, \bibinfo {author} {\bibfnamefont {S.~L.}\ \bibnamefont {Shapiro}},
  \bibinfo {author} {\bibfnamefont {Z.~B.}\ \bibnamefont {Etienne}}, \ and\
  \bibinfo {author} {\bibfnamefont {K.}~\bibnamefont {Taniguchi}},\ }\href
  {\doibase 10.1103/PhysRevD.78.024012} {\bibfield  {journal} {\bibinfo
  {journal} {Phys. Rev. D}\ }\textbf {\bibinfo {volume} {78}},\ \bibinfo
  {pages} {024012} (\bibinfo {year} {2008})},\ \Eprint
  {http://arxiv.org/abs/0803.4193} {arXiv:0803.4193 [astro-ph]} \BibitemShut
  {NoStop}%
\bibitem [{\citenamefont {{Bernuzzi}}\ \emph
  {et~al.}(2012{\natexlab{a}})\citenamefont {{Bernuzzi}}, \citenamefont
  {{Thierfelder}},\ and\ \citenamefont {{Br{\"u}gmann}}}]{Bernuzzi2011}%
  \BibitemOpen
  \bibfield  {author} {\bibinfo {author} {\bibfnamefont {S.}~\bibnamefont
  {{Bernuzzi}}}, \bibinfo {author} {\bibfnamefont {M.}~\bibnamefont
  {{Thierfelder}}}, \ and\ \bibinfo {author} {\bibfnamefont {B.}~\bibnamefont
  {{Br{\"u}gmann}}},\ }\href {\doibase 10.1103/PhysRevD.85.104030} {\bibfield
  {journal} {\bibinfo  {journal} {Phys. Rev. D}\ }\textbf {\bibinfo {volume}
  {85}},\ \bibinfo {eid} {104030} (\bibinfo {year} {2012}{\natexlab{a}})},\
  \Eprint {http://arxiv.org/abs/1109.3611} {arXiv:1109.3611 [gr-qc]}
  \BibitemShut {NoStop}%
\bibitem [{\citenamefont {{Margalit}}\ \emph {et~al.}(2015)\citenamefont
  {{Margalit}}, \citenamefont {{Metzger}},\ and\ \citenamefont
  {{Beloborodov}}}]{Margalit2015}%
  \BibitemOpen
  \bibfield  {author} {\bibinfo {author} {\bibfnamefont {B.}~\bibnamefont
  {{Margalit}}}, \bibinfo {author} {\bibfnamefont {B.~D.}\ \bibnamefont
  {{Metzger}}}, \ and\ \bibinfo {author} {\bibfnamefont {A.~M.}\ \bibnamefont
  {{Beloborodov}}},\ }\href {\doibase 10.1103/PhysRevLett.115.171101}
  {\bibfield  {journal} {\bibinfo  {journal} {Phys. Rev. Lett.}\ }\textbf
  {\bibinfo {volume} {115}},\ \bibinfo {eid} {171101} (\bibinfo {year}
  {2015})},\ \Eprint {http://arxiv.org/abs/1505.01842} {arXiv:1505.01842
  [astro-ph.HE]} \BibitemShut {NoStop}%
\bibitem [{\citenamefont {{Paczynski}}(1986)}]{Paczynski86}%
  \BibitemOpen
  \bibfield  {author} {\bibinfo {author} {\bibfnamefont {B.}~\bibnamefont
  {{Paczynski}}},\ }\href {\doibase 10.1086/184740} {\bibfield  {journal}
  {\bibinfo  {journal} {Astrophys. J. Lett.}\ }\textbf {\bibinfo {volume}
  {308}},\ \bibinfo {pages} {L43} (\bibinfo {year} {1986})}\BibitemShut
  {NoStop}%
\bibitem [{\citenamefont {{Palenzuela}}\ \emph
  {et~al.}(2013{\natexlab{a}})\citenamefont {{Palenzuela}}, \citenamefont
  {{Lehner}}, \citenamefont {{Ponce}}, \citenamefont {{Liebling}},
  \citenamefont {{Anderson}}, \citenamefont {{Neilsen}},\ and\ \citenamefont
  {{Motl}}}]{Palenzuela2013a}%
  \BibitemOpen
  \bibfield  {author} {\bibinfo {author} {\bibfnamefont {C.}~\bibnamefont
  {{Palenzuela}}}, \bibinfo {author} {\bibfnamefont {L.}~\bibnamefont
  {{Lehner}}}, \bibinfo {author} {\bibfnamefont {M.}~\bibnamefont {{Ponce}}},
  \bibinfo {author} {\bibfnamefont {S.~L.}\ \bibnamefont {{Liebling}}},
  \bibinfo {author} {\bibfnamefont {M.}~\bibnamefont {{Anderson}}}, \bibinfo
  {author} {\bibfnamefont {D.}~\bibnamefont {{Neilsen}}}, \ and\ \bibinfo
  {author} {\bibfnamefont {P.}~\bibnamefont {{Motl}}},\ }\href {\doibase
  10.1103/PhysRevLett.111.061105} {\bibfield  {journal} {\bibinfo  {journal}
  {Phys. Rev. Lett.}\ }\textbf {\bibinfo {volume} {111}},\ \bibinfo {eid}
  {061105} (\bibinfo {year} {2013}{\natexlab{a}})},\ \Eprint
  {http://arxiv.org/abs/1301.7074} {arXiv:1301.7074 [gr-qc]} \BibitemShut
  {NoStop}%
\bibitem [{\citenamefont {{Rezzolla}}\ \emph {et~al.}(2011)\citenamefont
  {{Rezzolla}}, \citenamefont {{Giacomazzo}}, \citenamefont {{Baiotti}},
  \citenamefont {{Granot}}, \citenamefont {{Kouveliotou}},\ and\ \citenamefont
  {{Aloy}}}]{Rezzolla:2011}%
  \BibitemOpen
  \bibfield  {author} {\bibinfo {author} {\bibfnamefont {L.}~\bibnamefont
  {{Rezzolla}}}, \bibinfo {author} {\bibfnamefont {B.}~\bibnamefont
  {{Giacomazzo}}}, \bibinfo {author} {\bibfnamefont {L.}~\bibnamefont
  {{Baiotti}}}, \bibinfo {author} {\bibfnamefont {J.}~\bibnamefont {{Granot}}},
  \bibinfo {author} {\bibfnamefont {C.}~\bibnamefont {{Kouveliotou}}}, \ and\
  \bibinfo {author} {\bibfnamefont {M.~A.}\ \bibnamefont {{Aloy}}},\ }\href
  {\doibase 10.1088/2041-8205/732/1/L6} {\bibfield  {journal} {\bibinfo
  {journal} {Astrophys. J. Letters}\ }\textbf {\bibinfo {volume} {732}},\
  \bibinfo {eid} {L6} (\bibinfo {year} {2011})},\ \Eprint
  {http://arxiv.org/abs/1101.4298} {arXiv:1101.4298 [astro-ph.HE]} \BibitemShut
  {NoStop}%
\bibitem [{\citenamefont {{Paschalidis}}\ \emph
  {et~al.}(2015{\natexlab{a}})\citenamefont {{Paschalidis}}, \citenamefont
  {{Ruiz}},\ and\ \citenamefont {{Shapiro}}}]{Paschalidis2014}%
  \BibitemOpen
  \bibfield  {author} {\bibinfo {author} {\bibfnamefont {V.}~\bibnamefont
  {{Paschalidis}}}, \bibinfo {author} {\bibfnamefont {M.}~\bibnamefont
  {{Ruiz}}}, \ and\ \bibinfo {author} {\bibfnamefont {S.~L.}\ \bibnamefont
  {{Shapiro}}},\ }\href {\doibase 10.1088/2041-8205/806/1/L14} {\bibfield
  {journal} {\bibinfo  {journal} {Astrophys. J. Lett.}\ }\textbf {\bibinfo
  {volume} {806}},\ \bibinfo {eid} {L14} (\bibinfo {year}
  {2015}{\natexlab{a}})},\ \Eprint {http://arxiv.org/abs/1410.7392}
  {arXiv:1410.7392 [astro-ph.HE]} \BibitemShut {NoStop}%
\bibitem [{\citenamefont {{Dionysopoulou}}\ \emph {et~al.}(2015)\citenamefont
  {{Dionysopoulou}}, \citenamefont {{Alic}},\ and\ \citenamefont
  {{Rezzolla}}}]{Dionysopoulou2015}%
  \BibitemOpen
  \bibfield  {author} {\bibinfo {author} {\bibfnamefont {K.}~\bibnamefont
  {{Dionysopoulou}}}, \bibinfo {author} {\bibfnamefont {D.}~\bibnamefont
  {{Alic}}}, \ and\ \bibinfo {author} {\bibfnamefont {L.}~\bibnamefont
  {{Rezzolla}}},\ }\href {\doibase 10.1103/PhysRevD.92.084064} {\bibfield
  {journal} {\bibinfo  {journal} {Phys. Rev. D}\ }\textbf {\bibinfo {volume}
  {92}},\ \bibinfo {eid} {084064} (\bibinfo {year} {2015})},\ \Eprint
  {http://arxiv.org/abs/1502.02021} {arXiv:1502.02021 [gr-qc]} \BibitemShut
  {NoStop}%
\bibitem [{\citenamefont {{Ruiz}}\ \emph {et~al.}(2016)\citenamefont {{Ruiz}},
  \citenamefont {{Lang}}, \citenamefont {{Paschalidis}},\ and\ \citenamefont
  {{Shapiro}}}]{Ruiz2016}%
  \BibitemOpen
  \bibfield  {author} {\bibinfo {author} {\bibfnamefont {M.}~\bibnamefont
  {{Ruiz}}}, \bibinfo {author} {\bibfnamefont {R.~N.}\ \bibnamefont {{Lang}}},
  \bibinfo {author} {\bibfnamefont {V.}~\bibnamefont {{Paschalidis}}}, \ and\
  \bibinfo {author} {\bibfnamefont {S.~L.}\ \bibnamefont {{Shapiro}}},\ }\href
  {\doibase 10.3847/2041-8205/824/1/L6} {\bibfield  {journal} {\bibinfo
  {journal} {Astrophys. J. Lett.}\ }\textbf {\bibinfo {volume} {824}},\
  \bibinfo {eid} {L6} (\bibinfo {year} {2016})},\ \Eprint
  {http://arxiv.org/abs/1604.02455} {arXiv:1604.02455 [astro-ph.HE]}
  \BibitemShut {NoStop}%
\bibitem [{\citenamefont {{Berger}}\ \emph {et~al.}(2013)\citenamefont
  {{Berger}}, \citenamefont {{Fong}},\ and\ \citenamefont
  {{Chornock}}}]{Berger2013}%
  \BibitemOpen
  \bibfield  {author} {\bibinfo {author} {\bibfnamefont {E.}~\bibnamefont
  {{Berger}}}, \bibinfo {author} {\bibfnamefont {W.}~\bibnamefont {{Fong}}}, \
  and\ \bibinfo {author} {\bibfnamefont {R.}~\bibnamefont {{Chornock}}},\
  }\href {\doibase 10.1088/2041-8205/774/2/L23} {\bibfield  {journal} {\bibinfo
   {journal} {Astrophys. J.}\ }\textbf {\bibinfo {volume} {774}},\ \bibinfo
  {eid} {L23} (\bibinfo {year} {2013})},\ \Eprint
  {http://arxiv.org/abs/1306.3960} {arXiv:1306.3960 [astro-ph.HE]} \BibitemShut
  {NoStop}%
\bibitem [{\citenamefont {{Tanvir}}\ \emph {et~al.}(2013)\citenamefont
  {{Tanvir}}, \citenamefont {{Levan}}, \citenamefont {{Fruchter}},
  \citenamefont {{Hjorth}}, \citenamefont {{Hounsell}}, \citenamefont
  {{Wiersema}},\ and\ \citenamefont {{Tunnicliffe}}}]{Tanvir2013}%
  \BibitemOpen
  \bibfield  {author} {\bibinfo {author} {\bibfnamefont {N.~R.}\ \bibnamefont
  {{Tanvir}}}, \bibinfo {author} {\bibfnamefont {A.~J.}\ \bibnamefont
  {{Levan}}}, \bibinfo {author} {\bibfnamefont {A.~S.}\ \bibnamefont
  {{Fruchter}}}, \bibinfo {author} {\bibfnamefont {J.}~\bibnamefont
  {{Hjorth}}}, \bibinfo {author} {\bibfnamefont {R.~A.}\ \bibnamefont
  {{Hounsell}}}, \bibinfo {author} {\bibfnamefont {K.}~\bibnamefont
  {{Wiersema}}}, \ and\ \bibinfo {author} {\bibfnamefont {R.~L.}\ \bibnamefont
  {{Tunnicliffe}}},\ }\href {\doibase 10.1038/nature12505} {\bibfield
  {journal} {\bibinfo  {journal} {Nature}\ }\textbf {\bibinfo {volume} {500}},\
  \bibinfo {pages} {547} (\bibinfo {year} {2013})},\ \Eprint
  {http://arxiv.org/abs/1306.4971} {arXiv:1306.4971 [astro-ph.HE]} \BibitemShut
  {NoStop}%
\bibitem [{\citenamefont {{Li}}\ and\ \citenamefont
  {{Paczy{\'n}ski}}(1998)}]{Li1998}%
  \BibitemOpen
  \bibfield  {author} {\bibinfo {author} {\bibfnamefont {L.-X.}\ \bibnamefont
  {{Li}}}\ and\ \bibinfo {author} {\bibfnamefont {B.}~\bibnamefont
  {{Paczy{\'n}ski}}},\ }\href {\doibase 10.1086/311680} {\bibfield  {journal}
  {\bibinfo  {journal} {Astrophys. J.}\ }\textbf {\bibinfo {volume} {507}},\
  \bibinfo {pages} {L59} (\bibinfo {year} {1998})},\ \Eprint
  {http://arxiv.org/abs/astro-ph/9807272} {astro-ph/9807272} \BibitemShut
  {NoStop}%
\bibitem [{\citenamefont {{Kulkarni}}(2005)}]{Kulkarni2005_macronova-term}%
  \BibitemOpen
  \bibfield  {author} {\bibinfo {author} {\bibfnamefont {S.~R.}\ \bibnamefont
  {{Kulkarni}}},\ }\href@noop {} {\bibfield  {journal} {\bibinfo  {journal}
  {astro-ph/0510256}\ } (\bibinfo {year} {2005})},\ \Eprint
  {http://arxiv.org/abs/astro-ph/0510256} {astro-ph/0510256} \BibitemShut
  {NoStop}%
\bibitem [{\citenamefont {{Metzger}}\ \emph {et~al.}(2010)\citenamefont
  {{Metzger}}, \citenamefont {{Mart{\'{\i}}nez-Pinedo}}, \citenamefont
  {{Darbha}}, \citenamefont {{Quataert}}, \citenamefont {{Arcones}},
  \citenamefont {{Kasen}}, \citenamefont {{Thomas}}, \citenamefont {{Nugent}},
  \citenamefont {{Panov}},\ and\ \citenamefont {{Zinner}}}]{Metzger:2010}%
  \BibitemOpen
  \bibfield  {author} {\bibinfo {author} {\bibfnamefont {B.~D.}\ \bibnamefont
  {{Metzger}}}, \bibinfo {author} {\bibfnamefont {G.}~\bibnamefont
  {{Mart{\'{\i}}nez-Pinedo}}}, \bibinfo {author} {\bibfnamefont
  {S.}~\bibnamefont {{Darbha}}}, \bibinfo {author} {\bibfnamefont
  {E.}~\bibnamefont {{Quataert}}}, \bibinfo {author} {\bibfnamefont
  {A.}~\bibnamefont {{Arcones}}}, \bibinfo {author} {\bibfnamefont
  {D.}~\bibnamefont {{Kasen}}}, \bibinfo {author} {\bibfnamefont
  {R.}~\bibnamefont {{Thomas}}}, \bibinfo {author} {\bibfnamefont
  {P.}~\bibnamefont {{Nugent}}}, \bibinfo {author} {\bibfnamefont {I.~V.}\
  \bibnamefont {{Panov}}}, \ and\ \bibinfo {author} {\bibfnamefont {N.~T.}\
  \bibnamefont {{Zinner}}},\ }\href {\doibase 10.1111/j.1365-2966.2010.16864.x}
  {\bibfield  {journal} {\bibinfo  {journal} {Mon. Not. R. Astron. Soc.}\
  }\textbf {\bibinfo {volume} {406}},\ \bibinfo {pages} {2650} (\bibinfo {year}
  {2010})},\ \Eprint {http://arxiv.org/abs/1001.5029} {arXiv:1001.5029
  [astro-ph.HE]} \BibitemShut {NoStop}%
\bibitem [{\citenamefont {{Yang}}\ \emph {et~al.}(2015)\citenamefont {{Yang}},
  \citenamefont {{Jin}}, \citenamefont {{Li}}, \citenamefont {{Covino}},
  \citenamefont {{Zheng}}, \citenamefont {{Hotokezaka}}, \citenamefont {{Fan}},
  \citenamefont {{Piran}},\ and\ \citenamefont {{Wei}}}]{Yang2015}%
  \BibitemOpen
  \bibfield  {author} {\bibinfo {author} {\bibfnamefont {B.}~\bibnamefont
  {{Yang}}}, \bibinfo {author} {\bibfnamefont {Z.-P.}\ \bibnamefont {{Jin}}},
  \bibinfo {author} {\bibfnamefont {X.}~\bibnamefont {{Li}}}, \bibinfo {author}
  {\bibfnamefont {S.}~\bibnamefont {{Covino}}}, \bibinfo {author}
  {\bibfnamefont {X.-Z.}\ \bibnamefont {{Zheng}}}, \bibinfo {author}
  {\bibfnamefont {K.}~\bibnamefont {{Hotokezaka}}}, \bibinfo {author}
  {\bibfnamefont {Y.-Z.}\ \bibnamefont {{Fan}}}, \bibinfo {author}
  {\bibfnamefont {T.}~\bibnamefont {{Piran}}}, \ and\ \bibinfo {author}
  {\bibfnamefont {D.-M.}\ \bibnamefont {{Wei}}},\ }\href {\doibase
  10.1038/ncomms8323} {\bibfield  {journal} {\bibinfo  {journal} {Nature
  Communications}\ }\textbf {\bibinfo {volume} {6}},\ \bibinfo {eid} {7323}
  (\bibinfo {year} {2015})},\ \Eprint {http://arxiv.org/abs/1503.07761}
  {arXiv:1503.07761 [astro-ph.HE]} \BibitemShut {NoStop}%
\bibitem [{\citenamefont {{Jin}}\ \emph {et~al.}(2016)\citenamefont {{Jin}},
  \citenamefont {{Hotokezaka}}, \citenamefont {{Li}}, \citenamefont {{Tanaka}},
  \citenamefont {{D'Avanzo}}, \citenamefont {{Fan}}, \citenamefont {{Covino}},
  \citenamefont {{Wei}},\ and\ \citenamefont {{Piran}}}]{Jin2016}%
  \BibitemOpen
  \bibfield  {author} {\bibinfo {author} {\bibfnamefont {Z.-P.}\ \bibnamefont
  {{Jin}}}, \bibinfo {author} {\bibfnamefont {K.}~\bibnamefont {{Hotokezaka}}},
  \bibinfo {author} {\bibfnamefont {X.}~\bibnamefont {{Li}}}, \bibinfo {author}
  {\bibfnamefont {M.}~\bibnamefont {{Tanaka}}}, \bibinfo {author}
  {\bibfnamefont {P.}~\bibnamefont {{D'Avanzo}}}, \bibinfo {author}
  {\bibfnamefont {Y.-Z.}\ \bibnamefont {{Fan}}}, \bibinfo {author}
  {\bibfnamefont {S.}~\bibnamefont {{Covino}}}, \bibinfo {author}
  {\bibfnamefont {D.-M.}\ \bibnamefont {{Wei}}}, \ and\ \bibinfo {author}
  {\bibfnamefont {T.}~\bibnamefont {{Piran}}},\ }\href {\doibase
  10.1038/ncomms12898} {\bibfield  {journal} {\bibinfo  {journal} {Nature
  Communications}\ }\textbf {\bibinfo {volume} {7}},\ \bibinfo {eid} {12898}
  (\bibinfo {year} {2016})},\ \Eprint {http://arxiv.org/abs/1603.07869}
  {arXiv:1603.07869 [astro-ph.HE]} \BibitemShut {NoStop}%
\bibitem [{\citenamefont {{Jin}}\ \emph {et~al.}(2015)\citenamefont {{Jin}},
  \citenamefont {{Li}}, \citenamefont {{Cano}}, \citenamefont {{Covino}},
  \citenamefont {{Fan}},\ and\ \citenamefont {{Wei}}}]{Jin2015}%
  \BibitemOpen
  \bibfield  {author} {\bibinfo {author} {\bibfnamefont {Z.-P.}\ \bibnamefont
  {{Jin}}}, \bibinfo {author} {\bibfnamefont {X.}~\bibnamefont {{Li}}},
  \bibinfo {author} {\bibfnamefont {Z.}~\bibnamefont {{Cano}}}, \bibinfo
  {author} {\bibfnamefont {S.}~\bibnamefont {{Covino}}}, \bibinfo {author}
  {\bibfnamefont {Y.-Z.}\ \bibnamefont {{Fan}}}, \ and\ \bibinfo {author}
  {\bibfnamefont {D.-M.}\ \bibnamefont {{Wei}}},\ }\href {\doibase
  10.1088/2041-8205/811/2/L22} {\bibfield  {journal} {\bibinfo  {journal}
  {Astrophys. J. Lett.}\ }\textbf {\bibinfo {volume} {811}},\ \bibinfo {eid}
  {L22} (\bibinfo {year} {2015})},\ \Eprint {http://arxiv.org/abs/1507.07206}
  {arXiv:1507.07206 [astro-ph.HE]} \BibitemShut {NoStop}%
\bibitem [{\citenamefont {{Wallner}}\ \emph {et~al.}(2015)\citenamefont
  {{Wallner}}, \citenamefont {{Faestermann}}, \citenamefont {{Feige}},
  \citenamefont {{Feldstein}}, \citenamefont {{Knie}}, \citenamefont
  {{Korschinek}}, \citenamefont {{Kutschera}}, \citenamefont {{Ofan}},
  \citenamefont {{Paul}}, \citenamefont {{Quinto}}, \citenamefont {{Rugel}},\
  and\ \citenamefont {{Steier}}}]{wallner:15}%
  \BibitemOpen
  \bibfield  {author} {\bibinfo {author} {\bibfnamefont {A.}~\bibnamefont
  {{Wallner}}}, \bibinfo {author} {\bibfnamefont {T.}~\bibnamefont
  {{Faestermann}}}, \bibinfo {author} {\bibfnamefont {J.}~\bibnamefont
  {{Feige}}}, \bibinfo {author} {\bibfnamefont {C.}~\bibnamefont
  {{Feldstein}}}, \bibinfo {author} {\bibfnamefont {K.}~\bibnamefont {{Knie}}},
  \bibinfo {author} {\bibfnamefont {G.}~\bibnamefont {{Korschinek}}}, \bibinfo
  {author} {\bibfnamefont {W.}~\bibnamefont {{Kutschera}}}, \bibinfo {author}
  {\bibfnamefont {A.}~\bibnamefont {{Ofan}}}, \bibinfo {author} {\bibfnamefont
  {M.}~\bibnamefont {{Paul}}}, \bibinfo {author} {\bibfnamefont
  {F.}~\bibnamefont {{Quinto}}}, \bibinfo {author} {\bibfnamefont
  {G.}~\bibnamefont {{Rugel}}}, \ and\ \bibinfo {author} {\bibfnamefont
  {P.}~\bibnamefont {{Steier}}},\ }\href {\doibase 10.1038/ncomms6956}
  {\bibfield  {journal} {\bibinfo  {journal} {Nature Communications}\ }\textbf
  {\bibinfo {volume} {6}},\ \bibinfo {eid} {5956} (\bibinfo {year} {2015})},\
  \Eprint {http://arxiv.org/abs/1509.08054} {arXiv:1509.08054 [astro-ph.SR]}
  \BibitemShut {NoStop}%
\bibitem [{\citenamefont {{Hotokezaka}}\ \emph
  {et~al.}(2015{\natexlab{a}})\citenamefont {{Hotokezaka}}, \citenamefont
  {{Piran}},\ and\ \citenamefont {{Paul}}}]{Hotokezaka:2015b}%
  \BibitemOpen
  \bibfield  {author} {\bibinfo {author} {\bibfnamefont {K.}~\bibnamefont
  {{Hotokezaka}}}, \bibinfo {author} {\bibfnamefont {T.}~\bibnamefont
  {{Piran}}}, \ and\ \bibinfo {author} {\bibfnamefont {M.}~\bibnamefont
  {{Paul}}},\ }\href@noop {} {\bibfield  {journal} {\bibinfo  {journal} {ArXiv
  e-prints}\ } (\bibinfo {year} {2015}{\natexlab{a}})},\ \Eprint
  {http://arxiv.org/abs/1510.00711} {arXiv:1510.00711 [astro-ph.HE]}
  \BibitemShut {NoStop}%
\bibitem [{\citenamefont {{Ji}}\ \emph {et~al.}(2015)\citenamefont {{Ji}},
  \citenamefont {{Frebel}}, \citenamefont {{Chiti}},\ and\ \citenamefont
  {{Simon}}}]{Ji:15}%
  \BibitemOpen
  \bibfield  {author} {\bibinfo {author} {\bibfnamefont {A.~P.}\ \bibnamefont
  {{Ji}}}, \bibinfo {author} {\bibfnamefont {A.}~\bibnamefont {{Frebel}}},
  \bibinfo {author} {\bibfnamefont {A.}~\bibnamefont {{Chiti}}}, \ and\
  \bibinfo {author} {\bibfnamefont {J.~D.}\ \bibnamefont {{Simon}}},\
  }\href@noop {} {\bibfield  {journal} {\bibinfo  {journal} {ArXiv e-prints}\ }
  (\bibinfo {year} {2015})},\ \Eprint {http://arxiv.org/abs/1512.01558}
  {arXiv:1512.01558} \BibitemShut {NoStop}%
\bibitem [{\citenamefont {Tolman}(1939)}]{Tolman39}%
  \BibitemOpen
  \bibfield  {author} {\bibinfo {author} {\bibfnamefont {R.~C.}\ \bibnamefont
  {Tolman}},\ }\href@noop {} {\bibfield  {journal} {\bibinfo  {journal} {Phys.
  Rev.}\ }\textbf {\bibinfo {volume} {55}},\ \bibinfo {pages} {364} (\bibinfo
  {year} {1939})}\BibitemShut {NoStop}%
\bibitem [{\citenamefont {Oppenheimer}\ and\ \citenamefont
  {Volkoff}(1939)}]{Oppenheimer39b}%
  \BibitemOpen
  \bibfield  {author} {\bibinfo {author} {\bibfnamefont {J.~R.}\ \bibnamefont
  {Oppenheimer}}\ and\ \bibinfo {author} {\bibfnamefont {G.}~\bibnamefont
  {Volkoff}},\ }\href@noop {} {\bibfield  {journal} {\bibinfo  {journal} {Phys.
  Rev.}\ }\textbf {\bibinfo {volume} {55}},\ \bibinfo {pages} {374} (\bibinfo
  {year} {1939})}\BibitemShut {NoStop}%
\bibitem [{\citenamefont {{Antoniadis}}\ \emph {et~al.}(2013)\citenamefont
  {{Antoniadis}}, \citenamefont {{Freire}}, \citenamefont {{Wex}},
  \citenamefont {{Tauris}}, \citenamefont {{Lynch}}, \citenamefont {{van
  Kerkwijk}}, \citenamefont {{Kramer}}, \citenamefont {{Bassa}}, \citenamefont
  {{Dhillon}}, \citenamefont {{Driebe}}, \citenamefont {{Hessels}},
  \citenamefont {{Kaspi}}, \citenamefont {{Kondratiev}}, \citenamefont
  {{Langer}}, \citenamefont {{Marsh}}, \citenamefont {{McLaughlin}},
  \citenamefont {{Pennucci}}, \citenamefont {{Ransom}}, \citenamefont
  {{Stairs}}, \citenamefont {{van Leeuwen}}, \citenamefont {{Verbiest}},\ and\
  \citenamefont {{Whelan}}}]{Antoniadis2013}%
  \BibitemOpen
  \bibfield  {author} {\bibinfo {author} {\bibfnamefont {J.}~\bibnamefont
  {{Antoniadis}}}, \bibinfo {author} {\bibfnamefont {P.~C.~C.}\ \bibnamefont
  {{Freire}}}, \bibinfo {author} {\bibfnamefont {N.}~\bibnamefont {{Wex}}},
  \bibinfo {author} {\bibfnamefont {T.~M.}\ \bibnamefont {{Tauris}}}, \bibinfo
  {author} {\bibfnamefont {R.~S.}\ \bibnamefont {{Lynch}}}, \bibinfo {author}
  {\bibfnamefont {M.~H.}\ \bibnamefont {{van Kerkwijk}}}, \bibinfo {author}
  {\bibfnamefont {M.}~\bibnamefont {{Kramer}}}, \bibinfo {author}
  {\bibfnamefont {C.}~\bibnamefont {{Bassa}}}, \bibinfo {author} {\bibfnamefont
  {V.~S.}\ \bibnamefont {{Dhillon}}}, \bibinfo {author} {\bibfnamefont
  {T.}~\bibnamefont {{Driebe}}}, \bibinfo {author} {\bibfnamefont {J.~W.~T.}\
  \bibnamefont {{Hessels}}}, \bibinfo {author} {\bibfnamefont {V.~M.}\
  \bibnamefont {{Kaspi}}}, \bibinfo {author} {\bibfnamefont {V.~I.}\
  \bibnamefont {{Kondratiev}}}, \bibinfo {author} {\bibfnamefont
  {N.}~\bibnamefont {{Langer}}}, \bibinfo {author} {\bibfnamefont {T.~R.}\
  \bibnamefont {{Marsh}}}, \bibinfo {author} {\bibfnamefont {M.~A.}\
  \bibnamefont {{McLaughlin}}}, \bibinfo {author} {\bibfnamefont {T.~T.}\
  \bibnamefont {{Pennucci}}}, \bibinfo {author} {\bibfnamefont {S.~M.}\
  \bibnamefont {{Ransom}}}, \bibinfo {author} {\bibfnamefont {I.~H.}\
  \bibnamefont {{Stairs}}}, \bibinfo {author} {\bibfnamefont {J.}~\bibnamefont
  {{van Leeuwen}}}, \bibinfo {author} {\bibfnamefont {J.~P.~W.}\ \bibnamefont
  {{Verbiest}}}, \ and\ \bibinfo {author} {\bibfnamefont {D.~G.}\ \bibnamefont
  {{Whelan}}},\ }\href {\doibase 10.1126/science.1233232} {\bibfield  {journal}
  {\bibinfo  {journal} {Science}\ }\textbf {\bibinfo {volume} {340}},\ \bibinfo
  {pages} {448} (\bibinfo {year} {2013})},\ \Eprint
  {http://arxiv.org/abs/1304.6875} {arXiv:1304.6875 [astro-ph.HE]} \BibitemShut
  {NoStop}%
\bibitem [{\citenamefont {{Demorest}}\ \emph {et~al.}(2010)\citenamefont
  {{Demorest}}, \citenamefont {{Pennucci}}, \citenamefont {{Ransom}},
  \citenamefont {{Roberts}},\ and\ \citenamefont {{Hessels}}}]{Demorest2010}%
  \BibitemOpen
  \bibfield  {author} {\bibinfo {author} {\bibfnamefont {P.~B.}\ \bibnamefont
  {{Demorest}}}, \bibinfo {author} {\bibfnamefont {T.}~\bibnamefont
  {{Pennucci}}}, \bibinfo {author} {\bibfnamefont {S.~M.}\ \bibnamefont
  {{Ransom}}}, \bibinfo {author} {\bibfnamefont {M.~S.~E.}\ \bibnamefont
  {{Roberts}}}, \ and\ \bibinfo {author} {\bibfnamefont {J.~W.~T.}\
  \bibnamefont {{Hessels}}},\ }\href {\doibase 10.1038/nature09466} {\bibfield
  {journal} {\bibinfo  {journal} {Nature}\ }\textbf {\bibinfo {volume} {467}},\
  \bibinfo {pages} {1081} (\bibinfo {year} {2010})},\ \Eprint
  {http://arxiv.org/abs/1010.5788} {arXiv:1010.5788 [astro-ph.HE]} \BibitemShut
  {NoStop}%
\bibitem [{\citenamefont {{Rezzolla}}\ \emph {et~al.}(2010)\citenamefont
  {{Rezzolla}}, \citenamefont {{Baiotti}}, \citenamefont {{Giacomazzo}},
  \citenamefont {{Link}},\ and\ \citenamefont {{Font}}}]{Rezzolla:2010}%
  \BibitemOpen
  \bibfield  {author} {\bibinfo {author} {\bibfnamefont {L.}~\bibnamefont
  {{Rezzolla}}}, \bibinfo {author} {\bibfnamefont {L.}~\bibnamefont
  {{Baiotti}}}, \bibinfo {author} {\bibfnamefont {B.}~\bibnamefont
  {{Giacomazzo}}}, \bibinfo {author} {\bibfnamefont {D.}~\bibnamefont
  {{Link}}}, \ and\ \bibinfo {author} {\bibfnamefont {J.~A.}\ \bibnamefont
  {{Font}}},\ }\href {\doibase 10.1088/0264-9381/27/11/114105} {\bibfield
  {journal} {\bibinfo  {journal} {Class. Quantum Grav.}\ }\textbf {\bibinfo
  {volume} {27}},\ \bibinfo {pages} {114105} (\bibinfo {year} {2010})},\
  \Eprint {http://arxiv.org/abs/1001.3074} {arXiv:1001.3074 [gr-qc]}
  \BibitemShut {NoStop}%
\bibitem [{\citenamefont {{Rezzolla}}\ and\ \citenamefont
  {{Zanotti}}(2013)}]{Rezzolla_book:2013}%
  \BibitemOpen
  \bibfield  {author} {\bibinfo {author} {\bibfnamefont {L.}~\bibnamefont
  {{Rezzolla}}}\ and\ \bibinfo {author} {\bibfnamefont {O.}~\bibnamefont
  {{Zanotti}}},\ }\href {\doibase 10.1093/acprof:oso/9780198528906.001.0001}
  {\emph {\bibinfo {title} {Relativistic Hydrodynamics}}}\ (\bibinfo
  {publisher} {Oxford University Press},\ \bibinfo {address} {Oxford, UK},\
  \bibinfo {year} {2013})\BibitemShut {NoStop}%
\bibitem [{\citenamefont {{Breu}}\ and\ \citenamefont
  {{Rezzolla}}(2016)}]{Breu2016}%
  \BibitemOpen
  \bibfield  {author} {\bibinfo {author} {\bibfnamefont {C.}~\bibnamefont
  {{Breu}}}\ and\ \bibinfo {author} {\bibfnamefont {L.}~\bibnamefont
  {{Rezzolla}}},\ }\href {\doibase 10.1093/mnras/stw575} {\bibfield  {journal}
  {\bibinfo  {journal} {Mon. Not. R. Astron. Soc.}\ }\textbf {\bibinfo {volume}
  {459}},\ \bibinfo {pages} {646} (\bibinfo {year} {2016})},\ \Eprint
  {http://arxiv.org/abs/1601.06083} {arXiv:1601.06083 [gr-qc]} \BibitemShut
  {NoStop}%
\bibitem [{\citenamefont {{Shibata}}\ \emph {et~al.}(2000)\citenamefont
  {{Shibata}}, \citenamefont {{Baumgarte}},\ and\ \citenamefont
  {{Shapiro}}}]{Shibata:2000jt}%
  \BibitemOpen
  \bibfield  {author} {\bibinfo {author} {\bibfnamefont {M.}~\bibnamefont
  {{Shibata}}}, \bibinfo {author} {\bibfnamefont {T.~W.}\ \bibnamefont
  {{Baumgarte}}}, \ and\ \bibinfo {author} {\bibfnamefont {S.~L.}\ \bibnamefont
  {{Shapiro}}},\ }\href {\doibase 10.1086/309525} {\bibfield  {journal}
  {\bibinfo  {journal} {Astrophys. J.}\ }\textbf {\bibinfo {volume} {542}},\
  \bibinfo {pages} {453} (\bibinfo {year} {2000})},\ \Eprint
  {http://arxiv.org/abs/astro-ph/0005378} {astro-ph/0005378} \BibitemShut
  {NoStop}%
\bibitem [{\citenamefont {{Baiotti}}\ \emph {et~al.}(2007)\citenamefont
  {{Baiotti}}, \citenamefont {{de Pietri}}, \citenamefont {{Manca}},\ and\
  \citenamefont {{Rezzolla}}}]{Baiotti06b}%
  \BibitemOpen
  \bibfield  {author} {\bibinfo {author} {\bibfnamefont {L.}~\bibnamefont
  {{Baiotti}}}, \bibinfo {author} {\bibfnamefont {R.}~\bibnamefont {{de
  Pietri}}}, \bibinfo {author} {\bibfnamefont {G.~M.}\ \bibnamefont {{Manca}}},
  \ and\ \bibinfo {author} {\bibfnamefont {L.}~\bibnamefont {{Rezzolla}}},\
  }\href {\doibase 10.1103/PhysRevD.75.044023} {\bibfield  {journal} {\bibinfo
  {journal} {Phys. Rev. D}\ }\textbf {\bibinfo {volume} {75}},\ \bibinfo {eid}
  {044023} (\bibinfo {year} {2007})},\ \Eprint
  {http://arxiv.org/abs/astro-ph/0609473} {astro-ph/0609473} \BibitemShut
  {NoStop}%
\bibitem [{\citenamefont {{Franci}}\ \emph
  {et~al.}(2013{\natexlab{a}})\citenamefont {{Franci}}, \citenamefont {{De
  Pietri}}, \citenamefont {{Dionysopoulou}},\ and\ \citenamefont
  {{Rezzolla}}}]{Franci2013}%
  \BibitemOpen
  \bibfield  {author} {\bibinfo {author} {\bibfnamefont {L.}~\bibnamefont
  {{Franci}}}, \bibinfo {author} {\bibfnamefont {R.}~\bibnamefont {{De
  Pietri}}}, \bibinfo {author} {\bibfnamefont {K.}~\bibnamefont
  {{Dionysopoulou}}}, \ and\ \bibinfo {author} {\bibfnamefont {L.}~\bibnamefont
  {{Rezzolla}}},\ }\href {\doibase 10.1088/1742-6596/470/1/012008} {\bibfield
  {journal} {\bibinfo  {journal} {Journal of Physics Conference Series}\
  }\textbf {\bibinfo {volume} {470}},\ \bibinfo {eid} {012008} (\bibinfo {year}
  {2013}{\natexlab{a}})},\ \Eprint {http://arxiv.org/abs/1309.6549}
  {arXiv:1309.6549 [gr-qc]} \BibitemShut {NoStop}%
\bibitem [{\citenamefont {{Siegel}}\ \emph {et~al.}(2013)\citenamefont
  {{Siegel}}, \citenamefont {{Ciolfi}}, \citenamefont {{Harte}},\ and\
  \citenamefont {{Rezzolla}}}]{Siegel2013}%
  \BibitemOpen
  \bibfield  {author} {\bibinfo {author} {\bibfnamefont {D.~M.}\ \bibnamefont
  {{Siegel}}}, \bibinfo {author} {\bibfnamefont {R.}~\bibnamefont {{Ciolfi}}},
  \bibinfo {author} {\bibfnamefont {A.~I.}\ \bibnamefont {{Harte}}}, \ and\
  \bibinfo {author} {\bibfnamefont {L.}~\bibnamefont {{Rezzolla}}},\ }\href
  {\doibase 10.1103/PhysRevD.87.121302} {\bibfield  {journal} {\bibinfo
  {journal} {Phys. Rev. D R}\ }\textbf {\bibinfo {volume} {87}},\ \bibinfo
  {eid} {121302} (\bibinfo {year} {2013})},\ \Eprint
  {http://arxiv.org/abs/1302.4368} {arXiv:1302.4368 [gr-qc]} \BibitemShut
  {NoStop}%
\bibitem [{\citenamefont {{Kiuchi}}\ \emph {et~al.}(2014)\citenamefont
  {{Kiuchi}}, \citenamefont {{Kyutoku}}, \citenamefont {{Sekiguchi}},
  \citenamefont {{Shibata}},\ and\ \citenamefont {{Wada}}}]{Kiuchi2014}%
  \BibitemOpen
  \bibfield  {author} {\bibinfo {author} {\bibfnamefont {K.}~\bibnamefont
  {{Kiuchi}}}, \bibinfo {author} {\bibfnamefont {K.}~\bibnamefont {{Kyutoku}}},
  \bibinfo {author} {\bibfnamefont {Y.}~\bibnamefont {{Sekiguchi}}}, \bibinfo
  {author} {\bibfnamefont {M.}~\bibnamefont {{Shibata}}}, \ and\ \bibinfo
  {author} {\bibfnamefont {T.}~\bibnamefont {{Wada}}},\ }\href {\doibase
  10.1103/PhysRevD.90.041502} {\bibfield  {journal} {\bibinfo  {journal} {Phys.
  Rev. D}\ }\textbf {\bibinfo {volume} {90}},\ \bibinfo {eid} {041502}
  (\bibinfo {year} {2014})},\ \Eprint {http://arxiv.org/abs/1407.2660}
  {arXiv:1407.2660 [astro-ph.HE]} \BibitemShut {NoStop}%
\bibitem [{\citenamefont {Stergioulas}(2003)}]{Stergioulas03}%
  \BibitemOpen
  \bibfield  {author} {\bibinfo {author} {\bibfnamefont {N.}~\bibnamefont
  {Stergioulas}},\ }\href@noop {} {\bibfield  {journal} {\bibinfo  {journal}
  {Living Rev. Relativ.}\ }\textbf {\bibinfo {volume} {6}},\ \bibinfo {pages}
  {3} (\bibinfo {year} {2003})}\BibitemShut {NoStop}%
\bibitem [{\citenamefont {{Ou}}\ and\ \citenamefont {{Tohline}}(2006)}]{Ou06}%
  \BibitemOpen
  \bibfield  {author} {\bibinfo {author} {\bibfnamefont {S.}~\bibnamefont
  {{Ou}}}\ and\ \bibinfo {author} {\bibfnamefont {J.~E.}\ \bibnamefont
  {{Tohline}}},\ }\href {\doibase 10.1086/507597} {\bibfield  {journal}
  {\bibinfo  {journal} {Astrophys. J.}\ }\textbf {\bibinfo {volume} {651}},\
  \bibinfo {pages} {1068} (\bibinfo {year} {2006})},\ \Eprint
  {http://arxiv.org/abs/astro-ph/0604099} {astro-ph/0604099} \BibitemShut
  {NoStop}%
\bibitem [{\citenamefont {{Corvino}}\ \emph {et~al.}(2010)\citenamefont
  {{Corvino}}, \citenamefont {{Rezzolla}}, \citenamefont {{Bernuzzi}},
  \citenamefont {{De Pietri}},\ and\ \citenamefont
  {{Giacomazzo}}}]{Corvino:2010}%
  \BibitemOpen
  \bibfield  {author} {\bibinfo {author} {\bibfnamefont {G.}~\bibnamefont
  {{Corvino}}}, \bibinfo {author} {\bibfnamefont {L.}~\bibnamefont
  {{Rezzolla}}}, \bibinfo {author} {\bibfnamefont {S.}~\bibnamefont
  {{Bernuzzi}}}, \bibinfo {author} {\bibfnamefont {R.}~\bibnamefont {{De
  Pietri}}}, \ and\ \bibinfo {author} {\bibfnamefont {B.}~\bibnamefont
  {{Giacomazzo}}},\ }\href {\doibase 10.1088/0264-9381/27/11/114104} {\bibfield
   {journal} {\bibinfo  {journal} {Class. Quantum Grav.}\ }\textbf {\bibinfo
  {volume} {27}},\ \bibinfo {pages} {114104} (\bibinfo {year} {2010})},\
  \Eprint {http://arxiv.org/abs/1001.5281} {arXiv:1001.5281 [gr-qc]}
  \BibitemShut {NoStop}%
\bibitem [{\citenamefont {{Anderson}}\ \emph
  {et~al.}(2008{\natexlab{b}})\citenamefont {{Anderson}}, \citenamefont
  {{Hirschmann}}, \citenamefont {{Lehner}}, \citenamefont {{Liebling}},
  \citenamefont {{Motl}}, \citenamefont {{Neilsen}}, \citenamefont
  {{Palenzuela}},\ and\ \citenamefont {{Tohline}}}]{Anderson2008}%
  \BibitemOpen
  \bibfield  {author} {\bibinfo {author} {\bibfnamefont {M.}~\bibnamefont
  {{Anderson}}}, \bibinfo {author} {\bibfnamefont {E.~W.}\ \bibnamefont
  {{Hirschmann}}}, \bibinfo {author} {\bibfnamefont {L.}~\bibnamefont
  {{Lehner}}}, \bibinfo {author} {\bibfnamefont {S.~L.}\ \bibnamefont
  {{Liebling}}}, \bibinfo {author} {\bibfnamefont {P.~M.}\ \bibnamefont
  {{Motl}}}, \bibinfo {author} {\bibfnamefont {D.}~\bibnamefont {{Neilsen}}},
  \bibinfo {author} {\bibfnamefont {C.}~\bibnamefont {{Palenzuela}}}, \ and\
  \bibinfo {author} {\bibfnamefont {J.~E.}\ \bibnamefont {{Tohline}}},\ }\href
  {\doibase 10.1103/PhysRevLett.100.191101} {\bibfield  {journal} {\bibinfo
  {journal} {Phys. Rev. Lett.}\ }\textbf {\bibinfo {volume} {100}},\ \bibinfo
  {eid} {191101} (\bibinfo {year} {2008}{\natexlab{b}})},\ \Eprint
  {http://arxiv.org/abs/0801.4387} {arXiv:0801.4387 [gr-qc]} \BibitemShut
  {NoStop}%
\bibitem [{\citenamefont {{Paschalidis}}\ \emph
  {et~al.}(2015{\natexlab{b}})\citenamefont {{Paschalidis}}, \citenamefont
  {{East}}, \citenamefont {{Pretorius}},\ and\ \citenamefont
  {{Shapiro}}}]{Paschalidis2015}%
  \BibitemOpen
  \bibfield  {author} {\bibinfo {author} {\bibfnamefont {V.}~\bibnamefont
  {{Paschalidis}}}, \bibinfo {author} {\bibfnamefont {W.~E.}\ \bibnamefont
  {{East}}}, \bibinfo {author} {\bibfnamefont {F.}~\bibnamefont {{Pretorius}}},
  \ and\ \bibinfo {author} {\bibfnamefont {S.~L.}\ \bibnamefont {{Shapiro}}},\
  }\href {\doibase 10.1103/PhysRevD.92.121502} {\bibfield  {journal} {\bibinfo
  {journal} {Phys. Rev. D}\ }\textbf {\bibinfo {volume} {92}},\ \bibinfo {eid}
  {121502} (\bibinfo {year} {2015}{\natexlab{b}})},\ \Eprint
  {http://arxiv.org/abs/1510.03432} {arXiv:1510.03432 [astro-ph.HE]}
  \BibitemShut {NoStop}%
\bibitem [{\citenamefont {{East}}\ \emph {et~al.}(2016)\citenamefont {{East}},
  \citenamefont {{Paschalidis}}, \citenamefont {{Pretorius}},\ and\
  \citenamefont {{Shapiro}}}]{East2016}%
  \BibitemOpen
  \bibfield  {author} {\bibinfo {author} {\bibfnamefont {W.~E.}\ \bibnamefont
  {{East}}}, \bibinfo {author} {\bibfnamefont {V.}~\bibnamefont
  {{Paschalidis}}}, \bibinfo {author} {\bibfnamefont {F.}~\bibnamefont
  {{Pretorius}}}, \ and\ \bibinfo {author} {\bibfnamefont {S.~L.}\ \bibnamefont
  {{Shapiro}}},\ }\href {\doibase 10.1103/PhysRevD.93.024011} {\bibfield
  {journal} {\bibinfo  {journal} {Phys. Rev. D}\ }\textbf {\bibinfo {volume}
  {93}},\ \bibinfo {eid} {024011} (\bibinfo {year} {2016})},\ \Eprint
  {http://arxiv.org/abs/1511.01093} {arXiv:1511.01093 [astro-ph.HE]}
  \BibitemShut {NoStop}%
\bibitem [{\citenamefont {{Radice}}\ \emph
  {et~al.}(2016{\natexlab{a}})\citenamefont {{Radice}}, \citenamefont
  {{Galeazzi}}, \citenamefont {{Lippuner}}, \citenamefont {{Roberts}},
  \citenamefont {{Ott}},\ and\ \citenamefont {{Rezzolla}}}]{Radice2016}%
  \BibitemOpen
  \bibfield  {author} {\bibinfo {author} {\bibfnamefont {D.}~\bibnamefont
  {{Radice}}}, \bibinfo {author} {\bibfnamefont {F.}~\bibnamefont
  {{Galeazzi}}}, \bibinfo {author} {\bibfnamefont {J.}~\bibnamefont
  {{Lippuner}}}, \bibinfo {author} {\bibfnamefont {L.~F.}\ \bibnamefont
  {{Roberts}}}, \bibinfo {author} {\bibfnamefont {C.~D.}\ \bibnamefont
  {{Ott}}}, \ and\ \bibinfo {author} {\bibfnamefont {L.}~\bibnamefont
  {{Rezzolla}}},\ }\href {\doibase 10.1093/mnras/stw1227} {\bibfield  {journal}
  {\bibinfo  {journal} {Mon. Not. R. Astron. Soc.}\ }\textbf {\bibinfo {volume}
  {460}},\ \bibinfo {pages} {3255} (\bibinfo {year} {2016}{\natexlab{a}})},\
  \Eprint {http://arxiv.org/abs/1601.02426} {arXiv:1601.02426 [astro-ph.HE]}
  \BibitemShut {NoStop}%
\bibitem [{\citenamefont {{Lehner}}\ \emph
  {et~al.}(2016{\natexlab{a}})\citenamefont {{Lehner}}, \citenamefont
  {{Liebling}}, \citenamefont {{Palenzuela}},\ and\ \citenamefont
  {{Motl}}}]{Lehner2016a}%
  \BibitemOpen
  \bibfield  {author} {\bibinfo {author} {\bibfnamefont {L.}~\bibnamefont
  {{Lehner}}}, \bibinfo {author} {\bibfnamefont {S.~L.}\ \bibnamefont
  {{Liebling}}}, \bibinfo {author} {\bibfnamefont {C.}~\bibnamefont
  {{Palenzuela}}}, \ and\ \bibinfo {author} {\bibfnamefont {P.}~\bibnamefont
  {{Motl}}},\ }\href@noop {} {\bibfield  {journal} {\bibinfo  {journal} {ArXiv
  e-prints}\ } (\bibinfo {year} {2016}{\natexlab{a}})},\ \Eprint
  {http://arxiv.org/abs/1605.02369} {arXiv:1605.02369 [gr-qc]} \BibitemShut
  {NoStop}%
\bibitem [{\citenamefont {{Kastaun}}\ and\ \citenamefont
  {{Galeazzi}}(2015)}]{Kastaun2014}%
  \BibitemOpen
  \bibfield  {author} {\bibinfo {author} {\bibfnamefont {W.}~\bibnamefont
  {{Kastaun}}}\ and\ \bibinfo {author} {\bibfnamefont {F.}~\bibnamefont
  {{Galeazzi}}},\ }\href {\doibase 10.1103/PhysRevD.91.064027} {\bibfield
  {journal} {\bibinfo  {journal} {Phys. Rev. D}\ }\textbf {\bibinfo {volume}
  {91}},\ \bibinfo {eid} {064027} (\bibinfo {year} {2015})},\ \Eprint
  {http://arxiv.org/abs/1411.7975} {arXiv:1411.7975 [gr-qc]} \BibitemShut
  {NoStop}%
\bibitem [{\citenamefont {{Rezzolla}}\ and\ \citenamefont
  {{Kumar}}(2015)}]{Rezzolla2014b}%
  \BibitemOpen
  \bibfield  {author} {\bibinfo {author} {\bibfnamefont {L.}~\bibnamefont
  {{Rezzolla}}}\ and\ \bibinfo {author} {\bibfnamefont {P.}~\bibnamefont
  {{Kumar}}},\ }\href {\doibase 10.1088/0004-637X/802/2/95} {\bibfield
  {journal} {\bibinfo  {journal} {Astrophys. J.}\ }\textbf {\bibinfo {volume}
  {802}},\ \bibinfo {eid} {95} (\bibinfo {year} {2015})},\ \Eprint
  {http://arxiv.org/abs/1410.8560} {arXiv:1410.8560 [astro-ph.HE]} \BibitemShut
  {NoStop}%
\bibitem [{\citenamefont {{Ravi}}\ and\ \citenamefont
  {{Lasky}}(2014)}]{Ravi2014}%
  \BibitemOpen
  \bibfield  {author} {\bibinfo {author} {\bibfnamefont {V.}~\bibnamefont
  {{Ravi}}}\ and\ \bibinfo {author} {\bibfnamefont {P.~D.}\ \bibnamefont
  {{Lasky}}},\ }\href {\doibase 10.1093/mnras/stu720} {\bibfield  {journal}
  {\bibinfo  {journal} {Mon. Not. R. Astron. Soc.}\ }\textbf {\bibinfo {volume}
  {441}},\ \bibinfo {pages} {2433} (\bibinfo {year} {2014})}\BibitemShut
  {NoStop}%
\bibitem [{\citenamefont {Alcubierre}(2008)}]{Alcubierre:2008}%
  \BibitemOpen
  \bibfield  {author} {\bibinfo {author} {\bibfnamefont {M.}~\bibnamefont
  {Alcubierre}},\ }\href {\doibase 10.1093/acprof:oso/9780199205677.001.0001}
  {\emph {\bibinfo {title} {Introduction to $3+1$ {N}umerical {R}elativity}}}\
  (\bibinfo  {publisher} {Oxford University Press},\ \bibinfo {address}
  {Oxford, UK},\ \bibinfo {year} {2008})\BibitemShut {NoStop}%
\bibitem [{\citenamefont {Bona}\ \emph {et~al.}(2009)\citenamefont {Bona},
  \citenamefont {Palenzuela-Luque},\ and\ \citenamefont
  {Bona-Casas}}]{Bona2009}%
  \BibitemOpen
  \bibfield  {author} {\bibinfo {author} {\bibfnamefont {C.}~\bibnamefont
  {Bona}}, \bibinfo {author} {\bibfnamefont {C.}~\bibnamefont
  {Palenzuela-Luque}}, \ and\ \bibinfo {author} {\bibfnamefont
  {C.}~\bibnamefont {Bona-Casas}},\ }\href
  {http://books.google.co.uk/books?id=KgPGHaCUaAYC} {\emph {\bibinfo {title}
  {Elements of Numerical Relativity and Relativistic Hydrodynamics: From
  Einstein's Equations to Astrophysical Simulations}}},\ Lecture Notes in
  Physics\ (\bibinfo  {publisher} {Springer},\ \bibinfo {address} {Berlin
  Heidelberg},\ \bibinfo {year} {2009})\BibitemShut {NoStop}%
\bibitem [{\citenamefont {{Baumgarte}}\ and\ \citenamefont
  {{Shapiro}}(2010)}]{Baumgarte2010}%
  \BibitemOpen
  \bibfield  {author} {\bibinfo {author} {\bibfnamefont {T.~W.}\ \bibnamefont
  {{Baumgarte}}}\ and\ \bibinfo {author} {\bibfnamefont {S.~L.}\ \bibnamefont
  {{Shapiro}}},\ }\href {\doibase 10.1017/cbo9781139193344} {\emph {\bibinfo
  {title} {Numerical Relativity: Solving Einstein's Equations on the Computer
  by Thomas W.~Baumgarte and Stuart L.~Shapiro.~Cambridge University Press,
  2010.~ISBN: 9780521514071}}}\ (\bibinfo  {publisher} {Cambridge University
  Press},\ \bibinfo {address} {Cambridge, UK},\ \bibinfo {year}
  {2010})\BibitemShut {NoStop}%
\bibitem [{\citenamefont {{Gourgoulhon}}(2012)}]{Gourgoulhon2012}%
  \BibitemOpen
  \bibfield  {author} {\bibinfo {author} {\bibfnamefont {E.}~\bibnamefont
  {{Gourgoulhon}}},\ }\href {\doibase 10.1007/978-3-642-24525-1} {\emph
  {\bibinfo {title} {Lecture Notes in Physics, Berlin Springer Verlag}}},\
  \bibinfo {series} {Lecture Notes in Physics, Berlin Springer Verlag}, Vol.\
  \bibinfo {volume} {846}\ (\bibinfo {year} {2012})\BibitemShut {NoStop}%
\bibitem [{\citenamefont {{Shibata}}(2016)}]{Shibata_book:2016}%
  \BibitemOpen
  \bibfield  {author} {\bibinfo {author} {\bibfnamefont {M.}~\bibnamefont
  {{Shibata}}},\ }\href {\doibase 10.1142/9692} {\emph {\bibinfo {title}
  {Numerical Relativity}}}\ (\bibinfo  {publisher} {World Scientific},\
  \bibinfo {address} {Singapore},\ \bibinfo {year} {2016})\BibitemShut
  {NoStop}%
\bibitem [{\citenamefont {Arnowitt}\ \emph {et~al.}(1959)\citenamefont
  {Arnowitt}, \citenamefont {Deser},\ and\ \citenamefont
  {Misner}}]{Arnowitt59}%
  \BibitemOpen
  \bibfield  {author} {\bibinfo {author} {\bibfnamefont {R.}~\bibnamefont
  {Arnowitt}}, \bibinfo {author} {\bibfnamefont {S.}~\bibnamefont {Deser}}, \
  and\ \bibinfo {author} {\bibfnamefont {C.~W.}\ \bibnamefont {Misner}},\
  }\href {\doibase 10.1103/physrev.116.1322} {\bibfield  {journal} {\bibinfo
  {journal} {Phys. Rev.}\ }\textbf {\bibinfo {volume} {116}},\ \bibinfo {pages}
  {1322} (\bibinfo {year} {1959})}\BibitemShut {NoStop}%
\bibitem [{\citenamefont {{Arnowitt}}\ \emph {et~al.}(2008)\citenamefont
  {{Arnowitt}}, \citenamefont {{Deser}},\ and\ \citenamefont
  {{Misner}}}]{Arnowitt62}%
  \BibitemOpen
  \bibfield  {author} {\bibinfo {author} {\bibfnamefont {R.}~\bibnamefont
  {{Arnowitt}}}, \bibinfo {author} {\bibfnamefont {S.}~\bibnamefont {{Deser}}},
  \ and\ \bibinfo {author} {\bibfnamefont {C.~W.}\ \bibnamefont {{Misner}}},\
  }\href {\doibase 10.1007/s10714-008-0661-1} {\bibfield  {journal} {\bibinfo
  {journal} {General Relativity and Gravitation}\ }\textbf {\bibinfo {volume}
  {40}},\ \bibinfo {pages} {1997} (\bibinfo {year} {2008})},\ \Eprint
  {http://arxiv.org/abs/gr-qc/0405109} {gr-qc/0405109} \BibitemShut {NoStop}%
\bibitem [{\citenamefont {{Nakamura}}\ \emph {et~al.}(1987)\citenamefont
  {{Nakamura}}, \citenamefont {{Oohara}},\ and\ \citenamefont
  {{Kojima}}}]{Nakamura87}%
  \BibitemOpen
  \bibfield  {author} {\bibinfo {author} {\bibfnamefont {T.}~\bibnamefont
  {{Nakamura}}}, \bibinfo {author} {\bibfnamefont {K.}~\bibnamefont
  {{Oohara}}}, \ and\ \bibinfo {author} {\bibfnamefont {Y.}~\bibnamefont
  {{Kojima}}},\ }\href {\doibase 10.1143/PTPS.90.1} {\bibfield  {journal}
  {\bibinfo  {journal} {Progress of Theoretical Physics Supplement}\ }\textbf
  {\bibinfo {volume} {90}},\ \bibinfo {pages} {1} (\bibinfo {year}
  {1987})}\BibitemShut {NoStop}%
\bibitem [{\citenamefont {{Shibata}}\ and\ \citenamefont
  {{Nakamura}}(1995)}]{Shibata95}%
  \BibitemOpen
  \bibfield  {author} {\bibinfo {author} {\bibfnamefont {M.}~\bibnamefont
  {{Shibata}}}\ and\ \bibinfo {author} {\bibfnamefont {T.}~\bibnamefont
  {{Nakamura}}},\ }\href {\doibase 10.1103/PhysRevD.52.5428} {\bibfield
  {journal} {\bibinfo  {journal} {Phys. Rev. D}\ }\textbf {\bibinfo {volume}
  {52}},\ \bibinfo {pages} {5428} (\bibinfo {year} {1995})}\BibitemShut
  {NoStop}%
\bibitem [{\citenamefont {{Baumgarte}}\ and\ \citenamefont
  {{Shapiro}}(1999)}]{Baumgarte99}%
  \BibitemOpen
  \bibfield  {author} {\bibinfo {author} {\bibfnamefont {T.~W.}\ \bibnamefont
  {{Baumgarte}}}\ and\ \bibinfo {author} {\bibfnamefont {S.~L.}\ \bibnamefont
  {{Shapiro}}},\ }\href {\doibase 10.1103/PhysRevD.59.024007} {\bibfield
  {journal} {\bibinfo  {journal} {Phys. Rev. D}\ }\textbf {\bibinfo {volume}
  {59}},\ \bibinfo {eid} {024007} (\bibinfo {year} {1999})},\ \Eprint
  {http://arxiv.org/abs/gr-qc/9810065} {gr-qc/9810065} \BibitemShut {NoStop}%
\bibitem [{\citenamefont {{Alcubierre}}\ \emph {et~al.}(2000)\citenamefont
  {{Alcubierre}}, \citenamefont {{Br{\"u}gmann}}, \citenamefont {{Dramlitsch}},
  \citenamefont {{Font}}, \citenamefont {{Papadopoulos}}, \citenamefont
  {{Seidel}}, \citenamefont {{Stergioulas}},\ and\ \citenamefont
  {{Takahashi}}}]{Alcubierre99d}%
  \BibitemOpen
  \bibfield  {author} {\bibinfo {author} {\bibfnamefont {M.}~\bibnamefont
  {{Alcubierre}}}, \bibinfo {author} {\bibfnamefont {B.}~\bibnamefont
  {{Br{\"u}gmann}}}, \bibinfo {author} {\bibfnamefont {T.}~\bibnamefont
  {{Dramlitsch}}}, \bibinfo {author} {\bibfnamefont {J.~A.}\ \bibnamefont
  {{Font}}}, \bibinfo {author} {\bibfnamefont {P.}~\bibnamefont
  {{Papadopoulos}}}, \bibinfo {author} {\bibfnamefont {E.}~\bibnamefont
  {{Seidel}}}, \bibinfo {author} {\bibfnamefont {N.}~\bibnamefont
  {{Stergioulas}}}, \ and\ \bibinfo {author} {\bibfnamefont {R.}~\bibnamefont
  {{Takahashi}}},\ }\href {\doibase 10.1103/PhysRevD.62.044034} {\bibfield
  {journal} {\bibinfo  {journal} {Phys. Rev. D}\ }\textbf {\bibinfo {volume}
  {62}},\ \bibinfo {eid} {044034} (\bibinfo {year} {2000})},\ \Eprint
  {http://arxiv.org/abs/gr-qc/0003071} {gr-qc/0003071} \BibitemShut {NoStop}%
\bibitem [{\citenamefont {{Alcubierre}}\ \emph {et~al.}(2001)\citenamefont
  {{Alcubierre}}, \citenamefont {{Br{\"u}gmann}}, \citenamefont {{Pollney}},
  \citenamefont {{Seidel}},\ and\ \citenamefont {{Takahashi}}}]{Alcubierre01a}%
  \BibitemOpen
  \bibfield  {author} {\bibinfo {author} {\bibfnamefont {M.}~\bibnamefont
  {{Alcubierre}}}, \bibinfo {author} {\bibfnamefont {B.}~\bibnamefont
  {{Br{\"u}gmann}}}, \bibinfo {author} {\bibfnamefont {D.}~\bibnamefont
  {{Pollney}}}, \bibinfo {author} {\bibfnamefont {E.}~\bibnamefont {{Seidel}}},
  \ and\ \bibinfo {author} {\bibfnamefont {R.}~\bibnamefont {{Takahashi}}},\
  }\href {\doibase 10.1103/PhysRevD.64.061501} {\bibfield  {journal} {\bibinfo
  {journal} {Phys. Rev. D}\ }\textbf {\bibinfo {volume} {64}},\ \bibinfo {eid}
  {061501} (\bibinfo {year} {2001})},\ \Eprint
  {http://arxiv.org/abs/gr-qc/0104020} {gr-qc/0104020} \BibitemShut {NoStop}%
\bibitem [{\citenamefont {{Alcubierre}}\ \emph {et~al.}(2003)\citenamefont
  {{Alcubierre}}, \citenamefont {{Br{\"u}gmann}}, \citenamefont {{Diener}},
  \citenamefont {{Koppitz}}, \citenamefont {{Pollney}}, \citenamefont
  {{Seidel}},\ and\ \citenamefont {{Takahashi}}}]{Alcubierre02a}%
  \BibitemOpen
  \bibfield  {author} {\bibinfo {author} {\bibfnamefont {M.}~\bibnamefont
  {{Alcubierre}}}, \bibinfo {author} {\bibfnamefont {B.}~\bibnamefont
  {{Br{\"u}gmann}}}, \bibinfo {author} {\bibfnamefont {P.}~\bibnamefont
  {{Diener}}}, \bibinfo {author} {\bibfnamefont {M.}~\bibnamefont {{Koppitz}}},
  \bibinfo {author} {\bibfnamefont {D.}~\bibnamefont {{Pollney}}}, \bibinfo
  {author} {\bibfnamefont {E.}~\bibnamefont {{Seidel}}}, \ and\ \bibinfo
  {author} {\bibfnamefont {R.}~\bibnamefont {{Takahashi}}},\ }\href {\doibase
  10.1103/PhysRevD.67.084023} {\bibfield  {journal} {\bibinfo  {journal} {Phys.
  Rev. D}\ }\textbf {\bibinfo {volume} {67}},\ \bibinfo {eid} {084023}
  (\bibinfo {year} {2003})},\ \Eprint {http://arxiv.org/abs/gr-qc/0206072}
  {gr-qc/0206072} \BibitemShut {NoStop}%
\bibitem [{\citenamefont {Mueller}\ and\ \citenamefont
  {Br{\"u}gmann}(2010)}]{Mueller09}%
  \BibitemOpen
  \bibfield  {author} {\bibinfo {author} {\bibfnamefont {D.}~\bibnamefont
  {Mueller}}\ and\ \bibinfo {author} {\bibfnamefont {B.}~\bibnamefont
  {Br{\"u}gmann}},\ }\href {\doibase 10.1088/0264-9381/27/11/114008} {\bibfield
   {journal} {\bibinfo  {journal} {Class. Quantum Grav.}\ }\textbf {\bibinfo
  {volume} {27}},\ \bibinfo {pages} {114008} (\bibinfo {year} {2010})},\
  \Eprint {http://arxiv.org/abs/0912.3125} {arXiv:0912.3125 [gr-qc]}
  \BibitemShut {NoStop}%
\bibitem [{\citenamefont {{Alic}}\ \emph {et~al.}(2010)\citenamefont {{Alic}},
  \citenamefont {{Rezzolla}}, \citenamefont {{Hinder}},\ and\ \citenamefont
  {{M{\"o}sta}}}]{Alic:2010}%
  \BibitemOpen
  \bibfield  {author} {\bibinfo {author} {\bibfnamefont {D.}~\bibnamefont
  {{Alic}}}, \bibinfo {author} {\bibfnamefont {L.}~\bibnamefont {{Rezzolla}}},
  \bibinfo {author} {\bibfnamefont {I.}~\bibnamefont {{Hinder}}}, \ and\
  \bibinfo {author} {\bibfnamefont {P.}~\bibnamefont {{M{\"o}sta}}},\ }\href
  {\doibase 10.1088/0264-9381/27/24/245023} {\bibfield  {journal} {\bibinfo
  {journal} {Class. Quantum Grav.}\ }\textbf {\bibinfo {volume} {27}},\
  \bibinfo {pages} {245023} (\bibinfo {year} {2010})},\ \Eprint
  {http://arxiv.org/abs/1008.2212} {arXiv:1008.2212 [gr-qc]} \BibitemShut
  {NoStop}%
\bibitem [{\citenamefont {{Alic}}\ \emph {et~al.}(2013)\citenamefont {{Alic}},
  \citenamefont {{Kastaun}},\ and\ \citenamefont {{Rezzolla}}}]{Alic2013}%
  \BibitemOpen
  \bibfield  {author} {\bibinfo {author} {\bibfnamefont {D.}~\bibnamefont
  {{Alic}}}, \bibinfo {author} {\bibfnamefont {W.}~\bibnamefont {{Kastaun}}}, \
  and\ \bibinfo {author} {\bibfnamefont {L.}~\bibnamefont {{Rezzolla}}},\
  }\href {\doibase 10.1103/PhysRevD.88.064049} {\bibfield  {journal} {\bibinfo
  {journal} {Phys. Rev. D}\ }\textbf {\bibinfo {volume} {88}},\ \bibinfo {eid}
  {064049} (\bibinfo {year} {2013})},\ \Eprint {http://arxiv.org/abs/1307.7391}
  {arXiv:1307.7391 [gr-qc]} \BibitemShut {NoStop}%
\bibitem [{\citenamefont {Bernuzzi}\ and\ \citenamefont
  {Hilditch}(2010)}]{Bernuzzi:2009ex}%
  \BibitemOpen
  \bibfield  {author} {\bibinfo {author} {\bibfnamefont {S.}~\bibnamefont
  {Bernuzzi}}\ and\ \bibinfo {author} {\bibfnamefont {D.}~\bibnamefont
  {Hilditch}},\ }\href {\doibase 10.1103/PhysRevD.81.084003} {\bibfield
  {journal} {\bibinfo  {journal} {Phys. Rev. D}\ }\textbf {\bibinfo {volume}
  {81}},\ \bibinfo {pages} {084003} (\bibinfo {year} {2010})},\ \Eprint
  {http://arxiv.org/abs/0912.2920} {arXiv:0912.2920 [gr-qc]} \BibitemShut
  {NoStop}%
\bibitem [{\citenamefont {Weyhausen}\ \emph {et~al.}(2012)\citenamefont
  {Weyhausen}, \citenamefont {Bernuzzi},\ and\ \citenamefont
  {Hilditch}}]{Weyhausen:2011cg}%
  \BibitemOpen
  \bibfield  {author} {\bibinfo {author} {\bibfnamefont {A.}~\bibnamefont
  {Weyhausen}}, \bibinfo {author} {\bibfnamefont {S.}~\bibnamefont {Bernuzzi}},
  \ and\ \bibinfo {author} {\bibfnamefont {D.}~\bibnamefont {Hilditch}},\
  }\href {\doibase 10.1103/PhysRevD.85.024038} {\bibfield  {journal} {\bibinfo
  {journal} {Phys. Rev.}\ }\textbf {\bibinfo {volume} {D85}},\ \bibinfo {pages}
  {024038} (\bibinfo {year} {2012})},\ \Eprint {http://arxiv.org/abs/1107.5539}
  {arXiv:1107.5539 [gr-qc]} \BibitemShut {NoStop}%
\bibitem [{\citenamefont {{Cao}}\ and\ \citenamefont
  {{Hilditch}}(2012)}]{Cao:2012}%
  \BibitemOpen
  \bibfield  {author} {\bibinfo {author} {\bibfnamefont {Z.}~\bibnamefont
  {{Cao}}}\ and\ \bibinfo {author} {\bibfnamefont {D.}~\bibnamefont
  {{Hilditch}}},\ }\href {\doibase 10.1103/PhysRevD.85.124032} {\bibfield
  {journal} {\bibinfo  {journal} {Phys. Rev. D}\ }\textbf {\bibinfo {volume}
  {85}},\ \bibinfo {eid} {124032} (\bibinfo {year} {2012})},\ \Eprint
  {http://arxiv.org/abs/1111.2177} {arXiv:1111.2177 [gr-qc]} \BibitemShut
  {NoStop}%
\bibitem [{\citenamefont {{Hilditch}}\ \emph {et~al.}(2013)\citenamefont
  {{Hilditch}}, \citenamefont {{Bernuzzi}}, \citenamefont {{Thierfelder}},
  \citenamefont {{Cao}}, \citenamefont {{Tichy}},\ and\ \citenamefont
  {{Br{\"u}gmann}}}]{Hilditch2012}%
  \BibitemOpen
  \bibfield  {author} {\bibinfo {author} {\bibfnamefont {D.}~\bibnamefont
  {{Hilditch}}}, \bibinfo {author} {\bibfnamefont {S.}~\bibnamefont
  {{Bernuzzi}}}, \bibinfo {author} {\bibfnamefont {M.}~\bibnamefont
  {{Thierfelder}}}, \bibinfo {author} {\bibfnamefont {Z.}~\bibnamefont
  {{Cao}}}, \bibinfo {author} {\bibfnamefont {W.}~\bibnamefont {{Tichy}}}, \
  and\ \bibinfo {author} {\bibfnamefont {B.}~\bibnamefont {{Br{\"u}gmann}}},\
  }\href {\doibase 10.1103/PhysRevD.88.084057} {\bibfield  {journal} {\bibinfo
  {journal} {Phys. Rev. D}\ }\textbf {\bibinfo {volume} {88}},\ \bibinfo {eid}
  {084057} (\bibinfo {year} {2013})},\ \Eprint {http://arxiv.org/abs/1212.2901}
  {arXiv:1212.2901 [gr-qc]} \BibitemShut {NoStop}%
\bibitem [{\citenamefont {{Bona}}\ \emph {et~al.}(2003)\citenamefont {{Bona}},
  \citenamefont {{Ledvinka}}, \citenamefont {{Palenzuela}},\ and\ \citenamefont
  {{{\v Z}{\'a}{\v c}ek}}}]{Bona:2003fj}%
  \BibitemOpen
  \bibfield  {author} {\bibinfo {author} {\bibfnamefont {C.}~\bibnamefont
  {{Bona}}}, \bibinfo {author} {\bibfnamefont {T.}~\bibnamefont {{Ledvinka}}},
  \bibinfo {author} {\bibfnamefont {C.}~\bibnamefont {{Palenzuela}}}, \ and\
  \bibinfo {author} {\bibfnamefont {M.}~\bibnamefont {{{\v Z}{\'a}{\v c}ek}}},\
  }\href {\doibase 10.1103/PhysRevD.67.104005} {\bibfield  {journal} {\bibinfo
  {journal} {Phys. Rev. D}\ }\textbf {\bibinfo {volume} {67}},\ \bibinfo {eid}
  {104005} (\bibinfo {year} {2003})},\ \Eprint
  {http://arxiv.org/abs/gr-qc/0302083} {gr-qc/0302083} \BibitemShut {NoStop}%
\bibitem [{\citenamefont {Bona}\ \emph {et~al.}(2010)\citenamefont {Bona},
  \citenamefont {Bona-Casas},\ and\ \citenamefont {Palenzuela}}]{Bona:2010is}%
  \BibitemOpen
  \bibfield  {author} {\bibinfo {author} {\bibfnamefont {C.}~\bibnamefont
  {Bona}}, \bibinfo {author} {\bibfnamefont {C.}~\bibnamefont {Bona-Casas}}, \
  and\ \bibinfo {author} {\bibfnamefont {C.}~\bibnamefont {Palenzuela}},\
  }\href {\doibase 10.1103/PhysRevD.82.124010} {\bibfield  {journal} {\bibinfo
  {journal} {Phys. Rev. D}\ }\textbf {\bibinfo {volume} {82}},\ \bibinfo
  {pages} {124010} (\bibinfo {year} {2010})},\ \Eprint
  {http://arxiv.org/abs/1008.0747} {arXiv:1008.0747 [gr-qc]} \BibitemShut
  {NoStop}%
\bibitem [{\citenamefont {Gundlach}\ \emph {et~al.}(2005)\citenamefont
  {Gundlach}, \citenamefont {Martin-Garcia}, \citenamefont {Calabrese},\ and\
  \citenamefont {Hinder}}]{Gundlach2005:constraint-damping}%
  \BibitemOpen
  \bibfield  {author} {\bibinfo {author} {\bibfnamefont {C.}~\bibnamefont
  {Gundlach}}, \bibinfo {author} {\bibfnamefont {J.~M.}\ \bibnamefont
  {Martin-Garcia}}, \bibinfo {author} {\bibfnamefont {G.}~\bibnamefont
  {Calabrese}}, \ and\ \bibinfo {author} {\bibfnamefont {I.}~\bibnamefont
  {Hinder}},\ }\href {\doibase 10.1088/0264-9381/22/17/025} {\bibfield
  {journal} {\bibinfo  {journal} {Class. Quantum Grav.}\ }\textbf {\bibinfo
  {volume} {22}},\ \bibinfo {pages} {3767} (\bibinfo {year} {2005})},\ \Eprint
  {http://arxiv.org/abs/gr-qc/0504114} {gr-qc/0504114} \BibitemShut {NoStop}%
\bibitem [{\citenamefont {{Bona}}\ \emph {et~al.}(2004)\citenamefont {{Bona}},
  \citenamefont {{Ledvinka}}, \citenamefont {{Palenzuela}},\ and\ \citenamefont
  {{{\v Z}{\'a}{\v c}ek}}}]{Bona:2003qn}%
  \BibitemOpen
  \bibfield  {author} {\bibinfo {author} {\bibfnamefont {C.}~\bibnamefont
  {{Bona}}}, \bibinfo {author} {\bibfnamefont {T.}~\bibnamefont {{Ledvinka}}},
  \bibinfo {author} {\bibfnamefont {C.}~\bibnamefont {{Palenzuela}}}, \ and\
  \bibinfo {author} {\bibfnamefont {M.}~\bibnamefont {{{\v Z}{\'a}{\v c}ek}}},\
  }\href {\doibase 10.1103/PhysRevD.69.064036} {\bibfield  {journal} {\bibinfo
  {journal} {Phys. Rev. D}\ }\textbf {\bibinfo {volume} {69}},\ \bibinfo {eid}
  {064036} (\bibinfo {year} {2004})},\ \Eprint
  {http://arxiv.org/abs/gr-qc/0307067} {gr-qc/0307067} \BibitemShut {NoStop}%
\bibitem [{\citenamefont {{Bona}}\ and\ \citenamefont
  {{Palenzuela}}(2004)}]{Bona:2004yp}%
  \BibitemOpen
  \bibfield  {author} {\bibinfo {author} {\bibfnamefont {C.}~\bibnamefont
  {{Bona}}}\ and\ \bibinfo {author} {\bibfnamefont {C.}~\bibnamefont
  {{Palenzuela}}},\ }\href {\doibase 10.1103/PhysRevD.69.104003} {\bibfield
  {journal} {\bibinfo  {journal} {Phys. Rev. D}\ }\textbf {\bibinfo {volume}
  {69}},\ \bibinfo {eid} {104003} (\bibinfo {year} {2004})},\ \Eprint
  {http://arxiv.org/abs/gr-qc/0401019} {gr-qc/0401019} \BibitemShut {NoStop}%
\bibitem [{\citenamefont {{Alic}}\ \emph {et~al.}(2009)\citenamefont {{Alic}},
  \citenamefont {{Bona}},\ and\ \citenamefont {{Bona-Casas}}}]{Alic:2009}%
  \BibitemOpen
  \bibfield  {author} {\bibinfo {author} {\bibfnamefont {D.}~\bibnamefont
  {{Alic}}}, \bibinfo {author} {\bibfnamefont {C.}~\bibnamefont {{Bona}}}, \
  and\ \bibinfo {author} {\bibfnamefont {C.}~\bibnamefont {{Bona-Casas}}},\
  }\href {\doibase 10.1103/PhysRevD.79.044026} {\bibfield  {journal} {\bibinfo
  {journal} {Phys. Rev. D}\ }\textbf {\bibinfo {volume} {79}},\ \bibinfo
  {pages} {044026} (\bibinfo {year} {2009})},\ \Eprint
  {http://arxiv.org/abs/0811.1691} {arXiv:0811.1691 [gr-qc]} \BibitemShut
  {NoStop}%
\bibitem [{\citenamefont {{Pretorius}}(2005)}]{Pretorius:2004jg}%
  \BibitemOpen
  \bibfield  {author} {\bibinfo {author} {\bibfnamefont {F.}~\bibnamefont
  {{Pretorius}}},\ }\href {\doibase 10.1088/0264-9381/22/2/014} {\bibfield
  {journal} {\bibinfo  {journal} {Class. Quantum Grav.}\ }\textbf {\bibinfo
  {volume} {22}},\ \bibinfo {pages} {425} (\bibinfo {year} {2005})},\ \Eprint
  {http://arxiv.org/abs/gr-qc/0407110} {gr-qc/0407110} \BibitemShut {NoStop}%
\bibitem [{\citenamefont {{Garfinkle}}(2002)}]{Garfinkle02}%
  \BibitemOpen
  \bibfield  {author} {\bibinfo {author} {\bibfnamefont {D.}~\bibnamefont
  {{Garfinkle}}},\ }\href {\doibase 10.1103/PhysRevD.65.044029} {\bibfield
  {journal} {\bibinfo  {journal} {Phys. Rev. D}\ }\textbf {\bibinfo {volume}
  {65}},\ \bibinfo {eid} {044029} (\bibinfo {year} {2002})},\ \Eprint
  {http://arxiv.org/abs/gr-qc/0110013} {gr-qc/0110013} \BibitemShut {NoStop}%
\bibitem [{\citenamefont {{Lindblom}}\ \emph {et~al.}(2006)\citenamefont
  {{Lindblom}}, \citenamefont {{Scheel}}, \citenamefont {{Kidder}},
  \citenamefont {{Owen}},\ and\ \citenamefont {{Rinne}}}]{Lindblom:2005gh}%
  \BibitemOpen
  \bibfield  {author} {\bibinfo {author} {\bibfnamefont {L.}~\bibnamefont
  {{Lindblom}}}, \bibinfo {author} {\bibfnamefont {M.~A.}\ \bibnamefont
  {{Scheel}}}, \bibinfo {author} {\bibfnamefont {L.~E.}\ \bibnamefont
  {{Kidder}}}, \bibinfo {author} {\bibfnamefont {R.}~\bibnamefont {{Owen}}}, \
  and\ \bibinfo {author} {\bibfnamefont {O.}~\bibnamefont {{Rinne}}},\ }\href
  {\doibase 10.1088/0264-9381/23/16/S09} {\bibfield  {journal} {\bibinfo
  {journal} {Class. Quantum Grav.}\ }\textbf {\bibinfo {volume} {23}},\
  \bibinfo {pages} {447} (\bibinfo {year} {2006})},\ \Eprint
  {http://arxiv.org/abs/gr-qc/0512093} {gr-qc/0512093} \BibitemShut {NoStop}%
\bibitem [{\citenamefont {Szil{\'a}gyi}\ \emph {et~al.}(2007)\citenamefont
  {Szil{\'a}gyi}, \citenamefont {Pollney}, \citenamefont {Rezzolla},
  \citenamefont {Thornburg},\ and\ \citenamefont {Winicour}}]{Szilagyi:2006qy}%
  \BibitemOpen
  \bibfield  {author} {\bibinfo {author} {\bibfnamefont {B.}~\bibnamefont
  {Szil{\'a}gyi}}, \bibinfo {author} {\bibfnamefont {D.}~\bibnamefont
  {Pollney}}, \bibinfo {author} {\bibfnamefont {L.}~\bibnamefont {Rezzolla}},
  \bibinfo {author} {\bibfnamefont {J.}~\bibnamefont {Thornburg}}, \ and\
  \bibinfo {author} {\bibfnamefont {J.}~\bibnamefont {Winicour}},\ }\href@noop
  {} {\bibfield  {journal} {\bibinfo  {journal} {Class. Quantum Grav.}\
  }\textbf {\bibinfo {volume} {24}},\ \bibinfo {pages} {S275} (\bibinfo {year}
  {2007})},\ \Eprint {http://arxiv.org/abs/gr-qc/0612150} {gr-qc/0612150}
  \BibitemShut {NoStop}%
\bibitem [{\citenamefont {{Sorkin}}\ and\ \citenamefont
  {{Choptuik}}(2010)}]{Sorkin2010}%
  \BibitemOpen
  \bibfield  {author} {\bibinfo {author} {\bibfnamefont {E.}~\bibnamefont
  {{Sorkin}}}\ and\ \bibinfo {author} {\bibfnamefont {M.~W.}\ \bibnamefont
  {{Choptuik}}},\ }\href {\doibase 10.1007/s10714-009-0905-8} {\bibfield
  {journal} {\bibinfo  {journal} {General Relativity and Gravitation}\ }\textbf
  {\bibinfo {volume} {42}},\ \bibinfo {pages} {1239} (\bibinfo {year}
  {2010})},\ \Eprint {http://arxiv.org/abs/0908.2500} {arXiv:0908.2500 [gr-qc]}
  \BibitemShut {NoStop}%
\bibitem [{\citenamefont {{Brown}}(2011)}]{Brown2011}%
  \BibitemOpen
  \bibfield  {author} {\bibinfo {author} {\bibfnamefont {J.~D.}\ \bibnamefont
  {{Brown}}},\ }\href {\doibase 10.1103/PhysRevD.84.124012} {\bibfield
  {journal} {\bibinfo  {journal} {Phys. Rev. D}\ }\textbf {\bibinfo {volume}
  {84}},\ \bibinfo {eid} {124012} (\bibinfo {year} {2011})},\ \Eprint
  {http://arxiv.org/abs/1109.1707} {arXiv:1109.1707 [gr-qc]} \BibitemShut
  {NoStop}%
\bibitem [{\citenamefont {{East}}\ \emph
  {et~al.}(2012{\natexlab{a}})\citenamefont {{East}}, \citenamefont
  {{Pretorius}},\ and\ \citenamefont {{Stephens}}}]{East2012b}%
  \BibitemOpen
  \bibfield  {author} {\bibinfo {author} {\bibfnamefont {W.~E.}\ \bibnamefont
  {{East}}}, \bibinfo {author} {\bibfnamefont {F.}~\bibnamefont {{Pretorius}}},
  \ and\ \bibinfo {author} {\bibfnamefont {B.~C.}\ \bibnamefont {{Stephens}}},\
  }\href {\doibase 10.1103/PhysRevD.85.124010} {\bibfield  {journal} {\bibinfo
  {journal} {Phys. Rev. D}\ }\textbf {\bibinfo {volume} {85}},\ \bibinfo {eid}
  {124010} (\bibinfo {year} {2012}{\natexlab{a}})},\ \Eprint
  {http://arxiv.org/abs/1112.3094} {arXiv:1112.3094 [gr-qc]} \BibitemShut
  {NoStop}%
\bibitem [{\citenamefont {{Haas}}\ \emph {et~al.}(2016)\citenamefont {{Haas}},
  \citenamefont {{Ott}}, \citenamefont {{Szilagyi}}, \citenamefont {{Kaplan}},
  \citenamefont {{Lippuner}}, \citenamefont {{Scheel}}, \citenamefont
  {{Barkett}}, \citenamefont {{Muhlberger}}, \citenamefont {{Dietrich}},
  \citenamefont {{Duez}}, \citenamefont {{Foucart}}, \citenamefont
  {{Pfeiffer}}, \citenamefont {{Kidder}},\ and\ \citenamefont
  {{Teukolsky}}}]{Haas2016}%
  \BibitemOpen
  \bibfield  {author} {\bibinfo {author} {\bibfnamefont {R.}~\bibnamefont
  {{Haas}}}, \bibinfo {author} {\bibfnamefont {C.~D.}\ \bibnamefont {{Ott}}},
  \bibinfo {author} {\bibfnamefont {B.}~\bibnamefont {{Szilagyi}}}, \bibinfo
  {author} {\bibfnamefont {J.~D.}\ \bibnamefont {{Kaplan}}}, \bibinfo {author}
  {\bibfnamefont {J.}~\bibnamefont {{Lippuner}}}, \bibinfo {author}
  {\bibfnamefont {M.~A.}\ \bibnamefont {{Scheel}}}, \bibinfo {author}
  {\bibfnamefont {K.}~\bibnamefont {{Barkett}}}, \bibinfo {author}
  {\bibfnamefont {C.~D.}\ \bibnamefont {{Muhlberger}}}, \bibinfo {author}
  {\bibfnamefont {T.}~\bibnamefont {{Dietrich}}}, \bibinfo {author}
  {\bibfnamefont {M.~D.}\ \bibnamefont {{Duez}}}, \bibinfo {author}
  {\bibfnamefont {F.}~\bibnamefont {{Foucart}}}, \bibinfo {author}
  {\bibfnamefont {H.~P.}\ \bibnamefont {{Pfeiffer}}}, \bibinfo {author}
  {\bibfnamefont {L.~E.}\ \bibnamefont {{Kidder}}}, \ and\ \bibinfo {author}
  {\bibfnamefont {S.~A.}\ \bibnamefont {{Teukolsky}}},\ }\href@noop {}
  {\bibfield  {journal} {\bibinfo  {journal} {ArXiv e-prints}\ } (\bibinfo
  {year} {2016})},\ \Eprint {http://arxiv.org/abs/1604.00782} {arXiv:1604.00782
  [gr-qc]} \BibitemShut {NoStop}%
\bibitem [{\citenamefont {{Isenberg}}(2008)}]{Isenberg08}%
  \BibitemOpen
  \bibfield  {author} {\bibinfo {author} {\bibfnamefont {J.~A.}\ \bibnamefont
  {{Isenberg}}},\ }\href {\doibase 10.1142/S0218271808011997} {\bibfield
  {journal} {\bibinfo  {journal} {International Journal of Modern Physics D}\
  }\textbf {\bibinfo {volume} {17}},\ \bibinfo {pages} {265} (\bibinfo {year}
  {2008})},\ \Eprint {http://arxiv.org/abs/arXiv:gr-qc/0702113}
  {arXiv:gr-qc/0702113} \BibitemShut {NoStop}%
\bibitem [{\citenamefont {{Wilson}}\ and\ \citenamefont
  {{Mathews}}(1989)}]{Wilson89}%
  \BibitemOpen
  \bibfield  {author} {\bibinfo {author} {\bibfnamefont {J.~R.}\ \bibnamefont
  {{Wilson}}}\ and\ \bibinfo {author} {\bibfnamefont {G.~J.}\ \bibnamefont
  {{Mathews}}},\ }in\ \href@noop {} {\emph {\bibinfo {booktitle} {Frontiers in
  Numerical Relativity}}},\ \bibinfo {editor} {edited by\ \bibinfo {editor}
  {\bibfnamefont {C.~R.}\ \bibnamefont {{Evans}}}, \bibinfo {editor}
  {\bibfnamefont {L.~S.}\ \bibnamefont {{Finn}}}, \ and\ \bibinfo {editor}
  {\bibfnamefont {D.~W.}\ \bibnamefont {{Hobill}}}}\ (\bibinfo {year} {1989})\
  pp.\ \bibinfo {pages} {306--314}\BibitemShut {NoStop}%
\bibitem [{\citenamefont {Mart{\'\i}}\ \emph {et~al.}(1991)\citenamefont
  {Mart{\'\i}}, \citenamefont {Ib{\'a}{\~n}ez},\ and\ \citenamefont
  {Miralles}}]{Marti91}%
  \BibitemOpen
  \bibfield  {author} {\bibinfo {author} {\bibfnamefont {J.~M.}\ \bibnamefont
  {Mart{\'\i}}}, \bibinfo {author} {\bibfnamefont {J.~M.}\ \bibnamefont
  {Ib{\'a}{\~n}ez}}, \ and\ \bibinfo {author} {\bibfnamefont {J.~A.}\
  \bibnamefont {Miralles}},\ }\href {\doibase 10.1103/physrevd.43.3794}
  {\bibfield  {journal} {\bibinfo  {journal} {Phys. Rev. D}\ }\textbf {\bibinfo
  {volume} {43}},\ \bibinfo {pages} {3794} (\bibinfo {year}
  {1991})}\BibitemShut {NoStop}%
\bibitem [{\citenamefont {Banyuls}\ \emph {et~al.}(1997)\citenamefont
  {Banyuls}, \citenamefont {Font}, \citenamefont {Ib{\'a}{\~n}ez},
  \citenamefont {Mart{\'\i}},\ and\ \citenamefont {Miralles}}]{Banyuls97}%
  \BibitemOpen
  \bibfield  {author} {\bibinfo {author} {\bibfnamefont {F.}~\bibnamefont
  {Banyuls}}, \bibinfo {author} {\bibfnamefont {J.~A.}\ \bibnamefont {Font}},
  \bibinfo {author} {\bibfnamefont {J.~M.}\ \bibnamefont {Ib{\'a}{\~n}ez}},
  \bibinfo {author} {\bibfnamefont {J.~M.}\ \bibnamefont {Mart{\'\i}}}, \ and\
  \bibinfo {author} {\bibfnamefont {J.~A.}\ \bibnamefont {Miralles}},\ }\href
  {\doibase 10.1086/303604} {\bibfield  {journal} {\bibinfo  {journal}
  {Astrophys. J.}\ }\textbf {\bibinfo {volume} {476}},\ \bibinfo {pages} {221}
  (\bibinfo {year} {1997})}\BibitemShut {NoStop}%
\bibitem [{\citenamefont {Ib{\'a}{\~n}ez}\ \emph {et~al.}(2001)\citenamefont
  {Ib{\'a}{\~n}ez}, \citenamefont {Aloy}, \citenamefont {Font}, \citenamefont
  {Mart{\'\i}}, \citenamefont {Miralles},\ and\ \citenamefont
  {Pons}}]{Ibanez01}%
  \BibitemOpen
  \bibfield  {author} {\bibinfo {author} {\bibfnamefont {J.}~\bibnamefont
  {Ib{\'a}{\~n}ez}}, \bibinfo {author} {\bibfnamefont {M.}~\bibnamefont
  {Aloy}}, \bibinfo {author} {\bibfnamefont {J.}~\bibnamefont {Font}}, \bibinfo
  {author} {\bibfnamefont {J.}~\bibnamefont {Mart{\'\i}}}, \bibinfo {author}
  {\bibfnamefont {J.}~\bibnamefont {Miralles}}, \ and\ \bibinfo {author}
  {\bibfnamefont {J.}~\bibnamefont {Pons}},\ }in\ \href {\doibase
  10.1007/978-1-4615-0663-8_48} {\emph {\bibinfo {booktitle} {Godunov methods:
  theory and applications}}},\ \bibinfo {editor} {edited by\ \bibinfo {editor}
  {\bibfnamefont {E.}~\bibnamefont {Toro}}}\ (\bibinfo  {publisher} {Kluwer
  Academic/Plenum Publishers},\ \bibinfo {address} {New York},\ \bibinfo {year}
  {2001})\BibitemShut {NoStop}%
\bibitem [{\citenamefont {Font}(2003)}]{Font03}%
  \BibitemOpen
  \bibfield  {author} {\bibinfo {author} {\bibfnamefont {J.~A.}\ \bibnamefont
  {Font}},\ }\href@noop {} {\bibfield  {journal} {\bibinfo  {journal} {Living
  Rev. Relativ.}\ }\textbf {\bibinfo {volume} {6}},\ \bibinfo {pages} {4}
  (\bibinfo {year} {2003})}\BibitemShut {NoStop}%
\bibitem [{\citenamefont {{Radice}}\ \emph
  {et~al.}(2014{\natexlab{a}})\citenamefont {{Radice}}, \citenamefont
  {{Rezzolla}},\ and\ \citenamefont {{Galeazzi}}}]{Radice2013c}%
  \BibitemOpen
  \bibfield  {author} {\bibinfo {author} {\bibfnamefont {D.}~\bibnamefont
  {{Radice}}}, \bibinfo {author} {\bibfnamefont {L.}~\bibnamefont
  {{Rezzolla}}}, \ and\ \bibinfo {author} {\bibfnamefont {F.}~\bibnamefont
  {{Galeazzi}}},\ }\href {\doibase 10.1088/0264-9381/31/7/075012} {\bibfield
  {journal} {\bibinfo  {journal} {Class. Quantum Grav.}\ }\textbf {\bibinfo
  {volume} {31}},\ \bibinfo {eid} {075012} (\bibinfo {year}
  {2014}{\natexlab{a}})},\ \Eprint {http://arxiv.org/abs/1312.5004}
  {arXiv:1312.5004 [gr-qc]} \BibitemShut {NoStop}%
\bibitem [{\citenamefont {{Palenzuela}}\ \emph
  {et~al.}(2009{\natexlab{a}})\citenamefont {{Palenzuela}}, \citenamefont
  {{Lehner}}, \citenamefont {{Reula}},\ and\ \citenamefont
  {{Rezzolla}}}]{Palenzuela:2008sf}%
  \BibitemOpen
  \bibfield  {author} {\bibinfo {author} {\bibfnamefont {C.}~\bibnamefont
  {{Palenzuela}}}, \bibinfo {author} {\bibfnamefont {L.}~\bibnamefont
  {{Lehner}}}, \bibinfo {author} {\bibfnamefont {O.}~\bibnamefont {{Reula}}}, \
  and\ \bibinfo {author} {\bibfnamefont {L.}~\bibnamefont {{Rezzolla}}},\
  }\href {\doibase 10.1111/j.1365-2966.2009.14454.x} {\bibfield  {journal}
  {\bibinfo  {journal} {Mon. Not. R. Astron. Soc.}\ }\textbf {\bibinfo {volume}
  {394}},\ \bibinfo {pages} {1727} (\bibinfo {year} {2009}{\natexlab{a}})},\
  \Eprint {http://arxiv.org/abs/0810.1838} {arXiv:0810.1838} \BibitemShut
  {NoStop}%
\bibitem [{\citenamefont {{Dionysopoulou}}\ \emph {et~al.}(2013)\citenamefont
  {{Dionysopoulou}}, \citenamefont {{Alic}}, \citenamefont {{Palenzuela}},
  \citenamefont {{Rezzolla}},\ and\ \citenamefont
  {{Giacomazzo}}}]{Dionysopoulou:2012pp}%
  \BibitemOpen
  \bibfield  {author} {\bibinfo {author} {\bibfnamefont {K.}~\bibnamefont
  {{Dionysopoulou}}}, \bibinfo {author} {\bibfnamefont {D.}~\bibnamefont
  {{Alic}}}, \bibinfo {author} {\bibfnamefont {C.}~\bibnamefont
  {{Palenzuela}}}, \bibinfo {author} {\bibfnamefont {L.}~\bibnamefont
  {{Rezzolla}}}, \ and\ \bibinfo {author} {\bibfnamefont {B.}~\bibnamefont
  {{Giacomazzo}}},\ }\href {\doibase 10.1103/PhysRevD.88.044020} {\bibfield
  {journal} {\bibinfo  {journal} {Phys. Rev. D}\ }\textbf {\bibinfo {volume}
  {88}},\ \bibinfo {eid} {044020} (\bibinfo {year} {2013})},\ \Eprint
  {http://arxiv.org/abs/1208.3487} {arXiv:1208.3487 [gr-qc]} \BibitemShut
  {NoStop}%
\bibitem [{\citenamefont {{Palenzuela}}(2013)}]{Palenzuela2013}%
  \BibitemOpen
  \bibfield  {author} {\bibinfo {author} {\bibfnamefont {C.}~\bibnamefont
  {{Palenzuela}}},\ }\href {\doibase 10.1093/mnras/stt311} {\bibfield
  {journal} {\bibinfo  {journal} {Mon. Not. R. Astron. Soc.}\ }\textbf
  {\bibinfo {volume} {431}},\ \bibinfo {pages} {1853} (\bibinfo {year}
  {2013})},\ \Eprint {http://arxiv.org/abs/1212.0130} {arXiv:1212.0130
  [astro-ph.HE]} \BibitemShut {NoStop}%
\bibitem [{\citenamefont {{Ant{\'o}n}}\ \emph {et~al.}(2006)\citenamefont
  {{Ant{\'o}n}}, \citenamefont {{Zanotti}}, \citenamefont {{Miralles}},
  \citenamefont {{Mart{\'{\i}}}}, \citenamefont {{Ib{\'a}{\~n}ez}},
  \citenamefont {{Font}},\ and\ \citenamefont {{Pons}}}]{Anton06}%
  \BibitemOpen
  \bibfield  {author} {\bibinfo {author} {\bibfnamefont {L.}~\bibnamefont
  {{Ant{\'o}n}}}, \bibinfo {author} {\bibfnamefont {O.}~\bibnamefont
  {{Zanotti}}}, \bibinfo {author} {\bibfnamefont {J.~A.}\ \bibnamefont
  {{Miralles}}}, \bibinfo {author} {\bibfnamefont {J.~M.}\ \bibnamefont
  {{Mart{\'{\i}}}}}, \bibinfo {author} {\bibfnamefont {J.~M.}\ \bibnamefont
  {{Ib{\'a}{\~n}ez}}}, \bibinfo {author} {\bibfnamefont {J.~A.}\ \bibnamefont
  {{Font}}}, \ and\ \bibinfo {author} {\bibfnamefont {J.~A.}\ \bibnamefont
  {{Pons}}},\ }\href {\doibase 10.1086/498238} {\bibfield  {journal} {\bibinfo
  {journal} {Astrophys. J.}\ }\textbf {\bibinfo {volume} {637}},\ \bibinfo
  {pages} {296} (\bibinfo {year} {2006})},\ \Eprint
  {http://arxiv.org/abs/astro-ph/0506063} {astro-ph/0506063} \BibitemShut
  {NoStop}%
\bibitem [{\citenamefont {{Giacomazzo}}\ and\ \citenamefont
  {{Rezzolla}}(2007)}]{Giacomazzo:2007ti}%
  \BibitemOpen
  \bibfield  {author} {\bibinfo {author} {\bibfnamefont {B.}~\bibnamefont
  {{Giacomazzo}}}\ and\ \bibinfo {author} {\bibfnamefont {L.}~\bibnamefont
  {{Rezzolla}}},\ }\href {\doibase 10.1088/0264-9381/24/12/S16} {\bibfield
  {journal} {\bibinfo  {journal} {Class. Quantum Grav.}\ }\textbf {\bibinfo
  {volume} {24}},\ \bibinfo {pages} {235} (\bibinfo {year} {2007})},\ \Eprint
  {http://arxiv.org/abs/gr-qc/0701109} {gr-qc/0701109} \BibitemShut {NoStop}%
\bibitem [{\citenamefont {{Duez}}\ \emph {et~al.}(2005)\citenamefont {{Duez}},
  \citenamefont {{Liu}}, \citenamefont {{Shapiro}},\ and\ \citenamefont
  {{Stephens}}}]{Duez05MHD0}%
  \BibitemOpen
  \bibfield  {author} {\bibinfo {author} {\bibfnamefont {M.~D.}\ \bibnamefont
  {{Duez}}}, \bibinfo {author} {\bibfnamefont {Y.~T.}\ \bibnamefont {{Liu}}},
  \bibinfo {author} {\bibfnamefont {S.~L.}\ \bibnamefont {{Shapiro}}}, \ and\
  \bibinfo {author} {\bibfnamefont {B.~C.}\ \bibnamefont {{Stephens}}},\ }\href
  {\doibase 10.1103/PhysRevD.72.024028} {\bibfield  {journal} {\bibinfo
  {journal} {Phys. Rev. D}\ }\textbf {\bibinfo {volume} {72}},\ \bibinfo {eid}
  {024028} (\bibinfo {year} {2005})},\ \Eprint
  {http://arxiv.org/abs/astro-ph/0503420} {astro-ph/0503420} \BibitemShut
  {NoStop}%
\bibitem [{\citenamefont {Pareschi}\ and\ \citenamefont
  {Russo}(2005)}]{pareschi_2005_ier}%
  \BibitemOpen
  \bibfield  {author} {\bibinfo {author} {\bibfnamefont {L.}~\bibnamefont
  {Pareschi}}\ and\ \bibinfo {author} {\bibfnamefont {G.}~\bibnamefont
  {Russo}},\ }\href {\doibase 10.1007/s10915-004-4636-4} {\bibfield  {journal}
  {\bibinfo  {journal} {Journal of Scientific Computing}\ }\textbf {\bibinfo
  {volume} {25}},\ \bibinfo {pages} {129} (\bibinfo {year} {2005})}\BibitemShut
  {NoStop}%
\bibitem [{\citenamefont {Wilson}\ and\ \citenamefont
  {Mathews}(1995)}]{Wilson95}%
  \BibitemOpen
  \bibfield  {author} {\bibinfo {author} {\bibfnamefont {J.~R.}\ \bibnamefont
  {Wilson}}\ and\ \bibinfo {author} {\bibfnamefont {G.~J.}\ \bibnamefont
  {Mathews}},\ }\href@noop {} {\bibfield  {journal} {\bibinfo  {journal} {Phys.
  Rev. Lett.}\ }\textbf {\bibinfo {volume} {75}},\ \bibinfo {pages} {4161}
  (\bibinfo {year} {1995})}\BibitemShut {NoStop}%
\bibitem [{\citenamefont {{Bonazzola}}\ \emph {et~al.}(1997)\citenamefont
  {{Bonazzola}}, \citenamefont {{Gourgoulhon}},\ and\ \citenamefont
  {{Marck}}}]{Bonazzola97}%
  \BibitemOpen
  \bibfield  {author} {\bibinfo {author} {\bibfnamefont {S.}~\bibnamefont
  {{Bonazzola}}}, \bibinfo {author} {\bibfnamefont {E.}~\bibnamefont
  {{Gourgoulhon}}}, \ and\ \bibinfo {author} {\bibfnamefont {J.-A.}\
  \bibnamefont {{Marck}}},\ }\href {\doibase 10.1103/PhysRevD.56.7740}
  {\bibfield  {journal} {\bibinfo  {journal} {Phys. Rev. D}\ }\textbf {\bibinfo
  {volume} {56}},\ \bibinfo {pages} {7740} (\bibinfo {year} {1997})},\ \Eprint
  {http://arxiv.org/abs/gr-qc/9710031} {gr-qc/9710031} \BibitemShut {NoStop}%
\bibitem [{\citenamefont {Marronetti}\ \emph {et~al.}(1998)\citenamefont
  {Marronetti}, \citenamefont {Mathews},\ and\ \citenamefont
  {Wilson}}]{Marronetti98}%
  \BibitemOpen
  \bibfield  {author} {\bibinfo {author} {\bibfnamefont {P.}~\bibnamefont
  {Marronetti}}, \bibinfo {author} {\bibfnamefont {G.~J.}\ \bibnamefont
  {Mathews}}, \ and\ \bibinfo {author} {\bibfnamefont {J.~R.}\ \bibnamefont
  {Wilson}},\ }\href@noop {} {\bibfield  {journal} {\bibinfo  {journal} {Phys.
  Rev. D}\ }\textbf {\bibinfo {volume} {58}},\ \bibinfo {pages} {107503}
  (\bibinfo {year} {1998})}\BibitemShut {NoStop}%
\bibitem [{\citenamefont {Baumgarte}\ \emph {et~al.}(1997)\citenamefont
  {Baumgarte}, \citenamefont {Cook}, \citenamefont {Scheel}, \citenamefont
  {Shapiro},\ and\ \citenamefont {Teukolsky}}]{Baumgarte97}%
  \BibitemOpen
  \bibfield  {author} {\bibinfo {author} {\bibfnamefont {T.~W.}\ \bibnamefont
  {Baumgarte}}, \bibinfo {author} {\bibfnamefont {G.~B.}\ \bibnamefont {Cook}},
  \bibinfo {author} {\bibfnamefont {M.~A.}\ \bibnamefont {Scheel}}, \bibinfo
  {author} {\bibfnamefont {S.~L.}\ \bibnamefont {Shapiro}}, \ and\ \bibinfo
  {author} {\bibfnamefont {S.~A.}\ \bibnamefont {Teukolsky}},\ }\href@noop {}
  {\bibfield  {journal} {\bibinfo  {journal} {Phys. Rev. Lett.}\ }\textbf
  {\bibinfo {volume} {79}},\ \bibinfo {pages} {1182} (\bibinfo {year}
  {1997})}\BibitemShut {NoStop}%
\bibitem [{\citenamefont {Marronetti}\ \emph {et~al.}(1999)\citenamefont
  {Marronetti}, \citenamefont {Mathews},\ and\ \citenamefont
  {Wilson}}]{Marronetti99}%
  \BibitemOpen
  \bibfield  {author} {\bibinfo {author} {\bibfnamefont {P.}~\bibnamefont
  {Marronetti}}, \bibinfo {author} {\bibfnamefont {G.~J.}\ \bibnamefont
  {Mathews}}, \ and\ \bibinfo {author} {\bibfnamefont {J.~R.}\ \bibnamefont
  {Wilson}},\ }\href@noop {} {\bibfield  {journal} {\bibinfo  {journal} {Phys.
  Rev. D}\ }\textbf {\bibinfo {volume} {60}},\ \bibinfo {pages} {087301}
  (\bibinfo {year} {1999})}\BibitemShut {NoStop}%
\bibitem [{\citenamefont {Bonazzola}\ \emph {et~al.}(1999)\citenamefont
  {Bonazzola}, \citenamefont {Gourgoulhon},\ and\ \citenamefont
  {Marck}}]{Bonazzola98b}%
  \BibitemOpen
  \bibfield  {author} {\bibinfo {author} {\bibfnamefont {S.}~\bibnamefont
  {Bonazzola}}, \bibinfo {author} {\bibfnamefont {E.}~\bibnamefont
  {Gourgoulhon}}, \ and\ \bibinfo {author} {\bibfnamefont {J.~A.}\ \bibnamefont
  {Marck}},\ }\href@noop {} {\bibfield  {journal} {\bibinfo  {journal} {Phys.
  Rev. Lett.}\ }\textbf {\bibinfo {volume} {82}},\ \bibinfo {pages} {892}
  (\bibinfo {year} {1999})},\ \Eprint {http://arxiv.org/abs/gr-qc/9810072}
  {gr-qc/9810072} \BibitemShut {NoStop}%
\bibitem [{\citenamefont {{Ury{\= u}}}\ and\ \citenamefont
  {{Eriguchi}}(2000)}]{Uryu00}%
  \BibitemOpen
  \bibfield  {author} {\bibinfo {author} {\bibfnamefont {K.}~\bibnamefont
  {{Ury{\= u}}}}\ and\ \bibinfo {author} {\bibfnamefont {Y.}~\bibnamefont
  {{Eriguchi}}},\ }\href {\doibase 10.1103/PhysRevD.61.124023} {\bibfield
  {journal} {\bibinfo  {journal} {Phys. Rev. D}\ }\textbf {\bibinfo {volume}
  {61}},\ \bibinfo {eid} {124023} (\bibinfo {year} {2000})},\ \Eprint
  {http://arxiv.org/abs/gr-qc/9908059} {gr-qc/9908059} \BibitemShut {NoStop}%
\bibitem [{\citenamefont {{Ury{\= u}}}\ \emph {et~al.}(2000)\citenamefont
  {{Ury{\= u}}}, \citenamefont {{Shibata}},\ and\ \citenamefont
  {{Eriguchi}}}]{Uryu00a}%
  \BibitemOpen
  \bibfield  {author} {\bibinfo {author} {\bibfnamefont {K.}~\bibnamefont
  {{Ury{\= u}}}}, \bibinfo {author} {\bibfnamefont {M.}~\bibnamefont
  {{Shibata}}}, \ and\ \bibinfo {author} {\bibfnamefont {Y.}~\bibnamefont
  {{Eriguchi}}},\ }\href {\doibase 10.1103/PhysRevD.62.104015} {\bibfield
  {journal} {\bibinfo  {journal} {Phys. Rev. D}\ }\textbf {\bibinfo {volume}
  {62}},\ \bibinfo {eid} {104015} (\bibinfo {year} {2000})},\ \Eprint
  {http://arxiv.org/abs/gr-qc/0007042} {gr-qc/0007042} \BibitemShut {NoStop}%
\bibitem [{\citenamefont {{Usui}}\ \emph {et~al.}(2000)\citenamefont {{Usui}},
  \citenamefont {{Ury{\={u}}}},\ and\ \citenamefont {{Eriguchi}}}]{Usui2000}%
  \BibitemOpen
  \bibfield  {author} {\bibinfo {author} {\bibfnamefont {F.}~\bibnamefont
  {{Usui}}}, \bibinfo {author} {\bibfnamefont {K.}~\bibnamefont
  {{Ury{\={u}}}}}, \ and\ \bibinfo {author} {\bibfnamefont {Y.}~\bibnamefont
  {{Eriguchi}}},\ }\href {\doibase 10.1103/PhysRevD.61.024039} {\bibfield
  {journal} {\bibinfo  {journal} {Phys. Rev. D}\ }\textbf {\bibinfo {volume}
  {61}},\ \bibinfo {eid} {024039} (\bibinfo {year} {2000})},\ \Eprint
  {http://arxiv.org/abs/gr-qc/9906102} {gr-qc/9906102} \BibitemShut {NoStop}%
\bibitem [{\citenamefont {{Gourgoulhon}}\ \emph {et~al.}(2001)\citenamefont
  {{Gourgoulhon}}, \citenamefont {{Grandcl{\'e}ment}}, \citenamefont
  {{Taniguchi}}, \citenamefont {{Marck}},\ and\ \citenamefont
  {{Bonazzola}}}]{Gourgoulhon-etal-2000:2ns-initial-data}%
  \BibitemOpen
  \bibfield  {author} {\bibinfo {author} {\bibfnamefont {E.}~\bibnamefont
  {{Gourgoulhon}}}, \bibinfo {author} {\bibfnamefont {P.}~\bibnamefont
  {{Grandcl{\'e}ment}}}, \bibinfo {author} {\bibfnamefont {K.}~\bibnamefont
  {{Taniguchi}}}, \bibinfo {author} {\bibfnamefont {J.-A.}\ \bibnamefont
  {{Marck}}}, \ and\ \bibinfo {author} {\bibfnamefont {S.}~\bibnamefont
  {{Bonazzola}}},\ }\href {\doibase 10.1103/PhysRevD.63.064029} {\bibfield
  {journal} {\bibinfo  {journal} {Phys. Rev. D}\ }\textbf {\bibinfo {volume}
  {63}},\ \bibinfo {eid} {064029} (\bibinfo {year} {2001})},\ \Eprint
  {http://arxiv.org/abs/gr-qc/0007028} {gr-qc/0007028} \BibitemShut {NoStop}%
\bibitem [{\citenamefont {{Taniguchi}}\ \emph {et~al.}(2001)\citenamefont
  {{Taniguchi}}, \citenamefont {{Gourgoulhon}},\ and\ \citenamefont
  {{Bonazzola}}}]{Taniguchi01}%
  \BibitemOpen
  \bibfield  {author} {\bibinfo {author} {\bibfnamefont {K.}~\bibnamefont
  {{Taniguchi}}}, \bibinfo {author} {\bibfnamefont {E.}~\bibnamefont
  {{Gourgoulhon}}}, \ and\ \bibinfo {author} {\bibfnamefont {S.}~\bibnamefont
  {{Bonazzola}}},\ }\href {\doibase 10.1103/PhysRevD.64.064012} {\bibfield
  {journal} {\bibinfo  {journal} {Phys. Rev. D}\ }\textbf {\bibinfo {volume}
  {64}},\ \bibinfo {eid} {064012} (\bibinfo {year} {2001})},\ \Eprint
  {http://arxiv.org/abs/gr-qc/0103041} {gr-qc/0103041} \BibitemShut {NoStop}%
\bibitem [{\citenamefont {{Taniguchi}}\ and\ \citenamefont
  {{Gourgoulhon}}(2002)}]{Taniguchi02b}%
  \BibitemOpen
  \bibfield  {author} {\bibinfo {author} {\bibfnamefont {K.}~\bibnamefont
  {{Taniguchi}}}\ and\ \bibinfo {author} {\bibfnamefont {E.}~\bibnamefont
  {{Gourgoulhon}}},\ }\href {\doibase 10.1103/PhysRevD.66.104019} {\bibfield
  {journal} {\bibinfo  {journal} {Phys. Rev. D}\ }\textbf {\bibinfo {volume}
  {66}},\ \bibinfo {eid} {104019} (\bibinfo {year} {2002})},\ \Eprint
  {http://arxiv.org/abs/gr-qc/0207098} {gr-qc/0207098} \BibitemShut {NoStop}%
\bibitem [{\citenamefont {{Taniguchi}}\ and\ \citenamefont
  {{Gourgoulhon}}(2003)}]{Taniguchi03}%
  \BibitemOpen
  \bibfield  {author} {\bibinfo {author} {\bibfnamefont {K.}~\bibnamefont
  {{Taniguchi}}}\ and\ \bibinfo {author} {\bibfnamefont {E.}~\bibnamefont
  {{Gourgoulhon}}},\ }\href {\doibase 10.1103/PhysRevD.68.124025} {\bibfield
  {journal} {\bibinfo  {journal} {Phys. Rev. D}\ }\textbf {\bibinfo {volume}
  {68}},\ \bibinfo {eid} {124025} (\bibinfo {year} {2003})},\ \Eprint
  {http://arxiv.org/abs/gr-qc/0309045} {gr-qc/0309045} \BibitemShut {NoStop}%
\bibitem [{\citenamefont {Bejger}\ \emph {et~al.}(2005)\citenamefont {Bejger},
  \citenamefont {Gondek-Rosi{\'n}ska}, \citenamefont {Gourgoulhon},
  \citenamefont {Haensel}, \citenamefont {Taniguchi},\ and\ \citenamefont
  {Zdunik}}]{Bejger05}%
  \BibitemOpen
  \bibfield  {author} {\bibinfo {author} {\bibfnamefont {M.}~\bibnamefont
  {Bejger}}, \bibinfo {author} {\bibfnamefont {D.}~\bibnamefont
  {Gondek-Rosi{\'n}ska}}, \bibinfo {author} {\bibfnamefont {E.}~\bibnamefont
  {Gourgoulhon}}, \bibinfo {author} {\bibfnamefont {P.}~\bibnamefont
  {Haensel}}, \bibinfo {author} {\bibfnamefont {K.}~\bibnamefont {Taniguchi}},
  \ and\ \bibinfo {author} {\bibfnamefont {J.~L.}\ \bibnamefont {Zdunik}},\
  }\href@noop {} {\bibfield  {journal} {\bibinfo  {journal} {Astron.
  Astrophys.}\ }\textbf {\bibinfo {volume} {431}},\ \bibinfo {pages} {297}
  (\bibinfo {year} {2005})}\BibitemShut {NoStop}%
\bibitem [{\citenamefont {Tsokaros}\ and\ \citenamefont
  {Ury{\={u}}}(2007)}]{Tsokaros2007}%
  \BibitemOpen
  \bibfield  {author} {\bibinfo {author} {\bibfnamefont {A.~A.}\ \bibnamefont
  {Tsokaros}}\ and\ \bibinfo {author} {\bibfnamefont {K.}~\bibnamefont
  {Ury{\={u}}}},\ }\href {\doibase 10.1103/PhysRevD.75.044026} {\bibfield
  {journal} {\bibinfo  {journal} {Phys. Rev. D}\ }\textbf {\bibinfo {volume}
  {75}},\ \bibinfo {pages} {044026} (\bibinfo {year} {2007})}\BibitemShut
  {NoStop}%
\bibitem [{\citenamefont {{Kiuchi}}\ \emph {et~al.}(2009)\citenamefont
  {{Kiuchi}}, \citenamefont {{Sekiguchi}}, \citenamefont {{Shibata}},\ and\
  \citenamefont {{Taniguchi}}}]{Kiuchi2009}%
  \BibitemOpen
  \bibfield  {author} {\bibinfo {author} {\bibfnamefont {K.}~\bibnamefont
  {{Kiuchi}}}, \bibinfo {author} {\bibfnamefont {Y.}~\bibnamefont
  {{Sekiguchi}}}, \bibinfo {author} {\bibfnamefont {M.}~\bibnamefont
  {{Shibata}}}, \ and\ \bibinfo {author} {\bibfnamefont {K.}~\bibnamefont
  {{Taniguchi}}},\ }\href {\doibase 10.1103/PhysRevD.80.064037} {\bibfield
  {journal} {\bibinfo  {journal} {Phys. Rev. D}\ }\textbf {\bibinfo {volume}
  {80}},\ \bibinfo {eid} {064037} (\bibinfo {year} {2009})},\ \Eprint
  {http://arxiv.org/abs/0904.4551} {arXiv:0904.4551 [gr-qc]} \BibitemShut
  {NoStop}%
\bibitem [{\citenamefont {{Taniguchi}}\ and\ \citenamefont
  {{Shibata}}(2010)}]{Taniguchi2010}%
  \BibitemOpen
  \bibfield  {author} {\bibinfo {author} {\bibfnamefont {K.}~\bibnamefont
  {{Taniguchi}}}\ and\ \bibinfo {author} {\bibfnamefont {M.}~\bibnamefont
  {{Shibata}}},\ }\href {\doibase 10.1088/0067-0049/188/1/187} {\bibfield
  {journal} {\bibinfo  {journal} {Astrophys. J., Supp.}\ }\textbf {\bibinfo
  {volume} {188}},\ \bibinfo {pages} {187} (\bibinfo {year} {2010})},\ \Eprint
  {http://arxiv.org/abs/1005.0958} {arXiv:1005.0958 [astro-ph.SR]} \BibitemShut
  {NoStop}%
\bibitem [{\citenamefont {Tsokaros}\ and\ \citenamefont
  {Ury{\={u}}}(2012)}]{Tsokaros2012}%
  \BibitemOpen
  \bibfield  {author} {\bibinfo {author} {\bibfnamefont {A.}~\bibnamefont
  {Tsokaros}}\ and\ \bibinfo {author} {\bibfnamefont {K.}~\bibnamefont
  {Ury{\={u}}}},\ }\href {\doibase 10.1007/s10665-012-9585-6} {\bibfield
  {journal} {\bibinfo  {journal} {Journal of Engineering Mathematics}\ }\textbf
  {\bibinfo {volume} {82}},\ \bibinfo {pages} {133} (\bibinfo {year}
  {2012})}\BibitemShut {NoStop}%
\bibitem [{\citenamefont {{Tsokaros}}\ \emph {et~al.}(2015)\citenamefont
  {{Tsokaros}}, \citenamefont {{Ury{\={u}}}},\ and\ \citenamefont
  {{Rezzolla}}}]{Tsokaros2015}%
  \BibitemOpen
  \bibfield  {author} {\bibinfo {author} {\bibfnamefont {A.}~\bibnamefont
  {{Tsokaros}}}, \bibinfo {author} {\bibfnamefont {K.}~\bibnamefont
  {{Ury{\={u}}}}}, \ and\ \bibinfo {author} {\bibfnamefont {L.}~\bibnamefont
  {{Rezzolla}}},\ }\href {\doibase 10.1103/PhysRevD.91.104030} {\bibfield
  {journal} {\bibinfo  {journal} {Phys. Rev. D}\ }\textbf {\bibinfo {volume}
  {91}},\ \bibinfo {eid} {104030} (\bibinfo {year} {2015})},\ \Eprint
  {http://arxiv.org/abs/1502.05674} {arXiv:1502.05674 [gr-qc]} \BibitemShut
  {NoStop}%
\bibitem [{LORENE()}]{lorene}%
  \BibitemOpen
  LORENE,\ \href {http://www.lorene.obspm.fr} {}\bibinfo {note} {Langage Objet
  pour la RElativit\'{e} Num\'{e}rique, \url{www.lorene.obspm.fr}}\BibitemShut
  {NoStop}%
\bibitem [{\citenamefont {Peters}\ and\ \citenamefont
  {Mathews}(1963)}]{Peters:1963ux}%
  \BibitemOpen
  \bibfield  {author} {\bibinfo {author} {\bibfnamefont {P.~C.}\ \bibnamefont
  {Peters}}\ and\ \bibinfo {author} {\bibfnamefont {J.}~\bibnamefont
  {Mathews}},\ }\href@noop {} {\bibfield  {journal} {\bibinfo  {journal} {Phys.
  Rev.}\ }\textbf {\bibinfo {volume} {131}},\ \bibinfo {pages} {435} (\bibinfo
  {year} {1963})}\BibitemShut {NoStop}%
\bibitem [{\citenamefont {{Turner}}(1977{\natexlab{a}})}]{Turner1977a}%
  \BibitemOpen
  \bibfield  {author} {\bibinfo {author} {\bibfnamefont {M.}~\bibnamefont
  {{Turner}}},\ }\href {\doibase 10.1086/155536} {\bibfield  {journal}
  {\bibinfo  {journal} {Astrophys. J.}\ }\textbf {\bibinfo {volume} {216}},\
  \bibinfo {pages} {914} (\bibinfo {year} {1977}{\natexlab{a}})}\BibitemShut
  {NoStop}%
\bibitem [{\citenamefont {{Turner}}(1977{\natexlab{b}})}]{Turner1977b}%
  \BibitemOpen
  \bibfield  {author} {\bibinfo {author} {\bibfnamefont {M.}~\bibnamefont
  {{Turner}}},\ }\href {\doibase 10.1086/155501} {\bibfield  {journal}
  {\bibinfo  {journal} {Astrophys. J.}\ }\textbf {\bibinfo {volume} {216}},\
  \bibinfo {pages} {610} (\bibinfo {year} {1977}{\natexlab{b}})}\BibitemShut
  {NoStop}%
\bibitem [{\citenamefont {{East}}\ and\ \citenamefont
  {{Pretorius}}(2012)}]{East2012c}%
  \BibitemOpen
  \bibfield  {author} {\bibinfo {author} {\bibfnamefont {W.~E.}\ \bibnamefont
  {{East}}}\ and\ \bibinfo {author} {\bibfnamefont {F.}~\bibnamefont
  {{Pretorius}}},\ }\href {\doibase 10.1088/2041-8205/760/1/L4} {\bibfield
  {journal} {\bibinfo  {journal} {Astrophys. J.}\ }\textbf {\bibinfo {volume}
  {760}},\ \bibinfo {eid} {L4} (\bibinfo {year} {2012})},\ \Eprint
  {http://arxiv.org/abs/1208.5279} {arXiv:1208.5279 [astro-ph.HE]} \BibitemShut
  {NoStop}%
\bibitem [{\citenamefont {{Gold}}\ \emph {et~al.}(2012)\citenamefont {{Gold}},
  \citenamefont {{Bernuzzi}}, \citenamefont {{Thierfelder}}, \citenamefont
  {{Br{\"u}gmann}},\ and\ \citenamefont {{Pretorius}}}]{Gold2012}%
  \BibitemOpen
  \bibfield  {author} {\bibinfo {author} {\bibfnamefont {R.}~\bibnamefont
  {{Gold}}}, \bibinfo {author} {\bibfnamefont {S.}~\bibnamefont {{Bernuzzi}}},
  \bibinfo {author} {\bibfnamefont {M.}~\bibnamefont {{Thierfelder}}}, \bibinfo
  {author} {\bibfnamefont {B.}~\bibnamefont {{Br{\"u}gmann}}}, \ and\ \bibinfo
  {author} {\bibfnamefont {F.}~\bibnamefont {{Pretorius}}},\ }\href {\doibase
  10.1103/PhysRevD.86.121501} {\bibfield  {journal} {\bibinfo  {journal} {Phys.
  Rev. D}\ }\textbf {\bibinfo {volume} {86}},\ \bibinfo {eid} {121501}
  (\bibinfo {year} {2012})},\ \Eprint {http://arxiv.org/abs/1109.5128}
  {arXiv:1109.5128 [gr-qc]} \BibitemShut {NoStop}%
\bibitem [{\citenamefont {{Rosswog}}\ \emph {et~al.}(2013)\citenamefont
  {{Rosswog}}, \citenamefont {{Piran}},\ and\ \citenamefont
  {{Nakar}}}]{Rosswog2013}%
  \BibitemOpen
  \bibfield  {author} {\bibinfo {author} {\bibfnamefont {S.}~\bibnamefont
  {{Rosswog}}}, \bibinfo {author} {\bibfnamefont {T.}~\bibnamefont {{Piran}}},
  \ and\ \bibinfo {author} {\bibfnamefont {E.}~\bibnamefont {{Nakar}}},\ }\href
  {\doibase 10.1093/mnras/sts708} {\bibfield  {journal} {\bibinfo  {journal}
  {Mon. Not. R. Astron. Soc.}\ }\textbf {\bibinfo {volume} {430}},\ \bibinfo
  {pages} {2585} (\bibinfo {year} {2013})},\ \Eprint
  {http://arxiv.org/abs/1204.6240} {arXiv:1204.6240 [astro-ph.HE]} \BibitemShut
  {NoStop}%
\bibitem [{\citenamefont {Tsatsin}\ and\ \citenamefont
  {Marronetti}(2013)}]{Tsatsin2013}%
  \BibitemOpen
  \bibfield  {author} {\bibinfo {author} {\bibfnamefont {P.}~\bibnamefont
  {Tsatsin}}\ and\ \bibinfo {author} {\bibfnamefont {P.}~\bibnamefont
  {Marronetti}},\ }\href {\doibase 10.1103/PhysRevD.88.064060} {\bibfield
  {journal} {\bibinfo  {journal} {Phys. Rev. D}\ }\textbf {\bibinfo {volume}
  {88}},\ \bibinfo {pages} {064060} (\bibinfo {year} {2013})}\BibitemShut
  {NoStop}%
\bibitem [{\citenamefont {Miller}\ \emph {et~al.}(2004)\citenamefont {Miller},
  \citenamefont {Gressman},\ and\ \citenamefont {Suen}}]{Miller03b}%
  \BibitemOpen
  \bibfield  {author} {\bibinfo {author} {\bibfnamefont {M.}~\bibnamefont
  {Miller}}, \bibinfo {author} {\bibfnamefont {P.}~\bibnamefont {Gressman}}, \
  and\ \bibinfo {author} {\bibfnamefont {W.-M.}\ \bibnamefont {Suen}},\
  }\href@noop {} {\bibfield  {journal} {\bibinfo  {journal} {Phys. Rev. D}\
  }\textbf {\bibinfo {volume} {69}},\ \bibinfo {pages} {064026} (\bibinfo
  {year} {2004})},\ \Eprint {http://arxiv.org/abs/gr-qc/0312030}
  {gr-qc/0312030} \BibitemShut {NoStop}%
\bibitem [{\citenamefont {Miller}(2004)}]{Miller03c}%
  \BibitemOpen
  \bibfield  {author} {\bibinfo {author} {\bibfnamefont {M.}~\bibnamefont
  {Miller}},\ }\href@noop {} {\bibfield  {journal} {\bibinfo  {journal} {Phys.
  Rev. D}\ }\textbf {\bibinfo {volume} {69}},\ \bibinfo {pages} {124013}
  (\bibinfo {year} {2004})},\ \bibinfo {note} {gr-qc/0305024}\BibitemShut
  {NoStop}%
\bibitem [{\citenamefont {{Baiotti}}\ \emph {et~al.}(2011)\citenamefont
  {{Baiotti}}, \citenamefont {{Damour}}, \citenamefont {{Giacomazzo}},
  \citenamefont {{Nagar}},\ and\ \citenamefont {{Rezzolla}}}]{Baiotti2011}%
  \BibitemOpen
  \bibfield  {author} {\bibinfo {author} {\bibfnamefont {L.}~\bibnamefont
  {{Baiotti}}}, \bibinfo {author} {\bibfnamefont {T.}~\bibnamefont {{Damour}}},
  \bibinfo {author} {\bibfnamefont {B.}~\bibnamefont {{Giacomazzo}}}, \bibinfo
  {author} {\bibfnamefont {A.}~\bibnamefont {{Nagar}}}, \ and\ \bibinfo
  {author} {\bibfnamefont {L.}~\bibnamefont {{Rezzolla}}},\ }\href {\doibase
  10.1103/PhysRevD.84.024017} {\bibfield  {journal} {\bibinfo  {journal} {Phys.
  Rev. D}\ }\textbf {\bibinfo {volume} {84}},\ \bibinfo {eid} {024017}
  (\bibinfo {year} {2011})},\ \Eprint {http://arxiv.org/abs/1103.3874}
  {arXiv:1103.3874 [gr-qc]} \BibitemShut {NoStop}%
\bibitem [{\citenamefont {{Bernuzzi}}\ \emph
  {et~al.}(2012{\natexlab{b}})\citenamefont {{Bernuzzi}}, \citenamefont
  {{Nagar}}, \citenamefont {{Thierfelder}},\ and\ \citenamefont
  {{Br{\"u}gmann}}}]{Bernuzzi2012}%
  \BibitemOpen
  \bibfield  {author} {\bibinfo {author} {\bibfnamefont {S.}~\bibnamefont
  {{Bernuzzi}}}, \bibinfo {author} {\bibfnamefont {A.}~\bibnamefont {{Nagar}}},
  \bibinfo {author} {\bibfnamefont {M.}~\bibnamefont {{Thierfelder}}}, \ and\
  \bibinfo {author} {\bibfnamefont {B.}~\bibnamefont {{Br{\"u}gmann}}},\ }\href
  {\doibase 10.1103/PhysRevD.86.044030} {\bibfield  {journal} {\bibinfo
  {journal} {Phys. Rev. D}\ }\textbf {\bibinfo {volume} {86}},\ \bibinfo {eid}
  {044030} (\bibinfo {year} {2012}{\natexlab{b}})},\ \Eprint
  {http://arxiv.org/abs/1205.3403} {arXiv:1205.3403 [gr-qc]} \BibitemShut
  {NoStop}%
\bibitem [{\citenamefont {{Hotokezaka}}\ \emph
  {et~al.}(2013{\natexlab{a}})\citenamefont {{Hotokezaka}}, \citenamefont
  {{Kyutoku}},\ and\ \citenamefont {{Shibata}}}]{Hotokezaka2013b}%
  \BibitemOpen
  \bibfield  {author} {\bibinfo {author} {\bibfnamefont {K.}~\bibnamefont
  {{Hotokezaka}}}, \bibinfo {author} {\bibfnamefont {K.}~\bibnamefont
  {{Kyutoku}}}, \ and\ \bibinfo {author} {\bibfnamefont {M.}~\bibnamefont
  {{Shibata}}},\ }\href {\doibase 10.1103/PhysRevD.87.044001} {\bibfield
  {journal} {\bibinfo  {journal} {Phys. Rev. D}\ }\textbf {\bibinfo {volume}
  {87}},\ \bibinfo {eid} {044001} (\bibinfo {year} {2013}{\natexlab{a}})},\
  \Eprint {http://arxiv.org/abs/1301.3555} {arXiv:1301.3555 [gr-qc]}
  \BibitemShut {NoStop}%
\bibitem [{\citenamefont {Ajith}\ \emph {et~al.}(2007)\citenamefont {Ajith}
  \emph {et~al.}}]{Ajith:2007qp}%
  \BibitemOpen
  \bibfield  {author} {\bibinfo {author} {\bibfnamefont {P.}~\bibnamefont
  {Ajith}} \emph {et~al.},\ }\href@noop {} {\bibfield  {journal} {\bibinfo
  {journal} {Class. Quantum Grav.}\ }\textbf {\bibinfo {volume} {24}},\
  \bibinfo {pages} {S689} (\bibinfo {year} {2007})},\ \Eprint
  {http://arxiv.org/abs/arXiv:0704.3764} {arXiv:0704.3764} \BibitemShut
  {NoStop}%
\bibitem [{\citenamefont {Ajith}\ \emph {et~al.}(2008)\citenamefont {Ajith},
  \citenamefont {Babak}, \citenamefont {Chen}, \citenamefont {Hewitson},
  \citenamefont {Krishnan}, \citenamefont {Whelan}, \citenamefont
  {Br{\"u}gman}, \citenamefont {Gonzalez}, \citenamefont {Hannam},
  \citenamefont {Husa}, \citenamefont {Koppitz}, \citenamefont {Pollney},
  \citenamefont {Rezzolla}, \citenamefont {Santamar{\'i}a}, \citenamefont
  {Sintes}, \citenamefont {Sperhake},\ and\ \citenamefont
  {Thornburg}}]{Ajith:2007kx:longal}%
  \BibitemOpen
  \bibfield  {author} {\bibinfo {author} {\bibfnamefont {P.}~\bibnamefont
  {Ajith}}, \bibinfo {author} {\bibfnamefont {S.}~\bibnamefont {Babak}},
  \bibinfo {author} {\bibfnamefont {Y.}~\bibnamefont {Chen}}, \bibinfo {author}
  {\bibfnamefont {M.}~\bibnamefont {Hewitson}}, \bibinfo {author}
  {\bibfnamefont {B.}~\bibnamefont {Krishnan}}, \bibinfo {author}
  {\bibfnamefont {J.~T.}\ \bibnamefont {Whelan}}, \bibinfo {author}
  {\bibfnamefont {B.}~\bibnamefont {Br{\"u}gman}}, \bibinfo {author}
  {\bibfnamefont {J.}~\bibnamefont {Gonzalez}}, \bibinfo {author}
  {\bibfnamefont {M.}~\bibnamefont {Hannam}}, \bibinfo {author} {\bibfnamefont
  {S.}~\bibnamefont {Husa}}, \bibinfo {author} {\bibfnamefont {M.}~\bibnamefont
  {Koppitz}}, \bibinfo {author} {\bibfnamefont {D.}~\bibnamefont {Pollney}},
  \bibinfo {author} {\bibfnamefont {L.}~\bibnamefont {Rezzolla}}, \bibinfo
  {author} {\bibfnamefont {L.}~\bibnamefont {Santamar{\'i}a}}, \bibinfo
  {author} {\bibfnamefont {A.~M.}\ \bibnamefont {Sintes}}, \bibinfo {author}
  {\bibfnamefont {U.}~\bibnamefont {Sperhake}}, \ and\ \bibinfo {author}
  {\bibfnamefont {J.}~\bibnamefont {Thornburg}},\ }\href@noop {} {\bibfield
  {journal} {\bibinfo  {journal} {Phys. Re. D}\ }\textbf {\bibinfo {volume}
  {77}},\ \bibinfo {pages} {104017} (\bibinfo {year} {2008})},\ \Eprint
  {http://arxiv.org/abs/arXiv:0710.2335} {arXiv:0710.2335} \BibitemShut
  {NoStop}%
\bibitem [{\citenamefont {{Bernuzzi}}\ \emph
  {et~al.}(2015{\natexlab{a}})\citenamefont {{Bernuzzi}}, \citenamefont
  {{Nagar}}, \citenamefont {{Dietrich}},\ and\ \citenamefont
  {{Damour}}}]{Bernuzzi2015}%
  \BibitemOpen
  \bibfield  {author} {\bibinfo {author} {\bibfnamefont {S.}~\bibnamefont
  {{Bernuzzi}}}, \bibinfo {author} {\bibfnamefont {A.}~\bibnamefont {{Nagar}}},
  \bibinfo {author} {\bibfnamefont {T.}~\bibnamefont {{Dietrich}}}, \ and\
  \bibinfo {author} {\bibfnamefont {T.}~\bibnamefont {{Damour}}},\ }\href
  {\doibase 10.1103/PhysRevLett.114.161103} {\bibfield  {journal} {\bibinfo
  {journal} {Phys. Rev. Lett.}\ }\textbf {\bibinfo {volume} {114}},\ \bibinfo
  {eid} {161103} (\bibinfo {year} {2015}{\natexlab{a}})},\ \Eprint
  {http://arxiv.org/abs/1412.4553} {arXiv:1412.4553 [gr-qc]} \BibitemShut
  {NoStop}%
\bibitem [{\citenamefont {{Read}}\ \emph {et~al.}(2013)\citenamefont {{Read}},
  \citenamefont {{Baiotti}}, \citenamefont {{Creighton}}, \citenamefont
  {{Friedman}}, \citenamefont {{Giacomazzo}}, \citenamefont {{Kyutoku}},
  \citenamefont {{Markakis}}, \citenamefont {{Rezzolla}}, \citenamefont
  {{Shibata}},\ and\ \citenamefont {{Taniguchi}}}]{Read2013}%
  \BibitemOpen
  \bibfield  {author} {\bibinfo {author} {\bibfnamefont {J.~S.}\ \bibnamefont
  {{Read}}}, \bibinfo {author} {\bibfnamefont {L.}~\bibnamefont {{Baiotti}}},
  \bibinfo {author} {\bibfnamefont {J.~D.~E.}\ \bibnamefont {{Creighton}}},
  \bibinfo {author} {\bibfnamefont {J.~L.}\ \bibnamefont {{Friedman}}},
  \bibinfo {author} {\bibfnamefont {B.}~\bibnamefont {{Giacomazzo}}}, \bibinfo
  {author} {\bibfnamefont {K.}~\bibnamefont {{Kyutoku}}}, \bibinfo {author}
  {\bibfnamefont {C.}~\bibnamefont {{Markakis}}}, \bibinfo {author}
  {\bibfnamefont {L.}~\bibnamefont {{Rezzolla}}}, \bibinfo {author}
  {\bibfnamefont {M.}~\bibnamefont {{Shibata}}}, \ and\ \bibinfo {author}
  {\bibfnamefont {K.}~\bibnamefont {{Taniguchi}}},\ }\href {\doibase
  10.1103/PhysRevD.88.044042} {\bibfield  {journal} {\bibinfo  {journal} {Phys.
  Rev. D}\ }\textbf {\bibinfo {volume} {88}},\ \bibinfo {eid} {044042}
  (\bibinfo {year} {2013})},\ \Eprint {http://arxiv.org/abs/1306.4065}
  {arXiv:1306.4065 [gr-qc]} \BibitemShut {NoStop}%
\bibitem [{\citenamefont {{Favata}}(2014)}]{Favata2014}%
  \BibitemOpen
  \bibfield  {author} {\bibinfo {author} {\bibfnamefont {M.}~\bibnamefont
  {{Favata}}},\ }\href {\doibase 10.1103/PhysRevLett.112.101101} {\bibfield
  {journal} {\bibinfo  {journal} {Phys. Rev. Lett.}\ }\textbf {\bibinfo
  {volume} {112}},\ \bibinfo {eid} {101101} (\bibinfo {year} {2014})},\ \Eprint
  {http://arxiv.org/abs/1310.8288} {arXiv:1310.8288 [gr-qc]} \BibitemShut
  {NoStop}%
\bibitem [{\citenamefont {{Yagi}}\ \emph {et~al.}(2014)\citenamefont {{Yagi}},
  \citenamefont {{Kyutoku}}, \citenamefont {{Pappas}}, \citenamefont
  {{Yunes}},\ and\ \citenamefont {{Apostolatos}}}]{Yagi2014}%
  \BibitemOpen
  \bibfield  {author} {\bibinfo {author} {\bibfnamefont {K.}~\bibnamefont
  {{Yagi}}}, \bibinfo {author} {\bibfnamefont {K.}~\bibnamefont {{Kyutoku}}},
  \bibinfo {author} {\bibfnamefont {G.}~\bibnamefont {{Pappas}}}, \bibinfo
  {author} {\bibfnamefont {N.}~\bibnamefont {{Yunes}}}, \ and\ \bibinfo
  {author} {\bibfnamefont {T.~A.}\ \bibnamefont {{Apostolatos}}},\ }\href
  {\doibase 10.1103/PhysRevD.89.124013} {\bibfield  {journal} {\bibinfo
  {journal} {Phys. Rev. D}\ }\textbf {\bibinfo {volume} {89}},\ \bibinfo {eid}
  {124013} (\bibinfo {year} {2014})},\ \Eprint {http://arxiv.org/abs/1403.6243}
  {arXiv:1403.6243 [gr-qc]} \BibitemShut {NoStop}%
\bibitem [{\citenamefont {{Wade}}\ \emph {et~al.}(2014)\citenamefont {{Wade}},
  \citenamefont {{Creighton}}, \citenamefont {{Ochsner}}, \citenamefont
  {{Lackey}}, \citenamefont {{Farr}}, \citenamefont {{Littenberg}},\ and\
  \citenamefont {{Raymond}}}]{Wade2014}%
  \BibitemOpen
  \bibfield  {author} {\bibinfo {author} {\bibfnamefont {L.}~\bibnamefont
  {{Wade}}}, \bibinfo {author} {\bibfnamefont {J.~D.~E.}\ \bibnamefont
  {{Creighton}}}, \bibinfo {author} {\bibfnamefont {E.}~\bibnamefont
  {{Ochsner}}}, \bibinfo {author} {\bibfnamefont {B.~D.}\ \bibnamefont
  {{Lackey}}}, \bibinfo {author} {\bibfnamefont {B.~F.}\ \bibnamefont
  {{Farr}}}, \bibinfo {author} {\bibfnamefont {T.~B.}\ \bibnamefont
  {{Littenberg}}}, \ and\ \bibinfo {author} {\bibfnamefont {V.}~\bibnamefont
  {{Raymond}}},\ }\href {\doibase 10.1103/PhysRevD.89.103012} {\bibfield
  {journal} {\bibinfo  {journal} {Phys. Rev. D}\ }\textbf {\bibinfo {volume}
  {89}},\ \bibinfo {eid} {103012} (\bibinfo {year} {2014})},\ \Eprint
  {http://arxiv.org/abs/1402.5156} {arXiv:1402.5156 [gr-qc]} \BibitemShut
  {NoStop}%
\bibitem [{\citenamefont {{Ury{\= u}}}\ \emph {et~al.}(2006)\citenamefont
  {{Ury{\= u}}}, \citenamefont {{Limousin}}, \citenamefont {{Friedman}},
  \citenamefont {{Gourgoulhon}},\ and\ \citenamefont {{Shibata}}}]{Uryu2006}%
  \BibitemOpen
  \bibfield  {author} {\bibinfo {author} {\bibfnamefont {K.}~\bibnamefont
  {{Ury{\= u}}}}, \bibinfo {author} {\bibfnamefont {F.}~\bibnamefont
  {{Limousin}}}, \bibinfo {author} {\bibfnamefont {J.~L.}\ \bibnamefont
  {{Friedman}}}, \bibinfo {author} {\bibfnamefont {E.}~\bibnamefont
  {{Gourgoulhon}}}, \ and\ \bibinfo {author} {\bibfnamefont {M.}~\bibnamefont
  {{Shibata}}},\ }\href {\doibase 10.1103/PhysRevLett.97.171101} {\bibfield
  {journal} {\bibinfo  {journal} {Phys. Rev. Lett.}\ }\textbf {\bibinfo
  {volume} {97}},\ \bibinfo {eid} {171101} (\bibinfo {year} {2006})},\ \Eprint
  {http://arxiv.org/abs/gr-qc/0511136} {gr-qc/0511136} \BibitemShut {NoStop}%
\bibitem [{\citenamefont {Ury{\= u}}\ \emph {et~al.}(2009)\citenamefont {Ury{\=
  u}}, \citenamefont {Limousin}, \citenamefont {Friedman}, \citenamefont
  {Gourgoulhon},\ and\ \citenamefont {Shibata}}]{Uryu:2009ye}%
  \BibitemOpen
  \bibfield  {author} {\bibinfo {author} {\bibfnamefont {K.}~\bibnamefont
  {Ury{\= u}}}, \bibinfo {author} {\bibfnamefont {F.}~\bibnamefont {Limousin}},
  \bibinfo {author} {\bibfnamefont {J.~L.}\ \bibnamefont {Friedman}}, \bibinfo
  {author} {\bibfnamefont {E.}~\bibnamefont {Gourgoulhon}}, \ and\ \bibinfo
  {author} {\bibfnamefont {M.}~\bibnamefont {Shibata}},\ }\href {\doibase
  10.1103/PhysRevD.80.124004} {\bibfield  {journal} {\bibinfo  {journal} {Phys.
  Rev. D}\ }\textbf {\bibinfo {volume} {80}},\ \bibinfo {pages} {124004}
  (\bibinfo {year} {2009})},\ \Eprint {http://arxiv.org/abs/0908.0579}
  {arXiv:0908.0579 [gr-qc]} \BibitemShut {NoStop}%
\bibitem [{\citenamefont {Isenberg}(1979)}]{Isenberg-1979}%
  \BibitemOpen
  \bibfield  {author} {\bibinfo {author} {\bibfnamefont {J.~A.}\ \bibnamefont
  {Isenberg}},\ }\emph {\bibinfo {title} {The Construction of Spacetimes from
  Initial Data}},\ \href@noop {} {Ph.D. thesis},\ \bibinfo  {school}
  {University of Maryland} (\bibinfo {year} {1979})\BibitemShut {NoStop}%
\bibitem [{\citenamefont {{Ury{\={u}}}}\ and\ \citenamefont
  {{Tsokaros}}(2012)}]{Uryu2012}%
  \BibitemOpen
  \bibfield  {author} {\bibinfo {author} {\bibfnamefont {K.}~\bibnamefont
  {{Ury{\={u}}}}}\ and\ \bibinfo {author} {\bibfnamefont {A.}~\bibnamefont
  {{Tsokaros}}},\ }\href {\doibase 10.1103/PhysRevD.85.064014} {\bibfield
  {journal} {\bibinfo  {journal} {Phys. Rev. D}\ }\textbf {\bibinfo {volume}
  {85}},\ \bibinfo {eid} {064014} (\bibinfo {year} {2012})},\ \Eprint
  {http://arxiv.org/abs/1108.3065} {arXiv:1108.3065 [gr-qc]} \BibitemShut
  {NoStop}%
\bibitem [{\citenamefont {{Baumgarte}}\ \emph {et~al.}(1998)\citenamefont
  {{Baumgarte}}, \citenamefont {{Cook}}, \citenamefont {{Scheel}},
  \citenamefont {{Shapiro}},\ and\ \citenamefont {{Teukolsky}}}]{Baumgarte98b}%
  \BibitemOpen
  \bibfield  {author} {\bibinfo {author} {\bibfnamefont {T.~W.}\ \bibnamefont
  {{Baumgarte}}}, \bibinfo {author} {\bibfnamefont {G.~B.}\ \bibnamefont
  {{Cook}}}, \bibinfo {author} {\bibfnamefont {M.~A.}\ \bibnamefont
  {{Scheel}}}, \bibinfo {author} {\bibfnamefont {S.~L.}\ \bibnamefont
  {{Shapiro}}}, \ and\ \bibinfo {author} {\bibfnamefont {S.~A.}\ \bibnamefont
  {{Teukolsky}}},\ }\href {\doibase 10.1103/PhysRevD.57.7299} {\bibfield
  {journal} {\bibinfo  {journal} {Phys. Rev. D}\ }\textbf {\bibinfo {volume}
  {57}},\ \bibinfo {pages} {7299} (\bibinfo {year} {1998})},\ \Eprint
  {http://arxiv.org/abs/gr-qc/9709026} {gr-qc/9709026} \BibitemShut {NoStop}%
\bibitem [{\citenamefont {{Bildsten}}\ and\ \citenamefont
  {{Cutler}}(1992)}]{Bildsten92}%
  \BibitemOpen
  \bibfield  {author} {\bibinfo {author} {\bibfnamefont {L.}~\bibnamefont
  {{Bildsten}}}\ and\ \bibinfo {author} {\bibfnamefont {C.}~\bibnamefont
  {{Cutler}}},\ }\href {\doibase 10.1086/171983} {\bibfield  {journal}
  {\bibinfo  {journal} {Astrophys. J.}\ }\textbf {\bibinfo {volume} {400}},\
  \bibinfo {pages} {175} (\bibinfo {year} {1992})}\BibitemShut {NoStop}%
\bibitem [{\citenamefont {{Asada}}(1998)}]{Asada1998}%
  \BibitemOpen
  \bibfield  {author} {\bibinfo {author} {\bibfnamefont {H.}~\bibnamefont
  {{Asada}}},\ }\href {\doibase 10.1103/PhysRevD.57.7292} {\bibfield  {journal}
  {\bibinfo  {journal} {Phys. Rev. D}\ }\textbf {\bibinfo {volume} {57}},\
  \bibinfo {pages} {7292} (\bibinfo {year} {1998})},\ \Eprint
  {http://arxiv.org/abs/gr-qc/9804003} {gr-qc/9804003} \BibitemShut {NoStop}%
\bibitem [{\citenamefont {{Teukolsky}}(1998)}]{Teukolsky98}%
  \BibitemOpen
  \bibfield  {author} {\bibinfo {author} {\bibfnamefont {S.~A.}\ \bibnamefont
  {{Teukolsky}}},\ }\href {\doibase 10.1086/306082} {\bibfield  {journal}
  {\bibinfo  {journal} {Astrophys. J.}\ }\textbf {\bibinfo {volume} {504}},\
  \bibinfo {pages} {442} (\bibinfo {year} {1998})},\ \Eprint
  {http://arxiv.org/abs/gr-qc/9803082} {gr-qc/9803082} \BibitemShut {NoStop}%
\bibitem [{\citenamefont {{Shibata}}(1998)}]{Shibata98}%
  \BibitemOpen
  \bibfield  {author} {\bibinfo {author} {\bibfnamefont {M.}~\bibnamefont
  {{Shibata}}},\ }\href {\doibase 10.1103/PhysRevD.58.024012} {\bibfield
  {journal} {\bibinfo  {journal} {Phys. Rev. D}\ }\textbf {\bibinfo {volume}
  {58}},\ \bibinfo {eid} {024012} (\bibinfo {year} {1998})},\ \Eprint
  {http://arxiv.org/abs/gr-qc/9803085} {gr-qc/9803085} \BibitemShut {NoStop}%
\bibitem [{\citenamefont {{Kyutoku}}\ \emph
  {et~al.}(2014{\natexlab{a}})\citenamefont {{Kyutoku}}, \citenamefont
  {{Shibata}},\ and\ \citenamefont {{Taniguchi}}}]{Kyutoku2014}%
  \BibitemOpen
  \bibfield  {author} {\bibinfo {author} {\bibfnamefont {K.}~\bibnamefont
  {{Kyutoku}}}, \bibinfo {author} {\bibfnamefont {M.}~\bibnamefont
  {{Shibata}}}, \ and\ \bibinfo {author} {\bibfnamefont {K.}~\bibnamefont
  {{Taniguchi}}},\ }\href {\doibase 10.1103/PhysRevD.90.064006} {\bibfield
  {journal} {\bibinfo  {journal} {Phys. Rev. D}\ }\textbf {\bibinfo {volume}
  {90}},\ \bibinfo {eid} {064006} (\bibinfo {year} {2014}{\natexlab{a}})},\
  \Eprint {http://arxiv.org/abs/1405.6207} {arXiv:1405.6207 [gr-qc]}
  \BibitemShut {NoStop}%
\bibitem [{\citenamefont {Pfeiffer}\ \emph {et~al.}(2007)\citenamefont
  {Pfeiffer} \emph {et~al.}}]{Pfeiffer:2007yz}%
  \BibitemOpen
  \bibfield  {author} {\bibinfo {author} {\bibfnamefont {H.~P.}\ \bibnamefont
  {Pfeiffer}} \emph {et~al.},\ }\href {\doibase 10.1088/0264-9381/24/12/S06}
  {\bibfield  {journal} {\bibinfo  {journal} {Class. Quantum Grav.}\ }\textbf
  {\bibinfo {volume} {24}},\ \bibinfo {pages} {S59} (\bibinfo {year} {2007})},\
  \Eprint {http://arxiv.org/abs/gr-qc/0702106} {arXiv:gr-qc/0702106}
  \BibitemShut {NoStop}%
\bibitem [{\citenamefont {{Hinder}}\ and\ \citenamefont
  {et~al.}(2013)}]{Hinder2013}%
  \BibitemOpen
  \bibfield  {author} {\bibinfo {author} {\bibfnamefont {I.}~\bibnamefont
  {{Hinder}}}\ and\ \bibinfo {author} {\bibnamefont {et~al.}},\ }\href
  {\doibase 10.1088/0264-9381/31/2/025012} {\bibfield  {journal} {\bibinfo
  {journal} {Class. Quantum Grav.}\ }\textbf {\bibinfo {volume} {31}},\
  \bibinfo {eid} {025012} (\bibinfo {year} {2013})},\ \Eprint
  {http://arxiv.org/abs/1307.5307} {arXiv:1307.5307 [gr-qc]} \BibitemShut
  {NoStop}%
\bibitem [{\citenamefont {{Moldenhauer}}\ \emph {et~al.}(2014)\citenamefont
  {{Moldenhauer}}, \citenamefont {{Markakis}}, \citenamefont
  {{Johnson-McDaniel}}, \citenamefont {{Tichy}},\ and\ \citenamefont
  {{Br{\"u}gmann}}}]{Moldenhauer2014}%
  \BibitemOpen
  \bibfield  {author} {\bibinfo {author} {\bibfnamefont {N.}~\bibnamefont
  {{Moldenhauer}}}, \bibinfo {author} {\bibfnamefont {C.~M.}\ \bibnamefont
  {{Markakis}}}, \bibinfo {author} {\bibfnamefont {N.~K.}\ \bibnamefont
  {{Johnson-McDaniel}}}, \bibinfo {author} {\bibfnamefont {W.}~\bibnamefont
  {{Tichy}}}, \ and\ \bibinfo {author} {\bibfnamefont {B.}~\bibnamefont
  {{Br{\"u}gmann}}},\ }\href {\doibase 10.1103/PhysRevD.90.084043} {\bibfield
  {journal} {\bibinfo  {journal} {Phys. Rev. D}\ }\textbf {\bibinfo {volume}
  {90}},\ \bibinfo {eid} {084043} (\bibinfo {year} {2014})},\ \Eprint
  {http://arxiv.org/abs/1408.4136} {arXiv:1408.4136 [gr-qc]} \BibitemShut
  {NoStop}%
\bibitem [{\citenamefont {{Os{\l}owski}}\ \emph {et~al.}(2011)\citenamefont
  {{Os{\l}owski}}, \citenamefont {{Bulik}}, \citenamefont
  {{Gondek-Rosi{\'n}ska}},\ and\ \citenamefont
  {{Belczy{\'n}ski}}}]{Oslowski2011}%
  \BibitemOpen
  \bibfield  {author} {\bibinfo {author} {\bibfnamefont {S.}~\bibnamefont
  {{Os{\l}owski}}}, \bibinfo {author} {\bibfnamefont {T.}~\bibnamefont
  {{Bulik}}}, \bibinfo {author} {\bibfnamefont {D.}~\bibnamefont
  {{Gondek-Rosi{\'n}ska}}}, \ and\ \bibinfo {author} {\bibfnamefont
  {K.}~\bibnamefont {{Belczy{\'n}ski}}},\ }\href {\doibase
  10.1111/j.1365-2966.2010.18147.x} {\bibfield  {journal} {\bibinfo  {journal}
  {Mon. Not. R. Astron. Soc.}\ }\textbf {\bibinfo {volume} {413}},\ \bibinfo
  {pages} {461} (\bibinfo {year} {2011})},\ \Eprint
  {http://arxiv.org/abs/0903.3538} {arXiv:0903.3538 [astro-ph.GA]} \BibitemShut
  {NoStop}%
\bibitem [{\citenamefont {{Kiel}}\ \emph {et~al.}(2010)\citenamefont {{Kiel}},
  \citenamefont {{Hurley}},\ and\ \citenamefont {{Bailes}}}]{Kiel2010}%
  \BibitemOpen
  \bibfield  {author} {\bibinfo {author} {\bibfnamefont {P.~D.}\ \bibnamefont
  {{Kiel}}}, \bibinfo {author} {\bibfnamefont {J.~R.}\ \bibnamefont
  {{Hurley}}}, \ and\ \bibinfo {author} {\bibfnamefont {M.}~\bibnamefont
  {{Bailes}}},\ }\href {\doibase 10.1111/j.1365-2966.2010.16717.x} {\bibfield
  {journal} {\bibinfo  {journal} {Mon. Not. R. Astron. Soc.}\ }\textbf
  {\bibinfo {volume} {406}},\ \bibinfo {pages} {656} (\bibinfo {year}
  {2010})},\ \Eprint {http://arxiv.org/abs/1004.0131} {arXiv:1004.0131
  [astro-ph.GA]} \BibitemShut {NoStop}%
\bibitem [{\citenamefont {{Tichy}}(2011)}]{Tichy11}%
  \BibitemOpen
  \bibfield  {author} {\bibinfo {author} {\bibfnamefont {W.}~\bibnamefont
  {{Tichy}}},\ }\href {\doibase 10.1103/PhysRevD.84.024041} {\bibfield
  {journal} {\bibinfo  {journal} {Phys. Rev. D}\ }\textbf {\bibinfo {volume}
  {84}},\ \bibinfo {eid} {024041} (\bibinfo {year} {2011})},\ \Eprint
  {http://arxiv.org/abs/1107.1440} {arXiv:1107.1440 [gr-qc]} \BibitemShut
  {NoStop}%
\bibitem [{\citenamefont {{Marronetti}}\ and\ \citenamefont
  {{Shapiro}}(2003)}]{Marronetti03}%
  \BibitemOpen
  \bibfield  {author} {\bibinfo {author} {\bibfnamefont {P.}~\bibnamefont
  {{Marronetti}}}\ and\ \bibinfo {author} {\bibfnamefont {S.~L.}\ \bibnamefont
  {{Shapiro}}},\ }\href {\doibase 10.1103/PhysRevD.68.104024} {\bibfield
  {journal} {\bibinfo  {journal} {Phys. Rev. D}\ }\textbf {\bibinfo {volume}
  {68}},\ \bibinfo {eid} {104024} (\bibinfo {year} {2003})},\ \Eprint
  {http://arxiv.org/abs/gr-qc/0306075} {gr-qc/0306075} \BibitemShut {NoStop}%
\bibitem [{\citenamefont {{Baumgarte}}\ and\ \citenamefont
  {{Shapiro}}(2009)}]{Baumgarte:2009}%
  \BibitemOpen
  \bibfield  {author} {\bibinfo {author} {\bibfnamefont {T.~W.}\ \bibnamefont
  {{Baumgarte}}}\ and\ \bibinfo {author} {\bibfnamefont {S.~L.}\ \bibnamefont
  {{Shapiro}}},\ }\href {\doibase 10.1103/PhysRevD.80.064009} {\bibfield
  {journal} {\bibinfo  {journal} {Phys. Rev. D}\ }\textbf {\bibinfo {volume}
  {80}},\ \bibinfo {eid} {064009} (\bibinfo {year} {2009})},\ \Eprint
  {http://arxiv.org/abs/0909.0952} {arXiv:0909.0952 [gr-qc]} \BibitemShut
  {NoStop}%
\bibitem [{\citenamefont {{Tichy}}(2012)}]{Tichy12}%
  \BibitemOpen
  \bibfield  {author} {\bibinfo {author} {\bibfnamefont {W.}~\bibnamefont
  {{Tichy}}},\ }\href {\doibase 10.1103/PhysRevD.86.064024} {\bibfield
  {journal} {\bibinfo  {journal} {Phys. Rev. D}\ }\textbf {\bibinfo {volume}
  {86}},\ \bibinfo {eid} {064024} (\bibinfo {year} {2012})},\ \Eprint
  {http://arxiv.org/abs/1209.5336} {arXiv:1209.5336 [gr-qc]} \BibitemShut
  {NoStop}%
\bibitem [{\citenamefont {{East}}\ \emph
  {et~al.}(2012{\natexlab{b}})\citenamefont {{East}}, \citenamefont
  {{Ramazano{\v g}lu}},\ and\ \citenamefont {{Pretorius}}}]{East2012d}%
  \BibitemOpen
  \bibfield  {author} {\bibinfo {author} {\bibfnamefont {W.~E.}\ \bibnamefont
  {{East}}}, \bibinfo {author} {\bibfnamefont {F.~M.}\ \bibnamefont
  {{Ramazano{\v g}lu}}}, \ and\ \bibinfo {author} {\bibfnamefont
  {F.}~\bibnamefont {{Pretorius}}},\ }\href {\doibase
  10.1103/PhysRevD.86.104053} {\bibfield  {journal} {\bibinfo  {journal} {Phys.
  Rev. D}\ }\textbf {\bibinfo {volume} {86}},\ \bibinfo {eid} {104053}
  (\bibinfo {year} {2012}{\natexlab{b}})},\ \Eprint
  {http://arxiv.org/abs/1208.3473} {arXiv:1208.3473 [gr-qc]} \BibitemShut
  {NoStop}%
\bibitem [{\citenamefont {{Kastaun}}\ \emph {et~al.}(2013)\citenamefont
  {{Kastaun}}, \citenamefont {{Galeazzi}}, \citenamefont {{Alic}},
  \citenamefont {{Rezzolla}},\ and\ \citenamefont {{Font}}}]{Kastaun2013}%
  \BibitemOpen
  \bibfield  {author} {\bibinfo {author} {\bibfnamefont {W.}~\bibnamefont
  {{Kastaun}}}, \bibinfo {author} {\bibfnamefont {F.}~\bibnamefont
  {{Galeazzi}}}, \bibinfo {author} {\bibfnamefont {D.}~\bibnamefont {{Alic}}},
  \bibinfo {author} {\bibfnamefont {L.}~\bibnamefont {{Rezzolla}}}, \ and\
  \bibinfo {author} {\bibfnamefont {J.~A.}\ \bibnamefont {{Font}}},\ }\href
  {\doibase 10.1103/PhysRevD.88.021501} {\bibfield  {journal} {\bibinfo
  {journal} {Phys. Rev. D}\ }\textbf {\bibinfo {volume} {88}},\ \bibinfo {eid}
  {021501} (\bibinfo {year} {2013})},\ \Eprint {http://arxiv.org/abs/1301.7348}
  {arXiv:1301.7348 [gr-qc]} \BibitemShut {NoStop}%
\bibitem [{\citenamefont {{Tacik}}\ \emph {et~al.}(2015)\citenamefont
  {{Tacik}}, \citenamefont {{Foucart}}, \citenamefont {{Pfeiffer}},
  \citenamefont {{Haas}}, \citenamefont {{Ossokine}}, \citenamefont {{Kaplan}},
  \citenamefont {{Muhlberger}}, \citenamefont {{Duez}}, \citenamefont
  {{Kidder}}, \citenamefont {{Scheel}},\ and\ \citenamefont
  {{Szil{\'a}gyi}}}]{Tacik15}%
  \BibitemOpen
  \bibfield  {author} {\bibinfo {author} {\bibfnamefont {N.}~\bibnamefont
  {{Tacik}}}, \bibinfo {author} {\bibfnamefont {F.}~\bibnamefont {{Foucart}}},
  \bibinfo {author} {\bibfnamefont {H.~P.}\ \bibnamefont {{Pfeiffer}}},
  \bibinfo {author} {\bibfnamefont {R.}~\bibnamefont {{Haas}}}, \bibinfo
  {author} {\bibfnamefont {S.}~\bibnamefont {{Ossokine}}}, \bibinfo {author}
  {\bibfnamefont {J.}~\bibnamefont {{Kaplan}}}, \bibinfo {author}
  {\bibfnamefont {C.}~\bibnamefont {{Muhlberger}}}, \bibinfo {author}
  {\bibfnamefont {M.~D.}\ \bibnamefont {{Duez}}}, \bibinfo {author}
  {\bibfnamefont {L.~E.}\ \bibnamefont {{Kidder}}}, \bibinfo {author}
  {\bibfnamefont {M.~A.}\ \bibnamefont {{Scheel}}}, \ and\ \bibinfo {author}
  {\bibfnamefont {B.}~\bibnamefont {{Szil{\'a}gyi}}},\ }\href {\doibase
  10.1103/PhysRevD.92.124012} {\bibfield  {journal} {\bibinfo  {journal} {Phys.
  Rev. D}\ }\textbf {\bibinfo {volume} {92}},\ \bibinfo {eid} {124012}
  (\bibinfo {year} {2015})},\ \Eprint {http://arxiv.org/abs/1508.06986}
  {arXiv:1508.06986 [gr-qc]} \BibitemShut {NoStop}%
\bibitem [{\citenamefont {Pfeiffer}\ \emph {et~al.}(2003)\citenamefont
  {Pfeiffer}, \citenamefont {Kidder}, \citenamefont {Scheel},\ and\
  \citenamefont {Teukolsky}}]{Pfeiffer:2002wt}%
  \BibitemOpen
  \bibfield  {author} {\bibinfo {author} {\bibfnamefont {H.~P.}\ \bibnamefont
  {Pfeiffer}}, \bibinfo {author} {\bibfnamefont {L.~E.}\ \bibnamefont
  {Kidder}}, \bibinfo {author} {\bibfnamefont {M.~A.}\ \bibnamefont {Scheel}},
  \ and\ \bibinfo {author} {\bibfnamefont {S.~A.}\ \bibnamefont {Teukolsky}},\
  }\href@noop {} {\bibfield  {journal} {\bibinfo  {journal} {Comput. Phys.
  Commun.}\ }\textbf {\bibinfo {volume} {152}},\ \bibinfo {pages} {253}
  (\bibinfo {year} {2003})},\ \Eprint {http://arxiv.org/abs/gr-qc/0202096}
  {gr-qc/0202096} \BibitemShut {NoStop}%
\bibitem [{\citenamefont {{Foucart}}\ \emph {et~al.}(2008)\citenamefont
  {{Foucart}}, \citenamefont {{Kidder}}, \citenamefont {{Pfeiffer}},\ and\
  \citenamefont {{Teukolsky}}}]{Foucart2008}%
  \BibitemOpen
  \bibfield  {author} {\bibinfo {author} {\bibfnamefont {F.}~\bibnamefont
  {{Foucart}}}, \bibinfo {author} {\bibfnamefont {L.~E.}\ \bibnamefont
  {{Kidder}}}, \bibinfo {author} {\bibfnamefont {H.~P.}\ \bibnamefont
  {{Pfeiffer}}}, \ and\ \bibinfo {author} {\bibfnamefont {S.~A.}\ \bibnamefont
  {{Teukolsky}}},\ }\href {\doibase 10.1103/PhysRevD.77.124051} {\bibfield
  {journal} {\bibinfo  {journal} {Phys. Rev. D}\ }\textbf {\bibinfo {volume}
  {77}},\ \bibinfo {eid} {124051} (\bibinfo {year} {2008})},\ \Eprint
  {http://arxiv.org/abs/0804.3787} {arXiv:0804.3787 [gr-qc]} \BibitemShut
  {NoStop}%
\bibitem [{\citenamefont {{Foucart}}\ \emph {et~al.}(2012)\citenamefont
  {{Foucart}}, \citenamefont {{Duez}}, \citenamefont {{Kidder}}, \citenamefont
  {{Scheel}}, \citenamefont {{Szilagyi}},\ and\ \citenamefont
  {{Teukolsky}}}]{Foucart2011}%
  \BibitemOpen
  \bibfield  {author} {\bibinfo {author} {\bibfnamefont {F.}~\bibnamefont
  {{Foucart}}}, \bibinfo {author} {\bibfnamefont {M.~D.}\ \bibnamefont
  {{Duez}}}, \bibinfo {author} {\bibfnamefont {L.~E.}\ \bibnamefont
  {{Kidder}}}, \bibinfo {author} {\bibfnamefont {M.~A.}\ \bibnamefont
  {{Scheel}}}, \bibinfo {author} {\bibfnamefont {B.}~\bibnamefont
  {{Szilagyi}}}, \ and\ \bibinfo {author} {\bibfnamefont {S.~A.}\ \bibnamefont
  {{Teukolsky}}},\ }\href {\doibase 10.1103/PhysRevD.85.044015} {\bibfield
  {journal} {\bibinfo  {journal} {Phys. Rev. D}\ }\textbf {\bibinfo {volume}
  {85}},\ \bibinfo {eid} {044015} (\bibinfo {year} {2012})},\ \Eprint
  {http://arxiv.org/abs/1111.1677} {arXiv:1111.1677 [gr-qc]} \BibitemShut
  {NoStop}%
\bibitem [{\citenamefont {Boyle}\ \emph {et~al.}(2007)\citenamefont {Boyle},
  \citenamefont {Barrow}, \citenamefont {Kidder}, \citenamefont {{Mrou\'e}},
  \citenamefont {Pfeiffer}, \citenamefont {Scheel}, \citenamefont {Cook},\ and\
  \citenamefont {Teukolsky}}]{Boyle:2007ft}%
  \BibitemOpen
  \bibfield  {author} {\bibinfo {author} {\bibfnamefont {M.}~\bibnamefont
  {Boyle}}, \bibinfo {author} {\bibfnamefont {D.~A.}\ \bibnamefont {Barrow}},
  \bibinfo {author} {\bibfnamefont {L.~E.}\ \bibnamefont {Kidder}}, \bibinfo
  {author} {\bibfnamefont {A.~H.}\ \bibnamefont {{Mrou\'e}}}, \bibinfo {author}
  {\bibfnamefont {H.~P.}\ \bibnamefont {Pfeiffer}}, \bibinfo {author}
  {\bibfnamefont {M.~A.}\ \bibnamefont {Scheel}}, \bibinfo {author}
  {\bibfnamefont {G.~B.}\ \bibnamefont {Cook}}, \ and\ \bibinfo {author}
  {\bibfnamefont {S.~A.}\ \bibnamefont {Teukolsky}},\ }\href {\doibase
  10.1103/PhysRevD.76.124038} {\bibfield  {journal} {\bibinfo  {journal} {Phys.
  Rev. D}\ }\textbf {\bibinfo {volume} {76}},\ \bibinfo {pages} {124038}
  (\bibinfo {year} {2007})},\ \Eprint {http://arxiv.org/abs/0710.0158}
  {arXiv:0710.0158 [gr-qc]} \BibitemShut {NoStop}%
\bibitem [{\citenamefont {{Buonanno}}\ \emph {et~al.}(2011)\citenamefont
  {{Buonanno}}, \citenamefont {{Kidder}}, \citenamefont {{Mrou{\'e}}},
  \citenamefont {{Pfeiffer}},\ and\ \citenamefont
  {{Taracchini}}}]{Buonanno2011}%
  \BibitemOpen
  \bibfield  {author} {\bibinfo {author} {\bibfnamefont {A.}~\bibnamefont
  {{Buonanno}}}, \bibinfo {author} {\bibfnamefont {L.~E.}\ \bibnamefont
  {{Kidder}}}, \bibinfo {author} {\bibfnamefont {A.~H.}\ \bibnamefont
  {{Mrou{\'e}}}}, \bibinfo {author} {\bibfnamefont {H.~P.}\ \bibnamefont
  {{Pfeiffer}}}, \ and\ \bibinfo {author} {\bibfnamefont {A.}~\bibnamefont
  {{Taracchini}}},\ }\href {\doibase 10.1103/PhysRevD.83.104034} {\bibfield
  {journal} {\bibinfo  {journal} {Phys. Rev. D}\ }\textbf {\bibinfo {volume}
  {83}},\ \bibinfo {eid} {104034} (\bibinfo {year} {2011})},\ \Eprint
  {http://arxiv.org/abs/1012.1549} {arXiv:1012.1549 [gr-qc]} \BibitemShut
  {NoStop}%
\bibitem [{\citenamefont {{Taniguchi}}\ \emph {et~al.}(2015)\citenamefont
  {{Taniguchi}}, \citenamefont {{Shibata}},\ and\ \citenamefont
  {{Buonanno}}}]{Taniguchi2015}%
  \BibitemOpen
  \bibfield  {author} {\bibinfo {author} {\bibfnamefont {K.}~\bibnamefont
  {{Taniguchi}}}, \bibinfo {author} {\bibfnamefont {M.}~\bibnamefont
  {{Shibata}}}, \ and\ \bibinfo {author} {\bibfnamefont {A.}~\bibnamefont
  {{Buonanno}}},\ }\href {\doibase 10.1103/PhysRevD.91.024033} {\bibfield
  {journal} {\bibinfo  {journal} {Phys. Rev. D}\ }\textbf {\bibinfo {volume}
  {91}},\ \bibinfo {eid} {024033} (\bibinfo {year} {2015})},\ \Eprint
  {http://arxiv.org/abs/1410.0738} {arXiv:1410.0738 [gr-qc]} \BibitemShut
  {NoStop}%
\bibitem [{\citenamefont {{Tsokaros}}\ \emph {et~al.}(2016)\citenamefont
  {{Tsokaros}}, \citenamefont {{Mundim}}, \citenamefont {{Galeazzi}},
  \citenamefont {{Rezzolla}},\ and\ \citenamefont {{Ury{\=
  u}}}}]{Tsokaros2016}%
  \BibitemOpen
  \bibfield  {author} {\bibinfo {author} {\bibfnamefont {A.}~\bibnamefont
  {{Tsokaros}}}, \bibinfo {author} {\bibfnamefont {B.~C.}\ \bibnamefont
  {{Mundim}}}, \bibinfo {author} {\bibfnamefont {F.}~\bibnamefont
  {{Galeazzi}}}, \bibinfo {author} {\bibfnamefont {L.}~\bibnamefont
  {{Rezzolla}}}, \ and\ \bibinfo {author} {\bibfnamefont {K.}~\bibnamefont
  {{Ury{\= u}}}},\ }\href@noop {} {\bibfield  {journal} {\bibinfo  {journal}
  {arXiv:1605.07205}\ } (\bibinfo {year} {2016})},\ \Eprint
  {http://arxiv.org/abs/1605.07205} {arXiv:1605.07205 [gr-qc]} \BibitemShut
  {NoStop}%
\bibitem [{\citenamefont {{Radice}}\ \emph
  {et~al.}(2014{\natexlab{b}})\citenamefont {{Radice}}, \citenamefont
  {{Rezzolla}},\ and\ \citenamefont {{Galeazzi}}}]{Radice2013b}%
  \BibitemOpen
  \bibfield  {author} {\bibinfo {author} {\bibfnamefont {D.}~\bibnamefont
  {{Radice}}}, \bibinfo {author} {\bibfnamefont {L.}~\bibnamefont
  {{Rezzolla}}}, \ and\ \bibinfo {author} {\bibfnamefont {F.}~\bibnamefont
  {{Galeazzi}}},\ }\href {\doibase 10.1093/mnrasl/slt137} {\bibfield  {journal}
  {\bibinfo  {journal} {Mon. Not. R. Astron. Soc. L.}\ }\textbf {\bibinfo
  {volume} {437}},\ \bibinfo {pages} {L46} (\bibinfo {year}
  {2014}{\natexlab{b}})},\ \Eprint {http://arxiv.org/abs/1306.6052}
  {arXiv:1306.6052 [gr-qc]} \BibitemShut {NoStop}%
\bibitem [{\citenamefont {{Radice}}\ \emph {et~al.}(2015)\citenamefont
  {{Radice}}, \citenamefont {{Rezzolla}},\ and\ \citenamefont
  {{Galeazzi}}}]{Radice2015}%
  \BibitemOpen
  \bibfield  {author} {\bibinfo {author} {\bibfnamefont {D.}~\bibnamefont
  {{Radice}}}, \bibinfo {author} {\bibfnamefont {L.}~\bibnamefont
  {{Rezzolla}}}, \ and\ \bibinfo {author} {\bibfnamefont {F.}~\bibnamefont
  {{Galeazzi}}},\ }\bibfield  {booktitle} {\emph {\bibinfo {booktitle}
  {Numerical Modeling of Space Plasma Flows ASTRONUM-2014}},\ }\href@noop {} {\
  \bibinfo {series} {Astronomical Society of the Pacific Conference Series},\
  \textbf {\bibinfo {volume} {498}},\ \bibinfo {pages} {121} (\bibinfo {year}
  {2015})},\ \Eprint {http://arxiv.org/abs/1502.00551} {arXiv:1502.00551
  [gr-qc]} \BibitemShut {NoStop}%
\bibitem [{\citenamefont {Nakamura}\ and\ \citenamefont
  {Oohara}(1998)}]{Nakamura99a}%
  \BibitemOpen
  \bibfield  {author} {\bibinfo {author} {\bibfnamefont {T.}~\bibnamefont
  {Nakamura}}\ and\ \bibinfo {author} {\bibfnamefont {K.}~\bibnamefont
  {Oohara}},\ }in\ \href@noop {} {\emph {\bibinfo {booktitle} {Numerical
  Astrophysics 1998 (NAP98) -- Proceedings}}}\ (\bibinfo {year} {1998})\
  \Eprint {http://arxiv.org/abs/gr-qc/9812054} {gr-qc/9812054} \BibitemShut
  {NoStop}%
\bibitem [{\citenamefont {Oohara}\ and\ \citenamefont
  {Nakamura}(1999)}]{Oohara:1999}%
  \BibitemOpen
  \bibfield  {author} {\bibinfo {author} {\bibfnamefont {K.-i.}\ \bibnamefont
  {Oohara}}\ and\ \bibinfo {author} {\bibfnamefont {T.}~\bibnamefont
  {Nakamura}},\ }\bibfield  {booktitle} {\emph {\bibinfo {booktitle} {{Black
  holes and gravitational waves: New eyes in the 21st century. Proceedings, 9th
  Yukawa International Seminar, Kyoto, Japan, June 28-July 2, 1999}}},\ }\href
  {\doibase 10.1143/PTPS.136.270} {\bibfield  {journal} {\bibinfo  {journal}
  {Prog. Theor. Phys. Suppl.}\ }\textbf {\bibinfo {volume} {136}},\ \bibinfo
  {pages} {270} (\bibinfo {year} {1999})},\ \Eprint
  {http://arxiv.org/abs/astro-ph/9912085} {arXiv:astro-ph/9912085 [astro-ph]}
  \BibitemShut {NoStop}%
\bibitem [{\citenamefont {{Shibata}}(1999{\natexlab{a}})}]{Shibata99c}%
  \BibitemOpen
  \bibfield  {author} {\bibinfo {author} {\bibfnamefont {M.}~\bibnamefont
  {{Shibata}}},\ }\href {\doibase 10.1103/PhysRevD.60.104052} {\bibfield
  {journal} {\bibinfo  {journal} {Phys. Rev. D}\ }\textbf {\bibinfo {volume}
  {60}},\ \bibinfo {eid} {104052} (\bibinfo {year} {1999}{\natexlab{a}})},\
  \Eprint {http://arxiv.org/abs/gr-qc/9908027} {gr-qc/9908027} \BibitemShut
  {NoStop}%
\bibitem [{\citenamefont {{Shibata}}\ and\ \citenamefont {{Ury{\=
  u}}}(2002)}]{Shibata02a}%
  \BibitemOpen
  \bibfield  {author} {\bibinfo {author} {\bibfnamefont {M.}~\bibnamefont
  {{Shibata}}}\ and\ \bibinfo {author} {\bibfnamefont {K.}~\bibnamefont
  {{Ury{\= u}}}},\ }\href {\doibase 10.1143/PTP.107.265} {\bibfield  {journal}
  {\bibinfo  {journal} {Progress of Theoretical Physics}\ }\textbf {\bibinfo
  {volume} {107}},\ \bibinfo {pages} {265} (\bibinfo {year} {2002})},\ \Eprint
  {http://arxiv.org/abs/gr-qc/0203037} {gr-qc/0203037} \BibitemShut {NoStop}%
\bibitem [{\citenamefont {{Shibata}}\ \emph {et~al.}(2003)\citenamefont
  {{Shibata}}, \citenamefont {{Taniguchi}},\ and\ \citenamefont {{Ury{\=
  u}}}}]{Shibata:2003ga}%
  \BibitemOpen
  \bibfield  {author} {\bibinfo {author} {\bibfnamefont {M.}~\bibnamefont
  {{Shibata}}}, \bibinfo {author} {\bibfnamefont {K.}~\bibnamefont
  {{Taniguchi}}}, \ and\ \bibinfo {author} {\bibfnamefont {K.}~\bibnamefont
  {{Ury{\= u}}}},\ }\href {\doibase 10.1103/PhysRevD.68.084020} {\bibfield
  {journal} {\bibinfo  {journal} {Phys. Rev. D}\ }\textbf {\bibinfo {volume}
  {68}},\ \bibinfo {eid} {084020} (\bibinfo {year} {2003})},\ \Eprint
  {http://arxiv.org/abs/gr-qc/0310030} {gr-qc/0310030} \BibitemShut {NoStop}%
\bibitem [{\citenamefont {{Stergioulas}}\ \emph {et~al.}(2004)\citenamefont
  {{Stergioulas}}, \citenamefont {{Apostolatos}},\ and\ \citenamefont
  {{Font}}}]{Stergioulas04}%
  \BibitemOpen
  \bibfield  {author} {\bibinfo {author} {\bibfnamefont {N.}~\bibnamefont
  {{Stergioulas}}}, \bibinfo {author} {\bibfnamefont {T.~A.}\ \bibnamefont
  {{Apostolatos}}}, \ and\ \bibinfo {author} {\bibfnamefont {J.~A.}\
  \bibnamefont {{Font}}},\ }\href {\doibase 10.1111/j.1365-2966.2004.07973.x}
  {\bibfield  {journal} {\bibinfo  {journal} {Mon. Not. R. Astron. Soc.}\
  }\textbf {\bibinfo {volume} {352}},\ \bibinfo {pages} {1089} (\bibinfo {year}
  {2004})},\ \Eprint {http://arxiv.org/abs/arXiv:astro-ph/0312648}
  {arXiv:astro-ph/0312648} \BibitemShut {NoStop}%
\bibitem [{\citenamefont {{Nagakura}}\ \emph {et~al.}(2014)\citenamefont
  {{Nagakura}}, \citenamefont {{Hotokezaka}}, \citenamefont {{Sekiguchi}},
  \citenamefont {{Shibata}},\ and\ \citenamefont {{Ioka}}}]{Nagakura2014}%
  \BibitemOpen
  \bibfield  {author} {\bibinfo {author} {\bibfnamefont {H.}~\bibnamefont
  {{Nagakura}}}, \bibinfo {author} {\bibfnamefont {K.}~\bibnamefont
  {{Hotokezaka}}}, \bibinfo {author} {\bibfnamefont {Y.}~\bibnamefont
  {{Sekiguchi}}}, \bibinfo {author} {\bibfnamefont {M.}~\bibnamefont
  {{Shibata}}}, \ and\ \bibinfo {author} {\bibfnamefont {K.}~\bibnamefont
  {{Ioka}}},\ }\href {\doibase 10.1088/2041-8205/784/2/L28} {\bibfield
  {journal} {\bibinfo  {journal} {Astrophys. J.}\ }\textbf {\bibinfo {volume}
  {784}},\ \bibinfo {eid} {L28} (\bibinfo {year} {2014})},\ \Eprint
  {http://arxiv.org/abs/1403.0956} {arXiv:1403.0956 [astro-ph.HE]} \BibitemShut
  {NoStop}%
\bibitem [{\citenamefont {Chandrasekhar}(1981)}]{Chandrasekhar81}%
  \BibitemOpen
  \bibfield  {author} {\bibinfo {author} {\bibfnamefont {S.}~\bibnamefont
  {Chandrasekhar}},\ }\href@noop {} {\emph {\bibinfo {title} {Hydrodynamic and
  hydromagnetic stability}}},\ Chandrasekhar81\ (\bibinfo  {publisher} {Dover
  Edition},\ \bibinfo {address} {New York, USA},\ \bibinfo {year}
  {1981})\BibitemShut {NoStop}%
\bibitem [{\citenamefont {{Bodo}}\ \emph {et~al.}(1994)\citenamefont {{Bodo}},
  \citenamefont {{Massaglia}}, \citenamefont {{Ferrari}},\ and\ \citenamefont
  {{Trussoni}}}]{Bodo1994}%
  \BibitemOpen
  \bibfield  {author} {\bibinfo {author} {\bibfnamefont {G.}~\bibnamefont
  {{Bodo}}}, \bibinfo {author} {\bibfnamefont {S.}~\bibnamefont {{Massaglia}}},
  \bibinfo {author} {\bibfnamefont {A.}~\bibnamefont {{Ferrari}}}, \ and\
  \bibinfo {author} {\bibfnamefont {E.}~\bibnamefont {{Trussoni}}},\
  }\href@noop {} {\bibfield  {journal} {\bibinfo  {journal} {Astron.
  Astrophys.}\ }\textbf {\bibinfo {volume} {283}},\ \bibinfo {pages} {655}
  (\bibinfo {year} {1994})}\BibitemShut {NoStop}%
\bibitem [{\citenamefont {{Price}}\ and\ \citenamefont
  {{Rosswog}}(2006)}]{Price06}%
  \BibitemOpen
  \bibfield  {author} {\bibinfo {author} {\bibfnamefont {D.~J.}\ \bibnamefont
  {{Price}}}\ and\ \bibinfo {author} {\bibfnamefont {S.}~\bibnamefont
  {{Rosswog}}},\ }\href {\doibase 10.1126/science.1125201} {\bibfield
  {journal} {\bibinfo  {journal} {Science}\ }\textbf {\bibinfo {volume}
  {312}},\ \bibinfo {pages} {719} (\bibinfo {year} {2006})},\ \Eprint
  {http://arxiv.org/abs/astro-ph/0603845} {astro-ph/0603845} \BibitemShut
  {NoStop}%
\bibitem [{\citenamefont {{Giacomazzo}}\ \emph {et~al.}(2011)\citenamefont
  {{Giacomazzo}}, \citenamefont {{Rezzolla}},\ and\ \citenamefont
  {{Baiotti}}}]{Giacomazzo2011b}%
  \BibitemOpen
  \bibfield  {author} {\bibinfo {author} {\bibfnamefont {B.}~\bibnamefont
  {{Giacomazzo}}}, \bibinfo {author} {\bibfnamefont {L.}~\bibnamefont
  {{Rezzolla}}}, \ and\ \bibinfo {author} {\bibfnamefont {L.}~\bibnamefont
  {{Baiotti}}},\ }\href {\doibase 10.1103/PhysRevD.83.044014} {\bibfield
  {journal} {\bibinfo  {journal} {Phys. Rev. D}\ }\textbf {\bibinfo {volume}
  {83}},\ \bibinfo {eid} {044014} (\bibinfo {year} {2011})},\ \Eprint
  {http://arxiv.org/abs/1009.2468} {arXiv:1009.2468 [gr-qc]} \BibitemShut
  {NoStop}%
\bibitem [{\citenamefont {{Neilsen}}\ \emph {et~al.}(2014)\citenamefont
  {{Neilsen}}, \citenamefont {{Liebling}}, \citenamefont {{Anderson}},
  \citenamefont {{Lehner}}, \citenamefont {{O'Connor}},\ and\ \citenamefont
  {{Palenzuela}}}]{Neilsen2014}%
  \BibitemOpen
  \bibfield  {author} {\bibinfo {author} {\bibfnamefont {D.~W.}\ \bibnamefont
  {{Neilsen}}}, \bibinfo {author} {\bibfnamefont {S.~L.}\ \bibnamefont
  {{Liebling}}}, \bibinfo {author} {\bibfnamefont {M.}~\bibnamefont
  {{Anderson}}}, \bibinfo {author} {\bibfnamefont {L.}~\bibnamefont
  {{Lehner}}}, \bibinfo {author} {\bibfnamefont {E.}~\bibnamefont
  {{O'Connor}}}, \ and\ \bibinfo {author} {\bibfnamefont {C.}~\bibnamefont
  {{Palenzuela}}},\ }\href {\doibase 10.1103/PhysRevD.89.104029} {\bibfield
  {journal} {\bibinfo  {journal} {Phys. Rev. D}\ }\textbf {\bibinfo {volume}
  {89}},\ \bibinfo {eid} {104029} (\bibinfo {year} {2014})},\ \Eprint
  {http://arxiv.org/abs/1403.3680} {arXiv:1403.3680 [gr-qc]} \BibitemShut
  {NoStop}%
\bibitem [{\citenamefont {{Rembiasz}}\ \emph {et~al.}(2016)\citenamefont
  {{Rembiasz}}, \citenamefont {{Guilet}}, \citenamefont {{Obergaulinger}},
  \citenamefont {{Cerd{\'a}-Dur{\'a}n}}, \citenamefont {{Aloy}},\ and\
  \citenamefont {{M{\"u}ller}}}]{Rembiasz2016}%
  \BibitemOpen
  \bibfield  {author} {\bibinfo {author} {\bibfnamefont {T.}~\bibnamefont
  {{Rembiasz}}}, \bibinfo {author} {\bibfnamefont {J.}~\bibnamefont
  {{Guilet}}}, \bibinfo {author} {\bibfnamefont {M.}~\bibnamefont
  {{Obergaulinger}}}, \bibinfo {author} {\bibfnamefont {P.}~\bibnamefont
  {{Cerd{\'a}-Dur{\'a}n}}}, \bibinfo {author} {\bibfnamefont {M.~A.}\
  \bibnamefont {{Aloy}}}, \ and\ \bibinfo {author} {\bibfnamefont
  {E.}~\bibnamefont {{M{\"u}ller}}},\ }\href {\doibase 10.1093/mnras/stw1201}
  {\bibfield  {journal} {\bibinfo  {journal} {Mon. Not. R. Astron. Soc.}\
  }\textbf {\bibinfo {volume} {460}},\ \bibinfo {pages} {3316} (\bibinfo {year}
  {2016})},\ \Eprint {http://arxiv.org/abs/1603.00466} {arXiv:1603.00466
  [astro-ph.SR]} \BibitemShut {NoStop}%
\bibitem [{\citenamefont {{Stergioulas}}\ \emph {et~al.}(2011)\citenamefont
  {{Stergioulas}}, \citenamefont {{Bauswein}}, \citenamefont {{Zagkouris}},\
  and\ \citenamefont {{Janka}}}]{Stergioulas2011b}%
  \BibitemOpen
  \bibfield  {author} {\bibinfo {author} {\bibfnamefont {N.}~\bibnamefont
  {{Stergioulas}}}, \bibinfo {author} {\bibfnamefont {A.}~\bibnamefont
  {{Bauswein}}}, \bibinfo {author} {\bibfnamefont {K.}~\bibnamefont
  {{Zagkouris}}}, \ and\ \bibinfo {author} {\bibfnamefont {H.-T.}\ \bibnamefont
  {{Janka}}},\ }\href {\doibase 10.1111/j.1365-2966.2011.19493.x} {\bibfield
  {journal} {\bibinfo  {journal} {Mon. Not. R. Astron. Soc.}\ }\textbf
  {\bibinfo {volume} {418}},\ \bibinfo {pages} {427} (\bibinfo {year}
  {2011})},\ \Eprint {http://arxiv.org/abs/1105.0368} {arXiv:1105.0368 [gr-qc]}
  \BibitemShut {NoStop}%
\bibitem [{\citenamefont {{Doneva}}\ \emph {et~al.}(2015)\citenamefont
  {{Doneva}}, \citenamefont {{Kokkotas}},\ and\ \citenamefont
  {{Pnigouras}}}]{Doneva2015}%
  \BibitemOpen
  \bibfield  {author} {\bibinfo {author} {\bibfnamefont {D.~D.}\ \bibnamefont
  {{Doneva}}}, \bibinfo {author} {\bibfnamefont {K.~D.}\ \bibnamefont
  {{Kokkotas}}}, \ and\ \bibinfo {author} {\bibfnamefont {P.}~\bibnamefont
  {{Pnigouras}}},\ }\href {\doibase 10.1103/PhysRevD.92.104040} {\bibfield
  {journal} {\bibinfo  {journal} {Phys. Rev. D}\ }\textbf {\bibinfo {volume}
  {92}},\ \bibinfo {eid} {104040} (\bibinfo {year} {2015})},\ \Eprint
  {http://arxiv.org/abs/1510.00673} {arXiv:1510.00673 [gr-qc]} \BibitemShut
  {NoStop}%
\bibitem [{\citenamefont {Shoemaker}(2009)}]{Shoemaker2009}%
  \BibitemOpen
  \bibfield  {author} {\bibinfo {author} {\bibfnamefont {D.}~\bibnamefont
  {Shoemaker}},\ }\href
  {https://dcc.ligo.org/cgi-bin/DocDB/ShowDocument?docid=2974} {\emph {\bibinfo
  {title} {Advanced {LIGO} anticipated sensitivity curves}}},\ \bibinfo {type}
  {Technical notes}\ \bibinfo {number} {LIGO-T0900288-v2}\ (\bibinfo
  {institution} {LIGO},\ \bibinfo {year} {2009})\ \bibinfo {note} {note: the
  high-power detuned model used in this paper is given in the data file
  \texttt{ZERO\_DET\_high\_P.txt}}\BibitemShut {NoStop}%
\bibitem [{\citenamefont {Punturo}\ \emph
  {et~al.}(2010{\natexlab{a}})\citenamefont {Punturo} \emph
  {et~al.}}]{Punturo:2010}%
  \BibitemOpen
  \bibfield  {author} {\bibinfo {author} {\bibfnamefont {M.}~\bibnamefont
  {Punturo}} \emph {et~al.},\ }\href {\doibase 10.1088/0264-9381/27/8/084007}
  {\bibfield  {journal} {\bibinfo  {journal} {Class. Quantum Grav.}\ }\textbf
  {\bibinfo {volume} {27}},\ \bibinfo {pages} {084007} (\bibinfo {year}
  {2010}{\natexlab{a}})}\BibitemShut {NoStop}%
\bibitem [{\citenamefont {{Santamar{\'{\i}}a}}\ \emph
  {et~al.}(2010)\citenamefont {{Santamar{\'{\i}}a}}, \citenamefont {{Ohme}},
  \citenamefont {{Ajith}}, \citenamefont {{Br{\"u}gmann}}, \citenamefont
  {{Dorband}}, \citenamefont {{Hannam}}, \citenamefont {{Husa}}, \citenamefont
  {{M{\"o}sta}}, \citenamefont {{Pollney}}, \citenamefont {{Reisswig}},
  \citenamefont {{Robinson}}, \citenamefont {{Seiler}},\ and\ \citenamefont
  {{Krishnan}}}]{Santamaria2010}%
  \BibitemOpen
  \bibfield  {author} {\bibinfo {author} {\bibfnamefont {L.}~\bibnamefont
  {{Santamar{\'{\i}}a}}}, \bibinfo {author} {\bibfnamefont {F.}~\bibnamefont
  {{Ohme}}}, \bibinfo {author} {\bibfnamefont {P.}~\bibnamefont {{Ajith}}},
  \bibinfo {author} {\bibfnamefont {B.}~\bibnamefont {{Br{\"u}gmann}}},
  \bibinfo {author} {\bibfnamefont {N.}~\bibnamefont {{Dorband}}}, \bibinfo
  {author} {\bibfnamefont {M.}~\bibnamefont {{Hannam}}}, \bibinfo {author}
  {\bibfnamefont {S.}~\bibnamefont {{Husa}}}, \bibinfo {author} {\bibfnamefont
  {P.}~\bibnamefont {{M{\"o}sta}}}, \bibinfo {author} {\bibfnamefont
  {D.}~\bibnamefont {{Pollney}}}, \bibinfo {author} {\bibfnamefont
  {C.}~\bibnamefont {{Reisswig}}}, \bibinfo {author} {\bibfnamefont {E.~L.}\
  \bibnamefont {{Robinson}}}, \bibinfo {author} {\bibfnamefont
  {J.}~\bibnamefont {{Seiler}}}, \ and\ \bibinfo {author} {\bibfnamefont
  {B.}~\bibnamefont {{Krishnan}}},\ }\href {\doibase
  10.1103/PhysRevD.82.064016} {\bibfield  {journal} {\bibinfo  {journal} {Phys.
  Rev. D}\ }\textbf {\bibinfo {volume} {82}},\ \bibinfo {eid} {064016}
  (\bibinfo {year} {2010})},\ \Eprint {http://arxiv.org/abs/1005.3306}
  {arXiv:1005.3306 [gr-qc]} \BibitemShut {NoStop}%
\bibitem [{\citenamefont {{Cutler}}\ and\ \citenamefont
  {{Flanagan}}(1994)}]{Cutler94}%
  \BibitemOpen
  \bibfield  {author} {\bibinfo {author} {\bibfnamefont {C.}~\bibnamefont
  {{Cutler}}}\ and\ \bibinfo {author} {\bibfnamefont {{\'E}.~E.}\ \bibnamefont
  {{Flanagan}}},\ }\href {\doibase 10.1103/PhysRevD.49.2658} {\bibfield
  {journal} {\bibinfo  {journal} {Phys. Rev. D}\ }\textbf {\bibinfo {volume}
  {49}},\ \bibinfo {pages} {2658} (\bibinfo {year} {1994})},\ \Eprint
  {http://arxiv.org/abs/gr-qc/9402014} {gr-qc/9402014} \BibitemShut {NoStop}%
\bibitem [{\citenamefont {{Del Pozzo}}\ \emph {et~al.}(2013)\citenamefont {{Del
  Pozzo}}, \citenamefont {{Li}}, \citenamefont {{Agathos}}, \citenamefont {{Van
  Den Broeck}},\ and\ \citenamefont {{Vitale}}}]{DelPozzo2013}%
  \BibitemOpen
  \bibfield  {author} {\bibinfo {author} {\bibfnamefont {W.}~\bibnamefont {{Del
  Pozzo}}}, \bibinfo {author} {\bibfnamefont {T.~G.~F.}\ \bibnamefont {{Li}}},
  \bibinfo {author} {\bibfnamefont {M.}~\bibnamefont {{Agathos}}}, \bibinfo
  {author} {\bibfnamefont {C.}~\bibnamefont {{Van Den Broeck}}}, \ and\
  \bibinfo {author} {\bibfnamefont {S.}~\bibnamefont {{Vitale}}},\ }\href
  {\doibase 10.1103/PhysRevLett.111.071101} {\bibfield  {journal} {\bibinfo
  {journal} {Phys. Rev. Lett.}\ }\textbf {\bibinfo {volume} {111}},\ \bibinfo
  {eid} {071101} (\bibinfo {year} {2013})},\ \Eprint
  {http://arxiv.org/abs/1307.8338} {arXiv:1307.8338 [gr-qc]} \BibitemShut
  {NoStop}%
\bibitem [{\citenamefont {{Lackey}}\ and\ \citenamefont
  {{Wade}}(2015)}]{Lackey2015}%
  \BibitemOpen
  \bibfield  {author} {\bibinfo {author} {\bibfnamefont {B.~D.}\ \bibnamefont
  {{Lackey}}}\ and\ \bibinfo {author} {\bibfnamefont {L.}~\bibnamefont
  {{Wade}}},\ }\href {\doibase 10.1103/PhysRevD.91.043002} {\bibfield
  {journal} {\bibinfo  {journal} {Phys. Rev. D}\ }\textbf {\bibinfo {volume}
  {91}},\ \bibinfo {eid} {043002} (\bibinfo {year} {2015})},\ \Eprint
  {http://arxiv.org/abs/1410.8866} {arXiv:1410.8866 [gr-qc]} \BibitemShut
  {NoStop}%
\bibitem [{\citenamefont {{Agathos}}\ \emph {et~al.}(2015)\citenamefont
  {{Agathos}}, \citenamefont {{Meidam}}, \citenamefont {{Del Pozzo}},
  \citenamefont {{Li}}, \citenamefont {{Tompitak}}, \citenamefont {{Veitch}},
  \citenamefont {{Vitale}},\ and\ \citenamefont {{Van Den
  Broeck}}}]{Agathos2015}%
  \BibitemOpen
  \bibfield  {author} {\bibinfo {author} {\bibfnamefont {M.}~\bibnamefont
  {{Agathos}}}, \bibinfo {author} {\bibfnamefont {J.}~\bibnamefont {{Meidam}}},
  \bibinfo {author} {\bibfnamefont {W.}~\bibnamefont {{Del Pozzo}}}, \bibinfo
  {author} {\bibfnamefont {T.~G.~F.}\ \bibnamefont {{Li}}}, \bibinfo {author}
  {\bibfnamefont {M.}~\bibnamefont {{Tompitak}}}, \bibinfo {author}
  {\bibfnamefont {J.}~\bibnamefont {{Veitch}}}, \bibinfo {author}
  {\bibfnamefont {S.}~\bibnamefont {{Vitale}}}, \ and\ \bibinfo {author}
  {\bibfnamefont {C.}~\bibnamefont {{Van Den Broeck}}},\ }\href {\doibase
  10.1103/PhysRevD.92.023012} {\bibfield  {journal} {\bibinfo  {journal} {Phys.
  Rev. D}\ }\textbf {\bibinfo {volume} {92}},\ \bibinfo {eid} {023012}
  (\bibinfo {year} {2015})},\ \Eprint {http://arxiv.org/abs/1503.05405}
  {arXiv:1503.05405 [gr-qc]} \BibitemShut {NoStop}%
\bibitem [{\citenamefont {{Read}}\ \emph
  {et~al.}(2009{\natexlab{a}})\citenamefont {{Read}}, \citenamefont {{Lackey}},
  \citenamefont {{Owen}},\ and\ \citenamefont {{Friedman}}}]{Read:2009a}%
  \BibitemOpen
  \bibfield  {author} {\bibinfo {author} {\bibfnamefont {J.~S.}\ \bibnamefont
  {{Read}}}, \bibinfo {author} {\bibfnamefont {B.~D.}\ \bibnamefont
  {{Lackey}}}, \bibinfo {author} {\bibfnamefont {B.~J.}\ \bibnamefont
  {{Owen}}}, \ and\ \bibinfo {author} {\bibfnamefont {J.~L.}\ \bibnamefont
  {{Friedman}}},\ }\href {\doibase 10.1103/PhysRevD.79.124032} {\bibfield
  {journal} {\bibinfo  {journal} {Phys. Rev. D}\ }\textbf {\bibinfo {volume}
  {79}},\ \bibinfo {eid} {124032} (\bibinfo {year} {2009}{\natexlab{a}})},\
  \Eprint {http://arxiv.org/abs/0812.2163} {arXiv:0812.2163} \BibitemShut
  {NoStop}%
\bibitem [{\citenamefont {{Bernuzzi}}\ \emph
  {et~al.}(2014{\natexlab{a}})\citenamefont {{Bernuzzi}}, \citenamefont
  {{Nagar}}, \citenamefont {{Balmelli}}, \citenamefont {{Dietrich}},\ and\
  \citenamefont {{Ujevic}}}]{Bernuzzi2014}%
  \BibitemOpen
  \bibfield  {author} {\bibinfo {author} {\bibfnamefont {S.}~\bibnamefont
  {{Bernuzzi}}}, \bibinfo {author} {\bibfnamefont {A.}~\bibnamefont {{Nagar}}},
  \bibinfo {author} {\bibfnamefont {S.}~\bibnamefont {{Balmelli}}}, \bibinfo
  {author} {\bibfnamefont {T.}~\bibnamefont {{Dietrich}}}, \ and\ \bibinfo
  {author} {\bibfnamefont {M.}~\bibnamefont {{Ujevic}}},\ }\href {\doibase
  10.1103/PhysRevLett.112.201101} {\bibfield  {journal} {\bibinfo  {journal}
  {Phys. Rev. Lett.}\ }\textbf {\bibinfo {volume} {112}},\ \bibinfo {eid}
  {201101} (\bibinfo {year} {2014}{\natexlab{a}})},\ \Eprint
  {http://arxiv.org/abs/1402.6244} {arXiv:1402.6244 [gr-qc]} \BibitemShut
  {NoStop}%
\bibitem [{\citenamefont {{Takami}}\ \emph {et~al.}(2015)\citenamefont
  {{Takami}}, \citenamefont {{Rezzolla}},\ and\ \citenamefont
  {{Baiotti}}}]{Takami2015}%
  \BibitemOpen
  \bibfield  {author} {\bibinfo {author} {\bibfnamefont {K.}~\bibnamefont
  {{Takami}}}, \bibinfo {author} {\bibfnamefont {L.}~\bibnamefont
  {{Rezzolla}}}, \ and\ \bibinfo {author} {\bibfnamefont {L.}~\bibnamefont
  {{Baiotti}}},\ }\href {\doibase 10.1103/PhysRevD.91.064001} {\bibfield
  {journal} {\bibinfo  {journal} {Phys. Rev. D}\ }\textbf {\bibinfo {volume}
  {91}},\ \bibinfo {eid} {064001} (\bibinfo {year} {2015})},\ \Eprint
  {http://arxiv.org/abs/1412.3240} {arXiv:1412.3240 [gr-qc]} \BibitemShut
  {NoStop}%
\bibitem [{\citenamefont {{Hotokezaka}}\ \emph
  {et~al.}(2016{\natexlab{a}})\citenamefont {{Hotokezaka}}, \citenamefont
  {{Kyutoku}}, \citenamefont {{Sekiguchi}},\ and\ \citenamefont
  {{Shibata}}}]{Hotokezaka2016}%
  \BibitemOpen
  \bibfield  {author} {\bibinfo {author} {\bibfnamefont {K.}~\bibnamefont
  {{Hotokezaka}}}, \bibinfo {author} {\bibfnamefont {K.}~\bibnamefont
  {{Kyutoku}}}, \bibinfo {author} {\bibfnamefont {Y.-i.}\ \bibnamefont
  {{Sekiguchi}}}, \ and\ \bibinfo {author} {\bibfnamefont {M.}~\bibnamefont
  {{Shibata}}},\ }\href@noop {} {\bibfield  {journal} {\bibinfo  {journal}
  {arXiv:1603.01286}\ } (\bibinfo {year} {2016}{\natexlab{a}})},\ \Eprint
  {http://arxiv.org/abs/1603.01286} {arXiv:1603.01286 [gr-qc]} \BibitemShut
  {NoStop}%
\bibitem [{\citenamefont {{Hotokezaka}}\ \emph
  {et~al.}(2015{\natexlab{b}})\citenamefont {{Hotokezaka}}, \citenamefont
  {{Kyutoku}}, \citenamefont {{Okawa}},\ and\ \citenamefont
  {{Shibata}}}]{Hotokezaka2015}%
  \BibitemOpen
  \bibfield  {author} {\bibinfo {author} {\bibfnamefont {K.}~\bibnamefont
  {{Hotokezaka}}}, \bibinfo {author} {\bibfnamefont {K.}~\bibnamefont
  {{Kyutoku}}}, \bibinfo {author} {\bibfnamefont {H.}~\bibnamefont {{Okawa}}},
  \ and\ \bibinfo {author} {\bibfnamefont {M.}~\bibnamefont {{Shibata}}},\
  }\href {\doibase 10.1103/PhysRevD.91.064060} {\bibfield  {journal} {\bibinfo
  {journal} {Phys. Rev. D}\ }\textbf {\bibinfo {volume} {91}},\ \bibinfo {eid}
  {064060} (\bibinfo {year} {2015}{\natexlab{b}})},\ \Eprint
  {http://arxiv.org/abs/1502.03457} {arXiv:1502.03457 [gr-qc]} \BibitemShut
  {NoStop}%
\bibitem [{\citenamefont {{Thierfelder}}\ \emph {et~al.}(2011)\citenamefont
  {{Thierfelder}}, \citenamefont {{Bernuzzi}},\ and\ \citenamefont
  {{Br{\"u}gmann}}}]{Thierfelder2011}%
  \BibitemOpen
  \bibfield  {author} {\bibinfo {author} {\bibfnamefont {M.}~\bibnamefont
  {{Thierfelder}}}, \bibinfo {author} {\bibfnamefont {S.}~\bibnamefont
  {{Bernuzzi}}}, \ and\ \bibinfo {author} {\bibfnamefont {B.}~\bibnamefont
  {{Br{\"u}gmann}}},\ }\href {\doibase 10.1103/PhysRevD.84.044012} {\bibfield
  {journal} {\bibinfo  {journal} {Phys. Rev. D}\ }\textbf {\bibinfo {volume}
  {84}},\ \bibinfo {eid} {044012} (\bibinfo {year} {2011})},\ \Eprint
  {http://arxiv.org/abs/1104.4751} {arXiv:1104.4751 [gr-qc]} \BibitemShut
  {NoStop}%
\bibitem [{\citenamefont {{Steiner}}\ \emph {et~al.}(2013)\citenamefont
  {{Steiner}}, \citenamefont {{Hempel}},\ and\ \citenamefont
  {{Fischer}}}]{Steiner2013}%
  \BibitemOpen
  \bibfield  {author} {\bibinfo {author} {\bibfnamefont {A.~W.}\ \bibnamefont
  {{Steiner}}}, \bibinfo {author} {\bibfnamefont {M.}~\bibnamefont {{Hempel}}},
  \ and\ \bibinfo {author} {\bibfnamefont {T.}~\bibnamefont {{Fischer}}},\
  }\href {\doibase 10.1088/0004-637X/774/1/17} {\bibfield  {journal} {\bibinfo
  {journal} {Astrophys. J.}\ }\textbf {\bibinfo {volume} {774}},\ \bibinfo
  {eid} {17} (\bibinfo {year} {2013})},\ \Eprint
  {http://arxiv.org/abs/1207.2184} {arXiv:1207.2184 [astro-ph.SR]} \BibitemShut
  {NoStop}%
\bibitem [{\citenamefont {{Banik}}\ \emph {et~al.}(2014)\citenamefont
  {{Banik}}, \citenamefont {{Hempel}},\ and\ \citenamefont
  {{Bandyopadhyay}}}]{Banik2014}%
  \BibitemOpen
  \bibfield  {author} {\bibinfo {author} {\bibfnamefont {S.}~\bibnamefont
  {{Banik}}}, \bibinfo {author} {\bibfnamefont {M.}~\bibnamefont {{Hempel}}}, \
  and\ \bibinfo {author} {\bibfnamefont {D.}~\bibnamefont {{Bandyopadhyay}}},\
  }\href {\doibase 10.1088/0067-0049/214/2/22} {\bibfield  {journal} {\bibinfo
  {journal} {Astrohys. J. Suppl.}\ }\textbf {\bibinfo {volume} {214}},\
  \bibinfo {eid} {22} (\bibinfo {year} {2014})},\ \Eprint
  {http://arxiv.org/abs/1404.6173} {arXiv:1404.6173 [astro-ph.HE]} \BibitemShut
  {NoStop}%
\bibitem [{\citenamefont {{Hempel}}\ \emph {et~al.}(2012)\citenamefont
  {{Hempel}}, \citenamefont {{Fischer}}, \citenamefont {{Schaffner-Bielich}},\
  and\ \citenamefont {{Liebend{\"o}rfer}}}]{Hempel2012}%
  \BibitemOpen
  \bibfield  {author} {\bibinfo {author} {\bibfnamefont {M.}~\bibnamefont
  {{Hempel}}}, \bibinfo {author} {\bibfnamefont {T.}~\bibnamefont {{Fischer}}},
  \bibinfo {author} {\bibfnamefont {J.}~\bibnamefont {{Schaffner-Bielich}}}, \
  and\ \bibinfo {author} {\bibfnamefont {M.}~\bibnamefont
  {{Liebend{\"o}rfer}}},\ }\href {\doibase 10.1088/0004-637X/748/1/70}
  {\bibfield  {journal} {\bibinfo  {journal} {Astrophys. J.}\ }\textbf
  {\bibinfo {volume} {748}},\ \bibinfo {eid} {70} (\bibinfo {year} {2012})},\
  \Eprint {http://arxiv.org/abs/1108.0848} {arXiv:1108.0848 [astro-ph.HE]}
  \BibitemShut {NoStop}%
\bibitem [{\citenamefont {Blanchet}(2006)}]{Blanchet06}%
  \BibitemOpen
  \bibfield  {author} {\bibinfo {author} {\bibfnamefont {L.}~\bibnamefont
  {Blanchet}},\ }\href@noop {} {\bibfield  {journal} {\bibinfo  {journal}
  {Living Rev. Relativ.}\ }\textbf {\bibinfo {volume} {9}},\ \bibinfo {pages}
  {4} (\bibinfo {year} {2006})}\BibitemShut {NoStop}%
\bibitem [{\citenamefont {{Poisson}}\ and\ \citenamefont
  {{Will}}(2014)}]{Poisson2014}%
  \BibitemOpen
  \bibfield  {author} {\bibinfo {author} {\bibfnamefont {E.}~\bibnamefont
  {{Poisson}}}\ and\ \bibinfo {author} {\bibfnamefont {C.~M.}\ \bibnamefont
  {{Will}}},\ }\href {\doibase 10.1017/cbo9781139507486} {\emph {\bibinfo
  {title} {Gravity, by Eric Poisson , Clifford M.~Will, Cambridge, UK:
  Cambridge University Press, 2014}}}\ (\bibinfo {year} {2014})\BibitemShut
  {NoStop}%
\bibitem [{\citenamefont {{Buonanno}}\ and\ \citenamefont
  {{Damour}}(1999)}]{Buonanno:1998gg}%
  \BibitemOpen
  \bibfield  {author} {\bibinfo {author} {\bibfnamefont {A.}~\bibnamefont
  {{Buonanno}}}\ and\ \bibinfo {author} {\bibfnamefont {T.}~\bibnamefont
  {{Damour}}},\ }\href {\doibase 10.1103/PhysRevD.59.084006} {\bibfield
  {journal} {\bibinfo  {journal} {Phys. Rev. D}\ }\textbf {\bibinfo {volume}
  {59}},\ \bibinfo {eid} {084006} (\bibinfo {year} {1999})},\ \Eprint
  {http://arxiv.org/abs/gr-qc/9811091} {gr-qc/9811091} \BibitemShut {NoStop}%
\bibitem [{\citenamefont {Buonanno}\ and\ \citenamefont
  {Damour}(2000)}]{Buonanno00a}%
  \BibitemOpen
  \bibfield  {author} {\bibinfo {author} {\bibfnamefont {A.}~\bibnamefont
  {Buonanno}}\ and\ \bibinfo {author} {\bibfnamefont {T.}~\bibnamefont
  {Damour}},\ }\href {\doibase 10.1103/PhysRevD.62.064015} {\bibfield
  {journal} {\bibinfo  {journal} {Phys. Rev. D}\ }\textbf {\bibinfo {volume}
  {62}},\ \bibinfo {pages} {064015} (\bibinfo {year} {2000})},\ \Eprint
  {http://arxiv.org/abs/gr-qc/0001013} {gr-qc/0001013} \BibitemShut {NoStop}%
\bibitem [{\citenamefont {Damour}\ \emph {et~al.}(2000)\citenamefont {Damour},
  \citenamefont {Jaranowski},\ and\ \citenamefont
  {Sch{\"a}fer}}]{Damour:1999cr}%
  \BibitemOpen
  \bibfield  {author} {\bibinfo {author} {\bibfnamefont {T.}~\bibnamefont
  {Damour}}, \bibinfo {author} {\bibfnamefont {P.}~\bibnamefont {Jaranowski}},
  \ and\ \bibinfo {author} {\bibfnamefont {G.}~\bibnamefont {Sch{\"a}fer}},\
  }\href@noop {} {\bibfield  {journal} {\bibinfo  {journal} {Phys. Rev. D}\
  }\textbf {\bibinfo {volume} {62}},\ \bibinfo {pages} {044024} (\bibinfo
  {year} {2000})},\ \Eprint {http://arxiv.org/abs/gr-qc/9912092}
  {gr-qc/9912092} \BibitemShut {NoStop}%
\bibitem [{\citenamefont {Damour}(2001)}]{Damour:2001tu}%
  \BibitemOpen
  \bibfield  {author} {\bibinfo {author} {\bibfnamefont {T.}~\bibnamefont
  {Damour}},\ }\href {\doibase 10.1103/PhysRevD.64.124013} {\bibfield
  {journal} {\bibinfo  {journal} {Phys. Rev. D}\ }\textbf {\bibinfo {volume}
  {64}},\ \bibinfo {pages} {124013} (\bibinfo {year} {2001})},\ \Eprint
  {http://arXiv.org/abs/gr-qc/0103018} {gr-qc/0103018} \BibitemShut {NoStop}%
\bibitem [{\citenamefont {{Hinderer}}(2008)}]{Hinderer08}%
  \BibitemOpen
  \bibfield  {author} {\bibinfo {author} {\bibfnamefont {T.}~\bibnamefont
  {{Hinderer}}},\ }\href {\doibase 10.1086/533487} {\bibfield  {journal}
  {\bibinfo  {journal} {Astrophys. J.}\ }\textbf {\bibinfo {volume} {677}},\
  \bibinfo {pages} {1216} (\bibinfo {year} {2008})},\ \Eprint
  {http://arxiv.org/abs/0711.2420} {arXiv:0711.2420} \BibitemShut {NoStop}%
\bibitem [{\citenamefont {{Damour}}\ and\ \citenamefont
  {{Nagar}}(2009)}]{Damour:2009}%
  \BibitemOpen
  \bibfield  {author} {\bibinfo {author} {\bibfnamefont {T.}~\bibnamefont
  {{Damour}}}\ and\ \bibinfo {author} {\bibfnamefont {A.}~\bibnamefont
  {{Nagar}}},\ }\href {\doibase 10.1103/PhysRevD.80.084035} {\bibfield
  {journal} {\bibinfo  {journal} {Phys. Rev. D}\ }\textbf {\bibinfo {volume}
  {80}},\ \bibinfo {pages} {084035} (\bibinfo {year} {2009})},\ \Eprint
  {http://arxiv.org/abs/0906.0096} {arXiv:0906.0096 [gr-qc]} \BibitemShut
  {NoStop}%
\bibitem [{\citenamefont {Binnington}\ and\ \citenamefont
  {Poisson}(2009)}]{Binnington:2009bb}%
  \BibitemOpen
  \bibfield  {author} {\bibinfo {author} {\bibfnamefont {T.}~\bibnamefont
  {Binnington}}\ and\ \bibinfo {author} {\bibfnamefont {E.}~\bibnamefont
  {Poisson}},\ }\href {\doibase 10.1103/PhysRevD.80.084018} {\bibfield
  {journal} {\bibinfo  {journal} {Phys. Rev. D}\ }\textbf {\bibinfo {volume}
  {80}},\ \bibinfo {pages} {084018} (\bibinfo {year} {2009})},\ \Eprint
  {http://arxiv.org/abs/0906.1366} {arXiv:0906.1366 [gr-qc]} \BibitemShut
  {NoStop}%
\bibitem [{\citenamefont {Hinderer}\ \emph {et~al.}(2010)\citenamefont
  {Hinderer}, \citenamefont {Lackey}, \citenamefont {Lang},\ and\ \citenamefont
  {Read}}]{Hinderer09}%
  \BibitemOpen
  \bibfield  {author} {\bibinfo {author} {\bibfnamefont {T.}~\bibnamefont
  {Hinderer}}, \bibinfo {author} {\bibfnamefont {B.~D.}\ \bibnamefont
  {Lackey}}, \bibinfo {author} {\bibfnamefont {R.~N.}\ \bibnamefont {Lang}}, \
  and\ \bibinfo {author} {\bibfnamefont {J.~S.}\ \bibnamefont {Read}},\ }\href
  {\doibase 10.1103/PhysRevD.81.123016} {\bibfield  {journal} {\bibinfo
  {journal} {Phys. Rev. D}\ }\textbf {\bibinfo {volume} {81}},\ \bibinfo
  {pages} {123016} (\bibinfo {year} {2010})},\ \Eprint
  {http://arxiv.org/abs/0911.3535} {arXiv:0911.3535 [astro-ph.HE]} \BibitemShut
  {NoStop}%
\bibitem [{\citenamefont {Damour}\ \emph {et~al.}(2012)\citenamefont {Damour},
  \citenamefont {Nagar},\ and\ \citenamefont {Villain}}]{Damour:2012}%
  \BibitemOpen
  \bibfield  {author} {\bibinfo {author} {\bibfnamefont {T.}~\bibnamefont
  {Damour}}, \bibinfo {author} {\bibfnamefont {A.}~\bibnamefont {Nagar}}, \
  and\ \bibinfo {author} {\bibfnamefont {L.}~\bibnamefont {Villain}},\ }\href
  {\doibase 10.1103/PhysRevD.85.123007} {\bibfield  {journal} {\bibinfo
  {journal} {Phys. Rev. D}\ }\textbf {\bibinfo {volume} {85}},\ \bibinfo
  {pages} {123007} (\bibinfo {year} {2012})}\BibitemShut {NoStop}%
\bibitem [{\citenamefont {{Damour}}\ and\ \citenamefont
  {{Nagar}}(2010)}]{Damour:2009wj}%
  \BibitemOpen
  \bibfield  {author} {\bibinfo {author} {\bibfnamefont {T.}~\bibnamefont
  {{Damour}}}\ and\ \bibinfo {author} {\bibfnamefont {A.}~\bibnamefont
  {{Nagar}}},\ }\href {\doibase 10.1103/PhysRevD.81.084016} {\bibfield
  {journal} {\bibinfo  {journal} {Phys. Rev. D}\ }\textbf {\bibinfo {volume}
  {81}},\ \bibinfo {pages} {084016} (\bibinfo {year} {2010})},\ \Eprint
  {http://arxiv.org/abs/0911.5041} {arXiv:0911.5041 [gr-qc]} \BibitemShut
  {NoStop}%
\bibitem [{\citenamefont {{Bini}}\ and\ \citenamefont
  {{Damour}}(2014{\natexlab{a}})}]{Bini2014a}%
  \BibitemOpen
  \bibfield  {author} {\bibinfo {author} {\bibfnamefont {D.}~\bibnamefont
  {{Bini}}}\ and\ \bibinfo {author} {\bibfnamefont {T.}~\bibnamefont
  {{Damour}}},\ }\href {\doibase 10.1103/PhysRevD.90.124037} {\bibfield
  {journal} {\bibinfo  {journal} {Phys. Rev. D}\ }\textbf {\bibinfo {volume}
  {90}},\ \bibinfo {eid} {124037} (\bibinfo {year} {2014}{\natexlab{a}})},\
  \Eprint {http://arxiv.org/abs/1409.6933} {arXiv:1409.6933 [gr-qc]}
  \BibitemShut {NoStop}%
\bibitem [{\citenamefont {{Bini}}\ and\ \citenamefont
  {{Damour}}(2014{\natexlab{b}})}]{Bini2014b}%
  \BibitemOpen
  \bibfield  {author} {\bibinfo {author} {\bibfnamefont {D.}~\bibnamefont
  {{Bini}}}\ and\ \bibinfo {author} {\bibfnamefont {T.}~\bibnamefont
  {{Damour}}},\ }\href {\doibase 10.1103/PhysRevD.89.064063} {\bibfield
  {journal} {\bibinfo  {journal} {Phys. Rev. D}\ }\textbf {\bibinfo {volume}
  {89}},\ \bibinfo {eid} {064063} (\bibinfo {year} {2014}{\natexlab{b}})},\
  \Eprint {http://arxiv.org/abs/1312.2503} {arXiv:1312.2503 [gr-qc]}
  \BibitemShut {NoStop}%
\bibitem [{\citenamefont {{Dolan}}\ \emph {et~al.}(2015)\citenamefont
  {{Dolan}}, \citenamefont {{Nolan}}, \citenamefont {{Ottewill}}, \citenamefont
  {{Warburton}},\ and\ \citenamefont {{Wardell}}}]{Dolan2015}%
  \BibitemOpen
  \bibfield  {author} {\bibinfo {author} {\bibfnamefont {S.~R.}\ \bibnamefont
  {{Dolan}}}, \bibinfo {author} {\bibfnamefont {P.}~\bibnamefont {{Nolan}}},
  \bibinfo {author} {\bibfnamefont {A.~C.}\ \bibnamefont {{Ottewill}}},
  \bibinfo {author} {\bibfnamefont {N.}~\bibnamefont {{Warburton}}}, \ and\
  \bibinfo {author} {\bibfnamefont {B.}~\bibnamefont {{Wardell}}},\ }\href
  {\doibase 10.1103/PhysRevD.91.023009} {\bibfield  {journal} {\bibinfo
  {journal} {Phys. Rev. D}\ }\textbf {\bibinfo {volume} {91}},\ \bibinfo {eid}
  {023009} (\bibinfo {year} {2015})},\ \Eprint {http://arxiv.org/abs/1406.4890}
  {arXiv:1406.4890 [gr-qc]} \BibitemShut {NoStop}%
\bibitem [{\citenamefont {Baiotti}\ \emph {et~al.}(2010)\citenamefont
  {Baiotti}, \citenamefont {Damour}, \citenamefont {Giacomazzo}, \citenamefont
  {Nagar},\ and\ \citenamefont {Rezzolla}}]{Baiotti:2010}%
  \BibitemOpen
  \bibfield  {author} {\bibinfo {author} {\bibfnamefont {L.}~\bibnamefont
  {Baiotti}}, \bibinfo {author} {\bibfnamefont {T.}~\bibnamefont {Damour}},
  \bibinfo {author} {\bibfnamefont {B.}~\bibnamefont {Giacomazzo}}, \bibinfo
  {author} {\bibfnamefont {A.}~\bibnamefont {Nagar}}, \ and\ \bibinfo {author}
  {\bibfnamefont {L.}~\bibnamefont {Rezzolla}},\ }\href {\doibase
  10.1103/PhysRevLett.105.261101} {\bibfield  {journal} {\bibinfo  {journal}
  {Phys. Rev. Lett.}\ }\textbf {\bibinfo {volume} {105}},\ \bibinfo {pages}
  {261101} (\bibinfo {year} {2010})},\ \Eprint {http://arxiv.org/abs/1009.0521}
  {arXiv:1009.0521 [gr-qc]} \BibitemShut {NoStop}%
\bibitem [{\citenamefont {{Hinderer}}\ \emph {et~al.}(2016)\citenamefont
  {{Hinderer}}, \citenamefont {{Taracchini}}, \citenamefont {{Foucart}},
  \citenamefont {{Buonanno}}, \citenamefont {{Steinhoff}}, \citenamefont
  {{Duez}}, \citenamefont {{Kidder}}, \citenamefont {{Pfeiffer}}, \citenamefont
  {{Scheel}}, \citenamefont {{Szilagyi}}, \citenamefont {{Hotokezaka}},
  \citenamefont {{Kyutoku}}, \citenamefont {{Shibata}},\ and\ \citenamefont
  {{Carpenter}}}]{Hinderer2016}%
  \BibitemOpen
  \bibfield  {author} {\bibinfo {author} {\bibfnamefont {T.}~\bibnamefont
  {{Hinderer}}}, \bibinfo {author} {\bibfnamefont {A.}~\bibnamefont
  {{Taracchini}}}, \bibinfo {author} {\bibfnamefont {F.}~\bibnamefont
  {{Foucart}}}, \bibinfo {author} {\bibfnamefont {A.}~\bibnamefont
  {{Buonanno}}}, \bibinfo {author} {\bibfnamefont {J.}~\bibnamefont
  {{Steinhoff}}}, \bibinfo {author} {\bibfnamefont {M.}~\bibnamefont {{Duez}}},
  \bibinfo {author} {\bibfnamefont {L.~E.}\ \bibnamefont {{Kidder}}}, \bibinfo
  {author} {\bibfnamefont {H.~P.}\ \bibnamefont {{Pfeiffer}}}, \bibinfo
  {author} {\bibfnamefont {M.~A.}\ \bibnamefont {{Scheel}}}, \bibinfo {author}
  {\bibfnamefont {B.}~\bibnamefont {{Szilagyi}}}, \bibinfo {author}
  {\bibfnamefont {K.}~\bibnamefont {{Hotokezaka}}}, \bibinfo {author}
  {\bibfnamefont {K.}~\bibnamefont {{Kyutoku}}}, \bibinfo {author}
  {\bibfnamefont {M.}~\bibnamefont {{Shibata}}}, \ and\ \bibinfo {author}
  {\bibfnamefont {C.~W.}\ \bibnamefont {{Carpenter}}},\ }\href@noop {}
  {\bibfield  {journal} {\bibinfo  {journal} {arXiv:1602.00599}\ } (\bibinfo
  {year} {2016})},\ \Eprint {http://arxiv.org/abs/1602.00599} {arXiv:1602.00599
  [gr-qc]} \BibitemShut {NoStop}%
\bibitem [{\citenamefont {{Messenger}}\ and\ \citenamefont
  {{Read}}(2012)}]{Messenger:2011}%
  \BibitemOpen
  \bibfield  {author} {\bibinfo {author} {\bibfnamefont {C.}~\bibnamefont
  {{Messenger}}}\ and\ \bibinfo {author} {\bibfnamefont {J.}~\bibnamefont
  {{Read}}},\ }\href {\doibase 10.1103/PhysRevLett.108.091101} {\bibfield
  {journal} {\bibinfo  {journal} {Phys. Rev. Lett.}\ }\textbf {\bibinfo
  {volume} {108}},\ \bibinfo {eid} {091101} (\bibinfo {year} {2012})},\ \Eprint
  {http://arxiv.org/abs/1107.5725} {arXiv:1107.5725 [gr-qc]} \BibitemShut
  {NoStop}%
\bibitem [{\citenamefont {Messenger}\ \emph {et~al.}(2014)\citenamefont
  {Messenger}, \citenamefont {Takami}, \citenamefont {Gossan}, \citenamefont
  {Rezzolla},\ and\ \citenamefont {Sathyaprakash}}]{Messenger2013}%
  \BibitemOpen
  \bibfield  {author} {\bibinfo {author} {\bibfnamefont {C.}~\bibnamefont
  {Messenger}}, \bibinfo {author} {\bibfnamefont {K.}~\bibnamefont {Takami}},
  \bibinfo {author} {\bibfnamefont {S.}~\bibnamefont {Gossan}}, \bibinfo
  {author} {\bibfnamefont {L.}~\bibnamefont {Rezzolla}}, \ and\ \bibinfo
  {author} {\bibfnamefont {B.~S.}\ \bibnamefont {Sathyaprakash}},\ }\href
  {\doibase 10.1103/PhysRevX.4.041004} {\bibfield  {journal} {\bibinfo
  {journal} {Phys. Rev. X}\ }\textbf {\bibinfo {volume} {4}},\ \bibinfo {pages}
  {041004} (\bibinfo {year} {2014})}\BibitemShut {NoStop}%
\bibitem [{\citenamefont {{Read}}\ \emph
  {et~al.}(2009{\natexlab{b}})\citenamefont {{Read}}, \citenamefont
  {{Markakis}}, \citenamefont {{Shibata}}, \citenamefont {{Ury{\= u}}},
  \citenamefont {{Creighton}},\ and\ \citenamefont {{Friedman}}}]{Read:2009b}%
  \BibitemOpen
  \bibfield  {author} {\bibinfo {author} {\bibfnamefont {J.~S.}\ \bibnamefont
  {{Read}}}, \bibinfo {author} {\bibfnamefont {C.}~\bibnamefont {{Markakis}}},
  \bibinfo {author} {\bibfnamefont {M.}~\bibnamefont {{Shibata}}}, \bibinfo
  {author} {\bibfnamefont {K.}~\bibnamefont {{Ury{\= u}}}}, \bibinfo {author}
  {\bibfnamefont {J.~D.~E.}\ \bibnamefont {{Creighton}}}, \ and\ \bibinfo
  {author} {\bibfnamefont {J.~L.}\ \bibnamefont {{Friedman}}},\ }\href
  {\doibase 10.1103/PhysRevD.79.124033} {\bibfield  {journal} {\bibinfo
  {journal} {Phys. Rev. D}\ }\textbf {\bibinfo {volume} {79}},\ \bibinfo {eid}
  {124033} (\bibinfo {year} {2009}{\natexlab{b}})},\ \Eprint
  {http://arxiv.org/abs/0901.3258} {arXiv:0901.3258 [gr-qc]} \BibitemShut
  {NoStop}%
\bibitem [{\citenamefont {{Barkett}}\ \emph {et~al.}(2016)\citenamefont
  {{Barkett}}, \citenamefont {{Scheel}}, \citenamefont {{Haas}}, \citenamefont
  {{Ott}}, \citenamefont {{Bernuzzi}}, \citenamefont {{Brown}}, \citenamefont
  {{Szil{\'a}gyi}}, \citenamefont {{Kaplan}}, \citenamefont {{Lippuner}},
  \citenamefont {{Muhlberger}}, \citenamefont {{Foucart}},\ and\ \citenamefont
  {{Duez}}}]{Barkett2016}%
  \BibitemOpen
  \bibfield  {author} {\bibinfo {author} {\bibfnamefont {K.}~\bibnamefont
  {{Barkett}}}, \bibinfo {author} {\bibfnamefont {M.~A.}\ \bibnamefont
  {{Scheel}}}, \bibinfo {author} {\bibfnamefont {R.}~\bibnamefont {{Haas}}},
  \bibinfo {author} {\bibfnamefont {C.~D.}\ \bibnamefont {{Ott}}}, \bibinfo
  {author} {\bibfnamefont {S.}~\bibnamefont {{Bernuzzi}}}, \bibinfo {author}
  {\bibfnamefont {D.~A.}\ \bibnamefont {{Brown}}}, \bibinfo {author}
  {\bibfnamefont {B.}~\bibnamefont {{Szil{\'a}gyi}}}, \bibinfo {author}
  {\bibfnamefont {J.~D.}\ \bibnamefont {{Kaplan}}}, \bibinfo {author}
  {\bibfnamefont {J.}~\bibnamefont {{Lippuner}}}, \bibinfo {author}
  {\bibfnamefont {C.~D.}\ \bibnamefont {{Muhlberger}}}, \bibinfo {author}
  {\bibfnamefont {F.}~\bibnamefont {{Foucart}}}, \ and\ \bibinfo {author}
  {\bibfnamefont {M.~D.}\ \bibnamefont {{Duez}}},\ }\href {\doibase
  10.1103/PhysRevD.93.044064} {\bibfield  {journal} {\bibinfo  {journal} {Phys.
  Rev. D}\ }\textbf {\bibinfo {volume} {93}},\ \bibinfo {eid} {044064}
  (\bibinfo {year} {2016})},\ \Eprint {http://arxiv.org/abs/1509.05782}
  {arXiv:1509.05782 [gr-qc]} \BibitemShut {NoStop}%
\bibitem [{\citenamefont {Campanelli}\ \emph {et~al.}(2006)\citenamefont
  {Campanelli}, \citenamefont {Lousto},\ and\ \citenamefont
  {Zlochower}}]{Campanelli:2006uy}%
  \BibitemOpen
  \bibfield  {author} {\bibinfo {author} {\bibfnamefont {M.}~\bibnamefont
  {Campanelli}}, \bibinfo {author} {\bibfnamefont {C.~O.}\ \bibnamefont
  {Lousto}}, \ and\ \bibinfo {author} {\bibfnamefont {Y.}~\bibnamefont
  {Zlochower}},\ }\href {\doibase 10.1103/PhysRevD.74.041501} {\bibfield
  {journal} {\bibinfo  {journal} {Phys. Rev. D}\ }\textbf {\bibinfo {volume}
  {74}},\ \bibinfo {pages} {041501} (\bibinfo {year} {2006})},\ \Eprint
  {http://arxiv.org/abs/gr-qc/0604012} {gr-qc/0604012} \BibitemShut {NoStop}%
\bibitem [{\citenamefont {{Pollney}}\ \emph {et~al.}(2007)\citenamefont
  {{Pollney}}, \citenamefont {{Reisswig}}, \citenamefont {{Rezzolla}},
  \citenamefont {{Szil{\'a}gyi}}, \citenamefont {{Ansorg}}, \citenamefont
  {{Deris}}, \citenamefont {{Diener}}, \citenamefont {{Dorband}}, \citenamefont
  {{Koppitz}}, \citenamefont {{Nagar}},\ and\ \citenamefont
  {{Schnetter}}}]{Pollney:2007ss}%
  \BibitemOpen
  \bibfield  {author} {\bibinfo {author} {\bibfnamefont {D.}~\bibnamefont
  {{Pollney}}}, \bibinfo {author} {\bibfnamefont {C.}~\bibnamefont
  {{Reisswig}}}, \bibinfo {author} {\bibfnamefont {L.}~\bibnamefont
  {{Rezzolla}}}, \bibinfo {author} {\bibfnamefont {B.}~\bibnamefont
  {{Szil{\'a}gyi}}}, \bibinfo {author} {\bibfnamefont {M.}~\bibnamefont
  {{Ansorg}}}, \bibinfo {author} {\bibfnamefont {B.}~\bibnamefont {{Deris}}},
  \bibinfo {author} {\bibfnamefont {P.}~\bibnamefont {{Diener}}}, \bibinfo
  {author} {\bibfnamefont {E.~N.}\ \bibnamefont {{Dorband}}}, \bibinfo {author}
  {\bibfnamefont {M.}~\bibnamefont {{Koppitz}}}, \bibinfo {author}
  {\bibfnamefont {A.}~\bibnamefont {{Nagar}}}, \ and\ \bibinfo {author}
  {\bibfnamefont {E.}~\bibnamefont {{Schnetter}}},\ }\href {\doibase
  10.1103/PhysRevD.76.124002} {\bibfield  {journal} {\bibinfo  {journal} {Phys.
  Rev. D}\ }\textbf {\bibinfo {volume} {76}},\ \bibinfo {eid} {124002}
  (\bibinfo {year} {2007})},\ \Eprint {http://arxiv.org/abs/0707.2559}
  {arXiv:0707.2559 [gr-qc]} \BibitemShut {NoStop}%
\bibitem [{\citenamefont {Hannam}\ \emph {et~al.}(2008)\citenamefont {Hannam},
  \citenamefont {Husa}, \citenamefont {Bruegmann},\ and\ \citenamefont
  {Gopakumar}}]{Hannam:2007wf}%
  \BibitemOpen
  \bibfield  {author} {\bibinfo {author} {\bibfnamefont {M.}~\bibnamefont
  {Hannam}}, \bibinfo {author} {\bibfnamefont {S.}~\bibnamefont {Husa}},
  \bibinfo {author} {\bibfnamefont {B.}~\bibnamefont {Bruegmann}}, \ and\
  \bibinfo {author} {\bibfnamefont {A.}~\bibnamefont {Gopakumar}},\ }\href
  {\doibase 10.1103/PhysRevD.78.104007} {\bibfield  {journal} {\bibinfo
  {journal} {Phys. Rev. D}\ }\textbf {\bibinfo {volume} {78}},\ \bibinfo
  {pages} {104007} (\bibinfo {year} {2008})},\ \Eprint
  {http://arxiv.org/abs/0712.3787} {arXiv:0712.3787 [gr-qc]} \BibitemShut
  {NoStop}%
\bibitem [{\citenamefont {{Bernuzzi}}\ \emph
  {et~al.}(2014{\natexlab{b}})\citenamefont {{Bernuzzi}}, \citenamefont
  {{Dietrich}}, \citenamefont {{Tichy}},\ and\ \citenamefont
  {{Br{\"u}gmann}}}]{Bernuzzi2013}%
  \BibitemOpen
  \bibfield  {author} {\bibinfo {author} {\bibfnamefont {S.}~\bibnamefont
  {{Bernuzzi}}}, \bibinfo {author} {\bibfnamefont {T.}~\bibnamefont
  {{Dietrich}}}, \bibinfo {author} {\bibfnamefont {W.}~\bibnamefont {{Tichy}}},
  \ and\ \bibinfo {author} {\bibfnamefont {B.}~\bibnamefont {{Br{\"u}gmann}}},\
  }\href {\doibase 10.1103/PhysRevD.89.104021} {\bibfield  {journal} {\bibinfo
  {journal} {Phys. Rev. D}\ }\textbf {\bibinfo {volume} {89}},\ \bibinfo {eid}
  {104021} (\bibinfo {year} {2014}{\natexlab{b}})},\ \Eprint
  {http://arxiv.org/abs/1311.4443} {arXiv:1311.4443 [gr-qc]} \BibitemShut
  {NoStop}%
\bibitem [{\citenamefont {{Bernuzzi}}\ \emph
  {et~al.}(2015{\natexlab{b}})\citenamefont {{Bernuzzi}}, \citenamefont
  {{Radice}}, \citenamefont {{Ott}}, \citenamefont {{Roberts}}, \citenamefont
  {{Moesta}},\ and\ \citenamefont {{Galeazzi}}}]{Bernuzzi2015b}%
  \BibitemOpen
  \bibfield  {author} {\bibinfo {author} {\bibfnamefont {S.}~\bibnamefont
  {{Bernuzzi}}}, \bibinfo {author} {\bibfnamefont {D.}~\bibnamefont
  {{Radice}}}, \bibinfo {author} {\bibfnamefont {C.~D.}\ \bibnamefont {{Ott}}},
  \bibinfo {author} {\bibfnamefont {L.~F.}\ \bibnamefont {{Roberts}}}, \bibinfo
  {author} {\bibfnamefont {P.}~\bibnamefont {{Moesta}}}, \ and\ \bibinfo
  {author} {\bibfnamefont {F.}~\bibnamefont {{Galeazzi}}},\ }\href@noop {}
  {\bibfield  {journal} {\bibinfo  {journal} {arXiv:1512.06397}\ } (\bibinfo
  {year} {2015}{\natexlab{b}})},\ \Eprint {http://arxiv.org/abs/1512.06397}
  {arXiv:1512.06397 [gr-qc]} \BibitemShut {NoStop}%
\bibitem [{\citenamefont {{Shibata}}\ \emph {et~al.}(2005)\citenamefont
  {{Shibata}}, \citenamefont {{Taniguchi}},\ and\ \citenamefont {{Ury{\=
  u}}}}]{Shibata05c}%
  \BibitemOpen
  \bibfield  {author} {\bibinfo {author} {\bibfnamefont {M.}~\bibnamefont
  {{Shibata}}}, \bibinfo {author} {\bibfnamefont {K.}~\bibnamefont
  {{Taniguchi}}}, \ and\ \bibinfo {author} {\bibfnamefont {K.}~\bibnamefont
  {{Ury{\= u}}}},\ }\href {\doibase 10.1103/PhysRevD.71.084021} {\bibfield
  {journal} {\bibinfo  {journal} {Phys. Rev. D}\ }\textbf {\bibinfo {volume}
  {71}},\ \bibinfo {eid} {084021} (\bibinfo {year} {2005})},\ \Eprint
  {http://arxiv.org/abs/gr-qc/0503119} {gr-qc/0503119} \BibitemShut {NoStop}%
\bibitem [{\citenamefont {{Shibata}}\ and\ \citenamefont
  {{Taniguchi}}(2006)}]{Shibata06a}%
  \BibitemOpen
  \bibfield  {author} {\bibinfo {author} {\bibfnamefont {M.}~\bibnamefont
  {{Shibata}}}\ and\ \bibinfo {author} {\bibfnamefont {K.}~\bibnamefont
  {{Taniguchi}}},\ }\href {\doibase 10.1103/PhysRevD.73.064027} {\bibfield
  {journal} {\bibinfo  {journal} {Phys. Rev. D}\ }\textbf {\bibinfo {volume}
  {73}},\ \bibinfo {eid} {064027} (\bibinfo {year} {2006})},\ \Eprint
  {http://arxiv.org/abs/astro-ph/0603145} {astro-ph/0603145} \BibitemShut
  {NoStop}%
\bibitem [{\citenamefont {{Yamamoto}}\ \emph {et~al.}(2008)\citenamefont
  {{Yamamoto}}, \citenamefont {{Shibata}},\ and\ \citenamefont
  {{Taniguchi}}}]{Yamamoto2008}%
  \BibitemOpen
  \bibfield  {author} {\bibinfo {author} {\bibfnamefont {T.}~\bibnamefont
  {{Yamamoto}}}, \bibinfo {author} {\bibfnamefont {M.}~\bibnamefont
  {{Shibata}}}, \ and\ \bibinfo {author} {\bibfnamefont {K.}~\bibnamefont
  {{Taniguchi}}},\ }\href {\doibase 10.1103/PhysRevD.78.064054} {\bibfield
  {journal} {\bibinfo  {journal} {Phys. Rev. D}\ }\textbf {\bibinfo {volume}
  {78}},\ \bibinfo {eid} {064054} (\bibinfo {year} {2008})},\ \Eprint
  {http://arxiv.org/abs/0806.4007} {arXiv:0806.4007 [gr-qc]} \BibitemShut
  {NoStop}%
\bibitem [{\citenamefont {{Kaplan}}\ \emph {et~al.}(2014)\citenamefont
  {{Kaplan}}, \citenamefont {{Ott}}, \citenamefont {{O'Connor}}, \citenamefont
  {{Kiuchi}}, \citenamefont {{Roberts}},\ and\ \citenamefont
  {{Duez}}}]{Kaplan2013}%
  \BibitemOpen
  \bibfield  {author} {\bibinfo {author} {\bibfnamefont {J.~D.}\ \bibnamefont
  {{Kaplan}}}, \bibinfo {author} {\bibfnamefont {C.~D.}\ \bibnamefont {{Ott}}},
  \bibinfo {author} {\bibfnamefont {E.~P.}\ \bibnamefont {{O'Connor}}},
  \bibinfo {author} {\bibfnamefont {K.}~\bibnamefont {{Kiuchi}}}, \bibinfo
  {author} {\bibfnamefont {L.}~\bibnamefont {{Roberts}}}, \ and\ \bibinfo
  {author} {\bibfnamefont {M.}~\bibnamefont {{Duez}}},\ }\href {\doibase
  10.1088/0004-637X/790/1/19} {\bibfield  {journal} {\bibinfo  {journal}
  {Astrophys. J.}\ }\textbf {\bibinfo {volume} {790}},\ \bibinfo {eid} {19}
  (\bibinfo {year} {2014})},\ \Eprint {http://arxiv.org/abs/1306.4034}
  {arXiv:1306.4034 [astro-ph.HE]} \BibitemShut {NoStop}%
\bibitem [{\citenamefont {Centrella}\ \emph {et~al.}(2001)\citenamefont
  {Centrella}, \citenamefont {New}, \citenamefont {Lowe},\ and\ \citenamefont
  {Brown}}]{Centrella:2001xp}%
  \BibitemOpen
  \bibfield  {author} {\bibinfo {author} {\bibfnamefont {J.~M.}\ \bibnamefont
  {Centrella}}, \bibinfo {author} {\bibfnamefont {K.~C.~B.}\ \bibnamefont
  {New}}, \bibinfo {author} {\bibfnamefont {L.~L.}\ \bibnamefont {Lowe}}, \
  and\ \bibinfo {author} {\bibfnamefont {J.~D.}\ \bibnamefont {Brown}},\
  }\href@noop {} {\bibfield  {journal} {\bibinfo  {journal} {Astrophys. J.}\
  }\textbf {\bibinfo {volume} {550}},\ \bibinfo {pages} {L193} (\bibinfo {year}
  {2001})},\ \Eprint {http://arxiv.org/abs/astro-ph/0010574} {astro-ph/0010574}
  \BibitemShut {NoStop}%
\bibitem [{\citenamefont {Watts}\ \emph {et~al.}(2005)\citenamefont {Watts},
  \citenamefont {Andersson},\ and\ \citenamefont {Jones}}]{Watts:2003nn}%
  \BibitemOpen
  \bibfield  {author} {\bibinfo {author} {\bibfnamefont {A.~L.}\ \bibnamefont
  {Watts}}, \bibinfo {author} {\bibfnamefont {N.}~\bibnamefont {Andersson}}, \
  and\ \bibinfo {author} {\bibfnamefont {D.~I.}\ \bibnamefont {Jones}},\
  }\href@noop {} {\bibfield  {journal} {\bibinfo  {journal} {Astrophys. J.}\
  }\textbf {\bibinfo {volume} {618}},\ \bibinfo {pages} {L37} (\bibinfo {year}
  {2005})},\ \Eprint {http://arxiv.org/abs/astro-ph/0309554} {astro-ph/0309554}
  \BibitemShut {NoStop}%
\bibitem [{\citenamefont {{Dietrich}}\ \emph
  {et~al.}(2015{\natexlab{a}})\citenamefont {{Dietrich}}, \citenamefont
  {{Moldenhauer}}, \citenamefont {{Johnson-McDaniel}}, \citenamefont
  {{Bernuzzi}}, \citenamefont {{Markakis}}, \citenamefont {{Br{\"u}gmann}},\
  and\ \citenamefont {{Tichy}}}]{Dietrich:2015b}%
  \BibitemOpen
  \bibfield  {author} {\bibinfo {author} {\bibfnamefont {T.}~\bibnamefont
  {{Dietrich}}}, \bibinfo {author} {\bibfnamefont {N.}~\bibnamefont
  {{Moldenhauer}}}, \bibinfo {author} {\bibfnamefont {N.~K.}\ \bibnamefont
  {{Johnson-McDaniel}}}, \bibinfo {author} {\bibfnamefont {S.}~\bibnamefont
  {{Bernuzzi}}}, \bibinfo {author} {\bibfnamefont {C.~M.}\ \bibnamefont
  {{Markakis}}}, \bibinfo {author} {\bibfnamefont {B.}~\bibnamefont
  {{Br{\"u}gmann}}}, \ and\ \bibinfo {author} {\bibfnamefont {W.}~\bibnamefont
  {{Tichy}}},\ }\href {\doibase 10.1103/PhysRevD.92.124007} {\bibfield
  {journal} {\bibinfo  {journal} {Phys. Rev. D}\ }\textbf {\bibinfo {volume}
  {92}},\ \bibinfo {eid} {124007} (\bibinfo {year} {2015}{\natexlab{a}})},\
  \Eprint {http://arxiv.org/abs/1507.07100} {arXiv:1507.07100 [gr-qc]}
  \BibitemShut {NoStop}%
\bibitem [{\citenamefont {{Radice}}\ \emph
  {et~al.}(2016{\natexlab{b}})\citenamefont {{Radice}}, \citenamefont
  {{Bernuzzi}},\ and\ \citenamefont {{Ott}}}]{Radice2016a}%
  \BibitemOpen
  \bibfield  {author} {\bibinfo {author} {\bibfnamefont {D.}~\bibnamefont
  {{Radice}}}, \bibinfo {author} {\bibfnamefont {S.}~\bibnamefont
  {{Bernuzzi}}}, \ and\ \bibinfo {author} {\bibfnamefont {C.~D.}\ \bibnamefont
  {{Ott}}},\ }\href@noop {} {\bibfield  {journal} {\bibinfo  {journal}
  {arXiv:1603.05726}\ } (\bibinfo {year} {2016}{\natexlab{b}})},\ \Eprint
  {http://arxiv.org/abs/1603.05726} {arXiv:1603.05726 [gr-qc]} \BibitemShut
  {NoStop}%
\bibitem [{\citenamefont {Camarda}\ \emph {et~al.}(2009)\citenamefont
  {Camarda}, \citenamefont {Anninos}, \citenamefont {Fragile},\ and\
  \citenamefont {Font}}]{Camarda:2009mk}%
  \BibitemOpen
  \bibfield  {author} {\bibinfo {author} {\bibfnamefont {K.~D.}\ \bibnamefont
  {Camarda}}, \bibinfo {author} {\bibfnamefont {P.}~\bibnamefont {Anninos}},
  \bibinfo {author} {\bibfnamefont {P.~C.}\ \bibnamefont {Fragile}}, \ and\
  \bibinfo {author} {\bibfnamefont {J.~A.}\ \bibnamefont {Font}},\ }\href
  {\doibase 10.1088/0004-637X/707/2/1610} {\bibfield  {journal} {\bibinfo
  {journal} {Astrophys. J.}\ }\textbf {\bibinfo {volume} {707}},\ \bibinfo
  {pages} {1610} (\bibinfo {year} {2009})},\ \Eprint
  {http://arxiv.org/abs/0911.0670} {arXiv:0911.0670 [astro-ph.SR]} \BibitemShut
  {NoStop}%
\bibitem [{\citenamefont {{Franci}}\ \emph
  {et~al.}(2013{\natexlab{b}})\citenamefont {{Franci}}, \citenamefont {{De
  Pietri}}, \citenamefont {{Dionysopoulou}},\ and\ \citenamefont
  {{Rezzolla}}}]{Franci2013b}%
  \BibitemOpen
  \bibfield  {author} {\bibinfo {author} {\bibfnamefont {L.}~\bibnamefont
  {{Franci}}}, \bibinfo {author} {\bibfnamefont {R.}~\bibnamefont {{De
  Pietri}}}, \bibinfo {author} {\bibfnamefont {K.}~\bibnamefont
  {{Dionysopoulou}}}, \ and\ \bibinfo {author} {\bibfnamefont {L.}~\bibnamefont
  {{Rezzolla}}},\ }\href {\doibase 10.1103/PhysRevD.88.104028} {\bibfield
  {journal} {\bibinfo  {journal} {Phys. Rev. D}\ }\textbf {\bibinfo {volume}
  {88}},\ \bibinfo {eid} {104028} (\bibinfo {year} {2013}{\natexlab{b}})},\
  \Eprint {http://arxiv.org/abs/1308.3989} {arXiv:1308.3989 [gr-qc]}
  \BibitemShut {NoStop}%
\bibitem [{\citenamefont {{Muhlberger}}\ \emph {et~al.}(2014)\citenamefont
  {{Muhlberger}}, \citenamefont {{Nouri}}, \citenamefont {{Duez}},
  \citenamefont {{Foucart}}, \citenamefont {{Kidder}}, \citenamefont {{Ott}},
  \citenamefont {{Scheel}}, \citenamefont {{Szil{\'a}gyi}},\ and\ \citenamefont
  {{Teukolsky}}}]{Muhlberger2014}%
  \BibitemOpen
  \bibfield  {author} {\bibinfo {author} {\bibfnamefont {C.~D.}\ \bibnamefont
  {{Muhlberger}}}, \bibinfo {author} {\bibfnamefont {F.~H.}\ \bibnamefont
  {{Nouri}}}, \bibinfo {author} {\bibfnamefont {M.~D.}\ \bibnamefont {{Duez}}},
  \bibinfo {author} {\bibfnamefont {F.}~\bibnamefont {{Foucart}}}, \bibinfo
  {author} {\bibfnamefont {L.~E.}\ \bibnamefont {{Kidder}}}, \bibinfo {author}
  {\bibfnamefont {C.~D.}\ \bibnamefont {{Ott}}}, \bibinfo {author}
  {\bibfnamefont {M.~A.}\ \bibnamefont {{Scheel}}}, \bibinfo {author}
  {\bibfnamefont {B.}~\bibnamefont {{Szil{\'a}gyi}}}, \ and\ \bibinfo {author}
  {\bibfnamefont {S.~A.}\ \bibnamefont {{Teukolsky}}},\ }\href {\doibase
  10.1103/PhysRevD.90.104014} {\bibfield  {journal} {\bibinfo  {journal} {Phys.
  Rev. D}\ }\textbf {\bibinfo {volume} {90}},\ \bibinfo {eid} {104014}
  (\bibinfo {year} {2014})},\ \Eprint {http://arxiv.org/abs/1405.2144}
  {arXiv:1405.2144 [astro-ph.HE]} \BibitemShut {NoStop}%
\bibitem [{\citenamefont {{Saijo}}\ \emph {et~al.}(2003)\citenamefont
  {{Saijo}}, \citenamefont {{Baumgarte}},\ and\ \citenamefont
  {{Shapiro}}}]{Saijo2003}%
  \BibitemOpen
  \bibfield  {author} {\bibinfo {author} {\bibfnamefont {M.}~\bibnamefont
  {{Saijo}}}, \bibinfo {author} {\bibfnamefont {T.~W.}\ \bibnamefont
  {{Baumgarte}}}, \ and\ \bibinfo {author} {\bibfnamefont {S.~L.}\ \bibnamefont
  {{Shapiro}}},\ }\href {\doibase 10.1086/377334} {\bibfield  {journal}
  {\bibinfo  {journal} {Astrophys. J.}\ }\textbf {\bibinfo {volume} {595}},\
  \bibinfo {pages} {352} (\bibinfo {year} {2003})},\ \Eprint
  {http://arxiv.org/abs/astro-ph/0302436} {astro-ph/0302436} \BibitemShut
  {NoStop}%
\bibitem [{\citenamefont {{Balbinski}}(1985)}]{Balbinski85b}%
  \BibitemOpen
  \bibfield  {author} {\bibinfo {author} {\bibfnamefont {E.}~\bibnamefont
  {{Balbinski}}},\ }\href@noop {} {\bibfield  {journal} {\bibinfo  {journal}
  {Mon. Not. R. Astron. Soc.}\ }\textbf {\bibinfo {volume} {216}},\ \bibinfo
  {pages} {897} (\bibinfo {year} {1985})}\BibitemShut {NoStop}%
\bibitem [{\citenamefont {{Luyten}}(1990)}]{Luyten:1990}%
  \BibitemOpen
  \bibfield  {author} {\bibinfo {author} {\bibfnamefont {P.~J.}\ \bibnamefont
  {{Luyten}}},\ }\href@noop {} {\bibfield  {journal} {\bibinfo  {journal} {Mon.
  Not. R. Astron. Soc.}\ }\textbf {\bibinfo {volume} {242}},\ \bibinfo {pages}
  {447} (\bibinfo {year} {1990})}\BibitemShut {NoStop}%
\bibitem [{\citenamefont {{Bauswein}}\ \emph {et~al.}(2014)\citenamefont
  {{Bauswein}}, \citenamefont {{Stergioulas}},\ and\ \citenamefont
  {{Janka}}}]{Bauswein2014}%
  \BibitemOpen
  \bibfield  {author} {\bibinfo {author} {\bibfnamefont {A.}~\bibnamefont
  {{Bauswein}}}, \bibinfo {author} {\bibfnamefont {N.}~\bibnamefont
  {{Stergioulas}}}, \ and\ \bibinfo {author} {\bibfnamefont {H.-T.}\
  \bibnamefont {{Janka}}},\ }\href {\doibase 10.1103/PhysRevD.90.023002}
  {\bibfield  {journal} {\bibinfo  {journal} {Phys. Rev. D}\ }\textbf {\bibinfo
  {volume} {90}},\ \bibinfo {eid} {023002} (\bibinfo {year} {2014})},\ \Eprint
  {http://arxiv.org/abs/1403.5301} {arXiv:1403.5301 [astro-ph.SR]} \BibitemShut
  {NoStop}%
\bibitem [{\citenamefont {{Takami}}\ \emph
  {et~al.}(2014{\natexlab{a}})\citenamefont {{Takami}}, \citenamefont
  {{Rezzolla}},\ and\ \citenamefont {{Baiotti}}}]{Takami:2014}%
  \BibitemOpen
  \bibfield  {author} {\bibinfo {author} {\bibfnamefont {K.}~\bibnamefont
  {{Takami}}}, \bibinfo {author} {\bibfnamefont {L.}~\bibnamefont
  {{Rezzolla}}}, \ and\ \bibinfo {author} {\bibfnamefont {L.}~\bibnamefont
  {{Baiotti}}},\ }\href {\doibase 10.1103/PhysRevLett.113.091104} {\bibfield
  {journal} {\bibinfo  {journal} {Phys. Rev. Lett.}\ }\textbf {\bibinfo
  {volume} {113}},\ \bibinfo {eid} {091104} (\bibinfo {year}
  {2014}{\natexlab{a}})},\ \Eprint {http://arxiv.org/abs/1403.5672}
  {arXiv:1403.5672 [gr-qc]} \BibitemShut {NoStop}%
\bibitem [{\citenamefont {{Bernuzzi}}\ \emph
  {et~al.}(2015{\natexlab{c}})\citenamefont {{Bernuzzi}}, \citenamefont
  {{Dietrich}},\ and\ \citenamefont {{Nagar}}}]{Bernuzzi2015a}%
  \BibitemOpen
  \bibfield  {author} {\bibinfo {author} {\bibfnamefont {S.}~\bibnamefont
  {{Bernuzzi}}}, \bibinfo {author} {\bibfnamefont {T.}~\bibnamefont
  {{Dietrich}}}, \ and\ \bibinfo {author} {\bibfnamefont {A.}~\bibnamefont
  {{Nagar}}},\ }\href {\doibase 10.1103/PhysRevLett.115.091101} {\bibfield
  {journal} {\bibinfo  {journal} {Phys. Rev. Lett.}\ }\textbf {\bibinfo
  {volume} {115}},\ \bibinfo {eid} {091101} (\bibinfo {year}
  {2015}{\natexlab{c}})},\ \Eprint {http://arxiv.org/abs/1504.01764}
  {arXiv:1504.01764 [gr-qc]} \BibitemShut {NoStop}%
\bibitem [{\citenamefont {{Maione}}\ \emph {et~al.}(2016)\citenamefont
  {{Maione}}, \citenamefont {{De Pietri}}, \citenamefont {{Feo}},\ and\
  \citenamefont {{L{\"o}ffler}}}]{Maione2016}%
  \BibitemOpen
  \bibfield  {author} {\bibinfo {author} {\bibfnamefont {F.}~\bibnamefont
  {{Maione}}}, \bibinfo {author} {\bibfnamefont {R.}~\bibnamefont {{De
  Pietri}}}, \bibinfo {author} {\bibfnamefont {A.}~\bibnamefont {{Feo}}}, \
  and\ \bibinfo {author} {\bibfnamefont {F.}~\bibnamefont {{L{\"o}ffler}}},\
  }\href@noop {} {\bibfield  {journal} {\bibinfo  {journal} {arXiv:1605.03424}\
  } (\bibinfo {year} {2016})},\ \Eprint {http://arxiv.org/abs/1605.03424}
  {arXiv:1605.03424 [gr-qc]} \BibitemShut {NoStop}%
\bibitem [{\citenamefont {{Andersson}}\ \emph {et~al.}(2011)\citenamefont
  {{Andersson}}, \citenamefont {{Ferrari}}, \citenamefont {{Jones}},
  \citenamefont {{Kokkotas}}, \citenamefont {{Krishnan}}, \citenamefont
  {{Read}}, \citenamefont {{Rezzolla}},\ and\ \citenamefont
  {{Zink}}}]{Andersson:2009yt}%
  \BibitemOpen
  \bibfield  {author} {\bibinfo {author} {\bibfnamefont {N.}~\bibnamefont
  {{Andersson}}}, \bibinfo {author} {\bibfnamefont {V.}~\bibnamefont
  {{Ferrari}}}, \bibinfo {author} {\bibfnamefont {D.~I.}\ \bibnamefont
  {{Jones}}}, \bibinfo {author} {\bibfnamefont {K.~D.}\ \bibnamefont
  {{Kokkotas}}}, \bibinfo {author} {\bibfnamefont {B.}~\bibnamefont
  {{Krishnan}}}, \bibinfo {author} {\bibfnamefont {J.~S.}\ \bibnamefont
  {{Read}}}, \bibinfo {author} {\bibfnamefont {L.}~\bibnamefont {{Rezzolla}}},
  \ and\ \bibinfo {author} {\bibfnamefont {B.}~\bibnamefont {{Zink}}},\ }\href
  {\doibase 10.1007/s10714-010-1059-4} {\bibfield  {journal} {\bibinfo
  {journal} {General Relativity and Gravitation}\ }\textbf {\bibinfo {volume}
  {43}},\ \bibinfo {pages} {409} (\bibinfo {year} {2011})},\ \Eprint
  {http://arxiv.org/abs/0912.0384} {arXiv:0912.0384 [astro-ph.SR]} \BibitemShut
  {NoStop}%
\bibitem [{\citenamefont {{Akmal}}\ \emph {et~al.}(1998)\citenamefont
  {{Akmal}}, \citenamefont {{Pandharipande}},\ and\ \citenamefont
  {{Ravenhall}}}]{Akmal1998a}%
  \BibitemOpen
  \bibfield  {author} {\bibinfo {author} {\bibfnamefont {A.}~\bibnamefont
  {{Akmal}}}, \bibinfo {author} {\bibfnamefont {V.~R.}\ \bibnamefont
  {{Pandharipande}}}, \ and\ \bibinfo {author} {\bibfnamefont {D.~G.}\
  \bibnamefont {{Ravenhall}}},\ }\href {\doibase 10.1103/PhysRevC.58.1804}
  {\bibfield  {journal} {\bibinfo  {journal} {Phys. Rev. C}\ }\textbf {\bibinfo
  {volume} {58}},\ \bibinfo {pages} {1804} (\bibinfo {year} {1998})},\ \Eprint
  {http://arxiv.org/abs/arXiv:hep-ph/9804388} {arXiv:hep-ph/9804388}
  \BibitemShut {NoStop}%
\bibitem [{\citenamefont {{Taranto}}\ \emph {et~al.}(2013)\citenamefont
  {{Taranto}}, \citenamefont {{Baldo}},\ and\ \citenamefont
  {{Burgio}}}]{Taranto2013}%
  \BibitemOpen
  \bibfield  {author} {\bibinfo {author} {\bibfnamefont {G.}~\bibnamefont
  {{Taranto}}}, \bibinfo {author} {\bibfnamefont {M.}~\bibnamefont {{Baldo}}},
  \ and\ \bibinfo {author} {\bibfnamefont {G.~F.}\ \bibnamefont {{Burgio}}},\
  }\href {\doibase 10.1103/PhysRevC.87.045803} {\bibfield  {journal} {\bibinfo
  {journal} {Phys. Rev. C}\ }\textbf {\bibinfo {volume} {87}},\ \bibinfo {eid}
  {045803} (\bibinfo {year} {2013})},\ \Eprint {http://arxiv.org/abs/1302.6882}
  {arXiv:1302.6882 [nucl-th]} \BibitemShut {NoStop}%
\bibitem [{\citenamefont {{Hotokezaka}}\ \emph {et~al.}(2011)\citenamefont
  {{Hotokezaka}}, \citenamefont {{Kyutoku}}, \citenamefont {{Okawa}},
  \citenamefont {{Shibata}},\ and\ \citenamefont {{Kiuchi}}}]{Hotokezaka2011}%
  \BibitemOpen
  \bibfield  {author} {\bibinfo {author} {\bibfnamefont {K.}~\bibnamefont
  {{Hotokezaka}}}, \bibinfo {author} {\bibfnamefont {K.}~\bibnamefont
  {{Kyutoku}}}, \bibinfo {author} {\bibfnamefont {H.}~\bibnamefont {{Okawa}}},
  \bibinfo {author} {\bibfnamefont {M.}~\bibnamefont {{Shibata}}}, \ and\
  \bibinfo {author} {\bibfnamefont {K.}~\bibnamefont {{Kiuchi}}},\ }\href
  {\doibase 10.1103/PhysRevD.83.124008} {\bibfield  {journal} {\bibinfo
  {journal} {Phys. Rev. D}\ }\textbf {\bibinfo {volume} {83}},\ \bibinfo {eid}
  {124008} (\bibinfo {year} {2011})},\ \Eprint {http://arxiv.org/abs/1105.4370}
  {arXiv:1105.4370 [astro-ph.HE]} \BibitemShut {NoStop}%
\bibitem [{\citenamefont {Witten}(1984)}]{Witten84}%
  \BibitemOpen
  \bibfield  {author} {\bibinfo {author} {\bibfnamefont {E.}~\bibnamefont
  {Witten}},\ }\href {\doibase 10.1103/physrevd.30.272} {\bibfield  {journal}
  {\bibinfo  {journal} {Phys. Rev. D}\ }\textbf {\bibinfo {volume} {30}},\
  \bibinfo {pages} {272} (\bibinfo {year} {1984})}\BibitemShut {NoStop}%
\bibitem [{\citenamefont {{Haensel}}\ \emph {et~al.}(1986)\citenamefont
  {{Haensel}}, \citenamefont {{Zdunik}},\ and\ \citenamefont
  {{Schaefer}}}]{Haensel86}%
  \BibitemOpen
  \bibfield  {author} {\bibinfo {author} {\bibfnamefont {P.}~\bibnamefont
  {{Haensel}}}, \bibinfo {author} {\bibfnamefont {J.~L.}\ \bibnamefont
  {{Zdunik}}}, \ and\ \bibinfo {author} {\bibfnamefont {R.}~\bibnamefont
  {{Schaefer}}},\ }\href@noop {} {\bibfield  {journal} {\bibinfo  {journal}
  {Astron. Astrophys.}\ }\textbf {\bibinfo {volume} {160}},\ \bibinfo {pages}
  {121} (\bibinfo {year} {1986})}\BibitemShut {NoStop}%
\bibitem [{\citenamefont {{Alcock}}\ \emph {et~al.}(1986)\citenamefont
  {{Alcock}}, \citenamefont {{Farhi}},\ and\ \citenamefont
  {{Olinto}}}]{Alcock86}%
  \BibitemOpen
  \bibfield  {author} {\bibinfo {author} {\bibfnamefont {C.}~\bibnamefont
  {{Alcock}}}, \bibinfo {author} {\bibfnamefont {E.}~\bibnamefont {{Farhi}}}, \
  and\ \bibinfo {author} {\bibfnamefont {A.}~\bibnamefont {{Olinto}}},\ }\href
  {\doibase 10.1086/164679} {\bibfield  {journal} {\bibinfo  {journal}
  {Astrophys. J.}\ }\textbf {\bibinfo {volume} {310}},\ \bibinfo {pages} {261}
  (\bibinfo {year} {1986})}\BibitemShut {NoStop}%
\bibitem [{\citenamefont {{Alford}}\ \emph {et~al.}(2005)\citenamefont
  {{Alford}}, \citenamefont {{Braby}}, \citenamefont {{Paris}},\ and\
  \citenamefont {{Reddy}}}]{Alford2005}%
  \BibitemOpen
  \bibfield  {author} {\bibinfo {author} {\bibfnamefont {M.}~\bibnamefont
  {{Alford}}}, \bibinfo {author} {\bibfnamefont {M.}~\bibnamefont {{Braby}}},
  \bibinfo {author} {\bibfnamefont {M.}~\bibnamefont {{Paris}}}, \ and\
  \bibinfo {author} {\bibfnamefont {S.}~\bibnamefont {{Reddy}}},\ }\href
  {\doibase 10.1086/430902} {\bibfield  {journal} {\bibinfo  {journal}
  {Astrophys. J.}\ }\textbf {\bibinfo {volume} {629}},\ \bibinfo {pages} {969}
  (\bibinfo {year} {2005})},\ \Eprint {http://arxiv.org/abs/nucl-th/0411016}
  {nucl-th/0411016} \BibitemShut {NoStop}%
\bibitem [{\citenamefont {Rikovska-Stone}\ \emph {et~al.}(2007)\citenamefont
  {Rikovska-Stone}, \citenamefont {Guichon}, \citenamefont {Matevosyan},\ and\
  \citenamefont {Thomas}}]{RikovskaStone:2006ta}%
  \BibitemOpen
  \bibfield  {author} {\bibinfo {author} {\bibfnamefont {J.}~\bibnamefont
  {Rikovska-Stone}}, \bibinfo {author} {\bibfnamefont {P.~A.}\ \bibnamefont
  {Guichon}}, \bibinfo {author} {\bibfnamefont {H.~H.}\ \bibnamefont
  {Matevosyan}}, \ and\ \bibinfo {author} {\bibfnamefont {A.~W.}\ \bibnamefont
  {Thomas}},\ }\href {\doibase 10.1016/j.nuclphysa.2007.05.011} {\bibfield
  {journal} {\bibinfo  {journal} {Nucl.Phys.}\ }\textbf {\bibinfo {volume}
  {A792}},\ \bibinfo {pages} {341} (\bibinfo {year} {2007})},\ \Eprint
  {http://arxiv.org/abs/nucl-th/0611030} {arXiv:nucl-th/0611030 [nucl-th]}
  \BibitemShut {NoStop}%
\bibitem [{\citenamefont {Weissenborn}\ \emph {et~al.}(2011)\citenamefont
  {Weissenborn}, \citenamefont {Sagert}, \citenamefont {Pagliara},
  \citenamefont {Hempel},\ and\ \citenamefont
  {Schaffner-Bielich}}]{Weissenborn:2011qu}%
  \BibitemOpen
  \bibfield  {author} {\bibinfo {author} {\bibfnamefont {S.}~\bibnamefont
  {Weissenborn}}, \bibinfo {author} {\bibfnamefont {I.}~\bibnamefont {Sagert}},
  \bibinfo {author} {\bibfnamefont {G.}~\bibnamefont {Pagliara}}, \bibinfo
  {author} {\bibfnamefont {M.}~\bibnamefont {Hempel}}, \ and\ \bibinfo {author}
  {\bibfnamefont {J.}~\bibnamefont {Schaffner-Bielich}},\ }\href {\doibase
  10.1088/2041-8205/740/1/L14} {\bibfield  {journal} {\bibinfo  {journal}
  {Astrophys.J.}\ }\textbf {\bibinfo {volume} {740}},\ \bibinfo {pages} {L14}
  (\bibinfo {year} {2011})},\ \Eprint {http://arxiv.org/abs/1102.2869}
  {arXiv:1102.2869 [astro-ph.HE]} \BibitemShut {NoStop}%
\bibitem [{\citenamefont {Bhowmick}\ \emph {et~al.}(2014)\citenamefont
  {Bhowmick}, \citenamefont {Bhattacharya}, \citenamefont {Bhattacharyya},\
  and\ \citenamefont {Gangopadhyay}}]{Bhowmick:2014pma}%
  \BibitemOpen
  \bibfield  {author} {\bibinfo {author} {\bibfnamefont {B.}~\bibnamefont
  {Bhowmick}}, \bibinfo {author} {\bibfnamefont {M.}~\bibnamefont
  {Bhattacharya}}, \bibinfo {author} {\bibfnamefont {A.}~\bibnamefont
  {Bhattacharyya}}, \ and\ \bibinfo {author} {\bibfnamefont {G.}~\bibnamefont
  {Gangopadhyay}},\ }\href {\doibase 10.1103/PhysRevC.89.065806} {\bibfield
  {journal} {\bibinfo  {journal} {Phys.Rev.}\ }\textbf {\bibinfo {volume}
  {C89}},\ \bibinfo {pages} {065806} (\bibinfo {year} {2014})},\ \Eprint
  {http://arxiv.org/abs/1403.0341} {arXiv:1403.0341 [nucl-th]} \BibitemShut
  {NoStop}%
\bibitem [{\citenamefont {{Bauswein}}\ \emph {et~al.}(2009)\citenamefont
  {{Bauswein}}, \citenamefont {{Janka}}, \citenamefont {{Oechslin}},
  \citenamefont {{Pagliara}}, \citenamefont {{Sagert}}, \citenamefont
  {{Schaffner-Bielich}}, \citenamefont {{Hohle}},\ and\ \citenamefont
  {{Neuh{\"a}user}}}]{Bauswein2009}%
  \BibitemOpen
  \bibfield  {author} {\bibinfo {author} {\bibfnamefont {A.}~\bibnamefont
  {{Bauswein}}}, \bibinfo {author} {\bibfnamefont {H.-T.}\ \bibnamefont
  {{Janka}}}, \bibinfo {author} {\bibfnamefont {R.}~\bibnamefont {{Oechslin}}},
  \bibinfo {author} {\bibfnamefont {G.}~\bibnamefont {{Pagliara}}}, \bibinfo
  {author} {\bibfnamefont {I.}~\bibnamefont {{Sagert}}}, \bibinfo {author}
  {\bibfnamefont {J.}~\bibnamefont {{Schaffner-Bielich}}}, \bibinfo {author}
  {\bibfnamefont {M.~M.}\ \bibnamefont {{Hohle}}}, \ and\ \bibinfo {author}
  {\bibfnamefont {R.}~\bibnamefont {{Neuh{\"a}user}}},\ }\href {\doibase
  10.1103/PhysRevLett.103.011101} {\bibfield  {journal} {\bibinfo  {journal}
  {Phys. Rev. Lett.}\ }\textbf {\bibinfo {volume} {103}},\ \bibinfo {eid}
  {011101} (\bibinfo {year} {2009})},\ \Eprint {http://arxiv.org/abs/0812.4248}
  {arXiv:0812.4248} \BibitemShut {NoStop}%
\bibitem [{\citenamefont {{Bauswein}}\ \emph
  {et~al.}(2010{\natexlab{a}})\citenamefont {{Bauswein}}, \citenamefont
  {{Oechslin}},\ and\ \citenamefont {{Janka}}}]{Bauswein2010}%
  \BibitemOpen
  \bibfield  {author} {\bibinfo {author} {\bibfnamefont {A.}~\bibnamefont
  {{Bauswein}}}, \bibinfo {author} {\bibfnamefont {R.}~\bibnamefont
  {{Oechslin}}}, \ and\ \bibinfo {author} {\bibfnamefont {H.-T.}\ \bibnamefont
  {{Janka}}},\ }\href {\doibase 10.1103/PhysRevD.81.024012} {\bibfield
  {journal} {\bibinfo  {journal} {Phys. Rev. D}\ }\textbf {\bibinfo {volume}
  {81}},\ \bibinfo {eid} {024012} (\bibinfo {year} {2010}{\natexlab{a}})},\
  \Eprint {http://arxiv.org/abs/0910.5169} {arXiv:0910.5169 [astro-ph.SR]}
  \BibitemShut {NoStop}%
\bibitem [{\citenamefont {{Farhi}}\ and\ \citenamefont
  {{Jaffe}}(1984)}]{Farhi1984}%
  \BibitemOpen
  \bibfield  {author} {\bibinfo {author} {\bibfnamefont {E.}~\bibnamefont
  {{Farhi}}}\ and\ \bibinfo {author} {\bibfnamefont {R.~L.}\ \bibnamefont
  {{Jaffe}}},\ }\href {\doibase 10.1103/PhysRevD.30.2379} {\bibfield  {journal}
  {\bibinfo  {journal} {Phys. Rev. D}\ }\textbf {\bibinfo {volume} {30}},\
  \bibinfo {pages} {2379} (\bibinfo {year} {1984})}\BibitemShut {NoStop}%
\bibitem [{\citenamefont {{Sekiguchi}}\ \emph
  {et~al.}(2011{\natexlab{a}})\citenamefont {{Sekiguchi}}, \citenamefont
  {{Kiuchi}}, \citenamefont {{Kyutoku}},\ and\ \citenamefont
  {{Shibata}}}]{Sekiguchi2011b}%
  \BibitemOpen
  \bibfield  {author} {\bibinfo {author} {\bibfnamefont {Y.}~\bibnamefont
  {{Sekiguchi}}}, \bibinfo {author} {\bibfnamefont {K.}~\bibnamefont
  {{Kiuchi}}}, \bibinfo {author} {\bibfnamefont {K.}~\bibnamefont {{Kyutoku}}},
  \ and\ \bibinfo {author} {\bibfnamefont {M.}~\bibnamefont {{Shibata}}},\
  }\href {\doibase 10.1103/PhysRevLett.107.211101} {\bibfield  {journal}
  {\bibinfo  {journal} {Phys. Rev. Lett.}\ }\textbf {\bibinfo {volume} {107}},\
  \bibinfo {eid} {211101} (\bibinfo {year} {2011}{\natexlab{a}})},\ \Eprint
  {http://arxiv.org/abs/1110.4442} {arXiv:1110.4442 [astro-ph.HE]} \BibitemShut
  {NoStop}%
\bibitem [{\citenamefont {{Kiuchi}}\ \emph
  {et~al.}(2012{\natexlab{a}})\citenamefont {{Kiuchi}}, \citenamefont
  {{Sekiguchi}}, \citenamefont {{Kyutoku}},\ and\ \citenamefont
  {{Shibata}}}]{Kiuchi2012}%
  \BibitemOpen
  \bibfield  {author} {\bibinfo {author} {\bibfnamefont {K.}~\bibnamefont
  {{Kiuchi}}}, \bibinfo {author} {\bibfnamefont {Y.}~\bibnamefont
  {{Sekiguchi}}}, \bibinfo {author} {\bibfnamefont {K.}~\bibnamefont
  {{Kyutoku}}}, \ and\ \bibinfo {author} {\bibfnamefont {M.}~\bibnamefont
  {{Shibata}}},\ }\href {\doibase 10.1088/0264-9381/29/12/124003} {\bibfield
  {journal} {\bibinfo  {journal} {Class. Quantum Grav.}\ }\textbf {\bibinfo
  {volume} {29}},\ \bibinfo {pages} {124003} (\bibinfo {year}
  {2012}{\natexlab{a}})},\ \Eprint {http://arxiv.org/abs/1206.0509}
  {arXiv:1206.0509 [astro-ph.HE]} \BibitemShut {NoStop}%
\bibitem [{\citenamefont {{Sekiguchi}}\ \emph {et~al.}(2012)\citenamefont
  {{Sekiguchi}}, \citenamefont {{Kiuchi}}, \citenamefont {{Kyutoku}},\ and\
  \citenamefont {{Shibata}}}]{Sekiguchi2012}%
  \BibitemOpen
  \bibfield  {author} {\bibinfo {author} {\bibfnamefont {Y.}~\bibnamefont
  {{Sekiguchi}}}, \bibinfo {author} {\bibfnamefont {K.}~\bibnamefont
  {{Kiuchi}}}, \bibinfo {author} {\bibfnamefont {K.}~\bibnamefont {{Kyutoku}}},
  \ and\ \bibinfo {author} {\bibfnamefont {M.}~\bibnamefont {{Shibata}}},\
  }\href {\doibase 10.1093/ptep/pts011} {\bibfield  {journal} {\bibinfo
  {journal} {Progress of Theoretical and Experimental Physics}\ }\textbf
  {\bibinfo {volume} {2012}},\ \bibinfo {eid} {01A304} (\bibinfo {year}
  {2012})},\ \Eprint {http://arxiv.org/abs/1206.5927} {arXiv:1206.5927
  [astro-ph.HE]} \BibitemShut {NoStop}%
\bibitem [{\citenamefont {{Kiuchi}}\ \emph
  {et~al.}(2012{\natexlab{b}})\citenamefont {{Kiuchi}}, \citenamefont
  {{Sekiguchi}}, \citenamefont {{Kyutoku}},\ and\ \citenamefont
  {{Shibata}}}]{Kiuchi2012a}%
  \BibitemOpen
  \bibfield  {author} {\bibinfo {author} {\bibfnamefont {K.}~\bibnamefont
  {{Kiuchi}}}, \bibinfo {author} {\bibfnamefont {Y.}~\bibnamefont
  {{Sekiguchi}}}, \bibinfo {author} {\bibfnamefont {K.}~\bibnamefont
  {{Kyutoku}}}, \ and\ \bibinfo {author} {\bibfnamefont {M.}~\bibnamefont
  {{Shibata}}},\ }in\ \href@noop {} {\emph {\bibinfo {booktitle} {Numerical
  Modeling of Space Plasma Slows (ASTRONUM 2011)}}},\ \bibinfo {series}
  {Astronomical Society of the Pacific Conference Series}, Vol.\ \bibinfo
  {volume} {459},\ \bibinfo {editor} {edited by\ \bibinfo {editor}
  {\bibfnamefont {N.~V.}\ \bibnamefont {{Pogorelov}}}, \bibinfo {editor}
  {\bibfnamefont {J.~A.}\ \bibnamefont {{Font}}}, \bibinfo {editor}
  {\bibfnamefont {E.}~\bibnamefont {{Audit}}}, \ and\ \bibinfo {editor}
  {\bibfnamefont {G.~P.}\ \bibnamefont {{Zank}}}}\ (\bibinfo {year} {2012})\
  p.~\bibinfo {pages} {85}\BibitemShut {NoStop}%
\bibitem [{\citenamefont {{Flanagan}}\ and\ \citenamefont
  {{Hinderer}}(2008)}]{Flanagan08}%
  \BibitemOpen
  \bibfield  {author} {\bibinfo {author} {\bibfnamefont {{\'E}.~{\'E}.}\
  \bibnamefont {{Flanagan}}}\ and\ \bibinfo {author} {\bibfnamefont
  {T.}~\bibnamefont {{Hinderer}}},\ }\href {\doibase
  10.1103/PhysRevD.77.021502} {\bibfield  {journal} {\bibinfo  {journal} {Phys.
  Rev. D}\ }\textbf {\bibinfo {volume} {77}},\ \bibinfo {pages} {021502}
  (\bibinfo {year} {2008})},\ \Eprint {http://arxiv.org/abs/0709.1915}
  {arXiv:0709.1915} \BibitemShut {NoStop}%
\bibitem [{\citenamefont {{Hotokezaka}}\ \emph
  {et~al.}(2013{\natexlab{b}})\citenamefont {{Hotokezaka}}, \citenamefont
  {{Kiuchi}}, \citenamefont {{Kyutoku}}, \citenamefont {{Muranushi}},
  \citenamefont {{Sekiguchi}}, \citenamefont {{Shibata}},\ and\ \citenamefont
  {{Taniguchi}}}]{Hotokezaka2013c}%
  \BibitemOpen
  \bibfield  {author} {\bibinfo {author} {\bibfnamefont {K.}~\bibnamefont
  {{Hotokezaka}}}, \bibinfo {author} {\bibfnamefont {K.}~\bibnamefont
  {{Kiuchi}}}, \bibinfo {author} {\bibfnamefont {K.}~\bibnamefont {{Kyutoku}}},
  \bibinfo {author} {\bibfnamefont {T.}~\bibnamefont {{Muranushi}}}, \bibinfo
  {author} {\bibfnamefont {Y.-i.}\ \bibnamefont {{Sekiguchi}}}, \bibinfo
  {author} {\bibfnamefont {M.}~\bibnamefont {{Shibata}}}, \ and\ \bibinfo
  {author} {\bibfnamefont {K.}~\bibnamefont {{Taniguchi}}},\ }\href {\doibase
  10.1103/PhysRevD.88.044026} {\bibfield  {journal} {\bibinfo  {journal} {Phys.
  Rev. D}\ }\textbf {\bibinfo {volume} {88}},\ \bibinfo {eid} {044026}
  (\bibinfo {year} {2013}{\natexlab{b}})},\ \Eprint
  {http://arxiv.org/abs/1307.5888} {arXiv:1307.5888 [astro-ph.HE]} \BibitemShut
  {NoStop}%
\bibitem [{url()}]{url:adLIGO_Sh_curve}%
  \BibitemOpen
  \href {https://dcc.ligo.org/LIGO-T0900288/public} {}\bibinfo {note} {Advanced
  LIGO anticipated sensitivity curves, LIGO Document No.
  T0900288-v3}\BibitemShut {NoStop}%
\bibitem [{\citenamefont {Punturo}\ \emph
  {et~al.}(2010{\natexlab{b}})\citenamefont {Punturo} \emph
  {et~al.}}]{Punturo2010b}%
  \BibitemOpen
  \bibfield  {author} {\bibinfo {author} {\bibfnamefont {M.}~\bibnamefont
  {Punturo}} \emph {et~al.},\ }\href {\doibase 10.1088/0264-9381/27/19/194002}
  {\bibfield  {journal} {\bibinfo  {journal} {Class. Quantum Grav.}\ }\textbf
  {\bibinfo {volume} {27}},\ \bibinfo {pages} {194002} (\bibinfo {year}
  {2010}{\natexlab{b}})}\BibitemShut {NoStop}%
\bibitem [{\citenamefont {Sathyaprakash}\ and\ \citenamefont
  {Schutz}(2009)}]{Sathyaprakash:2009xs}%
  \BibitemOpen
  \bibfield  {author} {\bibinfo {author} {\bibfnamefont {B.~S.}\ \bibnamefont
  {Sathyaprakash}}\ and\ \bibinfo {author} {\bibfnamefont {B.~F.}\ \bibnamefont
  {Schutz}},\ }\href@noop {} {\bibfield  {journal} {\bibinfo  {journal} {Living
  Rev. Relativ.}\ }\textbf {\bibinfo {volume} {12}},\ \bibinfo {pages} {2}
  (\bibinfo {year} {2009})},\ \Eprint {http://arxiv.org/abs/0903.0338}
  {arXiv:0903.0338 [gr-qc]} \BibitemShut {NoStop}%
\bibitem [{\citenamefont {{Kokkotas}}\ and\ \citenamefont
  {{Schmidt}}(1999)}]{Kokkotas99a}%
  \BibitemOpen
  \bibfield  {author} {\bibinfo {author} {\bibfnamefont {K.}~\bibnamefont
  {{Kokkotas}}}\ and\ \bibinfo {author} {\bibfnamefont {B.}~\bibnamefont
  {{Schmidt}}},\ }\href {\doibase 10.12942/lrr-1999-2} {\bibfield  {journal}
  {\bibinfo  {journal} {Living Rev. Relativity}\ }\textbf {\bibinfo {volume}
  {2}},\ \bibinfo {pages} {2} (\bibinfo {year} {1999})},\ \Eprint
  {http://arxiv.org/abs/gr-qc/9909058} {gr-qc/9909058} \BibitemShut {NoStop}%
\bibitem [{\citenamefont {{Oechslin}}\ and\ \citenamefont
  {{Janka}}(2007)}]{Oechslin07b}%
  \BibitemOpen
  \bibfield  {author} {\bibinfo {author} {\bibfnamefont {R.}~\bibnamefont
  {{Oechslin}}}\ and\ \bibinfo {author} {\bibfnamefont {H.-T.}\ \bibnamefont
  {{Janka}}},\ }\href {\doibase 10.1103/PhysRevLett.99.121102} {\bibfield
  {journal} {\bibinfo  {journal} {Phys. Rev. Lett.}\ }\textbf {\bibinfo
  {volume} {99}},\ \bibinfo {eid} {121102} (\bibinfo {year} {2007})},\ \Eprint
  {http://arxiv.org/abs/astro-ph/0702228} {astro-ph/0702228} \BibitemShut
  {NoStop}%
\bibitem [{\citenamefont {{Bauswein}}\ and\ \citenamefont
  {{Janka}}(2012)}]{Bauswein2012a}%
  \BibitemOpen
  \bibfield  {author} {\bibinfo {author} {\bibfnamefont {A.}~\bibnamefont
  {{Bauswein}}}\ and\ \bibinfo {author} {\bibfnamefont {H.-T.}\ \bibnamefont
  {{Janka}}},\ }\href {\doibase 10.1103/PhysRevLett.108.011101} {\bibfield
  {journal} {\bibinfo  {journal} {Phys. Rev. Lett.}\ }\textbf {\bibinfo
  {volume} {108}},\ \bibinfo {eid} {011101} (\bibinfo {year} {2012})},\ \Eprint
  {http://arxiv.org/abs/1106.1616} {arXiv:1106.1616 [astro-ph.SR]} \BibitemShut
  {NoStop}%
\bibitem [{\citenamefont {{Bauswein}}\ \emph {et~al.}(2012)\citenamefont
  {{Bauswein}}, \citenamefont {{Janka}}, \citenamefont {{Hebeler}},\ and\
  \citenamefont {{Schwenk}}}]{Bauswein2012}%
  \BibitemOpen
  \bibfield  {author} {\bibinfo {author} {\bibfnamefont {A.}~\bibnamefont
  {{Bauswein}}}, \bibinfo {author} {\bibfnamefont {H.-T.}\ \bibnamefont
  {{Janka}}}, \bibinfo {author} {\bibfnamefont {K.}~\bibnamefont {{Hebeler}}},
  \ and\ \bibinfo {author} {\bibfnamefont {A.}~\bibnamefont {{Schwenk}}},\
  }\href {\doibase 10.1103/PhysRevD.86.063001} {\bibfield  {journal} {\bibinfo
  {journal} {Phys. Rev. D}\ }\textbf {\bibinfo {volume} {86}},\ \bibinfo {eid}
  {063001} (\bibinfo {year} {2012})},\ \Eprint {http://arxiv.org/abs/1204.1888}
  {arXiv:1204.1888 [astro-ph.SR]} \BibitemShut {NoStop}%
\bibitem [{\citenamefont {{Clark}}\ \emph {et~al.}(2014)\citenamefont
  {{Clark}}, \citenamefont {{Bauswein}}, \citenamefont {{Cadonati}},
  \citenamefont {{Janka}}, \citenamefont {{Pankow}},\ and\ \citenamefont
  {{Stergioulas}}}]{Clark2014}%
  \BibitemOpen
  \bibfield  {author} {\bibinfo {author} {\bibfnamefont {J.}~\bibnamefont
  {{Clark}}}, \bibinfo {author} {\bibfnamefont {A.}~\bibnamefont {{Bauswein}}},
  \bibinfo {author} {\bibfnamefont {L.}~\bibnamefont {{Cadonati}}}, \bibinfo
  {author} {\bibfnamefont {H.-T.}\ \bibnamefont {{Janka}}}, \bibinfo {author}
  {\bibfnamefont {C.}~\bibnamefont {{Pankow}}}, \ and\ \bibinfo {author}
  {\bibfnamefont {N.}~\bibnamefont {{Stergioulas}}},\ }\href {\doibase
  10.1103/PhysRevD.90.062004} {\bibfield  {journal} {\bibinfo  {journal} {Phys.
  Rev. D}\ }\textbf {\bibinfo {volume} {90}},\ \bibinfo {eid} {062004}
  (\bibinfo {year} {2014})},\ \Eprint {http://arxiv.org/abs/1406.5444}
  {arXiv:1406.5444 [astro-ph.HE]} \BibitemShut {NoStop}%
\bibitem [{\citenamefont {{Bauswein}}\ and\ \citenamefont
  {{Stergioulas}}(2015)}]{Bauswein2015}%
  \BibitemOpen
  \bibfield  {author} {\bibinfo {author} {\bibfnamefont {A.}~\bibnamefont
  {{Bauswein}}}\ and\ \bibinfo {author} {\bibfnamefont {N.}~\bibnamefont
  {{Stergioulas}}},\ }\href {\doibase 10.1103/PhysRevD.91.124056} {\bibfield
  {journal} {\bibinfo  {journal} {Phys. Rev. D}\ }\textbf {\bibinfo {volume}
  {91}},\ \bibinfo {eid} {124056} (\bibinfo {year} {2015})},\ \Eprint
  {http://arxiv.org/abs/1502.03176} {arXiv:1502.03176 [astro-ph.SR]}
  \BibitemShut {NoStop}%
\bibitem [{\citenamefont {{Dietrich}}\ \emph
  {et~al.}(2015{\natexlab{b}})\citenamefont {{Dietrich}}, \citenamefont
  {{Bernuzzi}}, \citenamefont {{Ujevic}},\ and\ \citenamefont
  {{Br{\"u}gmann}}}]{Dietrich2015}%
  \BibitemOpen
  \bibfield  {author} {\bibinfo {author} {\bibfnamefont {T.}~\bibnamefont
  {{Dietrich}}}, \bibinfo {author} {\bibfnamefont {S.}~\bibnamefont
  {{Bernuzzi}}}, \bibinfo {author} {\bibfnamefont {M.}~\bibnamefont
  {{Ujevic}}}, \ and\ \bibinfo {author} {\bibfnamefont {B.}~\bibnamefont
  {{Br{\"u}gmann}}},\ }\href {\doibase 10.1103/PhysRevD.91.124041} {\bibfield
  {journal} {\bibinfo  {journal} {Phys. Rev. D}\ }\textbf {\bibinfo {volume}
  {91}},\ \bibinfo {eid} {124041} (\bibinfo {year} {2015}{\natexlab{b}})},\
  \Eprint {http://arxiv.org/abs/1504.01266} {arXiv:1504.01266 [gr-qc]}
  \BibitemShut {NoStop}%
\bibitem [{\citenamefont {{Foucart}}\ \emph {et~al.}(2016)\citenamefont
  {{Foucart}}, \citenamefont {{Haas}}, \citenamefont {{Duez}}, \citenamefont
  {{O'Connor}}, \citenamefont {{Ott}}, \citenamefont {{Roberts}}, \citenamefont
  {{Kidder}}, \citenamefont {{Lippuner}}, \citenamefont {{Pfeiffer}},\ and\
  \citenamefont {{Scheel}}}]{Foucart2015}%
  \BibitemOpen
  \bibfield  {author} {\bibinfo {author} {\bibfnamefont {F.}~\bibnamefont
  {{Foucart}}}, \bibinfo {author} {\bibfnamefont {R.}~\bibnamefont {{Haas}}},
  \bibinfo {author} {\bibfnamefont {M.~D.}\ \bibnamefont {{Duez}}}, \bibinfo
  {author} {\bibfnamefont {E.}~\bibnamefont {{O'Connor}}}, \bibinfo {author}
  {\bibfnamefont {C.~D.}\ \bibnamefont {{Ott}}}, \bibinfo {author}
  {\bibfnamefont {L.}~\bibnamefont {{Roberts}}}, \bibinfo {author}
  {\bibfnamefont {L.~E.}\ \bibnamefont {{Kidder}}}, \bibinfo {author}
  {\bibfnamefont {J.}~\bibnamefont {{Lippuner}}}, \bibinfo {author}
  {\bibfnamefont {H.~P.}\ \bibnamefont {{Pfeiffer}}}, \ and\ \bibinfo {author}
  {\bibfnamefont {M.~A.}\ \bibnamefont {{Scheel}}},\ }\href {\doibase
  10.1103/PhysRevD.93.044019} {\bibfield  {journal} {\bibinfo  {journal} {Phys.
  Rev. D}\ }\textbf {\bibinfo {volume} {93}},\ \bibinfo {eid} {044019}
  (\bibinfo {year} {2016})},\ \Eprint {http://arxiv.org/abs/1510.06398}
  {arXiv:1510.06398 [astro-ph.HE]} \BibitemShut {NoStop}%
\bibitem [{\citenamefont {{De Pietri}}\ \emph {et~al.}(2016)\citenamefont {{De
  Pietri}}, \citenamefont {{Feo}}, \citenamefont {{Maione}},\ and\
  \citenamefont {{L{\"o}ffler}}}]{DePietri2016}%
  \BibitemOpen
  \bibfield  {author} {\bibinfo {author} {\bibfnamefont {R.}~\bibnamefont {{De
  Pietri}}}, \bibinfo {author} {\bibfnamefont {A.}~\bibnamefont {{Feo}}},
  \bibinfo {author} {\bibfnamefont {F.}~\bibnamefont {{Maione}}}, \ and\
  \bibinfo {author} {\bibfnamefont {F.}~\bibnamefont {{L{\"o}ffler}}},\ }\href
  {\doibase 10.1103/PhysRevD.93.064047} {\bibfield  {journal} {\bibinfo
  {journal} {Phys. Rev. D}\ }\textbf {\bibinfo {volume} {93}},\ \bibinfo {eid}
  {064047} (\bibinfo {year} {2016})},\ \Eprint
  {http://arxiv.org/abs/1509.08804} {arXiv:1509.08804 [gr-qc]} \BibitemShut
  {NoStop}%
\bibitem [{\citenamefont {{Bauswein}}\ \emph {et~al.}(2016)\citenamefont
  {{Bauswein}}, \citenamefont {{Stergioulas}},\ and\ \citenamefont
  {{Janka}}}]{Bauswein2015b}%
  \BibitemOpen
  \bibfield  {author} {\bibinfo {author} {\bibfnamefont {A.}~\bibnamefont
  {{Bauswein}}}, \bibinfo {author} {\bibfnamefont {N.}~\bibnamefont
  {{Stergioulas}}}, \ and\ \bibinfo {author} {\bibfnamefont {H.-T.}\
  \bibnamefont {{Janka}}},\ }\href {\doibase 10.1140/epja/i2016-16056-7}
  {\bibfield  {journal} {\bibinfo  {journal} {European Physical Journal A}\
  }\textbf {\bibinfo {volume} {52}},\ \bibinfo {eid} {56} (\bibinfo {year}
  {2016})},\ \Eprint {http://arxiv.org/abs/1508.05493} {arXiv:1508.05493
  [astro-ph.HE]} \BibitemShut {NoStop}%
\bibitem [{\citenamefont {{Rezzolla}}\ and\ \citenamefont
  {{Takami}}(2016)}]{Rezzolla2016}%
  \BibitemOpen
  \bibfield  {author} {\bibinfo {author} {\bibfnamefont {L.}~\bibnamefont
  {{Rezzolla}}}\ and\ \bibinfo {author} {\bibfnamefont {K.}~\bibnamefont
  {{Takami}}},\ }\href {\doibase 10.1103/PhysRevD.93.124051} {\bibfield
  {journal} {\bibinfo  {journal} {Phys. Rev. D}\ }\textbf {\bibinfo {volume}
  {93}},\ \bibinfo {eid} {124051} (\bibinfo {year} {2016})},\ \Eprint
  {http://arxiv.org/abs/1604.00246} {arXiv:1604.00246 [gr-qc]} \BibitemShut
  {NoStop}%
\bibitem [{\citenamefont {{Bauswein}}\ \emph
  {et~al.}(2013{\natexlab{a}})\citenamefont {{Bauswein}}, \citenamefont
  {{Baumgarte}},\ and\ \citenamefont {{Janka}}}]{Bauswein2013}%
  \BibitemOpen
  \bibfield  {author} {\bibinfo {author} {\bibfnamefont {A.}~\bibnamefont
  {{Bauswein}}}, \bibinfo {author} {\bibfnamefont {T.~W.}\ \bibnamefont
  {{Baumgarte}}}, \ and\ \bibinfo {author} {\bibfnamefont {H.-T.}\ \bibnamefont
  {{Janka}}},\ }\href {\doibase 10.1103/PhysRevLett.111.131101} {\bibfield
  {journal} {\bibinfo  {journal} {Phys. Rev. Lett.}\ }\textbf {\bibinfo
  {volume} {111}},\ \bibinfo {eid} {131101} (\bibinfo {year}
  {2013}{\natexlab{a}})},\ \Eprint {http://arxiv.org/abs/1307.5191}
  {arXiv:1307.5191 [astro-ph.SR]} \BibitemShut {NoStop}%
\bibitem [{\citenamefont {{Takami}}\ \emph
  {et~al.}(2014{\natexlab{b}})\citenamefont {{Takami}}, \citenamefont
  {{Rezzolla}},\ and\ \citenamefont {{Baiotti}}}]{Takami2014}%
  \BibitemOpen
  \bibfield  {author} {\bibinfo {author} {\bibfnamefont {K.}~\bibnamefont
  {{Takami}}}, \bibinfo {author} {\bibfnamefont {L.}~\bibnamefont
  {{Rezzolla}}}, \ and\ \bibinfo {author} {\bibfnamefont {L.}~\bibnamefont
  {{Baiotti}}},\ }\href {\doibase 10.1103/PhysRevLett.113.091104} {\bibfield
  {journal} {\bibinfo  {journal} {Phys. Rev. Lett.}\ }\textbf {\bibinfo
  {volume} {113}},\ \bibinfo {eid} {091104} (\bibinfo {year}
  {2014}{\natexlab{b}})},\ \Eprint {http://arxiv.org/abs/1403.5672}
  {arXiv:1403.5672 [gr-qc]} \BibitemShut {NoStop}%
\bibitem [{\citenamefont {{Endrizzi}}\ \emph {et~al.}(2016)\citenamefont
  {{Endrizzi}}, \citenamefont {{Ciolfi}}, \citenamefont {{Giacomazzo}},
  \citenamefont {{Kastaun}},\ and\ \citenamefont {{Kawamura}}}]{Endrizzi2016}%
  \BibitemOpen
  \bibfield  {author} {\bibinfo {author} {\bibfnamefont {A.}~\bibnamefont
  {{Endrizzi}}}, \bibinfo {author} {\bibfnamefont {R.}~\bibnamefont
  {{Ciolfi}}}, \bibinfo {author} {\bibfnamefont {B.}~\bibnamefont
  {{Giacomazzo}}}, \bibinfo {author} {\bibfnamefont {W.}~\bibnamefont
  {{Kastaun}}}, \ and\ \bibinfo {author} {\bibfnamefont {T.}~\bibnamefont
  {{Kawamura}}},\ }\href@noop {} {\bibfield  {journal} {\bibinfo  {journal}
  {arXiv:1604.03445}\ } (\bibinfo {year} {2016})},\ \Eprint
  {http://arxiv.org/abs/1604.03445} {arXiv:1604.03445 [astro-ph.HE]}
  \BibitemShut {NoStop}%
\bibitem [{\citenamefont {{Shibata}}\ \emph
  {et~al.}(2014{\natexlab{a}})\citenamefont {{Shibata}}, \citenamefont
  {{Taniguchi}}, \citenamefont {{Okawa}},\ and\ \citenamefont
  {{Buonanno}}}]{Shibata2014}%
  \BibitemOpen
  \bibfield  {author} {\bibinfo {author} {\bibfnamefont {M.}~\bibnamefont
  {{Shibata}}}, \bibinfo {author} {\bibfnamefont {K.}~\bibnamefont
  {{Taniguchi}}}, \bibinfo {author} {\bibfnamefont {H.}~\bibnamefont
  {{Okawa}}}, \ and\ \bibinfo {author} {\bibfnamefont {A.}~\bibnamefont
  {{Buonanno}}},\ }\href {\doibase 10.1103/PhysRevD.89.084005} {\bibfield
  {journal} {\bibinfo  {journal} {Phys. Rev. D}\ }\textbf {\bibinfo {volume}
  {89}},\ \bibinfo {eid} {084005} (\bibinfo {year} {2014}{\natexlab{a}})},\
  \Eprint {http://arxiv.org/abs/1310.0627} {arXiv:1310.0627 [gr-qc]}
  \BibitemShut {NoStop}%
\bibitem [{\citenamefont {{Clark}}\ \emph {et~al.}(2016)\citenamefont
  {{Clark}}, \citenamefont {{Bauswein}}, \citenamefont {{Stergioulas}},\ and\
  \citenamefont {{Shoemaker}}}]{Clark2016}%
  \BibitemOpen
  \bibfield  {author} {\bibinfo {author} {\bibfnamefont {J.~A.}\ \bibnamefont
  {{Clark}}}, \bibinfo {author} {\bibfnamefont {A.}~\bibnamefont {{Bauswein}}},
  \bibinfo {author} {\bibfnamefont {N.}~\bibnamefont {{Stergioulas}}}, \ and\
  \bibinfo {author} {\bibfnamefont {D.}~\bibnamefont {{Shoemaker}}},\ }\href
  {\doibase 10.1088/0264-9381/33/8/085003} {\bibfield  {journal} {\bibinfo
  {journal} {Class. Quantum Grav.}\ }\textbf {\bibinfo {volume} {33}},\
  \bibinfo {eid} {085003} (\bibinfo {year} {2016})},\ \Eprint
  {http://arxiv.org/abs/1509.08522} {arXiv:1509.08522 [astro-ph.HE]}
  \BibitemShut {NoStop}%
\bibitem [{\citenamefont {Nakar}(2007)}]{Nakar:2007yr}%
  \BibitemOpen
  \bibfield  {author} {\bibinfo {author} {\bibfnamefont {E.}~\bibnamefont
  {Nakar}},\ }\href {\doibase 10.1016/j.physrep.2007.02.005} {\bibfield
  {journal} {\bibinfo  {journal} {Phys. Rep.}\ }\textbf {\bibinfo {volume}
  {442}},\ \bibinfo {pages} {166} (\bibinfo {year} {2007})},\ \Eprint
  {http://arxiv.org/abs/astro-ph/0701748} {arXiv:astro-ph/0701748} \BibitemShut
  {NoStop}%
\bibitem [{\citenamefont {Lee}\ and\ \citenamefont
  {Ramirez-Ruiz}(2007)}]{Lee:2007js}%
  \BibitemOpen
  \bibfield  {author} {\bibinfo {author} {\bibfnamefont {W.~H.}\ \bibnamefont
  {Lee}}\ and\ \bibinfo {author} {\bibfnamefont {E.}~\bibnamefont
  {Ramirez-Ruiz}},\ }\href {\doibase 10.1088/1367-2630/9/1/017} {\bibfield
  {journal} {\bibinfo  {journal} {New J. Phys.}\ }\textbf {\bibinfo {volume}
  {9}},\ \bibinfo {pages} {17} (\bibinfo {year} {2007})},\ \Eprint
  {http://arxiv.org/abs/astro-ph/0701874} {arXiv:astro-ph/0701874} \BibitemShut
  {NoStop}%
\bibitem [{\citenamefont {{Lattimer}}(2012)}]{Lattimer2012rev}%
  \BibitemOpen
  \bibfield  {author} {\bibinfo {author} {\bibfnamefont {J.~M.}\ \bibnamefont
  {{Lattimer}}},\ }\href {\doibase 10.1146/annurev-nucl-102711-095018}
  {\bibfield  {journal} {\bibinfo  {journal} {Annual Review of Nuclear and
  Particle Science}\ }\textbf {\bibinfo {volume} {62}},\ \bibinfo {pages} {485}
  (\bibinfo {year} {2012})}\BibitemShut {NoStop}%
\bibitem [{\citenamefont {{Rosswog}}\ \emph {et~al.}(2000)\citenamefont
  {{Rosswog}}, \citenamefont {{Davies}}, \citenamefont {{Thielemann}},\ and\
  \citenamefont {{Piran}}}]{Rosswog2000}%
  \BibitemOpen
  \bibfield  {author} {\bibinfo {author} {\bibfnamefont {S.}~\bibnamefont
  {{Rosswog}}}, \bibinfo {author} {\bibfnamefont {M.~B.}\ \bibnamefont
  {{Davies}}}, \bibinfo {author} {\bibfnamefont {F.-K.}\ \bibnamefont
  {{Thielemann}}}, \ and\ \bibinfo {author} {\bibfnamefont {T.}~\bibnamefont
  {{Piran}}},\ }\href@noop {} {\bibfield  {journal} {\bibinfo  {journal}
  {Astron. Astrophys.}\ }\textbf {\bibinfo {volume} {360}},\ \bibinfo {pages}
  {171} (\bibinfo {year} {2000})},\ \Eprint
  {http://arxiv.org/abs/astro-ph/0005550} {astro-ph/0005550} \BibitemShut
  {NoStop}%
\bibitem [{\citenamefont {{Oechslin}}\ and\ \citenamefont
  {{Janka}}(2006)}]{Oechslin06}%
  \BibitemOpen
  \bibfield  {author} {\bibinfo {author} {\bibfnamefont {R.}~\bibnamefont
  {{Oechslin}}}\ and\ \bibinfo {author} {\bibfnamefont {H.-T.}\ \bibnamefont
  {{Janka}}},\ }\href {\doibase 10.1111/j.1365-2966.2006.10238.x} {\bibfield
  {journal} {\bibinfo  {journal} {Mon. Not. R. Astron. Soc.}\ }\textbf
  {\bibinfo {volume} {368}},\ \bibinfo {pages} {1489} (\bibinfo {year}
  {2006})},\ \Eprint {http://arxiv.org/abs/astro-ph/0507099} {astro-ph/0507099}
  \BibitemShut {NoStop}%
\bibitem [{\citenamefont {{Kiuchi}}\ \emph {et~al.}(2010)\citenamefont
  {{Kiuchi}}, \citenamefont {{Sekiguchi}}, \citenamefont {{Shibata}},\ and\
  \citenamefont {{Taniguchi}}}]{Kiuchi2010}%
  \BibitemOpen
  \bibfield  {author} {\bibinfo {author} {\bibfnamefont {K.}~\bibnamefont
  {{Kiuchi}}}, \bibinfo {author} {\bibfnamefont {Y.}~\bibnamefont
  {{Sekiguchi}}}, \bibinfo {author} {\bibfnamefont {M.}~\bibnamefont
  {{Shibata}}}, \ and\ \bibinfo {author} {\bibfnamefont {K.}~\bibnamefont
  {{Taniguchi}}},\ }\href {\doibase 10.1103/PhysRevLett.104.141101} {\bibfield
  {journal} {\bibinfo  {journal} {Phys. Rev. Lett.}\ }\textbf {\bibinfo
  {volume} {104}},\ \bibinfo {eid} {141101} (\bibinfo {year} {2010})},\ \Eprint
  {http://arxiv.org/abs/1002.2689} {arXiv:1002.2689 [astro-ph.HE]} \BibitemShut
  {NoStop}%
\bibitem [{\citenamefont {Gonzalez}\ \emph {et~al.}(2007)\citenamefont
  {Gonzalez}, \citenamefont {Sperhake}, \citenamefont {Bruegmann},
  \citenamefont {Hannam},\ and\ \citenamefont {Husa}}]{Gonzalez:2006md}%
  \BibitemOpen
  \bibfield  {author} {\bibinfo {author} {\bibfnamefont {J.~A.}\ \bibnamefont
  {Gonzalez}}, \bibinfo {author} {\bibfnamefont {U.}~\bibnamefont {Sperhake}},
  \bibinfo {author} {\bibfnamefont {B.}~\bibnamefont {Bruegmann}}, \bibinfo
  {author} {\bibfnamefont {M.}~\bibnamefont {Hannam}}, \ and\ \bibinfo {author}
  {\bibfnamefont {S.}~\bibnamefont {Husa}},\ }\href@noop {} {\bibfield
  {journal} {\bibinfo  {journal} {Phys. Rev. Lett.}\ }\textbf {\bibinfo
  {volume} {98}},\ \bibinfo {pages} {091101} (\bibinfo {year} {2007})},\
  \Eprint {http://arxiv.org/abs/gr-qc/0610154} {gr-qc/0610154} \BibitemShut
  {NoStop}%
\bibitem [{\citenamefont {{Koppitz}}\ \emph {et~al.}(2007)\citenamefont
  {{Koppitz}}, \citenamefont {{Pollney}}, \citenamefont {{Reisswig}},
  \citenamefont {{Rezzolla}}, \citenamefont {{Thornburg}}, \citenamefont
  {{Diener}},\ and\ \citenamefont {{Schnetter}}}]{Koppitz-etal-2007aa}%
  \BibitemOpen
  \bibfield  {author} {\bibinfo {author} {\bibfnamefont {M.}~\bibnamefont
  {{Koppitz}}}, \bibinfo {author} {\bibfnamefont {D.}~\bibnamefont
  {{Pollney}}}, \bibinfo {author} {\bibfnamefont {C.}~\bibnamefont
  {{Reisswig}}}, \bibinfo {author} {\bibfnamefont {L.}~\bibnamefont
  {{Rezzolla}}}, \bibinfo {author} {\bibfnamefont {J.}~\bibnamefont
  {{Thornburg}}}, \bibinfo {author} {\bibfnamefont {P.}~\bibnamefont
  {{Diener}}}, \ and\ \bibinfo {author} {\bibfnamefont {E.}~\bibnamefont
  {{Schnetter}}},\ }\href {\doibase 10.1103/PhysRevLett.99.041102} {\bibfield
  {journal} {\bibinfo  {journal} {Phys. Rev. Lett.}\ }\textbf {\bibinfo
  {volume} {99}},\ \bibinfo {eid} {041102} (\bibinfo {year} {2007})},\ \Eprint
  {http://arxiv.org/abs/gr-qc/0701163} {gr-qc/0701163} \BibitemShut {NoStop}%
\bibitem [{\citenamefont {{Healy}}\ \emph {et~al.}(2014)\citenamefont
  {{Healy}}, \citenamefont {{Lousto}},\ and\ \citenamefont
  {{Zlochower}}}]{Healy2014}%
  \BibitemOpen
  \bibfield  {author} {\bibinfo {author} {\bibfnamefont {J.}~\bibnamefont
  {{Healy}}}, \bibinfo {author} {\bibfnamefont {C.~O.}\ \bibnamefont
  {{Lousto}}}, \ and\ \bibinfo {author} {\bibfnamefont {Y.}~\bibnamefont
  {{Zlochower}}},\ }\href {\doibase 10.1103/PhysRevD.90.104004} {\bibfield
  {journal} {\bibinfo  {journal} {Phys. Rev. D}\ }\textbf {\bibinfo {volume}
  {90}},\ \bibinfo {eid} {104004} (\bibinfo {year} {2014})},\ \Eprint
  {http://arxiv.org/abs/1406.7295} {arXiv:1406.7295 [gr-qc]} \BibitemShut
  {NoStop}%
\bibitem [{\citenamefont {{Webbink}}(1985)}]{Webbink:1985}%
  \BibitemOpen
  \bibfield  {author} {\bibinfo {author} {\bibfnamefont {R.~F.}\ \bibnamefont
  {{Webbink}}},\ }in\ \href {\doibase 10.1007/978-94-009-5335-2} {\emph
  {\bibinfo {booktitle} {Dynamics of Star Clusters}}},\ \bibinfo {series} {IAU
  Symposium}, Vol.\ \bibinfo {volume} {113},\ \bibinfo {editor} {edited by\
  \bibinfo {editor} {\bibnamefont {{J.~Goodman \& P.~Hut}}}}\ (\bibinfo {year}
  {1985})\ pp.\ \bibinfo {pages} {541--577}\BibitemShut {NoStop}%
\bibitem [{\citenamefont {{Giacomazzo}}\ \emph {et~al.}(2013)\citenamefont
  {{Giacomazzo}}, \citenamefont {{Perna}}, \citenamefont {{Rezzolla}},
  \citenamefont {{Troja}},\ and\ \citenamefont {{Lazzati}}}]{Giacomazzo2012b}%
  \BibitemOpen
  \bibfield  {author} {\bibinfo {author} {\bibfnamefont {B.}~\bibnamefont
  {{Giacomazzo}}}, \bibinfo {author} {\bibfnamefont {R.}~\bibnamefont
  {{Perna}}}, \bibinfo {author} {\bibfnamefont {L.}~\bibnamefont {{Rezzolla}}},
  \bibinfo {author} {\bibfnamefont {E.}~\bibnamefont {{Troja}}}, \ and\
  \bibinfo {author} {\bibfnamefont {D.}~\bibnamefont {{Lazzati}}},\ }\href
  {\doibase 10.1088/2041-8205/762/2/L18} {\bibfield  {journal} {\bibinfo
  {journal} {Astrophys. J.}\ }\textbf {\bibinfo {volume} {762}},\ \bibinfo
  {eid} {L18} (\bibinfo {year} {2013})},\ \Eprint
  {http://arxiv.org/abs/1210.8152} {arXiv:1210.8152 [astro-ph.HE]} \BibitemShut
  {NoStop}%
\bibitem [{\citenamefont {{Lehner}}\ \emph
  {et~al.}(2016{\natexlab{b}})\citenamefont {{Lehner}}, \citenamefont
  {{Liebling}}, \citenamefont {{Palenzuela}}, \citenamefont {{Caballero}},
  \citenamefont {{O'Connor}}, \citenamefont {{Anderson}},\ and\ \citenamefont
  {{Neilsen}}}]{Lehner2016}%
  \BibitemOpen
  \bibfield  {author} {\bibinfo {author} {\bibfnamefont {L.}~\bibnamefont
  {{Lehner}}}, \bibinfo {author} {\bibfnamefont {S.~L.}\ \bibnamefont
  {{Liebling}}}, \bibinfo {author} {\bibfnamefont {C.}~\bibnamefont
  {{Palenzuela}}}, \bibinfo {author} {\bibfnamefont {O.~L.}\ \bibnamefont
  {{Caballero}}}, \bibinfo {author} {\bibfnamefont {E.}~\bibnamefont
  {{O'Connor}}}, \bibinfo {author} {\bibfnamefont {M.}~\bibnamefont
  {{Anderson}}}, \ and\ \bibinfo {author} {\bibfnamefont {D.}~\bibnamefont
  {{Neilsen}}},\ }\href@noop {} {\bibfield  {journal} {\bibinfo  {journal}
  {arXiv:1603.00501}\ } (\bibinfo {year} {2016}{\natexlab{b}})},\ \Eprint
  {http://arxiv.org/abs/1603.00501} {arXiv:1603.00501 [gr-qc]} \BibitemShut
  {NoStop}%
\bibitem [{\citenamefont {{Sekiguchi}}\ \emph {et~al.}(2016)\citenamefont
  {{Sekiguchi}}, \citenamefont {{Kiuchi}}, \citenamefont {{Kyutoku}},
  \citenamefont {{Shibata}},\ and\ \citenamefont
  {{Taniguchi}}}]{Sekiguchi2016}%
  \BibitemOpen
  \bibfield  {author} {\bibinfo {author} {\bibfnamefont {Y.}~\bibnamefont
  {{Sekiguchi}}}, \bibinfo {author} {\bibfnamefont {K.}~\bibnamefont
  {{Kiuchi}}}, \bibinfo {author} {\bibfnamefont {K.}~\bibnamefont {{Kyutoku}}},
  \bibinfo {author} {\bibfnamefont {M.}~\bibnamefont {{Shibata}}}, \ and\
  \bibinfo {author} {\bibfnamefont {K.}~\bibnamefont {{Taniguchi}}},\
  }\href@noop {} {\bibfield  {journal} {\bibinfo  {journal} {ArXiv e-prints}\ }
  (\bibinfo {year} {2016})},\ \Eprint {http://arxiv.org/abs/1603.01918}
  {arXiv:1603.01918 [astro-ph.HE]} \BibitemShut {NoStop}%
\bibitem [{\citenamefont {{East}}\ \emph {et~al.}(2013)\citenamefont {{East}},
  \citenamefont {{McWilliams}}, \citenamefont {{Levin}},\ and\ \citenamefont
  {{Pretorius}}}]{East2013}%
  \BibitemOpen
  \bibfield  {author} {\bibinfo {author} {\bibfnamefont {W.~E.}\ \bibnamefont
  {{East}}}, \bibinfo {author} {\bibfnamefont {S.~T.}\ \bibnamefont
  {{McWilliams}}}, \bibinfo {author} {\bibfnamefont {J.}~\bibnamefont
  {{Levin}}}, \ and\ \bibinfo {author} {\bibfnamefont {F.}~\bibnamefont
  {{Pretorius}}},\ }\href {\doibase 10.1103/PhysRevD.87.043004} {\bibfield
  {journal} {\bibinfo  {journal} {Phys. Rev. D}\ }\textbf {\bibinfo {volume}
  {87}},\ \bibinfo {eid} {043004} (\bibinfo {year} {2013})},\ \Eprint
  {http://arxiv.org/abs/1212.0837} {arXiv:1212.0837 [gr-qc]} \BibitemShut
  {NoStop}%
\bibitem [{\citenamefont {{Harding}}\ and\ \citenamefont
  {{Lai}}(2006)}]{Harding2006}%
  \BibitemOpen
  \bibfield  {author} {\bibinfo {author} {\bibfnamefont {A.~K.}\ \bibnamefont
  {{Harding}}}\ and\ \bibinfo {author} {\bibfnamefont {D.}~\bibnamefont
  {{Lai}}},\ }\href {\doibase 10.1088/0034-4885/69/9/R03} {\bibfield  {journal}
  {\bibinfo  {journal} {Reports on Progress in Physics}\ }\textbf {\bibinfo
  {volume} {69}},\ \bibinfo {pages} {2631} (\bibinfo {year} {2006})},\ \Eprint
  {http://arxiv.org/abs/astro-ph/0606674} {astro-ph/0606674} \BibitemShut
  {NoStop}%
\bibitem [{\citenamefont {{Kiuchi}}\ \emph
  {et~al.}(2015{\natexlab{a}})\citenamefont {{Kiuchi}}, \citenamefont
  {{Cerd{\'a}-Dur{\'a}n}}, \citenamefont {{Kyutoku}}, \citenamefont
  {{Sekiguchi}},\ and\ \citenamefont {{Shibata}}}]{Kiuchi2015a}%
  \BibitemOpen
  \bibfield  {author} {\bibinfo {author} {\bibfnamefont {K.}~\bibnamefont
  {{Kiuchi}}}, \bibinfo {author} {\bibfnamefont {P.}~\bibnamefont
  {{Cerd{\'a}-Dur{\'a}n}}}, \bibinfo {author} {\bibfnamefont {K.}~\bibnamefont
  {{Kyutoku}}}, \bibinfo {author} {\bibfnamefont {Y.}~\bibnamefont
  {{Sekiguchi}}}, \ and\ \bibinfo {author} {\bibfnamefont {M.}~\bibnamefont
  {{Shibata}}},\ }\href {\doibase 10.1103/PhysRevD.92.124034} {\bibfield
  {journal} {\bibinfo  {journal} {Phys. Rev. D}\ }\textbf {\bibinfo {volume}
  {92}},\ \bibinfo {eid} {124034} (\bibinfo {year} {2015}{\natexlab{a}})},\
  \Eprint {http://arxiv.org/abs/1509.09205} {arXiv:1509.09205 [astro-ph.HE]}
  \BibitemShut {NoStop}%
\bibitem [{\citenamefont {{Velikhov}}(1959)}]{Velikhov1959}%
  \BibitemOpen
  \bibfield  {author} {\bibinfo {author} {\bibfnamefont {E.~P.}\ \bibnamefont
  {{Velikhov}}},\ }\href@noop {} {\bibfield  {journal} {\bibinfo  {journal}
  {Sov. Phys. JETP}\ }\textbf {\bibinfo {volume} {9}},\ \bibinfo {pages} {995}
  (\bibinfo {year} {1959})}\BibitemShut {NoStop}%
\bibitem [{\citenamefont {{Chandrasekhar}}(1960)}]{Chandrasekhar1960}%
  \BibitemOpen
  \bibfield  {author} {\bibinfo {author} {\bibfnamefont {S.}~\bibnamefont
  {{Chandrasekhar}}},\ }\href {\doibase 10.1073/pnas.46.2.253} {\bibfield
  {journal} {\bibinfo  {journal} {Proc. Natl. Acad. Sci.}\ }\textbf {\bibinfo
  {volume} {46}},\ \bibinfo {pages} {253} (\bibinfo {year} {1960})}\BibitemShut
  {NoStop}%
\bibitem [{\citenamefont {{Balbus}}\ and\ \citenamefont
  {{Hawley}}(1991)}]{Balbus1991}%
  \BibitemOpen
  \bibfield  {author} {\bibinfo {author} {\bibfnamefont {S.~A.}\ \bibnamefont
  {{Balbus}}}\ and\ \bibinfo {author} {\bibfnamefont {J.~F.}\ \bibnamefont
  {{Hawley}}},\ }\href {\doibase 10.1086/170270} {\bibfield  {journal}
  {\bibinfo  {journal} {Astrophys. J.}\ }\textbf {\bibinfo {volume} {376}},\
  \bibinfo {pages} {214} (\bibinfo {year} {1991})}\BibitemShut {NoStop}%
\bibitem [{\citenamefont {{Balbus}}\ and\ \citenamefont
  {{Hawley}}(1998)}]{BalbusHawley1998}%
  \BibitemOpen
  \bibfield  {author} {\bibinfo {author} {\bibfnamefont {S.~A.}\ \bibnamefont
  {{Balbus}}}\ and\ \bibinfo {author} {\bibfnamefont {J.~F.}\ \bibnamefont
  {{Hawley}}},\ }\href {\doibase 10.1103/RevModPhys.70.1} {\bibfield  {journal}
  {\bibinfo  {journal} {Rev. Mod. Phys.}\ }\textbf {\bibinfo {volume} {70}},\
  \bibinfo {pages} {1} (\bibinfo {year} {1998})}\BibitemShut {NoStop}%
\bibitem [{\citenamefont {{Komissarov}}(1999)}]{Komissarov1999}%
  \BibitemOpen
  \bibfield  {author} {\bibinfo {author} {\bibfnamefont {S.~S.}\ \bibnamefont
  {{Komissarov}}},\ }\href {\doibase 10.1046/j.1365-8711.1999.02244.x}
  {\bibfield  {journal} {\bibinfo  {journal} {Mon. Not. R. Astron. Soc.}\
  }\textbf {\bibinfo {volume} {303}},\ \bibinfo {pages} {343} (\bibinfo {year}
  {1999})}\BibitemShut {NoStop}%
\bibitem [{\citenamefont {{Koide}}\ \emph {et~al.}(1998)\citenamefont
  {{Koide}}, \citenamefont {{Shibata}},\ and\ \citenamefont
  {{Kudoh}}}]{Koide98}%
  \BibitemOpen
  \bibfield  {author} {\bibinfo {author} {\bibfnamefont {S.}~\bibnamefont
  {{Koide}}}, \bibinfo {author} {\bibfnamefont {K.}~\bibnamefont {{Shibata}}},
  \ and\ \bibinfo {author} {\bibfnamefont {T.}~\bibnamefont {{Kudoh}}},\ }\href
  {\doibase 10.1086/311204} {\bibfield  {journal} {\bibinfo  {journal}
  {Astrophys. J}\ }\textbf {\bibinfo {volume} {495}},\ \bibinfo {pages} {L63}
  (\bibinfo {year} {1998})}\BibitemShut {NoStop}%
\bibitem [{\citenamefont {{Del Zanna}}\ \emph {et~al.}(2003)\citenamefont {{Del
  Zanna}}, \citenamefont {{Bucciantini}},\ and\ \citenamefont
  {{Londrillo}}}]{DelZanna2003}%
  \BibitemOpen
  \bibfield  {author} {\bibinfo {author} {\bibfnamefont {L.}~\bibnamefont {{Del
  Zanna}}}, \bibinfo {author} {\bibfnamefont {N.}~\bibnamefont
  {{Bucciantini}}}, \ and\ \bibinfo {author} {\bibfnamefont {P.}~\bibnamefont
  {{Londrillo}}},\ }\href {\doibase 10.1051/0004-6361:20021641} {\bibfield
  {journal} {\bibinfo  {journal} {Astron. Astrophys.}\ }\textbf {\bibinfo
  {volume} {400}},\ \bibinfo {pages} {397} (\bibinfo {year} {2003})},\ \Eprint
  {http://arxiv.org/abs/arXiv:astro-ph/0210618} {arXiv:astro-ph/0210618}
  \BibitemShut {NoStop}%
\bibitem [{\citenamefont {Gammie}\ \emph {et~al.}(2003)\citenamefont {Gammie},
  \citenamefont {McKinney},\ and\ \citenamefont {T{\'o}th}}]{Gammie03}%
  \BibitemOpen
  \bibfield  {author} {\bibinfo {author} {\bibfnamefont {C.~F.}\ \bibnamefont
  {Gammie}}, \bibinfo {author} {\bibfnamefont {J.~C.}\ \bibnamefont
  {McKinney}}, \ and\ \bibinfo {author} {\bibfnamefont {G.}~\bibnamefont
  {T{\'o}th}},\ }\href@noop {} {\bibfield  {journal} {\bibinfo  {journal}
  {Astrophys. J.}\ }\textbf {\bibinfo {volume} {589}},\ \bibinfo {pages} {458}
  (\bibinfo {year} {2003})},\ \Eprint {http://arxiv.org/abs/astro-ph/0301509}
  {astro-ph/0301509} \BibitemShut {NoStop}%
\bibitem [{\citenamefont {Anninos}\ \emph {et~al.}(2005)\citenamefont
  {Anninos}, \citenamefont {Fragile},\ and\ \citenamefont
  {Salmonson}}]{Anninos05c}%
  \BibitemOpen
  \bibfield  {author} {\bibinfo {author} {\bibfnamefont {P.}~\bibnamefont
  {Anninos}}, \bibinfo {author} {\bibfnamefont {P.~C.}\ \bibnamefont
  {Fragile}}, \ and\ \bibinfo {author} {\bibfnamefont {J.~D.}\ \bibnamefont
  {Salmonson}},\ }\href {\doibase 10.1086} {\bibfield  {journal} {\bibinfo
  {journal} {Astrophys. J.}\ }\textbf {\bibinfo {volume} {635}},\ \bibinfo
  {pages} {723} (\bibinfo {year} {2005})}\BibitemShut {NoStop}%
\bibitem [{\citenamefont {{Del Zanna}}\ \emph {et~al.}(2007)\citenamefont {{Del
  Zanna}}, \citenamefont {{Zanotti}}, \citenamefont {{Bucciantini}},\ and\
  \citenamefont {{Londrillo}}}]{DelZanna2007}%
  \BibitemOpen
  \bibfield  {author} {\bibinfo {author} {\bibfnamefont {L.}~\bibnamefont {{Del
  Zanna}}}, \bibinfo {author} {\bibfnamefont {O.}~\bibnamefont {{Zanotti}}},
  \bibinfo {author} {\bibfnamefont {N.}~\bibnamefont {{Bucciantini}}}, \ and\
  \bibinfo {author} {\bibfnamefont {P.}~\bibnamefont {{Londrillo}}},\ }\href
  {\doibase 10.1051/0004-6361:20077093} {\bibfield  {journal} {\bibinfo
  {journal} {Astron. Astrophys.}\ }\textbf {\bibinfo {volume} {473}},\ \bibinfo
  {pages} {11} (\bibinfo {year} {2007})},\ \Eprint
  {http://arxiv.org/abs/0704.3206} {arXiv:0704.3206} \BibitemShut {NoStop}%
\bibitem [{\citenamefont {{Zink}}(2011)}]{Zink2011}%
  \BibitemOpen
  \bibfield  {author} {\bibinfo {author} {\bibfnamefont {B.}~\bibnamefont
  {{Zink}}},\ }\href@noop {} {\bibfield  {journal} {\bibinfo  {journal} {ArXiv
  e-prints}\ } (\bibinfo {year} {2011})},\ \Eprint
  {http://arxiv.org/abs/1102.5202} {arXiv:1102.5202 [gr-qc]} \BibitemShut
  {NoStop}%
\bibitem [{\citenamefont {{Shibata}}\ and\ \citenamefont
  {{Sekiguchi}}(2005)}]{Shibata05b}%
  \BibitemOpen
  \bibfield  {author} {\bibinfo {author} {\bibfnamefont {M.}~\bibnamefont
  {{Shibata}}}\ and\ \bibinfo {author} {\bibfnamefont {Y.-I.}\ \bibnamefont
  {{Sekiguchi}}},\ }\href {\doibase 10.1103/PhysRevD.72.044014} {\bibfield
  {journal} {\bibinfo  {journal} {Phys. Rev. D}\ }\textbf {\bibinfo {volume}
  {72}},\ \bibinfo {eid} {044014} (\bibinfo {year} {2005})},\ \Eprint
  {http://arxiv.org/abs/astro-ph/0507383} {astro-ph/0507383} \BibitemShut
  {NoStop}%
\bibitem [{\citenamefont {Neilsen}\ \emph {et~al.}(2006)\citenamefont
  {Neilsen}, \citenamefont {Hirschmann},\ and\ \citenamefont
  {Millward}}]{Neilsen2005}%
  \BibitemOpen
  \bibfield  {author} {\bibinfo {author} {\bibfnamefont {D.~W.}\ \bibnamefont
  {Neilsen}}, \bibinfo {author} {\bibfnamefont {E.~W.}\ \bibnamefont
  {Hirschmann}}, \ and\ \bibinfo {author} {\bibfnamefont {R.~S.}\ \bibnamefont
  {Millward}},\ }\href@noop {} {\bibfield  {journal} {\bibinfo  {journal}
  {Class. Quantum Grav.}\ }\textbf {\bibinfo {volume} {23}},\ \bibinfo {pages}
  {S505} (\bibinfo {year} {2006})}\BibitemShut {NoStop}%
\bibitem [{\citenamefont {{Farris}}\ \emph {et~al.}(2008)\citenamefont
  {{Farris}}, \citenamefont {{Li}}, \citenamefont {{Liu}},\ and\ \citenamefont
  {{Shapiro}}}]{Farris08}%
  \BibitemOpen
  \bibfield  {author} {\bibinfo {author} {\bibfnamefont {B.~D.}\ \bibnamefont
  {{Farris}}}, \bibinfo {author} {\bibfnamefont {T.~K.}\ \bibnamefont {{Li}}},
  \bibinfo {author} {\bibfnamefont {Y.~T.}\ \bibnamefont {{Liu}}}, \ and\
  \bibinfo {author} {\bibfnamefont {S.~L.}\ \bibnamefont {{Shapiro}}},\ }\href
  {\doibase 10.1103/PhysRevD.78.024023} {\bibfield  {journal} {\bibinfo
  {journal} {Phys. Rev. D}\ }\textbf {\bibinfo {volume} {78}},\ \bibinfo
  {pages} {024023} (\bibinfo {year} {2008})},\ \Eprint
  {http://arxiv.org/abs/0802.3210} {arXiv:0802.3210} \BibitemShut {NoStop}%
\bibitem [{\citenamefont {{M{\"o}sta}}\ \emph {et~al.}(2014)\citenamefont
  {{M{\"o}sta}}, \citenamefont {{Mundim}}, \citenamefont {{Faber}},
  \citenamefont {{Haas}}, \citenamefont {{Noble}}, \citenamefont {{Bode}},
  \citenamefont {{L{\"o}ffler}}, \citenamefont {{Ott}}, \citenamefont
  {{Reisswig}},\ and\ \citenamefont {{Schnetter}}}]{Moesta13_GRHydro}%
  \BibitemOpen
  \bibfield  {author} {\bibinfo {author} {\bibfnamefont {P.}~\bibnamefont
  {{M{\"o}sta}}}, \bibinfo {author} {\bibfnamefont {B.~C.}\ \bibnamefont
  {{Mundim}}}, \bibinfo {author} {\bibfnamefont {J.~A.}\ \bibnamefont
  {{Faber}}}, \bibinfo {author} {\bibfnamefont {R.}~\bibnamefont {{Haas}}},
  \bibinfo {author} {\bibfnamefont {S.~C.}\ \bibnamefont {{Noble}}}, \bibinfo
  {author} {\bibfnamefont {T.}~\bibnamefont {{Bode}}}, \bibinfo {author}
  {\bibfnamefont {F.}~\bibnamefont {{L{\"o}ffler}}}, \bibinfo {author}
  {\bibfnamefont {C.~D.}\ \bibnamefont {{Ott}}}, \bibinfo {author}
  {\bibfnamefont {C.}~\bibnamefont {{Reisswig}}}, \ and\ \bibinfo {author}
  {\bibfnamefont {E.}~\bibnamefont {{Schnetter}}},\ }\href {\doibase
  10.1088/0264-9381/31/1/015005} {\bibfield  {journal} {\bibinfo  {journal}
  {Classical and Quantum Gravity}\ }\textbf {\bibinfo {volume} {31}},\ \bibinfo
  {eid} {015005} (\bibinfo {year} {2014})},\ \Eprint
  {http://arxiv.org/abs/1304.5544} {arXiv:1304.5544 [gr-qc]} \BibitemShut
  {NoStop}%
\bibitem [{\citenamefont {{Biskamp}}(1986)}]{Biskamp1986}%
  \BibitemOpen
  \bibfield  {author} {\bibinfo {author} {\bibfnamefont {D.}~\bibnamefont
  {{Biskamp}}},\ }\href {\doibase 10.1063/1.865670} {\bibfield  {journal}
  {\bibinfo  {journal} {Physics of Fluids}\ }\textbf {\bibinfo {volume} {29}},\
  \bibinfo {pages} {1520} (\bibinfo {year} {1986})}\BibitemShut {NoStop}%
\bibitem [{\citenamefont {{Komissarov}}(2007)}]{Komissarov2007}%
  \BibitemOpen
  \bibfield  {author} {\bibinfo {author} {\bibfnamefont {S.~S.}\ \bibnamefont
  {{Komissarov}}},\ }\href {\doibase 10.1111/j.1365-2966.2007.12448.x}
  {\bibfield  {journal} {\bibinfo  {journal} {Mon. Not. R. Astron. Soc.}\
  }\textbf {\bibinfo {volume} {382}},\ \bibinfo {pages} {995} (\bibinfo {year}
  {2007})},\ \Eprint {http://arxiv.org/abs/0708.0323} {arXiv:0708.0323}
  \BibitemShut {NoStop}%
\bibitem [{\citenamefont {{Dumbser}}\ and\ \citenamefont
  {{Zanotti}}(2009)}]{Dumbser2009}%
  \BibitemOpen
  \bibfield  {author} {\bibinfo {author} {\bibfnamefont {M.}~\bibnamefont
  {{Dumbser}}}\ and\ \bibinfo {author} {\bibfnamefont {O.}~\bibnamefont
  {{Zanotti}}},\ }\href {\doibase 10.1016/j.jcp.2009.06.009} {\bibfield
  {journal} {\bibinfo  {journal} {Journal of Computational Physics}\ }\textbf
  {\bibinfo {volume} {228}},\ \bibinfo {pages} {6991} (\bibinfo {year}
  {2009})},\ \Eprint {http://arxiv.org/abs/0903.4832} {arXiv:0903.4832}
  \BibitemShut {NoStop}%
\bibitem [{\citenamefont {{Zenitani}}\ \emph {et~al.}(2010)\citenamefont
  {{Zenitani}}, \citenamefont {{Hesse}},\ and\ \citenamefont
  {{Klimas}}}]{Zenitani2010}%
  \BibitemOpen
  \bibfield  {author} {\bibinfo {author} {\bibfnamefont {S.}~\bibnamefont
  {{Zenitani}}}, \bibinfo {author} {\bibfnamefont {M.}~\bibnamefont {{Hesse}}},
  \ and\ \bibinfo {author} {\bibfnamefont {A.}~\bibnamefont {{Klimas}}},\
  }\href {\doibase 10.1088/2041-8205/716/2/L214} {\bibfield  {journal}
  {\bibinfo  {journal} {Astrophys. J. Lett.}\ }\textbf {\bibinfo {volume}
  {716}},\ \bibinfo {pages} {L214} (\bibinfo {year} {2010})},\ \Eprint
  {http://arxiv.org/abs/1005.4485} {arXiv:1005.4485 [astro-ph.HE]} \BibitemShut
  {NoStop}%
\bibitem [{\citenamefont {{Takamoto}}\ and\ \citenamefont
  {{Inoue}}(2011)}]{Takamoto2011b}%
  \BibitemOpen
  \bibfield  {author} {\bibinfo {author} {\bibfnamefont {M.}~\bibnamefont
  {{Takamoto}}}\ and\ \bibinfo {author} {\bibfnamefont {T.}~\bibnamefont
  {{Inoue}}},\ }\href {\doibase 10.1088/0004-637X/735/2/113} {\bibfield
  {journal} {\bibinfo  {journal} {Astrophys. J.}\ }\textbf {\bibinfo {volume}
  {735}},\ \bibinfo {eid} {113} (\bibinfo {year} {2011})},\ \Eprint
  {http://arxiv.org/abs/1105.5683} {arXiv:1105.5683 [astro-ph.HE]} \BibitemShut
  {NoStop}%
\bibitem [{\citenamefont {{Zanotti}}\ and\ \citenamefont
  {{Dumbser}}(2011)}]{Zanotti2011b}%
  \BibitemOpen
  \bibfield  {author} {\bibinfo {author} {\bibfnamefont {O.}~\bibnamefont
  {{Zanotti}}}\ and\ \bibinfo {author} {\bibfnamefont {M.}~\bibnamefont
  {{Dumbser}}},\ }\href {\doibase 10.1111/j.1365-2966.2011.19551.x} {\bibfield
  {journal} {\bibinfo  {journal} {Mon. Not. R. Astron. Soc.}\ }\textbf
  {\bibinfo {volume} {418}},\ \bibinfo {pages} {1004} (\bibinfo {year}
  {2011})},\ \Eprint {http://arxiv.org/abs/1103.5924} {arXiv:1103.5924
  [astro-ph.HE]} \BibitemShut {NoStop}%
\bibitem [{\citenamefont {{Bucciantini}}\ \emph {et~al.}(2012)\citenamefont
  {{Bucciantini}}, \citenamefont {{Metzger}}, \citenamefont {{Thompson}},\ and\
  \citenamefont {{Quataert}}}]{Bucciantini2012}%
  \BibitemOpen
  \bibfield  {author} {\bibinfo {author} {\bibfnamefont {N.}~\bibnamefont
  {{Bucciantini}}}, \bibinfo {author} {\bibfnamefont {B.~D.}\ \bibnamefont
  {{Metzger}}}, \bibinfo {author} {\bibfnamefont {T.~A.}\ \bibnamefont
  {{Thompson}}}, \ and\ \bibinfo {author} {\bibfnamefont {E.}~\bibnamefont
  {{Quataert}}},\ }\href {\doibase 10.1111/j.1365-2966.2011.19810.x} {\bibfield
   {journal} {\bibinfo  {journal} {Mon. Not. R. Astron. Soc.}\ }\textbf
  {\bibinfo {volume} {419}},\ \bibinfo {pages} {1537} (\bibinfo {year}
  {2012})},\ \Eprint {http://arxiv.org/abs/1106.4668} {arXiv:1106.4668
  [astro-ph.HE]} \BibitemShut {NoStop}%
\bibitem [{\citenamefont {{Palenzuela}}\ \emph
  {et~al.}(2013{\natexlab{b}})\citenamefont {{Palenzuela}}, \citenamefont
  {{Lehner}}, \citenamefont {{Liebling}}, \citenamefont {{Ponce}},
  \citenamefont {{Anderson}}, \citenamefont {{Neilsen}},\ and\ \citenamefont
  {{Motl}}}]{Palenzuela2013b}%
  \BibitemOpen
  \bibfield  {author} {\bibinfo {author} {\bibfnamefont {C.}~\bibnamefont
  {{Palenzuela}}}, \bibinfo {author} {\bibfnamefont {L.}~\bibnamefont
  {{Lehner}}}, \bibinfo {author} {\bibfnamefont {S.~L.}\ \bibnamefont
  {{Liebling}}}, \bibinfo {author} {\bibfnamefont {M.}~\bibnamefont {{Ponce}}},
  \bibinfo {author} {\bibfnamefont {M.}~\bibnamefont {{Anderson}}}, \bibinfo
  {author} {\bibfnamefont {D.}~\bibnamefont {{Neilsen}}}, \ and\ \bibinfo
  {author} {\bibfnamefont {P.}~\bibnamefont {{Motl}}},\ }\href {\doibase
  10.1103/PhysRevD.88.043011} {\bibfield  {journal} {\bibinfo  {journal} {Phys.
  Rev. D}\ }\textbf {\bibinfo {volume} {88}},\ \bibinfo {eid} {043011}
  (\bibinfo {year} {2013}{\natexlab{b}})},\ \Eprint
  {http://arxiv.org/abs/1307.7372} {arXiv:1307.7372 [gr-qc]} \BibitemShut
  {NoStop}%
\bibitem [{\citenamefont {{Ponce}}\ \emph {et~al.}(2014)\citenamefont
  {{Ponce}}, \citenamefont {{Palenzuela}}, \citenamefont {{Lehner}},\ and\
  \citenamefont {{Liebling}}}]{Ponce2014}%
  \BibitemOpen
  \bibfield  {author} {\bibinfo {author} {\bibfnamefont {M.}~\bibnamefont
  {{Ponce}}}, \bibinfo {author} {\bibfnamefont {C.}~\bibnamefont
  {{Palenzuela}}}, \bibinfo {author} {\bibfnamefont {L.}~\bibnamefont
  {{Lehner}}}, \ and\ \bibinfo {author} {\bibfnamefont {S.~L.}\ \bibnamefont
  {{Liebling}}},\ }\href {\doibase 10.1103/PhysRevD.90.044007} {\bibfield
  {journal} {\bibinfo  {journal} {Phys. Rev. D}\ }\textbf {\bibinfo {volume}
  {90}},\ \bibinfo {eid} {044007} (\bibinfo {year} {2014})},\ \Eprint
  {http://arxiv.org/abs/1404.0692} {arXiv:1404.0692 [gr-qc]} \BibitemShut
  {NoStop}%
\bibitem [{\citenamefont {{Radice}}\ and\ \citenamefont
  {{Rezzolla}}(2012)}]{Radice2012a}%
  \BibitemOpen
  \bibfield  {author} {\bibinfo {author} {\bibfnamefont {D.}~\bibnamefont
  {{Radice}}}\ and\ \bibinfo {author} {\bibfnamefont {L.}~\bibnamefont
  {{Rezzolla}}},\ }\href {\doibase 10.1051/0004-6361/201219735} {\bibfield
  {journal} {\bibinfo  {journal} {Astron. Astrophys.}\ }\textbf {\bibinfo
  {volume} {547}},\ \bibinfo {pages} {A26} (\bibinfo {year} {2012})},\ \Eprint
  {http://arxiv.org/abs/1206.6502} {arXiv:1206.6502 [astro-ph.IM]} \BibitemShut
  {NoStop}%
\bibitem [{\citenamefont {{Kiuchi}}\ \emph
  {et~al.}(2015{\natexlab{b}})\citenamefont {{Kiuchi}}, \citenamefont
  {{Sekiguchi}}, \citenamefont {{Kyutoku}}, \citenamefont {{Shibata}},
  \citenamefont {{Taniguchi}},\ and\ \citenamefont {{Wada}}}]{Kiuchi2015}%
  \BibitemOpen
  \bibfield  {author} {\bibinfo {author} {\bibfnamefont {K.}~\bibnamefont
  {{Kiuchi}}}, \bibinfo {author} {\bibfnamefont {Y.}~\bibnamefont
  {{Sekiguchi}}}, \bibinfo {author} {\bibfnamefont {K.}~\bibnamefont
  {{Kyutoku}}}, \bibinfo {author} {\bibfnamefont {M.}~\bibnamefont
  {{Shibata}}}, \bibinfo {author} {\bibfnamefont {K.}~\bibnamefont
  {{Taniguchi}}}, \ and\ \bibinfo {author} {\bibfnamefont {T.}~\bibnamefont
  {{Wada}}},\ }\href@noop {} {\bibfield  {journal} {\bibinfo  {journal}
  {arXiv:1506.06811}\ } (\bibinfo {year} {2015}{\natexlab{b}})},\ \Eprint
  {http://arxiv.org/abs/1506.06811} {arXiv:1506.06811 [astro-ph.HE]}
  \BibitemShut {NoStop}%
\bibitem [{\citenamefont {{Giacomazzo}}\ \emph {et~al.}(2015)\citenamefont
  {{Giacomazzo}}, \citenamefont {{Zrake}}, \citenamefont {{Duffell}},
  \citenamefont {{MacFadyen}},\ and\ \citenamefont
  {{Perna}}}]{Giacomazzo:2014b}%
  \BibitemOpen
  \bibfield  {author} {\bibinfo {author} {\bibfnamefont {B.}~\bibnamefont
  {{Giacomazzo}}}, \bibinfo {author} {\bibfnamefont {J.}~\bibnamefont
  {{Zrake}}}, \bibinfo {author} {\bibfnamefont {P.~C.}\ \bibnamefont
  {{Duffell}}}, \bibinfo {author} {\bibfnamefont {A.~I.}\ \bibnamefont
  {{MacFadyen}}}, \ and\ \bibinfo {author} {\bibfnamefont {R.}~\bibnamefont
  {{Perna}}},\ }\href {\doibase 10.1088/0004-637X/809/1/39} {\bibfield
  {journal} {\bibinfo  {journal} {Astrophys. J.}\ }\textbf {\bibinfo {volume}
  {809}},\ \bibinfo {eid} {39} (\bibinfo {year} {2015})},\ \Eprint
  {http://arxiv.org/abs/1410.0013} {arXiv:1410.0013 [astro-ph.HE]} \BibitemShut
  {NoStop}%
\bibitem [{\citenamefont {{Palenzuela}}\ \emph {et~al.}(2015)\citenamefont
  {{Palenzuela}}, \citenamefont {{Liebling}}, \citenamefont {{Neilsen}},
  \citenamefont {{Lehner}}, \citenamefont {{Caballero}}, \citenamefont
  {{O'Connor}},\ and\ \citenamefont {{Anderson}}}]{Palenzuela2015}%
  \BibitemOpen
  \bibfield  {author} {\bibinfo {author} {\bibfnamefont {C.}~\bibnamefont
  {{Palenzuela}}}, \bibinfo {author} {\bibfnamefont {S.~L.}\ \bibnamefont
  {{Liebling}}}, \bibinfo {author} {\bibfnamefont {D.}~\bibnamefont
  {{Neilsen}}}, \bibinfo {author} {\bibfnamefont {L.}~\bibnamefont {{Lehner}}},
  \bibinfo {author} {\bibfnamefont {O.~L.}\ \bibnamefont {{Caballero}}},
  \bibinfo {author} {\bibfnamefont {E.}~\bibnamefont {{O'Connor}}}, \ and\
  \bibinfo {author} {\bibfnamefont {M.}~\bibnamefont {{Anderson}}},\ }\href
  {\doibase 10.1103/PhysRevD.92.044045} {\bibfield  {journal} {\bibinfo
  {journal} {Phys. Rev. D}\ }\textbf {\bibinfo {volume} {92}},\ \bibinfo {eid}
  {044045} (\bibinfo {year} {2015})},\ \Eprint
  {http://arxiv.org/abs/1505.01607} {arXiv:1505.01607 [gr-qc]} \BibitemShut
  {NoStop}%
\bibitem [{\citenamefont {{Obergaulinger}}\ \emph {et~al.}(2010)\citenamefont
  {{Obergaulinger}}, \citenamefont {{Aloy}},\ and\ \citenamefont
  {{M{\"u}ller}}}]{Obergaulinger10}%
  \BibitemOpen
  \bibfield  {author} {\bibinfo {author} {\bibfnamefont {M.}~\bibnamefont
  {{Obergaulinger}}}, \bibinfo {author} {\bibfnamefont {M.~A.}\ \bibnamefont
  {{Aloy}}}, \ and\ \bibinfo {author} {\bibfnamefont {E.}~\bibnamefont
  {{M{\"u}ller}}},\ }\href {\doibase 10.1051/0004-6361/200913386} {\bibfield
  {journal} {\bibinfo  {journal} {Astron. Astrophys.}\ }\textbf {\bibinfo
  {volume} {515}},\ \bibinfo {eid} {A30} (\bibinfo {year} {2010})},\ \Eprint
  {http://arxiv.org/abs/1003.6031} {arXiv:1003.6031 [astro-ph.SR]} \BibitemShut
  {NoStop}%
\bibitem [{\citenamefont {Zrake}\ and\ \citenamefont
  {MacFadyen}(2011)}]{Zrake2011}%
  \BibitemOpen
  \bibfield  {author} {\bibinfo {author} {\bibfnamefont {J.}~\bibnamefont
  {Zrake}}\ and\ \bibinfo {author} {\bibfnamefont {A.}~\bibnamefont
  {MacFadyen}}\ }(\bibinfo {year} {2011})\ pp.\ \bibinfo {pages} {102--105},\
  \Eprint {http://arxiv.org/abs/1109.6294} {arXiv:1109.6294} \BibitemShut
  {NoStop}%
\bibitem [{\citenamefont {Zrake}\ and\ \citenamefont
  {MacFadyen}(2012)}]{Zrake2011a}%
  \BibitemOpen
  \bibfield  {author} {\bibinfo {author} {\bibfnamefont {J.}~\bibnamefont
  {Zrake}}\ and\ \bibinfo {author} {\bibfnamefont {A.~I.}\ \bibnamefont
  {MacFadyen}},\ }\href {\doibase 10.1088/0004-637X/744/1/32} {\bibfield
  {journal} {\bibinfo  {journal} {Astrophys. J.}\ }\textbf {\bibinfo {volume}
  {744}},\ \bibinfo {pages} {32} (\bibinfo {year} {2012})},\ \Eprint
  {http://arxiv.org/abs/1108.1991} {arXiv:1108.1991} \BibitemShut {NoStop}%
\bibitem [{\citenamefont {{Radice}}\ and\ \citenamefont
  {{Rezzolla}}(2013)}]{Radice2012b}%
  \BibitemOpen
  \bibfield  {author} {\bibinfo {author} {\bibfnamefont {D.}~\bibnamefont
  {{Radice}}}\ and\ \bibinfo {author} {\bibfnamefont {L.}~\bibnamefont
  {{Rezzolla}}},\ }\href {\doibase 10.1088/2041-8205/766/1/L10} {\bibfield
  {journal} {\bibinfo  {journal} {Astrophys. J.}\ }\textbf {\bibinfo {volume}
  {766}},\ \bibinfo {eid} {L10} (\bibinfo {year} {2013})},\ \Eprint
  {http://arxiv.org/abs/1209.2936} {arXiv:1209.2936 [astro-ph.HE]} \BibitemShut
  {NoStop}%
\bibitem [{\citenamefont {{Evans}}\ and\ \citenamefont
  {{Hawley}}(1988)}]{Evans1988}%
  \BibitemOpen
  \bibfield  {author} {\bibinfo {author} {\bibfnamefont {C.~R.}\ \bibnamefont
  {{Evans}}}\ and\ \bibinfo {author} {\bibfnamefont {J.~F.}\ \bibnamefont
  {{Hawley}}},\ }\href {\doibase 10.1086/166684} {\bibfield  {journal}
  {\bibinfo  {journal} {Astrophys. J.}\ }\textbf {\bibinfo {volume} {332}},\
  \bibinfo {pages} {659} (\bibinfo {year} {1988})}\BibitemShut {NoStop}%
\bibitem [{\citenamefont {{Balsara}}\ and\ \citenamefont
  {{Spicer}}(1999)}]{Balsara99}%
  \BibitemOpen
  \bibfield  {author} {\bibinfo {author} {\bibfnamefont {D.~S.}\ \bibnamefont
  {{Balsara}}}\ and\ \bibinfo {author} {\bibfnamefont {D.~S.}\ \bibnamefont
  {{Spicer}}},\ }\href {\doibase 10.1006/jcph.1998.6153} {\bibfield  {journal}
  {\bibinfo  {journal} {J. Comput. Phys.}\ }\textbf {\bibinfo {volume} {149}},\
  \bibinfo {pages} {270} (\bibinfo {year} {1999})}\BibitemShut {NoStop}%
\bibitem [{\citenamefont {Toth}(2000)}]{Toth2000}%
  \BibitemOpen
  \bibfield  {author} {\bibinfo {author} {\bibfnamefont {G.}~\bibnamefont
  {Toth}},\ }\href {\doibase 10.1006/jcph.2000.6519} {\bibfield  {journal}
  {\bibinfo  {journal} {J. Comput. Phys.}\ }\textbf {\bibinfo {volume} {161}},\
  \bibinfo {pages} {605} (\bibinfo {year} {2000})}\BibitemShut {NoStop}%
\bibitem [{\citenamefont {{Balsara}}(2001)}]{Balsara2001}%
  \BibitemOpen
  \bibfield  {author} {\bibinfo {author} {\bibfnamefont {D.}~\bibnamefont
  {{Balsara}}},\ }\href {\doibase 10.1086/318941} {\bibfield  {journal}
  {\bibinfo  {journal} {Astrophysical Journal Suppl. Series}\ }\textbf
  {\bibinfo {volume} {132}},\ \bibinfo {pages} {83} (\bibinfo {year}
  {2001})}\BibitemShut {NoStop}%
\bibitem [{\citenamefont {{Balsara}}\ \emph {et~al.}(2009)\citenamefont
  {{Balsara}}, \citenamefont {{Rumpf}}, \citenamefont {{Dumbser}},\ and\
  \citenamefont {{Munz}}}]{Balsara2009}%
  \BibitemOpen
  \bibfield  {author} {\bibinfo {author} {\bibfnamefont {D.~S.}\ \bibnamefont
  {{Balsara}}}, \bibinfo {author} {\bibfnamefont {T.}~\bibnamefont {{Rumpf}}},
  \bibinfo {author} {\bibfnamefont {M.}~\bibnamefont {{Dumbser}}}, \ and\
  \bibinfo {author} {\bibfnamefont {C.-D.}\ \bibnamefont {{Munz}}},\ }\href
  {\doibase 10.1016/j.jcp.2008.12.003} {\bibfield  {journal} {\bibinfo
  {journal} {Journal of Computational Physics}\ }\textbf {\bibinfo {volume}
  {228}},\ \bibinfo {pages} {2480} (\bibinfo {year} {2009})},\ \Eprint
  {http://arxiv.org/abs/0811.2200} {arXiv:0811.2200 [physics.comp-ph]}
  \BibitemShut {NoStop}%
\bibitem [{\citenamefont {{Dedner}}\ \emph {et~al.}(2002)\citenamefont
  {{Dedner}}, \citenamefont {{Kemm}}, \citenamefont {{Kr\"oner}}, \citenamefont
  {{Munz}}, \citenamefont {{Schnitzer}},\ and\ \citenamefont
  {{Wesenberg}}}]{Dedner:2002}%
  \BibitemOpen
  \bibfield  {author} {\bibinfo {author} {\bibfnamefont {A.}~\bibnamefont
  {{Dedner}}}, \bibinfo {author} {\bibfnamefont {F.}~\bibnamefont {{Kemm}}},
  \bibinfo {author} {\bibfnamefont {D.}~\bibnamefont {{Kr\"oner}}}, \bibinfo
  {author} {\bibfnamefont {C.~D.}\ \bibnamefont {{Munz}}}, \bibinfo {author}
  {\bibfnamefont {T.}~\bibnamefont {{Schnitzer}}}, \ and\ \bibinfo {author}
  {\bibfnamefont {M.}~\bibnamefont {{Wesenberg}}},\ }\href {\doibase
  10.1006/jcph.2001.6961} {\bibfield  {journal} {\bibinfo  {journal} {Journal
  of Computational Physics}\ }\textbf {\bibinfo {volume} {175}},\ \bibinfo
  {pages} {645} (\bibinfo {year} {2002})}\BibitemShut {NoStop}%
\bibitem [{\citenamefont {{Etienne}}\ \emph {et~al.}(2010)\citenamefont
  {{Etienne}}, \citenamefont {{Liu}},\ and\ \citenamefont
  {{Shapiro}}}]{Etienne:2010ui}%
  \BibitemOpen
  \bibfield  {author} {\bibinfo {author} {\bibfnamefont {Z.~B.}\ \bibnamefont
  {{Etienne}}}, \bibinfo {author} {\bibfnamefont {Y.~T.}\ \bibnamefont
  {{Liu}}}, \ and\ \bibinfo {author} {\bibfnamefont {S.~L.}\ \bibnamefont
  {{Shapiro}}},\ }\href {\doibase 10.1103/PhysRevD.82.084031} {\bibfield
  {journal} {\bibinfo  {journal} {Phys. Rev. D}\ }\textbf {\bibinfo {volume}
  {82}},\ \bibinfo {eid} {084031} (\bibinfo {year} {2010})},\ \Eprint
  {http://arxiv.org/abs/1007.2848} {arXiv:1007.2848 [astro-ph.HE]} \BibitemShut
  {NoStop}%
\bibitem [{\citenamefont {{Etienne}}\ \emph
  {et~al.}(2012{\natexlab{a}})\citenamefont {{Etienne}}, \citenamefont
  {{Paschalidis}}, \citenamefont {{Liu}},\ and\ \citenamefont
  {{Shapiro}}}]{Etienne2012a}%
  \BibitemOpen
  \bibfield  {author} {\bibinfo {author} {\bibfnamefont {Z.~B.}\ \bibnamefont
  {{Etienne}}}, \bibinfo {author} {\bibfnamefont {V.}~\bibnamefont
  {{Paschalidis}}}, \bibinfo {author} {\bibfnamefont {Y.~T.}\ \bibnamefont
  {{Liu}}}, \ and\ \bibinfo {author} {\bibfnamefont {S.~L.}\ \bibnamefont
  {{Shapiro}}},\ }\href {\doibase 10.1103/PhysRevD.85.024013} {\bibfield
  {journal} {\bibinfo  {journal} {Phys. Rev. D}\ }\textbf {\bibinfo {volume}
  {85}},\ \bibinfo {eid} {024013} (\bibinfo {year} {2012}{\natexlab{a}})},\
  \Eprint {http://arxiv.org/abs/1110.4633} {arXiv:1110.4633 [astro-ph.HE]}
  \BibitemShut {NoStop}%
\bibitem [{\citenamefont {{Farris}}\ \emph {et~al.}(2012)\citenamefont
  {{Farris}}, \citenamefont {{Gold}}, \citenamefont {{Paschalidis}},
  \citenamefont {{Etienne}},\ and\ \citenamefont {{Shapiro}}}]{Farris2012}%
  \BibitemOpen
  \bibfield  {author} {\bibinfo {author} {\bibfnamefont {B.~D.}\ \bibnamefont
  {{Farris}}}, \bibinfo {author} {\bibfnamefont {R.}~\bibnamefont {{Gold}}},
  \bibinfo {author} {\bibfnamefont {V.}~\bibnamefont {{Paschalidis}}}, \bibinfo
  {author} {\bibfnamefont {Z.~B.}\ \bibnamefont {{Etienne}}}, \ and\ \bibinfo
  {author} {\bibfnamefont {S.~L.}\ \bibnamefont {{Shapiro}}},\ }\href {\doibase
  10.1103/PhysRevLett.109.221102} {\bibfield  {journal} {\bibinfo  {journal}
  {Phys. Rev. Lett.}\ }\textbf {\bibinfo {volume} {109}},\ \bibinfo {eid}
  {221102} (\bibinfo {year} {2012})},\ \Eprint {http://arxiv.org/abs/1207.3354}
  {arXiv:1207.3354 [astro-ph.HE]} \BibitemShut {NoStop}%
\bibitem [{\citenamefont {{Giacomazzo}}\ \emph {et~al.}(2009)\citenamefont
  {{Giacomazzo}}, \citenamefont {{Rezzolla}},\ and\ \citenamefont
  {{Baiotti}}}]{Giacomazzo:2009mp}%
  \BibitemOpen
  \bibfield  {author} {\bibinfo {author} {\bibfnamefont {B.}~\bibnamefont
  {{Giacomazzo}}}, \bibinfo {author} {\bibfnamefont {L.}~\bibnamefont
  {{Rezzolla}}}, \ and\ \bibinfo {author} {\bibfnamefont {L.}~\bibnamefont
  {{Baiotti}}},\ }\href {\doibase 10.1111/j.1745-3933.2009.00745.x} {\bibfield
  {journal} {\bibinfo  {journal} {Mon. Not. R. Astron. Soc.}\ }\textbf
  {\bibinfo {volume} {399}},\ \bibinfo {pages} {L164} (\bibinfo {year}
  {2009})},\ \Eprint {http://arxiv.org/abs/0901.2722} {arXiv:0901.2722 [gr-qc]}
  \BibitemShut {NoStop}%
\bibitem [{\citenamefont {{Kiuchi}}\ \emph
  {et~al.}(2012{\natexlab{c}})\citenamefont {{Kiuchi}}, \citenamefont
  {{Kyutoku}},\ and\ \citenamefont {{Shibata}}}]{Kiuchi2012b}%
  \BibitemOpen
  \bibfield  {author} {\bibinfo {author} {\bibfnamefont {K.}~\bibnamefont
  {{Kiuchi}}}, \bibinfo {author} {\bibfnamefont {K.}~\bibnamefont {{Kyutoku}}},
  \ and\ \bibinfo {author} {\bibfnamefont {M.}~\bibnamefont {{Shibata}}},\
  }\href {\doibase 10.1103/PhysRevD.86.064008} {\bibfield  {journal} {\bibinfo
  {journal} {Phys. Rev. D}\ }\textbf {\bibinfo {volume} {86}},\ \bibinfo {eid}
  {064008} (\bibinfo {year} {2012}{\natexlab{c}})},\ \Eprint
  {http://arxiv.org/abs/1207.6444} {arXiv:1207.6444 [astro-ph.HE]} \BibitemShut
  {NoStop}%
\bibitem [{\citenamefont {{Kiuchi}}\ \emph {et~al.}(2013)\citenamefont
  {{Kiuchi}}, \citenamefont {{Kyutoku}}, \citenamefont {{Hotokezaka}},
  \citenamefont {{Sekiguchi}},\ and\ \citenamefont {{Shibata}}}]{Kiuchi2013}%
  \BibitemOpen
  \bibfield  {author} {\bibinfo {author} {\bibfnamefont {K.}~\bibnamefont
  {{Kiuchi}}}, \bibinfo {author} {\bibfnamefont {K.}~\bibnamefont {{Kyutoku}}},
  \bibinfo {author} {\bibfnamefont {K.}~\bibnamefont {{Hotokezaka}}}, \bibinfo
  {author} {\bibfnamefont {Y.}~\bibnamefont {{Sekiguchi}}}, \ and\ \bibinfo
  {author} {\bibfnamefont {M.}~\bibnamefont {{Shibata}}},\ }in\ \href@noop {}
  {\emph {\bibinfo {booktitle} {Numerical Modeling of Space Plasma Flows
  (ASTRONUM2012)}}},\ \bibinfo {series} {Astronomical Society of the Pacific
  Conference Series}, Vol.\ \bibinfo {volume} {474},\ \bibinfo {editor} {edited
  by\ \bibinfo {editor} {\bibfnamefont {N.~V.}\ \bibnamefont {{Pogorelov}}},
  \bibinfo {editor} {\bibfnamefont {E.}~\bibnamefont {{Audit}}}, \ and\
  \bibinfo {editor} {\bibfnamefont {G.~P.}\ \bibnamefont {{Zank}}}}\ (\bibinfo
  {year} {2013})\ p.~\bibinfo {pages} {84}\BibitemShut {NoStop}%
\bibitem [{\citenamefont {Glendenning}\ and\ \citenamefont
  {Moszkowski}(1991)}]{GlendenningMoszkowski91}%
  \BibitemOpen
  \bibfield  {author} {\bibinfo {author} {\bibfnamefont {N.~K.}\ \bibnamefont
  {Glendenning}}\ and\ \bibinfo {author} {\bibfnamefont {S.~A.}\ \bibnamefont
  {Moszkowski}},\ }\href {\doibase 10.1103/PhysRevLett.67.2414} {\bibfield
  {journal} {\bibinfo  {journal} {Phys. Rev. Lett.}\ }\textbf {\bibinfo
  {volume} {67}},\ \bibinfo {pages} {2414} (\bibinfo {year}
  {1991})}\BibitemShut {NoStop}%
\bibitem [{\citenamefont {{Zrake}}\ and\ \citenamefont
  {{MacFadyen}}(2013)}]{Zrake2013b}%
  \BibitemOpen
  \bibfield  {author} {\bibinfo {author} {\bibfnamefont {J.}~\bibnamefont
  {{Zrake}}}\ and\ \bibinfo {author} {\bibfnamefont {A.~I.}\ \bibnamefont
  {{MacFadyen}}},\ }\href {\doibase 10.1088/2041-8205/769/2/L29} {\bibfield
  {journal} {\bibinfo  {journal} {Astrophys. J.}\ }\textbf {\bibinfo {volume}
  {769}},\ \bibinfo {eid} {L29} (\bibinfo {year} {2013})},\ \Eprint
  {http://arxiv.org/abs/1303.1450} {arXiv:1303.1450 [astro-ph.HE]} \BibitemShut
  {NoStop}%
\bibitem [{\citenamefont {{Goldreich}}\ and\ \citenamefont
  {{Lynden-Bell}}(1969)}]{Goldreich1969a}%
  \BibitemOpen
  \bibfield  {author} {\bibinfo {author} {\bibfnamefont {P.}~\bibnamefont
  {{Goldreich}}}\ and\ \bibinfo {author} {\bibfnamefont {D.}~\bibnamefont
  {{Lynden-Bell}}},\ }\href {\doibase 10.1086/149947} {\bibfield  {journal}
  {\bibinfo  {journal} {Astrophys. J.}\ }\textbf {\bibinfo {volume} {156}},\
  \bibinfo {pages} {59} (\bibinfo {year} {1969})}\BibitemShut {NoStop}%
\bibitem [{\citenamefont {{Ponce}}\ \emph {et~al.}(2015)\citenamefont
  {{Ponce}}, \citenamefont {{Palenzuela}}, \citenamefont {{Barausse}},\ and\
  \citenamefont {{Lehner}}}]{Ponce2014a}%
  \BibitemOpen
  \bibfield  {author} {\bibinfo {author} {\bibfnamefont {M.}~\bibnamefont
  {{Ponce}}}, \bibinfo {author} {\bibfnamefont {C.}~\bibnamefont
  {{Palenzuela}}}, \bibinfo {author} {\bibfnamefont {E.}~\bibnamefont
  {{Barausse}}}, \ and\ \bibinfo {author} {\bibfnamefont {L.}~\bibnamefont
  {{Lehner}}},\ }\href {\doibase 10.1103/PhysRevD.91.084038} {\bibfield
  {journal} {\bibinfo  {journal} {Phys. Rev. D}\ }\textbf {\bibinfo {volume}
  {91}},\ \bibinfo {eid} {084038} (\bibinfo {year} {2015})},\ \Eprint
  {http://arxiv.org/abs/1410.0638} {arXiv:1410.0638 [gr-qc]} \BibitemShut
  {NoStop}%
\bibitem [{\citenamefont {{Aloy}}\ \emph {et~al.}(2012)\citenamefont {{Aloy}},
  \citenamefont {{Rezzolla}}, \citenamefont {{Giacomazzo}},\ and\ \citenamefont
  {{Obergaulinger}}}]{Aloy2012}%
  \BibitemOpen
  \bibfield  {author} {\bibinfo {author} {\bibfnamefont {M.~A.}\ \bibnamefont
  {{Aloy}}}, \bibinfo {author} {\bibfnamefont {L.}~\bibnamefont {{Rezzolla}}},
  \bibinfo {author} {\bibfnamefont {B.}~\bibnamefont {{Giacomazzo}}}, \ and\
  \bibinfo {author} {\bibfnamefont {M.}~\bibnamefont {{Obergaulinger}}},\ }in\
  \href@noop {} {\emph {\bibinfo {booktitle} {Numerical Modeling of Space
  Plasma Slows (ASTRONUM 2011)}}},\ \bibinfo {series} {Astronomical Society of
  the Pacific Conference Series}, Vol.\ \bibinfo {volume} {459},\ \bibinfo
  {editor} {edited by\ \bibinfo {editor} {\bibfnamefont {N.~V.}\ \bibnamefont
  {{Pogorelov}}}, \bibinfo {editor} {\bibfnamefont {J.~A.}\ \bibnamefont
  {{Font}}}, \bibinfo {editor} {\bibfnamefont {E.}~\bibnamefont {{Audit}}}, \
  and\ \bibinfo {editor} {\bibfnamefont {G.~P.}\ \bibnamefont {{Zank}}}}\
  (\bibinfo {year} {2012})\ p.~\bibinfo {pages} {49}\BibitemShut {NoStop}%
\bibitem [{\citenamefont {{Etienne}}\ \emph
  {et~al.}(2012{\natexlab{b}})\citenamefont {{Etienne}}, \citenamefont
  {{Paschalidis}},\ and\ \citenamefont {{Shapiro}}}]{Etienne2012b}%
  \BibitemOpen
  \bibfield  {author} {\bibinfo {author} {\bibfnamefont {Z.~B.}\ \bibnamefont
  {{Etienne}}}, \bibinfo {author} {\bibfnamefont {V.}~\bibnamefont
  {{Paschalidis}}}, \ and\ \bibinfo {author} {\bibfnamefont {S.~L.}\
  \bibnamefont {{Shapiro}}},\ }\href {\doibase 10.1103/PhysRevD.86.084026}
  {\bibfield  {journal} {\bibinfo  {journal} {Phys. Rev. D}\ }\textbf {\bibinfo
  {volume} {86}},\ \bibinfo {eid} {084026} (\bibinfo {year}
  {2012}{\natexlab{b}})},\ \Eprint {http://arxiv.org/abs/1209.1632}
  {arXiv:1209.1632 [astro-ph.HE]} \BibitemShut {NoStop}%
\bibitem [{\citenamefont {{Blandford}}\ and\ \citenamefont
  {{Znajek}}(1977)}]{Blandford1977}%
  \BibitemOpen
  \bibfield  {author} {\bibinfo {author} {\bibfnamefont {R.~D.}\ \bibnamefont
  {{Blandford}}}\ and\ \bibinfo {author} {\bibfnamefont {R.~L.}\ \bibnamefont
  {{Znajek}}},\ }\href@noop {} {\bibfield  {journal} {\bibinfo  {journal} {Mon.
  Not. R. Astron. Soc.}\ }\textbf {\bibinfo {volume} {179}},\ \bibinfo {pages}
  {433} (\bibinfo {year} {1977})}\BibitemShut {NoStop}%
\bibitem [{\citenamefont {Aloy}\ and\ \citenamefont
  {Rezzolla}(2006)}]{Aloy:2006rd}%
  \BibitemOpen
  \bibfield  {author} {\bibinfo {author} {\bibfnamefont {M.~A.}\ \bibnamefont
  {Aloy}}\ and\ \bibinfo {author} {\bibfnamefont {L.}~\bibnamefont
  {Rezzolla}},\ }\href@noop {} {\bibfield  {journal} {\bibinfo  {journal}
  {Astrophys. J.}\ }\textbf {\bibinfo {volume} {640}},\ \bibinfo {pages} {L115}
  (\bibinfo {year} {2006})}\BibitemShut {NoStop}%
\bibitem [{\citenamefont {{Uzdensky}}(2011)}]{Uzdensky2011}%
  \BibitemOpen
  \bibfield  {author} {\bibinfo {author} {\bibfnamefont {D.~A.}\ \bibnamefont
  {{Uzdensky}}},\ }\href {\doibase 10.1007/s11214-011-9744-5} {\bibfield
  {journal} {\bibinfo  {journal} {Space Science Reviews}\ }\textbf {\bibinfo
  {volume} {160}},\ \bibinfo {pages} {45} (\bibinfo {year} {2011})},\ \Eprint
  {http://arxiv.org/abs/1101.2472} {arXiv:1101.2472 [astro-ph.HE]} \BibitemShut
  {NoStop}%
\bibitem [{\citenamefont {{Gehrels}}\ and\ \citenamefont {{et
  al}.}(2004)}]{Gehrels_etal2004}%
  \BibitemOpen
  \bibfield  {author} {\bibinfo {author} {\bibfnamefont {N.}~\bibnamefont
  {{Gehrels}}}\ and\ \bibinfo {author} {\bibnamefont {{et al}.}},\ }\href
  {\doibase 10.1086/422091} {\bibfield  {journal} {\bibinfo  {journal}
  {Astrophys. J.}\ }\textbf {\bibinfo {volume} {611}},\ \bibinfo {pages} {1005}
  (\bibinfo {year} {2004})}\BibitemShut {NoStop}%
\bibitem [{\citenamefont {{Rowlinson}}\ \emph {et~al.}(2013)\citenamefont
  {{Rowlinson}}, \citenamefont {{O'Brien}}, \citenamefont {{Metzger}},
  \citenamefont {{Tanvir}},\ and\ \citenamefont {{Levan}}}]{Rowlinson2013}%
  \BibitemOpen
  \bibfield  {author} {\bibinfo {author} {\bibfnamefont {A.}~\bibnamefont
  {{Rowlinson}}}, \bibinfo {author} {\bibfnamefont {P.~T.}\ \bibnamefont
  {{O'Brien}}}, \bibinfo {author} {\bibfnamefont {B.~D.}\ \bibnamefont
  {{Metzger}}}, \bibinfo {author} {\bibfnamefont {N.~R.}\ \bibnamefont
  {{Tanvir}}}, \ and\ \bibinfo {author} {\bibfnamefont {A.~J.}\ \bibnamefont
  {{Levan}}},\ }\href {\doibase 10.1093/mnras/sts683} {\bibfield  {journal}
  {\bibinfo  {journal} {Mon. Not. R. Astron. Soc.}\ }\textbf {\bibinfo {volume}
  {430}},\ \bibinfo {pages} {1061} (\bibinfo {year} {2013})},\ \Eprint
  {http://arxiv.org/abs/1301.0629} {arXiv:1301.0629 [astro-ph.HE]} \BibitemShut
  {NoStop}%
\bibitem [{\citenamefont {{Gompertz}}\ \emph {et~al.}(2014)\citenamefont
  {{Gompertz}}, \citenamefont {{O'Brien}},\ and\ \citenamefont
  {{Wynn}}}]{Gompertz2013}%
  \BibitemOpen
  \bibfield  {author} {\bibinfo {author} {\bibfnamefont {B.~P.}\ \bibnamefont
  {{Gompertz}}}, \bibinfo {author} {\bibfnamefont {P.~T.}\ \bibnamefont
  {{O'Brien}}}, \ and\ \bibinfo {author} {\bibfnamefont {G.~A.}\ \bibnamefont
  {{Wynn}}},\ }\href {\doibase 10.1093/mnras/stt2165} {\bibfield  {journal}
  {\bibinfo  {journal} {Mon. Not. R. Astron. Soc.}\ }\textbf {\bibinfo {volume}
  {438}},\ \bibinfo {pages} {240} (\bibinfo {year} {2014})},\ \Eprint
  {http://arxiv.org/abs/1311.1505} {arXiv:1311.1505 [astro-ph.HE]} \BibitemShut
  {NoStop}%
\bibitem [{\citenamefont {{Hotokezaka}}\ \emph
  {et~al.}(2013{\natexlab{c}})\citenamefont {{Hotokezaka}}, \citenamefont
  {{Kiuchi}}, \citenamefont {{Kyutoku}}, \citenamefont {{Okawa}}, \citenamefont
  {{Sekiguchi}}, \citenamefont {{Shibata}},\ and\ \citenamefont
  {{Taniguchi}}}]{Hotokezaka2013}%
  \BibitemOpen
  \bibfield  {author} {\bibinfo {author} {\bibfnamefont {K.}~\bibnamefont
  {{Hotokezaka}}}, \bibinfo {author} {\bibfnamefont {K.}~\bibnamefont
  {{Kiuchi}}}, \bibinfo {author} {\bibfnamefont {K.}~\bibnamefont {{Kyutoku}}},
  \bibinfo {author} {\bibfnamefont {H.}~\bibnamefont {{Okawa}}}, \bibinfo
  {author} {\bibfnamefont {Y.-i.}\ \bibnamefont {{Sekiguchi}}}, \bibinfo
  {author} {\bibfnamefont {M.}~\bibnamefont {{Shibata}}}, \ and\ \bibinfo
  {author} {\bibfnamefont {K.}~\bibnamefont {{Taniguchi}}},\ }\href {\doibase
  10.1103/PhysRevD.87.024001} {\bibfield  {journal} {\bibinfo  {journal} {Phys.
  Rev. D}\ }\textbf {\bibinfo {volume} {87}},\ \bibinfo {eid} {024001}
  (\bibinfo {year} {2013}{\natexlab{c}})},\ \Eprint
  {http://arxiv.org/abs/1212.0905} {arXiv:1212.0905 [astro-ph.HE]} \BibitemShut
  {NoStop}%
\bibitem [{\citenamefont {{Zhang}}\ and\ \citenamefont
  {{M{\'e}sz{\'a}ros}}(2001)}]{Zhang2001}%
  \BibitemOpen
  \bibfield  {author} {\bibinfo {author} {\bibfnamefont {B.}~\bibnamefont
  {{Zhang}}}\ and\ \bibinfo {author} {\bibfnamefont {P.}~\bibnamefont
  {{M{\'e}sz{\'a}ros}}},\ }\href {\doibase 10.1086/320255} {\bibfield
  {journal} {\bibinfo  {journal} {Astrophys. J.}\ }\textbf {\bibinfo {volume}
  {552}},\ \bibinfo {pages} {L35} (\bibinfo {year} {2001})},\ \Eprint
  {http://arxiv.org/abs/astro-ph/0011133} {astro-ph/0011133} \BibitemShut
  {NoStop}%
\bibitem [{\citenamefont {{Gao}}\ and\ \citenamefont {{Fan}}(2006)}]{Gao2006}%
  \BibitemOpen
  \bibfield  {author} {\bibinfo {author} {\bibfnamefont {W.-H.}\ \bibnamefont
  {{Gao}}}\ and\ \bibinfo {author} {\bibfnamefont {Y.-Z.}\ \bibnamefont
  {{Fan}}},\ }\href {\doibase 10.1088/1009-9271/6/5/01} {\bibfield  {journal}
  {\bibinfo  {journal} {Chinese Journal of Astronomy and Astrophysics}\
  }\textbf {\bibinfo {volume} {6}},\ \bibinfo {pages} {513} (\bibinfo {year}
  {2006})},\ \Eprint {http://arxiv.org/abs/astro-ph/0512646} {astro-ph/0512646}
  \BibitemShut {NoStop}%
\bibitem [{\citenamefont {{Fan}}\ and\ \citenamefont {{Xu}}(2006)}]{Fan2006}%
  \BibitemOpen
  \bibfield  {author} {\bibinfo {author} {\bibfnamefont {Y.-Z.}\ \bibnamefont
  {{Fan}}}\ and\ \bibinfo {author} {\bibfnamefont {D.}~\bibnamefont {{Xu}}},\
  }\href {\doibase 10.1111/j.1745-3933.2006.00217.x} {\bibfield  {journal}
  {\bibinfo  {journal} {Mon. Not. R. Astron. Soc.}\ }\textbf {\bibinfo {volume}
  {372}},\ \bibinfo {pages} {L19} (\bibinfo {year} {2006})},\ \Eprint
  {http://arxiv.org/abs/astro-ph/0605445} {astro-ph/0605445} \BibitemShut
  {NoStop}%
\bibitem [{\citenamefont {{Metzger}}\ \emph {et~al.}(2008)\citenamefont
  {{Metzger}}, \citenamefont {{Quataert}},\ and\ \citenamefont
  {{Thompson}}}]{Metzger2008}%
  \BibitemOpen
  \bibfield  {author} {\bibinfo {author} {\bibfnamefont {B.~D.}\ \bibnamefont
  {{Metzger}}}, \bibinfo {author} {\bibfnamefont {E.}~\bibnamefont
  {{Quataert}}}, \ and\ \bibinfo {author} {\bibfnamefont {T.~A.}\ \bibnamefont
  {{Thompson}}},\ }\href {\doibase 10.1111/j.1365-2966.2008.12923.x} {\bibfield
   {journal} {\bibinfo  {journal} {Mon. Not. R. Astron. Soc.}\ }\textbf
  {\bibinfo {volume} {385}},\ \bibinfo {pages} {1455} (\bibinfo {year}
  {2008})},\ \Eprint {http://arxiv.org/abs/0712.1233} {arXiv:0712.1233}
  \BibitemShut {NoStop}%
\bibitem [{\citenamefont {{Metzger}}\ \emph {et~al.}(2011)\citenamefont
  {{Metzger}}, \citenamefont {{Giannios}}, \citenamefont {{Thompson}},
  \citenamefont {{Bucciantini}},\ and\ \citenamefont
  {{Quataert}}}]{Metzger:2011}%
  \BibitemOpen
  \bibfield  {author} {\bibinfo {author} {\bibfnamefont {B.~D.}\ \bibnamefont
  {{Metzger}}}, \bibinfo {author} {\bibfnamefont {D.}~\bibnamefont
  {{Giannios}}}, \bibinfo {author} {\bibfnamefont {T.~A.}\ \bibnamefont
  {{Thompson}}}, \bibinfo {author} {\bibfnamefont {N.}~\bibnamefont
  {{Bucciantini}}}, \ and\ \bibinfo {author} {\bibfnamefont {E.}~\bibnamefont
  {{Quataert}}},\ }\href {\doibase 10.1111/j.1365-2966.2011.18280.x} {\bibfield
   {journal} {\bibinfo  {journal} {Mon. Not. R. Astron. Soc.}\ }\textbf
  {\bibinfo {volume} {413}},\ \bibinfo {pages} {2031} (\bibinfo {year}
  {2011})},\ \Eprint {http://arxiv.org/abs/1012.0001} {arXiv:1012.0001
  [astro-ph.HE]} \BibitemShut {NoStop}%
\bibitem [{\citenamefont {{Giacomazzo}}\ and\ \citenamefont
  {{Perna}}(2013)}]{Giacomazzo2013}%
  \BibitemOpen
  \bibfield  {author} {\bibinfo {author} {\bibfnamefont {B.}~\bibnamefont
  {{Giacomazzo}}}\ and\ \bibinfo {author} {\bibfnamefont {R.}~\bibnamefont
  {{Perna}}},\ }\href {\doibase 10.1088/2041-8205/771/2/L26} {\bibfield
  {journal} {\bibinfo  {journal} {Astrophys. J.}\ }\textbf {\bibinfo {volume}
  {771}},\ \bibinfo {eid} {L26} (\bibinfo {year} {2013})},\ \Eprint
  {http://arxiv.org/abs/1306.1608} {arXiv:1306.1608 [astro-ph.HE]} \BibitemShut
  {NoStop}%
\bibitem [{\citenamefont {{Dall'Osso}}\ \emph {et~al.}(2015)\citenamefont
  {{Dall'Osso}}, \citenamefont {{Giacomazzo}}, \citenamefont {{Perna}},\ and\
  \citenamefont {{Stella}}}]{DallOsso2014}%
  \BibitemOpen
  \bibfield  {author} {\bibinfo {author} {\bibfnamefont {S.}~\bibnamefont
  {{Dall'Osso}}}, \bibinfo {author} {\bibfnamefont {B.}~\bibnamefont
  {{Giacomazzo}}}, \bibinfo {author} {\bibfnamefont {R.}~\bibnamefont
  {{Perna}}}, \ and\ \bibinfo {author} {\bibfnamefont {L.}~\bibnamefont
  {{Stella}}},\ }\href {\doibase 10.1088/0004-637X/798/1/25} {\bibfield
  {journal} {\bibinfo  {journal} {Astrophys. J.}\ }\textbf {\bibinfo {volume}
  {798}},\ \bibinfo {eid} {25} (\bibinfo {year} {2015})},\ \Eprint
  {http://arxiv.org/abs/1408.0013} {arXiv:1408.0013 [astro-ph.HE]} \BibitemShut
  {NoStop}%
\bibitem [{\citenamefont {{Ciolfi}}\ \emph {et~al.}(2010)\citenamefont
  {{Ciolfi}}, \citenamefont {{Ferrari}},\ and\ \citenamefont
  {{Gualtieri}}}]{Ciolfi2010}%
  \BibitemOpen
  \bibfield  {author} {\bibinfo {author} {\bibfnamefont {R.}~\bibnamefont
  {{Ciolfi}}}, \bibinfo {author} {\bibfnamefont {V.}~\bibnamefont {{Ferrari}}},
  \ and\ \bibinfo {author} {\bibfnamefont {L.}~\bibnamefont {{Gualtieri}}},\
  }\href {\doibase 10.1111/j.1365-2966.2010.16847.x} {\bibfield  {journal}
  {\bibinfo  {journal} {Mon. Not. R. Astron. Soc.}\ }\textbf {\bibinfo {volume}
  {406}},\ \bibinfo {pages} {2540} (\bibinfo {year} {2010})},\ \Eprint
  {http://arxiv.org/abs/1003.2148} {arXiv:1003.2148 [astro-ph.SR]} \BibitemShut
  {NoStop}%
\bibitem [{\citenamefont {{Frieben}}\ and\ \citenamefont
  {{Rezzolla}}(2012)}]{Frieben2012}%
  \BibitemOpen
  \bibfield  {author} {\bibinfo {author} {\bibfnamefont {J.}~\bibnamefont
  {{Frieben}}}\ and\ \bibinfo {author} {\bibfnamefont {L.}~\bibnamefont
  {{Rezzolla}}},\ }\href {\doibase 10.1111/j.1365-2966.2012.22027.x} {\bibfield
   {journal} {\bibinfo  {journal} {Mon. Not. R. Astron. Soc.}\ }\textbf
  {\bibinfo {volume} {427}},\ \bibinfo {pages} {3406} (\bibinfo {year}
  {2012})},\ \Eprint {http://arxiv.org/abs/1207.4035} {arXiv:1207.4035 [gr-qc]}
  \BibitemShut {NoStop}%
\bibitem [{\citenamefont {{Ciolfi}}\ and\ \citenamefont
  {{Rezzolla}}(2013)}]{Ciolfi2013}%
  \BibitemOpen
  \bibfield  {author} {\bibinfo {author} {\bibfnamefont {R.}~\bibnamefont
  {{Ciolfi}}}\ and\ \bibinfo {author} {\bibfnamefont {L.}~\bibnamefont
  {{Rezzolla}}},\ }\href {\doibase 10.1093/mnrasl/slt092} {\bibfield  {journal}
  {\bibinfo  {journal} {Mon. Not. R. Astron. Soc.}\ }\textbf {\bibinfo {volume}
  {435}},\ \bibinfo {pages} {L43} (\bibinfo {year} {2013})},\ \Eprint
  {http://arxiv.org/abs/1306.2803} {arXiv:1306.2803 [astro-ph.SR]} \BibitemShut
  {NoStop}%
\bibitem [{\citenamefont {Giacomazzo}\ \emph {et~al.}(2011)\citenamefont
  {Giacomazzo}, \citenamefont {Rezzolla},\ and\ \citenamefont
  {Baiotti}}]{Giacomazzo:2010}%
  \BibitemOpen
  \bibfield  {author} {\bibinfo {author} {\bibfnamefont {B.}~\bibnamefont
  {Giacomazzo}}, \bibinfo {author} {\bibfnamefont {L.}~\bibnamefont
  {Rezzolla}}, \ and\ \bibinfo {author} {\bibfnamefont {L.}~\bibnamefont
  {Baiotti}},\ }\href {\doibase 10.1103/PhysRevD.83.044014} {\bibfield
  {journal} {\bibinfo  {journal} {Phys. Rev. D}\ }\textbf {\bibinfo {volume}
  {83}},\ \bibinfo {pages} {044014} (\bibinfo {year} {2011})}\BibitemShut
  {NoStop}%
\bibitem [{\citenamefont {{Ciolfi}}\ and\ \citenamefont
  {{Siegel}}(2015)}]{Ciolfi2014}%
  \BibitemOpen
  \bibfield  {author} {\bibinfo {author} {\bibfnamefont {R.}~\bibnamefont
  {{Ciolfi}}}\ and\ \bibinfo {author} {\bibfnamefont {D.~M.}\ \bibnamefont
  {{Siegel}}},\ }\href {\doibase 10.1088/2041-8205/798/2/L36} {\bibfield
  {journal} {\bibinfo  {journal} {Astrophys. J.}\ }\textbf {\bibinfo {volume}
  {798}},\ \bibinfo {eid} {L36} (\bibinfo {year} {2015})},\ \Eprint
  {http://arxiv.org/abs/1411.2015} {arXiv:1411.2015 [astro-ph.HE]} \BibitemShut
  {NoStop}%
\bibitem [{\citenamefont {{Siegel}}\ \emph {et~al.}(2014)\citenamefont
  {{Siegel}}, \citenamefont {{Ciolfi}},\ and\ \citenamefont
  {{Rezzolla}}}]{Siegel2014}%
  \BibitemOpen
  \bibfield  {author} {\bibinfo {author} {\bibfnamefont {D.~M.}\ \bibnamefont
  {{Siegel}}}, \bibinfo {author} {\bibfnamefont {R.}~\bibnamefont {{Ciolfi}}},
  \ and\ \bibinfo {author} {\bibfnamefont {L.}~\bibnamefont {{Rezzolla}}},\
  }\href {\doibase 10.1088/2041-8205/785/1/L6} {\bibfield  {journal} {\bibinfo
  {journal} {Astrophys. J.}\ }\textbf {\bibinfo {volume} {785}},\ \bibinfo
  {eid} {L6} (\bibinfo {year} {2014})},\ \Eprint
  {http://arxiv.org/abs/1401.4544} {arXiv:1401.4544 [astro-ph.HE]} \BibitemShut
  {NoStop}%
\bibitem [{\citenamefont {{Metzger}}\ and\ \citenamefont
  {{Fern{\'a}ndez}}(2014)}]{Metzger2014}%
  \BibitemOpen
  \bibfield  {author} {\bibinfo {author} {\bibfnamefont {B.~D.}\ \bibnamefont
  {{Metzger}}}\ and\ \bibinfo {author} {\bibfnamefont {R.}~\bibnamefont
  {{Fern{\'a}ndez}}},\ }\href {\doibase 10.1093/mnras/stu802} {\bibfield
  {journal} {\bibinfo  {journal} {Mon. Not. R. Astron. Soc.}\ }\textbf
  {\bibinfo {volume} {441}},\ \bibinfo {pages} {3444} (\bibinfo {year}
  {2014})},\ \Eprint {http://arxiv.org/abs/1402.4803} {arXiv:1402.4803
  [astro-ph.HE]} \BibitemShut {NoStop}%
\bibitem [{\citenamefont {{Perego}}\ \emph {et~al.}(2014)\citenamefont
  {{Perego}}, \citenamefont {{Rosswog}}, \citenamefont {{Cabez{\'o}n}},
  \citenamefont {{Korobkin}}, \citenamefont {{K{\"a}ppeli}}, \citenamefont
  {{Arcones}},\ and\ \citenamefont {{Liebend{\"o}rfer}}}]{Perego2014}%
  \BibitemOpen
  \bibfield  {author} {\bibinfo {author} {\bibfnamefont {A.}~\bibnamefont
  {{Perego}}}, \bibinfo {author} {\bibfnamefont {S.}~\bibnamefont {{Rosswog}}},
  \bibinfo {author} {\bibfnamefont {R.~M.}\ \bibnamefont {{Cabez{\'o}n}}},
  \bibinfo {author} {\bibfnamefont {O.}~\bibnamefont {{Korobkin}}}, \bibinfo
  {author} {\bibfnamefont {R.}~\bibnamefont {{K{\"a}ppeli}}}, \bibinfo {author}
  {\bibfnamefont {A.}~\bibnamefont {{Arcones}}}, \ and\ \bibinfo {author}
  {\bibfnamefont {M.}~\bibnamefont {{Liebend{\"o}rfer}}},\ }\href {\doibase
  10.1093/mnras/stu1352} {\bibfield  {journal} {\bibinfo  {journal} {Mon. Not.
  R. Astron. Soc.}\ }\textbf {\bibinfo {volume} {443}},\ \bibinfo {pages}
  {3134} (\bibinfo {year} {2014})},\ \Eprint {http://arxiv.org/abs/1405.6730}
  {arXiv:1405.6730 [astro-ph.HE]} \BibitemShut {NoStop}%
\bibitem [{\citenamefont {{Ruffert}}\ and\ \citenamefont
  {{Janka}}(1999)}]{Ruffert99b}%
  \BibitemOpen
  \bibfield  {author} {\bibinfo {author} {\bibfnamefont {M.}~\bibnamefont
  {{Ruffert}}}\ and\ \bibinfo {author} {\bibfnamefont {H.-T.}\ \bibnamefont
  {{Janka}}},\ }\href@noop {} {\bibfield  {journal} {\bibinfo  {journal}
  {Astron. Astrophys.}\ }\textbf {\bibinfo {volume} {344}},\ \bibinfo {pages}
  {573} (\bibinfo {year} {1999})},\ \Eprint
  {http://arxiv.org/abs/astro-ph/9809280} {astro-ph/9809280} \BibitemShut
  {NoStop}%
\bibitem [{\citenamefont {{Aloy}}\ \emph {et~al.}(2005)\citenamefont {{Aloy}},
  \citenamefont {{Janka}},\ and\ \citenamefont {{M{\"u}ller}}}]{Aloy:2005}%
  \BibitemOpen
  \bibfield  {author} {\bibinfo {author} {\bibfnamefont {M.~A.}\ \bibnamefont
  {{Aloy}}}, \bibinfo {author} {\bibfnamefont {H.}~\bibnamefont {{Janka}}}, \
  and\ \bibinfo {author} {\bibfnamefont {E.}~\bibnamefont {{M{\"u}ller}}},\
  }\href {\doibase 10.1051/0004-6361:20041865} {\bibfield  {journal} {\bibinfo
  {journal} {Astron. Astrophys.}\ }\textbf {\bibinfo {volume} {436}},\ \bibinfo
  {pages} {273} (\bibinfo {year} {2005})},\ \Eprint
  {http://arxiv.org/abs/arXiv:astro-ph/0408291} {arXiv:astro-ph/0408291}
  \BibitemShut {NoStop}%
\bibitem [{\citenamefont {{Murguia-Berthier}}\ \emph
  {et~al.}(2014)\citenamefont {{Murguia-Berthier}}, \citenamefont {{Montes}},
  \citenamefont {{Ramirez-Ruiz}}, \citenamefont {{De Colle}},\ and\
  \citenamefont {{Lee}}}]{Murguia-Berthier2014}%
  \BibitemOpen
  \bibfield  {author} {\bibinfo {author} {\bibfnamefont {A.}~\bibnamefont
  {{Murguia-Berthier}}}, \bibinfo {author} {\bibfnamefont {G.}~\bibnamefont
  {{Montes}}}, \bibinfo {author} {\bibfnamefont {E.}~\bibnamefont
  {{Ramirez-Ruiz}}}, \bibinfo {author} {\bibfnamefont {F.}~\bibnamefont {{De
  Colle}}}, \ and\ \bibinfo {author} {\bibfnamefont {W.~H.}\ \bibnamefont
  {{Lee}}},\ }\href {\doibase 10.1088/2041-8205/788/1/L8} {\bibfield  {journal}
  {\bibinfo  {journal} {Astrophys. J.}\ }\textbf {\bibinfo {volume} {788}},\
  \bibinfo {eid} {L8} (\bibinfo {year} {2014})},\ \Eprint
  {http://arxiv.org/abs/1404.0383} {arXiv:1404.0383 [astro-ph.HE]} \BibitemShut
  {NoStop}%
\bibitem [{\citenamefont {{Troja}}\ \emph {et~al.}(2010)\citenamefont
  {{Troja}}, \citenamefont {{Rosswog}},\ and\ \citenamefont
  {{Gehrels}}}]{Troja2010}%
  \BibitemOpen
  \bibfield  {author} {\bibinfo {author} {\bibfnamefont {E.}~\bibnamefont
  {{Troja}}}, \bibinfo {author} {\bibfnamefont {S.}~\bibnamefont {{Rosswog}}},
  \ and\ \bibinfo {author} {\bibfnamefont {N.}~\bibnamefont {{Gehrels}}},\
  }\href {\doibase 10.1088/0004-637X/723/2/1711} {\bibfield  {journal}
  {\bibinfo  {journal} {Astrophys. J.}\ }\textbf {\bibinfo {volume} {723}},\
  \bibinfo {pages} {1711} (\bibinfo {year} {2010})},\ \Eprint
  {http://arxiv.org/abs/1009.1385} {arXiv:1009.1385 [astro-ph.HE]} \BibitemShut
  {NoStop}%
\bibitem [{\citenamefont {{Piran}}(2005)}]{Piran:2004ba}%
  \BibitemOpen
  \bibfield  {author} {\bibinfo {author} {\bibfnamefont {T.}~\bibnamefont
  {{Piran}}},\ }\href {\doibase 10.1103/RevModPhys.76.1143} {\bibfield
  {journal} {\bibinfo  {journal} {Reviews of Modern Physics}\ }\textbf
  {\bibinfo {volume} {76}},\ \bibinfo {pages} {1143} (\bibinfo {year}
  {2005})},\ \Eprint {http://arxiv.org/abs/astro-ph/0405503} {astro-ph/0405503}
  \BibitemShut {NoStop}%
\bibitem [{\citenamefont {{Just}}\ \emph {et~al.}(2016)\citenamefont {{Just}},
  \citenamefont {{Obergaulinger}}, \citenamefont {{Janka}}, \citenamefont
  {{Bauswein}},\ and\ \citenamefont {{Schwarz}}}]{Just2016}%
  \BibitemOpen
  \bibfield  {author} {\bibinfo {author} {\bibfnamefont {O.}~\bibnamefont
  {{Just}}}, \bibinfo {author} {\bibfnamefont {M.}~\bibnamefont
  {{Obergaulinger}}}, \bibinfo {author} {\bibfnamefont {H.-T.}\ \bibnamefont
  {{Janka}}}, \bibinfo {author} {\bibfnamefont {A.}~\bibnamefont {{Bauswein}}},
  \ and\ \bibinfo {author} {\bibfnamefont {N.}~\bibnamefont {{Schwarz}}},\
  }\href {\doibase 10.3847/2041-8205/816/2/L30} {\bibfield  {journal} {\bibinfo
   {journal} {Astrophys. J. Lett.}\ }\textbf {\bibinfo {volume} {816}},\
  \bibinfo {eid} {L30} (\bibinfo {year} {2016})},\ \Eprint
  {http://arxiv.org/abs/1510.04288} {arXiv:1510.04288 [astro-ph.HE]}
  \BibitemShut {NoStop}%
\bibitem [{\citenamefont {{Shibata}}\ \emph
  {et~al.}(2014{\natexlab{b}})\citenamefont {{Shibata}}, \citenamefont
  {{Nagakura}}, \citenamefont {{Sekiguchi}},\ and\ \citenamefont
  {{Yamada}}}]{Shibata2014a}%
  \BibitemOpen
  \bibfield  {author} {\bibinfo {author} {\bibfnamefont {M.}~\bibnamefont
  {{Shibata}}}, \bibinfo {author} {\bibfnamefont {H.}~\bibnamefont
  {{Nagakura}}}, \bibinfo {author} {\bibfnamefont {Y.}~\bibnamefont
  {{Sekiguchi}}}, \ and\ \bibinfo {author} {\bibfnamefont {S.}~\bibnamefont
  {{Yamada}}},\ }\href {\doibase 10.1103/PhysRevD.89.084073} {\bibfield
  {journal} {\bibinfo  {journal} {Phys. Rev. D}\ }\textbf {\bibinfo {volume}
  {89}},\ \bibinfo {eid} {084073} (\bibinfo {year}
  {2014}{\natexlab{b}})}\BibitemShut {NoStop}%
\bibitem [{\citenamefont {{Abdikamalov}}\ \emph {et~al.}(2012)\citenamefont
  {{Abdikamalov}}, \citenamefont {{Burrows}}, \citenamefont {{Ott}},
  \citenamefont {{L{\"o}ffler}}, \citenamefont {{O'Connor}}, \citenamefont
  {{Dolence}},\ and\ \citenamefont {{Schnetter}}}]{Abdikamalov12}%
  \BibitemOpen
  \bibfield  {author} {\bibinfo {author} {\bibfnamefont {E.}~\bibnamefont
  {{Abdikamalov}}}, \bibinfo {author} {\bibfnamefont {A.}~\bibnamefont
  {{Burrows}}}, \bibinfo {author} {\bibfnamefont {C.~D.}\ \bibnamefont
  {{Ott}}}, \bibinfo {author} {\bibfnamefont {F.}~\bibnamefont
  {{L{\"o}ffler}}}, \bibinfo {author} {\bibfnamefont {E.}~\bibnamefont
  {{O'Connor}}}, \bibinfo {author} {\bibfnamefont {J.~C.}\ \bibnamefont
  {{Dolence}}}, \ and\ \bibinfo {author} {\bibfnamefont {E.}~\bibnamefont
  {{Schnetter}}},\ }\href {\doibase 10.1088/0004-637X/755/2/111} {\bibfield
  {journal} {\bibinfo  {journal} {Astrophys. J.}\ }\textbf {\bibinfo {volume}
  {755}},\ \bibinfo {eid} {111} (\bibinfo {year} {2012})},\ \Eprint
  {http://arxiv.org/abs/1203.2915} {arXiv:1203.2915 [astro-ph.SR]} \BibitemShut
  {NoStop}%
\bibitem [{\citenamefont {{van Riper}}\ and\ \citenamefont
  {{Lattimer}}(1981)}]{vanRiper1981}%
  \BibitemOpen
  \bibfield  {author} {\bibinfo {author} {\bibfnamefont {K.~A.}\ \bibnamefont
  {{van Riper}}}\ and\ \bibinfo {author} {\bibfnamefont {J.~M.}\ \bibnamefont
  {{Lattimer}}},\ }\href {\doibase 10.1086/159285} {\bibfield  {journal}
  {\bibinfo  {journal} {Astrophys. J.}\ }\textbf {\bibinfo {volume} {249}},\
  \bibinfo {pages} {270} (\bibinfo {year} {1981})}\BibitemShut {NoStop}%
\bibitem [{\citenamefont {{Ruffert}}\ \emph {et~al.}(1996)\citenamefont
  {{Ruffert}}, \citenamefont {{Janka}},\ and\ \citenamefont
  {{Schaefer}}}]{Ruffert96b}%
  \BibitemOpen
  \bibfield  {author} {\bibinfo {author} {\bibfnamefont {M.}~\bibnamefont
  {{Ruffert}}}, \bibinfo {author} {\bibfnamefont {H.-T.}\ \bibnamefont
  {{Janka}}}, \ and\ \bibinfo {author} {\bibfnamefont {G.}~\bibnamefont
  {{Schaefer}}},\ }\href@noop {} {\bibfield  {journal} {\bibinfo  {journal}
  {Astron. Astrophys.}\ }\textbf {\bibinfo {volume} {311}},\ \bibinfo {pages}
  {532} (\bibinfo {year} {1996})},\ \Eprint
  {http://arxiv.org/abs/astro-ph/9509006} {astro-ph/9509006} \BibitemShut
  {NoStop}%
\bibitem [{\citenamefont {{Rosswog}}\ and\ \citenamefont
  {{Liebend{\"o}rfer}}(2003)}]{Rosswog:2003b}%
  \BibitemOpen
  \bibfield  {author} {\bibinfo {author} {\bibfnamefont {S.}~\bibnamefont
  {{Rosswog}}}\ and\ \bibinfo {author} {\bibfnamefont {M.}~\bibnamefont
  {{Liebend{\"o}rfer}}},\ }\href {\doibase 10.1046/j.1365-8711.2003.06579.x}
  {\bibfield  {journal} {\bibinfo  {journal} {Mon. Not. R. Astron. Soc.}\
  }\textbf {\bibinfo {volume} {342}},\ \bibinfo {pages} {673} (\bibinfo {year}
  {2003})},\ \Eprint {http://arxiv.org/abs/arXiv:astro-ph/0302301}
  {arXiv:astro-ph/0302301} \BibitemShut {NoStop}%
\bibitem [{\citenamefont {{O'Connor}}\ and\ \citenamefont
  {{Ott}}(2010)}]{OConnor10}%
  \BibitemOpen
  \bibfield  {author} {\bibinfo {author} {\bibfnamefont {E.}~\bibnamefont
  {{O'Connor}}}\ and\ \bibinfo {author} {\bibfnamefont {C.~D.}\ \bibnamefont
  {{Ott}}},\ }\href {\doibase 10.1088/0264-9381/27/11/114103} {\bibfield
  {journal} {\bibinfo  {journal} {Class. Quantum Grav.}\ }\textbf {\bibinfo
  {volume} {27}},\ \bibinfo {pages} {114103} (\bibinfo {year} {2010})},\
  \Eprint {http://arxiv.org/abs/0912.2393} {arXiv:0912.2393 [astro-ph.HE]}
  \BibitemShut {NoStop}%
\bibitem [{\citenamefont {{Galeazzi}}\ \emph {et~al.}(2013)\citenamefont
  {{Galeazzi}}, \citenamefont {{Kastaun}}, \citenamefont {{Rezzolla}},\ and\
  \citenamefont {{Font}}}]{Galeazzi2013}%
  \BibitemOpen
  \bibfield  {author} {\bibinfo {author} {\bibfnamefont {F.}~\bibnamefont
  {{Galeazzi}}}, \bibinfo {author} {\bibfnamefont {W.}~\bibnamefont
  {{Kastaun}}}, \bibinfo {author} {\bibfnamefont {L.}~\bibnamefont
  {{Rezzolla}}}, \ and\ \bibinfo {author} {\bibfnamefont {J.~A.}\ \bibnamefont
  {{Font}}},\ }\href {\doibase 10.1103/PhysRevD.88.064009} {\bibfield
  {journal} {\bibinfo  {journal} {Phys. Rev. D}\ }\textbf {\bibinfo {volume}
  {88}},\ \bibinfo {eid} {064009} (\bibinfo {year} {2013})},\ \Eprint
  {http://arxiv.org/abs/1306.4953} {arXiv:1306.4953 [gr-qc]} \BibitemShut
  {NoStop}%
\bibitem [{\citenamefont {{Thorne}}(1981)}]{Thorne1981}%
  \BibitemOpen
  \bibfield  {author} {\bibinfo {author} {\bibfnamefont {K.~S.}\ \bibnamefont
  {{Thorne}}},\ }\href {\doibase 10.1093/mnras/194.2.439} {\bibfield  {journal}
  {\bibinfo  {journal} {Mon. Not. R. Astron. Soc.}\ }\textbf {\bibinfo {volume}
  {194}},\ \bibinfo {pages} {439} (\bibinfo {year} {1981})}\BibitemShut
  {NoStop}%
\bibitem [{\citenamefont {{Rezzolla}}\ and\ \citenamefont
  {{Miller}}(1994)}]{Rezzolla1994}%
  \BibitemOpen
  \bibfield  {author} {\bibinfo {author} {\bibfnamefont {L.}~\bibnamefont
  {{Rezzolla}}}\ and\ \bibinfo {author} {\bibfnamefont {J.~C.}\ \bibnamefont
  {{Miller}}},\ }\href {\doibase 10.1088/0264-9381/11/7/018} {\bibfield
  {journal} {\bibinfo  {journal} {Class. Quantum Grav.}\ }\textbf {\bibinfo
  {volume} {11}},\ \bibinfo {pages} {1815} (\bibinfo {year} {1994})},\ \Eprint
  {http://arxiv.org/abs/arXiv:astro-ph/9406055} {arXiv:astro-ph/9406055}
  \BibitemShut {NoStop}%
\bibitem [{\citenamefont {{Shibata}}\ \emph
  {et~al.}(2011{\natexlab{a}})\citenamefont {{Shibata}}, \citenamefont
  {{Kiuchi}}, \citenamefont {{Sekiguchi}},\ and\ \citenamefont
  {{Suwa}}}]{Shibata2011}%
  \BibitemOpen
  \bibfield  {author} {\bibinfo {author} {\bibfnamefont {M.}~\bibnamefont
  {{Shibata}}}, \bibinfo {author} {\bibfnamefont {K.}~\bibnamefont {{Kiuchi}}},
  \bibinfo {author} {\bibfnamefont {Y.}~\bibnamefont {{Sekiguchi}}}, \ and\
  \bibinfo {author} {\bibfnamefont {Y.}~\bibnamefont {{Suwa}}},\ }\href@noop {}
  {\bibfield  {journal} {\bibinfo  {journal} {Progress of Theoretical Physics}\
  }\textbf {\bibinfo {volume} {125}},\ \bibinfo {pages} {1255} (\bibinfo {year}
  {2011}{\natexlab{a}})},\ \Eprint {http://arxiv.org/abs/1104.3937}
  {arXiv:1104.3937 [astro-ph.HE]} \BibitemShut {NoStop}%
\bibitem [{\citenamefont {{Shibata}}\ and\ \citenamefont
  {{Sekiguchi}}(2012)}]{Shibata2012}%
  \BibitemOpen
  \bibfield  {author} {\bibinfo {author} {\bibfnamefont {M.}~\bibnamefont
  {{Shibata}}}\ and\ \bibinfo {author} {\bibfnamefont {Y.}~\bibnamefont
  {{Sekiguchi}}},\ }\href@noop {} {\bibfield  {journal} {\bibinfo  {journal}
  {Progress of Theoretical Physics}\ }\textbf {\bibinfo {volume} {127}},\
  \bibinfo {pages} {535} (\bibinfo {year} {2012})},\ \Eprint
  {http://arxiv.org/abs/1206.5911} {arXiv:1206.5911 [astro-ph.HE]} \BibitemShut
  {NoStop}%
\bibitem [{\citenamefont {{Cardall}}\ \emph {et~al.}(2013)\citenamefont
  {{Cardall}}, \citenamefont {{Endeve}},\ and\ \citenamefont
  {{Mezzacappa}}}]{Cardall2013}%
  \BibitemOpen
  \bibfield  {author} {\bibinfo {author} {\bibfnamefont {C.~Y.}\ \bibnamefont
  {{Cardall}}}, \bibinfo {author} {\bibfnamefont {E.}~\bibnamefont {{Endeve}}},
  \ and\ \bibinfo {author} {\bibfnamefont {A.}~\bibnamefont {{Mezzacappa}}},\
  }\href {\doibase 10.1103/PhysRevD.88.023011} {\bibfield  {journal} {\bibinfo
  {journal} {Phys. Rev. D}\ }\textbf {\bibinfo {volume} {88}},\ \bibinfo {eid}
  {023011} (\bibinfo {year} {2013})},\ \Eprint {http://arxiv.org/abs/1305.0037}
  {arXiv:1305.0037 [astro-ph.HE]} \BibitemShut {NoStop}%
\bibitem [{\citenamefont {{Takahashi}}\ \emph {et~al.}(2013)\citenamefont
  {{Takahashi}}, \citenamefont {{Ohsuga}}, \citenamefont {{Sekiguchi}},
  \citenamefont {{Inoue}},\ and\ \citenamefont {{Tomida}}}]{Takahashi2013a}%
  \BibitemOpen
  \bibfield  {author} {\bibinfo {author} {\bibfnamefont {H.~R.}\ \bibnamefont
  {{Takahashi}}}, \bibinfo {author} {\bibfnamefont {K.}~\bibnamefont
  {{Ohsuga}}}, \bibinfo {author} {\bibfnamefont {Y.}~\bibnamefont
  {{Sekiguchi}}}, \bibinfo {author} {\bibfnamefont {T.}~\bibnamefont
  {{Inoue}}}, \ and\ \bibinfo {author} {\bibfnamefont {K.}~\bibnamefont
  {{Tomida}}},\ }\href {\doibase 10.1088/0004-637X/764/2/122} {\bibfield
  {journal} {\bibinfo  {journal} {Astrophys. J.}\ }\textbf {\bibinfo {volume}
  {764}},\ \bibinfo {eid} {122} (\bibinfo {year} {2013})},\ \Eprint
  {http://arxiv.org/abs/1212.4910} {arXiv:1212.4910 [astro-ph.HE]} \BibitemShut
  {NoStop}%
\bibitem [{\citenamefont {{Hotokezaka}}\ \emph
  {et~al.}(2013{\natexlab{d}})\citenamefont {{Hotokezaka}}, \citenamefont
  {{Kyutoku}}, \citenamefont {{Tanaka}}, \citenamefont {{Kiuchi}},
  \citenamefont {{Sekiguchi}}, \citenamefont {{Shibata}},\ and\ \citenamefont
  {{Wanajo}}}]{Hotokezaka2013d}%
  \BibitemOpen
  \bibfield  {author} {\bibinfo {author} {\bibfnamefont {K.}~\bibnamefont
  {{Hotokezaka}}}, \bibinfo {author} {\bibfnamefont {K.}~\bibnamefont
  {{Kyutoku}}}, \bibinfo {author} {\bibfnamefont {M.}~\bibnamefont {{Tanaka}}},
  \bibinfo {author} {\bibfnamefont {K.}~\bibnamefont {{Kiuchi}}}, \bibinfo
  {author} {\bibfnamefont {Y.}~\bibnamefont {{Sekiguchi}}}, \bibinfo {author}
  {\bibfnamefont {M.}~\bibnamefont {{Shibata}}}, \ and\ \bibinfo {author}
  {\bibfnamefont {S.}~\bibnamefont {{Wanajo}}},\ }\href {\doibase
  10.1088/2041-8205/778/1/L16} {\bibfield  {journal} {\bibinfo  {journal}
  {Astrophys. J.}\ }\textbf {\bibinfo {volume} {778}},\ \bibinfo {eid} {L16}
  (\bibinfo {year} {2013}{\natexlab{d}})},\ \Eprint
  {http://arxiv.org/abs/1310.1623} {arXiv:1310.1623 [astro-ph.HE]} \BibitemShut
  {NoStop}%
\bibitem [{\citenamefont {{Wanajo}}\ \emph {et~al.}(2014)\citenamefont
  {{Wanajo}}, \citenamefont {{Sekiguchi}}, \citenamefont {{Nishimura}},
  \citenamefont {{Kiuchi}}, \citenamefont {{Kyutoku}},\ and\ \citenamefont
  {{Shibata}}}]{Wanajo2014}%
  \BibitemOpen
  \bibfield  {author} {\bibinfo {author} {\bibfnamefont {S.}~\bibnamefont
  {{Wanajo}}}, \bibinfo {author} {\bibfnamefont {Y.}~\bibnamefont
  {{Sekiguchi}}}, \bibinfo {author} {\bibfnamefont {N.}~\bibnamefont
  {{Nishimura}}}, \bibinfo {author} {\bibfnamefont {K.}~\bibnamefont
  {{Kiuchi}}}, \bibinfo {author} {\bibfnamefont {K.}~\bibnamefont {{Kyutoku}}},
  \ and\ \bibinfo {author} {\bibfnamefont {M.}~\bibnamefont {{Shibata}}},\
  }\href {\doibase 10.1088/2041-8205/789/2/L39} {\bibfield  {journal} {\bibinfo
   {journal} {Astrophys. J.}\ }\textbf {\bibinfo {volume} {789}},\ \bibinfo
  {eid} {L39} (\bibinfo {year} {2014})},\ \Eprint
  {http://arxiv.org/abs/1402.7317} {arXiv:1402.7317 [astro-ph.SR]} \BibitemShut
  {NoStop}%
\bibitem [{\citenamefont {{Sekiguchi}}\ \emph {et~al.}(2015)\citenamefont
  {{Sekiguchi}}, \citenamefont {{Kiuchi}}, \citenamefont {{Kyutoku}},\ and\
  \citenamefont {{Shibata}}}]{Sekiguchi2015}%
  \BibitemOpen
  \bibfield  {author} {\bibinfo {author} {\bibfnamefont {Y.}~\bibnamefont
  {{Sekiguchi}}}, \bibinfo {author} {\bibfnamefont {K.}~\bibnamefont
  {{Kiuchi}}}, \bibinfo {author} {\bibfnamefont {K.}~\bibnamefont {{Kyutoku}}},
  \ and\ \bibinfo {author} {\bibfnamefont {M.}~\bibnamefont {{Shibata}}},\
  }\href {\doibase 10.1103/PhysRevD.91.064059} {\bibfield  {journal} {\bibinfo
  {journal} {Phys. Rev. D}\ }\textbf {\bibinfo {volume} {91}},\ \bibinfo {eid}
  {064059} (\bibinfo {year} {2015})},\ \Eprint
  {http://arxiv.org/abs/1502.06660} {arXiv:1502.06660 [astro-ph.HE]}
  \BibitemShut {NoStop}%
\bibitem [{\citenamefont {{Sekiguchi}}(2010)}]{Sekiguchi2010}%
  \BibitemOpen
  \bibfield  {author} {\bibinfo {author} {\bibfnamefont {Y.}~\bibnamefont
  {{Sekiguchi}}},\ }\href {\doibase 10.1143/ptp.124.331} {\bibfield  {journal}
  {\bibinfo  {journal} {Progress of Theoretical Physics}\ }\textbf {\bibinfo
  {volume} {124}},\ \bibinfo {pages} {331} (\bibinfo {year} {2010})},\ \Eprint
  {http://arxiv.org/abs/1009.3320} {arXiv:1009.3320 [astro-ph.HE]} \BibitemShut
  {NoStop}%
\bibitem [{\citenamefont {{Sekiguchi}}\ and\ \citenamefont
  {{Shibata}}(2011)}]{Sekiguchi2011a}%
  \BibitemOpen
  \bibfield  {author} {\bibinfo {author} {\bibfnamefont {Y.}~\bibnamefont
  {{Sekiguchi}}}\ and\ \bibinfo {author} {\bibfnamefont {M.}~\bibnamefont
  {{Shibata}}},\ }\href {\doibase 10.1088/0004-637X/737/1/6} {\bibfield
  {journal} {\bibinfo  {journal} {Astrophys. J.}\ }\textbf {\bibinfo {volume}
  {737}},\ \bibinfo {eid} {6} (\bibinfo {year} {2011})},\ \Eprint
  {http://arxiv.org/abs/1009.5303} {arXiv:1009.5303 [astro-ph.HE]} \BibitemShut
  {NoStop}%
\bibitem [{\citenamefont {{Foucart}}\ \emph {et~al.}(2015)\citenamefont
  {{Foucart}}, \citenamefont {{O'Connor}}, \citenamefont {{Roberts}},
  \citenamefont {{Duez}}, \citenamefont {{Haas}}, \citenamefont {{Kidder}},
  \citenamefont {{Ott}}, \citenamefont {{Pfeiffer}}, \citenamefont {{Scheel}},\
  and\ \citenamefont {{Szilagyi}}}]{Foucart2015a}%
  \BibitemOpen
  \bibfield  {author} {\bibinfo {author} {\bibfnamefont {F.}~\bibnamefont
  {{Foucart}}}, \bibinfo {author} {\bibfnamefont {E.}~\bibnamefont
  {{O'Connor}}}, \bibinfo {author} {\bibfnamefont {L.}~\bibnamefont
  {{Roberts}}}, \bibinfo {author} {\bibfnamefont {M.~D.}\ \bibnamefont
  {{Duez}}}, \bibinfo {author} {\bibfnamefont {R.}~\bibnamefont {{Haas}}},
  \bibinfo {author} {\bibfnamefont {L.~E.}\ \bibnamefont {{Kidder}}}, \bibinfo
  {author} {\bibfnamefont {C.~D.}\ \bibnamefont {{Ott}}}, \bibinfo {author}
  {\bibfnamefont {H.~P.}\ \bibnamefont {{Pfeiffer}}}, \bibinfo {author}
  {\bibfnamefont {M.~A.}\ \bibnamefont {{Scheel}}}, \ and\ \bibinfo {author}
  {\bibfnamefont {B.}~\bibnamefont {{Szilagyi}}},\ }\href {\doibase
  10.1103/PhysRevD.91.124021} {\bibfield  {journal} {\bibinfo  {journal} {Phys.
  Rev. D}\ }\textbf {\bibinfo {volume} {91}},\ \bibinfo {eid} {124021}
  (\bibinfo {year} {2015})},\ \Eprint {http://arxiv.org/abs/1502.04146}
  {arXiv:1502.04146 [astro-ph.HE]} \BibitemShut {NoStop}%
\bibitem [{\citenamefont {McClarren}\ and\ \citenamefont
  {Hauck}(2010)}]{McClarren10}%
  \BibitemOpen
  \bibfield  {author} {\bibinfo {author} {\bibfnamefont {R.~G.}\ \bibnamefont
  {McClarren}}\ and\ \bibinfo {author} {\bibfnamefont {C.~D.}\ \bibnamefont
  {Hauck}},\ }\href {\doibase 10.1016/j.jcp.2010.03.043} {\bibfield  {journal}
  {\bibinfo  {journal} {J. Comput. Phys.}\ }\textbf {\bibinfo {volume} {229}},\
  \bibinfo {pages} {5597} (\bibinfo {year} {2010})}\BibitemShut {NoStop}%
\bibitem [{\citenamefont {{Mezzacappa}}\ and\ \citenamefont
  {{Bruenn}}(1993)}]{Mezzacappa1993}%
  \BibitemOpen
  \bibfield  {author} {\bibinfo {author} {\bibfnamefont {A.}~\bibnamefont
  {{Mezzacappa}}}\ and\ \bibinfo {author} {\bibfnamefont {S.~W.}\ \bibnamefont
  {{Bruenn}}},\ }\href {\doibase 10.1086/172394} {\bibfield  {journal}
  {\bibinfo  {journal} {Astrophys. J.}\ }\textbf {\bibinfo {volume} {405}},\
  \bibinfo {pages} {637} (\bibinfo {year} {1993})}\BibitemShut {NoStop}%
\bibitem [{\citenamefont {{Scheck}}\ \emph {et~al.}(2006)\citenamefont
  {{Scheck}}, \citenamefont {{Kifonidis}}, \citenamefont {{Janka}},\ and\
  \citenamefont {{M{\"u}ller}}}]{Scheck2006}%
  \BibitemOpen
  \bibfield  {author} {\bibinfo {author} {\bibfnamefont {L.}~\bibnamefont
  {{Scheck}}}, \bibinfo {author} {\bibfnamefont {K.}~\bibnamefont
  {{Kifonidis}}}, \bibinfo {author} {\bibfnamefont {H.-T.}\ \bibnamefont
  {{Janka}}}, \ and\ \bibinfo {author} {\bibfnamefont {E.}~\bibnamefont
  {{M{\"u}ller}}},\ }\href {\doibase 10.1051/0004-6361:20064855} {\bibfield
  {journal} {\bibinfo  {journal} {Astron. Astrophys.}\ }\textbf {\bibinfo
  {volume} {457}},\ \bibinfo {pages} {963} (\bibinfo {year} {2006})},\ \Eprint
  {http://arxiv.org/abs/arXiv:astro-ph/0601302} {arXiv:astro-ph/0601302}
  \BibitemShut {NoStop}%
\bibitem [{\citenamefont {{Ott}}\ \emph {et~al.}(2008)\citenamefont {{Ott}},
  \citenamefont {{Burrows}}, \citenamefont {{Dessart}},\ and\ \citenamefont
  {{Livne}}}]{Ott08}%
  \BibitemOpen
  \bibfield  {author} {\bibinfo {author} {\bibfnamefont {C.~D.}\ \bibnamefont
  {{Ott}}}, \bibinfo {author} {\bibfnamefont {A.}~\bibnamefont {{Burrows}}},
  \bibinfo {author} {\bibfnamefont {L.}~\bibnamefont {{Dessart}}}, \ and\
  \bibinfo {author} {\bibfnamefont {E.}~\bibnamefont {{Livne}}},\ }\href
  {\doibase 10.1086/591440} {\bibfield  {journal} {\bibinfo  {journal}
  {Astrophys. J.}\ }\textbf {\bibinfo {volume} {685}},\ \bibinfo {pages} {1069}
  (\bibinfo {year} {2008})},\ \Eprint {http://arxiv.org/abs/0804.0239}
  {arXiv:0804.0239} \BibitemShut {NoStop}%
\bibitem [{\citenamefont {{Liebend{\"o}rfer}}\ \emph
  {et~al.}(2009)\citenamefont {{Liebend{\"o}rfer}}, \citenamefont
  {{Whitehouse}},\ and\ \citenamefont {{Fischer}}}]{Liebendoerfer2009}%
  \BibitemOpen
  \bibfield  {author} {\bibinfo {author} {\bibfnamefont {M.}~\bibnamefont
  {{Liebend{\"o}rfer}}}, \bibinfo {author} {\bibfnamefont {S.~C.}\ \bibnamefont
  {{Whitehouse}}}, \ and\ \bibinfo {author} {\bibfnamefont {T.}~\bibnamefont
  {{Fischer}}},\ }\href {\doibase 10.1088/0004-637X/698/2/1174} {\bibfield
  {journal} {\bibinfo  {journal} {Astrophys. J.}\ }\textbf {\bibinfo {volume}
  {698}},\ \bibinfo {pages} {1174} (\bibinfo {year} {2009})},\ \Eprint
  {http://arxiv.org/abs/0711.2929} {arXiv:0711.2929} \BibitemShut {NoStop}%
\bibitem [{\citenamefont {{Sumiyoshi}}\ and\ \citenamefont
  {{Yamada}}(2012)}]{Sumiyoshi:12}%
  \BibitemOpen
  \bibfield  {author} {\bibinfo {author} {\bibfnamefont {K.}~\bibnamefont
  {{Sumiyoshi}}}\ and\ \bibinfo {author} {\bibfnamefont {S.}~\bibnamefont
  {{Yamada}}},\ }\href {\doibase 10.1088/0067-0049/199/1/17} {\bibfield
  {journal} {\bibinfo  {journal} {Astrophys. J., Supp.}\ }\textbf {\bibinfo
  {volume} {199}},\ \bibinfo {eid} {17} (\bibinfo {year} {2012})},\ \Eprint
  {http://arxiv.org/abs/1201.2244} {arXiv:1201.2244 [astro-ph.HE]} \BibitemShut
  {NoStop}%
\bibitem [{\citenamefont {{Ruffert}}\ \emph {et~al.}(1997)\citenamefont
  {{Ruffert}}, \citenamefont {{Janka}}, \citenamefont {{Takahashi}},\ and\
  \citenamefont {{Schaefer}}}]{Ruffert97}%
  \BibitemOpen
  \bibfield  {author} {\bibinfo {author} {\bibfnamefont {M.}~\bibnamefont
  {{Ruffert}}}, \bibinfo {author} {\bibfnamefont {H.-T.}\ \bibnamefont
  {{Janka}}}, \bibinfo {author} {\bibfnamefont {K.}~\bibnamefont
  {{Takahashi}}}, \ and\ \bibinfo {author} {\bibfnamefont {G.}~\bibnamefont
  {{Schaefer}}},\ }\href@noop {} {\bibfield  {journal} {\bibinfo  {journal}
  {Astron. Astrophys.}\ }\textbf {\bibinfo {volume} {319}},\ \bibinfo {pages}
  {122} (\bibinfo {year} {1997})},\ \Eprint
  {http://arxiv.org/abs/astro-ph/9606181} {astro-ph/9606181} \BibitemShut
  {NoStop}%
\bibitem [{\citenamefont {{Ruffert}}\ and\ \citenamefont
  {{Janka}}(2001)}]{Ruffert01}%
  \BibitemOpen
  \bibfield  {author} {\bibinfo {author} {\bibfnamefont {M.}~\bibnamefont
  {{Ruffert}}}\ and\ \bibinfo {author} {\bibfnamefont {H.-T.}\ \bibnamefont
  {{Janka}}},\ }\href {\doibase 10.1051/0004-6361:20011453} {\bibfield
  {journal} {\bibinfo  {journal} {Astron. Astrophys.}\ }\textbf {\bibinfo
  {volume} {380}},\ \bibinfo {pages} {544} (\bibinfo {year} {2001})},\ \Eprint
  {http://arxiv.org/abs/astro-ph/0106229} {astro-ph/0106229} \BibitemShut
  {NoStop}%
\bibitem [{\citenamefont {{Rosswog}}\ \emph {et~al.}(2003)\citenamefont
  {{Rosswog}}, \citenamefont {{Ramirez-Ruiz}},\ and\ \citenamefont
  {{Davies}}}]{Rosswog:2003}%
  \BibitemOpen
  \bibfield  {author} {\bibinfo {author} {\bibfnamefont {S.}~\bibnamefont
  {{Rosswog}}}, \bibinfo {author} {\bibfnamefont {E.}~\bibnamefont
  {{Ramirez-Ruiz}}}, \ and\ \bibinfo {author} {\bibfnamefont {M.~B.}\
  \bibnamefont {{Davies}}},\ }\href {\doibase 10.1046/j.1365-2966.2003.07032.x}
  {\bibfield  {journal} {\bibinfo  {journal} {Mon. Not. R. Astron. Soc.}\
  }\textbf {\bibinfo {volume} {345}},\ \bibinfo {pages} {1077} (\bibinfo {year}
  {2003})},\ \Eprint {http://arxiv.org/abs/arXiv:astro-ph/0110180}
  {arXiv:astro-ph/0110180} \BibitemShut {NoStop}%
\bibitem [{\citenamefont {{Sekiguchi}}\ \emph
  {et~al.}(2011{\natexlab{b}})\citenamefont {{Sekiguchi}}, \citenamefont
  {{Kiuchi}}, \citenamefont {{Kyutoku}},\ and\ \citenamefont
  {{Shibata}}}]{Sekiguchi2011}%
  \BibitemOpen
  \bibfield  {author} {\bibinfo {author} {\bibfnamefont {Y.}~\bibnamefont
  {{Sekiguchi}}}, \bibinfo {author} {\bibfnamefont {K.}~\bibnamefont
  {{Kiuchi}}}, \bibinfo {author} {\bibfnamefont {K.}~\bibnamefont {{Kyutoku}}},
  \ and\ \bibinfo {author} {\bibfnamefont {M.}~\bibnamefont {{Shibata}}},\
  }\href {\doibase 10.1103/PhysRevLett.107.051102} {\bibfield  {journal}
  {\bibinfo  {journal} {Phys. Rev. Lett.}\ }\textbf {\bibinfo {volume} {107}},\
  \bibinfo {eid} {051102} (\bibinfo {year} {2011}{\natexlab{b}})},\ \Eprint
  {http://arxiv.org/abs/1105.2125} {arXiv:1105.2125 [gr-qc]} \BibitemShut
  {NoStop}%
\bibitem [{\citenamefont {{Korobkin}}\ \emph {et~al.}(2012)\citenamefont
  {{Korobkin}}, \citenamefont {{Rosswog}}, \citenamefont {{Arcones}},\ and\
  \citenamefont {{Winteler}}}]{Korobkin2012}%
  \BibitemOpen
  \bibfield  {author} {\bibinfo {author} {\bibfnamefont {O.}~\bibnamefont
  {{Korobkin}}}, \bibinfo {author} {\bibfnamefont {S.}~\bibnamefont
  {{Rosswog}}}, \bibinfo {author} {\bibfnamefont {A.}~\bibnamefont
  {{Arcones}}}, \ and\ \bibinfo {author} {\bibfnamefont {C.}~\bibnamefont
  {{Winteler}}},\ }\href {\doibase 10.1111/j.1365-2966.2012.21859.x} {\bibfield
   {journal} {\bibinfo  {journal} {Mon. Not. R. Astron. Soc.}\ }\textbf
  {\bibinfo {volume} {426}},\ \bibinfo {pages} {1940} (\bibinfo {year}
  {2012})},\ \Eprint {http://arxiv.org/abs/1206.2379} {arXiv:1206.2379
  [astro-ph.SR]} \BibitemShut {NoStop}%
\bibitem [{\citenamefont {{Rosswog}}\ \emph {et~al.}(2014)\citenamefont
  {{Rosswog}}, \citenamefont {{Korobkin}}, \citenamefont {{Arcones}},
  \citenamefont {{Thielemann}},\ and\ \citenamefont {{Piran}}}]{Rosswog2014a}%
  \BibitemOpen
  \bibfield  {author} {\bibinfo {author} {\bibfnamefont {S.}~\bibnamefont
  {{Rosswog}}}, \bibinfo {author} {\bibfnamefont {O.}~\bibnamefont
  {{Korobkin}}}, \bibinfo {author} {\bibfnamefont {A.}~\bibnamefont
  {{Arcones}}}, \bibinfo {author} {\bibfnamefont {F.-K.}\ \bibnamefont
  {{Thielemann}}}, \ and\ \bibinfo {author} {\bibfnamefont {T.}~\bibnamefont
  {{Piran}}},\ }\href {\doibase 10.1093/mnras/stt2502} {\bibfield  {journal}
  {\bibinfo  {journal} {Mon. Not. R. Astron. Soc.}\ }\textbf {\bibinfo {volume}
  {439}},\ \bibinfo {pages} {744} (\bibinfo {year} {2014})},\ \Eprint
  {http://arxiv.org/abs/1307.2939} {arXiv:1307.2939 [astro-ph.HE]} \BibitemShut
  {NoStop}%
\bibitem [{\citenamefont {{Oechslin}}\ \emph {et~al.}(2007)\citenamefont
  {{Oechslin}}, \citenamefont {{Janka}},\ and\ \citenamefont
  {{Marek}}}]{Oechslin07a}%
  \BibitemOpen
  \bibfield  {author} {\bibinfo {author} {\bibfnamefont {R.}~\bibnamefont
  {{Oechslin}}}, \bibinfo {author} {\bibfnamefont {H.-T.}\ \bibnamefont
  {{Janka}}}, \ and\ \bibinfo {author} {\bibfnamefont {A.}~\bibnamefont
  {{Marek}}},\ }\href {\doibase 10.1051/0004-6361:20066682} {\bibfield
  {journal} {\bibinfo  {journal} {Astron. Astrophys.}\ }\textbf {\bibinfo
  {volume} {467}},\ \bibinfo {pages} {395} (\bibinfo {year} {2007})},\ \Eprint
  {http://arxiv.org/abs/astro-ph/0611047} {astro-ph/0611047} \BibitemShut
  {NoStop}%
\bibitem [{\citenamefont {{Bauswein}}\ \emph
  {et~al.}(2013{\natexlab{b}})\citenamefont {{Bauswein}}, \citenamefont
  {{Goriely}},\ and\ \citenamefont {{Janka}}}]{Bauswein2013b}%
  \BibitemOpen
  \bibfield  {author} {\bibinfo {author} {\bibfnamefont {A.}~\bibnamefont
  {{Bauswein}}}, \bibinfo {author} {\bibfnamefont {S.}~\bibnamefont
  {{Goriely}}}, \ and\ \bibinfo {author} {\bibfnamefont {H.-T.}\ \bibnamefont
  {{Janka}}},\ }\href {\doibase 10.1088/0004-637X/773/1/78} {\bibfield
  {journal} {\bibinfo  {journal} {Astrophys. J.}\ }\textbf {\bibinfo {volume}
  {773}},\ \bibinfo {eid} {78} (\bibinfo {year} {2013}{\natexlab{b}})},\
  \Eprint {http://arxiv.org/abs/1302.6530} {arXiv:1302.6530 [astro-ph.SR]}
  \BibitemShut {NoStop}%
\bibitem [{\citenamefont {{Arnould}}\ \emph {et~al.}(2007)\citenamefont
  {{Arnould}}, \citenamefont {{Goriely}},\ and\ \citenamefont
  {{Takahashi}}}]{Arnould2007}%
  \BibitemOpen
  \bibfield  {author} {\bibinfo {author} {\bibfnamefont {M.}~\bibnamefont
  {{Arnould}}}, \bibinfo {author} {\bibfnamefont {S.}~\bibnamefont
  {{Goriely}}}, \ and\ \bibinfo {author} {\bibfnamefont {K.}~\bibnamefont
  {{Takahashi}}},\ }\href {\doibase 10.1016/j.physrep.2007.06.002} {\bibfield
  {journal} {\bibinfo  {journal} {Physics Reports}\ }\textbf {\bibinfo {volume}
  {450}},\ \bibinfo {pages} {97} (\bibinfo {year} {2007})},\ \Eprint
  {http://arxiv.org/abs/0705.4512} {arXiv:0705.4512} \BibitemShut {NoStop}%
\bibitem [{\citenamefont {Rosswog}(2005)}]{Rosswog05}%
  \BibitemOpen
  \bibfield  {author} {\bibinfo {author} {\bibfnamefont {S.}~\bibnamefont
  {Rosswog}},\ }\href {\doibase 10.1086/497062} {\bibfield  {journal} {\bibinfo
   {journal} {Astrophys. J.}\ }\textbf {\bibinfo {volume} {634}},\ \bibinfo
  {pages} {1202} (\bibinfo {year} {2005})},\ \Eprint
  {http://arxiv.org/abs/astro-ph/0508138} {astro-ph/0508138} \BibitemShut
  {NoStop}%
\bibitem [{\citenamefont {{Roberts}}\ \emph {et~al.}(2011)\citenamefont
  {{Roberts}}, \citenamefont {{Kasen}}, \citenamefont {{Lee}},\ and\
  \citenamefont {{Ramirez-Ruiz}}}]{Roberts2011}%
  \BibitemOpen
  \bibfield  {author} {\bibinfo {author} {\bibfnamefont {L.~F.}\ \bibnamefont
  {{Roberts}}}, \bibinfo {author} {\bibfnamefont {D.}~\bibnamefont {{Kasen}}},
  \bibinfo {author} {\bibfnamefont {W.~H.}\ \bibnamefont {{Lee}}}, \ and\
  \bibinfo {author} {\bibfnamefont {E.}~\bibnamefont {{Ramirez-Ruiz}}},\ }\href
  {\doibase 10.1088/2041-8205/736/1/L21} {\bibfield  {journal} {\bibinfo
  {journal} {Astrophys. J. Lett.}\ }\textbf {\bibinfo {volume} {736}},\
  \bibinfo {eid} {L21} (\bibinfo {year} {2011})},\ \Eprint
  {http://arxiv.org/abs/1104.5504} {arXiv:1104.5504 [astro-ph.HE]} \BibitemShut
  {NoStop}%
\bibitem [{\citenamefont {{Rosswog}}(2013)}]{Rosswog2013a}%
  \BibitemOpen
  \bibfield  {author} {\bibinfo {author} {\bibfnamefont {S.}~\bibnamefont
  {{Rosswog}}},\ }\href {\doibase 10.1098/rsta.2012.0272} {\bibfield  {journal}
  {\bibinfo  {journal} {Royal Society of London Philosophical Transactions
  Series A}\ }\textbf {\bibinfo {volume} {371}},\ \bibinfo {pages} {20272}
  (\bibinfo {year} {2013})},\ \Eprint {http://arxiv.org/abs/1210.6549}
  {arXiv:1210.6549 [astro-ph.HE]} \BibitemShut {NoStop}%
\bibitem [{\citenamefont {{Kyutoku}}\ \emph
  {et~al.}(2014{\natexlab{b}})\citenamefont {{Kyutoku}}, \citenamefont
  {{Ioka}},\ and\ \citenamefont {{Shibata}}}]{Kyutoku2012}%
  \BibitemOpen
  \bibfield  {author} {\bibinfo {author} {\bibfnamefont {K.}~\bibnamefont
  {{Kyutoku}}}, \bibinfo {author} {\bibfnamefont {K.}~\bibnamefont {{Ioka}}}, \
  and\ \bibinfo {author} {\bibfnamefont {M.}~\bibnamefont {{Shibata}}},\ }\href
  {\doibase 10.1093/mnrasl/slt128} {\bibfield  {journal} {\bibinfo  {journal}
  {Mon. Not. R. Astron.Soc.}\ }\textbf {\bibinfo {volume} {437}},\ \bibinfo
  {pages} {L6} (\bibinfo {year} {2014}{\natexlab{b}})},\ \Eprint
  {http://arxiv.org/abs/1209.5747} {arXiv:1209.5747 [astro-ph.HE]} \BibitemShut
  {NoStop}%
\bibitem [{\citenamefont {{Dessart}}\ \emph {et~al.}(2009)\citenamefont
  {{Dessart}}, \citenamefont {{Ott}}, \citenamefont {{Burrows}}, \citenamefont
  {{Rosswog}},\ and\ \citenamefont {{Livne}}}]{Dessart2009}%
  \BibitemOpen
  \bibfield  {author} {\bibinfo {author} {\bibfnamefont {L.}~\bibnamefont
  {{Dessart}}}, \bibinfo {author} {\bibfnamefont {C.~D.}\ \bibnamefont
  {{Ott}}}, \bibinfo {author} {\bibfnamefont {A.}~\bibnamefont {{Burrows}}},
  \bibinfo {author} {\bibfnamefont {S.}~\bibnamefont {{Rosswog}}}, \ and\
  \bibinfo {author} {\bibfnamefont {E.}~\bibnamefont {{Livne}}},\ }\href
  {\doibase 10.1088/0004-637X/690/2/1681} {\bibfield  {journal} {\bibinfo
  {journal} {Astrophys. J.}\ }\textbf {\bibinfo {volume} {690}},\ \bibinfo
  {pages} {1681} (\bibinfo {year} {2009})},\ \Eprint
  {http://arxiv.org/abs/0806.4380} {arXiv:0806.4380} \BibitemShut {NoStop}%
\bibitem [{\citenamefont {{Just}}\ \emph {et~al.}(2015)\citenamefont {{Just}},
  \citenamefont {{Bauswein}}, \citenamefont {{Pulpillo}}, \citenamefont
  {{Goriely}},\ and\ \citenamefont {{Janka}}}]{Just2015}%
  \BibitemOpen
  \bibfield  {author} {\bibinfo {author} {\bibfnamefont {O.}~\bibnamefont
  {{Just}}}, \bibinfo {author} {\bibfnamefont {A.}~\bibnamefont {{Bauswein}}},
  \bibinfo {author} {\bibfnamefont {R.~A.}\ \bibnamefont {{Pulpillo}}},
  \bibinfo {author} {\bibfnamefont {S.}~\bibnamefont {{Goriely}}}, \ and\
  \bibinfo {author} {\bibfnamefont {H.-T.}\ \bibnamefont {{Janka}}},\ }\href
  {\doibase 10.1093/mnras/stv009} {\bibfield  {journal} {\bibinfo  {journal}
  {Mon. Not. R. Astron. Soc.}\ }\textbf {\bibinfo {volume} {448}},\ \bibinfo
  {pages} {541} (\bibinfo {year} {2015})},\ \Eprint
  {http://arxiv.org/abs/1406.2687} {arXiv:1406.2687 [astro-ph.SR]} \BibitemShut
  {NoStop}%
\bibitem [{\citenamefont {{Shibata}}\ \emph
  {et~al.}(2011{\natexlab{b}})\citenamefont {{Shibata}}, \citenamefont
  {{Suwa}}, \citenamefont {{Kiuchi}},\ and\ \citenamefont
  {{Ioka}}}]{Shibata2011b}%
  \BibitemOpen
  \bibfield  {author} {\bibinfo {author} {\bibfnamefont {M.}~\bibnamefont
  {{Shibata}}}, \bibinfo {author} {\bibfnamefont {Y.}~\bibnamefont {{Suwa}}},
  \bibinfo {author} {\bibfnamefont {K.}~\bibnamefont {{Kiuchi}}}, \ and\
  \bibinfo {author} {\bibfnamefont {K.}~\bibnamefont {{Ioka}}},\ }\href
  {\doibase 10.1088/2041-8205/734/2/L36} {\bibfield  {journal} {\bibinfo
  {journal} {Astrophys. J.l}\ }\textbf {\bibinfo {volume} {734}},\ \bibinfo
  {eid} {L36} (\bibinfo {year} {2011}{\natexlab{b}})},\ \Eprint
  {http://arxiv.org/abs/1105.3302} {arXiv:1105.3302 [astro-ph.HE]} \BibitemShut
  {NoStop}%
\bibitem [{\citenamefont {{Goriely}}\ \emph {et~al.}(2011)\citenamefont
  {{Goriely}}, \citenamefont {{Bauswein}},\ and\ \citenamefont
  {{Janka}}}]{Goriely2011}%
  \BibitemOpen
  \bibfield  {author} {\bibinfo {author} {\bibfnamefont {S.}~\bibnamefont
  {{Goriely}}}, \bibinfo {author} {\bibfnamefont {A.}~\bibnamefont
  {{Bauswein}}}, \ and\ \bibinfo {author} {\bibfnamefont {H.-T.}\ \bibnamefont
  {{Janka}}},\ }\href {\doibase 10.1088/2041-8205/738/2/L32} {\bibfield
  {journal} {\bibinfo  {journal} {Astrophys. J.}\ }\textbf {\bibinfo {volume}
  {738}},\ \bibinfo {eid} {L32} (\bibinfo {year} {2011})},\ \Eprint
  {http://arxiv.org/abs/1107.0899} {arXiv:1107.0899 [astro-ph.SR]} \BibitemShut
  {NoStop}%
\bibitem [{\citenamefont {{Piran}}\ \emph {et~al.}(2013)\citenamefont
  {{Piran}}, \citenamefont {{Nakar}},\ and\ \citenamefont
  {{Rosswog}}}]{Piran2013}%
  \BibitemOpen
  \bibfield  {author} {\bibinfo {author} {\bibfnamefont {T.}~\bibnamefont
  {{Piran}}}, \bibinfo {author} {\bibfnamefont {E.}~\bibnamefont {{Nakar}}}, \
  and\ \bibinfo {author} {\bibfnamefont {S.}~\bibnamefont {{Rosswog}}},\ }\href
  {\doibase 10.1093/mnras/stt037} {\bibfield  {journal} {\bibinfo  {journal}
  {Mon. Not. R. Astron. Soc.}\ }\textbf {\bibinfo {volume} {430}},\ \bibinfo
  {pages} {2121} (\bibinfo {year} {2013})},\ \Eprint
  {http://arxiv.org/abs/1204.6242} {arXiv:1204.6242 [astro-ph.HE]} \BibitemShut
  {NoStop}%
\bibitem [{\citenamefont {{Grossman}}\ \emph {et~al.}(2014)\citenamefont
  {{Grossman}}, \citenamefont {{Korobkin}}, \citenamefont {{Rosswog}},\ and\
  \citenamefont {{Piran}}}]{Grossman2014}%
  \BibitemOpen
  \bibfield  {author} {\bibinfo {author} {\bibfnamefont {D.}~\bibnamefont
  {{Grossman}}}, \bibinfo {author} {\bibfnamefont {O.}~\bibnamefont
  {{Korobkin}}}, \bibinfo {author} {\bibfnamefont {S.}~\bibnamefont
  {{Rosswog}}}, \ and\ \bibinfo {author} {\bibfnamefont {T.}~\bibnamefont
  {{Piran}}},\ }\href {\doibase 10.1093/mnras/stt2503} {\bibfield  {journal}
  {\bibinfo  {journal} {Mon. Not. R. Astron. Soc.}\ }\textbf {\bibinfo {volume}
  {439}},\ \bibinfo {pages} {757} (\bibinfo {year} {2014})},\ \Eprint
  {http://arxiv.org/abs/1307.2943} {arXiv:1307.2943 [astro-ph.HE]} \BibitemShut
  {NoStop}%
\bibitem [{\citenamefont {{Metzger}}\ \emph {et~al.}(2015)\citenamefont
  {{Metzger}}, \citenamefont {{Bauswein}}, \citenamefont {{Goriely}},\ and\
  \citenamefont {{Kasen}}}]{Metzger2015}%
  \BibitemOpen
  \bibfield  {author} {\bibinfo {author} {\bibfnamefont {B.~D.}\ \bibnamefont
  {{Metzger}}}, \bibinfo {author} {\bibfnamefont {A.}~\bibnamefont
  {{Bauswein}}}, \bibinfo {author} {\bibfnamefont {S.}~\bibnamefont
  {{Goriely}}}, \ and\ \bibinfo {author} {\bibfnamefont {D.}~\bibnamefont
  {{Kasen}}},\ }\href {\doibase 10.1093/mnras/stu2225} {\bibfield  {journal}
  {\bibinfo  {journal} {Mon. Not. R. Astron. Soc.}\ }\textbf {\bibinfo {volume}
  {446}},\ \bibinfo {pages} {1115} (\bibinfo {year} {2015})},\ \Eprint
  {http://arxiv.org/abs/1409.0544} {arXiv:1409.0544 [astro-ph.HE]} \BibitemShut
  {NoStop}%
\bibitem [{\citenamefont {{Goriely}}\ \emph {et~al.}(2015)\citenamefont
  {{Goriely}}, \citenamefont {{Bauswein}}, \citenamefont {{Just}},
  \citenamefont {{Pllumbi}},\ and\ \citenamefont {{Janka}}}]{Goriely2015}%
  \BibitemOpen
  \bibfield  {author} {\bibinfo {author} {\bibfnamefont {S.}~\bibnamefont
  {{Goriely}}}, \bibinfo {author} {\bibfnamefont {A.}~\bibnamefont
  {{Bauswein}}}, \bibinfo {author} {\bibfnamefont {O.}~\bibnamefont {{Just}}},
  \bibinfo {author} {\bibfnamefont {E.}~\bibnamefont {{Pllumbi}}}, \ and\
  \bibinfo {author} {\bibfnamefont {H.-T.}\ \bibnamefont {{Janka}}},\ }\href
  {\doibase 10.1093/mnras/stv1526} {\bibfield  {journal} {\bibinfo  {journal}
  {Mon. Not. R. Astron. Soc.}\ }\textbf {\bibinfo {volume} {452}},\ \bibinfo
  {pages} {3894} (\bibinfo {year} {2015})},\ \Eprint
  {http://arxiv.org/abs/1504.04377} {arXiv:1504.04377 [astro-ph.SR]}
  \BibitemShut {NoStop}%
\bibitem [{\citenamefont {{Kyutoku}}\ \emph {et~al.}(2015)\citenamefont
  {{Kyutoku}}, \citenamefont {{Ioka}}, \citenamefont {{Okawa}}, \citenamefont
  {{Shibata}},\ and\ \citenamefont {{Taniguchi}}}]{Kyutoku2015}%
  \BibitemOpen
  \bibfield  {author} {\bibinfo {author} {\bibfnamefont {K.}~\bibnamefont
  {{Kyutoku}}}, \bibinfo {author} {\bibfnamefont {K.}~\bibnamefont {{Ioka}}},
  \bibinfo {author} {\bibfnamefont {H.}~\bibnamefont {{Okawa}}}, \bibinfo
  {author} {\bibfnamefont {M.}~\bibnamefont {{Shibata}}}, \ and\ \bibinfo
  {author} {\bibfnamefont {K.}~\bibnamefont {{Taniguchi}}},\ }\href {\doibase
  10.1103/PhysRevD.92.044028} {\bibfield  {journal} {\bibinfo  {journal} {Phys.
  Rev. D}\ }\textbf {\bibinfo {volume} {92}},\ \bibinfo {eid} {044028}
  (\bibinfo {year} {2015})},\ \Eprint {http://arxiv.org/abs/1502.05402}
  {arXiv:1502.05402 [astro-ph.HE]} \BibitemShut {NoStop}%
\bibitem [{\citenamefont {{Tanaka}}\ and\ \citenamefont
  {{Hotokezaka}}(2013)}]{Tanaka2013}%
  \BibitemOpen
  \bibfield  {author} {\bibinfo {author} {\bibfnamefont {M.}~\bibnamefont
  {{Tanaka}}}\ and\ \bibinfo {author} {\bibfnamefont {K.}~\bibnamefont
  {{Hotokezaka}}},\ }\href {\doibase 10.1088/0004-637X/775/2/113} {\bibfield
  {journal} {\bibinfo  {journal} {Astrophys. J.}\ }\textbf {\bibinfo {volume}
  {775}},\ \bibinfo {eid} {113} (\bibinfo {year} {2013})},\ \Eprint
  {http://arxiv.org/abs/1306.3742} {arXiv:1306.3742 [astro-ph.HE]} \BibitemShut
  {NoStop}%
\bibitem [{\citenamefont {{Metzger}}\ and\ \citenamefont
  {{Berger}}(2012)}]{Metzger2012}%
  \BibitemOpen
  \bibfield  {author} {\bibinfo {author} {\bibfnamefont {B.~D.}\ \bibnamefont
  {{Metzger}}}\ and\ \bibinfo {author} {\bibfnamefont {E.}~\bibnamefont
  {{Berger}}},\ }\href {\doibase 10.1088/0004-637X/746/1/48} {\bibfield
  {journal} {\bibinfo  {journal} {Astrophys. J.}\ }\textbf {\bibinfo {volume}
  {746}},\ \bibinfo {eid} {48} (\bibinfo {year} {2012})},\ \Eprint
  {http://arxiv.org/abs/1108.6056} {arXiv:1108.6056 [astro-ph.HE]} \BibitemShut
  {NoStop}%
\bibitem [{\citenamefont {{Kisaka}}\ \emph {et~al.}(2015)\citenamefont
  {{Kisaka}}, \citenamefont {{Ioka}},\ and\ \citenamefont
  {{Nakamura}}}]{Kisaka2015}%
  \BibitemOpen
  \bibfield  {author} {\bibinfo {author} {\bibfnamefont {S.}~\bibnamefont
  {{Kisaka}}}, \bibinfo {author} {\bibfnamefont {K.}~\bibnamefont {{Ioka}}}, \
  and\ \bibinfo {author} {\bibfnamefont {T.}~\bibnamefont {{Nakamura}}},\
  }\href {\doibase 10.1088/2041-8205/809/1/L8} {\bibfield  {journal} {\bibinfo
  {journal} {Astrophys. J. Lett.}\ }\textbf {\bibinfo {volume} {809}},\
  \bibinfo {eid} {L8} (\bibinfo {year} {2015})},\ \Eprint
  {http://arxiv.org/abs/1506.02030} {arXiv:1506.02030 [astro-ph.HE]}
  \BibitemShut {NoStop}%
\bibitem [{\citenamefont {{Rosswog}}\ and\ \citenamefont
  {{Price}}(2007)}]{Rosswog2007}%
  \BibitemOpen
  \bibfield  {author} {\bibinfo {author} {\bibfnamefont {S.}~\bibnamefont
  {{Rosswog}}}\ and\ \bibinfo {author} {\bibfnamefont {D.}~\bibnamefont
  {{Price}}},\ }\href {\doibase 10.1111/j.1365-2966.2007.11984.x} {\bibfield
  {journal} {\bibinfo  {journal} {Mon. Not. R. Astron. Soc.}\ }\textbf
  {\bibinfo {volume} {379}},\ \bibinfo {pages} {915} (\bibinfo {year}
  {2007})},\ \Eprint {http://arxiv.org/abs/0705.1441} {arXiv:0705.1441}
  \BibitemShut {NoStop}%
\bibitem [{\citenamefont {{Hotokezaka}}\ \emph
  {et~al.}(2016{\natexlab{b}})\citenamefont {{Hotokezaka}}, \citenamefont
  {{Wanajo}}, \citenamefont {{Tanaka}}, \citenamefont {{Bamba}}, \citenamefont
  {{Terada}},\ and\ \citenamefont {{Piran}}}]{Hotokezaka2016b}%
  \BibitemOpen
  \bibfield  {author} {\bibinfo {author} {\bibfnamefont {K.}~\bibnamefont
  {{Hotokezaka}}}, \bibinfo {author} {\bibfnamefont {S.}~\bibnamefont
  {{Wanajo}}}, \bibinfo {author} {\bibfnamefont {M.}~\bibnamefont {{Tanaka}}},
  \bibinfo {author} {\bibfnamefont {A.}~\bibnamefont {{Bamba}}}, \bibinfo
  {author} {\bibfnamefont {Y.}~\bibnamefont {{Terada}}}, \ and\ \bibinfo
  {author} {\bibfnamefont {T.}~\bibnamefont {{Piran}}},\ }\href {\doibase
  10.1093/mnras/stw404} {\bibfield  {journal} {\bibinfo  {journal} {Mon. Not.
  R. Astron. Soc.}\ }\textbf {\bibinfo {volume} {459}},\ \bibinfo {pages} {35}
  (\bibinfo {year} {2016}{\natexlab{b}})},\ \Eprint
  {http://arxiv.org/abs/1511.05580} {arXiv:1511.05580 [astro-ph.HE]}
  \BibitemShut {NoStop}%
\bibitem [{\citenamefont {{Barnes}}\ \emph {et~al.}(2016)\citenamefont
  {{Barnes}}, \citenamefont {{Kasen}}, \citenamefont {{Wu}},\ and\
  \citenamefont {{Mart'inez-Pinedo}}}]{Barnes2016}%
  \BibitemOpen
  \bibfield  {author} {\bibinfo {author} {\bibfnamefont {J.}~\bibnamefont
  {{Barnes}}}, \bibinfo {author} {\bibfnamefont {D.}~\bibnamefont {{Kasen}}},
  \bibinfo {author} {\bibfnamefont {M.-R.}\ \bibnamefont {{Wu}}}, \ and\
  \bibinfo {author} {\bibfnamefont {G.}~\bibnamefont {{Mart'inez-Pinedo}}},\
  }\href@noop {} {\bibfield  {journal} {\bibinfo  {journal} {ArXiv e-prints}\ }
  (\bibinfo {year} {2016})},\ \Eprint {http://arxiv.org/abs/1605.07218}
  {arXiv:1605.07218 [astro-ph.HE]} \BibitemShut {NoStop}%
\bibitem [{\citenamefont {{Nakar}}\ and\ \citenamefont
  {{Piran}}(2011)}]{Nakar2011}%
  \BibitemOpen
  \bibfield  {author} {\bibinfo {author} {\bibfnamefont {E.}~\bibnamefont
  {{Nakar}}}\ and\ \bibinfo {author} {\bibfnamefont {T.}~\bibnamefont
  {{Piran}}},\ }\href {\doibase 10.1038/nature10365} {\bibfield  {journal}
  {\bibinfo  {journal} {Nature}\ }\textbf {\bibinfo {volume} {478}},\ \bibinfo
  {pages} {82} (\bibinfo {year} {2011})},\ \Eprint
  {http://arxiv.org/abs/1102.1020} {arXiv:1102.1020 [astro-ph.HE]} \BibitemShut
  {NoStop}%
\bibitem [{\citenamefont {{Hotokezaka}}\ and\ \citenamefont
  {{Piran}}(2015)}]{Hotokezaka2015MNRAS}%
  \BibitemOpen
  \bibfield  {author} {\bibinfo {author} {\bibfnamefont {K.}~\bibnamefont
  {{Hotokezaka}}}\ and\ \bibinfo {author} {\bibfnamefont {T.}~\bibnamefont
  {{Piran}}},\ }\href {\doibase 10.1093/mnras/stv620} {\bibfield  {journal}
  {\bibinfo  {journal} {Mon. Not. R. Astron. Soc.}\ }\textbf {\bibinfo {volume}
  {450}},\ \bibinfo {pages} {1430} (\bibinfo {year} {2015})},\ \Eprint
  {http://arxiv.org/abs/1501.01986} {arXiv:1501.01986 [astro-ph.HE]}
  \BibitemShut {NoStop}%
\bibitem [{\citenamefont {{Stone}}\ \emph {et~al.}(2008)\citenamefont
  {{Stone}}, \citenamefont {{Gardiner}}, \citenamefont {{Teuben}},
  \citenamefont {{Hawley}},\ and\ \citenamefont {{Simon}}}]{Stone2008}%
  \BibitemOpen
  \bibfield  {author} {\bibinfo {author} {\bibfnamefont {J.~M.}\ \bibnamefont
  {{Stone}}}, \bibinfo {author} {\bibfnamefont {T.~A.}\ \bibnamefont
  {{Gardiner}}}, \bibinfo {author} {\bibfnamefont {P.}~\bibnamefont
  {{Teuben}}}, \bibinfo {author} {\bibfnamefont {J.~F.}\ \bibnamefont
  {{Hawley}}}, \ and\ \bibinfo {author} {\bibfnamefont {J.~B.}\ \bibnamefont
  {{Simon}}},\ }\href {\doibase 10.1086/588755} {\bibfield  {journal} {\bibinfo
   {journal} {Astrophys. J.s}\ }\textbf {\bibinfo {volume} {178}},\ \bibinfo
  {pages} {137} (\bibinfo {year} {2008})},\ \Eprint
  {http://arxiv.org/abs/0804.0402} {arXiv:0804.0402} \BibitemShut {NoStop}%
\bibitem [{\citenamefont {{White}}\ and\ \citenamefont
  {{Stone}}(2015)}]{White2015}%
  \BibitemOpen
  \bibfield  {author} {\bibinfo {author} {\bibfnamefont {C.~J.}\ \bibnamefont
  {{White}}}\ and\ \bibinfo {author} {\bibfnamefont {J.~M.}\ \bibnamefont
  {{Stone}}},\ }\href@noop {} {\bibfield  {journal} {\bibinfo  {journal} {ArXiv
  e-prints}\ } (\bibinfo {year} {2015})},\ \Eprint
  {http://arxiv.org/abs/1511.00943} {arXiv:1511.00943 [astro-ph.HE]}
  \BibitemShut {NoStop}%
\bibitem [{\citenamefont {{Almgren}}\ \emph {et~al.}(2010)\citenamefont
  {{Almgren}}, \citenamefont {{Beckner}}, \citenamefont {{Bell}}, \citenamefont
  {{Day}}, \citenamefont {{Howell}}, \citenamefont {{Joggerst}}, \citenamefont
  {{Lijewski}}, \citenamefont {{Nonaka}}, \citenamefont {{Singer}},\ and\
  \citenamefont {{Zingale}}}]{Almgren2010}%
  \BibitemOpen
  \bibfield  {author} {\bibinfo {author} {\bibfnamefont {A.~S.}\ \bibnamefont
  {{Almgren}}}, \bibinfo {author} {\bibfnamefont {V.~E.}\ \bibnamefont
  {{Beckner}}}, \bibinfo {author} {\bibfnamefont {J.~B.}\ \bibnamefont
  {{Bell}}}, \bibinfo {author} {\bibfnamefont {M.~S.}\ \bibnamefont {{Day}}},
  \bibinfo {author} {\bibfnamefont {L.~H.}\ \bibnamefont {{Howell}}}, \bibinfo
  {author} {\bibfnamefont {C.~C.}\ \bibnamefont {{Joggerst}}}, \bibinfo
  {author} {\bibfnamefont {M.~J.}\ \bibnamefont {{Lijewski}}}, \bibinfo
  {author} {\bibfnamefont {A.}~\bibnamefont {{Nonaka}}}, \bibinfo {author}
  {\bibfnamefont {M.}~\bibnamefont {{Singer}}}, \ and\ \bibinfo {author}
  {\bibfnamefont {M.}~\bibnamefont {{Zingale}}},\ }\href {\doibase
  10.1088/0004-637X/715/2/1221} {\bibfield  {journal} {\bibinfo  {journal}
  {Astrophys. J.}\ }\textbf {\bibinfo {volume} {715}},\ \bibinfo {pages} {1221}
  (\bibinfo {year} {2010})},\ \Eprint {http://arxiv.org/abs/1005.0114}
  {arXiv:1005.0114 [astro-ph.IM]} \BibitemShut {NoStop}%
\bibitem [{\citenamefont {{Collins}}\ \emph {et~al.}(2010)\citenamefont
  {{Collins}}, \citenamefont {{Xu}}, \citenamefont {{Norman}}, \citenamefont
  {{Li}},\ and\ \citenamefont {{Li}}}]{Collins2010}%
  \BibitemOpen
  \bibfield  {author} {\bibinfo {author} {\bibfnamefont {D.~C.}\ \bibnamefont
  {{Collins}}}, \bibinfo {author} {\bibfnamefont {H.}~\bibnamefont {{Xu}}},
  \bibinfo {author} {\bibfnamefont {M.~L.}\ \bibnamefont {{Norman}}}, \bibinfo
  {author} {\bibfnamefont {H.}~\bibnamefont {{Li}}}, \ and\ \bibinfo {author}
  {\bibfnamefont {S.}~\bibnamefont {{Li}}},\ }\href {\doibase
  10.1088/0067-0049/186/2/308} {\bibfield  {journal} {\bibinfo  {journal}
  {Astrophys. J.s}\ }\textbf {\bibinfo {volume} {186}},\ \bibinfo {pages} {308}
  (\bibinfo {year} {2010})},\ \Eprint {http://arxiv.org/abs/0902.2594}
  {arXiv:0902.2594 [astro-ph.IM]} \BibitemShut {NoStop}%
\bibitem [{\citenamefont {{Fryxell}}\ \emph {et~al.}(2000)\citenamefont
  {{Fryxell}}, \citenamefont {{Olson}}, \citenamefont {{Ricker}}, \citenamefont
  {{Timmes}}, \citenamefont {{Zingale}}, \citenamefont {{Lamb}}, \citenamefont
  {{MacNeice}}, \citenamefont {{Rosner}}, \citenamefont {{Truran}},\ and\
  \citenamefont {{Tufo}}}]{Fryxell2000}%
  \BibitemOpen
  \bibfield  {author} {\bibinfo {author} {\bibfnamefont {B.}~\bibnamefont
  {{Fryxell}}}, \bibinfo {author} {\bibfnamefont {K.}~\bibnamefont {{Olson}}},
  \bibinfo {author} {\bibfnamefont {P.}~\bibnamefont {{Ricker}}}, \bibinfo
  {author} {\bibfnamefont {F.~X.}\ \bibnamefont {{Timmes}}}, \bibinfo {author}
  {\bibfnamefont {M.}~\bibnamefont {{Zingale}}}, \bibinfo {author}
  {\bibfnamefont {D.~Q.}\ \bibnamefont {{Lamb}}}, \bibinfo {author}
  {\bibfnamefont {P.}~\bibnamefont {{MacNeice}}}, \bibinfo {author}
  {\bibfnamefont {R.}~\bibnamefont {{Rosner}}}, \bibinfo {author}
  {\bibfnamefont {J.~W.}\ \bibnamefont {{Truran}}}, \ and\ \bibinfo {author}
  {\bibfnamefont {H.}~\bibnamefont {{Tufo}}},\ }\href {\doibase 10.1086/317361}
  {\bibfield  {journal} {\bibinfo  {journal} {Astrophys. J.s}\ }\textbf
  {\bibinfo {volume} {131}},\ \bibinfo {pages} {273} (\bibinfo {year}
  {2000})}\BibitemShut {NoStop}%
\bibitem [{\citenamefont {{Shibata}}\ \emph {et~al.}(2008)\citenamefont
  {{Shibata}}, \citenamefont {{Okawa}},\ and\ \citenamefont
  {{Yamamoto}}}]{Shibata2008}%
  \BibitemOpen
  \bibfield  {author} {\bibinfo {author} {\bibfnamefont {M.}~\bibnamefont
  {{Shibata}}}, \bibinfo {author} {\bibfnamefont {H.}~\bibnamefont {{Okawa}}},
  \ and\ \bibinfo {author} {\bibfnamefont {T.}~\bibnamefont {{Yamamoto}}},\
  }\href {\doibase 10.1103/PhysRevD.78.101501} {\bibfield  {journal} {\bibinfo
  {journal} {Phys. Rev. D}\ }\textbf {\bibinfo {volume} {78}},\ \bibinfo {eid}
  {101501} (\bibinfo {year} {2008})},\ \Eprint {http://arxiv.org/abs/0810.4735}
  {arXiv:0810.4735 [gr-qc]} \BibitemShut {NoStop}%
\bibitem [{\citenamefont {{Pannarale}}\ \emph {et~al.}(2013)\citenamefont
  {{Pannarale}}, \citenamefont {{Berti}}, \citenamefont {{Kyutoku}},\ and\
  \citenamefont {{Shibata}}}]{Pannarale2013a}%
  \BibitemOpen
  \bibfield  {author} {\bibinfo {author} {\bibfnamefont {F.}~\bibnamefont
  {{Pannarale}}}, \bibinfo {author} {\bibfnamefont {E.}~\bibnamefont
  {{Berti}}}, \bibinfo {author} {\bibfnamefont {K.}~\bibnamefont {{Kyutoku}}},
  \ and\ \bibinfo {author} {\bibfnamefont {M.}~\bibnamefont {{Shibata}}},\
  }\href {\doibase 10.1103/PhysRevD.88.084011} {\bibfield  {journal} {\bibinfo
  {journal} {Phys. Rev. D}\ }\textbf {\bibinfo {volume} {88}},\ \bibinfo {eid}
  {084011} (\bibinfo {year} {2013})},\ \Eprint {http://arxiv.org/abs/1307.5111}
  {arXiv:1307.5111 [gr-qc]} \BibitemShut {NoStop}%
\bibitem [{\citenamefont {{Shibata}}(1999{\natexlab{b}})}]{Shibata99a}%
  \BibitemOpen
  \bibfield  {author} {\bibinfo {author} {\bibfnamefont {M.}~\bibnamefont
  {{Shibata}}},\ }\href {\doibase 10.1143/PTP.101.1199} {\bibfield  {journal}
  {\bibinfo  {journal} {Progress of Theoretical Physics}\ }\textbf {\bibinfo
  {volume} {101}},\ \bibinfo {pages} {1199} (\bibinfo {year}
  {1999}{\natexlab{b}})},\ \Eprint {http://arxiv.org/abs/gr-qc/9905058}
  {gr-qc/9905058} \BibitemShut {NoStop}%
\bibitem [{\citenamefont {{Baiotti}}\ \emph {et~al.}(2005)\citenamefont
  {{Baiotti}}, \citenamefont {{Hawke}}, \citenamefont {{Montero}},
  \citenamefont {{L{\"o}ffler}}, \citenamefont {{Rezzolla}}, \citenamefont
  {{Stergioulas}}, \citenamefont {{Font}},\ and\ \citenamefont
  {{Seidel}}}]{Baiotti04}%
  \BibitemOpen
  \bibfield  {author} {\bibinfo {author} {\bibfnamefont {L.}~\bibnamefont
  {{Baiotti}}}, \bibinfo {author} {\bibfnamefont {I.}~\bibnamefont {{Hawke}}},
  \bibinfo {author} {\bibfnamefont {P.~J.}\ \bibnamefont {{Montero}}}, \bibinfo
  {author} {\bibfnamefont {F.}~\bibnamefont {{L{\"o}ffler}}}, \bibinfo {author}
  {\bibfnamefont {L.}~\bibnamefont {{Rezzolla}}}, \bibinfo {author}
  {\bibfnamefont {N.}~\bibnamefont {{Stergioulas}}}, \bibinfo {author}
  {\bibfnamefont {J.~A.}\ \bibnamefont {{Font}}}, \ and\ \bibinfo {author}
  {\bibfnamefont {E.}~\bibnamefont {{Seidel}}},\ }\href {\doibase
  10.1103/PhysRevD.71.024035} {\bibfield  {journal} {\bibinfo  {journal} {Phys.
  Rev. D}\ }\textbf {\bibinfo {volume} {71}},\ \bibinfo {eid} {024035}
  (\bibinfo {year} {2005})},\ \Eprint {http://arxiv.org/abs/gr-qc/0403029}
  {gr-qc/0403029} \BibitemShut {NoStop}%
\bibitem [{\citenamefont {{Palenzuela}}\ \emph
  {et~al.}(2009{\natexlab{b}})\citenamefont {{Palenzuela}}, \citenamefont
  {{Anderson}}, \citenamefont {{Lehner}}, \citenamefont {{Liebling}},\ and\
  \citenamefont {{Neilsen}}}]{Palenzuela2009b}%
  \BibitemOpen
  \bibfield  {author} {\bibinfo {author} {\bibfnamefont {C.}~\bibnamefont
  {{Palenzuela}}}, \bibinfo {author} {\bibfnamefont {M.}~\bibnamefont
  {{Anderson}}}, \bibinfo {author} {\bibfnamefont {L.}~\bibnamefont
  {{Lehner}}}, \bibinfo {author} {\bibfnamefont {S.~L.}\ \bibnamefont
  {{Liebling}}}, \ and\ \bibinfo {author} {\bibfnamefont {D.}~\bibnamefont
  {{Neilsen}}},\ }\href {\doibase 10.1103/PhysRevLett.103.081101} {\bibfield
  {journal} {\bibinfo  {journal} {Phys. Rev. Lett.}\ }\textbf {\bibinfo
  {volume} {103}},\ \bibinfo {pages} {081101} (\bibinfo {year}
  {2009}{\natexlab{b}})},\ \Eprint {http://arxiv.org/abs/0905.1121}
  {arXiv:0905.1121 [astro-ph.HE]} \BibitemShut {NoStop}%
\bibitem [{\citenamefont {Yo}\ \emph {et~al.}(2002)\citenamefont {Yo},
  \citenamefont {Baumgarte},\ and\ \citenamefont {Shapiro}}]{Yo02a}%
  \BibitemOpen
  \bibfield  {author} {\bibinfo {author} {\bibfnamefont {H.-J.}\ \bibnamefont
  {Yo}}, \bibinfo {author} {\bibfnamefont {T.~W.}\ \bibnamefont {Baumgarte}}, \
  and\ \bibinfo {author} {\bibfnamefont {S.~L.}\ \bibnamefont {Shapiro}},\
  }\href@noop {} {\bibfield  {journal} {\bibinfo  {journal} {Phys. Rev. D}\
  }\textbf {\bibinfo {volume} {66}},\ \bibinfo {pages} {084026} (\bibinfo
  {year} {2002})}\BibitemShut {NoStop}%
\bibitem [{\citenamefont {Duez}\ \emph {et~al.}(2003)\citenamefont {Duez},
  \citenamefont {Marronetti}, \citenamefont {Shapiro},\ and\ \citenamefont
  {Baumgarte}}]{Duez:2002bn}%
  \BibitemOpen
  \bibfield  {author} {\bibinfo {author} {\bibfnamefont {M.~D.}\ \bibnamefont
  {Duez}}, \bibinfo {author} {\bibfnamefont {P.}~\bibnamefont {Marronetti}},
  \bibinfo {author} {\bibfnamefont {S.~L.}\ \bibnamefont {Shapiro}}, \ and\
  \bibinfo {author} {\bibfnamefont {T.~W.}\ \bibnamefont {Baumgarte}},\
  }\href@noop {} {\bibfield  {journal} {\bibinfo  {journal} {Phys. Rev. D}\
  }\textbf {\bibinfo {volume} {67}},\ \bibinfo {pages} {024004} (\bibinfo
  {year} {2003})},\ \Eprint {http://arxiv.org/abs/gr-qc/0209102}
  {gr-qc/0209102} \BibitemShut {NoStop}%
\bibitem [{\citenamefont {{Faber}}\ \emph {et~al.}(2007)\citenamefont
  {{Faber}}, \citenamefont {{Baumgarte}}, \citenamefont {{Etienne}},
  \citenamefont {{Shapiro}},\ and\ \citenamefont {{Taniguchi}}}]{Faber2007}%
  \BibitemOpen
  \bibfield  {author} {\bibinfo {author} {\bibfnamefont {J.~A.}\ \bibnamefont
  {{Faber}}}, \bibinfo {author} {\bibfnamefont {T.~W.}\ \bibnamefont
  {{Baumgarte}}}, \bibinfo {author} {\bibfnamefont {Z.~B.}\ \bibnamefont
  {{Etienne}}}, \bibinfo {author} {\bibfnamefont {S.~L.}\ \bibnamefont
  {{Shapiro}}}, \ and\ \bibinfo {author} {\bibfnamefont {K.}~\bibnamefont
  {{Taniguchi}}},\ }\href {\doibase 10.1103/PhysRevD.76.104021} {\bibfield
  {journal} {\bibinfo  {journal} {Phys. Rev. D}\ }\textbf {\bibinfo {volume}
  {76}},\ \bibinfo {eid} {104021} (\bibinfo {year} {2007})},\ \Eprint
  {http://arxiv.org/abs/0708.2436} {arXiv:0708.2436 [gr-qc]} \BibitemShut
  {NoStop}%
\bibitem [{\citenamefont {{Etienne}}\ \emph {et~al.}(2015)\citenamefont
  {{Etienne}}, \citenamefont {{Paschalidis}}, \citenamefont {{Haas}},
  \citenamefont {{M{\"o}sta}},\ and\ \citenamefont {{Shapiro}}}]{Etienne2015}%
  \BibitemOpen
  \bibfield  {author} {\bibinfo {author} {\bibfnamefont {Z.~B.}\ \bibnamefont
  {{Etienne}}}, \bibinfo {author} {\bibfnamefont {V.}~\bibnamefont
  {{Paschalidis}}}, \bibinfo {author} {\bibfnamefont {R.}~\bibnamefont
  {{Haas}}}, \bibinfo {author} {\bibfnamefont {P.}~\bibnamefont {{M{\"o}sta}}},
  \ and\ \bibinfo {author} {\bibfnamefont {S.~L.}\ \bibnamefont {{Shapiro}}},\
  }\href {\doibase 10.1088/0264-9381/32/17/175009} {\bibfield  {journal}
  {\bibinfo  {journal} {Class. Quantum Grav.}\ }\textbf {\bibinfo {volume}
  {32}},\ \bibinfo {eid} {175009} (\bibinfo {year} {2015})},\ \Eprint
  {http://arxiv.org/abs/1501.07276} {arXiv:1501.07276 [astro-ph.HE]}
  \BibitemShut {NoStop}%
\bibitem [{\citenamefont {{L{\"o}ffler}}\ \emph {et~al.}(2012)\citenamefont
  {{L{\"o}ffler}}, \citenamefont {{Faber}}, \citenamefont {{Bentivegna}},
  \citenamefont {{Bode}}, \citenamefont {{Diener}}, \citenamefont {{Haas}},
  \citenamefont {{Hinder}}, \citenamefont {{Mundim}}, \citenamefont {{Ott}},
  \citenamefont {{Schnetter}}, \citenamefont {{Allen}}, \citenamefont
  {{Campanelli}},\ and\ \citenamefont {{Laguna}}}]{loeffler_2011_et}%
  \BibitemOpen
  \bibfield  {author} {\bibinfo {author} {\bibfnamefont {F.}~\bibnamefont
  {{L{\"o}ffler}}}, \bibinfo {author} {\bibfnamefont {J.}~\bibnamefont
  {{Faber}}}, \bibinfo {author} {\bibfnamefont {E.}~\bibnamefont
  {{Bentivegna}}}, \bibinfo {author} {\bibfnamefont {T.}~\bibnamefont
  {{Bode}}}, \bibinfo {author} {\bibfnamefont {P.}~\bibnamefont {{Diener}}},
  \bibinfo {author} {\bibfnamefont {R.}~\bibnamefont {{Haas}}}, \bibinfo
  {author} {\bibfnamefont {I.}~\bibnamefont {{Hinder}}}, \bibinfo {author}
  {\bibfnamefont {B.~C.}\ \bibnamefont {{Mundim}}}, \bibinfo {author}
  {\bibfnamefont {C.~D.}\ \bibnamefont {{Ott}}}, \bibinfo {author}
  {\bibfnamefont {E.}~\bibnamefont {{Schnetter}}}, \bibinfo {author}
  {\bibfnamefont {G.}~\bibnamefont {{Allen}}}, \bibinfo {author} {\bibfnamefont
  {M.}~\bibnamefont {{Campanelli}}}, \ and\ \bibinfo {author} {\bibfnamefont
  {P.}~\bibnamefont {{Laguna}}},\ }\href {\doibase
  10.1088/0264-9381/29/11/115001} {\bibfield  {journal} {\bibinfo  {journal}
  {Class. Quantum Grav.}\ }\textbf {\bibinfo {volume} {29}},\ \bibinfo {eid}
  {115001} (\bibinfo {year} {2012})},\ \Eprint {http://arxiv.org/abs/1111.3344}
  {arXiv:1111.3344 [gr-qc]} \BibitemShut {NoStop}%
\bibitem [{\citenamefont {{Zilh{\~a}o}}\ and\ \citenamefont
  {{L{\"o}ffler}}(2013)}]{ET2013}%
  \BibitemOpen
  \bibfield  {author} {\bibinfo {author} {\bibfnamefont {M.}~\bibnamefont
  {{Zilh{\~a}o}}}\ and\ \bibinfo {author} {\bibfnamefont {F.}~\bibnamefont
  {{L{\"o}ffler}}},\ }\href {\doibase 10.1142/S0217751X13400149} {\bibfield
  {journal} {\bibinfo  {journal} {International Journal of Modern Physics A}\
  }\textbf {\bibinfo {volume} {28}},\ \bibinfo {eid} {1340014} (\bibinfo {year}
  {2013})},\ \Eprint {http://arxiv.org/abs/1305.5299} {arXiv:1305.5299 [gr-qc]}
  \BibitemShut {NoStop}%
\bibitem [{ein()}]{einsteintoolkitweb}%
  \BibitemOpen
  \href@noop {} {}\bibinfo {note} {The Einstein Toolkit Consortium:
  \url{einsteintoolkit.org}}\BibitemShut {NoStop}%
\bibitem [{Whisky()}]{whisky-web}%
  \BibitemOpen
  Whisky,\ \href@noop {} {}\bibinfo {note} {EU Network GR Hydrodynamics Code:
  \url{www.whiskycode.org}}\BibitemShut {NoStop}%
\bibitem [{\citenamefont {{Br{\"u}gmann}}(1999)}]{Bruegmann97}%
  \BibitemOpen
  \bibfield  {author} {\bibinfo {author} {\bibfnamefont {B.}~\bibnamefont
  {{Br{\"u}gmann}}},\ }\href {\doibase 10.1142/S0218271899000080} {\bibfield
  {journal} {\bibinfo  {journal} {International Journal of Modern Physics D}\
  }\textbf {\bibinfo {volume} {8}},\ \bibinfo {pages} {85} (\bibinfo {year}
  {1999})},\ \Eprint {http://arxiv.org/abs/gr-qc/9708035} {gr-qc/9708035}
  \BibitemShut {NoStop}%
\bibitem [{\citenamefont {{Br{\"u}gmann}}\ \emph {et~al.}(2004)\citenamefont
  {{Br{\"u}gmann}}, \citenamefont {{Tichy}},\ and\ \citenamefont
  {{Jansen}}}]{Bruegmann:2003aw}%
  \BibitemOpen
  \bibfield  {author} {\bibinfo {author} {\bibfnamefont {B.}~\bibnamefont
  {{Br{\"u}gmann}}}, \bibinfo {author} {\bibfnamefont {W.}~\bibnamefont
  {{Tichy}}}, \ and\ \bibinfo {author} {\bibfnamefont {N.}~\bibnamefont
  {{Jansen}}},\ }\href {\doibase 10.1103/PhysRevLett.92.211101} {\bibfield
  {journal} {\bibinfo  {journal} {Phys. Rev. Lett.}\ }\textbf {\bibinfo
  {volume} {92}},\ \bibinfo {eid} {211101} (\bibinfo {year} {2004})},\ \Eprint
  {http://arxiv.org/abs/gr-qc/0312112} {gr-qc/0312112} \BibitemShut {NoStop}%
\bibitem [{\citenamefont {Br{\"u}gmann}\ \emph {et~al.}(2008)\citenamefont
  {Br{\"u}gmann}, \citenamefont {Gonz{\'a}lez}, \citenamefont {Hannam},
  \citenamefont {Husa}, \citenamefont {Sperhake},\ and\ \citenamefont
  {Tichy}}]{Bruegmann:2006at}%
  \BibitemOpen
  \bibfield  {author} {\bibinfo {author} {\bibfnamefont {B.}~\bibnamefont
  {Br{\"u}gmann}}, \bibinfo {author} {\bibfnamefont {J.~A.}\ \bibnamefont
  {Gonz{\'a}lez}}, \bibinfo {author} {\bibfnamefont {M.}~\bibnamefont
  {Hannam}}, \bibinfo {author} {\bibfnamefont {S.}~\bibnamefont {Husa}},
  \bibinfo {author} {\bibfnamefont {U.}~\bibnamefont {Sperhake}}, \ and\
  \bibinfo {author} {\bibfnamefont {W.}~\bibnamefont {Tichy}},\ }\href@noop {}
  {\bibfield  {journal} {\bibinfo  {journal} {Phys. Rev. D}\ }\textbf {\bibinfo
  {volume} {77}},\ \bibinfo {pages} {024027} (\bibinfo {year} {2008})},\
  \bibinfo {note} {gr-qc/0610128},\ \Eprint
  {http://arxiv.org/abs/gr-qc/0610128} {gr-qc/0610128} \BibitemShut {NoStop}%
\bibitem [{\citenamefont {{East}}\ \emph
  {et~al.}(2012{\natexlab{c}})\citenamefont {{East}}, \citenamefont
  {{Pretorius}},\ and\ \citenamefont {{Stephens}}}]{East2012b0}%
  \BibitemOpen
  \bibfield  {author} {\bibinfo {author} {\bibfnamefont {W.~E.}\ \bibnamefont
  {{East}}}, \bibinfo {author} {\bibfnamefont {F.}~\bibnamefont {{Pretorius}}},
  \ and\ \bibinfo {author} {\bibfnamefont {B.~C.}\ \bibnamefont {{Stephens}}},\
  }\href {\doibase 10.1103/PhysRevD.85.124009} {\bibfield  {journal} {\bibinfo
  {journal} {Phys. Rev. D}\ }\textbf {\bibinfo {volume} {85}},\ \bibinfo {eid}
  {124009} (\bibinfo {year} {2012}{\natexlab{c}})},\ \Eprint
  {http://arxiv.org/abs/1111.3055} {arXiv:1111.3055 [astro-ph.HE]} \BibitemShut
  {NoStop}%
\bibitem [{\citenamefont {{East}}\ \emph {et~al.}(2015)\citenamefont {{East}},
  \citenamefont {{Paschalidis}},\ and\ \citenamefont {{Pretorius}}}]{East2015}%
  \BibitemOpen
  \bibfield  {author} {\bibinfo {author} {\bibfnamefont {W.~E.}\ \bibnamefont
  {{East}}}, \bibinfo {author} {\bibfnamefont {V.}~\bibnamefont
  {{Paschalidis}}}, \ and\ \bibinfo {author} {\bibfnamefont {F.}~\bibnamefont
  {{Pretorius}}},\ }\href {\doibase 10.1088/2041-8205/807/1/L3} {\bibfield
  {journal} {\bibinfo  {journal} {Astrophys. J. Lett.}\ }\textbf {\bibinfo
  {volume} {807}},\ \bibinfo {eid} {L3} (\bibinfo {year} {2015})},\ \Eprint
  {http://arxiv.org/abs/1503.07171} {arXiv:1503.07171 [astro-ph.HE]}
  \BibitemShut {NoStop}%
\bibitem [{\citenamefont {Duez}\ \emph {et~al.}(2008)\citenamefont {Duez} \emph
  {et~al.}}]{Duez:2008rb}%
  \BibitemOpen
  \bibfield  {author} {\bibinfo {author} {\bibfnamefont {M.~D.}\ \bibnamefont
  {Duez}} \emph {et~al.},\ }\href {\doibase 10.1103/PhysRevD.78.104015}
  {\bibfield  {journal} {\bibinfo  {journal} {Phys. Rev. D}\ }\textbf {\bibinfo
  {volume} {78}},\ \bibinfo {pages} {104015} (\bibinfo {year} {2008})},\
  \Eprint {http://arxiv.org/abs/0809.0002} {arXiv:0809.0002 [gr-qc]}
  \BibitemShut {NoStop}%
\bibitem [{\citenamefont {{Foucart}}\ \emph {et~al.}(2011)\citenamefont
  {{Foucart}}, \citenamefont {{Duez}}, \citenamefont {{Kidder}},\ and\
  \citenamefont {{Teukolsky}}}]{Foucart2010}%
  \BibitemOpen
  \bibfield  {author} {\bibinfo {author} {\bibfnamefont {F.}~\bibnamefont
  {{Foucart}}}, \bibinfo {author} {\bibfnamefont {M.~D.}\ \bibnamefont
  {{Duez}}}, \bibinfo {author} {\bibfnamefont {L.~E.}\ \bibnamefont
  {{Kidder}}}, \ and\ \bibinfo {author} {\bibfnamefont {S.~A.}\ \bibnamefont
  {{Teukolsky}}},\ }\href {\doibase 10.1103/PhysRevD.83.024005} {\bibfield
  {journal} {\bibinfo  {journal} {Phys. Rev. D}\ }\textbf {\bibinfo {volume}
  {83}},\ \bibinfo {eid} {024005} (\bibinfo {year} {2011})},\ \Eprint
  {http://arxiv.org/abs/1007.4203} {arXiv:1007.4203 [astro-ph.HE]} \BibitemShut
  {NoStop}%
\bibitem [{\citenamefont {{Foucart}}(2012)}]{Foucart2012}%
  \BibitemOpen
  \bibfield  {author} {\bibinfo {author} {\bibfnamefont {F.}~\bibnamefont
  {{Foucart}}},\ }\href {\doibase 10.1103/PhysRevD.86.124007} {\bibfield
  {journal} {\bibinfo  {journal} {Phys. Rev. D}\ }\textbf {\bibinfo {volume}
  {86}},\ \bibinfo {eid} {124007} (\bibinfo {year} {2012})},\ \Eprint
  {http://arxiv.org/abs/1207.6304} {arXiv:1207.6304 [astro-ph.HE]} \BibitemShut
  {NoStop}%
\bibitem [{\citenamefont {{Foucart}}\ \emph {et~al.}(2013)\citenamefont
  {{Foucart}}, \citenamefont {{Deaton}}, \citenamefont {{Duez}}, \citenamefont
  {{Kidder}}, \citenamefont {{MacDonald}}, \citenamefont {{Ott}}, \citenamefont
  {{Pfeiffer}}, \citenamefont {{Scheel}}, \citenamefont {{Szilagyi}},\ and\
  \citenamefont {{Teukolsky}}}]{Foucart2013a}%
  \BibitemOpen
  \bibfield  {author} {\bibinfo {author} {\bibfnamefont {F.}~\bibnamefont
  {{Foucart}}}, \bibinfo {author} {\bibfnamefont {M.~B.}\ \bibnamefont
  {{Deaton}}}, \bibinfo {author} {\bibfnamefont {M.~D.}\ \bibnamefont
  {{Duez}}}, \bibinfo {author} {\bibfnamefont {L.~E.}\ \bibnamefont
  {{Kidder}}}, \bibinfo {author} {\bibfnamefont {I.}~\bibnamefont
  {{MacDonald}}}, \bibinfo {author} {\bibfnamefont {C.~D.}\ \bibnamefont
  {{Ott}}}, \bibinfo {author} {\bibfnamefont {H.~P.}\ \bibnamefont
  {{Pfeiffer}}}, \bibinfo {author} {\bibfnamefont {M.~A.}\ \bibnamefont
  {{Scheel}}}, \bibinfo {author} {\bibfnamefont {B.}~\bibnamefont
  {{Szilagyi}}}, \ and\ \bibinfo {author} {\bibfnamefont {S.~A.}\ \bibnamefont
  {{Teukolsky}}},\ }\href {\doibase 10.1103/PhysRevD.87.084006} {\bibfield
  {journal} {\bibinfo  {journal} {Phys. Rev. D}\ }\textbf {\bibinfo {volume}
  {87}},\ \bibinfo {eid} {084006} (\bibinfo {year} {2013})},\ \Eprint
  {http://arxiv.org/abs/1212.4810} {arXiv:1212.4810 [gr-qc]} \BibitemShut
  {NoStop}%
\bibitem [{\citenamefont {{Scheel}}\ \emph {et~al.}(2009)\citenamefont
  {{Scheel}}, \citenamefont {{Boyle}}, \citenamefont {{Chu}}, \citenamefont
  {{Kidder}}, \citenamefont {{Matthews}},\ and\ \citenamefont
  {{Pfeiffer}}}]{Scheel:2008rj}%
  \BibitemOpen
  \bibfield  {author} {\bibinfo {author} {\bibfnamefont {M.~A.}\ \bibnamefont
  {{Scheel}}}, \bibinfo {author} {\bibfnamefont {M.}~\bibnamefont {{Boyle}}},
  \bibinfo {author} {\bibfnamefont {T.}~\bibnamefont {{Chu}}}, \bibinfo
  {author} {\bibfnamefont {L.~E.}\ \bibnamefont {{Kidder}}}, \bibinfo {author}
  {\bibfnamefont {K.~D.}\ \bibnamefont {{Matthews}}}, \ and\ \bibinfo {author}
  {\bibfnamefont {H.~P.}\ \bibnamefont {{Pfeiffer}}},\ }\href {\doibase
  10.1103/PhysRevD.79.024003} {\bibfield  {journal} {\bibinfo  {journal} {Phys.
  Rev. D}\ }\textbf {\bibinfo {volume} {79}},\ \bibinfo {eid} {024003}
  (\bibinfo {year} {2009})},\ \Eprint {http://arxiv.org/abs/0810.1767}
  {arXiv:0810.1767 [gr-qc]} \BibitemShut {NoStop}%
\bibitem [{\citenamefont {{Lovelace}}\ \emph {et~al.}(2011)\citenamefont
  {{Lovelace}}, \citenamefont {{Scheel}},\ and\ \citenamefont
  {{Szil{\'a}gyi}}}]{Lovelace2011}%
  \BibitemOpen
  \bibfield  {author} {\bibinfo {author} {\bibfnamefont {G.}~\bibnamefont
  {{Lovelace}}}, \bibinfo {author} {\bibfnamefont {M.~A.}\ \bibnamefont
  {{Scheel}}}, \ and\ \bibinfo {author} {\bibfnamefont {B.}~\bibnamefont
  {{Szil{\'a}gyi}}},\ }\href {\doibase 10.1103/PhysRevD.83.024010} {\bibfield
  {journal} {\bibinfo  {journal} {Phys. Rev. D}\ }\textbf {\bibinfo {volume}
  {83}},\ \bibinfo {eid} {024010} (\bibinfo {year} {2011})},\ \Eprint
  {http://arxiv.org/abs/1010.2777} {arXiv:1010.2777 [gr-qc]} \BibitemShut
  {NoStop}%
\bibitem [{\citenamefont {{Buchman}}\ \emph {et~al.}(2012)\citenamefont
  {{Buchman}}, \citenamefont {{Pfeiffer}}, \citenamefont {{Scheel}},\ and\
  \citenamefont {{Szil{\'a}gyi}}}]{Buchman2012}%
  \BibitemOpen
  \bibfield  {author} {\bibinfo {author} {\bibfnamefont {L.~T.}\ \bibnamefont
  {{Buchman}}}, \bibinfo {author} {\bibfnamefont {H.~P.}\ \bibnamefont
  {{Pfeiffer}}}, \bibinfo {author} {\bibfnamefont {M.~A.}\ \bibnamefont
  {{Scheel}}}, \ and\ \bibinfo {author} {\bibfnamefont {B.}~\bibnamefont
  {{Szil{\'a}gyi}}},\ }\href {\doibase 10.1103/PhysRevD.86.084033} {\bibfield
  {journal} {\bibinfo  {journal} {Phys. Rev. D}\ }\textbf {\bibinfo {volume}
  {86}},\ \bibinfo {eid} {084033} (\bibinfo {year} {2012})},\ \Eprint
  {http://arxiv.org/abs/1206.3015} {arXiv:1206.3015 [gr-qc]} \BibitemShut
  {NoStop}%
\bibitem [{\citenamefont {{Ossokine}}\ \emph {et~al.}(2013)\citenamefont
  {{Ossokine}}, \citenamefont {{Kidder}},\ and\ \citenamefont
  {{Pfeiffer}}}]{Ossokine2013}%
  \BibitemOpen
  \bibfield  {author} {\bibinfo {author} {\bibfnamefont {S.}~\bibnamefont
  {{Ossokine}}}, \bibinfo {author} {\bibfnamefont {L.~E.}\ \bibnamefont
  {{Kidder}}}, \ and\ \bibinfo {author} {\bibfnamefont {H.~P.}\ \bibnamefont
  {{Pfeiffer}}},\ }\href {\doibase 10.1103/PhysRevD.88.084031} {\bibfield
  {journal} {\bibinfo  {journal} {Phys. Rev. D}\ }\textbf {\bibinfo {volume}
  {88}},\ \bibinfo {eid} {084031} (\bibinfo {year} {2013})},\ \Eprint
  {http://arxiv.org/abs/1304.3067} {arXiv:1304.3067 [gr-qc]} \BibitemShut
  {NoStop}%
\bibitem [{\citenamefont {{Rosswog}}(2010)}]{rosswog_2010_csr}%
  \BibitemOpen
  \bibfield  {author} {\bibinfo {author} {\bibfnamefont {S.}~\bibnamefont
  {{Rosswog}}},\ }\href {\doibase 10.1016/j.jcp.2010.08.002} {\bibfield
  {journal} {\bibinfo  {journal} {Journal of Computational Physics}\ }\textbf
  {\bibinfo {volume} {229}},\ \bibinfo {pages} {8591} (\bibinfo {year}
  {2010})},\ \Eprint {http://arxiv.org/abs/0907.4890} {arXiv:0907.4890
  [astro-ph.HE]} \BibitemShut {NoStop}%
\bibitem [{\citenamefont {{Oechslin}}\ \emph {et~al.}(2002)\citenamefont
  {{Oechslin}}, \citenamefont {{Rosswog}},\ and\ \citenamefont
  {{Thielemann}}}]{Oechslin02}%
  \BibitemOpen
  \bibfield  {author} {\bibinfo {author} {\bibfnamefont {R.}~\bibnamefont
  {{Oechslin}}}, \bibinfo {author} {\bibfnamefont {S.}~\bibnamefont
  {{Rosswog}}}, \ and\ \bibinfo {author} {\bibfnamefont {F.-K.}\ \bibnamefont
  {{Thielemann}}},\ }\href {\doibase 10.1103/PhysRevD.65.103005} {\bibfield
  {journal} {\bibinfo  {journal} {Phys. Rev. D}\ }\textbf {\bibinfo {volume}
  {65}},\ \bibinfo {eid} {103005} (\bibinfo {year} {2002})},\ \Eprint
  {http://arxiv.org/abs/gr-qc/0111005} {gr-qc/0111005} \BibitemShut {NoStop}%
\bibitem [{\citenamefont {{Bauswein}}\ \emph
  {et~al.}(2010{\natexlab{b}})\citenamefont {{Bauswein}}, \citenamefont
  {{Janka}},\ and\ \citenamefont {{Oechslin}}}]{Bauswein:2010dn}%
  \BibitemOpen
  \bibfield  {author} {\bibinfo {author} {\bibfnamefont {A.}~\bibnamefont
  {{Bauswein}}}, \bibinfo {author} {\bibfnamefont {H.}~\bibnamefont {{Janka}}},
  \ and\ \bibinfo {author} {\bibfnamefont {R.}~\bibnamefont {{Oechslin}}},\
  }\href {\doibase 10.1103/PhysRevD.82.084043} {\bibfield  {journal} {\bibinfo
  {journal} {Phys. Rev. D}\ }\textbf {\bibinfo {volume} {82}},\ \bibinfo
  {pages} {084043} (\bibinfo {year} {2010}{\natexlab{b}})},\ \Eprint
  {http://arxiv.org/abs/1006.3315} {arXiv:1006.3315 [astro-ph.SR]} \BibitemShut
  {NoStop}%
\bibitem [{\citenamefont {{Baiotti}}\ \emph {et~al.}(2010)\citenamefont
  {{Baiotti}}, \citenamefont {{Shibata}},\ and\ \citenamefont
  {{Yamamoto}}}]{Baiotti:2010ka}%
  \BibitemOpen
  \bibfield  {author} {\bibinfo {author} {\bibfnamefont {L.}~\bibnamefont
  {{Baiotti}}}, \bibinfo {author} {\bibfnamefont {M.}~\bibnamefont
  {{Shibata}}}, \ and\ \bibinfo {author} {\bibfnamefont {T.}~\bibnamefont
  {{Yamamoto}}},\ }\href {\doibase 10.1103/PhysRevD.82.064015} {\bibfield
  {journal} {\bibinfo  {journal} {Phys. Rev. D}\ }\textbf {\bibinfo {volume}
  {82}},\ \bibinfo {eid} {064015} (\bibinfo {year} {2010})},\ \Eprint
  {http://arxiv.org/abs/1007.1754} {arXiv:1007.1754 [gr-qc]} \BibitemShut
  {NoStop}%
\bibitem [{\citenamefont {Vines}\ \emph {et~al.}(2011)\citenamefont {Vines},
  \citenamefont {Flanagan},\ and\ \citenamefont {Hinderer}}]{Vines:2010ca}%
  \BibitemOpen
  \bibfield  {author} {\bibinfo {author} {\bibfnamefont {J.}~\bibnamefont
  {Vines}}, \bibinfo {author} {\bibfnamefont {E.~E.}\ \bibnamefont {Flanagan}},
  \ and\ \bibinfo {author} {\bibfnamefont {T.}~\bibnamefont {Hinderer}},\
  }\href {\doibase 10.1103/PhysRevD.83.084051} {\bibfield  {journal} {\bibinfo
  {journal} {Phys. Rev. D}\ }\textbf {\bibinfo {volume} {83}},\ \bibinfo
  {pages} {084051} (\bibinfo {year} {2011})},\ \Eprint
  {http://arxiv.org/abs/1101.1673} {arXiv:1101.1673 [gr-qc]} \BibitemShut
  {NoStop}%
\bibitem [{\citenamefont {{Pannarale}}\ \emph {et~al.}(2011)\citenamefont
  {{Pannarale}}, \citenamefont {{Rezzolla}}, \citenamefont {{Ohme}},\ and\
  \citenamefont {{Read}}}]{Pannarale2011}%
  \BibitemOpen
  \bibfield  {author} {\bibinfo {author} {\bibfnamefont {F.}~\bibnamefont
  {{Pannarale}}}, \bibinfo {author} {\bibfnamefont {L.}~\bibnamefont
  {{Rezzolla}}}, \bibinfo {author} {\bibfnamefont {F.}~\bibnamefont {{Ohme}}},
  \ and\ \bibinfo {author} {\bibfnamefont {J.~S.}\ \bibnamefont {{Read}}},\
  }\href {\doibase 10.1103/PhysRevD.84.104017} {\bibfield  {journal} {\bibinfo
  {journal} {Phys. Rev. D}\ }\textbf {\bibinfo {volume} {84}},\ \bibinfo {eid}
  {104017} (\bibinfo {year} {2011})},\ \Eprint {http://arxiv.org/abs/1103.3526}
  {arXiv:1103.3526 [astro-ph.HE]} \BibitemShut {NoStop}%
\bibitem [{\citenamefont {Maselli}\ \emph {et~al.}(2012)\citenamefont
  {Maselli}, \citenamefont {Gualtieri}, \citenamefont {Pannarale},\ and\
  \citenamefont {Ferrari}}]{Maselli2012}%
  \BibitemOpen
  \bibfield  {author} {\bibinfo {author} {\bibfnamefont {A.}~\bibnamefont
  {Maselli}}, \bibinfo {author} {\bibfnamefont {L.}~\bibnamefont {Gualtieri}},
  \bibinfo {author} {\bibfnamefont {F.}~\bibnamefont {Pannarale}}, \ and\
  \bibinfo {author} {\bibfnamefont {V.}~\bibnamefont {Ferrari}},\ }\href
  {\doibase 10.1103/PhysRevD.86.044032} {\bibfield  {journal} {\bibinfo
  {journal} {Phys. Rev. D}\ }\textbf {\bibinfo {volume} {86}},\ \bibinfo
  {pages} {044032} (\bibinfo {year} {2012})}\BibitemShut {NoStop}%
\bibitem [{\citenamefont {Baiotti}\ \emph {et~al.}(2003)\citenamefont
  {Baiotti}, \citenamefont {Hawke}, \citenamefont {Montero},\ and\
  \citenamefont {Rezzolla}}]{Baiotti03a}%
  \BibitemOpen
  \bibfield  {author} {\bibinfo {author} {\bibfnamefont {L.}~\bibnamefont
  {Baiotti}}, \bibinfo {author} {\bibfnamefont {I.}~\bibnamefont {Hawke}},
  \bibinfo {author} {\bibfnamefont {P.}~\bibnamefont {Montero}}, \ and\
  \bibinfo {author} {\bibfnamefont {L.}~\bibnamefont {Rezzolla}},\ }in\
  \href@noop {} {\emph {\bibinfo {booktitle} {Computational Astrophysics in
  Italy: Methods and Tools}}},\ Vol.~\bibinfo {volume} {1},\ \bibinfo {editor}
  {edited by\ \bibinfo {editor} {\bibfnamefont {R.}~\bibnamefont
  {Capuzzo-Dolcetta}}}\ (\bibinfo  {publisher} {MSAIt},\ \bibinfo {address}
  {Trieste},\ \bibinfo {year} {2003})\ p.\ \bibinfo {pages} {210}\BibitemShut
  {NoStop}%
\bibitem [{\citenamefont {Suresh}\ and\ \citenamefont
  {Huynh}(1997)}]{suresh_1997_amp}%
  \BibitemOpen
  \bibfield  {author} {\bibinfo {author} {\bibfnamefont {A.}~\bibnamefont
  {Suresh}}\ and\ \bibinfo {author} {\bibfnamefont {H.~T.}\ \bibnamefont
  {Huynh}},\ }\href {\doibase DOI: 10.1006/jcph.1997.5745} {\bibfield
  {journal} {\bibinfo  {journal} {Journal of Computational Physics}\ }\textbf
  {\bibinfo {volume} {136}},\ \bibinfo {pages} {83} (\bibinfo {year}
  {1997})}\BibitemShut {NoStop}%
\bibitem [{mcl()}]{mclachlanweb}%
  \BibitemOpen
  \href {http://www.cct.lsu.edu/~eschnett/McLachlan/index.html} {}\bibinfo
  {note} {{McLachlan}, a Public {BSSN} Code}\BibitemShut {NoStop}%
\bibitem [{\citenamefont {Shu}(1997)}]{Shu97}%
  \BibitemOpen
  \bibfield  {author} {\bibinfo {author} {\bibfnamefont {C.~W.}\ \bibnamefont
  {Shu}},\ }\href
  {http://ntrs.nasa.gov/archive/nasa/casi.ntrs.nasa.gov/19980007543\_1998045663.pdf}
  {\emph {\bibinfo {title} {{E}ssentially non-oscillatory and weighted
  essentially non-oscillatory schemes for hyperbolic conservation laws}}},\
  \bibinfo {type} {Lecture notes}\ \bibinfo {number} {ICASE Report 97-65; NASA
  CR-97-206253}\ (\bibinfo  {institution} {NASA Langley Research Center},\
  \bibinfo {year} {1997})\BibitemShut {NoStop}%
\bibitem [{\citenamefont {Zlochower}\ \emph {et~al.}(2012)\citenamefont
  {Zlochower}, \citenamefont {Ponce},\ and\ \citenamefont
  {Lousto}}]{Zlochower2012}%
  \BibitemOpen
  \bibfield  {author} {\bibinfo {author} {\bibfnamefont {Y.}~\bibnamefont
  {Zlochower}}, \bibinfo {author} {\bibfnamefont {M.}~\bibnamefont {Ponce}}, \
  and\ \bibinfo {author} {\bibfnamefont {C.~O.}\ \bibnamefont {Lousto}},\
  }\href {\doibase 10.1103/PhysRevD.86.104056} {\bibfield  {journal} {\bibinfo
  {journal} {Phys. Rev. D}\ }\textbf {\bibinfo {volume} {86}},\ \bibinfo
  {pages} {104056} (\bibinfo {year} {2012})}\BibitemShut {NoStop}%
\bibitem [{\citenamefont {{Bernuzzi}}\ and\ \citenamefont
  {{Dietrich}}(2016)}]{Bernuzzi2016}%
  \BibitemOpen
  \bibfield  {author} {\bibinfo {author} {\bibfnamefont {S.}~\bibnamefont
  {{Bernuzzi}}}\ and\ \bibinfo {author} {\bibfnamefont {T.}~\bibnamefont
  {{Dietrich}}},\ }\href@noop {} {\bibfield  {journal} {\bibinfo  {journal}
  {arXiv:1604.07999}\ } (\bibinfo {year} {2016})},\ \Eprint
  {http://arxiv.org/abs/1604.07999} {arXiv:1604.07999 [gr-qc]} \BibitemShut
  {NoStop}%
\bibitem [{\citenamefont {{Cockburn}}\ and\ \citenamefont
  {{Shu}}(1989)}]{Cockburn1989a}%
  \BibitemOpen
  \bibfield  {author} {\bibinfo {author} {\bibfnamefont {B.}~\bibnamefont
  {{Cockburn}}}\ and\ \bibinfo {author} {\bibfnamefont {C.}~\bibnamefont
  {{Shu}}},\ }\href@noop {} {\bibfield  {journal} {\bibinfo  {journal} {Math.
  Comp.}\ }\textbf {\bibinfo {volume} {52}},\ \bibinfo {pages} {411} (\bibinfo
  {year} {1989})}\BibitemShut {NoStop}%
\bibitem [{\citenamefont {{Cockburn}}\ \emph {et~al.}(1990)\citenamefont
  {{Cockburn}}, \citenamefont {{How}},\ and\ \citenamefont
  {{Shu}}}]{Cockburn1990}%
  \BibitemOpen
  \bibfield  {author} {\bibinfo {author} {\bibfnamefont {B.}~\bibnamefont
  {{Cockburn}}}, \bibinfo {author} {\bibfnamefont {S.}~\bibnamefont {{How}}}, \
  and\ \bibinfo {author} {\bibfnamefont {C.}~\bibnamefont {{Shu}}},\
  }\href@noop {} {\bibfield  {journal} {\bibinfo  {journal} {Math. Comp.}\
  }\textbf {\bibinfo {volume} {54}},\ \bibinfo {pages} {545} (\bibinfo {year}
  {1990})}\BibitemShut {NoStop}%
\bibitem [{\citenamefont {{Cockburn}}(1998)}]{Cockburn1998}%
  \BibitemOpen
  \bibfield  {author} {\bibinfo {author} {\bibfnamefont {B.}~\bibnamefont
  {{Cockburn}}},\ }\href {\doibase 10.1006/jcph.1998.5892} {\bibfield
  {journal} {\bibinfo  {journal} {Journal of Computational Physics}\ }\textbf
  {\bibinfo {volume} {141}},\ \bibinfo {pages} {199} (\bibinfo {year}
  {1998})}\BibitemShut {NoStop}%
\bibitem [{\citenamefont {{Radice}}\ and\ \citenamefont
  {{Rezzolla}}(2011)}]{Radice2011}%
  \BibitemOpen
  \bibfield  {author} {\bibinfo {author} {\bibfnamefont {D.}~\bibnamefont
  {{Radice}}}\ and\ \bibinfo {author} {\bibfnamefont {L.}~\bibnamefont
  {{Rezzolla}}},\ }\href {\doibase 10.1103/PhysRevD.84.024010} {\bibfield
  {journal} {\bibinfo  {journal} {Phys. Rev. D}\ }\textbf {\bibinfo {volume}
  {84}},\ \bibinfo {eid} {024010} (\bibinfo {year} {2011})},\ \Eprint
  {http://arxiv.org/abs/1103.2426} {arXiv:1103.2426 [gr-qc]} \BibitemShut
  {NoStop}%
\bibitem [{\citenamefont {{Bugner}}\ \emph {et~al.}(2016)\citenamefont
  {{Bugner}}, \citenamefont {{Dietrich}}, \citenamefont {{Bernuzzi}},
  \citenamefont {{Weyhausen}},\ and\ \citenamefont
  {{Br{\"u}gmann}}}]{Bugner2015}%
  \BibitemOpen
  \bibfield  {author} {\bibinfo {author} {\bibfnamefont {M.}~\bibnamefont
  {{Bugner}}}, \bibinfo {author} {\bibfnamefont {T.}~\bibnamefont
  {{Dietrich}}}, \bibinfo {author} {\bibfnamefont {S.}~\bibnamefont
  {{Bernuzzi}}}, \bibinfo {author} {\bibfnamefont {A.}~\bibnamefont
  {{Weyhausen}}}, \ and\ \bibinfo {author} {\bibfnamefont {B.}~\bibnamefont
  {{Br{\"u}gmann}}},\ }\href {\doibase 10.1103/PhysRevD.94.084004} {\bibfield
  {journal} {\bibinfo  {journal} {Phys. Rev. D}\ }\textbf {\bibinfo {volume}
  {94}},\ \bibinfo {eid} {084004} (\bibinfo {year} {2016})},\ \Eprint
  {http://arxiv.org/abs/1508.07147} {arXiv:1508.07147 [gr-qc]} \BibitemShut
  {NoStop}%
\bibitem [{\citenamefont {{Zanotti}}\ and\ \citenamefont
  {{Dumbser}}(2015)}]{Zanotti2015}%
  \BibitemOpen
  \bibfield  {author} {\bibinfo {author} {\bibfnamefont {O.}~\bibnamefont
  {{Zanotti}}}\ and\ \bibinfo {author} {\bibfnamefont {M.}~\bibnamefont
  {{Dumbser}}},\ }\href {\doibase 10.1016/j.cpc.2014.11.015} {\bibfield
  {journal} {\bibinfo  {journal} {Computer Physics Communications}\ }\textbf
  {\bibinfo {volume} {188}},\ \bibinfo {pages} {110} (\bibinfo {year}
  {2015})},\ \Eprint {http://arxiv.org/abs/1312.7784} {arXiv:1312.7784
  [astro-ph.HE]} \BibitemShut {NoStop}%
\bibitem [{\citenamefont {{Zanotti}}\ \emph {et~al.}(2015)\citenamefont
  {{Zanotti}}, \citenamefont {{Fambri}},\ and\ \citenamefont
  {{Dumbser}}}]{Zanotti2015b}%
  \BibitemOpen
  \bibfield  {author} {\bibinfo {author} {\bibfnamefont {O.}~\bibnamefont
  {{Zanotti}}}, \bibinfo {author} {\bibfnamefont {F.}~\bibnamefont {{Fambri}}},
  \ and\ \bibinfo {author} {\bibfnamefont {M.}~\bibnamefont {{Dumbser}}},\
  }\href {\doibase 10.1093/mnras/stv1510} {\bibfield  {journal} {\bibinfo
  {journal} {Mon. Not. R. Astron. Soc.}\ }\textbf {\bibinfo {volume} {452}},\
  \bibinfo {pages} {3010} (\bibinfo {year} {2015})},\ \Eprint
  {http://arxiv.org/abs/1504.07458} {arXiv:1504.07458 [astro-ph.HE]}
  \BibitemShut {NoStop}%
\bibitem [{\citenamefont {Galeazzi}(2008)}]{galeazzi_master}%
  \BibitemOpen
  \bibfield  {author} {\bibinfo {author} {\bibfnamefont {F.}~\bibnamefont
  {Galeazzi}},\ }\emph {\bibinfo {title} {Modelling fluid interfaces in
  numerical relativistic hydrodynamics}},\ \href@noop {} {Master's thesis},\
  \bibinfo  {school} {Universit{\`{a}} degli studi di Padova} (\bibinfo {year}
  {2008})\BibitemShut {NoStop}%
\bibitem [{\citenamefont {Kastaun}(2006)}]{kastaun_2006_hrs}%
  \BibitemOpen
  \bibfield  {author} {\bibinfo {author} {\bibfnamefont {W.}~\bibnamefont
  {Kastaun}},\ }\href {\doibase 10.1103/PhysRevD.74.124024} {\bibfield
  {journal} {\bibinfo  {journal} {Phys. Rev. D}\ }\textbf {\bibinfo {volume}
  {74}},\ \bibinfo {pages} {124024} (\bibinfo {year} {2006})}\BibitemShut
  {NoStop}%
\bibitem [{\citenamefont {{Millmore}}\ and\ \citenamefont
  {{Hawke}}(2010)}]{Millmore2010}%
  \BibitemOpen
  \bibfield  {author} {\bibinfo {author} {\bibfnamefont {S.~T.}\ \bibnamefont
  {{Millmore}}}\ and\ \bibinfo {author} {\bibfnamefont {I.}~\bibnamefont
  {{Hawke}}},\ }\href {\doibase 10.1088/0264-9381/27/1/015007} {\bibfield
  {journal} {\bibinfo  {journal} {Class. Quantum Grav.}\ }\textbf {\bibinfo
  {volume} {27}},\ \bibinfo {pages} {015007} (\bibinfo {year} {2010})},\
  \Eprint {http://arxiv.org/abs/0909.4217} {arXiv:0909.4217 [gr-qc]}
  \BibitemShut {NoStop}%
\bibitem [{\citenamefont {{Hu}}\ \emph {et~al.}(2013)\citenamefont {{Hu}},
  \citenamefont {{Adams}},\ and\ \citenamefont {{Shu}}}]{Hu2013}%
  \BibitemOpen
  \bibfield  {author} {\bibinfo {author} {\bibfnamefont {X.~Y.}\ \bibnamefont
  {{Hu}}}, \bibinfo {author} {\bibfnamefont {N.~A.}\ \bibnamefont {{Adams}}}, \
  and\ \bibinfo {author} {\bibfnamefont {C.-W.}\ \bibnamefont {{Shu}}},\ }\href
  {\doibase 10.1016/j.jcp.2013.01.024} {\bibfield  {journal} {\bibinfo
  {journal} {Journal of Computational Physics}\ }\textbf {\bibinfo {volume}
  {242}},\ \bibinfo {pages} {169} (\bibinfo {year} {2013})},\ \Eprint
  {http://arxiv.org/abs/1203.1540} {arXiv:1203.1540 [physics.flu-dyn]}
  \BibitemShut {NoStop}%
\bibitem [{\citenamefont {Berger}\ and\ \citenamefont
  {Colella}(1989)}]{Berger89}%
  \BibitemOpen
  \bibfield  {author} {\bibinfo {author} {\bibfnamefont {M.~J.}\ \bibnamefont
  {Berger}}\ and\ \bibinfo {author} {\bibfnamefont {P.}~\bibnamefont
  {Colella}},\ }\href {\doibase 10.1016/0021-9991(89)90035-1} {\bibfield
  {journal} {\bibinfo  {journal} {J. Comput. Phys.}\ }\textbf {\bibinfo
  {volume} {82}},\ \bibinfo {pages} {64} (\bibinfo {year} {1989})}\BibitemShut
  {NoStop}%
\bibitem [{\citenamefont {{Reisswig}}\ \emph {et~al.}(2013)\citenamefont
  {{Reisswig}}, \citenamefont {{Haas}}, \citenamefont {{Ott}}, \citenamefont
  {{Abdikamalov}}, \citenamefont {{M{\"o}sta}}, \citenamefont {{Pollney}},\
  and\ \citenamefont {{Schnetter}}}]{Reisswig2012b}%
  \BibitemOpen
  \bibfield  {author} {\bibinfo {author} {\bibfnamefont {C.}~\bibnamefont
  {{Reisswig}}}, \bibinfo {author} {\bibfnamefont {R.}~\bibnamefont {{Haas}}},
  \bibinfo {author} {\bibfnamefont {C.~D.}\ \bibnamefont {{Ott}}}, \bibinfo
  {author} {\bibfnamefont {E.}~\bibnamefont {{Abdikamalov}}}, \bibinfo {author}
  {\bibfnamefont {P.}~\bibnamefont {{M{\"o}sta}}}, \bibinfo {author}
  {\bibfnamefont {D.}~\bibnamefont {{Pollney}}}, \ and\ \bibinfo {author}
  {\bibfnamefont {E.}~\bibnamefont {{Schnetter}}},\ }\href {\doibase
  10.1103/PhysRevD.87.064023} {\bibfield  {journal} {\bibinfo  {journal} {Phys.
  Rev. D}\ }\textbf {\bibinfo {volume} {87}},\ \bibinfo {eid} {064023}
  (\bibinfo {year} {2013})},\ \Eprint {http://arxiv.org/abs/1212.1191}
  {arXiv:1212.1191 [astro-ph.HE]} \BibitemShut {NoStop}%
\bibitem [{\citenamefont {Bishop}\ and\ \citenamefont
  {Rezzolla}(2016)}]{Bishop2016}%
  \BibitemOpen
  \bibfield  {author} {\bibinfo {author} {\bibfnamefont {N.~T.}\ \bibnamefont
  {Bishop}}\ and\ \bibinfo {author} {\bibfnamefont {L.}~\bibnamefont
  {Rezzolla}},\ }\href@noop {} {\bibfield  {journal} {\bibinfo  {journal}
  {Living Rev. Relativity}\ } (\bibinfo {year} {2016})}\BibitemShut {NoStop}%
\bibitem [{\citenamefont {{Bishop}}\ \emph {et~al.}(1997)\citenamefont
  {{Bishop}}, \citenamefont {{G{\'o}mez}}, \citenamefont {{Lehner}},
  \citenamefont {{Maharaj}},\ and\ \citenamefont {{Winicour}}}]{Bishop97b}%
  \BibitemOpen
  \bibfield  {author} {\bibinfo {author} {\bibfnamefont {N.~T.}\ \bibnamefont
  {{Bishop}}}, \bibinfo {author} {\bibfnamefont {R.}~\bibnamefont
  {{G{\'o}mez}}}, \bibinfo {author} {\bibfnamefont {L.}~\bibnamefont
  {{Lehner}}}, \bibinfo {author} {\bibfnamefont {M.}~\bibnamefont {{Maharaj}}},
  \ and\ \bibinfo {author} {\bibfnamefont {J.}~\bibnamefont {{Winicour}}},\
  }\href {\doibase 10.1103/PhysRevD.56.6298} {\bibfield  {journal} {\bibinfo
  {journal} {Phys. Rev. D}\ }\textbf {\bibinfo {volume} {56}},\ \bibinfo
  {pages} {6298} (\bibinfo {year} {1997})},\ \Eprint
  {http://arxiv.org/abs/gr-qc/9708065} {gr-qc/9708065} \BibitemShut {NoStop}%
\bibitem [{\citenamefont {{Eardley}}(1975)}]{Eardley1975a}%
  \BibitemOpen
  \bibfield  {author} {\bibinfo {author} {\bibfnamefont {D.~M.}\ \bibnamefont
  {{Eardley}}},\ }\href {\doibase 10.1086/181744} {\bibfield  {journal}
  {\bibinfo  {journal} {Astrophys. J.}\ }\textbf {\bibinfo {volume} {196}},\
  \bibinfo {pages} {L59} (\bibinfo {year} {1975})}\BibitemShut {NoStop}%
\bibitem [{\citenamefont {{Will}}(1993)}]{Will92}%
  \BibitemOpen
  \bibfield  {author} {\bibinfo {author} {\bibfnamefont {C.~M.}\ \bibnamefont
  {{Will}}},\ }\href {\doibase 10.1017/cbo9780511564246} {\emph {\bibinfo
  {title} {Theory and Experiment in Gravitational Physics, by Clifford M.~Will,
  pp.~396.~ISBN 0521439736.~Cambridge, UK: Cambridge University Press, March
  1993.}}}\ (\bibinfo {year} {1993})\BibitemShut {NoStop}%
\bibitem [{\citenamefont {{Alsing}}\ \emph {et~al.}(2012)\citenamefont
  {{Alsing}}, \citenamefont {{Berti}}, \citenamefont {{Will}},\ and\
  \citenamefont {{Zaglauer}}}]{Alsing2012}%
  \BibitemOpen
  \bibfield  {author} {\bibinfo {author} {\bibfnamefont {J.}~\bibnamefont
  {{Alsing}}}, \bibinfo {author} {\bibfnamefont {E.}~\bibnamefont {{Berti}}},
  \bibinfo {author} {\bibfnamefont {C.~M.}\ \bibnamefont {{Will}}}, \ and\
  \bibinfo {author} {\bibfnamefont {H.}~\bibnamefont {{Zaglauer}}},\ }\href
  {\doibase 10.1103/PhysRevD.85.064041} {\bibfield  {journal} {\bibinfo
  {journal} {Phys. Rev. D}\ }\textbf {\bibinfo {volume} {85}},\ \bibinfo {eid}
  {064041} (\bibinfo {year} {2012})},\ \Eprint {http://arxiv.org/abs/1112.4903}
  {arXiv:1112.4903 [gr-qc]} \BibitemShut {NoStop}%
\bibitem [{\citenamefont {{Mirshekari}}\ and\ \citenamefont
  {{Will}}(2013)}]{Mirshekari2013}%
  \BibitemOpen
  \bibfield  {author} {\bibinfo {author} {\bibfnamefont {S.}~\bibnamefont
  {{Mirshekari}}}\ and\ \bibinfo {author} {\bibfnamefont {C.~M.}\ \bibnamefont
  {{Will}}},\ }\href {\doibase 10.1103/PhysRevD.87.084070} {\bibfield
  {journal} {\bibinfo  {journal} {Phys. Rev. D}\ }\textbf {\bibinfo {volume}
  {87}},\ \bibinfo {eid} {084070} (\bibinfo {year} {2013})},\ \Eprint
  {http://arxiv.org/abs/1301.4680} {arXiv:1301.4680 [gr-qc]} \BibitemShut
  {NoStop}%
\bibitem [{\citenamefont {{Barausse}}\ \emph {et~al.}(2013)\citenamefont
  {{Barausse}}, \citenamefont {{Palenzuela}}, \citenamefont {{Ponce}},\ and\
  \citenamefont {{Lehner}}}]{Barausse2013}%
  \BibitemOpen
  \bibfield  {author} {\bibinfo {author} {\bibfnamefont {E.}~\bibnamefont
  {{Barausse}}}, \bibinfo {author} {\bibfnamefont {C.}~\bibnamefont
  {{Palenzuela}}}, \bibinfo {author} {\bibfnamefont {M.}~\bibnamefont
  {{Ponce}}}, \ and\ \bibinfo {author} {\bibfnamefont {L.}~\bibnamefont
  {{Lehner}}},\ }\href {\doibase 10.1103/PhysRevD.87.081506} {\bibfield
  {journal} {\bibinfo  {journal} {Phys. Rev. D}\ }\textbf {\bibinfo {volume}
  {87}},\ \bibinfo {eid} {081506} (\bibinfo {year} {2013})},\ \Eprint
  {http://arxiv.org/abs/1212.5053} {arXiv:1212.5053 [gr-qc]} \BibitemShut
  {NoStop}%
\bibitem [{\citenamefont {{Palenzuela}}\ \emph {et~al.}(2014)\citenamefont
  {{Palenzuela}}, \citenamefont {{Barausse}}, \citenamefont {{Ponce}},\ and\
  \citenamefont {{Lehner}}}]{Palenzuela2014}%
  \BibitemOpen
  \bibfield  {author} {\bibinfo {author} {\bibfnamefont {C.}~\bibnamefont
  {{Palenzuela}}}, \bibinfo {author} {\bibfnamefont {E.}~\bibnamefont
  {{Barausse}}}, \bibinfo {author} {\bibfnamefont {M.}~\bibnamefont {{Ponce}}},
  \ and\ \bibinfo {author} {\bibfnamefont {L.}~\bibnamefont {{Lehner}}},\
  }\href {\doibase 10.1103/PhysRevD.89.044024} {\bibfield  {journal} {\bibinfo
  {journal} {Phys. Rev. D}\ }\textbf {\bibinfo {volume} {89}},\ \bibinfo {eid}
  {044024} (\bibinfo {year} {2014})},\ \Eprint {http://arxiv.org/abs/1310.4481}
  {arXiv:1310.4481 [gr-qc]} \BibitemShut {NoStop}%
\bibitem [{\citenamefont {{Sampson}}\ \emph {et~al.}(2014)\citenamefont
  {{Sampson}}, \citenamefont {{Yunes}}, \citenamefont {{Cornish}},
  \citenamefont {{Ponce}}, \citenamefont {{Barausse}}, \citenamefont {{Klein}},
  \citenamefont {{Palenzuela}},\ and\ \citenamefont {{Lehner}}}]{Sampson2014}%
  \BibitemOpen
  \bibfield  {author} {\bibinfo {author} {\bibfnamefont {L.}~\bibnamefont
  {{Sampson}}}, \bibinfo {author} {\bibfnamefont {N.}~\bibnamefont {{Yunes}}},
  \bibinfo {author} {\bibfnamefont {N.}~\bibnamefont {{Cornish}}}, \bibinfo
  {author} {\bibfnamefont {M.}~\bibnamefont {{Ponce}}}, \bibinfo {author}
  {\bibfnamefont {E.}~\bibnamefont {{Barausse}}}, \bibinfo {author}
  {\bibfnamefont {A.}~\bibnamefont {{Klein}}}, \bibinfo {author} {\bibfnamefont
  {C.}~\bibnamefont {{Palenzuela}}}, \ and\ \bibinfo {author} {\bibfnamefont
  {L.}~\bibnamefont {{Lehner}}},\ }\href {\doibase 10.1103/PhysRevD.90.124091}
  {\bibfield  {journal} {\bibinfo  {journal} {Phys. Rev. D}\ }\textbf {\bibinfo
  {volume} {90}},\ \bibinfo {eid} {124091} (\bibinfo {year} {2014})},\ \Eprint
  {http://arxiv.org/abs/1407.7038} {arXiv:1407.7038 [gr-qc]} \BibitemShut
  {NoStop}%
\bibitem [{\citenamefont {{Wagoner}}(1970)}]{Wagoner1970}%
  \BibitemOpen
  \bibfield  {author} {\bibinfo {author} {\bibfnamefont {R.~V.}\ \bibnamefont
  {{Wagoner}}},\ }\href {\doibase 10.1103/PhysRevD.1.3209} {\bibfield
  {journal} {\bibinfo  {journal} {Phys. Rev. D}\ }\textbf {\bibinfo {volume}
  {1}},\ \bibinfo {pages} {3209} (\bibinfo {year} {1970})}\BibitemShut
  {NoStop}%
\bibitem [{\citenamefont {{Nordtvedt}}(1970)}]{Nordtvedt1970}%
  \BibitemOpen
  \bibfield  {author} {\bibinfo {author} {\bibfnamefont {K.}~\bibnamefont
  {{Nordtvedt}}, \bibfnamefont {Jr.}},\ }\href {\doibase 10.1086/150607}
  {\bibfield  {journal} {\bibinfo  {journal} {Astrophys. J.}\ }\textbf
  {\bibinfo {volume} {161}},\ \bibinfo {pages} {1059} (\bibinfo {year}
  {1970})}\BibitemShut {NoStop}%
\bibitem [{\citenamefont {{Damour}}\ and\ \citenamefont
  {{Esposito-Farese}}(1992)}]{Damour1992}%
  \BibitemOpen
  \bibfield  {author} {\bibinfo {author} {\bibfnamefont {T.}~\bibnamefont
  {{Damour}}}\ and\ \bibinfo {author} {\bibfnamefont {G.}~\bibnamefont
  {{Esposito-Farese}}},\ }\href {\doibase 10.1088/0264-9381/9/9/015} {\bibfield
   {journal} {\bibinfo  {journal} {Class. Quantum Grav.}\ }\textbf {\bibinfo
  {volume} {9}},\ \bibinfo {pages} {2093} (\bibinfo {year} {1992})}\BibitemShut
  {NoStop}%
\bibitem [{\citenamefont {{Kramer}}\ and\ \citenamefont
  {{Wex}}(2009)}]{Kramer2009}%
  \BibitemOpen
  \bibfield  {author} {\bibinfo {author} {\bibfnamefont {M.}~\bibnamefont
  {{Kramer}}}\ and\ \bibinfo {author} {\bibfnamefont {N.}~\bibnamefont
  {{Wex}}},\ }\href {\doibase 10.1088/0264-9381/26/7/073001} {\bibfield
  {journal} {\bibinfo  {journal} {Class. Quantum Grav.}\ }\textbf {\bibinfo
  {volume} {26}},\ \bibinfo {eid} {073001} (\bibinfo {year}
  {2009})}\BibitemShut {NoStop}%
\bibitem [{\citenamefont {{Freire}}\ \emph {et~al.}(2012)\citenamefont
  {{Freire}}, \citenamefont {{Wex}}, \citenamefont {{Esposito-Far{\`e}se}},
  \citenamefont {{Verbiest}}, \citenamefont {{Bailes}}, \citenamefont
  {{Jacoby}}, \citenamefont {{Kramer}}, \citenamefont {{Stairs}}, \citenamefont
  {{Antoniadis}},\ and\ \citenamefont {{Janssen}}}]{Freire2012}%
  \BibitemOpen
  \bibfield  {author} {\bibinfo {author} {\bibfnamefont {P.~C.~C.}\
  \bibnamefont {{Freire}}}, \bibinfo {author} {\bibfnamefont {N.}~\bibnamefont
  {{Wex}}}, \bibinfo {author} {\bibfnamefont {G.}~\bibnamefont
  {{Esposito-Far{\`e}se}}}, \bibinfo {author} {\bibfnamefont {J.~P.~W.}\
  \bibnamefont {{Verbiest}}}, \bibinfo {author} {\bibfnamefont
  {M.}~\bibnamefont {{Bailes}}}, \bibinfo {author} {\bibfnamefont {B.~A.}\
  \bibnamefont {{Jacoby}}}, \bibinfo {author} {\bibfnamefont {M.}~\bibnamefont
  {{Kramer}}}, \bibinfo {author} {\bibfnamefont {I.~H.}\ \bibnamefont
  {{Stairs}}}, \bibinfo {author} {\bibfnamefont {J.}~\bibnamefont
  {{Antoniadis}}}, \ and\ \bibinfo {author} {\bibfnamefont {G.~H.}\
  \bibnamefont {{Janssen}}},\ }\href {\doibase
  10.1111/j.1365-2966.2012.21253.x} {\bibfield  {journal} {\bibinfo  {journal}
  {Mon. Not. R. Astron. Soc.}\ }\textbf {\bibinfo {volume} {423}},\ \bibinfo
  {pages} {3328} (\bibinfo {year} {2012})},\ \Eprint
  {http://arxiv.org/abs/1205.1450} {arXiv:1205.1450 [astro-ph.GA]} \BibitemShut
  {NoStop}%
\bibitem [{\citenamefont {{Babusci}}\ \emph {et~al.}(2001)\citenamefont
  {{Babusci}}, \citenamefont {{Baiotti}}, \citenamefont {{Fucito}},\ and\
  \citenamefont {{Nagar}}}]{Babusci01}%
  \BibitemOpen
  \bibfield  {author} {\bibinfo {author} {\bibfnamefont {D.}~\bibnamefont
  {{Babusci}}}, \bibinfo {author} {\bibfnamefont {L.}~\bibnamefont
  {{Baiotti}}}, \bibinfo {author} {\bibfnamefont {F.}~\bibnamefont {{Fucito}}},
  \ and\ \bibinfo {author} {\bibfnamefont {A.}~\bibnamefont {{Nagar}}},\ }\href
  {\doibase 10.1103/PhysRevD.64.062001} {\bibfield  {journal} {\bibinfo
  {journal} {Phys. Rev. D}\ }\textbf {\bibinfo {volume} {64}},\ \bibinfo {eid}
  {062001} (\bibinfo {year} {2001})},\ \Eprint
  {http://arxiv.org/abs/gr-qc/0105028} {gr-qc/0105028} \BibitemShut {NoStop}%
\bibitem [{\citenamefont {{Damour}}\ and\ \citenamefont
  {{Esposito-Farese}}(1993)}]{Damour1993}%
  \BibitemOpen
  \bibfield  {author} {\bibinfo {author} {\bibfnamefont {T.}~\bibnamefont
  {{Damour}}}\ and\ \bibinfo {author} {\bibfnamefont {G.}~\bibnamefont
  {{Esposito-Farese}}},\ }\href {\doibase 10.1103/PhysRevLett.70.2220}
  {\bibfield  {journal} {\bibinfo  {journal} {Phys. Rev. Lett.}\ }\textbf
  {\bibinfo {volume} {70}},\ \bibinfo {pages} {2220} (\bibinfo {year}
  {1993})}\BibitemShut {NoStop}%
\bibitem [{\citenamefont {{Ramazano{\v g}lu}}\ and\ \citenamefont
  {{Pretorius}}(2016)}]{Ramazanoglu2016}%
  \BibitemOpen
  \bibfield  {author} {\bibinfo {author} {\bibfnamefont {F.~M.}\ \bibnamefont
  {{Ramazano{\v g}lu}}}\ and\ \bibinfo {author} {\bibfnamefont
  {F.}~\bibnamefont {{Pretorius}}},\ }\href {\doibase
  10.1103/PhysRevD.93.064005} {\bibfield  {journal} {\bibinfo  {journal} {Phys.
  Rev. D}\ }\textbf {\bibinfo {volume} {93}},\ \bibinfo {eid} {064005}
  (\bibinfo {year} {2016})},\ \Eprint {http://arxiv.org/abs/1601.07475}
  {arXiv:1601.07475 [gr-qc]} \BibitemShut {NoStop}%
\bibitem [{\citenamefont {{Sennett}}\ and\ \citenamefont
  {{Buonanno}}(2016)}]{Sennet2016}%
  \BibitemOpen
  \bibfield  {author} {\bibinfo {author} {\bibfnamefont {N.}~\bibnamefont
  {{Sennett}}}\ and\ \bibinfo {author} {\bibfnamefont {A.}~\bibnamefont
  {{Buonanno}}},\ }\href {\doibase 10.1103/PhysRevD.93.124004} {\bibfield
  {journal} {\bibinfo  {journal} {Phys. Rev. D}\ }\textbf {\bibinfo {volume}
  {93}},\ \bibinfo {eid} {124004} (\bibinfo {year} {2016})},\ \Eprint
  {http://arxiv.org/abs/1603.03300} {arXiv:1603.03300 [gr-qc]} \BibitemShut
  {NoStop}%
\bibitem [{\citenamefont {Thorne}(1972)}]{Thorne72a}%
  \BibitemOpen
  \bibfield  {author} {\bibinfo {author} {\bibfnamefont {K.}~\bibnamefont
  {Thorne}},\ }in\ \href {\doibase 10.1016/0550-3213(73)90607-x} {\emph
  {\bibinfo {booktitle} {Magic Without Magic: John Archibald Wheeler}}},\
  \bibinfo {editor} {edited by\ \bibinfo {editor} {\bibfnamefont
  {J.}~\bibnamefont {Klauder}}}\ (\bibinfo  {publisher} {Freeman},\ \bibinfo
  {address} {San Francisco},\ \bibinfo {year} {1972})\ p.\ \bibinfo {pages}
  {231}\BibitemShut {NoStop}%
\bibitem [{\citenamefont {{Eardley}}\ and\ \citenamefont
  {{Giddings}}(2002)}]{Eardley_2002}%
  \BibitemOpen
  \bibfield  {author} {\bibinfo {author} {\bibfnamefont {D.~M.}\ \bibnamefont
  {{Eardley}}}\ and\ \bibinfo {author} {\bibfnamefont {S.~B.}\ \bibnamefont
  {{Giddings}}},\ }\href {\doibase 10.1103/PhysRevD.66.044011} {\bibfield
  {journal} {\bibinfo  {journal} {Phys. Rev. D}\ }\textbf {\bibinfo {volume}
  {66}},\ \bibinfo {eid} {044011} (\bibinfo {year} {2002})},\ \Eprint
  {http://arxiv.org/abs/arXiv:gr-qc/0201034} {arXiv:gr-qc/0201034} \BibitemShut
  {NoStop}%
\bibitem [{\citenamefont {{Choptuik}}\ and\ \citenamefont
  {{Pretorius}}(2010)}]{Choptuik:2010a}%
  \BibitemOpen
  \bibfield  {author} {\bibinfo {author} {\bibfnamefont {M.~W.}\ \bibnamefont
  {{Choptuik}}}\ and\ \bibinfo {author} {\bibfnamefont {F.}~\bibnamefont
  {{Pretorius}}},\ }\href {\doibase 10.1103/PhysRevLett.104.111101} {\bibfield
  {journal} {\bibinfo  {journal} {Phys. Rev. Lett.}\ }\textbf {\bibinfo
  {volume} {104}},\ \bibinfo {eid} {111101} (\bibinfo {year} {2010})},\ \Eprint
  {http://arxiv.org/abs/0908.1780} {arXiv:0908.1780 [gr-qc]} \BibitemShut
  {NoStop}%
\bibitem [{\citenamefont {{Antoniadis}}\ \emph {et~al.}(1998)\citenamefont
  {{Antoniadis}}, \citenamefont {{Arkani-Hamed}}, \citenamefont
  {{Dimopoulos}},\ and\ \citenamefont {{Dvali}}}]{Antoniadis1998}%
  \BibitemOpen
  \bibfield  {author} {\bibinfo {author} {\bibfnamefont {I.}~\bibnamefont
  {{Antoniadis}}}, \bibinfo {author} {\bibfnamefont {N.}~\bibnamefont
  {{Arkani-Hamed}}}, \bibinfo {author} {\bibfnamefont {S.}~\bibnamefont
  {{Dimopoulos}}}, \ and\ \bibinfo {author} {\bibfnamefont {G.}~\bibnamefont
  {{Dvali}}},\ }\href {\doibase 10.1016/S0370-2693(98)00860-0} {\bibfield
  {journal} {\bibinfo  {journal} {Physics Letters B}\ }\textbf {\bibinfo
  {volume} {436}},\ \bibinfo {pages} {257} (\bibinfo {year} {1998})},\ \Eprint
  {http://arxiv.org/abs/hep-ph/9804398} {hep-ph/9804398} \BibitemShut {NoStop}%
\bibitem [{\citenamefont {{Argyres}}\ \emph {et~al.}(1998)\citenamefont
  {{Argyres}}, \citenamefont {{Dimopoulos}},\ and\ \citenamefont
  {{March-Russell}}}]{Argyres_1998}%
  \BibitemOpen
  \bibfield  {author} {\bibinfo {author} {\bibfnamefont {P.~C.}\ \bibnamefont
  {{Argyres}}}, \bibinfo {author} {\bibfnamefont {S.}~\bibnamefont
  {{Dimopoulos}}}, \ and\ \bibinfo {author} {\bibfnamefont {J.}~\bibnamefont
  {{March-Russell}}},\ }\href {\doibase 10.1016/S0370-2693(98)01184-8}
  {\bibfield  {journal} {\bibinfo  {journal} {Physics Letters B}\ }\textbf
  {\bibinfo {volume} {441}},\ \bibinfo {pages} {96} (\bibinfo {year} {1998})},\
  \Eprint {http://arxiv.org/abs/arXiv:hep-th/9808138} {arXiv:hep-th/9808138}
  \BibitemShut {NoStop}%
\bibitem [{\citenamefont {{Yoo}}\ \emph {et~al.}(2010)\citenamefont {{Yoo}},
  \citenamefont {{Ishihara}}, \citenamefont {{Kimura}},\ and\ \citenamefont
  {{Tanzawa}}}]{Yoo_2010}%
  \BibitemOpen
  \bibfield  {author} {\bibinfo {author} {\bibfnamefont {C.-M.}\ \bibnamefont
  {{Yoo}}}, \bibinfo {author} {\bibfnamefont {H.}~\bibnamefont {{Ishihara}}},
  \bibinfo {author} {\bibfnamefont {M.}~\bibnamefont {{Kimura}}}, \ and\
  \bibinfo {author} {\bibfnamefont {S.}~\bibnamefont {{Tanzawa}}},\ }\href
  {\doibase 10.1103/PhysRevD.81.024020} {\bibfield  {journal} {\bibinfo
  {journal} {Phys. Rev. D}\ }\textbf {\bibinfo {volume} {81}},\ \bibinfo {eid}
  {024020} (\bibinfo {year} {2010})},\ \Eprint {http://arxiv.org/abs/0906.0689}
  {arXiv:0906.0689 [gr-qc]} \BibitemShut {NoStop}%
\bibitem [{\citenamefont {{Yoshino}}\ and\ \citenamefont
  {{Nambu}}(2003)}]{Yoshino_2003}%
  \BibitemOpen
  \bibfield  {author} {\bibinfo {author} {\bibfnamefont {H.}~\bibnamefont
  {{Yoshino}}}\ and\ \bibinfo {author} {\bibfnamefont {Y.}~\bibnamefont
  {{Nambu}}},\ }\href {\doibase 10.1103/PhysRevD.67.024009} {\bibfield
  {journal} {\bibinfo  {journal} {Phys. Rev. D}\ }\textbf {\bibinfo {volume}
  {67}},\ \bibinfo {eid} {024009} (\bibinfo {year} {2003})},\ \Eprint
  {http://arxiv.org/abs/arXiv:gr-qc/0209003} {arXiv:gr-qc/0209003} \BibitemShut
  {NoStop}%
\bibitem [{\citenamefont {{Dimopoulos}}\ and\ \citenamefont
  {{Landsberg}}(2001)}]{Dimopoulos_2001}%
  \BibitemOpen
  \bibfield  {author} {\bibinfo {author} {\bibfnamefont {S.}~\bibnamefont
  {{Dimopoulos}}}\ and\ \bibinfo {author} {\bibfnamefont {G.}~\bibnamefont
  {{Landsberg}}},\ }\href {\doibase 10.1103/PhysRevLett.87.161602} {\bibfield
  {journal} {\bibinfo  {journal} {Phys. Rev. Lett.}\ }\textbf {\bibinfo
  {volume} {87}},\ \bibinfo {eid} {161602} (\bibinfo {year} {2001})},\ \Eprint
  {http://arxiv.org/abs/arXiv:hep-ph/0106295} {arXiv:hep-ph/0106295}
  \BibitemShut {NoStop}%
\bibitem [{\citenamefont {{Feng}}\ and\ \citenamefont
  {{Shapere}}(2002)}]{Feng2002}%
  \BibitemOpen
  \bibfield  {author} {\bibinfo {author} {\bibfnamefont {J.~L.}\ \bibnamefont
  {{Feng}}}\ and\ \bibinfo {author} {\bibfnamefont {A.~D.}\ \bibnamefont
  {{Shapere}}},\ }\href {\doibase 10.1103/PhysRevLett.88.021303} {\bibfield
  {journal} {\bibinfo  {journal} {Phys. Rev. Lett.}\ }\textbf {\bibinfo
  {volume} {88}},\ \bibinfo {eid} {021303} (\bibinfo {year} {2002})},\ \Eprint
  {http://arxiv.org/abs/hep-ph/0109106} {hep-ph/0109106} \BibitemShut {NoStop}%
\bibitem [{\citenamefont {{Chatrchyan}}\ \emph {et~al.}(2013)\citenamefont
  {{Chatrchyan}}, \citenamefont {{Khachatryan}}, \citenamefont {{Sirunyan}},
  \citenamefont {{Tumasyan}}, \citenamefont {{Adam}}, \citenamefont
  {{Bergauer}}, \citenamefont {{Dragicevic}}, \citenamefont {{Er{\"o}}},
  \citenamefont {{Fabjan}}, \citenamefont {{Friedl}},\ and\ \citenamefont
  {et~al.}}]{Chatrchyan2013}%
  \BibitemOpen
  \bibfield  {author} {\bibinfo {author} {\bibfnamefont {S.}~\bibnamefont
  {{Chatrchyan}}}, \bibinfo {author} {\bibfnamefont {V.}~\bibnamefont
  {{Khachatryan}}}, \bibinfo {author} {\bibfnamefont {A.~M.}\ \bibnamefont
  {{Sirunyan}}}, \bibinfo {author} {\bibfnamefont {A.}~\bibnamefont
  {{Tumasyan}}}, \bibinfo {author} {\bibfnamefont {W.}~\bibnamefont {{Adam}}},
  \bibinfo {author} {\bibfnamefont {T.}~\bibnamefont {{Bergauer}}}, \bibinfo
  {author} {\bibfnamefont {M.}~\bibnamefont {{Dragicevic}}}, \bibinfo {author}
  {\bibfnamefont {J.}~\bibnamefont {{Er{\"o}}}}, \bibinfo {author}
  {\bibfnamefont {C.}~\bibnamefont {{Fabjan}}}, \bibinfo {author}
  {\bibfnamefont {M.}~\bibnamefont {{Friedl}}}, \ and\ \bibinfo {author}
  {\bibnamefont {et~al.}},\ }\href {\doibase 10.1007/JHEP07(2013)178}
  {\bibfield  {journal} {\bibinfo  {journal} {Journal of High Energy Physics}\
  }\textbf {\bibinfo {volume} {7}},\ \bibinfo {eid} {178} (\bibinfo {year}
  {2013})},\ \Eprint {http://arxiv.org/abs/1303.5338} {arXiv:1303.5338
  [hep-ex]} \BibitemShut {NoStop}%
\bibitem [{\citenamefont {{Rezzolla}}\ and\ \citenamefont
  {{Takami}}(2013)}]{Rezzolla2013}%
  \BibitemOpen
  \bibfield  {author} {\bibinfo {author} {\bibfnamefont {L.}~\bibnamefont
  {{Rezzolla}}}\ and\ \bibinfo {author} {\bibfnamefont {K.}~\bibnamefont
  {{Takami}}},\ }\href {\doibase 10.1088/0264-9381/30/1/012001} {\bibfield
  {journal} {\bibinfo  {journal} {Class. Quantum Grav.}\ }\textbf {\bibinfo
  {volume} {30}},\ \bibinfo {pages} {012001} (\bibinfo {year} {2013})},\
  \Eprint {http://arxiv.org/abs/1209.6138} {arXiv:1209.6138 [gr-qc]}
  \BibitemShut {NoStop}%
\bibitem [{\citenamefont {{East}}\ and\ \citenamefont
  {{Pretorius}}(2013)}]{East2012}%
  \BibitemOpen
  \bibfield  {author} {\bibinfo {author} {\bibfnamefont {W.~E.}\ \bibnamefont
  {{East}}}\ and\ \bibinfo {author} {\bibfnamefont {F.}~\bibnamefont
  {{Pretorius}}},\ }\href {\doibase 10.1103/PhysRevLett.110.101101} {\bibfield
  {journal} {\bibinfo  {journal} {Phys. Rev. Lett.}\ }\textbf {\bibinfo
  {volume} {110}},\ \bibinfo {eid} {101101} (\bibinfo {year} {2013})},\ \Eprint
  {http://arxiv.org/abs/1210.0443} {arXiv:1210.0443 [gr-qc]} \BibitemShut
  {NoStop}%
\bibitem [{\citenamefont {{Kellerman}}\ \emph {et~al.}(2008)\citenamefont
  {{Kellerman}}, \citenamefont {{Baiotti}}, \citenamefont {{Giacomazzo}},\ and\
  \citenamefont {{Rezzolla}}}]{Kellermann:08a}%
  \BibitemOpen
  \bibfield  {author} {\bibinfo {author} {\bibfnamefont {T.}~\bibnamefont
  {{Kellerman}}}, \bibinfo {author} {\bibfnamefont {L.}~\bibnamefont
  {{Baiotti}}}, \bibinfo {author} {\bibfnamefont {B.}~\bibnamefont
  {{Giacomazzo}}}, \ and\ \bibinfo {author} {\bibfnamefont {L.}~\bibnamefont
  {{Rezzolla}}},\ }\href {\doibase 10.1088/0264-9381/25/22/225007} {\bibfield
  {journal} {\bibinfo  {journal} {Class. Quantum Grav.}\ }\textbf {\bibinfo
  {volume} {25}},\ \bibinfo {pages} {225007} (\bibinfo {year} {2008})},\
  \Eprint {http://arxiv.org/abs/0811.0938} {arXiv:0811.0938} \BibitemShut
  {NoStop}%
\bibitem [{\citenamefont {{Kellerman}}\ \emph {et~al.}(2010)\citenamefont
  {{Kellerman}}, \citenamefont {{Rezzolla}},\ and\ \citenamefont
  {{Radice}}}]{Kellermann:10}%
  \BibitemOpen
  \bibfield  {author} {\bibinfo {author} {\bibfnamefont {T.}~\bibnamefont
  {{Kellerman}}}, \bibinfo {author} {\bibfnamefont {L.}~\bibnamefont
  {{Rezzolla}}}, \ and\ \bibinfo {author} {\bibfnamefont {D.}~\bibnamefont
  {{Radice}}},\ }\href {\doibase 10.1088/0264-9381/27/23/235016} {\bibfield
  {journal} {\bibinfo  {journal} {Class. Quantum Grav.}\ }\textbf {\bibinfo
  {volume} {27}},\ \bibinfo {pages} {235016} (\bibinfo {year} {2010})},\
  \Eprint {http://arxiv.org/abs/1007.2797} {arXiv:1007.2797 [gr-qc]}
  \BibitemShut {NoStop}%
\bibitem [{\citenamefont {{Radice}}\ \emph {et~al.}(2010)\citenamefont
  {{Radice}}, \citenamefont {{Rezzolla}},\ and\ \citenamefont
  {{Kellerman}}}]{Radice:10}%
  \BibitemOpen
  \bibfield  {author} {\bibinfo {author} {\bibfnamefont {D.}~\bibnamefont
  {{Radice}}}, \bibinfo {author} {\bibfnamefont {L.}~\bibnamefont
  {{Rezzolla}}}, \ and\ \bibinfo {author} {\bibfnamefont {T.}~\bibnamefont
  {{Kellerman}}},\ }\href {\doibase 10.1088/0264-9381/27/23/235015} {\bibfield
  {journal} {\bibinfo  {journal} {Class. Quantum Grav.}\ }\textbf {\bibinfo
  {volume} {27}},\ \bibinfo {pages} {235015} (\bibinfo {year} {2010})},\
  \Eprint {http://arxiv.org/abs/1007.2809} {arXiv:1007.2809 [gr-qc]}
  \BibitemShut {NoStop}%
\bibitem [{\citenamefont {Noble}(2003)}]{noble_2003_nsr}%
  \BibitemOpen
  \bibfield  {author} {\bibinfo {author} {\bibfnamefont {S.~C.}\ \bibnamefont
  {Noble}},\ }\emph {\bibinfo {title} {A Numerical Study of Relativistic Fluid
  Collapse}},\ \href
  {http://www.citebase.org/abstract?id=oai:arXiv.org:gr-qc/0310116} {Ph.D.
  thesis},\ \bibinfo  {school} {University of Texas at Austin} (\bibinfo {year}
  {2003}),\ \Eprint {http://arxiv.org/abs/gr-qc/0310116v1} {gr-qc/0310116v1}
  \BibitemShut {NoStop}%
\bibitem [{\citenamefont {{Noble}}\ and\ \citenamefont
  {{Choptuik}}(2008)}]{Noble08a}%
  \BibitemOpen
  \bibfield  {author} {\bibinfo {author} {\bibfnamefont {S.~C.}\ \bibnamefont
  {{Noble}}}\ and\ \bibinfo {author} {\bibfnamefont {M.~W.}\ \bibnamefont
  {{Choptuik}}},\ }\href {\doibase 10.1103/PhysRevD.78.064059} {\bibfield
  {journal} {\bibinfo  {journal} {Phys. Rev. D}\ }\textbf {\bibinfo {volume}
  {78}},\ \bibinfo {pages} {064059} (\bibinfo {year} {2008})},\ \Eprint
  {http://arxiv.org/abs/0709.3527} {arXiv:0709.3527} \BibitemShut {NoStop}%
\bibitem [{\citenamefont {{Noble}}\ and\ \citenamefont
  {{Choptuik}}(2016)}]{Noble2016}%
  \BibitemOpen
  \bibfield  {author} {\bibinfo {author} {\bibfnamefont {S.~C.}\ \bibnamefont
  {{Noble}}}\ and\ \bibinfo {author} {\bibfnamefont {M.~W.}\ \bibnamefont
  {{Choptuik}}},\ }\href {\doibase 10.1103/PhysRevD.93.024015} {\bibfield
  {journal} {\bibinfo  {journal} {Phys. Rev. D}\ }\textbf {\bibinfo {volume}
  {93}},\ \bibinfo {eid} {024015} (\bibinfo {year} {2016})},\ \Eprint
  {http://arxiv.org/abs/1512.02999} {arXiv:1512.02999 [gr-qc]} \BibitemShut
  {NoStop}%
\bibitem [{\citenamefont {{Gundlach}}(2003)}]{Gundlach-2003-critical-review}%
  \BibitemOpen
  \bibfield  {author} {\bibinfo {author} {\bibfnamefont {C.}~\bibnamefont
  {{Gundlach}}},\ }\href {\doibase 10.1016/S0370-1573(02)00560-4} {\bibfield
  {journal} {\bibinfo  {journal} {Physics Reports}\ }\textbf {\bibinfo {volume}
  {376}},\ \bibinfo {pages} {339} (\bibinfo {year} {2003})},\ \Eprint
  {http://arxiv.org/abs/gr-qc/0210101} {gr-qc/0210101} \BibitemShut {NoStop}%
\end{thebibliography}%




\end{document}